\documentclass[fleqn]{2023SCGE}
\usepackage{graphicx}
\usepackage{mwe,tikz,ulem}
\usepackage[percent]{overpic}
\usepackage{graphicx}
\usepackage{xspace}
\usepackage[numbers]{natbib}
\usepackage[utf8x]{inputenc}

\usepackage[gen]{eurosym}
\usepackage{microtype}
\usepackage{color}
\usepackage{txfonts}
\usepackage{makeidx}
\usepackage[columns=1]{idxlayout}

\newcommand*\degr{\ensuremath{^\circ}}

\newcommand{\apj}{ApJ\xspace}
\newcommand{\apjl}{ApJ\xspace}
\newcommand{\apjs}{ApJ Supp.\xspace}
\newcommand{\mnras}{MNRAS\xspace}
\newcommand{\aap}{A\&A\xspace}

\newcommand{\nat}{Nature\xspace}
\newcommand{\araa}{ARAA\xspace}

\newcommand{\ssr}{Space Sci.\ Rev.\xspace}
\newcommand{\nar}{New Astron.\xspace}
\newcommand{\pasj}{PASJ\xspace}
\newcommand{\prd}{Phys.\ Rev.\ D\xspace}
\newcommand{\procspie}{Proc.\ SPIE\xspace}

\titleformat{\paragraph}
{\normalfont\normalsize\itshape}{\theparagraph}{1em}{}
\titlespacing*{\paragraph}
{0pt}{3.25ex plus 1ex minus .2ex}{1.5ex plus .2ex}

\begin{document}
\ensubject{subject}

%%%%%%%%%%%%%%%%%%%%%%%%%%%%%%%%%%%%%%%%%%%%%%%%%%%%%%%
%%% Authors do not modify the information below
%%% ????????????????
%%% ??????????, ????????????{}, ???????????????????
%Letter to the Editor??Article%??????
\ArticleType{Article}%??Article
\SpecialTopic{SPECIAL TOPIC: }%???????
\Year{ }
\Month{ }
\Vol{ }
\No{ }
\DOI{ }
\ArtNo{ }
%\ReceiveDate{January 11, 2016}
%\AcceptDate{April 6, 2016}
%\OnlineDate{January 1, 2016}
%%%%%%%%%%%%%%%%%%%%%%%%%%%%%%%%%%%%%%%%%%%%%%%%%%%%%%%

%%% title: ????
%%%   \title{}{title for citation}
\title{Physics of Strong Magnetism with eXTP}{eXTP WG3}
\author[1]{\\ Mingyu Ge$^{*}$}{}
\author[2,3]{Long Ji$^{*}$}{}
\author[4]{Roberto Taverna$^{*}$}{}
\author[5]{Sergey Tsygankov$^{*}$}{}
\author[1]{Yanjun Xu}{}
\author[6,1,32]{Andrea Santangelo}{}
\author[7]{\\ Silvia Zane}{}
\author[1]{Shuang-Nan Zhang}{}
\author[1]{Hua Feng}{}
\author[8]{Wei Chen}{}
\author[9]{Quan Cheng}{}
\author[10]{Xian Hou}{}
\author[11]{\\ Matteo Imbrogno}{}
\author[11]{Gian Luca Israel}{}
\author[7]{Ruth Kelly}{}
\author[6]{Ling-Da Kong}{}
\author[8]{\\ Kuan Liu}{}
\author[12]{Alexander Mushtukov}{}
\author[5]{Juri Poutanen}{}
\author[6]{ Valery Suleimanov}{}
\author[1]{Lian Tao}{}
\author[13,14,15]{Hao Tong}{}
\author[4,7]{\\ Roberto Turolla}{}
\author[16]{Weihua Wang}{}
\author[1]{Wentao Ye}{}
\author[1,17]{ Qing-Chang Zhao}{}
\author[18]{Nabil Brice}{}
\author[19]{Jinjun Geng}{}
\author[20,21]{\\Lin Lin}{}
\author[22]{Wei-Yang Wang}{}
\author[8]{Fei Xie}{}
\author[1]{Shao-Lin Xiong}{}
\author[1]{Shu Zhang}{}
\author[20,21]{Yucong Fu}{}
\author[23]{Dong Lai}{}
\author[24,25]{\\ Jian Li}{}
\author[1]{Pan-Ping Li}{}
\author[1]{Xiaobo Li}{}
\author[26]{Xinyu Li}{}
\author[6]{Honghui Liu}{}
\author[27]{Jiren Liu}{}
\author[1]{Jingqiang Peng}{}
\author[1]{\\ Qingcang Shui}{}
\author[6]{Youli Tuo}{}
\author[13,14,15]{Hongguang Wang}{}
\author[28]{Wei Wang}{}
\author[29]{\\ Shanshan Weng}{}
\author[1]{Yuan You}{}
\author[9,30]{Xiaoping Zheng}{}
\author[31,32]{Xia Zhou}{}

\address[1]{State Key Laboratory of Particle Astrophysics, Institute of High Energy Physics, Chinese Academy of Sciences, Beijing 100049, China. gemy@ihep.ac.cn}
\address[2]{School of Physics and Astronomy, Sun Yat-sen University, Zhuhai 519082, China. jilong@mail.sysu.edu.cn}
\address[3]{CSST Science Center for the Guangdong-Hong Kong-Macau Greater Bay Area, DaXue Road 2, 519082, Zhuhai, China}
\address[4]{Department of Physics and Astronomy, University of Padova; Via Marzolo 8, Padova, I35131, Italy. roberto.taverna@unipd.it}
\address[5]{Department of Physics and Astronomy, University of Turku, Turku FI-20014, Finland. sergey.tsygankov@utu.fi}
\address[6]{Institut f\"{u}r Astronomie und Astrophysik, Kepler Center for Astro and Particle Physics, Eberhard Karls Universit\"{a}t Tübingen, Sand 1, 72076 T\"{u}bingen, Germany}
\address[7]{Mullard Space Science Laboratory, University College London, Holmbury St Mary, Dorking, Surrey RH5 6NT, UK}
\address[8]{Guangxi Key Laboratory for Relativistic Astrophysics, School of Physical Science and Technology, Guangxi University, Nanning 530004, China}
\address[9]{Institute of Astrophysics, Central China Normal University, Wuhan 430079, China}
\address[10]{\ \ Yunnan Observatories, Chinese Academy of Sciences, Kunming 650216, China}
\address[11]{\ \ INAF – Osservatorio Astronomico di Roma, Via Frascati 33, I-00078 Monte Porzio Catone, (RM), Italy}
\address[12]{\ \ Astrophysics, Department of Physics, University of Oxford, Denys Wilkinson Building, Keble Road, Oxford OX1 3RH, UK}
\address[13]{\ \ Department of Astronomy, School of Physics and Materials Science, Guangzhou University, Guangzhou 510006, China}
\address[14]{\ \ Key Laboratory of Astronomical Observation and Technology of Guangdong Higher Education Institutes, Guangzhou 510006, China}
\address[15]{\ \ Great Bay Area, National Astronomical Data Center, Guangzhou 510006, China}
\address[16]{\ \ Department of Physics, College of Mathematics and Physics, Wenzhou University, Wenzhou 325035, China}
\address[17]{\ \ University of Chinese Academy of Sciences, Chinese Academy of Sciences, Beijing 100049, China}
\address[18]{Centre for Astrophysics Research, University of Hertfordshire, College Lane, Hatfield, Hertfordshire, AL10 9AB, UK}
\address[19]{\ \ Purple Mountain Observatory, Chinese Academy of Sciences, Nanjing 210023, China}
\address[20]{\ \ Institute for Frontiers in Astronomy and Astrophysics, Beijing Normal University, Beijing 102206, China}
\address[21]{\ \ School of Physics and Astronomy, Beijing Normal University, Beijing 100875, China}
\address[22]{\ \ School of Astronomy and Space Science, University of Chinese Academy of Sciences, Beijing 100049, China}
\address[23]{\ \ Cornell Center for Astrophysics and Planetary Science, Department of Astronomy, Cornell University, Ithaca, NY 14853, USA}
\address[24]{\ \ Department of Astronomy, University of Science and Technology of China, Hefei 230026, China}
\address[25]{\ \ School of Astronomy and Space Science, University of Science and Technology of China, Hefei 230026,  China}
\address[26]{\ \ Department of Astronomy, Tsinghua University, Beijing 100084, China}
\address[27]{\ \ School of Physical Science and Technology, Southwest Jiaotong University , Chengdu Sichuan 611756, China}
\address[28]{\ \ Department of Astronomy, School of Physics and Technology, Wuhan University, Wuhan 430072, China}
\address[29]{\ \ Department of Physics and Institute of Theoretical Physics, Nanjing Normal University, Nanjing, 210023, Jiangsu, China}
\address[30]{\ \ Department of Astronomy, Huazhong University of Science and Technology, Wuhan 430074, China}
\address[31]{\ \ Xinjiang Astronomical Observatory, Chinese Academy of Sciences, Urumqi 830011, Xinjiang, China}
\address[32]{\ \ Xinjiang Key Laboratory of Radio Astrophysics, Urumqi 830011, Xinjiang, China}

%\address[32]{Center for Astronomy and Astrophysics, Center for Field Theory and Particle Physics, and \\ Department of Physics, Fudan University, Shanghai 200438, China}

%\title{Physics and Astrophysics of Strong Magnetic Field systems with eXTP}{eXTP WG3}
%%% Corresponding author: ???????
%%%   \author[number]{Full name}{{email@xxx.com}}
%%% General author: ???????
%%%   \author[number]{Full name}{}
%\author[1]{First Author}{{zz@scichina.org}}%
%\author[2]{Second author}{email@xxx.com}
%\author[1]{Zhongxing ZHANG}{}%\protect\\???????
%\author[1]{Liyuan LIU}{}%
%\author[1]{Nanjian WU}{}

%%% Author information for page head. ??????????
%%% ??????????????, ??????????author???
\AuthorMark{ }%\authorcr????????

%%% Authors for citation. ????????????????
%%% ??????????????, ??????????author???

\AuthorCitation{}

%%% Address. ???
%%%   \address[number]{Address, City {\rm Postcode}, Country}
%\address[1]{Department, University, City, Postal code, Country;}
%\address[2]{For example: Institute of Mechanics, Chinese Academy of Sciences, Beijing 100190, China}

%%% Abstract. ??
\abstract{ In this paper we present the science potential of the enhanced X-ray Timing and Polarimetry (eXTP) mission, in its new configuration, for studies of strongly magnetized compact objects. We discuss the scientific potential of eXTP for quantum electrodynamic (QED) studies, especially leveraging on the recent observations made with the NASA IXPE mission. Given eXTP's unique combination of timing, spectroscopy, and polarimetry, we focus on the perspectives for physics and astrophysics studies of strongly magnetized compact objects, such as magnetars and accreting X-ray pulsars. Developed by an international Consortium led by the Institute of High Energy Physics of the Chinese Academy of Sciences, the eXTP mission is expected to launch in early 2030.
}

%%% Keywords. ?????


\maketitle

%\tableofcontents%?????

%%%%%%%%%%%%%%%%%%%%%%%%%%%%%%%%%%%%%%%%%%%%%%%%%%%%%%%
%%% The main text. ???????
%???????????????????\cref{fig1}
%\twocolumn\onecolumn
%%%%%%%%%%%%%%%%%%%%%%%%%%%%%%%%%%%%%%%%%%%%%%%%%%%%%%%
\begin{multicols}{2}

%%%%%%%%%%%%%%%%%%%%%%%%%%%%%%%%%%

%\pagebreak
%\setcounter{section}{0}

\section{Introduction}\label{sec:Introduction}
The enhanced X-ray Timing and Polarimetry mission (eXTP) is a science mission designed to study the state of matter under extreme conditions of density, gravity, and magnetism \cite{2016SPIE.9905E..1QZ,2019SCPMA..6229502Z}. In the new baseline, eXTP will be launched in early 2030. The payload of the mission consists of three main instruments: the Spectroscopic Focusing Array (SFA), the Polarimetry Focusing Array (PFA) and the Wide-band and Wide-field Camera (W2C). Here we provide a brief introduction to the scientific payloads. For a detailed description and comparison with other missions, please refer to the white paper of the whole eXTP project \cite{WP1}.

The SFA consists of five SFA-T (where T denotes Timing) X-ray focusing telescopes covering the energy range $0.5$--$10\, \mathrm{keV}$, featuring a total effective area of $2750\,{\rm cm^2}$ at $1.5\, \mathrm{keV}$ and $1670\,{\rm cm^2}$ at $6\,\mathrm{keV}$. The designed angular resolution of the SFA is $\le 1^\prime$ (HPD)  with a $18^{\prime}$ field of view (FoV). The SFA-T are equipped with silicon-drift detectors (SDDs), which combine good spectral resolution ($\sim$ 180~eV at 1.5~keV) with very short dead time and a high time resolution of $10\,{\mu\mathrm {s}}$. They are therefore well-suited for studies of compact objects emitting X-rays at the shortest time scales and even for the bright X-ray sources up to $10^{-7}$\,$\rm{erg\,cm^{-2}\, s^{-1}}$\cite{2019SCPMA..6229502Z}. The SFA array also includes one unit of the SFA-I (where I signifies Imaging) telescope equipped with pn-CCD detectors (p-n junction charged coupled device), to enhance imaging capabilities, which would supply strong upper limits to detect weak and extended sources, though suffering the pile-up effects when observing bright sources. The expected FoV of SFA-I is $18^\prime \times 18^\prime$. Therefore, the overall sensitivity of SFA could reach around $3.3\times 10^{-15}\,{\rm erg\,cm^{-2}\,s^{-1}}$ for an exposure time of 1 million seconds ($\mathrm{Ms}$). Since it is not excluded that the SFA might in the end include six SFA-T units, simulations presented here have taken this possibility into consideration. 

\begin{figure*}
    \centering
    \includegraphics[width=0.7\linewidth]{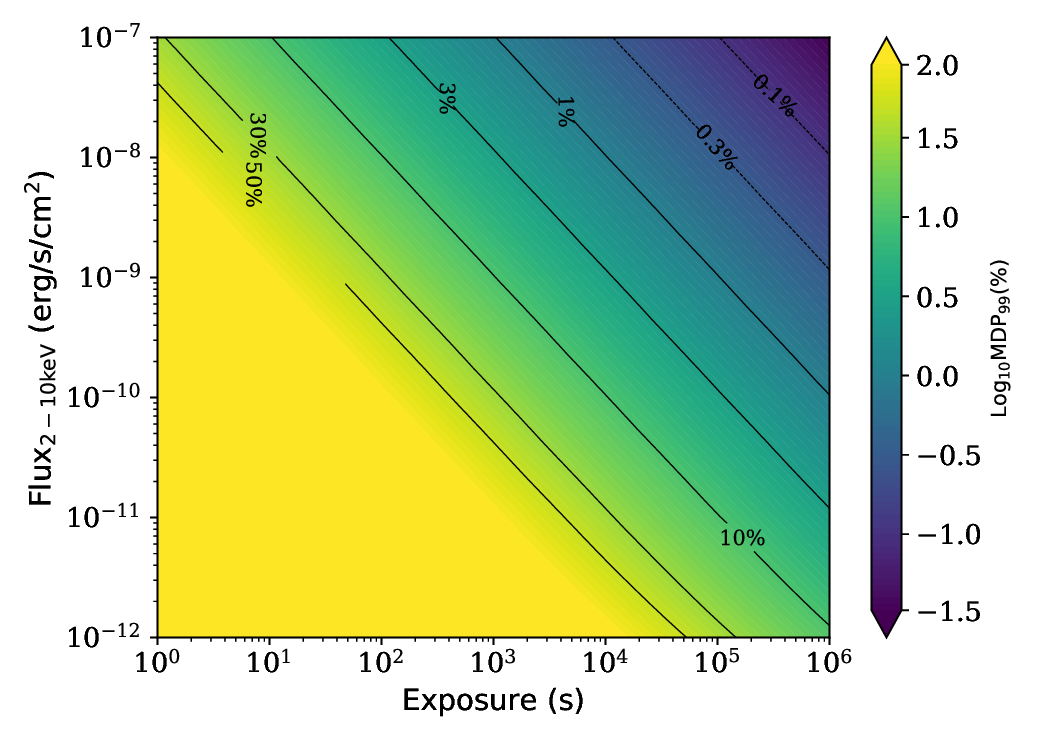}
    \caption{The estimated minimum detectable polarization at the 99\% confidence level achievable by eXTP/PFA. The calculations are performed assuming a powerlaw spectral shape with a photon index of 2.1 and considering different 2-10 keV flux levels and exposures.}
    \label{fig:MDP}
\end{figure*}

The PFA features three identical telescopes, with an angular resolution better than $30^{\prime\prime}$ (HPD) in a $9.8^{\prime} \times 9.8^{\prime}$ FoV, and a total effective area of $250\,{\mathrm{ cm^{2}}}$ at $3\, \mathrm{keV}$ (considering the detector efficiency). Polarization measurements are carried out by gas pixel detectors (GPDs) working at 2--8\,keV with an expected energy resolution of 20\% at 6\,keV and a time resolution better than $10\,{\mathrm {\mu{s}}}$ \cite{2001Natur.411..662C, 2003NIMPA.510..176B,2007NIMPA.579..853B,2013NIMPA.720..173B,2019SCPMA..6229502Z}. The instrument reaches an expected minimum detectable polarization (MDP) at $99\%$ confidence level ($\mathrm{MDP}_{99}$) of about $2\%$ in $1\,\mathrm {Ms}$ for a milliCrab-like source as shown in Figure \ref{fig:MDP}. In the current MDP calculation, the background is neglected under the assumption that the source is significantly brighter than the background. When the background cannot be neglected for weak sources (0.01 mCrab, $\sim 2\times 10^{-13}~\rm erg~cm^{-2}~s^{-1}$), the MDP changes from 20\% to 29\% with an exposure of 1\,Ms, which is inadequate for confident measurements.

The W2C is a secondary instrument of the science payload, featuring a coded mask camera with a FoV of approximately {1500} {square degrees} (Full-Width Zero Response, FWZR). The instrument achieves a sensitivity of $4\times 10^{-7}\,{\rm erg\, cm^{-2}\,s^{-1}}$ (1\,s exposure) across the 30–600\,keV energy range, with an angular resolution of $20^{\prime}$ and an energy resolution better than $30\%$ at 60\,keV.

It is fascinating to study the behavior of matter and radiation in strong magnetic fields, especially at strengths near or even far beyond the quantum critical field $B_\mathrm{q}\simeq4.4\times10^{13}\,\mathrm{G}$. Magnetars (observationally identified as anomalous X-ray pulsars, AXPs, and soft-gamma repeaters, SGRs) and pulsars in high-mass X-ray binaries supply astrophysical laboratories to reveal the nature of strongly magnetized sources. In the previous eXTP white paper dedicated to strong magnetism science, many topics have been discussed, such as the potential of the mission to investigate quantum electrodynamic (QED) effects, or the bursting activity and accretion mechanisms in neutron stars \cite[NSs,][]{2019SCPMA..6229505S}. However, since that publication, the design of the eXTP mission has evolved. Moreover, from the scientific point of view the situation underwent a rapid change, due to the operation of Polar-light and IXPE which achieved polarization measurements in X-rays, and collected data that lead to a quantum leap in our understanding of strongly magnetized sources \cite{2020NatAs...4..511F,2023ApJ...944L..27Z,2023ApJ...954...88T,2024MNRAS.52712219H,2024Galax..12....6T,2024ApJ...961..106P,2024arXiv241215811R,2024arXiv241216036S,2025ApJ...985L..35S,2022AAS...24024604T,2022tsra.confE.141T}. 

With its much larger effective area, combined with a unique and unprecedented suite of scientific instruments onboard, eXTP offers, for the first time, the possibility to make simultaneous high resolution spectral, timing, and polarimetric observations of cosmic sources in the $0.5$--$10\, \mathrm{keV}$ energy range. In this white paper we present an updated discussion of the science potential for studies of strongly magnetized objects,  based on the optimized current design of eXTP, and on the knowledge accumulated based on the recent IXPE findings.

\section{Ultra-strong magnetic fields}

\subsection{QED} \label{subsec:QED}
According to the theory of quantum electrodynamics, strong magnetic fields $\boldsymbol{B}$, in excess of the so-called quantum critical field $B_\mathrm{q}\approx4.4\times10^{13}\,\mathrm{G}$, make the vacuum dichroic and birefringent \cite{1936ZPhy...98..714H,1936DVS...14..5}. As a result, the polarization vector of a wave propagating in vacuum must be either  parallel or perpendicular to the $\boldsymbol{k}$-$\boldsymbol{B}$ plane (with $\boldsymbol{k}$ being the wave vector). These two normal modes of polarization are referred to as ordinary (O) and extraordinary (X), respectively \cite[][]{1978SvAL....4..117G}. For $B\lesssim B_\mathrm{q}$ the difference $\Delta n$ in the refraction index between the O- and X-mode photons is 
\begin{eqnarray}
    \Delta n=\frac{\alpha}{30\pi}\left(\frac{B}{B_\mathrm{q}}\right)^2\,,
\end{eqnarray}
where $\alpha$ is the fine-structure constant \cite[][]{1997JPhA...30.6485H}. For typical magnetic fields that can be achieved in laboratory ($B\lesssim 10^5\ \mathrm{G}$) $\Delta n$ turns out to be exceedingly small and challenging to measure, and this is the reason why so far only upper limits have been set on vacuum birefringence despite being $\sim$90 years since this phenomenon was first theorized \cite[][]{2020PhR...871....1E}.

\begin{figure*}
    \centering
    \includegraphics[width=1\linewidth]{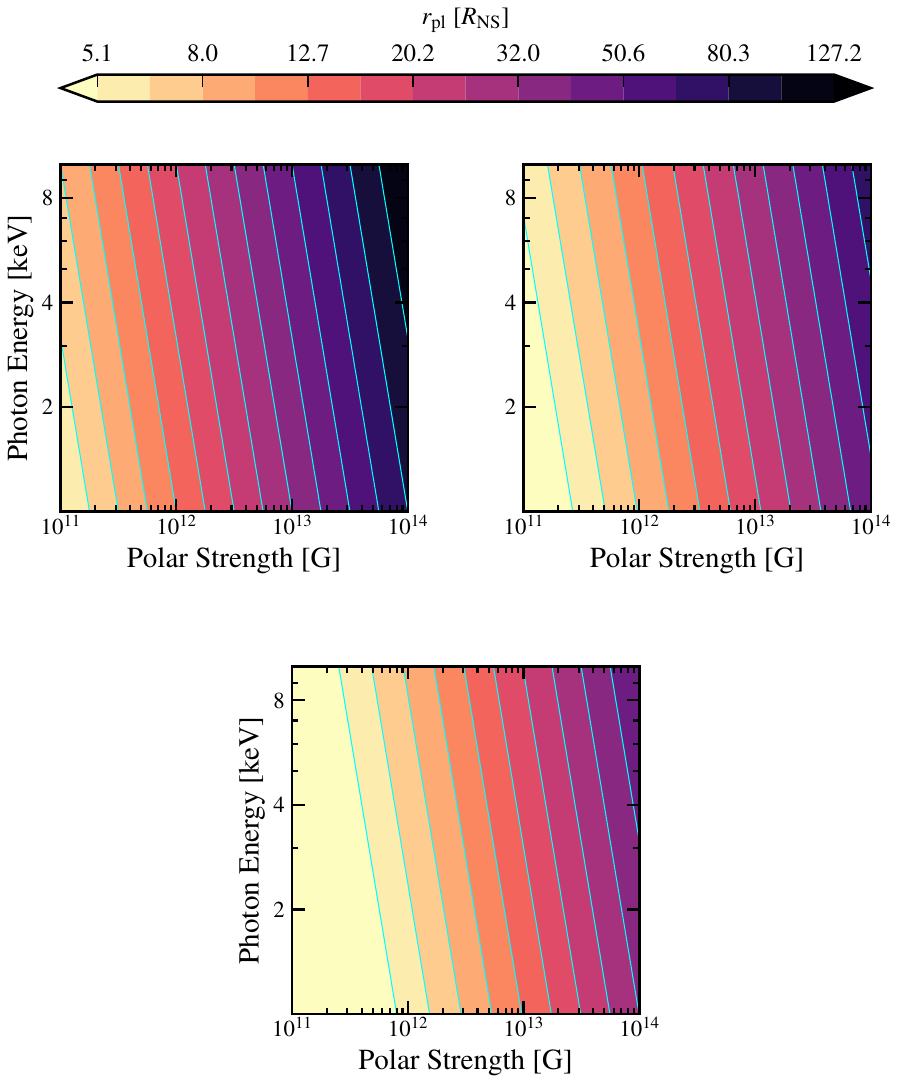}
    \caption{Contour plots of the polarization-limiting radius for a pure dipole (top left, see equation \ref{eqn:rpl}), dipole $+$ quadrupole $3$ times as strong at the surface (top right), dipole $+$ octupole $9$ times as strong at the surface (bottom), as a function of the (total) magnetic field strength at the surface and of the photon energy.}
    \label{fig:rpl_vs_B_E}
\end{figure*}

NSs possess magnetic fields far stronger than those of any other cosmic source,  with a typical strength of $\approx 10^{12}\ \mathrm{G}$ or higher. As such, they offer an alternative pathway to test vacuum birefringence by studying the properties of radiation coming from (or near to) their surface. There, the length scale $\ell_\mathrm{B}$ over which the direction and strength of the magnetic field change is much larger than the typical scale of variation of the wave polarization direction, $\ell_\mathrm{E}$ \cite[][]{2011ApJ...730..131F}. As a consequence, the polarization vector rotates adiabatically around $\boldsymbol{k}$, maintaining a fixed orientation with respect to the (local) magnetic field. In this way, despite the change in the magnetic field direction along the photon trajectory, photons polarized in the O (X) mode remain in the O (X) mode up to a distance from the NS that roughly corresponds to the so-called polarization-limiting (or adiabatic) radius \cite[][]{2003MNRAS.342..134H,2015MNRAS.454.3254T}. The exact expression of this quantity depends on the radial scaling of the magnetic field, and, for instance, under assumption of a dipolar field is given by 
\begin{equation} \label{eqn:rpl}
    r_\mathrm{pl}\approx80\left(\frac{B_\mathrm{p}}{10^{14}\,\mathrm{G}}\right)^{2/5}\left(\frac{E}{1\,\mathrm{keV}}\right)^{1/5}\left(\frac{R_\mathrm{NS}}{10\,\mathrm{km}}\right)^{1/5}R_\mathrm{NS}\, ,
\end{equation}
where $B_\mathrm{p}$ is the star magnetic field strength at the pole, $R_\mathrm{NS}$ is the NS radius, and $E$ is the photon energy. An important point is that, independently of the complexity of the magnetic field topology near the star surface, according to the most commonly-accepted scenario the magnetic field can be reasonably approximated as a pure dipole at such a large distance from the star surface (any multipolar, component would have substantially decayed). Therefore, escaping radiation crosses $r_\mathrm{pl}$ at different positions, where, however, the orientation of the magnetic field is roughly the same. As a result, the polarization state of the observed radiation is expected to closely follow the O/X-mode ratio at the emission, allowing one to gain insight into the emission mechanism that produced the signal. In contrast, if QED effects were not present around the star, the direction of the polarization vectors would not change as radiation travels to the observer, while the magnetic field does. This implies that the normal polarization modes would be washed out, thus substantially reducing the observed polarization degree (PD), even for photons originally emitted as polarized at $100\%$ \cite[][]{2015MNRAS.454.3254T,2016MNRAS.459.3585G}.

If, on the other hand, the magnetic field at the surface is dominated by higher-older multipoles, the polarization-limiting radius may assume values different from those provided by equation \ref{eqn:rpl}. By way of example, Fig. \ref{fig:rpl_vs_B_E} shows the behavior of $r_\mathrm{pl}$ as a function of the (total) surface magnetic field and the photon energy for a pure dipole, dipole $+$ quadrupole $3$ times as strong at the surface, and dipole $+$ octupole $9$ times as strong at the surface. Generally, the polarization-limiting radius tends to decrease when the field is dominated by higher-order multipoles. Nevertheless, at typical magnetar field strengths ($\gtrsim10^{14}\,\mathrm{G}$) and for photon energies within the eXTP PDA working band ($2$--$8\,\mathrm{keV}$), $r_\mathrm{pl}$ attains quite large values ($\gtrsim 50\,R_\mathrm{NS}$ in the quadrupolar configuration and $\gtrsim 30\,R_\mathrm{NS}$ in the octupolar one). Despite the presence of multipolar components changes $r_\mathrm{pl}$, at such large distances from the star the dipole is anyway dominant. In this situation, a substantial difference in the polarization degree with or without considering QED effects (as described above) is still expected. However, clear effects on the polarization angle behavior, as a result of the more complex magnetic field topology, should be visible (as discussed in detail in the following).

Among NSs, magnetars, with magnetic fields in the range $10^{14}$--$10^{15}\,\mathrm{G}$ \cite[i.e. $100$--$1000$ times stronger than those of the most common rotation-powered pulsars,][]{2014ApJS..212....6O} and emission peaked in the X-rays \cite[][]{2015RPPh...78k6901T}, are the best targets for observing this QED effect at work. In these sources, $r_\mathrm{pl}$ is about $100\,R_\mathrm{NS}$ (as shown by equation \ref{eqn:rpl}), so that the expected polarization pattern closely follows that at the emission. The same considerations hold also for other classes of strongly magnetized NSs, like the X-ray dim isolated neutron stars (for which $B\approx10^{13}\,\mathrm{G}$). 

\begin{figure*}
    \centering
    \includegraphics[width=0.6\linewidth]{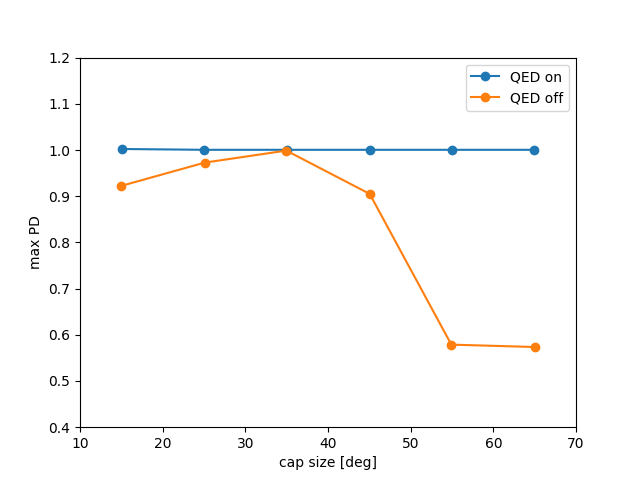}
    \caption{Polarization degree of radiation emerging from a standard (magnetized) cooling atmosphere, with $B_\mathrm{p}=10^{14}\,\mathrm{G}$ and $T=0.5\,\mathrm{keV}$ \cite[as that modeled in][]{2020MNRAS.492.5057T}, covering a cap on the star surface at one of the magnetic poles, for different values of the semi-aperture angular size of the cap (15\degr--65\degr, step 10\degr). For each cap size,  the plot shows the maximum value of the phase-averaged polarization degree that can be observed at Earth, among all possible orientations of the observer's line-of-sight and of the star's magnetic-dipole axis with respect to the star rotation axis. The cyan (orange) line with filled circles refers to models computed with (without) considering vacuum birefringence.}
    \label{fig:qedonoff_cap}
\end{figure*}

Nevertheless, the difference expected in the polarization signal when computed with and without taking into account vacuum birefringence tends to reduce as the emitting region shrinks. In fact, if photons originate from a  very small, localized spot on the star surface, the mutual orientation of their polarization vectors with respect to the star's magnetic field is basically the same even at the emission point (because the magnetic field is almost uniform across the spot), and the polarization modes are not changing much along the photon path. Hence, in this case, the intrinsic polarization degree (where for intrinsic we mean the polarization degree at the emission point, so at the star surface) should not be much different from the observed one, regardless of the presence of QED effects \cite[][]{2009MNRAS.399.1523V}. This is shown in Fig. \ref{fig:qedonoff_cap}, where the observed degree of polarization is plotted as a function of the size of the emitting region, for a computation made either with (QED-on) or without (QED-off) accounting for QED effects (in this example, radiation is assumed to be intrinsically polarized at about $100\%$ in the X mode). So, for the detection of a high polarization degree to be a proof of vacuum birefringence, observations should target highly magnetized NSs for which the X-ray radiation originates from a sufficiently large emission region (typically $\gtrsim 12\%$ of the star surface), which implies a low pulsed fraction. 

\begin{figure*}
    \centering
    \includegraphics[width=1\linewidth]{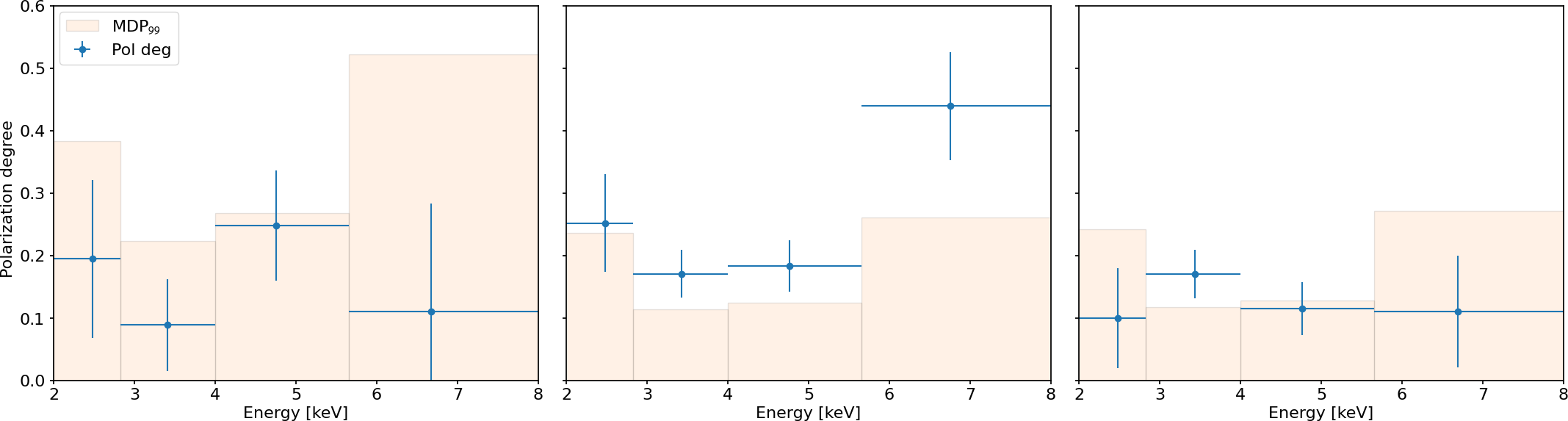}
    \caption{Simulated $1\,\mathrm{Ms}$ observations of polarization degree (blue points with error bars) from SGR 1806$-$20, performed using the {\sc ixpeobssim} suite \cite[][]{2022SoftX..1901194B} with IXPE (left-hand panel) and eXTP (central panel) response functions. Emission is assumed to come from two equatorial spots of the star condensed surface, reprocessed by RCS in the inner magnetosphere. %I used a particular geometrical configuration, such that the same simulation performed using IXPE response functions gives a result compatible with that discussed in \cite[][]{2023ApJ...954...88T}. 
    Another eXTP simulation, obtained for the same model but neglecting vacuum birefringence effects (QED-off), is shown for comparison in the right-hand panel. 
%    QED vacuum birefringence effects have been included (QED on) in the left-hand and central panels, while they are neglected (QED off) in the right-hand panel. 
In each plot, the shaded histogram shows the corresponding $\mathrm{MDP}_\mathrm{99}$.}
    \label{fig:1806}
\end{figure*}
\begin{figure*}
    \centering
    \includegraphics[width=0.8\linewidth]{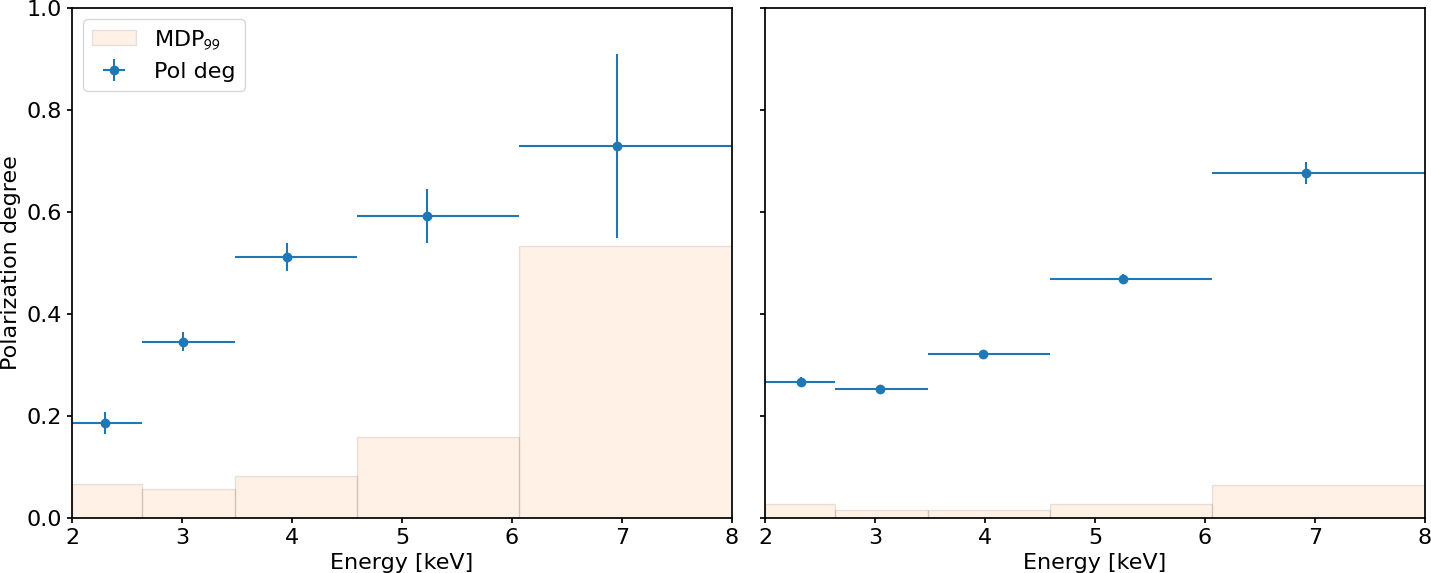}
    \caption{Left panel: Phase-integrated, energy-dependent polarization degree (blue points with error bars) as observed by IXPE from the AXP 1RXS J170849.0$-$400910 \cite[][]{2023ApJ...944L..27Z}. Right panel: Same for a simulated eXTP observation obtained using the {\sc ixpeobssim} suite, assuming the source emitting from a hotter polar spot covered by a magnetized atmosphere and a warmer equatorial region with the condensed-surface exposed (see text for details), for a $1\,\mathrm{Ms}$ observation time. 
    In each plot, the shaded histogram shows the corresponding $\mathrm{MDP}_\mathrm{99}$ and errors are given at $1\sigma$}
    \label{fig:1708IXPEvseXTP}
\end{figure*}

Transient magnetars (hereafter TMs), if observed relatively soon after the onset of an outburst event, satisfy this requirement (see Section \ref{subsubsec:bursts} for further details); but also some of the known persistent magnetars do. Among the magnetars observed so far by IXPE, the only one exhibiting a quite low pulsed fraction, $\approx 5\%$, is SGR 1806$-$20 \cite[][]{2023ApJ...954...88T}. However, despite the long exposure time ($\approx 1\,\mathrm{Ms}$), the low flux level, $\approx4\times10^{-12}\,\mathrm{erg}\,\mathrm{cm}^{-2}\,\mathrm{s}^{-1}$ in the $0.5$--$10\,\mathrm{keV}$ range prevented a secure measurement of the polarization properties with IXPE.  In this respect, eXTP, with its increased effective area, promises a much better sensitivity, ensuring a secure detection with similar or shorter exposure times. An example is shown in Fig.~\ref{fig:1806}, where we report the expected degree of polarization for a $1\ \mathrm {Ms}$ simulated observation of SGR 1806$-$20 made with eXTP, compared to an IXPE observation of the same duration. The calculation is based on the same model discussed in \cite[][]{2023ApJ...954...88T}, i.e. thermal emission from a condensed surface region located along the magnetic equator, reprocessed by resonant Compton scattering (RCS) in the inner magnetosphere (see \S\ref{subsec:atmocond}). The plots clearly show that an observation performed by eXTP (central panel) will lead to a significant improvement of the IXPE measurement (left panel), allowing for a signal detection above $\mathrm{MDP}_{99}$ in all the energy bins across the entire $2$--$8\,\mathrm{keV}$ band, for the model self-consistently computed considering QED effects ( passing from a level of significance $\approx 99\%$ to $>99.99\%$ for a $1\,\mathrm{Ms}$ exposure time). This will provide a confirmation of the theoretical interpretation of the emission, which could not be firmly obtained with IXPE because the polarization was measured only in a single energy bin at $99\%$ confidence level. In order to show how eXTP polarization measurements may be used to test vacuum birefringence, we performed an analogous simulation without including QED effects. As the right panel of Fig.~\ref{fig:1806} shows, the polarization in this case is expected to be much lower than in the previous case in all energy bins.

To better illustrate the improvement granted by eXTP on magnetar polarization measurements, Fig. \ref{fig:1708IXPEvseXTP} shows the energy-dependent polarization degree as measured by IXPE for the AXP 1RXS J170849.0$-$400910 \cite[see][]{2023ApJ...944L..27Z} compared with a simulated observation by eXTP obtained for the same observing time ($1\,\mathrm{Ms}$) starting from a theoretical model which gives a similar behavior of the polarization properties as a function of the energy\footnote{The model, discussed in \citet{2023ApJ...944L..27Z}, assumes the neutron star endowed with a magnetic field with polar strength $5\times10^{14}\,\mathrm{G}$, with emission from a polar cap covered by a magnetized atmosphere (with tempereature $\approx0.8\,\mathrm{keV}$ and semiaperture $\approx 10^\circ$) and from an equatorial belt with the condensed surface exposed (with temperature $\approx0.6\,\mathrm{keV}$, colatitudinal aperture $\approx30^\circ$ and covering $270^\circ$ in azimuth).}. In the latter case, a remarkable reduction in $\mathrm{MDP}_{99}$ and errors is evident, confirming that eXTP will enable significantly improved results with respect to IXPE for the same observation time or, alternatively, ensure no loss of significance even with reduced exposure times.

An alternative way to probe whether vacuum birefringence is indeed at work around magnetars is by looking at how the polarization properties of the observed X-ray emission change with the star's rotational phase. IXPE observations have shown that, for persistent magnetars, the polarization degree and angle exhibit different phase-dependent behaviors \cite{2022Sci...378..646T,2023ApJ...944L..27Z,2024MNRAS.52712219H,2024ApJ...961..106P}, with the former following the light curve and the latter oscillating according to the rotating vector model \cite[RVM;][]{1969ApL.....3..225R,2020A&A...641A.166P}. The behavior of the polarization degree can be explained if the polarization signal originates close to the star surface, and in such a case it will give insight into the emission geometry. Instead, we investigate the behavior of the polarization angle, which is completely unexpected for persistent magnetars, given that the RVM assumes dipolar magnetic fields, like those expected in rotation-powered pulsars. In magnetars, instead, the magnetic field close to the surface is likely far from being dipolar. The only way out of this conundrum is to invoke vacuum birefringence. If the polarization-limiting radius is large enough (as in magnetars, see equation \ref{eqn:rpl}), the pattern observed at infinity in the polarization degree reflects that of the emission near the surface, while the observed polarization vector mimics the magnetic field topology at $r_\mathrm{pl}\gg R_\mathrm{NS}$, where only the dipolar component is important. The capability of eXTP to make phase-resolved observations of persistent magnetars with a large accuracy, and  to expand the number of observable targets including other classes of sources,  is crucial to reaffirm and consolidate this conclusion, which is currently based on few preliminary results achieved up to now with IXPE.  

The situation would be different assuming more complex magnetic field topologies, with stronger multipolar components superimposed on the (weaker) dipole field (as in the examples provided in Fig. \ref{fig:rpl_vs_B_E}). In this cases, however, a substantial deviation of the phase-dependent polarization angle from the RVM is expected, as the axial-symmetry of the dipolar configuration would be definitively broken\footnote{ This holds true as long as the dipole and multipolar components do not share the same magnetic axis, which is, however, a quite unlikely situation for young sources such as magnetars \cite[see e.g.][]{2021ApJ...914..118D}.}. The same can be expected for alternative scenarios assuming, for example, nonlinear expansions of quantum-electrodynamics (like, e.g., within the Born-Infeld scheme discussed in \cite[][]{2025MNRAS.539.3655S}). In this respect, the better sensitivity of the eXTP detectors with respect to previous instrumentation will be certainly capable of resolving such deviations in future observations, allowing for a more thorough understanding of the magnetic field topology of magnetar sources.

\subsection{Spectral appearance}
Building upon the comprehensive discussion in the preceding section, which thoroughly addressed the main tests of QED effects attainable with X-ray polarimetry, the ensuing section will delve into how the forthcoming eXTP measurements can help in characterizing the intrinsic physical properties of highly magnetized cosmic sources like magnetars and X-ray pulsars. 
%*** RT please check if this transitional paragraph is fine for you.}
\subsubsection{Cyclotron resonant scattering features} \label{subsubsec:crsf}

\paragraph{Proton CRSFs in magnetar spectra}
Some of the currently known magnetar sources, such as SGR 0418$+$5729 and AXP 1E 2259$+$586, exhibit absorption features at X-ray energies, with peculiar variability across the rotational phase. SGR 0418$+$5729 entered an outburst phase in 2009, when the X-ray flux increased to $\approx5\times10^{-12}\,\mathrm{erg\,cm}^{-2}\,\mathrm{s}^{-1}$, and at that time an absorption line was detected in the spectrum by XMM-Newton, RXTE, and Swift, with a line energy in the $1$--$5\,\mathrm{keV}$ range \cite[][]{2013Natur.500..312T}. An absorption line was also visible in the spectrum of AXP 1E 2259$+$586, in two distinct sets of XMM-Newton observations, performed in 2002 and 2014, centered at an energy of $\approx0.7\,\mathrm{keV}$ \cite[][]{2019A&A...626A..39P}. The presence of this spectral line has been further confirmed by a joint XMM-Newton and IXPE observation performed in 2023 \cite[][]{2024MNRAS.52712219H}. Furthermore, spectral lines with similar properties have also been observed in the AXP Swift J1822.3$-$1606 \cite[][]{2016MNRAS.456.4145R}, as well as in some X-ray dim isolated neutron stars \cite[XDINSs,][]{2015ApJ...807L..20B,2017MNRAS.468.2975B}.

In some cases the absorption features were phase-dependent, leading to their interpretation in terms of proton cyclotron resonant scattering features (CRSFs) produced by a baryon-loaded bundle of magnetic field lines, confined to a very small portion of the star magnetosphere very close to the surface. Photons originating from an underlying hotspot on the star's surface are partly intercepted by the charged particles (mainly protons) streaming along the magnetic loop. During this process, those photons with energy equal to the proton cyclotron energy (in the particle frame) are resonantly scattered out of the observer's line of sight, resulting in a well-defined, phase-dependent absorption feature, as observed in the spectra. 
The proton cyclotron resonance is expected to occur at an energy $E_\mathrm{p}\approx0.44 B_{14}\,\mathrm{keV}$ (with $B_{14}$ the magnetic field in units of $10^{14}\,\mathrm{G}$); hence, from the energy at which the observed line is centered one can infer the local strength of the star magnetic field. Interestingly, for SGR 0418$+$5729 and AXP 1E 2259$+$586, the local magnetic field is found to be $\approx10^{15}\,\mathrm{G}$ and $3\times10^{14}$~G, respectively, both of which are much higher than the dipolar values inferred from the spin-down measurement in these sources.

\begin{figure*}
\center
\includegraphics[width=0.95\textwidth]{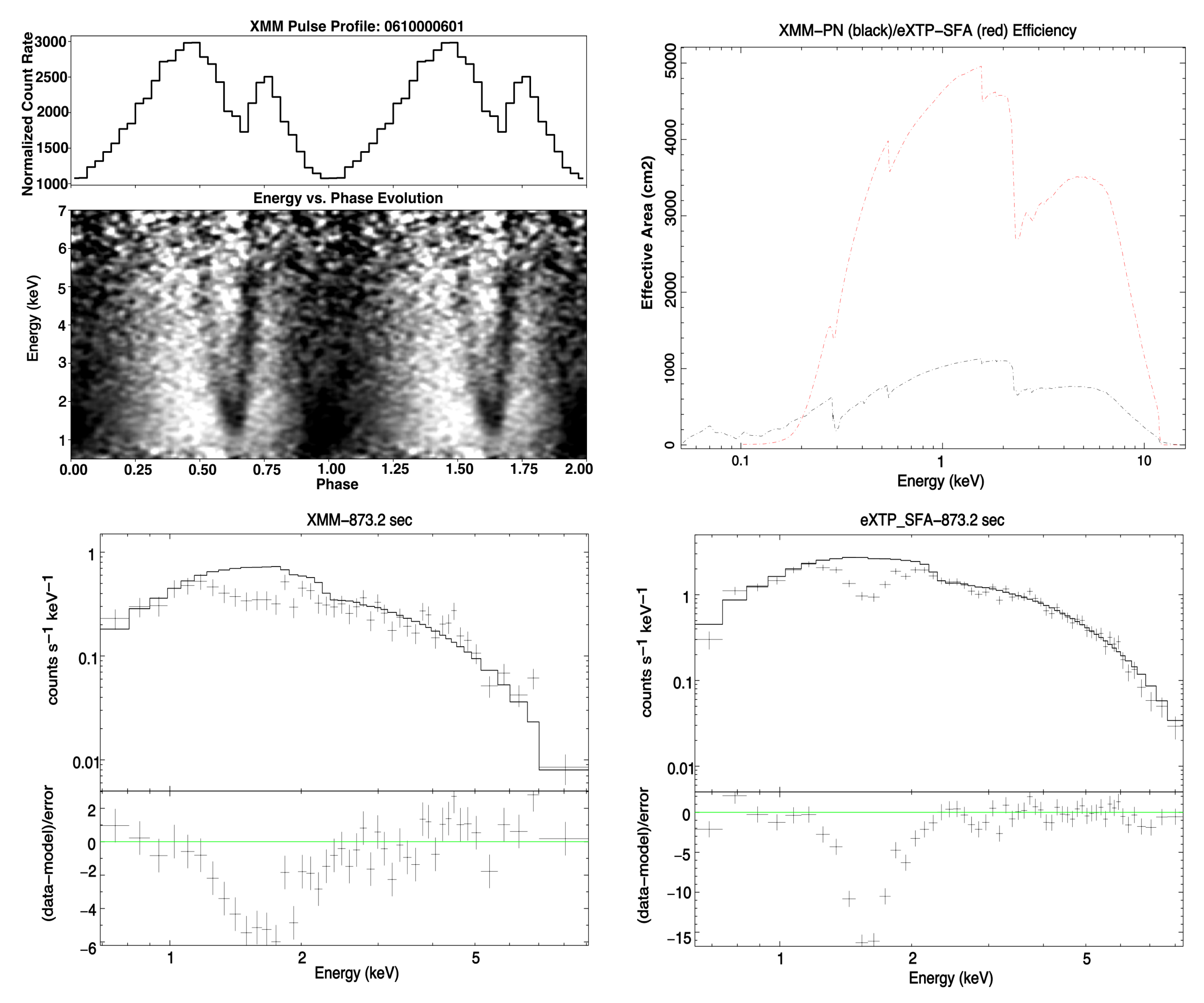}
\caption{This figure presents the observational and simulated results from XMM-PN and eXTP SFA for SGR 0418+5729. The top-left panel shows the pulse profile in the 0.5–10 keV energy range observed by XMM-PN, with the energy-phase evolution diagram plotted below, where the grayscale intensity represents spectral variations over the pulse phase and energy. The top-right panel compares the effective areas of the XMM-PN detector (black) and eXTP-SFA (red). The bottom-left panel displays the X-ray spectrum in phase 0.62--0.64 obtained from XMM-PN with an exposure time of 873k seconds, with the corresponding residuals shown in the lower sub-panel. The bottom-right panel presents the simulated X-ray spectrum obtained from eXTP-SFA for the same exposure time and spectral model, with residuals plotted in the lower sub-panel for comparison.}
\label{fig:CRSF-simu-SGR0418}
\end{figure*}

Figure~\ref{fig:CRSF-simu-SGR0418} shows the simulated $0.5$--$10\,\mathrm{keV}$ spectrum of SGR 0418$+$5729 (in the phase range $0.62$--$0.64$) obtained with six SFA-T for an exposure time of $\approx 870\,\mathrm{ks}$, compared to an analogous observation of XMM-Newton reported in \cite{2013Natur.500..312T}.  In both cases, the cyclotron feature is not included in the spectral model used for the fitting; therefore, its presence is evident in the residuals. It is clear that eXTP would detect the same feature at a much higher significance within the same exposure time.

The IXPE observation of 1E 2259$+$586 allowed one to measure the X-ray polarization properties of such an absorption feature for the first time, revealing a peculiar behavior of the degree and angle of polarization as a function of the rotational phase. In particular, the degree of polarization attains the largest values ($\approx 30\%$) at the minimum of the pulse profile, while it drops below the detection threshold in the rising phases of the light curve. Consequently, the polarization angle oscillates following the RVM model. However, poor statistics make it difficult to determine if a $90\degr$ swing
(indicative of a polarization mode switch) occurs. \citet{2024MNRAS.52712219H} found that these results are compatible with the reprocessing of photons via RCS in a localized magnetic loop. However, due to the low flux of the source, the IXPE observation was not sufficient to constrain the observed quantities to more than $\sim2.5\sigma$, even with an exposure time $\approx1.2\,\mathrm{Ms}$. The strong increase in the effective area ensured by eXTP, as well as the possibility of performing spectral and polarization observations at the same time from the same source, promise to increase the sensitivity for targets such as AXP 1E 2259$+$586. Such a systematic spectro-polarimetric study of CRSFs in magnetars (and, possibly, in other classes of highly magnetized NSs, such as the XDINSs) will help to identify the nature of the lines, potentially allowing one to map the surface magnetic field strength and topology in these sources.  Transient magnetars, during outburst phases, may also exhibit similar lines in their spectra, In this case, their study will allow one to resolve the untwisting of the magnetic loops above the heated surface region during the outburst decay, providing an independent test of the magnetar scenario (see section \ref{subsubsec:bursts}).
\ \\

\paragraph{Electron CRSFs in X-ray pulsars spectra}
For X-ray pulsars (XRPs), CRSFs due to resonant scattering onto electrons are considered the most direct method for measuring the magnetic field strength in the emission region, that is, in the vicinity of the NS surface. Currently, approximately 40 XRPs are known to exhibit absorption features in their spectra \citep{2019A&A...622A..61S,Mushtukov2022}. However, the vast majority of known CRSFs are located in the hard X-ray band, above $\sim$20~keV. This bias arises in part due to the complex nature of XRP spectra below 10 keV, where multiple spectral components, including the soft excess, iron emission line, iron edge, and hard continuum, overlap, leading to strong degeneracies that complicate the identification of additional absorption features. As a result, the measured distribution of magnetic field strengths in XRPs is skewed toward higher values.
Another challenge in CRSF detection is their potential phase dependence. In some cases, CRSFs appear only within a narrow range of spin phases, making them difficult to detect in phase-averaged spectra. However, several CRSFs below $\sim$10~keV have been discovered \citep{2015MNRAS.452.2490D,2019ApJ...883L..11M,2021ApJ...915L..27M}, some of which are phase-transient, a finding that highlights the need for detailed phase-resolved spectral analysis. Conducting such an analysis requires exceptionally high photon statistics, which will be provided by eXTP/SFA.
A key advantage of eXTP is its ability to perform simultaneous energy-resolved polarimetry with the PFA instrument in the same energy band. This unique combination will enable an unambiguous identification of CRSFs in soft X-rays by detecting the characteristic 90 degree swing in the polarization position angle, expected due to mode conversion at the electron-cyclotron resonance \citep{Mushtukov21,SokolovaLapa21}. This polarization signature will break degeneracies caused by complex spectral continua, providing a robust confirmation of CRSFs and offering new insights into the structure of magnetized atmospheres and accretion columns.
Futhermore, CRSF energies provide a direct probe of the local magnetic field strength, which is sensitive to both the height and latitude of the accretion region and hence crucial for constraining the surface magnetic topology of neutron stars. While the correlation between higher-energy CRSFs ($>$ 10 keV) and accretion rate, driven largely by changes in the accretion column height, is well-established in accreting pulsars \citep{2019A&A...622A..61S, Mushtukov2022}, the behavior of low-energy ($<$ 10 keV) CRSFs remains unexplored. As CRSF energies trace emission height \citep{Becker2012} and phase-resolved polarization measurements inform on the pulsar geometry \citep{Poutanen2024}, simultaneous spectroscopy and polarimetry are therefore key to capturing how magnetic field configurations and accretion geometries evolve with accretion rate and time. The unprecedented sensitivity and photon statistics of instruments like SFA and PFA will uniquely enable the first measurements of low-energy CRSF evolution and polarization in accreting pulsars, magnetars and PULXs.
Such observations hold the potential to fundamentally advance our understanding of the physics of extreme magnetic fields and accretion processes in neutron stars.

\subsubsection{Magnetized atmosphere and magnetic condensation}
\label{subsec:atmocond}
There is no doubt that the surfaces of accreting NSs in close binary systems are covered with gaseous envelopes. As a result, the emergent radiation originates from the upper layers of this envelope, which can be called the atmosphere \citep[see, for example, the atmospheres of X-ray bursters][]{1986ApJ...306..170L}.
However, the question of the presence of gaseous atmospheres also around isolated NSs remained open for a long time.  
Observational data support the idea that emergent radiation may be formed in an atmosphere of at least one class of isolated NSs, the central compact objects (CCOs), young sources still located near the center of their supernova remnants \citep{2004IAUS..218..239P}. Probably, the envelopes of these NSs have formed as a result of fallback accretion just after their birth \cite[][]{2017JPhCS.932a2006D}.

The problem of modeling the transport of radiation in magnetized NS atmospheres has been extensively investigated in the literature. The first self-consistent models for fully ionized atmospheres with weak magnetic fields $B \approx  10^{12}-10^{13}$ G were presented by \citet{1992A&A...266..313S} and \citet{1994A&A...289..837P}, and then extended and refined, either improving microphysics or considering different scenarios, by many authors \cite[][]{2000ApJ...537..387Z,2003ApJ...599.1293H,2004ApJ...612.1034P,2007MNRAS.377..905M,2007MNRAS.375..821H,2009A&A...500..891S,2012ApJ...751...15S}. 

The atmospheres of more highly magnetized NSs ($B\gtrsim 10^{14}$~G are more challenging to model, but have been studied in connection with magnetars to explain the thermal blackbody-like component detected in their emission spectra \cite[see e.g.][see also \citealt{2014PhyU...57..735P} for a review]{2001MNRAS.327.1081H,2001ApJ...563..276O,2001ApJ...560..384Z,vAL06}.

Instead of using Stokes parameters, the standard approach consists in solving a system of coupled radiative transfer equations for the specific intensities of ordinary and extraordinary photons. The opacities relative to these two polarization modes differ at photon energies below the electron cyclotron energy 
\begin{equation}
    E_{\rm cyc}=  \hbar \frac{eB}{m_e c} = 11.6 B_{12} \ \rm keV.
\end{equation} 
where $\hbar$, $c$, $e$ and $m_e$ are the reduced Plank constant, light speed, charge and mass of an electron while $B$ and $B_{12}$ denote magnetic field and magnetic strength in units of $10^{12}$\,Gauss.
The main effect is suppression of the X-mode opacity by a factor $(E/E_{\rm cyc})^2$ \citep{1979PhRvD..19.1684V}. Therefore, a typical feature of magnetic atmospheres for which $kT_{\rm eff} \ll E_{\rm cyc}$ is the presence of two photospheres, one for X-mode and the other for O-mode photons. The X-mode photosphere is located deeper in the atmosphere, and therefore, at least if the NSs is modeled as a passive cooler without heat deposition in its external layers,  its temperature is higher than that of the O-mode photosphere. As a result, such an atmosphere should radiate mainly in the X-mode, and its emergent radiation should be almost completely linearly polarized. 

However, this conclusion may change as the QED vacuum polarization effects are taken into account (see \S\ref{subsec:QED}).  For a given magnetic field strength, a so-called ``vacuum resonance`` can  occur at a depth in the atmosphere where vacuum and plasma contributions to the dielectric tensor balance, and such depth depends on the plasma density,  photon energy and direction.  The evolution of polarization in crossing such a resonance is poorly understood, and several studies indicate that complete or partial mode conversion can occur \cite[][]{2003MNRAS.338..233H,2006RPPh...69.2631H,1979JETP...49..741P}. 

If mode conversion occurs, it may affect the emergent polarization in a way that depends on the relative positions of the X and O photospheres and of the vacuum resonance. For those photons for which the vacuum resonance is located deeper than the X-mode photosphere, the polarization remains the same as in the absence of vacuum polarization and expected to be X-mode dominated. If, instead, the vacuum resonance is at a density lower than that of the photosphere of both polarization modes, the outcome may differ. Since X-mode photons decouple at higher densities (i.e. deeper in the atmosphere), there will be more X-mode photons than O-mode photons as radiation crosses the resonance. Here, a large fraction of X-mode photons convert into O-mode photons and, with both modes decoupled, this results in radiation being O-mode-dominated at the observer. If the resonance lies in between the two photospheres, with the X-mode photosphere deeper inside than the resonance, the flux before the resonance is dominated by X-mode photons, while O-mode ones prevail once the resonance is crossed. However, at this depth, the X-mode photons are free to travel to the observer, whereas the O-mode ones are still trapped and continue to interact, potentially changing into the X-mode, until they eventually reach the O-mode photosphere, resulting in a spectrum that remains dominated by X-mode photons. These phantomatic effects have a profound effect on the polarization properties of emergent radiation \cite[][]{2024MNRAS.528.3927K}, and eXTP will be crucial in probing them.  The effects of mode conversion may also potential suppress strong proton cyclotron absorption spectral lines \cite[][]{vAL06}.  

However, the possibility that the radiating surface of magnetars is described by a classical semi-infinite atmosphere without its own energy sources is still strongly debated \citep[see e.g.][]{2020MNRAS.492.5057T}. Possibly, energy can be released throughout the entire thickness of the atmosphere as a result of magnetic field dissipation.  Heat deposition in the outer layers by returning magnetospheric currents can also largely affect polarization \cite[][]{2019MNRAS.483..599G,2024MNRAS.534.1355K}.  
eXTP has the revolutionary capability to observe the polarization of thermal radiation from a large number of magnetars, extending the preliminary finding by IXPE, and this will unveil the properties of the radiating surface responsible for the thermal component in the spectra of magnetars.

In the presence of a strong magnetic field, the NS atmosphere may turn into a condensed state. This occurs for sufficiently low temperatures, at which atoms, elongated along the direction of $\boldsymbol{B}$, form molecular chains via covalent bonding \cite[][]{1997ApJ...491..270L,2001RvMP...73..629L,2004ApJ...603..265T,2012A&A...546A.121P}. Such elongated atoms settle on the star's surface in a solid/liquid phase, leaving the NS surface exposed (a phenomenon called ``magnetic condensation''). There is in fact theoretical and observational evidence that this may occur in some of the colder and more magnetized NSs, such as magnetars (see the next section and references therein) and possibly
the XDINSs (the so-called ``Magnificent Seven'') \citep{2007MNRAS.375..821H,2024ApJ...969...53B}. The discrimination between atmospheric and condensed surface emission can be easily performed by eXTP, since low polarization degree is expected for radiation coming from a condensed surface \cite[$\lesssim20\%$, dominated by either O- or X-mode photons according to the geometry of view,][]{2016MNRAS.459.3585G,2020MNRAS.492.5057T}.
Another exciting potentiality of eXTP would be the detection of polarization of X-ray emission from CCOs, since it could resolve the problem of the chemical composition of the atmospheres of some of these non-pulsating objects,
such as those in the supernova remnants Cas A \citep{2009ApJ...703..910P} and HESS J1731--347 \citep{2015A&A...573A..53K}. When these sources have been modelled by making the simple  assumption that  temperature is uniform on the surface, some authors obtained  the conclusion that their atmospheres are made of carbon \citep{2009Natur.462...71H}, and, in the case of HESS J1731--347, this led to the estimate of an extremely low mass of the NS, $\sim 0.8 M_\odot$ \citep{2022NatAs...6.1444D}. On the other hand, the assumption of constant temperature is probably unrealistic, in particular in presence of high magnetic fields: the presence of magnetised envelopes can effectively introduce anisotropies in the heat distribution and lead to complex thermal maps \cite{1999A&A...346..345P,2015SSRv..191..239P}.  In fact, an alternative and intriguing scenario is that these NSs host instead very high magnetar-like magnetic fields buried by fall-back accretion \citep[see e.g.][]{2014ApJ...790...94B}, in which case their surfaces may exhibit hot spots covered by hydrogen atmospheres. In that case, the absence of pulsations may be attributed to a very unfavorable viewing geometry \citep{2017A&A...600A..43S}. As discussed in \citep{2023A&A...673A..15S},
for carbon atmospheres the expected polarization is not detectable, while
detecting even a modest degree of polarization (up to a few percent for the CCO in HESS~J1731-347, in the 2-8 keV band)  
would support the hot spot hydrogen atmosphere hypothesis. These authors estimated that measuring this
polarization with IXPE required a long exposure time of a few million of seconds, while eXTP, with his enhanced sensitivity, would be excellently placed to make the detection.

\subsubsection{Spectral-continuum polarimetry in magnetars}
Observations of persistent magnetar sources show that soft X-ray spectra ($0.5$--$10\,\mathrm{keV}$) are characterized by the superposition of a thermal component, i.e. a blackbody with temperature $T\approx0.5\,\mathrm{keV}$, and a power-law tail, characterized by a spectral index $\Gamma$ typically between $2$ and $4$ \cite[][]{2008A&ARv..15..225M,2015RPPh...78k6901T}. The most viable interpretation of this phenomenology is given by the twisted-magnetosphere model \cite[][]{Thompson2002}. The internal magnetic field of magnetars is highly wounded \cite[][]{1992ApJ...392L...9D}. Magnetic stresses acting on the crust induce plastic deformations, and this results in a transfer of helicity from the internal to the external magnetic field. The external field becomes twisted and non-potential, and this forces charge carriers (typically electrons and positrons) to stream along the closed field lines, filling the inner magnetosphere with (relativistic) particles. Thermal photons coming from the star surface may interact with particles via RCS, which is responsible for the power-law tails observed in the spectra at soft X-ray energies \cite[][]{2008MNRAS.386.1527N}. 

The nature of the thermal component remains an open problem. Given the ultra-strong magnetic fields of these objects, emission from the star surface may either emerge from a magnetized atmosphere (possibly bombarded by the returning magnetospheric currents) or from a ``bare'' condensed surface. Despite the fact that the spectrum emitted in the two cases is expected to be not much different (and close to a blackbody distribution), polarization properties are poles apart \cite[][]{2020MNRAS.492.5057T}. As mentioned above, a much lower polarization degree is characteristic of radiation from a condensed surface, compared to the case of a magnetized atmosphere, for which a high polarization degree is generally expected, even when QED mode conversion is accounted for. Because for magnetar-like magnetic fields ($\gtrsim10^{14}\,\mathrm{G}$), the energy at which mode switching occurs is outside the $2$--$8\,\mathrm{keV}$ band, the working energy range for GPD-based polarimeters like IXPE and eXTP \cite[][]{2024MNRAS.528.3927K}.

Polarization measurements of magnetar persistent emission are therefore crucial to understand the physical nature of the observed radiation. This has already been achieved in the past 3 years by IXPE, which has observed five persistent magnetars. The low polarization degree ($5$--$20\%$) measured in 4U 0142$+$61 \cite[][]{2022Sci...378..646T}, 1RXS J170849.0$-$400910 \cite[][]{2023ApJ...944L..27Z}, 1E 2259$+$589 \cite[][]{2024MNRAS.52712219H,2024ApJ...961..106P} and 1E 1841$+$045 \cite[][]{2024arXiv241215811R,2024arXiv241216036S} suggests that condensed surface emission likely dominates the spectrum of these objects at soft X-ray energies ($2$--$3\,\mathrm{keV}$). In the case of 4U 0142$+$61, the polarization fraction was found to decrease with energy up to $4$--$5\,\mathrm{keV}$, where it becomes compatible with zero, then increases again, reaching $\approx35\%$ at the high-energy end of the instrumental range. Correspondingly, the polarization angle is observed to swing by $90\degr$ indeed at around $4$--$5\,\mathrm{keV}$ supporting a scenario in which low-energy (blackbody) and high-energy (power-law) spectral components are characterized by two different normal modes. These findings agree with the twisted magnetosphere model. In fact, photons reprocessed by RCS in the star magnetosphere are expected to emerge predominantly polarized in the X-mode, at a level close to $33\%$ \cite[][]{2008MNRAS.386.1527N,2014MNRAS.438.1686T,2020MNRAS.492.5057T}, consistent with the observed $6$--$8\,\mathrm{keV}$ value within the errors. Thermal photons should be emitted by the star condensed surface, being polarized at a lower polarization degree and, likely, in the O-mode. A similar behavior (i.e., small polarization at low energies and a possible $90\degr$ swing of the polarization angle) has also been suggested for 1E 2259$+$586 \cite[][]{2024MNRAS.52712219H}, although low counting statistics prevented any firm conclusion.

An alternative explanation for the properties observed in 4U 0142$+$61 has been proposed: photons are reprocessed in a magnetized atmosphere, and the low polarization degree is produced by partial mode conversion at the vacuum resonance \cite[][]{2023PNAS..12016534L}. In this way, the $90\degr$ swing of the polarization angle can be explained without invoking magnetospheric reprocessing via RCS. In this respect, new observations with eXTP may help in determining the most reliable model, both increasing the sensitivity with respect to previous IXPE measurements and assessing a larger number of targets, with fluxes too low for a significant detection with current X-ray polarimeters.

Different behaviors have been observed in the AXPs 1RXS J170849.0$-$400910 and 1E 1841$-$045. Although the relatively low degree of polarization at low energies ($2$--$4\,\mathrm{keV}$) still argues in favor of a condensed surface origin, it increases monotonically with energy, up to $60$--$80\%$ at $6$--$8\,\mathrm{keV}$. Moreover, contrary to the case of 4U 0142$+$61, the polarization angle is basically constant throughout the $2$--$8\,\mathrm{keV}$ range. In 1E 1841$-$045, a simultaneous observation with NuSTAR allowed us to associate the high degree of polarization at high energies with a power-law component, which still dominates at $\approx 6\,\mathrm{keV}$ \cite[][]{2024arXiv241215811R,2024arXiv241216036S}. Such a broad-band analysis was not possible for 1RXS J170849.0$-$400910, making it difficult to understand whether the observed polarization properties can only be attributed to surface / magnetospheric emission. Once more, eXTP observations of these sources (with both the SFA and PFA instruments) will provide invaluable insight, helping to fully characterize the spectral continuum of magnetar sources.

\subsection{Magnetic field structure}

\subsubsection{RVM and beyond}
The RVM can explain the phase-dependent behavior of the polarization angle observed in radio \cite{Johnson2023} and X-ray pulsars, both accreting \cite{Heyl_herX-1} and isolated ones \cite{2022Sci...378..646T,2023ApJ...944L..27Z,2024MNRAS.52712219H}. In fact, as discussed above, QED effects should ensure that the polarization angle as a function of the rotational phase follows the RVM also if emission comes from an extended region on (or close to) the star's surface. 

Using the RVM, one can obtain important information about the source geometry, mainly the magnetic obliquity $\alpha$ (i.e. the angle between the magnetic axis and rotational axis) and the inclination angle $\zeta$ between the line of sight and rotational axis, as well as the position angle $\psi_0$ of the rotational axis projected onto the plane of sky. In each phase, the observer sees different position angles (PAs, position angle of the rotating ``vector'' relative to the meridian plane defined by the rotational axis and line of sight). Therefore, compared to the case of a NS endowed with a dipolar magnetic field, the RVM will be modified if (i) the ``vector" intrinsically changes with time (e.g. due to forced or free precession) or (ii)  the field topology deviates from a dipole, e.g. because of the presence of a toroidal component (a twist). A variable PA is often seen in the case of magnetar radio emission\cite{Kramer2007,Levin2012}. PAs that can not be modeled by the simple form of RVM are also reported in pulsars and FRBs \cite{Johnson2023,Liu2025}. They all point to a dynamic magnetosphere with a magnetic field geometry more complex than the simple dipole field.

\begin{figure*}
\center
\includegraphics[width=0.5\textwidth,angle=270]{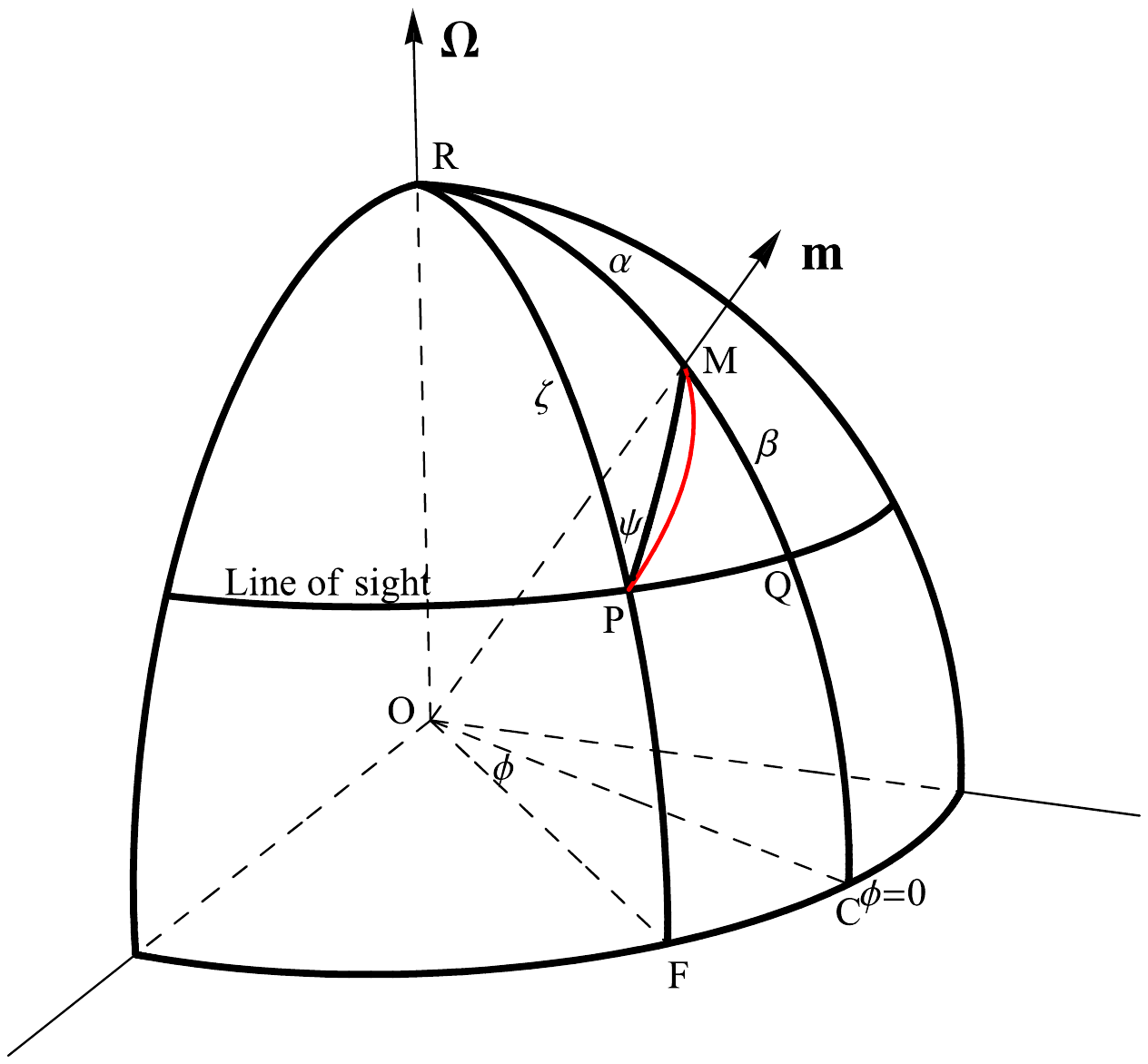}
\caption{Geometry of a rotating dipole. At the observational point, the magnetic axis makes a position angle $\psi$ with the meridian plane defined by the rotational axis and line of sight. A twisted dipole field is shown is red (Figure 1 in \cite{Tong2021}).}
\label{fig_gdipole}
\end{figure*}

The geometry of a rotating dipole field is shown in Fig.~\ref{fig_gdipole}. Using simple arguments from  spherical trigonometry, it is possible to obtain PA as a function of phase \citep[see, e.g.,][]{2020A&A...641A.166P}
\begin{equation}\label{eqn_RVM}
    \tan \psi = \frac{\sin\alpha \sin\phi}{\sin\zeta \cos\alpha - \cos\zeta \sin\alpha \cos\phi}.
\end{equation}
The labeling of different angles here ($\alpha$ and $\zeta$) follows that of pulsar astronomy, where the RVM is firstly proposed and widely used. In modeling the observations, the rotational phase angle $\phi$ is replaced by $\phi- \phi_0$ and the PA $\psi$ by $\psi-\psi_0$, where $\psi_0$ is the position angle at phase $\phi_0$, chosen such that it corresponds to the phase of the meridian plane defined by the rotational and magnetic axis. 

\paragraph{Change of the geometry due to precession} 
Actually in eq. (\ref{eqn_RVM}), except for the factor $(\sin\zeta, \ \cos\zeta)$, the remaining factors are $(\sin\alpha \cos\phi, \ \sin\alpha \sin\phi, \cos\alpha)$. It is the unit vector along the magnetic axis. Therefore, when the NS is precessing, the relative position of the magnetic axis can also be calculated by: (i) finding the direction of the magnetic axis in the body frame of the NS and (ii) finding the direction of the magnetic axis in the inertial frame using the coordinate transformation of rigid body kinematics (described by three Eulerian angles). Then the corresponding RVM for a precessing NS can be found, which is similar to equation~(\ref{eqn_RVM}) but more complicated \cite{Heyl_herX-1}. 

In practice, a simpler and straightforward approach can be used by fitting the observed polarization angle with the standard RVM model at different epochs. Then, different sets of RVM parameters $(\alpha,\ \zeta,\ \phi_0,\ \psi_0)$ will be obtained. In the case of forced precession, the rotation axis of the NS experiences a significant change, due to gravity effects produced by a companion star or a fallback disc. Then, the magnetic axis is dragged together with the spin, so that $\alpha$ remains constant. In contrast, free precession arises because of an aspherical deformation of the star, possibly produced by the strong $B$-field. In this case the rotation axis is practically fixed in space (which is approximately coincide with the star's angular momentum), so that $\zeta$ and $\psi_0$ constant, while the inclination $\alpha$ of the magnetic axis varies in time, and a consequence also $\phi_0$ changes. 

\begin{figure*}
    \centering
    \includegraphics[width=0.9\linewidth]{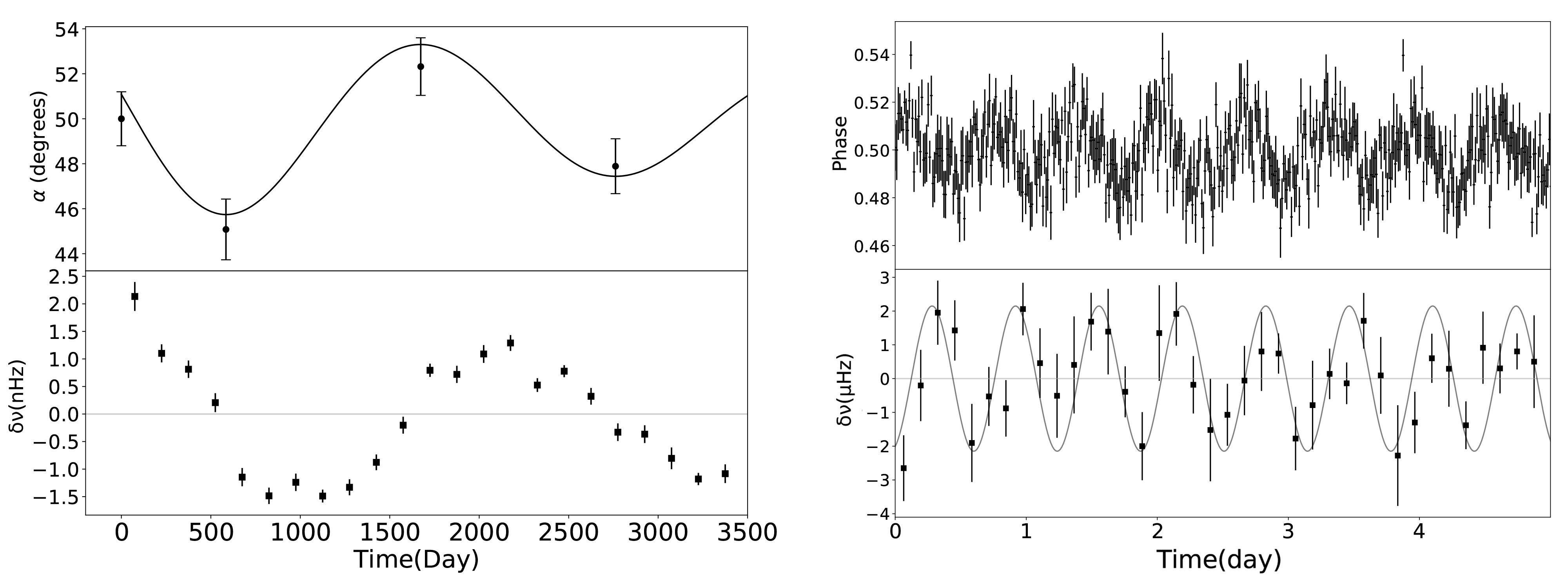}
    
    \caption{A simulation to probe long-term and short-term precession together with RVM and timing results for 4U 0142+61. In the left figure, upper panel shows the inclination angle measurement based on RVM method with exposure 1.6\,Ms. The frequency oscillation measurements induced by precession are plotted in the bottom panel. In the right figure, upper panel shows timing residuals with assumption precession period of 55\,ks and amplitude of 0.1\,s considering the energy range 3--10\,keV with net exposure 400\,ks. The frequency oscillation measurements induced by precession is shown in the bottom panel.
    }
    \label{fig:rvm}
\end{figure*}

We simulate the possible precession signal with the model prediction which could be measured by RVM and timing results as shown in Fig.~\ref{fig:rvm}. The oscillation signal could obviously be detected from the timing results and the inclination angle obtained from the RVM fitting with the exposure of 1.6\,Ms.

\paragraph{Change of field line geometry due to the twist of the magnetic field}
The difference between magnetars and normal radio pulsars is not only their magnetic field strength, but also their field topology. Magnetars, in fact, may have twisted dipolar fields and even twisted multipolar fields, although the geometry of the large-scale field line is dominated by the twisted dipole field at some distance from the surface \cite{Thompson2002,Beloborodov2009,Tong2019}. 

For a self-similar twisted dipole field, the magnetic field is proportional to $r^{-(2+n)}$, where the radial index $n$ takes values between $0$ (for the limiting situation of a split monopole) and $1$ (corresponding to a purely dipolar field). The twist grows as magnetic helicity is transferred from the tangled internal field to the external one, causing a slippage of crustal platelets and injecting magnetic energy into the upper crust. This is believed to power magnetar outbursts. However, once implanted, the twist must decay to provide the currents needed to support it \cite{Thompson2002,Beloborodov2009,Tong2019}. 

For a twisted magnetic field, the RVM will also be modified. According to the ``rotating vector model for magnetars", the RVM curve is modified with respect to the shape expected for purely dipolar fields (see equation \ref{eqn_RVM}), being stretched over a wider range of values and shifted in rotational phase \cite{2015MNRAS.454.3254T,Tong2021}. This translates into a monotonic change of $\psi_0$ and $\phi_0$ over time, according to the direction in which the magnetic field lines are twisted. Previous radio observations of magnetars found some hints of an untwisting magnetic field \cite{Levin2012}. 

\begin{figure*}
    \centering
    \includegraphics[width=0.6\linewidth]{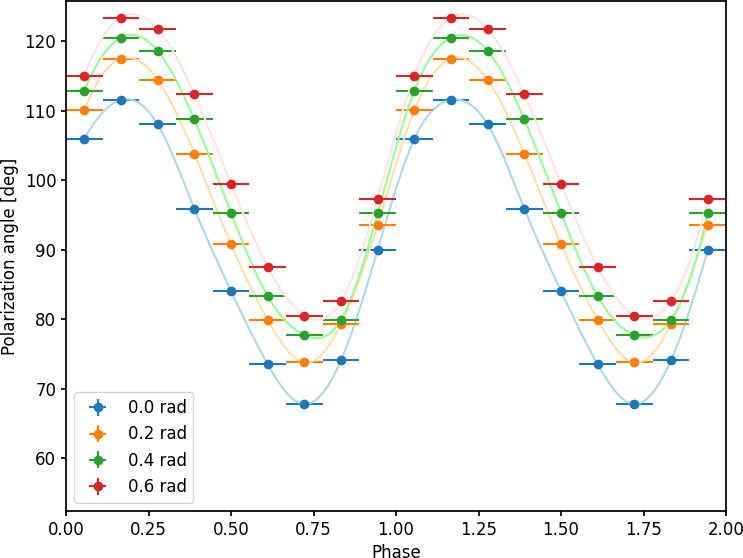}
    \caption{Phase-dependent polarization angle behaviors obtained for different $1\,\mathrm{Ms}$ eXTP simulated observations, reproducing different stage of a magnetar outburst, characterized by different values of the $1$--$10\,\mathrm{keV}$ unabsorbed flux with $1$--$10\,\mathrm{keV}$ unabsorbed flux: $10\times10^{-10}$ (red), $8\times10^{-11}$ (green), $6\times10^{-11}$ (orange) and $4\times10^{-11}\,\mathrm{erg\,cm^{-2}\,s^{-1}}$ (cyan). The star magnetic field is assumed to be self-similar and globally twisted, with different values of the radial index $n$ \cite[see][]{Thompson2002} for each outburst phase, i.e. $n=0.96$ (corresponding to a global twist angle $\Delta\phi_{\mathrm{N-S}}=0.6\,\mathrm{rad}$), $0.98$ ($\Delta\phi_{\mathrm{N-S}}=0.4\,\mathrm{rad}$), $0.99$ ($\Delta\phi_{\mathrm{N-S}}=0.2\,\mathrm{rad}$) and $1.00$ ($\Delta\phi_{\mathrm{N-S}}=0\,\mathrm{rad}$, dipolar field), respectively. Radiation is assumed to follow a blackbody distribution at a temperature $0.5\,\mathrm{keV}$, constant across the surface, $100\%$ polarized in the X-mode. All the simulations are performed using the {\sc ixpeobssim} suite \cite[][]{2022SoftX..1901194B} with the eXTP response files.  Notice that the error bars on the polarization degree are within the marker size.
    }
    \label{fig:rvm2}
\end{figure*}

If X-ray polarization observations of magnetars are performed during an outburst, a monotonically evolving PA (especially $\psi_0$ and $\phi_0$) is expected. This is shown in Fig.~\ref{fig:rvm2}, where the polarization angle resulting from different $1\,\mathrm{Ms}$ eXTP simulations is shown \cite[see][]{2015MNRAS.454.3254T,Tong2021}. Four stages of an outburst have been considered, characterized by different values of the flux $1$--$10\,\mathrm{keV}$ (unabsorbed) and of the radial index $n$ of the globally twisted field \cite[see][]{Thompson2002}. For simplicity, the source emission is modeled assuming a BB distribution (at a temperature $0.5\,\mathrm{keV}$, constant across the surface), $100\%$ polarized in the X-mode. According to the results shown in Fig.~\ref{fig:rvm2}, the deviation from the classical RVM for different magnetic field twists can be clearly resolved by eXTP observations with sufficiently long exposure. 

\subsubsection{Precession of Neutron Stars}
In the freely precessing-magnetar scenario, the measured modulation period $P_{\rm m}$ is considered to be the magnetar's precession period $P_{\rm p}$, which can be theoretically determined as 
\begin{equation}
\begin{aligned}
  P_{\rm p} \simeq P/|\epsilon_{\rm B}|.
\end{aligned}
\label{epsilon}
\end{equation}
where $P$ and $\epsilon_{\rm B}$ are respectively the magnetar's spin period and magnetically induced ellipticity. Here, depending on the configuration of internal fields (i.e., whether the internal fields are poloidal-dominated or toroidal-dominated), the ellipticity can be either $\epsilon_{\rm B}>0$ or $\epsilon_{\rm B}<0$. Assuming that the magnetar has a possible ellipticity $|\epsilon_{\rm B}|\sim10^{-4}$ as inferred from the magnetars 4U 0142$+$61, 1E 1547.0-5408, SGR 1900$+$14, and SGR 1806-20 \cite{2014PhRvL.112q1102M,2021MNRAS.502.2266M,2021ApJ...923...63M,2024PASJ...76..688M}, and a typical spin period $P\sim1-12$ s \cite{2014ApJS..212....6O}, then it would have a precession period $P_{\rm p}\sim10^4-10^5$ s. If the magnetar's precession amplitude of spin period is around sub-second as observed for the above four magnetars in hard X-ray band \cite{2014PhRvL.112q1102M,2021MNRAS.502.2266M,2021ApJ...923...63M,2024PASJ...76..688M}, our simulated results show that the periodic modulation in its magnetic tilt angle could be detected by the eXTP through X-ray timing observations as shown in Fig.~\ref{fig:rvm} with net exposure of 400\,ks and precession period of 55\,ks with energy range 3-10\,keV, which could be merged with other observations.

We also expect that detection of free precession of magnetars using eXTP may provide us valuable information on the following two open issues: (i) whether free precession of magnetars could indeed be damped (namely, decays with time); (ii) the specific viscous mechanisms that damp free precession of magnetars. For the first issue, polarization measurements of the radio emission of the magnetar XTE J1810-197 have recently shown that its free precession was damped on a time scale of
months \cite{Desvignes2024}. However, X-ray polarization observations of this magnetar is still lacking and it is unclear whether similar evidence could be found in X-ray band. Furthermore, for the above four magnetars that may be freely precessing \cite{2014PhRvL.112q1102M,2021MNRAS.502.2266M,2021ApJ...923...63M,2024PASJ...76..688M}, no observational evidence showing that their free precessions are damped. As a result, sophisticated X-ray polarization observations of magnetars by eXTP will help to figure out this issue.    

The second issue could be addressed if free precession of magnetars is found and their magnetic tilt angles are measured simultaneously. The damping of the free precession of magnetars is mainly due to the mutual frictions between the superfluid neutrons and other particles in the interior of the NS, which can be characterized by the number of precession cycles $\xi$ \cite{2002PhRvD..66h4025C,2009MNRAS.398.1869D,2019PhRvD..99h3011C,2024SCPMA..6729513Y}. Using the method proposed in \cite{2019PhRvD..99h3011C,2023RAA....23e5020H}, we may set constraints on $\xi$ of some magnetars and ordinary young pulsars when their magnetic tilt angles are accurately measured. From the value of $\xi$, we can grasp the specific viscous mechanisms that damp the free precession of NS \cite{2009MNRAS.398.1869D,2019PhRvD..99h3011C,2024SCPMA..6729513Y}. Finally, combining the measured precession periods with surface thermal emissions of magnetars, their internal magnetic fields can also be probed \cite{2025PhRvD.111b3038S}. Investigation of the magnetic fields of magnetars is also important for understanding gravitational wave radiation from magnetars \cite{2024EPJC...84.1043Y}. 

\subsection{Magnetically driven activities}

\subsubsection{Glitches}
The typical pulsar glitch is an abrupt increase in spin frequency ($\Delta\nu$), with the fractional glitch size $\Delta\nu/\nu\sim 10^{-11}-10^{-5}$, and in many cases the glitches are accompanied by a step increase ($\Delta\dot\nu$) in the absolute value of spin down rate ($\dot\nu$), with $\Delta\dot\nu/\dot\nu\sim 10^{-5}$--$10^{-2}$. Theoretically, there are mainly two models for the glitch phenomenon, the starquake model~\citep{1969Natur.223..597R, 1971AnPhy..66..816B,2003ApJ...596L..59X} and the superfluid vortex model~\citep{1969Natur.224..673B, 1975Natur.256...25A}. For a recent review on the glitch phenomenon, see ~\cite{2022RPPh...85l6901A, 2022Univ....8..641Z}.

In recent years, much progress has been made on observations of the glitch phenomenon, among which are radiative changes associated with glitches in magnetars and rotation-powered pulsars, spin-down glitch (also named anti-glitch) in magnetars and rotation-powered pulsars, delayed spin-up glitches in young pulsars, and the association of glitch/spin-down glitch-fast radio bursts.

The spin-down glitch is characterized by a sudden and permanent decrease in spin frequency of pulsars within short timescales, and it should be distinguished from the case of over-recovery glitches (for instance, the PSR J1522--5735 glitch~\cite{2013ApJ...779L..11P}) or net spin-down due to an enhanced post-glitch spin-down rate (for instance, glitch in the high magnetic field pulsar PSR J1119--6127~\citep{2018MNRAS.480.3584D,2018ApJ...869..180A}). The first spin-down glitch was observed in the magnetar 1E 2259+586, and it was accompanied by a X-ray burst, flux enhancement in $2$--$10~{\rm keV}$, pulse profile change, and a period of enhanced spin-down rate~\citep{2013Natur.497..591A}. 1E 2259+586 exhibited another spin-down glitch of a similar size to the previous one in April 2019, but this spin-down glitch was radiatively silent~\citep{2020ApJ...896L..42Y}. Later, another spin-down glitch was identified in the anomalous X-ray pulsar 1E 1841--045, with no significant variations in the pulsed X-ray emission of the $0.5$--$10~{\rm keV}$ flux~\citep{2014MNRAS.440.2916S}. Similarly, three spin-down glitches were found in the ultraluminous accreting pulsar NGC 300 ULX-1, without detection of pulse profile or spectral changes coincident with these glitches~\citep{2019ApJ...879..130R}. Recently, a radiatively quiet spin-down glitch was detected in the rotation-powered pulsar (RPP) PSR B0540-69~\cite{2024ApJ...967L..13T}, extending the spin-down glitch phenomenon to a wider pulsar population.

The physical origin of spin-down glitch is still unclear, and we do not know whether the radiation-loud and radiation-quiet spin-down glitches share the same physical origin. Based on the detection or non-detection of pulsed X-ray flux enhancement and pulse profile changes, the spin-down glitch could theoretically be of internal or external origin. The first group includes theories that transfer angular momentum from the crust to the superfluid component. Generally, the faster-rotating superfluid component transfers angular momentum to the slower-rotating crust. However, in some cases, this process could be reversed. It is proposed that the fraction of normal and superfluid components in NSs depends not only on the temperature but also on their spin lag, and changes in spin lag will result in matter transformation between superfluid component and normal component in pulsars with high internal temperature and strong toroidal magnetic fields, resulting in an abrupt spin-down glitch in some cases~\cite{2014ApJ...797L...4K}. It is also  explored that the possibility of vortices migrating inward instead of outward to explain spin-down glitches in the accreting pulsar NGC 300 ULX-1~\cite{2022MNRAS.514..863H}. ~\cite{2024ApJ...977..243Z} proposed the radially inward transportation of vortex lines due to broken plate motion following crustquakes.

The second group includes theories of collisions between NSs and asteroids/comets~\cite{2014ApJ...782L..20H}, enhanced particle wind braking~\cite{1999ApJ...525L.125H,
2014ApJ...784...86T}, partial opening of the magnetosphere~\cite{2013arXiv1306.2264L}, pulsar shape reconfiguration due to the decay of internal toroidal magnetic field~\cite{2015MNRAS.449L..73G,2015ApJ...807L..27M}, et al.

The biggest differences between the first and second groups lie in the existence and the level of X-ray flux enhancement. Theoretically, little flux variation is expected for internal models, while large flux variations are expected in the magnetospheric models, and these flux variations will also be correlated with the timing variations during the spin-down glitch event. Taking spin-down glitches in 1E 2259 + 586 as examples, to distinguish whether spin-down glitches originate from the interior or from the external of this pulsar, the lower limit of detectable variations in X-ray flux should be of the order of $\sim 10^{-13}~\rm{erg~s^{-1}~cm^{-2}}$~\cite{2014ApJ...784...86T}. 
For eXTP, observations with an exposure time of $10~\rm{ks}$ will satisfy this requirement.
Therefore, the eXTP program is promising in setting constraints on the mechanism of spin-down glitch through X-ray observations.

In RPPs, only rare cases of clear correlations between glitches and emission changes such as pulse profile and spectral characteristics have been found in pulsars such as PSRs J0742-2822, B1822-09, and J0835-4510~\cite{2025A&A...694A.124Z,2022ApJ...931..103L,2018Natur.556..219P}, suggesting an internal origin for most glitches and the existence of the angular momentum reservoir in RPPs~\citep{1999PhRvL..83.3362L}. The theory of how the internal vortex motion result in emission changes in RPPs is still establishing~\cite{2025A&A...694A.124Z}. 

However, in magnetars, a large fraction of glitches or spin-down glitches are associated with burst and/or outburst behaviors~\cite{2014ApJ...784...37D,2015MNRAS.449..933A,2019AN....340..340H}, and the magnitude of magnetar glitches ($\Delta\nu$) falls at the far end among all pulsar glitches in the histogram of $\Delta\nu$ versus normalized counts~\cite{2017A&A...608A.131F}. Why magnetar glitches behave differently from RPPs glitches or whether magnetar glitches and RPPs glitches share the same physical origin is an open question. It is proposed that large-scale crust failures due to vortex pinning stresses in a spinning down NS could give rise to glitches and bursts of X-ray and gamma-ray due to the release of elastic energy~\cite{1991ApJ...382..587R}. \cite{2019MNRAS.488.5887S} proposed that the hall drift of the strong and highly multipolar crustal magnetic fields in young magnetars may generate magnetic stress and induce frequent failures, which result in glitches and possibly fast radio bursts (FRBs). 

Starquakes in magnetars might trigger FRBs as revealed by some theoretical and statistical works \cite{2018ApJ...852..140W,2019ApJ...879....4W,2019MNRAS.488.5887S,2022ApJ...928...53W,2023MNRAS.526.2795T}. The energy functions of FRBs has a universal break which is similar to that of the frequency–magnitude relationship of earthquakes. Such break and the change in power-law indices across this transition can be explained by magnetar starquake models \citep{2025ApJ...979L..42W}. The energy released during the starquakes is estimated to support X-ray bursts via resonant inverse Compton scattering \cite{2021ApJ...919...89Y}. 
The quake model predicts that FRBs will be associated with glitches and quasi-periodic oscillations (QPOs), which are expected to be found in the associated X-ray bursts \cite{2022ApJ...931...56L}. Currently, several FRB-like bursts from SGR J1935+2154 are found to be associated with the source spin changes. 
\cite{2024RAA....24a5016G} found a glitch candidate before FRB 200428, the fast rise of this glitch was followed by a delayed spin-up rise, similar to that observed in the Crab pulsar~\cite{2018MNRAS.478.3832S}. It is reported that two neighboring glitches within a time interval of approximately nine hours with great details, and the two glitches bracket an FRB on 2022 October 14~\cite{2024Natur.626..500H}. Rapid spin down phase was observed between the two glitches. Previously, \cite{2023NatAs...7..339Y} reported a spin-down glitch around 2020 October 5 that occurred before three FRB-like bursts, and the observational cadence seems to be much lower than that in~\cite{2024Natur.626..500H}. The possibility that this spin-down glitch is actually a glitch followed by an enhanced spin-down rate (just as what was observed by \citet{2024Natur.626..500H}) is worthy of consideration. Note that, apart from starquakes, other models such as tidal capture of an asteroid by a magnetar could also result in the observed spin changes and FRB-like bursts simultaneously~\cite{2023MNRAS.523.2732W}. Given the apparent association between glitches/spin-down glitches and galactic FRBs, quick responses and high-cadence observations around the epoch of FRBs should be conducted. If the galactic FRB and glitch/spin-down glitch association is confirmed to be a necessity rather than a coincidence by eXTP, then a comprehensive research on the physical origin of these glitches/spin-down glitches through flux, pulse profile, polarization position angle variations and so on may shed light on the physical origin of galactic FRBs.

\begin{figure*}
    \centering
    \includegraphics[width=0.6\linewidth]{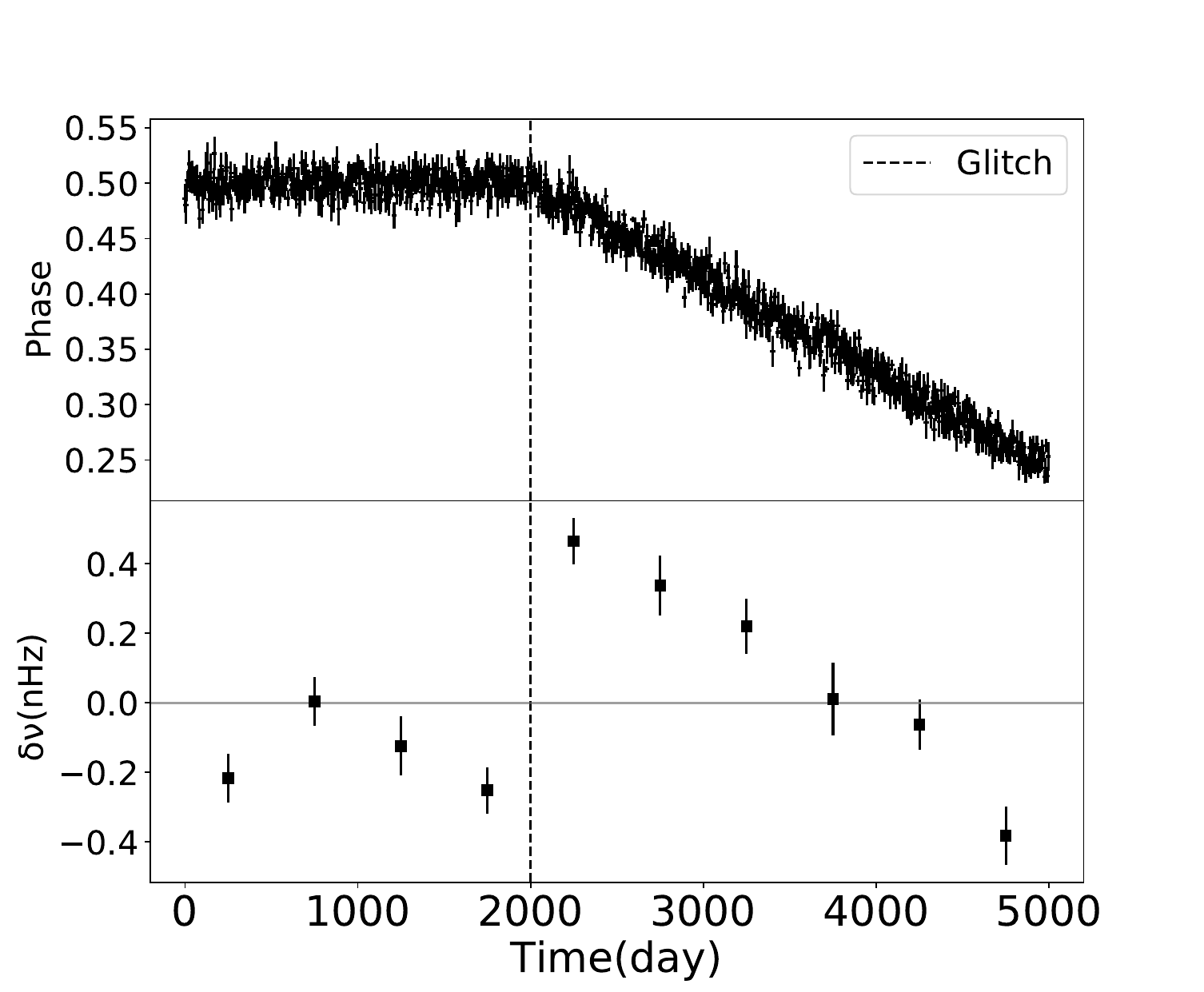}
    \caption{A simulation to probe the glitch event based on eXTP-SFA observations. Upper panel shows the timing residuals with the pre-glitch spin parameters. The spin frequency and frequency derivative are plotted in the middle and bottom panels. The vertical dashed line represent the simulated glitch epoch in three panels.
    }
    \label{fig:glitch}
\end{figure*}

We have simulated the possible glitch signal to study the connection between burst activities and glitch behaviors as revealed by previous studies~\cite{2021MNRAS.507.2208W}. We selected the anomalous X-ray pulsar 4U 0142-61 as the target for observational simulation. This pulsar exhibits a broad single-peak pulse profile. Its flux within the 0.5-10\,keV energy range is approximately $0.035\, {\rm counts\,cm^{-2}\,s^{-1}} $\cite{2002ApJ...568L..31J}. By performing an analysis with its energy spectrum and the effective area of eXTP, when eXTP conducts an observation of 4U 0142-61 for 1\,ks, a timing accuracy of approximately $5\times 10^{-3}$ in phase can be achieved. With an observation frequency of obtaining one TOA every five days, the spin frequency (\(\nu\)) and its first derivative (\(\dot{\nu}\)) can be extracted by utilizing the TEMPO2 software to fit the TOAs accumulated within each 100\,day \cite{2006MNRAS.369..655H}. As shown in Fig.~\ref{fig:glitch}, the minimum detectable amplitude of a glitch by eXTP is $\Delta{\rm \nu}/\nu\sim10^{-9}$ with the typical exposure. The detection capability of eXTP will help confirm whether all galactic FRBs detected in the future are associated with glitches.

\subsubsection{Bursting activity and outburst phases} \label{subsubsec:bursts}

\paragraph{Bursting activity and outbursts in magnetar sources}
One of the most distinctive behaviors of magnetars is the occurrence of events in which huge amounts of energy are released over different timescales, which, in many cases, led to the discovery of new magnetar sources. Persistent magnetars are characterized by the emission of short ($0.1$--$1\,\mathrm{s}$) bursts, with luminosities in the range $10^{39}$--$10^{41}\,\mathrm{erg}\,\mathrm{s}^{-1}$, and intermediate ($1$--$40\,\mathrm{s}$) flares, with peak luminosities ranging from $10^{41}$ to $10^{43}\,\mathrm{erg\,s}^{-1}$ \cite[][]{2015RPPh...78k6901T}. The rarest giant flares, with luminosities up to $\approx10^{47}\,\mathrm{erg\,s}^{-1}$ in the short ($0.1$--$0.2\,\mathrm{s}$) initial spike and decreasing down to $\approx10^{44}\,\mathrm{erg\,s}^{-1}$ in the $\sim100\,\mathrm{s}$ pulsating tail (modulated at the spin frequency of the star), have been definitively observed from only three magnetars \cite[SGR 0526$-$66, SGR 1900$+$14 and SGR 1806$-$20, see][]{1979Natur.282..587M,1999Natur.397...41H,2005Natur.434.1098H,2005Natur.434.1107P}. Several magnetar giant flare candidates have been reported from nearby galaxies, such as GRB 200415A in NGC 253 \citep{Svinkin2021Natur}. Conversely, outbursts are typically identified in transient sources following the detection of X-ray bursts by the wide-field instruments on-board such as Swift, Fermi, GECAM missions. These events are characterized by a sudden flux enhancement (by a factor of $\sim100$--$1000$), followed by a return to quiescence over timescales ranging from months to several years \cite[][]{CotiZelati2018}. In particular, the recent launch of the Einstein Probe and SVOM missions has significantly enhanced our ability to detect new magnetar outbursts \citep{Yuan2025EP,Wei2016SVOM}.
With its large FOV, accurate localization and coverage of the hard X-ray band, eXTP/W2C will be powerful to monitor the burst activity of known magnetars and discover new magnetar sources either in our galaxy or nearby galaxies.

Currently, a comprehensive and self-consistent model to explain the triggering mechanisms of magnetar bursts and flares remains elusive. One proposed mechanism involves electron acceleration along the star's twisted magnetic field lines, which undergo rapid reconnection within the magnetosphere. This process generates a cascade of electron-positron pairs and gamma-ray photons \cite[][]{1989BAAS...21..768S}. Alternatively, particularly for intermediate and giant flares, another model suggests that crustal displacements, driven by the intense internal magnetic field of the star, may induce a rapid rearrangement of external field lines, injecting an Alfvén pulse into the magnetosphere \cite[][]{1995MNRAS.275..255T}. These Alfvén waves may remain trapped within the closed field-line region, dissipating into an electron-positron pair plasma (the so-called ``trapped fireball''), which is predicted to produce the observed pulsating tails in the afterglow phase. Simple radiative transfer calculations \cite[][]{2016MNRAS.461..877V,2017MNRAS.469.3610T} have shown that this latter scenario effectively reproduces the spectral properties of magnetar intermediate flares, for which the $0.2$--$100\,\mathrm{keV}$ spectra are well fitted by two thermal components \cite[][]{2008ApJ...685.1114I}. Furthermore, this model predicts a high degree of polarization (exceeding $\approx80\%$) and a constant polarization angle in the soft X-ray band \cite[][]{2017MNRAS.469.3610T}, a measurement well within the capabilities of eXTP (see Fig.~\ref{fig:MDP}), despite the short duration of these events. 

However, the discovered correlation between the temperature and size of the emitting surface of the two blackbody components observed in the spectra of intermediate flares has also suggested the possibility that one of these blackbody is populated by ordinary photons and the other by extraordinary photons, emerging from their respective photospheres \cite[][]{2008ApJ...685.1114I}. In this scenario, the expected net polarization degree would be significantly lower than $80\%$, due to the competing contributions of the two normal modes. Instead, energy-resolved polarimetric observations may detect the polarization of the two components separately and, therefore, reveal a 90$^\circ$ swing of the polarization angle between the hot and the cold blackbody. 
Consequently, polarimetric observations with eXTP are crucial to disentangling these two scenarios and will provide unprecedented insights into the trigger mechanisms of magnetar bursts and flares.

Observing an outburst with eXTP, with its capability of performing simultaneous phase-resolved spectral and polarization measurements, will enable the extraction of extensive information regarding aspects such as the activation mechanism of these phenomena, the energy source driving their long-term emission, and the potential correlations with the emission of short bursts. At the onset of the event, a spectral hardening is usually observed, followed by a slow decay phase in which the X-ray spectral shape softens \cite{2011ASSP...21..247R,2014ApJS..212....6O,2015SSRv..191..315M,2017ApJS..231....8E,2025ApJ...980...99Y}. %If outbursts are indeed related to local heat deposition in the outer crust, caused by the release of internal magnetic energy or by the external magnetic field line untwisting, a very peculiar polarization pattern is expected to be associated with the changing spectral shape, allowing us to gain insight into both the magnetic field topology and the physical state of the surface during the event.
In the spectral observations carried out so far, radiation near the peak of an outburst is typically observed to be emitted from large regions, where, due to heat deposition, temperatures are expected to be relatively higher than those of the rest of the surface. This situation represents the most favorable condition to probe the effects of vacuum birefringence (see \S\ref{subsec:QED}). In fact, a sufficiently high surface temperature in the emitting region would prevent magnetic condensation of the gaseous NS atmosphere to occur, so that the intrinsic polarization of emitted radiation can reasonably be expected to be very high. Moreover, if vacuum birefringence were not effective, the observed polarization degree of radiation coming from such an extended and hot cap should be much lower than the intrinsic polarization at the emission, due to the geometrical depolarization caused by tangled magnetic field topology at the surface (see Fig.~\ref{fig:qedonoff_cap}). Hence, measuring a high degree of polarization at the onset of the outburst would provide long-awaited evidence of vacuum birefringence. Furthermore, as the emitting cap shrinks in the decay phase, an increase of the observed polarization degree should be observed if QED effects were not at work, due to geometrical depolarization being less effective for emitting region sizes smaller than $\approx40\degr$ (see again Fig.~\ref{fig:qedonoff_cap}). Thanks to the increased effective area and much shorter repointing time compared to those of IXPE, eXTP can map both spectral and polarization properties during a typical magnetar outburst, from the peak phase to the afterglow, so as to put a definitive end to the long-standing problem of the observational test of vacuum birefringence. 

\begin{figure*}
    \centering
    \includegraphics[width=1\linewidth]{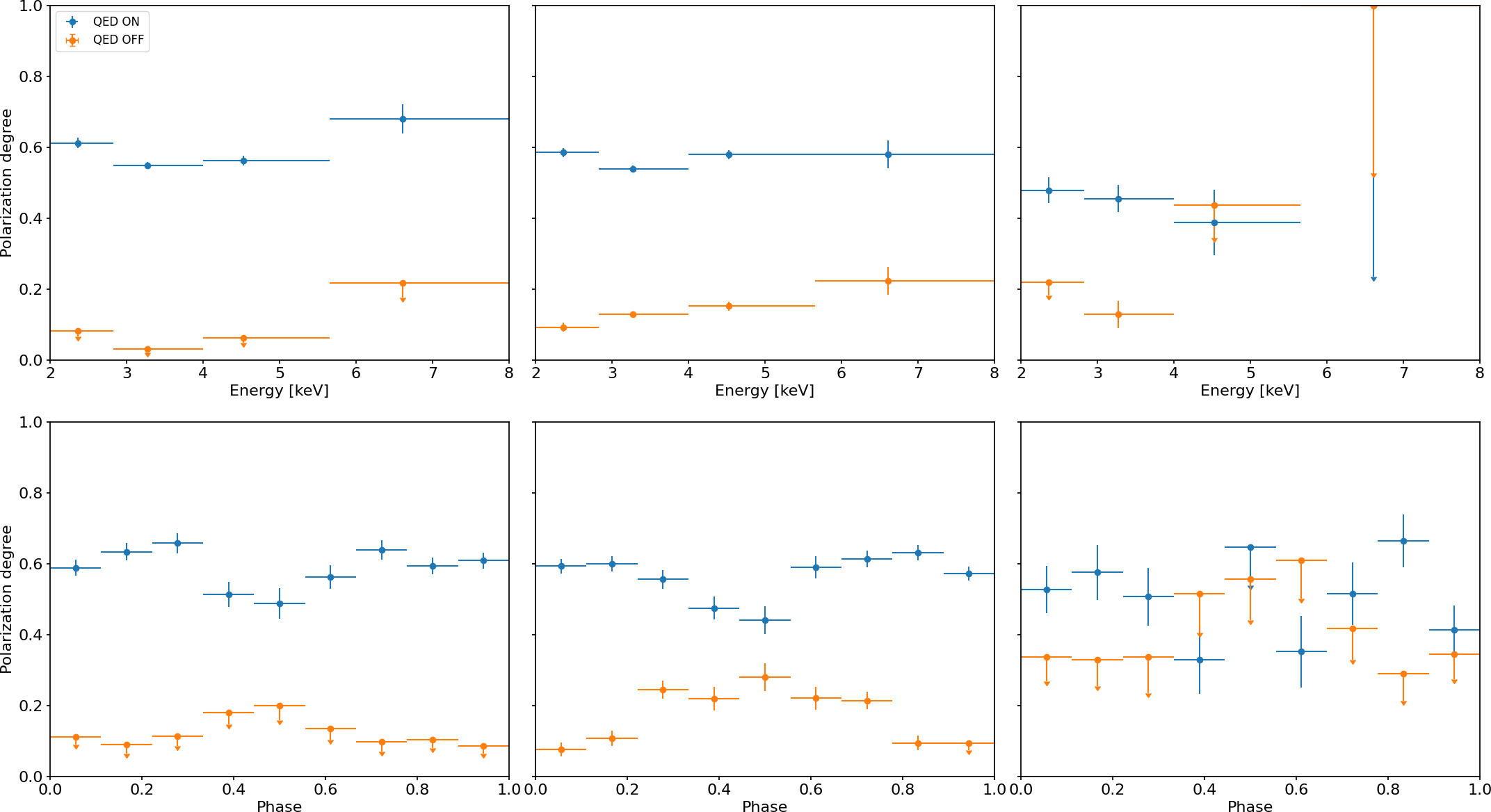}
    \caption{Simulated eXTP observations of a typical magnetar outburst (using the AXP XTE J1810$-$197 as template source) for three different $1$--$10\,\mathrm{keV}$ flux levels, $1\times10^{-10}$ (left), $7\times10^{-11}$ (center) and $1\times10^{-11}\,\mathrm{erg\,cm^{-2}\,s^{-1}}$ (right), with exposure times $t_\mathrm{exp}=100$, $200$ and $300\,\mathrm{ks}$, respectively (see text for details). The plots show the phase-integrated, energy-dependent (top row) and the energy-integrated (over the $2$--$8\,\mathrm{keV}$ range), phase-dependent (bottom row) polarization degree, modeled in both the QED-on (blue) and QED-off (orange) regimes. In the energy/phase bins where the polarization degree is below the corresponding $\mathrm{MDP}_{99}$, the $3\sigma$ upper limit is reported (marked by a downward arrow). All the simulations are performed using the {\sc ixpeobssim} suite \cite[][]{2022SoftX..1901194B} with the eXTP response functions.}
    \label{fig:outburstsimulation}
\end{figure*}

Fig.~\ref{fig:outburstsimulation} shows a simulation of an eXTP observation of a typical magnetar outburst from the template source XTE J1810$-$197 \cite[][]{2019ApJ...874L..25G,2021MNRAS.504.5244B}. The simulated observation consists of three distinct stages, for a total exposure time of $t=600\,\mathrm{ks}$. In the first stage (with exposure time $t_1=100\,\mathrm{ks}$), blackbody emission polarized in the X-mode at a $70\%$ level (reproducing emission from a magnetized gaseous layer, see \S\ref{subsec:atmocond}) is assumed to originate from a hot cap located at the magnetic pole of the star, with temperature $kT_\mathrm{h}=0.9\,\mathrm{keV}$ and semi-aperture $\theta_\mathrm{h}=20\degr$, concentric to a secondary warm cap, with temperature $kT_\mathrm{w}=0.4\,\mathrm{keV}$ and semi-aperture $\theta_\mathrm{w}=45\degr$. The $1$--$10\,\mathrm{keV}$ flux in this stage is assumed to be $10^{-10}\,\mathrm{erg\,cm}^{-2}\,\mathrm{s}^{-1}$. In the second stage (exposure time $t_2=200\,\mathrm{ks}$), the shrinking of the emitting region is simulated by maintaining the same polarization properties of emitted radiation, as well as hotter and warmer cap temperatures, but with semi-apertures $\theta_\mathrm{h}=10\degr$ and $\theta_\mathrm{w}=20\degr$, respectively. The 1--10\,keV flux is now $7\times10^{-11}\,\mathrm{erg\,cm}^{-2}\,\mathrm{s}^{-1}$. The final stage (with exposure time $t_3=300\,\mathrm{ks}$) mimics the afterglow of the outburst, with only a warm cap ($kT_\mathrm{w}=0.4\,\mathrm{keV}$) with semi-aperture $\theta_\mathrm{w}=20\degr$, $1$--$10\,\mathrm{keV}$ flux of $10^{-11}\,\mathrm{erg\,cm}^{-2}\,\mathrm{s}^{-1}$ and the same intrinsic polarization properties for emitted radiation. In all cases, the rest of the surface is assumed to emit unpolarized blackbody radiation, characterized by a temperature $kT_\mathrm{c}=0.15\,\mathrm{keV}$, to reproduce emission from a condensed surface (see again \S\ref{subsec:atmocond}). 

The phase- and energy-dependent polarization degrees, computed with (QED-on) and without (QED-off) vacuum birefringence effects, are clearly statistically distinguishable during the onset. In particular, no significant detection is expected in the QED-off case; however, even considering the $3\sigma$ upper limits, the degree of polarization of the QED-on simulation cannot be recovered in any bin. As expected, the polarization observed during the shrinking phase should increase, reaching values above $\mathrm{MDP}_{99}$, in the QED-off case, while no significant variations are visible in the QED-on case. Finally, simulations show how eXTP is able to provide a significant polarization detection even for low fluxes ($\lesssim10^{-11}\,\mathrm{erg\,cm}^{-2}\,\mathrm{s}^{-1}$) in the QED-on case, while no detectable signal should be observed in the afterglow if vacuum birefringence effects were absent. 
\ \\

\paragraph{FRB-associated X-ray burst} 
The association of FRB 200428 with a peculiar nonthermal X-ray burst from the Galactic magnetar SGR J1935+2154 \cite{Bochenek20,Andersen2020Natur} suggests that they originate from the same explosive event under extreme magnetic environments~\cite{LCK21,Mereghetti20,Tavani21,Ridnaia21}. The spectrum of the X-ray burst is characterized by the higher cut-off energy with a steeper power-law index compared to the rest of non-FRB-association bursts in the same bursting episode, indicating that the emission of FRB-association X-ray burst is more comptonized by the plasma. If the emission region of the X-ray burst is close to the magnetar surface, QED effects may play a crucial role in shaping the spectrum of those bursts. Specifically, the synchrotron radiation cooling of particles and the electron–positron pairs production via the single-photon conversion process would significantly affect the spectrum of nonthermal particles accelerated during magnetic reconnections, hence the observed radiation spectrum~\cite{Xieyu23}.
In such a reconnection-origin for X-ray burst, the polarimetric measurements would enable us to learn the configuration of twisted magnetic field lines and test the scenario. 
As the QED effect is sensitive to the magnetic strength (or distance to the NS surface equivalently), more association events detected by eXTP could help to constrain the emission radius of FRB-associated X-ray bursts.

%%%%%%%%%%%%%%%%%%%%%%%%%%%%%%%%%%%%
\begin{figure*}
\centering
\includegraphics[width=0.6\linewidth]{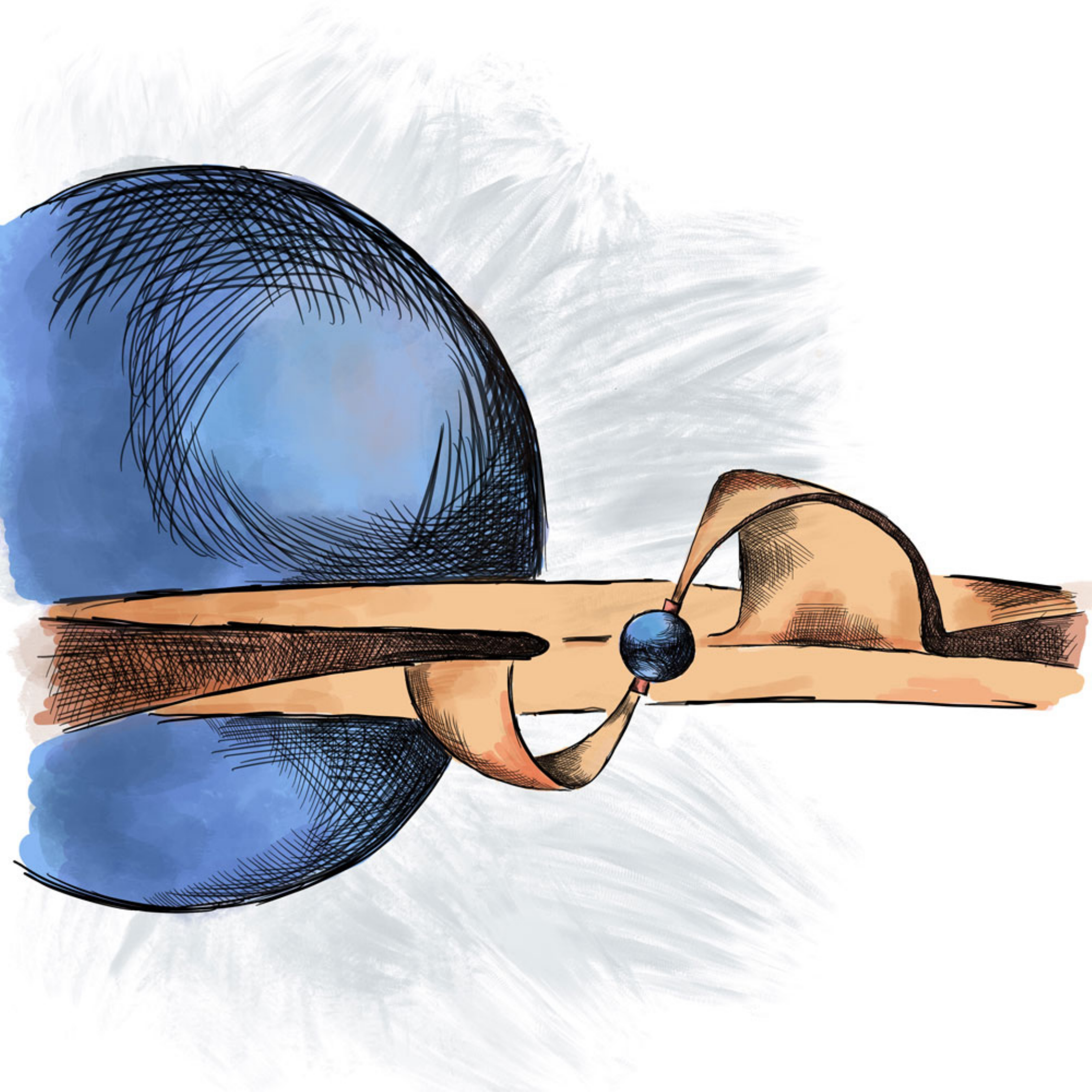}
\caption{Schematic illustration of accretion geometry in X-ray pulsars. Adopted from \citet{Tsygankov_etal_2022_cenx-3}.}
\label{fig:XRP}
\end{figure*}
%%%%%%%%%%%%%%%%%%%%%%%%%%%%%%%%%%%%

\section{Accretion under strong magnetic fields}

Although magnetars represent one of the most extreme manifestations of strong magnetic fields in NSs, accreting X-ray pulsars (XRPs) offer a crucial avenue for exploring the physics of interaction of plasma with highly magnetized NSs. 
Unlike magnetars, XRPs are fueled by accretion, a fundamental astrophysical process in which matter accumulates onto a central compact object.
In XRPs, where the NS has a strong magnetic field on the order of $B\sim10^{12-13}$,G, the accretion process is fundamentally shaped by magnetic forces, leading to a wide variety of observational phenomena \citep[for reviews, see][]{Mushtukov2022,Weng2024}. 
The interaction of accreting plasma with a strong magnetic field gives rise to a highly complex geometrical structure in these systems (see Fig.~\ref{fig:XRP}). 
At the magnetospheric radius, where the ram pressure of the accretion disk matches the magnetic pressure, infalling matter is captured by the field lines and channeled toward the magnetic poles, where its energy is converted into X-ray emission. 
The structure of the emission region is determined by the mass accretion rate, manifesting itself either as a localized hot spot or as a vertically extended accretion column for luminosities below and above $10^{37}$~erg~s$^{-1}$, respectively \citep{Basko1976,Mushtukov2015,Becker2012A&A...544A.123B}. 
The strong dependence of the scattering cross section on the magnetic field gives rise to highly anisotropic radiation patterns (commonly known as `beam patterns'), leading to intricate pulse profiles that reflect the complex interplay between the geometric and physical properties of XRPs. 
Unfortunately, this degeneracy has so far prevented the development of a reliable method for reconstructing the geometric configuration of such systems solely from their pulse profiles. 
However, the strong anisotropy in the scattering cross section induced by the magnetic field was expected to produce a high degree of polarization in the emission of XRPs \citep{Meszaros1988}, which, in turn, is sensitive to the geometry of the emission region.

The recent launch of the \textit{Imaging X-ray polarimetry Explorer} \cite[IXPE;][]{Weisskopf2022}, which was expected to resolve the aforementioned degeneracy, has instead revealed significant discrepancies between theoretical predictions and the observed polarization properties of XRPs \citep[see review by][]{Poutanen2024}. This provides strong motivation for the next-generation X-ray observatory, eXTP, which offers a significantly larger effective area for both timing and polarimetric studies.

%%%%%%%%%%%%%%%%%%%%%%%%%%%%%%%%%%%%%
\subsection{Geometry of the system}
%%%%%%%%%%%%%%%%%%%%%%%%%%%%%%%%%%%%%%

\subsubsection{Rotating vector model and NS orientation}
Due to the effect of vacuum birefringence, as discussed in Sect.~\ref{subsec:QED} for magnetars, the direction of photon polarization aligns with the local magnetic field geometry until it reaches the adiabatic radius. Even in XRPs, where the magnetic field is 1–2 orders of magnitude weaker than in magnetars, this radius extends to 20–30 NS radii, where the dipole component of the magnetic field dominates. This justifies the use of the RVM to determine the geometric configuration of the XRP and its orientation relative to the observer. 

\begin{table*}
\caption{A list of X-ray pulsars that have been observed by {\it IXPE}. The periods, orbital periods and cyclotron lines are adopted from \cite{Neumann2023,2019A&A...622A..61S}. The geometry information obtained from the RVM model is taken from \cite{Poutanen2024}.} 
	\centering
    \small
	\begin{tabular}{c|cccccc}
		\hline
		Source Name            & Period (s) & Orbital Period (d) & $E_{\rm cyc}$ (keV) &   $\zeta$ (deg)        & $\alpha$ (deg)       & IXPE  Reference\\
		\hline
		Her X-1                &  1.24      &     1.7            &  37                &  $25^{+24}_{-20}$, 90 $\pm$ 30 & $3.7^{+2.6}_{-1.9}$, $16.3^{+3.5}_{-4.1}$   & \citep{Doroshenko_etal_2022_herx-1,Garg_herx-1,Zhao_herx-1,Heyl_herX-1} \\
		Cen X-3	               &  4.8       &     2.09           &  28                 &      70.2 (fixed)      & 16.4 $\pm$ 1.3       &  \citep{Tsygankov_etal_2022_cenx-3} \\
		GRO J1008$-$57         &  93.5      &     249.5          &  78                 &       130 $\pm$ 3      &  74 $\pm$ 2          &  \citep{Tsygankov_groj1008} \\
		4U 1626$-$67	       &  7.7     &     0.02875          & 37, 61?             &          -             &      -               & \citep{Marshall_4U1627} \\
		X Persei	           &   837.67   &     250.3          &  29                 &       162 $\pm$ 12     &   90 $\pm$ 15        & \citep{Mushtukov_xpeisei} \\			     
		Vela X-1	           &   283      &     8.96           &  25, 53             &           -            &          -           & \citep{Forsblom_velaX-1,Forsblom2025} \\			     
		EXO 2030+375           &   41.31    &     46.02          & 36/63?              &      $128^{+8}_{-6}$   &    $60^{+5}_{-6}$    & \citep{Malacaria_exo2030} \\			
		GX 301$-$2             &   696.0     &     41.59          & 50                  &       135 $\pm$ 17     &   43 $\pm$ 12        &   \citep{Suleimanov_gx301} \\		
		RX J0440.9+4431        &  202.5       &     155.0          & 32?                 &        108 $\pm$ 2     &       48 $\pm$ 1     &  \citep{Victor_etal_2023,Zhao_RXJ0440} \\	
		Swift J0243.6+6124     &   9.87      &     28.3           &  146                &      $25^{+8}_{-17}$   &      $77^{+8}_{-29}$ &  \citep{SwiftJ0243_Majumder,SwiftJ0243_Poutanen} \\	
		SMC X-1                &    0.717     &     3.892          &  -                  &      $91^{+42}_{-41}$  &  $13^{+7}_{-6}$      &  \citep{SMCX-1_Forsblom} \\	
		4U 1538$-$52           &    526.42     &     3.73           &  22                 &      $77^{+50}_{-32}$  &   $30^{+16}_{-13}$   &  \citep{Loktev25}  \\		
            4U 1907+09             &    438.0    &     8.37          &   18, 36            &          -             &         -            &   \citep{ZhouML2025}           \\    
		\hline
	\end{tabular}
\label{tab_XPs}
\end{table*}

By now, a sample of about a dozen XRPs has been observed with IXPE (see Table~\ref{tab_XPs}).
%: Her X-1 \citep{Doroshenko_etal_2022_herx-1,Garg_herx-1,Zhao_herx-1,Heyl_herX-1}, Cen X-3 \citep{Tsygankov_etal_2022_cenx-3}, GRO J1008$-$57 \citep{Tsygankov_groj1008}, 4U 1626$-$67 \citep{Marshall_4U1627}, X Persei \citep{Mushtukov_xpeisei}, Vela X-1 \citep{Forsblom_velaX-1,Forsblom2025}, EXO 2030+375 \citep{Malacaria_exo2030}, GX 301$-$2 \citep{Suleimanov_gx301}, RX J0440.9+4431/LS V +44 17 \citep{Victor_etal_2023,Zhao_RXJ0440}, Swift J0243.6+6124 \citep{SwiftJ0243_Majumder,SwiftJ0243_Poutanen},  SMC X-1 \citep{SMCX-1_Forsblom}, and 4U~1538$-$52 \citep{Loktev25}. 
Although the PD may exhibit a complex evolution pattern over the pulse phase, the PA variations in most XRPs follow a much simpler trend, enabling the application of the RVM. This has enabled independent determination of the geometric configuration of at least the brightest XRPs, including the discovery of nearly aligned and nearly orthogonal rotators \citep{Poutanen2024}.

%%%%%%%%%%%%%%%%%%%%%%%%%%%%%%%%%%%%%
\begin{figure*}
\centering
\includegraphics[width=0.9\textwidth]{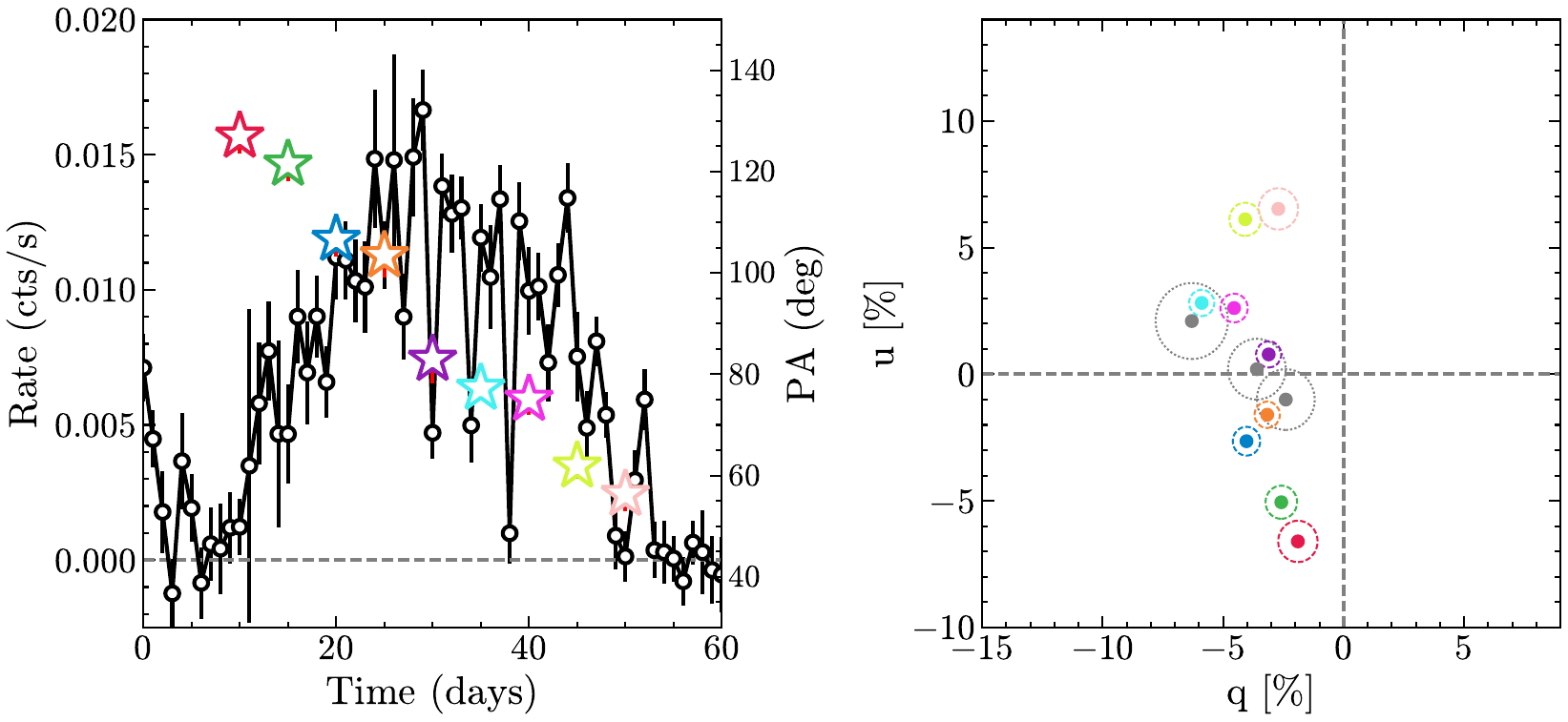}
\caption{\textit{Left panel}: Swift/BAT one-day averaged light curve of SMC X-1 during one super-orbital cycle (black circles), with the simulated PA at each segment of the cycle indicated by colored stars.  
\textit{Right panel}: the \(q\)-\(u\) plane for each super-orbital segment, with IXPE results shown in gray. The eXTP simulation uses the same exposure time as in \citet{SMCX-1_Forsblom}. Based on the results from \citet{SMCX-1_Forsblom}, a linear evolution of PA is assumed in this simulation.}
\label{fig:smcx_1_extp_simu}
\end{figure*}
%%%%%%%%%%%%%%%%%%%%%%%%%%%%%%%%%%%%%

Interestingly, it was found that the RVM parameters may change with time. In Her X-1, the geometrical parameters (particularly magnetic obliquity) changed significantly between the 'main-on' and 'short-on' states. A free precession of the NS crust is proposed to explain this modulation of geometrical parameters with a 35-day super-orbital period \citep{Heyl_herX-1}. At the same time, in two transient XRPs, RX J0440.9+4431 and Swift J0243.6+6124, the derived geometric parameters also changed unexpectedly and significantly within a very short period. An additional unpulsed but strongly polarized emission component of still unknown origin has been suggested to explain this behavior. It may arise from X-rays scattered by outflows launched from the inner regions of the accretion disk or the magnetospheric accretion flow \citep{Nitindala25}.

Although IXPE has made significant progress in determining the geometry of XRPs, it has also raised several fundamental questions, which require an instrument with far more advanced technical capabilities. 
One of the scientific goals that can be achieved with the improved sensitivity of eXTP is the expansion of the sample of XRPs with known geometry. This will enable the verification of models of NS spin evolution within the context of binary systems. Enhanced scheduling flexibility will also allow for the monitoring of bright outbursts from transient XRPs to determine the physical origin of the unpulsed emission component. Finally, significantly better count statistics will enable us to demonstrate the applicability of the RVM to such systems, rigorously examine potential discrepancies, and further refine our understanding of the polarization mechanisms in XRPs.\\

\subsubsection{Precession in accreting neutron stars}
Some accreting XRPs (e.g., Her X-1, SMC X-1, and LMC X-4) exhibit periodic modulations on timescales that significantly exceed their orbital periods, a phenomenon typically referred to as super-orbital periodicity. Among these sources, Her X-1 is one of the best known, displaying a 35-day super-orbital variability that has been attributed to the precession of a warped accretion disk. In addition to the flux modulation, its pulse profiles show systematic changes with super-orbital phase, which may be explained by the free precession of the NS. If the spin axis does not align with a symmetry axis of the NS crust, free precession is expected to occur even in the absence of a net torque over the precession period. This scenario is further supported by recent research by \citet{Heyl_herX-1} using X-ray polarimetry data obtained with IXPE. In these observations, the angle between the spin axis and the magnetic dipole differs significantly between the Main-on and Short-on states, a variation that can be explained by free precession of the NS. Moreover, a change in the position angle of the spin axis on the sky may indicate a forced precession occurring on a year-long timescale. 

%%%%%%%%%%%%%%%%%%%%%%%%%%%%%%%%%%%%%
\begin{figure*}
    \centering
     \includegraphics[width=0.9\linewidth]{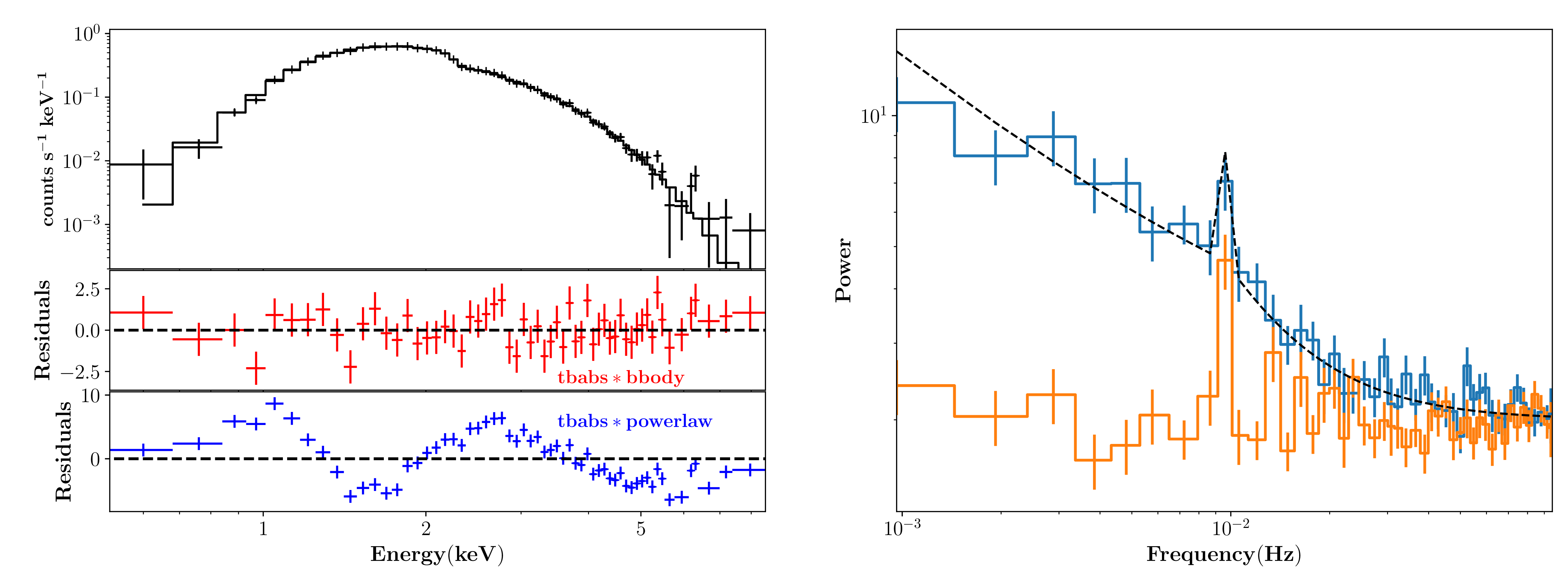}
    \caption{\textit{Left panel}: A 50\,ks eXTP/SFA simulation of the energy spectrum of an XRP in the quiescent state, assuming an absorbed blackbody model with a temperature of 0.5\,keV and 0.5$-$10\,keV flux of $\rm 10^{-12}\,erg\,s^{-1}\,cm^{-2}$ \citep{2016A&A...593A..16T}, which corresponds to a source with a luminosity of $\rm 4\times10^{33}\,erg\,s^{-1}$ at a distance of 6\,kpc. The bottom panels show residuals to the best-fit {``tbabs*bbody"} and {``tbabs*powerlaw"} models. 
    \textit{Right panel}: A 50\,ks eXTP/SFA simulation of the power spectrum for a source with a 0.5$-$10\,keV flux of $\rm 10^{-12}\,erg\,s^{-1}\,cm^{-2}$. The shapes of both the energy spectrum and the power density spectrum (dashed line) were adopted from a faint state of 1A 0535+26 \citep{Doroshenko2020b} (blue histogram). The red noise and the peak around 0.01\,Hz caused by pulsations are clearly detectable. We also present another scenario where no accretion occurs, and therefore the red noise is absent at low frequencies (orange histogram).
    }
    \label{fig:XRP_quiescence}
\end{figure*}
%%%%%%%%%%%%%%%%%%%%%%%%%%%%%%%%%%%%%

Although Her X-1 serves as a reference for studying such mechanisms, the currently available low-statistic data for fainter systems such as SMC X-1 remain insufficient to definitively confirm the presence of free precession \citep{SMCX-1_Forsblom}. In fact, \citet{SMCX-1_Forsblom} reported only a possible variation in PA over the super-orbital phase of this source. To assess eXTP's ability to trace polarimetric properties over the super-orbital cycle of faint XRPs, we performed a simulation of the expected signal from SMC X-1 using eXTP response files. As shown in Fig.~\ref{fig:smcx_1_extp_simu}, eXTP can achieve significantly improved statistics with the same exposure time as IXPE. With its enhanced capabilities, eXTP will be able to conduct polarimetric observations across different super-orbital phases for an extended sample of XRPs exhibiting this type of variability, a critical step in understanding its physical mechanisms.

%%%%%%%%%%%%%%%%%%%%%%%%%%%%%%%%%%%%%
\subsection{Low accretion rate: nature of emission and physics of the atmosphere}
%%%%%%%%%%%%%%%%%%%%%%%%%%%%%%%%%%%%%%

Observations of accreting NSs at very low luminosities require instruments with a high effective area. eXTP, with its high sensitivity in both spectral and polarimetric capabilities, will play a crucial role in verifying several fundamental concepts related to the low-luminosity states of XRPs.

\subsubsection{Nature of emission at very low luminosities}
One of the long-standing challenges in the study of XRPs is understanding the nature of their emission in the quiescent state. This question is closely related to the verification of the proposed centrifugal inhibition of accretion, which is expected to occur at very low mass accretion rates onto highly magnetized NSs \citep[the so-called `propeller effect';][]{Illarionov1975}. 
The propeller effect relies on the dipole magnetic field, which decays more slowly compared to the multipole field on a large scale. As a result, it provides a straightforward approach to confirm multipole fields by comparing the magnetic field inferred by the propeller effect and other measurements, e.g., CRSFs generated close to the NS surface \citep{Kong2022ApJ...933L...3K}.
Depending on whether this effect occurs in nature, different sources of quiescent luminosity have been considered in the literature, with thermal cooling of the NS and continued low-level accretion being the most probable explanations \citep[see, e.g.,][]{2017MNRAS.470..126T}. The key observable that can help distinguish between these two scenarios is the shape of the energy spectrum in this state: a soft, nearly thermal spectrum is expected in the absence of accretion, while a much harder spectrum is expected if accretion continues.

In the two transient XRPs, 4U~0115+63 and V~0332+53, a significant softening of the energy spectrum was observed after their possible transition to the propeller state \citep{2016A&A...593A..16T}. At the same time, the authors could not rule out softening of the continuum without its transformation to the thermal one. Indeed, some softening was later found in other XRPs (GX~304$-$1 and A~0535+26) when their luminosity dropped to $10^{34}$--$10^{35}$~erg~s$^{-1}$, independent of the propeller effect  \citep{2019MNRAS.483L.144T,2019MNRAS.487L..30T}. This softening, however, is evident only at energies below $\sim10$~keV, while at higher energies  a second hump in the spectrum is clearly observed, indicating the ongoing accretion. Therefore, our ability to distinguish between the propeller state and ongoing accretion ultimately depends on our capability to differentiate between thermal and nonthermal spectra of XRPs in quiescence.
Unfortunately, the limited sensitivity of existing X-ray telescopes has so far prevented the collection of high-quality spectra of XRPs in the propeller (quiescent) state. However, this science will be possible to perform with eXTP/SFA. 

%%%%%%%%%%%%%%%%%%%%%%%%%%%%%%%%%%%%%%
\begin{figure*}
    \centering
    \includegraphics[width=0.5\linewidth]{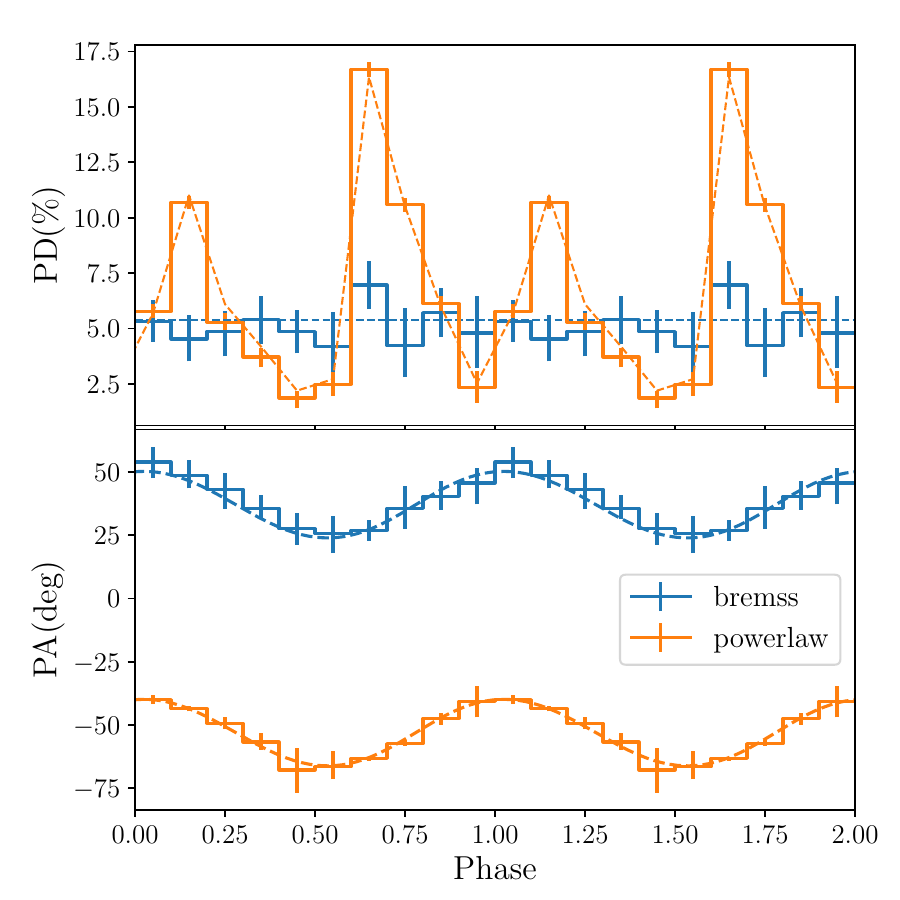}
    \caption{A simulation of phase-resolved spectro-polarimetric results for Vela X-1, where each data point corresponds to a 100\,ks exposure and the phase-averaged flux in the 2$-$8\,keV energy range is $\rm 1.8\times10^{-9}\,erg\,s^{-1}\,cm^{-2}$. 
    This simulation assumed two spectral components: a bremsstrahlung and a power law, each characterized by distinct PDs and a 90\degr\ difference in their PAs. The phase-resolved spectral shape, as well as PDs and PAs of the power-law component (dashed lines), were adopted from \citet{Forsblom2025}. The PD of the bremsstrahlung component was assumed to be phase-independent. The results demonstrate that eXTP is capable of precisely constraining the PDs and PAs of both spectral components across different rotational phases. 
    }
    \label{fig:two_components}
\end{figure*}
%%%%%%%%%%%%%%%%%%%%%%%%%%%%%%%%%%%%%%

To demonstrate this, we conducted a 50\,ks eXTP/SFA simulation of energy spectrum of an XRP in the quiescent state, assuming an absorbed blackbody model described by \texttt{tbabs*bbody} with a temperature of 0.5\,keV (see left panel in Fig.~\ref{fig:XRP_quiescence}). The flux level of the simulated source in 0.5--10\,keV was set to $10^{-12}$\,erg\,s$^{-1}$\,cm$^{-2}$, which corresponds to a source with a luminosity of $\rm 4\times10^{33}\,erg\,s^{-1}$ at a distance of 6\,kpc \citep{2016A&A...593A..16T}. We then fitted the simulated spectrum using both the input model and another model \texttt{tbabs*powerlaw}. The fittings led to distinctly different residuals, indicating that eXTP/SFA is capable of effectively distinguishing between these different spectral shapes even when the source is in a quiescence state.

Another signature of the accretion process is the presence of red noise in the source's power density spectrum (PDS), as demonstrated for several types of objects \citep{Doroshenko2020b}. Therefore, PDSs obtained by eXTP/SFA can also help distinguish between cooling NS and accreting at a low level. The instrumental capabilities for this purpose are illustrated in right panel of Fig.~\ref{fig:XRP_quiescence} where a 50\,ks simulation of the power spectrum for a source with a flux of $\rm 10^{-12}\,erg\,s^{-1}\,cm^{-2}$ in 0.5--10\,keV is shown. The shapes of both the energy spectrum and the power spectrum (dashed line) in the accreting state were adopted from a faint state of 1A~0535+26 \citep{Doroshenko2020b}. The red noise and the peak around 0.01\,Hz caused by pulsations can be evidently detected. We also presents another scenario where no accretion occurs and therefore the red noise at low frequencies is absent.\\

\subsubsection{Physics of the NS atmosphere and low PD in XRPs}

One of the most striking and unexpected results of IXPE is the discovery of PD values in XRPs much lower than any theoretical predictions \citep{Poutanen2024}. One possible explanation is incorrect assumptions about the physical conditions in the radiating regions of highly magnetized NSs made in these models.

Let us consider here XRPs with luminosities below critical one.  At low  mass accretion rates (\(\dot{M} \lesssim 10^{17}\,{\rm g\,s^{-1}}\)), the accretion flow does not have possibility to slow down significantly under radiation pressure, and accretion columns supported by radiation pressure do not form. 
Instead, the kinetic energy of the freely falling ion and electron flow is spent on heating the upper layers of the atmosphere due to Coulomb collisions and excitation of electrons to the upper Landau levels. 
As a result, the upper layers of the atmosphere are heated to temperatures of several tens of keV, approximately to the depth of the free 
path of protons.

There are no exact self-consistent models of such atmospheres \citep[see, however,][]{Suleimanov18, SokolovaLapa21, Mushtukov21}. 
However, it is possible to estimate the radiation properties of such an atmosphere using a toy model in the first approximation. 
In such a model, the magnetized pure hydrogen atmosphere can be represented as two isothermal plasma layers, an upper hot  ($kT\sim$10--100\,keV) with an exponential temperature drop to a lower cold  ($kT \sim 1$\,keV) semi-infinite layer \citep[see details in][]{Forsblom2025}. 
Assuming that these layers are in hydrostatic equilibrium in the gravitational field of a typical NS, we can calculate the number densities of protons and electrons, the opacity coefficients, and consider the radiative transport in two polarization modes in the resulting model atmosphere.

The spectrum of the emerging radiation from such an atmosphere, to the first approximation, consists of two components. 
A relatively cold semi-infinite layer emits an approximately blackbody, distorted by the influence of electron scattering.
The second component is represented by the radiation of the optically thin plasma from the upper heated layer. 
It dominates over the first component at high photon energies due to its higher temperature, and at sufficiently low photon energies, where its optical thickness becomes significant. 

Initially, the spectrum of the second component is dominated by radiation in the extraordinary mode (X-mode), since radiation in the ordinary mode (O-mode) is more significantly diluted by electron scattering.
However, the influence of virtual electron-positron pairs in the presence of a strong magnetic field (vacuum polarization) can significantly change the polarization of the emerging radiation due to mode conversion during passage through the vacuum resonance (a layer of the atmosphere in which the influence of plasma and virtual particles on photons of a given energy becomes the same, see details in  Sect.~\ref{subsec:QED}). 

The depth of the atmosphere at which the vacuum resonance occurs for a given magnetic field depends not only on the photon energy but also on the plasma density \citep[see details in][]{vAL06}. 
For low-energy photons, the vacuum resonance occurs at low plasma density in the upper optically thin layers of the atmosphere. 
In this case, a simple exchange of fluxes between the O- and X-modes takes place, and the PA of the emergent radiation rotates by 90\degr. 
Vacuum resonance for high-energy photons occurs in dense optically thick layers of the atmosphere and does not affect the polarization of the emergent radiation.

However, for photons with energies between these extreme values, a situation may arise where the vacuum resonance leads to approximate equality of the fluxes in the modes. 
For magnetic field values typical for XRPs ($\sim 10^{12}$\,G) and a heated atmospheric layer thickness of about 1~g\,cm$^{-2}$, and the vacuum resonance occurs in the transition layer between the heated and cold atmospheric layers with a very steep density gradient. 
This leads to the appearance of a fairly wide band in the spectrum (a few keVs), where the fluxes of the emergent radiation are almost equal to each other. 
This model can potentially explain the low polarization of the radiation observed with IXPE in a number of XRPs \citep{Poutanen2024}.

The presence of vacuum resonance  in the highly magnetized accretion-heated atmosphere could lead to more interesting effects.  
For example, \citet{Forsblom_velaX-1} performed an energy-resolved polarimetric analysis of Vela X-1 IXPE data and discovered a $\sim 90\degr$ swing in the PA between 2--3 and 5--8\,keV bands.  
Two potential scenarios have been proposed to account for this phenomenon: (1) the presence of two distinct spectral components, each exhibiting polarized emission with different values of the PA; and (2) the vacuum resonance in the NS atmosphere with an overheated upper layer \citep{Forsblom2025}. 
In the latter case, the parameters of the toy model with a overheated upper atmosphere can be selected in such a way that the transition from the predominance of the thermal component of the cold part of the atmosphere to the optically thin radiation of the overheated atmosphere occurs between photon energies of 2-5 keV. In this case, a rotation of the plane of polarization by 90\degr ~can be reproduced, see details in \citep{Forsblom2025}.

At the same time, the nature of the PA swings in Vela X-1 and some other XRPs is still not fully understood. The large effective area of eXTP enables us to perform a phase-resolved polarimetric and spectral study to distinguish between the two possibilities mentioned above. If the PA swing results from the vacuum resonance, the PA difference between the two spectral components would be approximately $90\degr$ at all pulse phases (see the simulation in Fig.~\ref{fig:two_components}).

%%%%%%%%%%%%%%%%%%%%%%%%%%%%%%%%%%%%%%%%%
%\subsection{Moderate accretion rate}
\subsection{Critical luminosity and geometry of emission region}
%%%%%%%%%%%%%%%%%%%%%%%%%%%%%%%%%%%%%%% 

The geometrical configuration of the emission regions at the magnetic poles of the NS in XRPs depends on the mass accretion rate. 
The so-called critical luminosity, which separates two accretion regimes, is a function of the magnetic field strength and for a typical XRP has a value around  10$^{37}$ erg~s$^{-1}$ \citep{Basko1976,Becker2012,Mushtukov2015}. 
Below this luminosity (the sub-critical regime), matter falls directly onto the NS surface, and the emission region forms a hot spot at the magnetic pole. 
When the luminosity exceeds the critical value, radiation pressure from the hot spots becomes significant, slowing the infalling matter (the super-critical regime).
As a result, a radiation-dominated shock forms some distance above the NS surface, below which lies an optically thick region, known as the accretion column \citep{Basko1976, Becker2012, Mushtukov2015}. 
The appearance of the accretion column is expected to lead to substantial changes in both the spectral shape and the pulse profile, as the radiation transfer from the polar caps is strongly influenced by the configuration of the emission region. 
The most compelling evidence for the existence of an accretion column was observed in V~0332+53 and 1A~0535+262, where the CRSF energy anti-correlates with luminosity during the bright super-critical regime. 
This has been interpreted as an increase in the height of the accretion columns as the sources become more luminous \citep{Doroshenko2017,Kong2021}.

In observations, transitions through the critical luminosity have been suggested for several transient XRPs based on changes in their pulse profiles and the hardness evolution with luminosity. This provides an indirect method to estimate the magnetic field of a NS \citep[e.g.,][]{WangPJ2022,Postnov2015,Reig2013}. 
Around the critical luminosity phase-resolved polarization properties of XRPs are expected to change dramatically \citep{Meszaros1988}. These changes can be precisely measured with eXTP/PFA during observations of a transient XRP. The accurate measurement of the critical luminosity would therefore provide a crucial test for the entire paradigm of accretion onto highly magnetized NSs.

High count statistics is  crucial for accurately determining the structure of emitting regions while accounting for all possible complexities. For instance, \citet{2013ApJ...777..115P} demonstrated that in the case of super-critical accretion, a significant fraction of the emission from the accretion column is intercepted by the NS surface and reflected to the observer, the so-called `reflection model'. As a result, the beam function and the polarimetric properties of this reflected component are strongly modified. Therefore, a comprehensive study of simultaneous timing, spectral, and polarimetric variations, enabled by the unprecedented combination of the large effective areas of eXTP's SFA and PFA detectors, will provide valuable new insights into the radiation processes near the magnetic poles of NSs.

\subsection{Extreme accretion: PULXs}

First detected by the \textit{Einstein} mission \citep{Fabbiano1989}, ultraluminous X-ray sources (ULXs) are bright, off-nuclear point sources shining at luminosities $L_\mathrm{X}\gtrsim10^{39}$\,erg\,s$^{-1}$, that is above the Eddington luminosity 
\begin{equation}
L_\mathrm{Edd} \simeq 1.3\times10^{38} \frac{M}{M_\odot} \rm \,erg\,s^{-1}
\end{equation} 
of a 10\,M$_\odot$ black hole \citep[BH; see reviews by][]{Kaaret2017,Fabrika2021,King2023,Pinto2023}. 
We know now of about 2000 ULXs \citep{Kovlakas2020_ULXcat,Walton2022catalogue,Tranin2024_cat}. 
Initially thought to be powered by intermediate-mass BHs ($M_\mathrm{BH}\simeq10^2-10^6$\,M$_\odot$) accreting at sub-Eddington rates \citep[see e.g.][]{Colbert1999}, since the early 2000s a few pieces of evidence pointed towards super-Eddington accretion in X-ray binaries (XRBs) \citep{King2001,Poutanen2007,Zampieri2009}. 
First of all, they are characterized by peculiar spectral states, different from those of Galactic XRBs \citep{Gladstone2009,Sutton2013}. 
Secondly, the energy spectra of a few ULXs have shown the presence of strong outflow winds, a key prediction of the super-Eddington scenario \citep{Pinto2016,Kosec2021}. 
The most striking evidence in support of this scenario, however, came in 2014, when coherent pulsations were detected in the X-ray flux of M82~X-2, clearly stating that this source is an accreting NS shining at luminosities $L_\mathrm{X}>10L_\mathrm{Edd}$, the first pulsating ULX \citep[PULX][]{Bachetti2014}. 

Since then, 5 more extragalactic PULXs emitting at luminosities far exceeding Eddington limit for a NS have been identified \citep{Bachetti2014,Furst2016,Israel2017,Israel2017a,Carpano2018,Sathyaprakash2019,RodriguezCastillo2020}, and a few more accreting NSs have reached similar luminosities at least once \citep[see][and references therein]{King2023}. Their spin periods are characterized by the presence of huge spin-up rates ($-10^{-10}\,\mathrm{s}\,\mathrm{s}^{-1}\lesssim\dot{P}\lesssim-10^{-11}\,\mathrm{s}\,\mathrm{s}^{-1}$) that support the super-Eddington nature of these sources \citep[see e.g.][for the case of NGC\,5907 ULX-1]{Israel2017a}. 
The super-Eddington nature is also supported by the magnitude of the orbital decay of M82 X-2 \citep{Bachetti2022}. It is interesting to note that M82 X-2 was found to show a long-term spin-down trend \citep{LiuJR24}.
PULXs are characterized by hard spectra \citep{Pintore2017,Gurpide2021a}, practically indistinguishable from the X-ray spectra of many ULXs \citep[see e.g.][]{Walton2018}, suggesting a much larger population. 
Finally, at least 2 PULXs have shown mHz QPOs with similar properties in the past years \citep{Feng2010,Imbrogno2024}.

How PULXs can reach such high luminosities is still an open question. 
A (partial) solution could lie in the magnetic field of the accreting NSs powering these systems. 
Particularly intense dipolar magnetic fields ($B\gtrsim10^{14}-10^{15}$\,G) could help overcome the Eddington limit \citep[see, e.g.,][]{DallOsso2015,Mushtukov2015MNRAS.454.2539M,Tong2015}, but at the observed spin periods ($P\simeq0.4-18$\,s) PULXs should constantly be in the propeller regime \citep{Illarionov1975}. 
In the case of NGC\,5907 ULX-1, it has been argued that a less intense dipolar component ($B_\mathrm{d}\sim10^{12}-10^{13}$\,G) accompanied by a strong multipolar component near the NS surface ($B_\mathrm{m}\sim10^{14}$\,G) could explain its extreme luminosities $L_\mathrm{X}\sim10^{41}$\,erg\,s$^{-1}\simeq10^3 L_\mathrm{Edd}$ \citep{Israel2017a}. 
A pronounced geometrical beaming of the X-ray radiation has also been invoked to explain the PULX luminosities \citep{King2009,King2017,King2020}, but although some degree of beaming must be present, an unrealistically high value of $b\sim1/100$ would be required to explain NGC\,5907 ULX-1 luminosities \citep[see again][]{Israel2017a}.

The large effective area and excellent timing resolution of the SFA, together with the polarimetric capabilities of the PFA onboard \textit{eXTP}, will significantly enhance our ability to study PULXs. The high count rate will enable sensitive searches for pulsations in faint or rapidly varying sources, facilitating the discovery of new PULXs and improving our understanding of their population properties. In addition, the superb time resolution of the SFA may aid in detecting fast-rotating PULXs—e.g. potential millisecond PULXs—given the high spin-up rates. The improved photon statistics also allow for detailed investigation of their physical characteristics, including constraints on magnetic field strengths. In particular, the detection of proton CRSFs may offer a direct probe of extreme magnetic fields in these systems \citep{2018NatAs...2..312B}. In addition, the superb effective area and good timing resolution of SFA also allow a comprehensive investigation on the timing properties evolution across the transition between a gas-dominated thin disk and a radiation-dominated thick disk as did in Swift J0243.6+6124 \citep{Doroshenko2020}.
PFA will, for the first time, enable polarization studies of extragalactic PULXs, which are currently beyond the reach of existing instruments, e.g. IXPE. To date, polarization measurements have only been achieved for Galactic ULX candidates, such as Swift J0243.6+6124 \citep{SwiftJ0243_Poutanen}, a PULX candidate with an estimated magnetic field of $\sim10^{13}$\,G \citep{Kong2022ApJ...933L...3K}. Observations of this source have revealed phase-dependent variations in the polarization angle which, after subtracting the contribution from an additional component, are well described by the RVM, which is caused by the effects of vacuum birefringence in strong magnetic fields.
With its high sensitivity and polarimetric capabilities, \textit{eXTP} will open a new window into the emission geometry and magnetic fields of PULXs, providing unique insights into their physics and placing them in the broader context of accreting XRPs.

\section{Conclusion}\label{sec:Conclusion}
The eXTP mission, scheduled for launch in 2030, will revolutionize the study of strongly magnetized compact objects by combining unprecedented timing, spectral, and polarimetric capabilities. Focused on extreme magnetic fields (up to \(10^{15}\) G), eXTP aims to directly test quantum electrodynamics (QED) effects, such as vacuum birefringence, through phase-resolved X-ray polarimetry of magnetars and accreting pulsars. By leveraging its high-sensitivity SFA and PFA, eXTP will probe magnetar surface emission mechanisms (condensed vs. atmospheric), track magnetic field evolution during outbursts, and resolve proton cyclotron absorption features to map ultra-strong surface fields. For accreting X-ray pulsars, eXTP will decode accretion column geometry and super-Eddington processes in PULXs through energy-resolved polarization and timing. W2C enables rapid detection of magnetar bursts and FRB-associated X-ray flares, while timing studies of glitches and precession will constrain neutron star interiors. With 5–10 higher sensitivity than current missions, eXTP’s multi-instrument synergy will advance our understanding of QED in extreme fields, magnetospheric dynamics, and accretion physics, bridging quantum-scale effects to astrophysical phenomena.

In addition, a potential configuration has been proposed, in which one SFA detector would be replaced by an imaging detector pnCCD (named SFA-I, togethor with the focusing mirror) \cite{2023ExA....55..603C}. The effective area and the energy resolution of SFA-I are comparable to those of the SFA unit, but its spatial resolution is much higher, significantly enhancing eXTP's imaging capability. It will provide improved measurements for weak and extended sources. In this white paper, we only focus on point sources, for which SFA-I is mainly helpful for more precisely estimating the background when sources are faint. Conversely, for bright sources, e.g., in high luminosity states or during bursts, SFA-I may suffer from the pile-up effect. However, in this case, the remaining SFA units will be still sufficient for collecting enough photons for spectral and timing studies proposed in this white paper. Overall, regardless of which configuration is ultimately implemented,  the scientific goals related to strongly magnetic objects are not expected to be affected.

%%%%%%%%%%%%%%%%%%%%%%%%%%%%%%%%%%%%%%%%%%%%%%%%%%%%%%%
%%% Acknowledgements. ??
%%%%%%%%%%%%%%%%%%%%%%%%%%%%%%%%%%%%%%%%%%%%%%%%%%%%%%%

\emph{Acknowledgements} 

This work is supported by China's  Space Origins Exploration Program and the National Key R\&D Program of China (2021YFA0718500). We acknowledge the support by the National Natural Science Foundation of China (Grant Nos. 12373051, and 12273028, 12173103 and 12261141691, and 12003009). RTav and RTur acknowledge the support by the Italian MUR (grant PRIN 2022 - 2022LWPEXW grant, “An X-ray view of compact objects in polarized light”, CUP C53D23001180006). VS thanks the German Research Foundation (DFG) grant WE 1312/59-1 for financial support. AAM thanks UKRI Stephen Hawking fellowship. RK aknowledges the support by UKRI STFC under the grant ST/W507891/1. S-NZ is supported by the National Natural Science Foundation of China (No. 12333007), the International Partnership Program of Chinese Academy of Sciences (No.113111KYSB20190020) and the Strategic Priority Research Program of the Chinese Academy of Sciences (No. XDA15020100). FX is supported by National Natural Science Foundation of China (grant No. 12373041 and No. 12422306), and Bagui Scholars Program (XF). NB is supported by STFC (grant code: ST/Y001060/1).

%%%%%%%%%%%%%%%%%%%%%%%%%%%%%%%%%%%%%%%%%%%%%%%%%%%%%%%
%%% Conflict of interest. ????????????
%%%%%%%%%%%%%%%%%%%%%%%%%%%%%%%%%%%%%%%%%%%%%%%%%%%%%%%
\textbf{Interest Conflict}: The authors declare that they have no conflict of interest.

%%%%%%%%%%%%%%%%%%%%%%%%%%%%%%%%%%%%%%%%%%%%%%%%%%%%%%%
%%% Supplements. ????????, ????
%%%%%%%%%%%%%%%%%%%%%%%%%%%%%%%%%%%%%%%%%%%%%%%%%%%%%%%
%\Supplements{}

%%%%%%%%%%%%%%%%%%%%%%%%%%%%%%%%%%%%%%%%%%%%%%%%%%%%%%%
%%% Reference section. ??????
%%% citation in the content using "some words~\cite{1,2}".
%%% ~ is needed to make the reference number is on the same line with the word before it.
%%%%%%%%%%%%%%%%%%%%%%%%%%%%%%%%%%%%%%%%%%%%%%%%%%%%%%%】
%\bibliographystyle{mnras}
\bibliographystyle{apsrev4-1} 
\bibliography{refs}

%merlin.mbs apsrev4-1.bst 2010-07-25 4.21a (PWD, AO, DPC) hacked
%Control: key (0)
%Control: author (72) initials jnrlst
%Control: editor formatted (1) identically to author
%Control: production of article title (-1) disabled
%Control: page (0) single
%Control: year (1) truncated
%Control: production of eprint (0) enabled
\begin{thebibliography}{270}%
\makeatletter
\providecommand \@ifxundefined [1]{%
 \@ifx{#1\undefined}
}%
\providecommand \@ifnum [1]{%
 \ifnum #1\expandafter \@firstoftwo
 \else \expandafter \@secondoftwo
 \fi
}%
\providecommand \@ifx [1]{%
 \ifx #1\expandafter \@firstoftwo
 \else \expandafter \@secondoftwo
 \fi
}%
\providecommand \natexlab [1]{#1}%
\providecommand \enquote  [1]{``#1''}%
\providecommand \bibnamefont  [1]{#1}%
\providecommand \bibfnamefont [1]{#1}%
\providecommand \citenamefont [1]{#1}%
\providecommand \href@noop [0]{\@secondoftwo}%
\providecommand \href [0]{\begingroup \@sanitize@url \@href}%
\providecommand \@href[1]{\@@startlink{#1}\@@href}%
\providecommand \@@href[1]{\endgroup#1\@@endlink}%
\providecommand \@sanitize@url [0]{\catcode `\\12\catcode `\$12\catcode `\&12\catcode `\#12\catcode `\^12\catcode `\_12\catcode `\%12\relax}%
\providecommand \@@startlink[1]{}%
\providecommand \@@endlink[0]{}%
\providecommand \url  [0]{\begingroup\@sanitize@url \@url }%
\providecommand \@url [1]{\endgroup\@href {#1}{\urlprefix }}%
\providecommand \urlprefix  [0]{URL }%
\providecommand \Eprint [0]{\href }%
\providecommand \doibase [0]{http://dx.doi.org/}%
\providecommand \selectlanguage [0]{\@gobble}%
\providecommand \bibinfo  [0]{\@secondoftwo}%
\providecommand \bibfield  [0]{\@secondoftwo}%
\providecommand \translation [1]{[#1]}%
\providecommand \BibitemOpen [0]{}%
\providecommand \bibitemStop [0]{}%
\providecommand \bibitemNoStop [0]{.\EOS\space}%
\providecommand \EOS [0]{\spacefactor3000\relax}%
\providecommand \BibitemShut  [1]{\csname bibitem#1\endcsname}%
\let\auto@bib@innerbib\@empty
%</preamble>
\bibitem [{\citenamefont {{Zhang}}\ \emph {et~al.}(2016)\citenamefont {{Zhang}}, \citenamefont {{Feroci}}, \citenamefont {{Santangelo}}, \citenamefont {{Dong}}, \citenamefont {{Feng}}, \citenamefont {{Lu}}, \citenamefont {{Nandra}}, \citenamefont {{Wang}}, \citenamefont {{Zhang}}, \citenamefont {{Bozzo}}, \citenamefont {{Brandt}}, \citenamefont {{De Rosa}}, \citenamefont {{Gou}}, \citenamefont {{Hernanz}}, \citenamefont {{van der Klis}}, \citenamefont {{Li}}, \citenamefont {{Liu}}, \citenamefont {{Orleanski}}, \citenamefont {{Pareschi}}, \citenamefont {{Pohl}}, \citenamefont {{Poutanen}}, \citenamefont {{Qu}}, \citenamefont {{Schanne}}, \citenamefont {{Stella}}, \citenamefont {{Uttley}}, \citenamefont {{Watts}}, \citenamefont {{Xu}}, \citenamefont {{Yu}}, \citenamefont {{in 't Zand}}, \citenamefont {{Zane}}, \citenamefont {{Alvarez}}, \citenamefont {{Amati}}, \citenamefont {{Baldini}}, \citenamefont {{Bambi}}, \citenamefont {{Basso}}, \citenamefont {{Bhattacharyya S.}}, \citenamefont {{}}, \citenamefont
  {{Belloni}}, \citenamefont {{Bellutti}}, \citenamefont {{Bianchi}}, \citenamefont {{Brez}}, \citenamefont {{Bursa}}, \citenamefont {{Burwitz}}, \citenamefont {{Budtz-J{\o}rgensen}}, \citenamefont {{Caiazzo}}, \citenamefont {{Campana}}, \citenamefont {{Cao}}, \citenamefont {{Casella}}, \citenamefont {{Chen}}, \citenamefont {{Chen}}, \citenamefont {{Chen}}, \citenamefont {{Chen}}, \citenamefont {{Chen}}, \citenamefont {{Chen}}, \citenamefont {{Civitani}}, \citenamefont {{Coti Zelati}}, \citenamefont {{Cui}}, \citenamefont {{Cui}}, \citenamefont {{Dai}}, \citenamefont {{Del Monte}}, \citenamefont {{de Martino}}, \citenamefont {{Di Cosimo}}, \citenamefont {{Diebold}}, \citenamefont {{Dovciak}}, \citenamefont {{Donnarumma}}, \citenamefont {{Doroshenko}}, \citenamefont {{Esposito}}, \citenamefont {{Evangelista}}, \citenamefont {{Favre}}, \citenamefont {{Friedrich}}, \citenamefont {{Fuschino}}, \citenamefont {{Galvez}}, \citenamefont {{Gao}}, \citenamefont {{Ge}}, \citenamefont {{Gevin}}, \citenamefont {{Goetz}},
  \citenamefont {{Han}}, \citenamefont {{Heyl}}, \citenamefont {{Horak}}, \citenamefont {{Hu}}, \citenamefont {{Huang}}, \citenamefont {{Huang}}, \citenamefont {{Hudec}}, \citenamefont {{Huppenkothen}}, \citenamefont {{Israel}}, \citenamefont {{Ingram}}, \citenamefont {{Karas}}, \citenamefont {{Karelin}}, \citenamefont {{Jenke}}, \citenamefont {{Ji}}, \citenamefont {{Korpela}}, \citenamefont {{Kunneriath}}, \citenamefont {{Labanti}}, \citenamefont {{Li}}, \citenamefont {{Li}}, \citenamefont {{Li}}, \citenamefont {{Liang}}, \citenamefont {{Limousin}}, \citenamefont {{Lin}}, \citenamefont {{Ling}}, \citenamefont {{Liu}}, \citenamefont {{Liu}}, \citenamefont {{Liu}}, \citenamefont {{Lu}}, \citenamefont {{Lund}}, \citenamefont {{Lai}}, \citenamefont {{Luo}}, \citenamefont {{Luo}}, \citenamefont {{Ma}}, \citenamefont {{Mahmoodifar}}, \citenamefont {{Marisaldi}}, \citenamefont {{Martindale}}, \citenamefont {{Meidinger}}, \citenamefont {{Men}}, \citenamefont {{Michalska}}, \citenamefont {{Mignani}}, \citenamefont
  {{Minuti}}, \citenamefont {{Motta}}, \citenamefont {{Muleri}}, \citenamefont {{Neilsen}}, \citenamefont {{Orlandini}}, \citenamefont {{Pan}}, \citenamefont {{Patruno}}, \citenamefont {{Perinati}}, \citenamefont {{Picciotto}}, \citenamefont {{Piemonte}}, \citenamefont {{Pinchera}}, \citenamefont {{Rachevski A.}}, \citenamefont {{Rapisarda}}, \citenamefont {{Rea}}, \citenamefont {{Rossi}}, \citenamefont {{Rubini}}, \citenamefont {{Sala}}, \citenamefont {{Shu}}, \citenamefont {{Sgro}}, \citenamefont {{Shen}}, \citenamefont {{Soffitta}}, \citenamefont {{Song}}, \citenamefont {{Spandre}}, \citenamefont {{Stratta}}, \citenamefont {{Strohmayer}}, \citenamefont {{Sun}}, \citenamefont {{Svoboda}}, \citenamefont {{Tagliaferri}}, \citenamefont {{Tenzer}}, \citenamefont {{Hong}}, \citenamefont {{Taverna}}, \citenamefont {{Torok}}, \citenamefont {{Turolla}}, \citenamefont {{Vacchi}}, \citenamefont {{Wang}}, \citenamefont {{Walton}}, \citenamefont {{Wang}}, \citenamefont {{Wang}}, \citenamefont {{Wang}}, \citenamefont
  {{Wang}}, \citenamefont {{Weng}}, \citenamefont {{Wilms}}, \citenamefont {{Winter}}, \citenamefont {{Wu}}, \citenamefont {{Wu}}, \citenamefont {{Xiong}}, \citenamefont {{Xu}}, \citenamefont {{Xue}}, \citenamefont {{Yan}}, \citenamefont {{Yang}}, \citenamefont {{Yang}}, \citenamefont {{Yang}}, \citenamefont {{Yuan}}, \citenamefont {{Yuan}}, \citenamefont {{Yuan}}, \citenamefont {{Zampa}}, \citenamefont {{Zampa}}, \citenamefont {{Zdziarski}}, \citenamefont {{Zhang}}, \citenamefont {{Zhang}}, \citenamefont {{Zhang}}, \citenamefont {{Zhang}}, \citenamefont {{Zhang}}, \citenamefont {{Zhang}}, \citenamefont {{Zheng}}, \citenamefont {{Zhou}},\ and\ \citenamefont {{Zhou X.~L.}}}]{2016SPIE.9905E..1QZ}%
  \BibitemOpen
  \bibfield  {author} {\bibinfo {author} {\bibfnamefont {S.~N.}\ \bibnamefont {{Zhang}}}, \bibinfo {author} {\bibfnamefont {M.}~\bibnamefont {{Feroci}}}, \bibinfo {author} {\bibfnamefont {A.}~\bibnamefont {{Santangelo}}}, \bibinfo {author} {\bibfnamefont {Y.~W.}\ \bibnamefont {{Dong}}}, \bibinfo {author} {\bibfnamefont {H.}~\bibnamefont {{Feng}}}, \bibinfo {author} {\bibfnamefont {F.~J.}\ \bibnamefont {{Lu}}}, \bibinfo {author} {\bibfnamefont {K.}~\bibnamefont {{Nandra}}}, \bibinfo {author} {\bibfnamefont {Z.~S.}\ \bibnamefont {{Wang}}}, \bibinfo {author} {\bibfnamefont {S.}~\bibnamefont {{Zhang}}}, \bibinfo {author} {\bibfnamefont {E.}~\bibnamefont {{Bozzo}}}, \bibinfo {author} {\bibfnamefont {S.}~\bibnamefont {{Brandt}}}, \bibinfo {author} {\bibfnamefont {A.}~\bibnamefont {{De Rosa}}}, \bibinfo {author} {\bibfnamefont {L.~J.}\ \bibnamefont {{Gou}}}, \bibinfo {author} {\bibfnamefont {M.}~\bibnamefont {{Hernanz}}}, \bibinfo {author} {\bibfnamefont {M.}~\bibnamefont {{van der Klis}}}, \bibinfo {author}
  {\bibfnamefont {X.~D.}\ \bibnamefont {{Li}}}, \bibinfo {author} {\bibfnamefont {Y.}~\bibnamefont {{Liu}}}, \bibinfo {author} {\bibfnamefont {P.}~\bibnamefont {{Orleanski}}}, \bibinfo {author} {\bibfnamefont {G.}~\bibnamefont {{Pareschi}}}, \bibinfo {author} {\bibfnamefont {M.}~\bibnamefont {{Pohl}}}, \bibinfo {author} {\bibfnamefont {J.}~\bibnamefont {{Poutanen}}}, \bibinfo {author} {\bibfnamefont {J.~L.}\ \bibnamefont {{Qu}}}, \bibinfo {author} {\bibfnamefont {S.}~\bibnamefont {{Schanne}}}, \bibinfo {author} {\bibfnamefont {L.}~\bibnamefont {{Stella}}}, \bibinfo {author} {\bibfnamefont {P.}~\bibnamefont {{Uttley}}}, \bibinfo {author} {\bibfnamefont {A.}~\bibnamefont {{Watts}}}, \bibinfo {author} {\bibfnamefont {R.~X.}\ \bibnamefont {{Xu}}}, \bibinfo {author} {\bibfnamefont {W.~F.}\ \bibnamefont {{Yu}}}, \bibinfo {author} {\bibfnamefont {J.~J.~M.}\ \bibnamefont {{in 't Zand}}}, \bibinfo {author} {\bibfnamefont {S.}~\bibnamefont {{Zane}}}, \bibinfo {author} {\bibfnamefont {L.}~\bibnamefont {{Alvarez}}},
  \bibinfo {author} {\bibfnamefont {L.}~\bibnamefont {{Amati}}}, \bibinfo {author} {\bibfnamefont {L.}~\bibnamefont {{Baldini}}}, \bibinfo {author} {\bibfnamefont {C.}~\bibnamefont {{Bambi}}}, \bibinfo {author} {\bibfnamefont {S.}~\bibnamefont {{Basso}}}, \bibinfo {author} {\bibnamefont {{Bhattacharyya S.}}}, \bibinfo {author} {\bibfnamefont {B.}~\bibnamefont {{}}, \bibfnamefont {R.}}, \bibinfo {author} {\bibfnamefont {T.}~\bibnamefont {{Belloni}}}, \bibinfo {author} {\bibfnamefont {P.}~\bibnamefont {{Bellutti}}}, \bibinfo {author} {\bibfnamefont {S.}~\bibnamefont {{Bianchi}}}, \bibinfo {author} {\bibfnamefont {A.}~\bibnamefont {{Brez}}}, \bibinfo {author} {\bibfnamefont {M.}~\bibnamefont {{Bursa}}}, \bibinfo {author} {\bibfnamefont {V.}~\bibnamefont {{Burwitz}}}, \bibinfo {author} {\bibfnamefont {C.}~\bibnamefont {{Budtz-J{\o}rgensen}}}, \bibinfo {author} {\bibfnamefont {I.}~\bibnamefont {{Caiazzo}}}, \bibinfo {author} {\bibfnamefont {R.}~\bibnamefont {{Campana}}}, \bibinfo {author} {\bibfnamefont
  {X.}~\bibnamefont {{Cao}}}, \bibinfo {author} {\bibfnamefont {P.}~\bibnamefont {{Casella}}}, \bibinfo {author} {\bibfnamefont {C.~Y.}\ \bibnamefont {{Chen}}}, \bibinfo {author} {\bibfnamefont {L.}~\bibnamefont {{Chen}}}, \bibinfo {author} {\bibfnamefont {T.}~\bibnamefont {{Chen}}}, \bibinfo {author} {\bibfnamefont {Y.}~\bibnamefont {{Chen}}}, \bibinfo {author} {\bibfnamefont {Y.}~\bibnamefont {{Chen}}}, \bibinfo {author} {\bibfnamefont {Y.~P.}\ \bibnamefont {{Chen}}}, \bibinfo {author} {\bibfnamefont {M.}~\bibnamefont {{Civitani}}}, \bibinfo {author} {\bibfnamefont {F.}~\bibnamefont {{Coti Zelati}}}, \bibinfo {author} {\bibfnamefont {W.}~\bibnamefont {{Cui}}}, \bibinfo {author} {\bibfnamefont {W.~W.}\ \bibnamefont {{Cui}}}, \bibinfo {author} {\bibfnamefont {Z.~G.}\ \bibnamefont {{Dai}}}, \bibinfo {author} {\bibfnamefont {E.}~\bibnamefont {{Del Monte}}}, \bibinfo {author} {\bibfnamefont {D.}~\bibnamefont {{de Martino}}}, \bibinfo {author} {\bibfnamefont {S.}~\bibnamefont {{Di Cosimo}}}, \bibinfo {author}
  {\bibfnamefont {S.}~\bibnamefont {{Diebold}}}, \bibinfo {author} {\bibfnamefont {M.}~\bibnamefont {{Dovciak}}}, \bibinfo {author} {\bibfnamefont {I.}~\bibnamefont {{Donnarumma}}}, \bibinfo {author} {\bibfnamefont {V.}~\bibnamefont {{Doroshenko}}}, \bibinfo {author} {\bibfnamefont {P.}~\bibnamefont {{Esposito}}}, \bibinfo {author} {\bibfnamefont {Y.}~\bibnamefont {{Evangelista}}}, \bibinfo {author} {\bibfnamefont {Y.}~\bibnamefont {{Favre}}}, \bibinfo {author} {\bibfnamefont {P.}~\bibnamefont {{Friedrich}}}, \bibinfo {author} {\bibfnamefont {F.}~\bibnamefont {{Fuschino}}}, \bibinfo {author} {\bibfnamefont {J.~L.}\ \bibnamefont {{Galvez}}}, \bibinfo {author} {\bibfnamefont {Z.~L.}\ \bibnamefont {{Gao}}}, \bibinfo {author} {\bibfnamefont {M.~Y.}\ \bibnamefont {{Ge}}}, \bibinfo {author} {\bibfnamefont {O.}~\bibnamefont {{Gevin}}}, \bibinfo {author} {\bibfnamefont {D.}~\bibnamefont {{Goetz}}}, \bibinfo {author} {\bibfnamefont {D.~W.}\ \bibnamefont {{Han}}}, \bibinfo {author} {\bibfnamefont {J.}~\bibnamefont
  {{Heyl}}}, \bibinfo {author} {\bibfnamefont {J.}~\bibnamefont {{Horak}}}, \bibinfo {author} {\bibfnamefont {W.}~\bibnamefont {{Hu}}}, \bibinfo {author} {\bibfnamefont {F.}~\bibnamefont {{Huang}}}, \bibinfo {author} {\bibfnamefont {Q.~S.}\ \bibnamefont {{Huang}}}, \bibinfo {author} {\bibfnamefont {R.}~\bibnamefont {{Hudec}}}, \bibinfo {author} {\bibfnamefont {D.}~\bibnamefont {{Huppenkothen}}}, \bibinfo {author} {\bibfnamefont {G.~L.}\ \bibnamefont {{Israel}}}, \bibinfo {author} {\bibfnamefont {A.}~\bibnamefont {{Ingram}}}, \bibinfo {author} {\bibfnamefont {V.}~\bibnamefont {{Karas}}}, \bibinfo {author} {\bibfnamefont {D.}~\bibnamefont {{Karelin}}}, \bibinfo {author} {\bibfnamefont {P.~A.}\ \bibnamefont {{Jenke}}}, \bibinfo {author} {\bibfnamefont {L.}~\bibnamefont {{Ji}}}, \bibinfo {author} {\bibfnamefont {S.}~\bibnamefont {{Korpela}}}, \bibinfo {author} {\bibfnamefont {D.}~\bibnamefont {{Kunneriath}}}, \bibinfo {author} {\bibfnamefont {C.}~\bibnamefont {{Labanti}}}, \bibinfo {author} {\bibfnamefont
  {G.}~\bibnamefont {{Li}}}, \bibinfo {author} {\bibfnamefont {X.}~\bibnamefont {{Li}}}, \bibinfo {author} {\bibfnamefont {Z.~S.}\ \bibnamefont {{Li}}}, \bibinfo {author} {\bibfnamefont {E.~W.}\ \bibnamefont {{Liang}}}, \bibinfo {author} {\bibfnamefont {O.}~\bibnamefont {{Limousin}}}, \bibinfo {author} {\bibfnamefont {L.}~\bibnamefont {{Lin}}}, \bibinfo {author} {\bibfnamefont {Z.~X.}\ \bibnamefont {{Ling}}}, \bibinfo {author} {\bibfnamefont {H.~B.}\ \bibnamefont {{Liu}}}, \bibinfo {author} {\bibfnamefont {H.~W.}\ \bibnamefont {{Liu}}}, \bibinfo {author} {\bibfnamefont {Z.}~\bibnamefont {{Liu}}}, \bibinfo {author} {\bibfnamefont {B.}~\bibnamefont {{Lu}}}, \bibinfo {author} {\bibfnamefont {N.}~\bibnamefont {{Lund}}}, \bibinfo {author} {\bibfnamefont {D.}~\bibnamefont {{Lai}}}, \bibinfo {author} {\bibfnamefont {B.}~\bibnamefont {{Luo}}}, \bibinfo {author} {\bibfnamefont {T.}~\bibnamefont {{Luo}}}, \bibinfo {author} {\bibfnamefont {B.}~\bibnamefont {{Ma}}}, \bibinfo {author} {\bibfnamefont {S.}~\bibnamefont
  {{Mahmoodifar}}}, \bibinfo {author} {\bibfnamefont {M.}~\bibnamefont {{Marisaldi}}}, \bibinfo {author} {\bibfnamefont {A.}~\bibnamefont {{Martindale}}}, \bibinfo {author} {\bibfnamefont {N.}~\bibnamefont {{Meidinger}}}, \bibinfo {author} {\bibfnamefont {Y.}~\bibnamefont {{Men}}}, \bibinfo {author} {\bibfnamefont {M.}~\bibnamefont {{Michalska}}}, \bibinfo {author} {\bibfnamefont {R.}~\bibnamefont {{Mignani}}}, \bibinfo {author} {\bibfnamefont {M.}~\bibnamefont {{Minuti}}}, \bibinfo {author} {\bibfnamefont {S.}~\bibnamefont {{Motta}}}, \bibinfo {author} {\bibfnamefont {F.}~\bibnamefont {{Muleri}}}, \bibinfo {author} {\bibfnamefont {J.}~\bibnamefont {{Neilsen}}}, \bibinfo {author} {\bibfnamefont {M.}~\bibnamefont {{Orlandini}}}, \bibinfo {author} {\bibfnamefont {A.~T.}\ \bibnamefont {{Pan}}}, \bibinfo {author} {\bibfnamefont {A.}~\bibnamefont {{Patruno}}}, \bibinfo {author} {\bibfnamefont {E.}~\bibnamefont {{Perinati}}}, \bibinfo {author} {\bibfnamefont {A.}~\bibnamefont {{Picciotto}}}, \bibinfo {author}
  {\bibfnamefont {C.}~\bibnamefont {{Piemonte}}}, \bibinfo {author} {\bibfnamefont {M.}~\bibnamefont {{Pinchera}}}, \bibinfo {author} {\bibnamefont {{Rachevski A.}}}, \bibinfo {author} {\bibfnamefont {M.}~\bibnamefont {{Rapisarda}}}, \bibinfo {author} {\bibfnamefont {N.}~\bibnamefont {{Rea}}}, \bibinfo {author} {\bibfnamefont {E.~M.~R.}\ \bibnamefont {{Rossi}}}, \bibinfo {author} {\bibfnamefont {A.}~\bibnamefont {{Rubini}}}, \bibinfo {author} {\bibfnamefont {G.}~\bibnamefont {{Sala}}}, \bibinfo {author} {\bibfnamefont {X.~W.}\ \bibnamefont {{Shu}}}, \bibinfo {author} {\bibfnamefont {C.}~\bibnamefont {{Sgro}}}, \bibinfo {author} {\bibfnamefont {Z.~X.}\ \bibnamefont {{Shen}}}, \bibinfo {author} {\bibfnamefont {P.}~\bibnamefont {{Soffitta}}}, \bibinfo {author} {\bibfnamefont {L.}~\bibnamefont {{Song}}}, \bibinfo {author} {\bibfnamefont {G.}~\bibnamefont {{Spandre}}}, \bibinfo {author} {\bibfnamefont {G.}~\bibnamefont {{Stratta}}}, \bibinfo {author} {\bibfnamefont {T.~E.}\ \bibnamefont {{Strohmayer}}}, \bibinfo
  {author} {\bibfnamefont {L.}~\bibnamefont {{Sun}}}, \bibinfo {author} {\bibfnamefont {J.}~\bibnamefont {{Svoboda}}}, \bibinfo {author} {\bibfnamefont {G.}~\bibnamefont {{Tagliaferri}}}, \bibinfo {author} {\bibfnamefont {G.}~\bibnamefont {{Tenzer}}}, \bibinfo {author} {\bibfnamefont {T.}~\bibnamefont {{Hong}}}, \bibinfo {author} {\bibfnamefont {R.}~\bibnamefont {{Taverna}}}, \bibinfo {author} {\bibfnamefont {G.}~\bibnamefont {{Torok}}}, \bibinfo {author} {\bibfnamefont {R.}~\bibnamefont {{Turolla}}}, \bibinfo {author} {\bibfnamefont {S.}~\bibnamefont {{Vacchi}}}, \bibinfo {author} {\bibfnamefont {J.}~\bibnamefont {{Wang}}}, \bibinfo {author} {\bibfnamefont {D.}~\bibnamefont {{Walton}}}, \bibinfo {author} {\bibfnamefont {K.}~\bibnamefont {{Wang}}}, \bibinfo {author} {\bibfnamefont {J.~F.}\ \bibnamefont {{Wang}}}, \bibinfo {author} {\bibfnamefont {R.~J.}\ \bibnamefont {{Wang}}}, \bibinfo {author} {\bibfnamefont {Y.~F.}\ \bibnamefont {{Wang}}}, \bibinfo {author} {\bibfnamefont {S.~S.}\ \bibnamefont {{Weng}}},
  \bibinfo {author} {\bibfnamefont {J.}~\bibnamefont {{Wilms}}}, \bibinfo {author} {\bibfnamefont {B.}~\bibnamefont {{Winter}}}, \bibinfo {author} {\bibfnamefont {X.}~\bibnamefont {{Wu}}}, \bibinfo {author} {\bibfnamefont {X.~F.}\ \bibnamefont {{Wu}}}, \bibinfo {author} {\bibfnamefont {S.~L.}\ \bibnamefont {{Xiong}}}, \bibinfo {author} {\bibfnamefont {Y.~P.}\ \bibnamefont {{Xu}}}, \bibinfo {author} {\bibfnamefont {Y.~Q.}\ \bibnamefont {{Xue}}}, \bibinfo {author} {\bibfnamefont {Z.}~\bibnamefont {{Yan}}}, \bibinfo {author} {\bibfnamefont {S.}~\bibnamefont {{Yang}}}, \bibinfo {author} {\bibfnamefont {X.}~\bibnamefont {{Yang}}}, \bibinfo {author} {\bibfnamefont {Y.~J.}\ \bibnamefont {{Yang}}}, \bibinfo {author} {\bibfnamefont {F.}~\bibnamefont {{Yuan}}}, \bibinfo {author} {\bibfnamefont {W.~M.}\ \bibnamefont {{Yuan}}}, \bibinfo {author} {\bibfnamefont {Y.~F.}\ \bibnamefont {{Yuan}}}, \bibinfo {author} {\bibfnamefont {G.}~\bibnamefont {{Zampa}}}, \bibinfo {author} {\bibfnamefont {N.}~\bibnamefont {{Zampa}}},
  \bibinfo {author} {\bibfnamefont {A.}~\bibnamefont {{Zdziarski}}}, \bibinfo {author} {\bibfnamefont {C.}~\bibnamefont {{Zhang}}}, \bibinfo {author} {\bibfnamefont {C.~L.}\ \bibnamefont {{Zhang}}}, \bibinfo {author} {\bibfnamefont {L.}~\bibnamefont {{Zhang}}}, \bibinfo {author} {\bibfnamefont {X.}~\bibnamefont {{Zhang}}}, \bibinfo {author} {\bibfnamefont {Z.}~\bibnamefont {{Zhang}}}, \bibinfo {author} {\bibfnamefont {W.~D.}\ \bibnamefont {{Zhang}}}, \bibinfo {author} {\bibfnamefont {S.~J.}\ \bibnamefont {{Zheng}}}, \bibinfo {author} {\bibfnamefont {P.}~\bibnamefont {{Zhou}}}, \ and\ \bibinfo {author} {\bibnamefont {{Zhou X.~L.}}},\ }in\ \href {\doibase 10.1117/12.2232034} {\emph {\bibinfo {booktitle} {Space Telescopes and Instrumentation 2016: Ultraviolet to Gamma Ray}}},\ \bibinfo {series} {\procspie}, Vol.\ \bibinfo {volume} {9905},\ \bibinfo {editor} {edited by\ \bibinfo {editor} {\bibfnamefont {J.-W.~A.}\ \bibnamefont {{den Herder}}}, \bibinfo {editor} {\bibfnamefont {T.}~\bibnamefont {{Takahashi}}}, \
  and\ \bibinfo {editor} {\bibfnamefont {M.}~\bibnamefont {{Bautz}}}}\ (\bibinfo {year} {2016})\ p.\ \bibinfo {pages} {99051Q},\ \Eprint {http://arxiv.org/abs/1607.08823} {arXiv:1607.08823 [astro-ph.IM]} \BibitemShut {NoStop}%
\bibitem [{\citenamefont {{Zhang}}\ \emph {et~al.}(2019)\citenamefont {{Zhang}}, \citenamefont {{Santangelo}}, \citenamefont {{Feroci}}, \citenamefont {{Xu}}, \citenamefont {{Lu}}, \citenamefont {{Chen}}, \citenamefont {{Feng}}, \citenamefont {{Zhang}}, \citenamefont {{Brandt}}, \citenamefont {{Hernanz}}, \citenamefont {{Baldini}}, \citenamefont {{Bozzo}}, \citenamefont {{Campana}}, \citenamefont {{De Rosa}}, \citenamefont {{Dong}}, \citenamefont {{Evangelista}}, \citenamefont {{Karas}}, \citenamefont {{Meidinger}}, \citenamefont {{Meuris}}, \citenamefont {{Nandra}}, \citenamefont {{Pan}}, \citenamefont {{Pareschi}}, \citenamefont {{Orleanski}}, \citenamefont {{Huang}}, \citenamefont {{Schanne}}, \citenamefont {{Sironi}}, \citenamefont {{Spiga}}, \citenamefont {{Svoboda}}, \citenamefont {{Tagliaferri}}, \citenamefont {{Tenzer}}, \citenamefont {{Vacchi}}, \citenamefont {{Zane}}, \citenamefont {{Walton}}, \citenamefont {{Wang}}, \citenamefont {{Winter}}, \citenamefont {{Wu}}, \citenamefont {{in't Zand}},
  \citenamefont {{Ahangarianabhari}}, \citenamefont {{Ambrosi}}, \citenamefont {{Ambrosino}}, \citenamefont {{Barbera}}, \citenamefont {{Basso}}, \citenamefont {{Bayer}}, \citenamefont {{Bellazzini}}, \citenamefont {{Bellutti}}, \citenamefont {{Bertucci}}, \citenamefont {{Bertuccio}}, \citenamefont {{Borghi}}, \citenamefont {{Cao}}, \citenamefont {{Cadoux}}, \citenamefont {{Campana}}, \citenamefont {{Ceraudo}}, \citenamefont {{Chen}}, \citenamefont {{Chen}}, \citenamefont {{Chevenez}}, \citenamefont {{Civitani}}, \citenamefont {{Cui}}, \citenamefont {{Cui}}, \citenamefont {{Dauser}}, \citenamefont {{Del Monte}}, \citenamefont {{Di Cosimo}}, \citenamefont {{Diebold}}, \citenamefont {{Doroshenko}}, \citenamefont {{Dovciak}}, \citenamefont {{Du}}, \citenamefont {{Ducci}}, \citenamefont {{Fan}}, \citenamefont {{Favre}}, \citenamefont {{Fuschino}}, \citenamefont {{G{\'a}lvez}}, \citenamefont {{Gao}}, \citenamefont {{Ge}}, \citenamefont {{Gevin}}, \citenamefont {{Grassi}}, \citenamefont {{Gu}}, \citenamefont
  {{Gu}}, \citenamefont {{Han}}, \citenamefont {{Hong}}, \citenamefont {{Hu}}, \citenamefont {{Ji}}, \citenamefont {{Jia}}, \citenamefont {{Jiang}}, \citenamefont {{Kennedy}}, \citenamefont {{Kreykenbohm}}, \citenamefont {{Kuvvetli}}, \citenamefont {{Labanti}}, \citenamefont {{Latronico}}, \citenamefont {{Li}}, \citenamefont {{Li}}, \citenamefont {{Li}}, \citenamefont {{Li}}, \citenamefont {{Li}}, \citenamefont {{Limousin}}, \citenamefont {{Liu}}, \citenamefont {{Liu}}, \citenamefont {{Lu}}, \citenamefont {{Luo}}, \citenamefont {{Macera}}, \citenamefont {{Malcovati}}, \citenamefont {{Martindale}}, \citenamefont {{Michalska}}, \citenamefont {{Meng}}, \citenamefont {{Minuti}}, \citenamefont {{Morbidini}}, \citenamefont {{Muleri}}, \citenamefont {{Paltani}}, \citenamefont {{Perinati}}, \citenamefont {{Picciotto}}, \citenamefont {{Piemonte}}, \citenamefont {{Qu}}, \citenamefont {{Rachevski}}, \citenamefont {{Rashevskaya}}, \citenamefont {{Rodriguez}}, \citenamefont {{Schanz}}, \citenamefont {{Shen}},
  \citenamefont {{Sheng}}, \citenamefont {{Song}}, \citenamefont {{Song}}, \citenamefont {{Sgro}}, \citenamefont {{Sun}}, \citenamefont {{Tan}}, \citenamefont {{Uttley}}, \citenamefont {{Wang}}, \citenamefont {{Wang}}, \citenamefont {{Wang}}, \citenamefont {{Wang}}, \citenamefont {{Wang}}, \citenamefont {{Wang}}, \citenamefont {{Watts}}, \citenamefont {{Wen}}, \citenamefont {{Wilms}}, \citenamefont {{Xiong}}, \citenamefont {{Yang}}, \citenamefont {{Yang}}, \citenamefont {{Yang}}, \citenamefont {{Yu}}, \citenamefont {{Zhang}}, \citenamefont {{Zampa}}, \citenamefont {{Zampa}}, \citenamefont {{Zdziarski}}, \citenamefont {{Zhang}}, \citenamefont {{Zhang}}, \citenamefont {{Zhang}}, \citenamefont {{Zhang}}, \citenamefont {{Zhang}}, \citenamefont {{Zhang}}, \citenamefont {{Zhang}}, \citenamefont {{Zhang}}, \citenamefont {{Zhao}}, \citenamefont {{Zheng}}, \citenamefont {{Zhou}}, \citenamefont {{Zorzi}},\ and\ \citenamefont {{Zwart}}}]{2019SCPMA..6229502Z}%
  \BibitemOpen
  \bibfield  {author} {\bibinfo {author} {\bibfnamefont {S.}~\bibnamefont {{Zhang}}}, \bibinfo {author} {\bibfnamefont {A.}~\bibnamefont {{Santangelo}}}, \bibinfo {author} {\bibfnamefont {M.}~\bibnamefont {{Feroci}}}, \bibinfo {author} {\bibfnamefont {Y.}~\bibnamefont {{Xu}}}, \bibinfo {author} {\bibfnamefont {F.}~\bibnamefont {{Lu}}}, \bibinfo {author} {\bibfnamefont {Y.}~\bibnamefont {{Chen}}}, \bibinfo {author} {\bibfnamefont {H.}~\bibnamefont {{Feng}}}, \bibinfo {author} {\bibfnamefont {S.}~\bibnamefont {{Zhang}}}, \bibinfo {author} {\bibfnamefont {S.}~\bibnamefont {{Brandt}}}, \bibinfo {author} {\bibfnamefont {M.}~\bibnamefont {{Hernanz}}}, \bibinfo {author} {\bibfnamefont {L.}~\bibnamefont {{Baldini}}}, \bibinfo {author} {\bibfnamefont {E.}~\bibnamefont {{Bozzo}}}, \bibinfo {author} {\bibfnamefont {R.}~\bibnamefont {{Campana}}}, \bibinfo {author} {\bibfnamefont {A.}~\bibnamefont {{De Rosa}}}, \bibinfo {author} {\bibfnamefont {Y.}~\bibnamefont {{Dong}}}, \bibinfo {author} {\bibfnamefont {Y.}~\bibnamefont
  {{Evangelista}}}, \bibinfo {author} {\bibfnamefont {V.}~\bibnamefont {{Karas}}}, \bibinfo {author} {\bibfnamefont {N.}~\bibnamefont {{Meidinger}}}, \bibinfo {author} {\bibfnamefont {A.}~\bibnamefont {{Meuris}}}, \bibinfo {author} {\bibfnamefont {K.}~\bibnamefont {{Nandra}}}, \bibinfo {author} {\bibfnamefont {T.}~\bibnamefont {{Pan}}}, \bibinfo {author} {\bibfnamefont {G.}~\bibnamefont {{Pareschi}}}, \bibinfo {author} {\bibfnamefont {P.}~\bibnamefont {{Orleanski}}}, \bibinfo {author} {\bibfnamefont {Q.}~\bibnamefont {{Huang}}}, \bibinfo {author} {\bibfnamefont {S.}~\bibnamefont {{Schanne}}}, \bibinfo {author} {\bibfnamefont {G.}~\bibnamefont {{Sironi}}}, \bibinfo {author} {\bibfnamefont {D.}~\bibnamefont {{Spiga}}}, \bibinfo {author} {\bibfnamefont {J.}~\bibnamefont {{Svoboda}}}, \bibinfo {author} {\bibfnamefont {G.}~\bibnamefont {{Tagliaferri}}}, \bibinfo {author} {\bibfnamefont {C.}~\bibnamefont {{Tenzer}}}, \bibinfo {author} {\bibfnamefont {A.}~\bibnamefont {{Vacchi}}}, \bibinfo {author} {\bibfnamefont
  {S.}~\bibnamefont {{Zane}}}, \bibinfo {author} {\bibfnamefont {D.}~\bibnamefont {{Walton}}}, \bibinfo {author} {\bibfnamefont {Z.}~\bibnamefont {{Wang}}}, \bibinfo {author} {\bibfnamefont {B.}~\bibnamefont {{Winter}}}, \bibinfo {author} {\bibfnamefont {X.}~\bibnamefont {{Wu}}}, \bibinfo {author} {\bibfnamefont {J.~J.~M.}\ \bibnamefont {{in't Zand}}}, \bibinfo {author} {\bibfnamefont {M.}~\bibnamefont {{Ahangarianabhari}}}, \bibinfo {author} {\bibfnamefont {G.}~\bibnamefont {{Ambrosi}}}, \bibinfo {author} {\bibfnamefont {F.}~\bibnamefont {{Ambrosino}}}, \bibinfo {author} {\bibfnamefont {M.}~\bibnamefont {{Barbera}}}, \bibinfo {author} {\bibfnamefont {S.}~\bibnamefont {{Basso}}}, \bibinfo {author} {\bibfnamefont {J.}~\bibnamefont {{Bayer}}}, \bibinfo {author} {\bibfnamefont {R.}~\bibnamefont {{Bellazzini}}}, \bibinfo {author} {\bibfnamefont {P.}~\bibnamefont {{Bellutti}}}, \bibinfo {author} {\bibfnamefont {B.}~\bibnamefont {{Bertucci}}}, \bibinfo {author} {\bibfnamefont {G.}~\bibnamefont {{Bertuccio}}},
  \bibinfo {author} {\bibfnamefont {G.}~\bibnamefont {{Borghi}}}, \bibinfo {author} {\bibfnamefont {X.}~\bibnamefont {{Cao}}}, \bibinfo {author} {\bibfnamefont {F.}~\bibnamefont {{Cadoux}}}, \bibinfo {author} {\bibfnamefont {R.}~\bibnamefont {{Campana}}}, \bibinfo {author} {\bibfnamefont {F.}~\bibnamefont {{Ceraudo}}}, \bibinfo {author} {\bibfnamefont {T.}~\bibnamefont {{Chen}}}, \bibinfo {author} {\bibfnamefont {Y.}~\bibnamefont {{Chen}}}, \bibinfo {author} {\bibfnamefont {J.}~\bibnamefont {{Chevenez}}}, \bibinfo {author} {\bibfnamefont {M.}~\bibnamefont {{Civitani}}}, \bibinfo {author} {\bibfnamefont {W.}~\bibnamefont {{Cui}}}, \bibinfo {author} {\bibfnamefont {W.}~\bibnamefont {{Cui}}}, \bibinfo {author} {\bibfnamefont {T.}~\bibnamefont {{Dauser}}}, \bibinfo {author} {\bibfnamefont {E.}~\bibnamefont {{Del Monte}}}, \bibinfo {author} {\bibfnamefont {S.}~\bibnamefont {{Di Cosimo}}}, \bibinfo {author} {\bibfnamefont {S.}~\bibnamefont {{Diebold}}}, \bibinfo {author} {\bibfnamefont {V.}~\bibnamefont
  {{Doroshenko}}}, \bibinfo {author} {\bibfnamefont {M.}~\bibnamefont {{Dovciak}}}, \bibinfo {author} {\bibfnamefont {Y.}~\bibnamefont {{Du}}}, \bibinfo {author} {\bibfnamefont {L.}~\bibnamefont {{Ducci}}}, \bibinfo {author} {\bibfnamefont {Q.}~\bibnamefont {{Fan}}}, \bibinfo {author} {\bibfnamefont {Y.}~\bibnamefont {{Favre}}}, \bibinfo {author} {\bibfnamefont {F.}~\bibnamefont {{Fuschino}}}, \bibinfo {author} {\bibfnamefont {J.~L.}\ \bibnamefont {{G{\'a}lvez}}}, \bibinfo {author} {\bibfnamefont {M.}~\bibnamefont {{Gao}}}, \bibinfo {author} {\bibfnamefont {M.}~\bibnamefont {{Ge}}}, \bibinfo {author} {\bibfnamefont {O.}~\bibnamefont {{Gevin}}}, \bibinfo {author} {\bibfnamefont {M.}~\bibnamefont {{Grassi}}}, \bibinfo {author} {\bibfnamefont {Q.}~\bibnamefont {{Gu}}}, \bibinfo {author} {\bibfnamefont {Y.}~\bibnamefont {{Gu}}}, \bibinfo {author} {\bibfnamefont {D.}~\bibnamefont {{Han}}}, \bibinfo {author} {\bibfnamefont {B.}~\bibnamefont {{Hong}}}, \bibinfo {author} {\bibfnamefont {W.}~\bibnamefont {{Hu}}},
  \bibinfo {author} {\bibfnamefont {L.}~\bibnamefont {{Ji}}}, \bibinfo {author} {\bibfnamefont {S.}~\bibnamefont {{Jia}}}, \bibinfo {author} {\bibfnamefont {W.}~\bibnamefont {{Jiang}}}, \bibinfo {author} {\bibfnamefont {T.}~\bibnamefont {{Kennedy}}}, \bibinfo {author} {\bibfnamefont {I.}~\bibnamefont {{Kreykenbohm}}}, \bibinfo {author} {\bibfnamefont {I.}~\bibnamefont {{Kuvvetli}}}, \bibinfo {author} {\bibfnamefont {C.}~\bibnamefont {{Labanti}}}, \bibinfo {author} {\bibfnamefont {L.}~\bibnamefont {{Latronico}}}, \bibinfo {author} {\bibfnamefont {G.}~\bibnamefont {{Li}}}, \bibinfo {author} {\bibfnamefont {M.}~\bibnamefont {{Li}}}, \bibinfo {author} {\bibfnamefont {X.}~\bibnamefont {{Li}}}, \bibinfo {author} {\bibfnamefont {W.}~\bibnamefont {{Li}}}, \bibinfo {author} {\bibfnamefont {Z.}~\bibnamefont {{Li}}}, \bibinfo {author} {\bibfnamefont {O.}~\bibnamefont {{Limousin}}}, \bibinfo {author} {\bibfnamefont {H.}~\bibnamefont {{Liu}}}, \bibinfo {author} {\bibfnamefont {X.}~\bibnamefont {{Liu}}}, \bibinfo {author}
  {\bibfnamefont {B.}~\bibnamefont {{Lu}}}, \bibinfo {author} {\bibfnamefont {T.}~\bibnamefont {{Luo}}}, \bibinfo {author} {\bibfnamefont {D.}~\bibnamefont {{Macera}}}, \bibinfo {author} {\bibfnamefont {P.}~\bibnamefont {{Malcovati}}}, \bibinfo {author} {\bibfnamefont {A.}~\bibnamefont {{Martindale}}}, \bibinfo {author} {\bibfnamefont {M.}~\bibnamefont {{Michalska}}}, \bibinfo {author} {\bibfnamefont {B.}~\bibnamefont {{Meng}}}, \bibinfo {author} {\bibfnamefont {M.}~\bibnamefont {{Minuti}}}, \bibinfo {author} {\bibfnamefont {A.}~\bibnamefont {{Morbidini}}}, \bibinfo {author} {\bibfnamefont {F.}~\bibnamefont {{Muleri}}}, \bibinfo {author} {\bibfnamefont {S.}~\bibnamefont {{Paltani}}}, \bibinfo {author} {\bibfnamefont {E.}~\bibnamefont {{Perinati}}}, \bibinfo {author} {\bibfnamefont {A.}~\bibnamefont {{Picciotto}}}, \bibinfo {author} {\bibfnamefont {C.}~\bibnamefont {{Piemonte}}}, \bibinfo {author} {\bibfnamefont {J.}~\bibnamefont {{Qu}}}, \bibinfo {author} {\bibfnamefont {A.}~\bibnamefont {{Rachevski}}},
  \bibinfo {author} {\bibfnamefont {I.}~\bibnamefont {{Rashevskaya}}}, \bibinfo {author} {\bibfnamefont {J.}~\bibnamefont {{Rodriguez}}}, \bibinfo {author} {\bibfnamefont {T.}~\bibnamefont {{Schanz}}}, \bibinfo {author} {\bibfnamefont {Z.}~\bibnamefont {{Shen}}}, \bibinfo {author} {\bibfnamefont {L.}~\bibnamefont {{Sheng}}}, \bibinfo {author} {\bibfnamefont {J.}~\bibnamefont {{Song}}}, \bibinfo {author} {\bibfnamefont {L.}~\bibnamefont {{Song}}}, \bibinfo {author} {\bibfnamefont {C.}~\bibnamefont {{Sgro}}}, \bibinfo {author} {\bibfnamefont {L.}~\bibnamefont {{Sun}}}, \bibinfo {author} {\bibfnamefont {Y.}~\bibnamefont {{Tan}}}, \bibinfo {author} {\bibfnamefont {P.}~\bibnamefont {{Uttley}}}, \bibinfo {author} {\bibfnamefont {B.}~\bibnamefont {{Wang}}}, \bibinfo {author} {\bibfnamefont {D.}~\bibnamefont {{Wang}}}, \bibinfo {author} {\bibfnamefont {G.}~\bibnamefont {{Wang}}}, \bibinfo {author} {\bibfnamefont {J.}~\bibnamefont {{Wang}}}, \bibinfo {author} {\bibfnamefont {L.}~\bibnamefont {{Wang}}}, \bibinfo
  {author} {\bibfnamefont {Y.}~\bibnamefont {{Wang}}}, \bibinfo {author} {\bibfnamefont {A.~L.}\ \bibnamefont {{Watts}}}, \bibinfo {author} {\bibfnamefont {X.}~\bibnamefont {{Wen}}}, \bibinfo {author} {\bibfnamefont {J.}~\bibnamefont {{Wilms}}}, \bibinfo {author} {\bibfnamefont {S.}~\bibnamefont {{Xiong}}}, \bibinfo {author} {\bibfnamefont {J.}~\bibnamefont {{Yang}}}, \bibinfo {author} {\bibfnamefont {S.}~\bibnamefont {{Yang}}}, \bibinfo {author} {\bibfnamefont {Y.}~\bibnamefont {{Yang}}}, \bibinfo {author} {\bibfnamefont {N.}~\bibnamefont {{Yu}}}, \bibinfo {author} {\bibfnamefont {W.}~\bibnamefont {{Zhang}}}, \bibinfo {author} {\bibfnamefont {G.}~\bibnamefont {{Zampa}}}, \bibinfo {author} {\bibfnamefont {N.}~\bibnamefont {{Zampa}}}, \bibinfo {author} {\bibfnamefont {A.~A.}\ \bibnamefont {{Zdziarski}}}, \bibinfo {author} {\bibfnamefont {A.}~\bibnamefont {{Zhang}}}, \bibinfo {author} {\bibfnamefont {C.}~\bibnamefont {{Zhang}}}, \bibinfo {author} {\bibfnamefont {F.}~\bibnamefont {{Zhang}}}, \bibinfo {author}
  {\bibfnamefont {L.}~\bibnamefont {{Zhang}}}, \bibinfo {author} {\bibfnamefont {T.}~\bibnamefont {{Zhang}}}, \bibinfo {author} {\bibfnamefont {Y.}~\bibnamefont {{Zhang}}}, \bibinfo {author} {\bibfnamefont {X.}~\bibnamefont {{Zhang}}}, \bibinfo {author} {\bibfnamefont {Z.}~\bibnamefont {{Zhang}}}, \bibinfo {author} {\bibfnamefont {B.}~\bibnamefont {{Zhao}}}, \bibinfo {author} {\bibfnamefont {S.}~\bibnamefont {{Zheng}}}, \bibinfo {author} {\bibfnamefont {Y.}~\bibnamefont {{Zhou}}}, \bibinfo {author} {\bibfnamefont {N.}~\bibnamefont {{Zorzi}}}, \ and\ \bibinfo {author} {\bibfnamefont {J.~F.}\ \bibnamefont {{Zwart}}},\ }\href {\doibase 10.1007/s11433-018-9309-2} {\bibfield  {journal} {\bibinfo  {journal} {Science China Physics, Mechanics, and Astronomy}\ }\textbf {\bibinfo {volume} {62}},\ \bibinfo {eid} {29502} (\bibinfo {year} {2019})},\ \Eprint {http://arxiv.org/abs/1812.04020} {arXiv:1812.04020 [astro-ph.IM]} \BibitemShut {NoStop}%
\bibitem [{\citenamefont {{Zhang}}\ and\ \citenamefont {{Santangelo}}(2025)}]{WP1}%
  \BibitemOpen
  \bibfield  {author} {\bibinfo {author} {\bibfnamefont {S.~N.}\ \bibnamefont {{Zhang}}}\ and\ \bibinfo {author} {\bibfnamefont {A.~e.~a.}\ \bibnamefont {{Santangelo}}},\ }\href@noop {} {\bibfield  {journal} {\bibinfo  {journal} {SCPMA}\ }\textbf {\bibinfo {volume} {xx}},\ \bibinfo {pages} {xx} (\bibinfo {year} {2025})},\ \Eprint {http://arxiv.org/abs/xx} {arXiv:xx [xx]} \BibitemShut {NoStop}%
\bibitem [{\citenamefont {{Costa}}\ \emph {et~al.}(2001)\citenamefont {{Costa}}, \citenamefont {{Soffitta}}, \citenamefont {{Bellazzini}}, \citenamefont {{Brez}}, \citenamefont {{Lumb}},\ and\ \citenamefont {{Spandre}}}]{2001Natur.411..662C}%
  \BibitemOpen
  \bibfield  {author} {\bibinfo {author} {\bibfnamefont {E.}~\bibnamefont {{Costa}}}, \bibinfo {author} {\bibfnamefont {P.}~\bibnamefont {{Soffitta}}}, \bibinfo {author} {\bibfnamefont {R.}~\bibnamefont {{Bellazzini}}}, \bibinfo {author} {\bibfnamefont {A.}~\bibnamefont {{Brez}}}, \bibinfo {author} {\bibfnamefont {N.}~\bibnamefont {{Lumb}}}, \ and\ \bibinfo {author} {\bibfnamefont {G.}~\bibnamefont {{Spandre}}},\ }\href {\doibase 10.1038/35079508} {\bibfield  {journal} {\bibinfo  {journal} {\nat}\ }\textbf {\bibinfo {volume} {411}},\ \bibinfo {pages} {662} (\bibinfo {year} {2001})},\ \Eprint {http://arxiv.org/abs/astro-ph/0107486} {arXiv:astro-ph/0107486 [astro-ph]} \BibitemShut {NoStop}%
\bibitem [{\citenamefont {{Bellazzini}}\ \emph {et~al.}(2003)\citenamefont {{Bellazzini}}, \citenamefont {{Baldini}}, \citenamefont {{Brez}}, \citenamefont {{Costa}}, \citenamefont {{Latronico}}, \citenamefont {{Omodei}}, \citenamefont {{Soffitta}},\ and\ \citenamefont {{Spandre}}}]{2003NIMPA.510..176B}%
  \BibitemOpen
  \bibfield  {author} {\bibinfo {author} {\bibfnamefont {R.}~\bibnamefont {{Bellazzini}}}, \bibinfo {author} {\bibfnamefont {L.}~\bibnamefont {{Baldini}}}, \bibinfo {author} {\bibfnamefont {A.}~\bibnamefont {{Brez}}}, \bibinfo {author} {\bibfnamefont {E.}~\bibnamefont {{Costa}}}, \bibinfo {author} {\bibfnamefont {L.}~\bibnamefont {{Latronico}}}, \bibinfo {author} {\bibfnamefont {N.}~\bibnamefont {{Omodei}}}, \bibinfo {author} {\bibfnamefont {P.}~\bibnamefont {{Soffitta}}}, \ and\ \bibinfo {author} {\bibfnamefont {G.}~\bibnamefont {{Spandre}}},\ }\href {\doibase 10.1016/S0168-9002(03)01695-4} {\bibfield  {journal} {\bibinfo  {journal} {Nuclear Instruments and Methods in Physics Research A}\ }\textbf {\bibinfo {volume} {510}},\ \bibinfo {pages} {176} (\bibinfo {year} {2003})}\BibitemShut {NoStop}%
\bibitem [{\citenamefont {{Bellazzini}}\ \emph {et~al.}(2007)\citenamefont {{Bellazzini}}, \citenamefont {{Spandre}}, \citenamefont {{Minuti}}, \citenamefont {{Baldini}}, \citenamefont {{Brez}}, \citenamefont {{Latronico}}, \citenamefont {{Omodei}}, \citenamefont {{Razzano}}, \citenamefont {{Massai}}, \citenamefont {{Pesce-Rollins}}, \citenamefont {{Sgr{\'o}}}, \citenamefont {{Costa}}, \citenamefont {{Soffitta}}, \citenamefont {{Sipila}},\ and\ \citenamefont {{Lempinen}}}]{2007NIMPA.579..853B}%
  \BibitemOpen
  \bibfield  {author} {\bibinfo {author} {\bibfnamefont {R.}~\bibnamefont {{Bellazzini}}}, \bibinfo {author} {\bibfnamefont {G.}~\bibnamefont {{Spandre}}}, \bibinfo {author} {\bibfnamefont {M.}~\bibnamefont {{Minuti}}}, \bibinfo {author} {\bibfnamefont {L.}~\bibnamefont {{Baldini}}}, \bibinfo {author} {\bibfnamefont {A.}~\bibnamefont {{Brez}}}, \bibinfo {author} {\bibfnamefont {L.}~\bibnamefont {{Latronico}}}, \bibinfo {author} {\bibfnamefont {N.}~\bibnamefont {{Omodei}}}, \bibinfo {author} {\bibfnamefont {M.}~\bibnamefont {{Razzano}}}, \bibinfo {author} {\bibfnamefont {M.~M.}\ \bibnamefont {{Massai}}}, \bibinfo {author} {\bibfnamefont {M.}~\bibnamefont {{Pesce-Rollins}}}, \bibinfo {author} {\bibfnamefont {C.}~\bibnamefont {{Sgr{\'o}}}}, \bibinfo {author} {\bibfnamefont {E.}~\bibnamefont {{Costa}}}, \bibinfo {author} {\bibfnamefont {P.}~\bibnamefont {{Soffitta}}}, \bibinfo {author} {\bibfnamefont {H.}~\bibnamefont {{Sipila}}}, \ and\ \bibinfo {author} {\bibfnamefont {E.}~\bibnamefont {{Lempinen}}},\ }\href
  {\doibase 10.1016/j.nima.2007.05.304} {\bibfield  {journal} {\bibinfo  {journal} {Nuclear Instruments and Methods in Physics Research A}\ }\textbf {\bibinfo {volume} {579}},\ \bibinfo {pages} {853} (\bibinfo {year} {2007})},\ \Eprint {http://arxiv.org/abs/astro-ph/0611512} {arXiv:astro-ph/0611512 [astro-ph]} \BibitemShut {NoStop}%
\bibitem [{\citenamefont {{Bellazzini}}\ \emph {et~al.}(2013)\citenamefont {{Bellazzini}}, \citenamefont {{Brez}}, \citenamefont {{Costa}}, \citenamefont {{Minuti}}, \citenamefont {{Muleri}}, \citenamefont {{Pinchera}}, \citenamefont {{Rubini}}, \citenamefont {{Soffitta}},\ and\ \citenamefont {{Spandre}}}]{2013NIMPA.720..173B}%
  \BibitemOpen
  \bibfield  {author} {\bibinfo {author} {\bibfnamefont {R.}~\bibnamefont {{Bellazzini}}}, \bibinfo {author} {\bibfnamefont {A.}~\bibnamefont {{Brez}}}, \bibinfo {author} {\bibfnamefont {E.}~\bibnamefont {{Costa}}}, \bibinfo {author} {\bibfnamefont {M.}~\bibnamefont {{Minuti}}}, \bibinfo {author} {\bibfnamefont {F.}~\bibnamefont {{Muleri}}}, \bibinfo {author} {\bibfnamefont {M.}~\bibnamefont {{Pinchera}}}, \bibinfo {author} {\bibfnamefont {A.}~\bibnamefont {{Rubini}}}, \bibinfo {author} {\bibfnamefont {P.}~\bibnamefont {{Soffitta}}}, \ and\ \bibinfo {author} {\bibfnamefont {G.}~\bibnamefont {{Spandre}}},\ }\href {\doibase 10.1016/j.nima.2012.12.006} {\bibfield  {journal} {\bibinfo  {journal} {Nuclear Instruments and Methods in Physics Research A}\ }\textbf {\bibinfo {volume} {720}},\ \bibinfo {pages} {173} (\bibinfo {year} {2013})}\BibitemShut {NoStop}%
\bibitem [{\citenamefont {{Santangelo}}\ \emph {et~al.}(2019)\citenamefont {{Santangelo}}, \citenamefont {{Zane}}, \citenamefont {{Feng}}, \citenamefont {{Xu}}, \citenamefont {{Doroshenko}}, \citenamefont {{Bozzo}}, \citenamefont {{Caiazzo}}, \citenamefont {{Coti Zelati}}, \citenamefont {{Esposito}}, \citenamefont {{Gonz{\'a}lez-Caniulef}}, \citenamefont {{Heyl}}, \citenamefont {{Huppenkothen}}, \citenamefont {{Israel}}, \citenamefont {{Li}}, \citenamefont {{Lin}}, \citenamefont {{Mignani}}, \citenamefont {{Rea}}, \citenamefont {{Orlandini}}, \citenamefont {{Taverna}}, \citenamefont {{Tong}}, \citenamefont {{Turolla}}, \citenamefont {{Baglio}}, \citenamefont {{Bernardini}}, \citenamefont {{Bucciantini}}, \citenamefont {{Feroci}}, \citenamefont {{F{\"u}rst}}, \citenamefont {{G{\"o}{\u{g}}{\"u}{\c{s}}}}, \citenamefont {{G{\"u}ng{\"o}r}}, \citenamefont {{Ji}}, \citenamefont {{Lu}}, \citenamefont {{Manousakis}}, \citenamefont {{Mereghetti}}, \citenamefont {{Mikusincova}}, \citenamefont {{Paul}}, \citenamefont
  {{Prescod-Weinstein}}, \citenamefont {{Younes}}, \citenamefont {{Tiengo}}, \citenamefont {{Xu}}, \citenamefont {{Watts}}, \citenamefont {{Zhang}},\ and\ \citenamefont {{Zhan}}}]{2019SCPMA..6229505S}%
  \BibitemOpen
  \bibfield  {author} {\bibinfo {author} {\bibfnamefont {A.}~\bibnamefont {{Santangelo}}}, \bibinfo {author} {\bibfnamefont {S.}~\bibnamefont {{Zane}}}, \bibinfo {author} {\bibfnamefont {H.}~\bibnamefont {{Feng}}}, \bibinfo {author} {\bibfnamefont {R.}~\bibnamefont {{Xu}}}, \bibinfo {author} {\bibfnamefont {V.}~\bibnamefont {{Doroshenko}}}, \bibinfo {author} {\bibfnamefont {E.}~\bibnamefont {{Bozzo}}}, \bibinfo {author} {\bibfnamefont {I.}~\bibnamefont {{Caiazzo}}}, \bibinfo {author} {\bibfnamefont {F.}~\bibnamefont {{Coti Zelati}}}, \bibinfo {author} {\bibfnamefont {P.}~\bibnamefont {{Esposito}}}, \bibinfo {author} {\bibfnamefont {D.}~\bibnamefont {{Gonz{\'a}lez-Caniulef}}}, \bibinfo {author} {\bibfnamefont {J.}~\bibnamefont {{Heyl}}}, \bibinfo {author} {\bibfnamefont {D.}~\bibnamefont {{Huppenkothen}}}, \bibinfo {author} {\bibfnamefont {G.}~\bibnamefont {{Israel}}}, \bibinfo {author} {\bibfnamefont {Z.}~\bibnamefont {{Li}}}, \bibinfo {author} {\bibfnamefont {L.}~\bibnamefont {{Lin}}}, \bibinfo {author}
  {\bibfnamefont {R.}~\bibnamefont {{Mignani}}}, \bibinfo {author} {\bibfnamefont {N.}~\bibnamefont {{Rea}}}, \bibinfo {author} {\bibfnamefont {M.}~\bibnamefont {{Orlandini}}}, \bibinfo {author} {\bibfnamefont {R.}~\bibnamefont {{Taverna}}}, \bibinfo {author} {\bibfnamefont {H.}~\bibnamefont {{Tong}}}, \bibinfo {author} {\bibfnamefont {R.}~\bibnamefont {{Turolla}}}, \bibinfo {author} {\bibfnamefont {C.}~\bibnamefont {{Baglio}}}, \bibinfo {author} {\bibfnamefont {F.}~\bibnamefont {{Bernardini}}}, \bibinfo {author} {\bibfnamefont {N.}~\bibnamefont {{Bucciantini}}}, \bibinfo {author} {\bibfnamefont {M.}~\bibnamefont {{Feroci}}}, \bibinfo {author} {\bibfnamefont {F.}~\bibnamefont {{F{\"u}rst}}}, \bibinfo {author} {\bibfnamefont {E.}~\bibnamefont {{G{\"o}{\u{g}}{\"u}{\c{s}}}}}, \bibinfo {author} {\bibfnamefont {C.}~\bibnamefont {{G{\"u}ng{\"o}r}}}, \bibinfo {author} {\bibfnamefont {L.}~\bibnamefont {{Ji}}}, \bibinfo {author} {\bibfnamefont {F.}~\bibnamefont {{Lu}}}, \bibinfo {author} {\bibfnamefont
  {A.}~\bibnamefont {{Manousakis}}}, \bibinfo {author} {\bibfnamefont {S.}~\bibnamefont {{Mereghetti}}}, \bibinfo {author} {\bibfnamefont {R.}~\bibnamefont {{Mikusincova}}}, \bibinfo {author} {\bibfnamefont {B.}~\bibnamefont {{Paul}}}, \bibinfo {author} {\bibfnamefont {C.}~\bibnamefont {{Prescod-Weinstein}}}, \bibinfo {author} {\bibfnamefont {G.}~\bibnamefont {{Younes}}}, \bibinfo {author} {\bibfnamefont {A.}~\bibnamefont {{Tiengo}}}, \bibinfo {author} {\bibfnamefont {Y.}~\bibnamefont {{Xu}}}, \bibinfo {author} {\bibfnamefont {A.}~\bibnamefont {{Watts}}}, \bibinfo {author} {\bibfnamefont {S.}~\bibnamefont {{Zhang}}}, \ and\ \bibinfo {author} {\bibfnamefont {S.-N.}\ \bibnamefont {{Zhan}}},\ }\href {\doibase 10.1007/s11433-018-9234-3} {\bibfield  {journal} {\bibinfo  {journal} {Science China Physics, Mechanics, and Astronomy}\ }\textbf {\bibinfo {volume} {62}},\ \bibinfo {eid} {29505} (\bibinfo {year} {2019})},\ \Eprint {http://arxiv.org/abs/1812.04460} {arXiv:1812.04460 [astro-ph.HE]} \BibitemShut {NoStop}%
\bibitem [{\citenamefont {{Feng}}\ \emph {et~al.}(2020)\citenamefont {{Feng}}, \citenamefont {{Li}}, \citenamefont {{Long}}, \citenamefont {{Bellazzini}}, \citenamefont {{Costa}}, \citenamefont {{Wu}}, \citenamefont {{Huang}}, \citenamefont {{Jiang}}, \citenamefont {{Minuti}}, \citenamefont {{Wang}}, \citenamefont {{Xu}}, \citenamefont {{Yang}}, \citenamefont {{Baldini}}, \citenamefont {{Citraro}}, \citenamefont {{Nasimi}}, \citenamefont {{Soffitta}}, \citenamefont {{Muleri}}, \citenamefont {{Jung}}, \citenamefont {{Yu}}, \citenamefont {{Jin}}, \citenamefont {{Zeng}}, \citenamefont {{An}}, \citenamefont {{Brez}}, \citenamefont {{Latronico}}, \citenamefont {{Sgro}}, \citenamefont {{Spandre}},\ and\ \citenamefont {{Pinchera}}}]{2020NatAs...4..511F}%
  \BibitemOpen
  \bibfield  {author} {\bibinfo {author} {\bibfnamefont {H.}~\bibnamefont {{Feng}}}, \bibinfo {author} {\bibfnamefont {H.}~\bibnamefont {{Li}}}, \bibinfo {author} {\bibfnamefont {X.}~\bibnamefont {{Long}}}, \bibinfo {author} {\bibfnamefont {R.}~\bibnamefont {{Bellazzini}}}, \bibinfo {author} {\bibfnamefont {E.}~\bibnamefont {{Costa}}}, \bibinfo {author} {\bibfnamefont {Q.}~\bibnamefont {{Wu}}}, \bibinfo {author} {\bibfnamefont {J.}~\bibnamefont {{Huang}}}, \bibinfo {author} {\bibfnamefont {W.}~\bibnamefont {{Jiang}}}, \bibinfo {author} {\bibfnamefont {M.}~\bibnamefont {{Minuti}}}, \bibinfo {author} {\bibfnamefont {W.}~\bibnamefont {{Wang}}}, \bibinfo {author} {\bibfnamefont {R.}~\bibnamefont {{Xu}}}, \bibinfo {author} {\bibfnamefont {D.}~\bibnamefont {{Yang}}}, \bibinfo {author} {\bibfnamefont {L.}~\bibnamefont {{Baldini}}}, \bibinfo {author} {\bibfnamefont {S.}~\bibnamefont {{Citraro}}}, \bibinfo {author} {\bibfnamefont {H.}~\bibnamefont {{Nasimi}}}, \bibinfo {author} {\bibfnamefont {P.}~\bibnamefont
  {{Soffitta}}}, \bibinfo {author} {\bibfnamefont {F.}~\bibnamefont {{Muleri}}}, \bibinfo {author} {\bibfnamefont {A.}~\bibnamefont {{Jung}}}, \bibinfo {author} {\bibfnamefont {J.}~\bibnamefont {{Yu}}}, \bibinfo {author} {\bibfnamefont {G.}~\bibnamefont {{Jin}}}, \bibinfo {author} {\bibfnamefont {M.}~\bibnamefont {{Zeng}}}, \bibinfo {author} {\bibfnamefont {P.}~\bibnamefont {{An}}}, \bibinfo {author} {\bibfnamefont {A.}~\bibnamefont {{Brez}}}, \bibinfo {author} {\bibfnamefont {L.}~\bibnamefont {{Latronico}}}, \bibinfo {author} {\bibfnamefont {C.}~\bibnamefont {{Sgro}}}, \bibinfo {author} {\bibfnamefont {G.}~\bibnamefont {{Spandre}}}, \ and\ \bibinfo {author} {\bibfnamefont {M.}~\bibnamefont {{Pinchera}}},\ }\href {\doibase 10.1038/s41550-020-1088-1} {\bibfield  {journal} {\bibinfo  {journal} {Nature Astronomy}\ }\textbf {\bibinfo {volume} {4}},\ \bibinfo {pages} {511} (\bibinfo {year} {2020})},\ \Eprint {http://arxiv.org/abs/2011.05487} {arXiv:2011.05487 [astro-ph.HE]} \BibitemShut {NoStop}%
\bibitem [{\citenamefont {{Zane}}\ \emph {et~al.}(2023)\citenamefont {{Zane}}, \citenamefont {{Taverna}}, \citenamefont {{Gonz{\'a}lez-Caniulef}}, \citenamefont {{Muleri}}, \citenamefont {{Turolla}}, \citenamefont {{Heyl}}, \citenamefont {{Uchiyama}}, \citenamefont {{Ng}}, \citenamefont {{Tamagawa}}, \citenamefont {{Caiazzo}}, \citenamefont {{Di Lalla}}, \citenamefont {{Marshall}}, \citenamefont {{Bachetti}}, \citenamefont {{La Monaca}}, \citenamefont {{Gau}}, \citenamefont {{Di Marco}}, \citenamefont {{Baldini}}, \citenamefont {{Negro}}, \citenamefont {{Omodei}}, \citenamefont {{Rankin}}, \citenamefont {{Matt}}, \citenamefont {{Pavlov}}, \citenamefont {{Kitaguchi}}, \citenamefont {{Krawczynski}}, \citenamefont {{Kislat}}, \citenamefont {{Kelly}}, \citenamefont {{Agudo}}, \citenamefont {{Antonelli}}, \citenamefont {{Baumgartner}}, \citenamefont {{Bellazzini}}, \citenamefont {{Bianchi}}, \citenamefont {{Bongiorno}}, \citenamefont {{Bonino}}, \citenamefont {{Brez}}, \citenamefont {{Bucciantini}}, \citenamefont
  {{Capitanio}}, \citenamefont {{Castellano}}, \citenamefont {{Cavazzuti}}, \citenamefont {{Chen}}, \citenamefont {{Ciprini}}, \citenamefont {{Costa}}, \citenamefont {{De Rosa}}, \citenamefont {{Del Monte}}, \citenamefont {{Di Gesu}}, \citenamefont {{Donnarumma}}, \citenamefont {{Doroshenko}}, \citenamefont {{Dov{\v{c}}iak}}, \citenamefont {{Ehlert}}, \citenamefont {{Enoto}}, \citenamefont {{Evangelista}}, \citenamefont {{Fabiani}}, \citenamefont {{Ferrazzoli}}, \citenamefont {{Garcia}}, \citenamefont {{Gunji}}, \citenamefont {{Hayashida}}, \citenamefont {{Iwakiri}}, \citenamefont {{Jorstad}}, \citenamefont {{Kaaret}}, \citenamefont {{Karas}}, \citenamefont {{Kolodziejczak}}, \citenamefont {{Latronico}}, \citenamefont {{Liodakis}}, \citenamefont {{Maldera}}, \citenamefont {{Manfreda}}, \citenamefont {{Marin}}, \citenamefont {{Marinucci}}, \citenamefont {{Marscher}}, \citenamefont {{Massaro}}, \citenamefont {{Mitsuishi}}, \citenamefont {{Mizuno}}, \citenamefont {{Ng}}, \citenamefont {{O'Dell}}, \citenamefont
  {{Oppedisano}}, \citenamefont {{Papitto}}, \citenamefont {{Peirson}}, \citenamefont {{Perri}}, \citenamefont {{Pesce-Rollins}}, \citenamefont {{Petrucci}}, \citenamefont {{Pilia}}, \citenamefont {{Possenti}}, \citenamefont {{Poutanen}}, \citenamefont {{Puccetti}}, \citenamefont {{Ramsey}}, \citenamefont {{Ratheesh}}, \citenamefont {{Roberts}}, \citenamefont {{Romani}}, \citenamefont {{Sgr{\'o}}}, \citenamefont {{Slane}}, \citenamefont {{Soffitta}}, \citenamefont {{Spandre}}, \citenamefont {{Swartz}}, \citenamefont {{Tavecchio}}, \citenamefont {{Tawara}}, \citenamefont {{Tennant}}, \citenamefont {{Thomas}}, \citenamefont {{Tombesi}}, \citenamefont {{Trois}}, \citenamefont {{Tsygankov}}, \citenamefont {{Vink}}, \citenamefont {{Weisskopf}}, \citenamefont {{Wu}},\ and\ \citenamefont {{Xie}}}]{2023ApJ...944L..27Z}%
  \BibitemOpen
  \bibfield  {author} {\bibinfo {author} {\bibfnamefont {S.}~\bibnamefont {{Zane}}}, \bibinfo {author} {\bibfnamefont {R.}~\bibnamefont {{Taverna}}}, \bibinfo {author} {\bibfnamefont {D.}~\bibnamefont {{Gonz{\'a}lez-Caniulef}}}, \bibinfo {author} {\bibfnamefont {F.}~\bibnamefont {{Muleri}}}, \bibinfo {author} {\bibfnamefont {R.}~\bibnamefont {{Turolla}}}, \bibinfo {author} {\bibfnamefont {J.}~\bibnamefont {{Heyl}}}, \bibinfo {author} {\bibfnamefont {K.}~\bibnamefont {{Uchiyama}}}, \bibinfo {author} {\bibfnamefont {M.}~\bibnamefont {{Ng}}}, \bibinfo {author} {\bibfnamefont {T.}~\bibnamefont {{Tamagawa}}}, \bibinfo {author} {\bibfnamefont {I.}~\bibnamefont {{Caiazzo}}}, \bibinfo {author} {\bibfnamefont {N.}~\bibnamefont {{Di Lalla}}}, \bibinfo {author} {\bibfnamefont {H.~L.}\ \bibnamefont {{Marshall}}}, \bibinfo {author} {\bibfnamefont {M.}~\bibnamefont {{Bachetti}}}, \bibinfo {author} {\bibfnamefont {F.}~\bibnamefont {{La Monaca}}}, \bibinfo {author} {\bibfnamefont {E.}~\bibnamefont {{Gau}}}, \bibinfo {author}
  {\bibfnamefont {A.}~\bibnamefont {{Di Marco}}}, \bibinfo {author} {\bibfnamefont {L.}~\bibnamefont {{Baldini}}}, \bibinfo {author} {\bibfnamefont {M.}~\bibnamefont {{Negro}}}, \bibinfo {author} {\bibfnamefont {N.}~\bibnamefont {{Omodei}}}, \bibinfo {author} {\bibfnamefont {J.}~\bibnamefont {{Rankin}}}, \bibinfo {author} {\bibfnamefont {G.}~\bibnamefont {{Matt}}}, \bibinfo {author} {\bibfnamefont {G.~G.}\ \bibnamefont {{Pavlov}}}, \bibinfo {author} {\bibfnamefont {T.}~\bibnamefont {{Kitaguchi}}}, \bibinfo {author} {\bibfnamefont {H.}~\bibnamefont {{Krawczynski}}}, \bibinfo {author} {\bibfnamefont {F.}~\bibnamefont {{Kislat}}}, \bibinfo {author} {\bibfnamefont {R.}~\bibnamefont {{Kelly}}}, \bibinfo {author} {\bibfnamefont {I.}~\bibnamefont {{Agudo}}}, \bibinfo {author} {\bibfnamefont {L.~A.}\ \bibnamefont {{Antonelli}}}, \bibinfo {author} {\bibfnamefont {W.~H.}\ \bibnamefont {{Baumgartner}}}, \bibinfo {author} {\bibfnamefont {R.}~\bibnamefont {{Bellazzini}}}, \bibinfo {author} {\bibfnamefont {S.}~\bibnamefont
  {{Bianchi}}}, \bibinfo {author} {\bibfnamefont {S.~D.}\ \bibnamefont {{Bongiorno}}}, \bibinfo {author} {\bibfnamefont {R.}~\bibnamefont {{Bonino}}}, \bibinfo {author} {\bibfnamefont {A.}~\bibnamefont {{Brez}}}, \bibinfo {author} {\bibfnamefont {N.}~\bibnamefont {{Bucciantini}}}, \bibinfo {author} {\bibfnamefont {F.}~\bibnamefont {{Capitanio}}}, \bibinfo {author} {\bibfnamefont {S.}~\bibnamefont {{Castellano}}}, \bibinfo {author} {\bibfnamefont {E.}~\bibnamefont {{Cavazzuti}}}, \bibinfo {author} {\bibfnamefont {C.-T.}\ \bibnamefont {{Chen}}}, \bibinfo {author} {\bibfnamefont {S.}~\bibnamefont {{Ciprini}}}, \bibinfo {author} {\bibfnamefont {E.}~\bibnamefont {{Costa}}}, \bibinfo {author} {\bibfnamefont {A.}~\bibnamefont {{De Rosa}}}, \bibinfo {author} {\bibfnamefont {E.}~\bibnamefont {{Del Monte}}}, \bibinfo {author} {\bibfnamefont {L.}~\bibnamefont {{Di Gesu}}}, \bibinfo {author} {\bibfnamefont {I.}~\bibnamefont {{Donnarumma}}}, \bibinfo {author} {\bibfnamefont {V.}~\bibnamefont {{Doroshenko}}}, \bibinfo
  {author} {\bibfnamefont {M.}~\bibnamefont {{Dov{\v{c}}iak}}}, \bibinfo {author} {\bibfnamefont {S.~R.}\ \bibnamefont {{Ehlert}}}, \bibinfo {author} {\bibfnamefont {T.}~\bibnamefont {{Enoto}}}, \bibinfo {author} {\bibfnamefont {Y.}~\bibnamefont {{Evangelista}}}, \bibinfo {author} {\bibfnamefont {S.}~\bibnamefont {{Fabiani}}}, \bibinfo {author} {\bibfnamefont {R.}~\bibnamefont {{Ferrazzoli}}}, \bibinfo {author} {\bibfnamefont {J.~A.}\ \bibnamefont {{Garcia}}}, \bibinfo {author} {\bibfnamefont {S.}~\bibnamefont {{Gunji}}}, \bibinfo {author} {\bibfnamefont {K.}~\bibnamefont {{Hayashida}}}, \bibinfo {author} {\bibfnamefont {W.}~\bibnamefont {{Iwakiri}}}, \bibinfo {author} {\bibfnamefont {S.~G.}\ \bibnamefont {{Jorstad}}}, \bibinfo {author} {\bibfnamefont {P.}~\bibnamefont {{Kaaret}}}, \bibinfo {author} {\bibfnamefont {V.}~\bibnamefont {{Karas}}}, \bibinfo {author} {\bibfnamefont {J.~J.}\ \bibnamefont {{Kolodziejczak}}}, \bibinfo {author} {\bibfnamefont {L.}~\bibnamefont {{Latronico}}}, \bibinfo {author}
  {\bibfnamefont {I.}~\bibnamefont {{Liodakis}}}, \bibinfo {author} {\bibfnamefont {S.}~\bibnamefont {{Maldera}}}, \bibinfo {author} {\bibfnamefont {A.}~\bibnamefont {{Manfreda}}}, \bibinfo {author} {\bibfnamefont {F.}~\bibnamefont {{Marin}}}, \bibinfo {author} {\bibfnamefont {A.}~\bibnamefont {{Marinucci}}}, \bibinfo {author} {\bibfnamefont {A.~P.}\ \bibnamefont {{Marscher}}}, \bibinfo {author} {\bibfnamefont {F.}~\bibnamefont {{Massaro}}}, \bibinfo {author} {\bibfnamefont {I.}~\bibnamefont {{Mitsuishi}}}, \bibinfo {author} {\bibfnamefont {T.}~\bibnamefont {{Mizuno}}}, \bibinfo {author} {\bibfnamefont {C.~Y.}\ \bibnamefont {{Ng}}}, \bibinfo {author} {\bibfnamefont {S.~L.}\ \bibnamefont {{O'Dell}}}, \bibinfo {author} {\bibfnamefont {C.}~\bibnamefont {{Oppedisano}}}, \bibinfo {author} {\bibfnamefont {A.}~\bibnamefont {{Papitto}}}, \bibinfo {author} {\bibfnamefont {A.~L.}\ \bibnamefont {{Peirson}}}, \bibinfo {author} {\bibfnamefont {M.}~\bibnamefont {{Perri}}}, \bibinfo {author} {\bibfnamefont {M.}~\bibnamefont
  {{Pesce-Rollins}}}, \bibinfo {author} {\bibfnamefont {P.-O.}\ \bibnamefont {{Petrucci}}}, \bibinfo {author} {\bibfnamefont {M.}~\bibnamefont {{Pilia}}}, \bibinfo {author} {\bibfnamefont {A.}~\bibnamefont {{Possenti}}}, \bibinfo {author} {\bibfnamefont {J.}~\bibnamefont {{Poutanen}}}, \bibinfo {author} {\bibfnamefont {S.}~\bibnamefont {{Puccetti}}}, \bibinfo {author} {\bibfnamefont {B.~D.}\ \bibnamefont {{Ramsey}}}, \bibinfo {author} {\bibfnamefont {A.}~\bibnamefont {{Ratheesh}}}, \bibinfo {author} {\bibfnamefont {O.~J.}\ \bibnamefont {{Roberts}}}, \bibinfo {author} {\bibfnamefont {R.~W.}\ \bibnamefont {{Romani}}}, \bibinfo {author} {\bibfnamefont {C.}~\bibnamefont {{Sgr{\'o}}}}, \bibinfo {author} {\bibfnamefont {P.}~\bibnamefont {{Slane}}}, \bibinfo {author} {\bibfnamefont {P.}~\bibnamefont {{Soffitta}}}, \bibinfo {author} {\bibfnamefont {G.}~\bibnamefont {{Spandre}}}, \bibinfo {author} {\bibfnamefont {D.~A.}\ \bibnamefont {{Swartz}}}, \bibinfo {author} {\bibfnamefont {F.}~\bibnamefont {{Tavecchio}}},
  \bibinfo {author} {\bibfnamefont {Y.}~\bibnamefont {{Tawara}}}, \bibinfo {author} {\bibfnamefont {A.~F.}\ \bibnamefont {{Tennant}}}, \bibinfo {author} {\bibfnamefont {N.~E.}\ \bibnamefont {{Thomas}}}, \bibinfo {author} {\bibfnamefont {F.}~\bibnamefont {{Tombesi}}}, \bibinfo {author} {\bibfnamefont {A.}~\bibnamefont {{Trois}}}, \bibinfo {author} {\bibfnamefont {S.~S.}\ \bibnamefont {{Tsygankov}}}, \bibinfo {author} {\bibfnamefont {J.}~\bibnamefont {{Vink}}}, \bibinfo {author} {\bibfnamefont {M.~C.}\ \bibnamefont {{Weisskopf}}}, \bibinfo {author} {\bibfnamefont {K.}~\bibnamefont {{Wu}}}, \ and\ \bibinfo {author} {\bibfnamefont {F.}~\bibnamefont {{Xie}}},\ }\href {\doibase 10.3847/2041-8213/acb703} {\bibfield  {journal} {\bibinfo  {journal} {\apjl}\ }\textbf {\bibinfo {volume} {944}},\ \bibinfo {eid} {L27} (\bibinfo {year} {2023})},\ \Eprint {http://arxiv.org/abs/2301.12919} {arXiv:2301.12919 [astro-ph.HE]} \BibitemShut {NoStop}%
\bibitem [{\citenamefont {{Turolla}}\ \emph {et~al.}(2023)\citenamefont {{Turolla}}, \citenamefont {{Taverna}}, \citenamefont {{Israel}}, \citenamefont {{Muleri}}, \citenamefont {{Zane}}, \citenamefont {{Bachetti}}, \citenamefont {{Heyl}}, \citenamefont {{Di Marco}}, \citenamefont {{Gau}}, \citenamefont {{Krawczynski}}, \citenamefont {{Ng}}, \citenamefont {{Possenti}}, \citenamefont {{Poutanen}}, \citenamefont {{Baldini}}, \citenamefont {{Matt}}, \citenamefont {{Negro}}, \citenamefont {{Agudo}}, \citenamefont {{Antonelli}}, \citenamefont {{Baumgartner}}, \citenamefont {{Bellazzini}}, \citenamefont {{Bianchi}}, \citenamefont {{Bongiorno}}, \citenamefont {{Bonino}}, \citenamefont {{Brez}}, \citenamefont {{Bucciantini}}, \citenamefont {{Capitanio}}, \citenamefont {{Castellano}}, \citenamefont {{Cavazzuti}}, \citenamefont {{Chen}}, \citenamefont {{Ciprini}}, \citenamefont {{Costa}}, \citenamefont {{De Rosa}}, \citenamefont {{Del Monte}}, \citenamefont {{Di Gesu}}, \citenamefont {{Di Lalla}}, \citenamefont
  {{Donnarumma}}, \citenamefont {{Doroshenko}}, \citenamefont {{Dov{\v{c}}iak}}, \citenamefont {{Ehlert}}, \citenamefont {{Enoto}}, \citenamefont {{Evangelista}}, \citenamefont {{Fabiani}}, \citenamefont {{Ferrazzoli}}, \citenamefont {{Garcia}}, \citenamefont {{Gunji}}, \citenamefont {{Hayashida}}, \citenamefont {{Iwakiri}}, \citenamefont {{Jorstad}}, \citenamefont {{Kaaret}}, \citenamefont {{Karas}}, \citenamefont {{Kislat}}, \citenamefont {{Kitaguchi}}, \citenamefont {{Kolodziejczak}}, \citenamefont {{La Monaca}}, \citenamefont {{Latronico}}, \citenamefont {{Liodakis}}, \citenamefont {{Maldera}}, \citenamefont {{Manfreda}}, \citenamefont {{Marin}}, \citenamefont {{Marinucci}}, \citenamefont {{Marscher}}, \citenamefont {{Marshall}}, \citenamefont {{Massaro}}, \citenamefont {{Mitsuishi}}, \citenamefont {{Mizuno}}, \citenamefont {{Ng}}, \citenamefont {{O'Dell}}, \citenamefont {{Omodei}}, \citenamefont {{Oppedisano}}, \citenamefont {{Papitto}}, \citenamefont {{Pavlov}}, \citenamefont {{Peirson}}, \citenamefont
  {{Perri}}, \citenamefont {{Pesce-Rollins}}, \citenamefont {{Petrucci}}, \citenamefont {{Pilia}}, \citenamefont {{Puccetti}}, \citenamefont {{Ramsey}}, \citenamefont {{Rankin}}, \citenamefont {{Ratheesh}}, \citenamefont {{Roberts}}, \citenamefont {{Romani}}, \citenamefont {{Sgr{\'o}}}, \citenamefont {{Slane}}, \citenamefont {{Soffitta}}, \citenamefont {{Spandre}}, \citenamefont {{Swartz}}, \citenamefont {{Tamagawa}}, \citenamefont {{Tavecchio}}, \citenamefont {{Tawara}}, \citenamefont {{Tennant}}, \citenamefont {{Thomas}}, \citenamefont {{Tombesi}}, \citenamefont {{Trois}}, \citenamefont {{Tsygankov}}, \citenamefont {{Vink}}, \citenamefont {{Weisskopf}}, \citenamefont {{Wu}},\ and\ \citenamefont {{Xie}}}]{2023ApJ...954...88T}%
  \BibitemOpen
  \bibfield  {author} {\bibinfo {author} {\bibfnamefont {R.}~\bibnamefont {{Turolla}}}, \bibinfo {author} {\bibfnamefont {R.}~\bibnamefont {{Taverna}}}, \bibinfo {author} {\bibfnamefont {G.~L.}\ \bibnamefont {{Israel}}}, \bibinfo {author} {\bibfnamefont {F.}~\bibnamefont {{Muleri}}}, \bibinfo {author} {\bibfnamefont {S.}~\bibnamefont {{Zane}}}, \bibinfo {author} {\bibfnamefont {M.}~\bibnamefont {{Bachetti}}}, \bibinfo {author} {\bibfnamefont {J.}~\bibnamefont {{Heyl}}}, \bibinfo {author} {\bibfnamefont {A.}~\bibnamefont {{Di Marco}}}, \bibinfo {author} {\bibfnamefont {E.}~\bibnamefont {{Gau}}}, \bibinfo {author} {\bibfnamefont {H.}~\bibnamefont {{Krawczynski}}}, \bibinfo {author} {\bibfnamefont {M.}~\bibnamefont {{Ng}}}, \bibinfo {author} {\bibfnamefont {A.}~\bibnamefont {{Possenti}}}, \bibinfo {author} {\bibfnamefont {J.}~\bibnamefont {{Poutanen}}}, \bibinfo {author} {\bibfnamefont {L.}~\bibnamefont {{Baldini}}}, \bibinfo {author} {\bibfnamefont {G.}~\bibnamefont {{Matt}}}, \bibinfo {author} {\bibfnamefont
  {M.}~\bibnamefont {{Negro}}}, \bibinfo {author} {\bibfnamefont {I.}~\bibnamefont {{Agudo}}}, \bibinfo {author} {\bibfnamefont {L.~A.}\ \bibnamefont {{Antonelli}}}, \bibinfo {author} {\bibfnamefont {W.~H.}\ \bibnamefont {{Baumgartner}}}, \bibinfo {author} {\bibfnamefont {R.}~\bibnamefont {{Bellazzini}}}, \bibinfo {author} {\bibfnamefont {S.}~\bibnamefont {{Bianchi}}}, \bibinfo {author} {\bibfnamefont {S.~D.}\ \bibnamefont {{Bongiorno}}}, \bibinfo {author} {\bibfnamefont {R.}~\bibnamefont {{Bonino}}}, \bibinfo {author} {\bibfnamefont {A.}~\bibnamefont {{Brez}}}, \bibinfo {author} {\bibfnamefont {N.}~\bibnamefont {{Bucciantini}}}, \bibinfo {author} {\bibfnamefont {F.}~\bibnamefont {{Capitanio}}}, \bibinfo {author} {\bibfnamefont {S.}~\bibnamefont {{Castellano}}}, \bibinfo {author} {\bibfnamefont {E.}~\bibnamefont {{Cavazzuti}}}, \bibinfo {author} {\bibfnamefont {C.-T.}\ \bibnamefont {{Chen}}}, \bibinfo {author} {\bibfnamefont {S.}~\bibnamefont {{Ciprini}}}, \bibinfo {author} {\bibfnamefont {E.}~\bibnamefont
  {{Costa}}}, \bibinfo {author} {\bibfnamefont {A.}~\bibnamefont {{De Rosa}}}, \bibinfo {author} {\bibfnamefont {E.}~\bibnamefont {{Del Monte}}}, \bibinfo {author} {\bibfnamefont {L.}~\bibnamefont {{Di Gesu}}}, \bibinfo {author} {\bibfnamefont {N.}~\bibnamefont {{Di Lalla}}}, \bibinfo {author} {\bibfnamefont {I.}~\bibnamefont {{Donnarumma}}}, \bibinfo {author} {\bibfnamefont {V.}~\bibnamefont {{Doroshenko}}}, \bibinfo {author} {\bibfnamefont {M.}~\bibnamefont {{Dov{\v{c}}iak}}}, \bibinfo {author} {\bibfnamefont {S.~R.}\ \bibnamefont {{Ehlert}}}, \bibinfo {author} {\bibfnamefont {T.}~\bibnamefont {{Enoto}}}, \bibinfo {author} {\bibfnamefont {Y.}~\bibnamefont {{Evangelista}}}, \bibinfo {author} {\bibfnamefont {S.}~\bibnamefont {{Fabiani}}}, \bibinfo {author} {\bibfnamefont {R.}~\bibnamefont {{Ferrazzoli}}}, \bibinfo {author} {\bibfnamefont {J.~A.}\ \bibnamefont {{Garcia}}}, \bibinfo {author} {\bibfnamefont {S.}~\bibnamefont {{Gunji}}}, \bibinfo {author} {\bibfnamefont {K.}~\bibnamefont {{Hayashida}}}, \bibinfo
  {author} {\bibfnamefont {W.}~\bibnamefont {{Iwakiri}}}, \bibinfo {author} {\bibfnamefont {S.~G.}\ \bibnamefont {{Jorstad}}}, \bibinfo {author} {\bibfnamefont {P.}~\bibnamefont {{Kaaret}}}, \bibinfo {author} {\bibfnamefont {V.}~\bibnamefont {{Karas}}}, \bibinfo {author} {\bibfnamefont {F.}~\bibnamefont {{Kislat}}}, \bibinfo {author} {\bibfnamefont {T.}~\bibnamefont {{Kitaguchi}}}, \bibinfo {author} {\bibfnamefont {J.~J.}\ \bibnamefont {{Kolodziejczak}}}, \bibinfo {author} {\bibfnamefont {F.}~\bibnamefont {{La Monaca}}}, \bibinfo {author} {\bibfnamefont {L.}~\bibnamefont {{Latronico}}}, \bibinfo {author} {\bibfnamefont {I.}~\bibnamefont {{Liodakis}}}, \bibinfo {author} {\bibfnamefont {S.}~\bibnamefont {{Maldera}}}, \bibinfo {author} {\bibfnamefont {A.}~\bibnamefont {{Manfreda}}}, \bibinfo {author} {\bibfnamefont {F.}~\bibnamefont {{Marin}}}, \bibinfo {author} {\bibfnamefont {A.}~\bibnamefont {{Marinucci}}}, \bibinfo {author} {\bibfnamefont {A.~P.}\ \bibnamefont {{Marscher}}}, \bibinfo {author} {\bibfnamefont
  {H.~L.}\ \bibnamefont {{Marshall}}}, \bibinfo {author} {\bibfnamefont {F.}~\bibnamefont {{Massaro}}}, \bibinfo {author} {\bibfnamefont {I.}~\bibnamefont {{Mitsuishi}}}, \bibinfo {author} {\bibfnamefont {T.}~\bibnamefont {{Mizuno}}}, \bibinfo {author} {\bibfnamefont {C.~Y.}\ \bibnamefont {{Ng}}}, \bibinfo {author} {\bibfnamefont {S.~L.}\ \bibnamefont {{O'Dell}}}, \bibinfo {author} {\bibfnamefont {N.}~\bibnamefont {{Omodei}}}, \bibinfo {author} {\bibfnamefont {C.}~\bibnamefont {{Oppedisano}}}, \bibinfo {author} {\bibfnamefont {A.}~\bibnamefont {{Papitto}}}, \bibinfo {author} {\bibfnamefont {G.~G.}\ \bibnamefont {{Pavlov}}}, \bibinfo {author} {\bibfnamefont {A.~L.}\ \bibnamefont {{Peirson}}}, \bibinfo {author} {\bibfnamefont {M.}~\bibnamefont {{Perri}}}, \bibinfo {author} {\bibfnamefont {M.}~\bibnamefont {{Pesce-Rollins}}}, \bibinfo {author} {\bibfnamefont {P.-O.}\ \bibnamefont {{Petrucci}}}, \bibinfo {author} {\bibfnamefont {M.}~\bibnamefont {{Pilia}}}, \bibinfo {author} {\bibfnamefont {S.}~\bibnamefont
  {{Puccetti}}}, \bibinfo {author} {\bibfnamefont {B.~D.}\ \bibnamefont {{Ramsey}}}, \bibinfo {author} {\bibfnamefont {J.}~\bibnamefont {{Rankin}}}, \bibinfo {author} {\bibfnamefont {A.}~\bibnamefont {{Ratheesh}}}, \bibinfo {author} {\bibfnamefont {O.~J.}\ \bibnamefont {{Roberts}}}, \bibinfo {author} {\bibfnamefont {R.~W.}\ \bibnamefont {{Romani}}}, \bibinfo {author} {\bibfnamefont {C.}~\bibnamefont {{Sgr{\'o}}}}, \bibinfo {author} {\bibfnamefont {P.}~\bibnamefont {{Slane}}}, \bibinfo {author} {\bibfnamefont {P.}~\bibnamefont {{Soffitta}}}, \bibinfo {author} {\bibfnamefont {G.}~\bibnamefont {{Spandre}}}, \bibinfo {author} {\bibfnamefont {D.~A.}\ \bibnamefont {{Swartz}}}, \bibinfo {author} {\bibfnamefont {T.}~\bibnamefont {{Tamagawa}}}, \bibinfo {author} {\bibfnamefont {F.}~\bibnamefont {{Tavecchio}}}, \bibinfo {author} {\bibfnamefont {Y.}~\bibnamefont {{Tawara}}}, \bibinfo {author} {\bibfnamefont {A.~F.}\ \bibnamefont {{Tennant}}}, \bibinfo {author} {\bibfnamefont {N.~E.}\ \bibnamefont {{Thomas}}}, \bibinfo
  {author} {\bibfnamefont {F.}~\bibnamefont {{Tombesi}}}, \bibinfo {author} {\bibfnamefont {A.}~\bibnamefont {{Trois}}}, \bibinfo {author} {\bibfnamefont {S.~S.}\ \bibnamefont {{Tsygankov}}}, \bibinfo {author} {\bibfnamefont {J.}~\bibnamefont {{Vink}}}, \bibinfo {author} {\bibfnamefont {M.~C.}\ \bibnamefont {{Weisskopf}}}, \bibinfo {author} {\bibfnamefont {K.}~\bibnamefont {{Wu}}}, \ and\ \bibinfo {author} {\bibfnamefont {F.}~\bibnamefont {{Xie}}},\ }\href {\doibase 10.3847/1538-4357/aced05} {\bibfield  {journal} {\bibinfo  {journal} {\apj}\ }\textbf {\bibinfo {volume} {954}},\ \bibinfo {eid} {88} (\bibinfo {year} {2023})},\ \Eprint {http://arxiv.org/abs/2308.01238} {arXiv:2308.01238 [astro-ph.HE]} \BibitemShut {NoStop}%
\bibitem [{\citenamefont {{Heyl}}\ \emph {et~al.}(2024{\natexlab{a}})\citenamefont {{Heyl}}, \citenamefont {{Taverna}}, \citenamefont {{Turolla}}, \citenamefont {{Israel}}, \citenamefont {{Ng}}, \citenamefont {{K{\i}rm{\i}z{\i}bayrak}}, \citenamefont {{Gonz{\'a}lez-Caniulef}}, \citenamefont {{Caiazzo}}, \citenamefont {{Zane}}, \citenamefont {{Ehlert}}, \citenamefont {{Negro}}, \citenamefont {{Agudo}}, \citenamefont {{Antonelli}}, \citenamefont {{Bachetti}}, \citenamefont {{Baldini}}, \citenamefont {{Baumgartner}}, \citenamefont {{Bellazzini}}, \citenamefont {{Bianchi}}, \citenamefont {{Bongiorno}}, \citenamefont {{Bonino}}, \citenamefont {{Brez}}, \citenamefont {{Bucciantini}}, \citenamefont {{Capitanio}}, \citenamefont {{Castellano}}, \citenamefont {{Cavazzuti}}, \citenamefont {{Chen}}, \citenamefont {{Ciprini}}, \citenamefont {{Costa}}, \citenamefont {{De Rosa}}, \citenamefont {{Del Monte}}, \citenamefont {{Di Gesu}}, \citenamefont {{Di Lalla}}, \citenamefont {{Di Marco}}, \citenamefont {{Donnarumma}},
  \citenamefont {{Doroshenko}}, \citenamefont {{Dov{\v{c}}iak}}, \citenamefont {{Enoto}}, \citenamefont {{Evangelista}}, \citenamefont {{Fabiani}}, \citenamefont {{Ferrazzoli}}, \citenamefont {{Garcia}}, \citenamefont {{Gunji}}, \citenamefont {{Hayashida}}, \citenamefont {{Iwakiri}}, \citenamefont {{Jorstad}}, \citenamefont {{Kaaret}}, \citenamefont {{Karas}}, \citenamefont {{Kislat}}, \citenamefont {{Kitaguchi}}, \citenamefont {{Kolodziejczak}}, \citenamefont {{Krawczynski}}, \citenamefont {{Monaca}}, \citenamefont {{Latronico}}, \citenamefont {{Liodakis}}, \citenamefont {{Maldera}}, \citenamefont {{Manfreda}}, \citenamefont {{Marin}}, \citenamefont {{Marinucci}}, \citenamefont {{Marscher}}, \citenamefont {{Marshall}}, \citenamefont {{Massaro}}, \citenamefont {{Matt}}, \citenamefont {{Mitsuishi}}, \citenamefont {{Mizuno}}, \citenamefont {{Muleri}}, \citenamefont {{Ng}}, \citenamefont {{O'Dell}}, \citenamefont {{Omodei}}, \citenamefont {{Oppedisano}}, \citenamefont {{Papitto}}, \citenamefont {{Pavlov}},
  \citenamefont {{Peirson}}, \citenamefont {{Perri}}, \citenamefont {{Pesce-Rollins}}, \citenamefont {{Petrucci}}, \citenamefont {{Pilia}}, \citenamefont {{Possenti}}, \citenamefont {{Poutanen}}, \citenamefont {{Puccetti}}, \citenamefont {{Ramsey}}, \citenamefont {{Rankin}}, \citenamefont {{Ratheesh}}, \citenamefont {{Roberts}}, \citenamefont {{Romani}}, \citenamefont {{Sgr{\`o}}}, \citenamefont {{Slane}}, \citenamefont {{Soffitta}}, \citenamefont {{Spandre}}, \citenamefont {{Swartz}}, \citenamefont {{Tamagawa}}, \citenamefont {{Tavecchio}}, \citenamefont {{Tawara}}, \citenamefont {{Tennant}}, \citenamefont {{Thomas}}, \citenamefont {{Tombesi}}, \citenamefont {{Trois}}, \citenamefont {{Tsygankov}}, \citenamefont {{Vink}}, \citenamefont {{Weisskopf}}, \citenamefont {{Wu}},\ and\ \citenamefont {{Xie}}}]{2024MNRAS.52712219H}%
  \BibitemOpen
  \bibfield  {author} {\bibinfo {author} {\bibfnamefont {J.}~\bibnamefont {{Heyl}}}, \bibinfo {author} {\bibfnamefont {R.}~\bibnamefont {{Taverna}}}, \bibinfo {author} {\bibfnamefont {R.}~\bibnamefont {{Turolla}}}, \bibinfo {author} {\bibfnamefont {G.~L.}\ \bibnamefont {{Israel}}}, \bibinfo {author} {\bibfnamefont {M.}~\bibnamefont {{Ng}}}, \bibinfo {author} {\bibfnamefont {D.}~\bibnamefont {{K{\i}rm{\i}z{\i}bayrak}}}, \bibinfo {author} {\bibfnamefont {D.}~\bibnamefont {{Gonz{\'a}lez-Caniulef}}}, \bibinfo {author} {\bibfnamefont {I.}~\bibnamefont {{Caiazzo}}}, \bibinfo {author} {\bibfnamefont {S.}~\bibnamefont {{Zane}}}, \bibinfo {author} {\bibfnamefont {S.~R.}\ \bibnamefont {{Ehlert}}}, \bibinfo {author} {\bibfnamefont {M.}~\bibnamefont {{Negro}}}, \bibinfo {author} {\bibfnamefont {I.}~\bibnamefont {{Agudo}}}, \bibinfo {author} {\bibfnamefont {L.~A.}\ \bibnamefont {{Antonelli}}}, \bibinfo {author} {\bibfnamefont {M.}~\bibnamefont {{Bachetti}}}, \bibinfo {author} {\bibfnamefont {L.}~\bibnamefont {{Baldini}}},
  \bibinfo {author} {\bibfnamefont {W.~H.}\ \bibnamefont {{Baumgartner}}}, \bibinfo {author} {\bibfnamefont {R.}~\bibnamefont {{Bellazzini}}}, \bibinfo {author} {\bibfnamefont {S.}~\bibnamefont {{Bianchi}}}, \bibinfo {author} {\bibfnamefont {S.~D.}\ \bibnamefont {{Bongiorno}}}, \bibinfo {author} {\bibfnamefont {R.}~\bibnamefont {{Bonino}}}, \bibinfo {author} {\bibfnamefont {A.}~\bibnamefont {{Brez}}}, \bibinfo {author} {\bibfnamefont {N.}~\bibnamefont {{Bucciantini}}}, \bibinfo {author} {\bibfnamefont {F.}~\bibnamefont {{Capitanio}}}, \bibinfo {author} {\bibfnamefont {S.}~\bibnamefont {{Castellano}}}, \bibinfo {author} {\bibfnamefont {E.}~\bibnamefont {{Cavazzuti}}}, \bibinfo {author} {\bibfnamefont {C.-T.}\ \bibnamefont {{Chen}}}, \bibinfo {author} {\bibfnamefont {S.}~\bibnamefont {{Ciprini}}}, \bibinfo {author} {\bibfnamefont {E.}~\bibnamefont {{Costa}}}, \bibinfo {author} {\bibfnamefont {A.}~\bibnamefont {{De Rosa}}}, \bibinfo {author} {\bibfnamefont {E.}~\bibnamefont {{Del Monte}}}, \bibinfo {author}
  {\bibfnamefont {L.}~\bibnamefont {{Di Gesu}}}, \bibinfo {author} {\bibfnamefont {N.}~\bibnamefont {{Di Lalla}}}, \bibinfo {author} {\bibfnamefont {A.}~\bibnamefont {{Di Marco}}}, \bibinfo {author} {\bibfnamefont {I.}~\bibnamefont {{Donnarumma}}}, \bibinfo {author} {\bibfnamefont {V.}~\bibnamefont {{Doroshenko}}}, \bibinfo {author} {\bibfnamefont {M.}~\bibnamefont {{Dov{\v{c}}iak}}}, \bibinfo {author} {\bibfnamefont {T.}~\bibnamefont {{Enoto}}}, \bibinfo {author} {\bibfnamefont {Y.}~\bibnamefont {{Evangelista}}}, \bibinfo {author} {\bibfnamefont {S.}~\bibnamefont {{Fabiani}}}, \bibinfo {author} {\bibfnamefont {R.}~\bibnamefont {{Ferrazzoli}}}, \bibinfo {author} {\bibfnamefont {J.~A.}\ \bibnamefont {{Garcia}}}, \bibinfo {author} {\bibfnamefont {S.}~\bibnamefont {{Gunji}}}, \bibinfo {author} {\bibfnamefont {K.}~\bibnamefont {{Hayashida}}}, \bibinfo {author} {\bibfnamefont {W.}~\bibnamefont {{Iwakiri}}}, \bibinfo {author} {\bibfnamefont {S.~G.}\ \bibnamefont {{Jorstad}}}, \bibinfo {author} {\bibfnamefont
  {P.}~\bibnamefont {{Kaaret}}}, \bibinfo {author} {\bibfnamefont {V.}~\bibnamefont {{Karas}}}, \bibinfo {author} {\bibfnamefont {F.}~\bibnamefont {{Kislat}}}, \bibinfo {author} {\bibfnamefont {T.}~\bibnamefont {{Kitaguchi}}}, \bibinfo {author} {\bibfnamefont {J.~J.}\ \bibnamefont {{Kolodziejczak}}}, \bibinfo {author} {\bibfnamefont {H.}~\bibnamefont {{Krawczynski}}}, \bibinfo {author} {\bibfnamefont {F.~L.}\ \bibnamefont {{Monaca}}}, \bibinfo {author} {\bibfnamefont {L.}~\bibnamefont {{Latronico}}}, \bibinfo {author} {\bibfnamefont {I.}~\bibnamefont {{Liodakis}}}, \bibinfo {author} {\bibfnamefont {S.}~\bibnamefont {{Maldera}}}, \bibinfo {author} {\bibfnamefont {A.}~\bibnamefont {{Manfreda}}}, \bibinfo {author} {\bibfnamefont {F.}~\bibnamefont {{Marin}}}, \bibinfo {author} {\bibfnamefont {A.}~\bibnamefont {{Marinucci}}}, \bibinfo {author} {\bibfnamefont {A.~P.}\ \bibnamefont {{Marscher}}}, \bibinfo {author} {\bibfnamefont {H.~L.}\ \bibnamefont {{Marshall}}}, \bibinfo {author} {\bibfnamefont {F.}~\bibnamefont
  {{Massaro}}}, \bibinfo {author} {\bibfnamefont {G.}~\bibnamefont {{Matt}}}, \bibinfo {author} {\bibfnamefont {I.}~\bibnamefont {{Mitsuishi}}}, \bibinfo {author} {\bibfnamefont {T.}~\bibnamefont {{Mizuno}}}, \bibinfo {author} {\bibfnamefont {F.}~\bibnamefont {{Muleri}}}, \bibinfo {author} {\bibfnamefont {C.~Y.}\ \bibnamefont {{Ng}}}, \bibinfo {author} {\bibfnamefont {S.~L.}\ \bibnamefont {{O'Dell}}}, \bibinfo {author} {\bibfnamefont {N.}~\bibnamefont {{Omodei}}}, \bibinfo {author} {\bibfnamefont {C.}~\bibnamefont {{Oppedisano}}}, \bibinfo {author} {\bibfnamefont {A.}~\bibnamefont {{Papitto}}}, \bibinfo {author} {\bibfnamefont {G.~G.}\ \bibnamefont {{Pavlov}}}, \bibinfo {author} {\bibfnamefont {A.~L.}\ \bibnamefont {{Peirson}}}, \bibinfo {author} {\bibfnamefont {M.}~\bibnamefont {{Perri}}}, \bibinfo {author} {\bibfnamefont {M.}~\bibnamefont {{Pesce-Rollins}}}, \bibinfo {author} {\bibfnamefont {P.-O.}\ \bibnamefont {{Petrucci}}}, \bibinfo {author} {\bibfnamefont {M.}~\bibnamefont {{Pilia}}}, \bibinfo {author}
  {\bibfnamefont {A.}~\bibnamefont {{Possenti}}}, \bibinfo {author} {\bibfnamefont {J.}~\bibnamefont {{Poutanen}}}, \bibinfo {author} {\bibfnamefont {S.}~\bibnamefont {{Puccetti}}}, \bibinfo {author} {\bibfnamefont {B.~D.}\ \bibnamefont {{Ramsey}}}, \bibinfo {author} {\bibfnamefont {J.}~\bibnamefont {{Rankin}}}, \bibinfo {author} {\bibfnamefont {A.}~\bibnamefont {{Ratheesh}}}, \bibinfo {author} {\bibfnamefont {O.~J.}\ \bibnamefont {{Roberts}}}, \bibinfo {author} {\bibfnamefont {R.~W.}\ \bibnamefont {{Romani}}}, \bibinfo {author} {\bibfnamefont {C.}~\bibnamefont {{Sgr{\`o}}}}, \bibinfo {author} {\bibfnamefont {P.}~\bibnamefont {{Slane}}}, \bibinfo {author} {\bibfnamefont {P.}~\bibnamefont {{Soffitta}}}, \bibinfo {author} {\bibfnamefont {G.}~\bibnamefont {{Spandre}}}, \bibinfo {author} {\bibfnamefont {D.~A.}\ \bibnamefont {{Swartz}}}, \bibinfo {author} {\bibfnamefont {T.}~\bibnamefont {{Tamagawa}}}, \bibinfo {author} {\bibfnamefont {F.}~\bibnamefont {{Tavecchio}}}, \bibinfo {author} {\bibfnamefont
  {Y.}~\bibnamefont {{Tawara}}}, \bibinfo {author} {\bibfnamefont {A.~F.}\ \bibnamefont {{Tennant}}}, \bibinfo {author} {\bibfnamefont {N.~E.}\ \bibnamefont {{Thomas}}}, \bibinfo {author} {\bibfnamefont {F.}~\bibnamefont {{Tombesi}}}, \bibinfo {author} {\bibfnamefont {A.}~\bibnamefont {{Trois}}}, \bibinfo {author} {\bibfnamefont {S.~S.}\ \bibnamefont {{Tsygankov}}}, \bibinfo {author} {\bibfnamefont {J.}~\bibnamefont {{Vink}}}, \bibinfo {author} {\bibfnamefont {M.~C.}\ \bibnamefont {{Weisskopf}}}, \bibinfo {author} {\bibfnamefont {K.}~\bibnamefont {{Wu}}}, \ and\ \bibinfo {author} {\bibfnamefont {F.}~\bibnamefont {{Xie}}},\ }\href {\doibase 10.1093/mnras/stad3680} {\bibfield  {journal} {\bibinfo  {journal} {\mnras}\ }\textbf {\bibinfo {volume} {527}},\ \bibinfo {pages} {12219} (\bibinfo {year} {2024}{\natexlab{a}})},\ \Eprint {http://arxiv.org/abs/2311.03637} {arXiv:2311.03637 [astro-ph.HE]} \BibitemShut {NoStop}%
\bibitem [{\citenamefont {{Taverna}}\ and\ \citenamefont {{Turolla}}(2024)}]{2024Galax..12....6T}%
  \BibitemOpen
  \bibfield  {author} {\bibinfo {author} {\bibfnamefont {R.}~\bibnamefont {{Taverna}}}\ and\ \bibinfo {author} {\bibfnamefont {R.}~\bibnamefont {{Turolla}}},\ }\href {\doibase 10.3390/galaxies12010006} {\bibfield  {journal} {\bibinfo  {journal} {Galaxies}\ }\textbf {\bibinfo {volume} {12}},\ \bibinfo {eid} {6} (\bibinfo {year} {2024})},\ \Eprint {http://arxiv.org/abs/2402.05622} {arXiv:2402.05622 [astro-ph.HE]} \BibitemShut {NoStop}%
\bibitem [{\citenamefont {{Peng}}\ \emph {et~al.}(2024)\citenamefont {{Peng}}, \citenamefont {{Ge}}, \citenamefont {{Weng}}, \citenamefont {{Zhao}}, \citenamefont {{Ye}}, \citenamefont {{Zhang}}, \citenamefont {{Qi}},\ and\ \citenamefont {{Tuo}}}]{2024ApJ...961..106P}%
  \BibitemOpen
  \bibfield  {author} {\bibinfo {author} {\bibfnamefont {H.-L.}\ \bibnamefont {{Peng}}}, \bibinfo {author} {\bibfnamefont {M.-Y.}\ \bibnamefont {{Ge}}}, \bibinfo {author} {\bibfnamefont {S.-S.}\ \bibnamefont {{Weng}}}, \bibinfo {author} {\bibfnamefont {Q.-C.}\ \bibnamefont {{Zhao}}}, \bibinfo {author} {\bibfnamefont {W.-T.}\ \bibnamefont {{Ye}}}, \bibinfo {author} {\bibfnamefont {L.}~\bibnamefont {{Zhang}}}, \bibinfo {author} {\bibfnamefont {L.-Q.}\ \bibnamefont {{Qi}}}, \ and\ \bibinfo {author} {\bibfnamefont {Y.-L.}\ \bibnamefont {{Tuo}}},\ }\href {\doibase 10.3847/1538-4357/ad1512} {\bibfield  {journal} {\bibinfo  {journal} {\apj}\ }\textbf {\bibinfo {volume} {961}},\ \bibinfo {eid} {106} (\bibinfo {year} {2024})}\BibitemShut {NoStop}%
\bibitem [{\citenamefont {{Rigoselli}}\ \emph {et~al.}(2025)\citenamefont {{Rigoselli}}, \citenamefont {{Taverna}}, \citenamefont {{Mereghetti}}, \citenamefont {{Turolla}}, \citenamefont {{Israel}}, \citenamefont {{Zane}}, \citenamefont {{Marra}}, \citenamefont {{Muleri}}, \citenamefont {{Borghese}}, \citenamefont {{Coti Zelati}}, \citenamefont {{De Grandis}}, \citenamefont {{Imbrogno}}, \citenamefont {{Kelly}}, \citenamefont {{Esposito}},\ and\ \citenamefont {{Rea}}}]{2024arXiv241215811R}%
  \BibitemOpen
  \bibfield  {author} {\bibinfo {author} {\bibfnamefont {M.}~\bibnamefont {{Rigoselli}}}, \bibinfo {author} {\bibfnamefont {R.}~\bibnamefont {{Taverna}}}, \bibinfo {author} {\bibfnamefont {S.}~\bibnamefont {{Mereghetti}}}, \bibinfo {author} {\bibfnamefont {R.}~\bibnamefont {{Turolla}}}, \bibinfo {author} {\bibfnamefont {G.~L.}\ \bibnamefont {{Israel}}}, \bibinfo {author} {\bibfnamefont {S.}~\bibnamefont {{Zane}}}, \bibinfo {author} {\bibfnamefont {L.}~\bibnamefont {{Marra}}}, \bibinfo {author} {\bibfnamefont {F.}~\bibnamefont {{Muleri}}}, \bibinfo {author} {\bibfnamefont {A.}~\bibnamefont {{Borghese}}}, \bibinfo {author} {\bibfnamefont {F.}~\bibnamefont {{Coti Zelati}}}, \bibinfo {author} {\bibfnamefont {D.}~\bibnamefont {{De Grandis}}}, \bibinfo {author} {\bibfnamefont {M.}~\bibnamefont {{Imbrogno}}}, \bibinfo {author} {\bibfnamefont {R.~M.~E.}\ \bibnamefont {{Kelly}}}, \bibinfo {author} {\bibfnamefont {P.}~\bibnamefont {{Esposito}}}, \ and\ \bibinfo {author} {\bibfnamefont {N.}~\bibnamefont {{Rea}}},\
  }\href {\doibase 10.3847/2041-8213/adbffb} {\bibfield  {journal} {\bibinfo  {journal} {\apjl}\ }\textbf {\bibinfo {volume} {985}},\ \bibinfo {eid} {L34} (\bibinfo {year} {2025})}\BibitemShut {NoStop}%
\bibitem [{\citenamefont {{Stewart}}\ \emph {et~al.}(2025{\natexlab{a}})\citenamefont {{Stewart}}, \citenamefont {{Younes}}, \citenamefont {{Harding}}, \citenamefont {{Wadiasingh}}, \citenamefont {{Baring}}, \citenamefont {{Negro}}, \citenamefont {{Strohmayer}}, \citenamefont {{Ho}}, \citenamefont {{Ng}}, \citenamefont {{Arzoumanian}}, \citenamefont {{Thi}}, \citenamefont {{Di Lalla}}, \citenamefont {{Enoto}}, \citenamefont {{Gendreau}}, \citenamefont {{Hu}}, \citenamefont {{van Kooten}}, \citenamefont {{Kouveliotou}},\ and\ \citenamefont {{McEwen}}}]{2024arXiv241216036S}%
  \BibitemOpen
  \bibfield  {author} {\bibinfo {author} {\bibfnamefont {R.}~\bibnamefont {{Stewart}}}, \bibinfo {author} {\bibfnamefont {G.~A.}\ \bibnamefont {{Younes}}}, \bibinfo {author} {\bibfnamefont {A.~K.}\ \bibnamefont {{Harding}}}, \bibinfo {author} {\bibfnamefont {Z.}~\bibnamefont {{Wadiasingh}}}, \bibinfo {author} {\bibfnamefont {M.~G.}\ \bibnamefont {{Baring}}}, \bibinfo {author} {\bibfnamefont {M.}~\bibnamefont {{Negro}}}, \bibinfo {author} {\bibfnamefont {T.~E.}\ \bibnamefont {{Strohmayer}}}, \bibinfo {author} {\bibfnamefont {W.~C.~G.}\ \bibnamefont {{Ho}}}, \bibinfo {author} {\bibfnamefont {M.}~\bibnamefont {{Ng}}}, \bibinfo {author} {\bibfnamefont {Z.}~\bibnamefont {{Arzoumanian}}}, \bibinfo {author} {\bibfnamefont {H.~D.}\ \bibnamefont {{Thi}}}, \bibinfo {author} {\bibfnamefont {N.}~\bibnamefont {{Di Lalla}}}, \bibinfo {author} {\bibfnamefont {T.}~\bibnamefont {{Enoto}}}, \bibinfo {author} {\bibfnamefont {K.}~\bibnamefont {{Gendreau}}}, \bibinfo {author} {\bibfnamefont {C.-P.}\ \bibnamefont {{Hu}}}, \bibinfo
  {author} {\bibfnamefont {A.}~\bibnamefont {{van Kooten}}}, \bibinfo {author} {\bibfnamefont {C.}~\bibnamefont {{Kouveliotou}}}, \ and\ \bibinfo {author} {\bibfnamefont {A.}~\bibnamefont {{McEwen}}},\ }\href {\doibase 10.3847/2041-8213/adbffa} {\bibfield  {journal} {\bibinfo  {journal} {\apjl}\ }\textbf {\bibinfo {volume} {985}},\ \bibinfo {eid} {L35} (\bibinfo {year} {2025}{\natexlab{a}})}\BibitemShut {NoStop}%
\bibitem [{\citenamefont {{Stewart}}\ \emph {et~al.}(2025{\natexlab{b}})\citenamefont {{Stewart}}, \citenamefont {{Younes}}, \citenamefont {{Harding}}, \citenamefont {{Wadiasingh}}, \citenamefont {{Baring}}, \citenamefont {{Negro}}, \citenamefont {{Strohmayer}}, \citenamefont {{Ho}}, \citenamefont {{Ng}}, \citenamefont {{Arzoumanian}}, \citenamefont {{Thi}}, \citenamefont {{Di Lalla}}, \citenamefont {{Enoto}}, \citenamefont {{Gendreau}}, \citenamefont {{Hu}}, \citenamefont {{van Kooten}}, \citenamefont {{Kouveliotou}},\ and\ \citenamefont {{McEwen}}}]{2025ApJ...985L..35S}%
  \BibitemOpen
  \bibfield  {author} {\bibinfo {author} {\bibfnamefont {R.}~\bibnamefont {{Stewart}}}, \bibinfo {author} {\bibfnamefont {G.~A.}\ \bibnamefont {{Younes}}}, \bibinfo {author} {\bibfnamefont {A.~K.}\ \bibnamefont {{Harding}}}, \bibinfo {author} {\bibfnamefont {Z.}~\bibnamefont {{Wadiasingh}}}, \bibinfo {author} {\bibfnamefont {M.~G.}\ \bibnamefont {{Baring}}}, \bibinfo {author} {\bibfnamefont {M.}~\bibnamefont {{Negro}}}, \bibinfo {author} {\bibfnamefont {T.~E.}\ \bibnamefont {{Strohmayer}}}, \bibinfo {author} {\bibfnamefont {W.~C.~G.}\ \bibnamefont {{Ho}}}, \bibinfo {author} {\bibfnamefont {M.}~\bibnamefont {{Ng}}}, \bibinfo {author} {\bibfnamefont {Z.}~\bibnamefont {{Arzoumanian}}}, \bibinfo {author} {\bibfnamefont {H.~D.}\ \bibnamefont {{Thi}}}, \bibinfo {author} {\bibfnamefont {N.}~\bibnamefont {{Di Lalla}}}, \bibinfo {author} {\bibfnamefont {T.}~\bibnamefont {{Enoto}}}, \bibinfo {author} {\bibfnamefont {K.}~\bibnamefont {{Gendreau}}}, \bibinfo {author} {\bibfnamefont {C.-P.}\ \bibnamefont {{Hu}}}, \bibinfo
  {author} {\bibfnamefont {A.}~\bibnamefont {{van Kooten}}}, \bibinfo {author} {\bibfnamefont {C.}~\bibnamefont {{Kouveliotou}}}, \ and\ \bibinfo {author} {\bibfnamefont {A.}~\bibnamefont {{McEwen}}},\ }\href {\doibase 10.3847/2041-8213/adbffa} {\bibfield  {journal} {\bibinfo  {journal} {\apjl}\ }\textbf {\bibinfo {volume} {985}},\ \bibinfo {eid} {L35} (\bibinfo {year} {2025}{\natexlab{b}})},\ \Eprint {http://arxiv.org/abs/2412.16036} {arXiv:2412.16036 [astro-ph.HE]} \BibitemShut {NoStop}%
\bibitem [{\citenamefont {{Taverna}}(2022{\natexlab{a}})}]{2022AAS...24024604T}%
  \BibitemOpen
  \bibfield  {author} {\bibinfo {author} {\bibfnamefont {R.}~\bibnamefont {{Taverna}}},\ }in\ \href@noop {} {\emph {\bibinfo {booktitle} {American Astronomical Society Meeting \#240}}},\ \bibinfo {series} {American Astronomical Society Meeting Abstracts}, Vol.\ \bibinfo {volume} {240}\ (\bibinfo {year} {2022})\ p.\ \bibinfo {pages} {246.04}\BibitemShut {NoStop}%
\bibitem [{\citenamefont {{Taverna}}(2022{\natexlab{b}})}]{2022tsra.confE.141T}%
  \BibitemOpen
  \bibfield  {author} {\bibinfo {author} {\bibfnamefont {R.}~\bibnamefont {{Taverna}}},\ }in\ \href@noop {} {\emph {\bibinfo {booktitle} {31st Texas Symposium on Relativistic Astrophysics}}}\ (\bibinfo {year} {2022})\ p.\ \bibinfo {pages} {141}\BibitemShut {NoStop}%
\bibitem [{\citenamefont {{Heisenberg}}\ and\ \citenamefont {{Euler}}(1936)}]{1936ZPhy...98..714H}%
  \BibitemOpen
  \bibfield  {author} {\bibinfo {author} {\bibfnamefont {W.}~\bibnamefont {{Heisenberg}}}\ and\ \bibinfo {author} {\bibfnamefont {H.}~\bibnamefont {{Euler}}},\ }\href {\doibase 10.1007/BF01343663} {\bibfield  {journal} {\bibinfo  {journal} {Zeitschrift fur Physik}\ }\textbf {\bibinfo {volume} {98}},\ \bibinfo {pages} {714} (\bibinfo {year} {1936})}\BibitemShut {NoStop}%
\bibitem [{\citenamefont {{Weisskopf}}(1936)}]{1936DVS...14..5}%
  \BibitemOpen
  \bibfield  {author} {\bibinfo {author} {\bibfnamefont {V.~S.}\ \bibnamefont {{Weisskopf}}},\ }\href@noop {} {\bibfield  {journal} {\bibinfo  {journal} {Det Kgl. Danske Videnskabernes Selskab. Mathematisk-fysiske Meddelelser}\ }\textbf {\bibinfo {volume} {14}},\ \bibinfo {pages} {1} (\bibinfo {year} {1936})}\BibitemShut {NoStop}%
\bibitem [{\citenamefont {{Gnedin}}\ \emph {et~al.}(1978)\citenamefont {{Gnedin}}, \citenamefont {{Pavlov}},\ and\ \citenamefont {{Shibanov}}}]{1978SvAL....4..117G}%
  \BibitemOpen
  \bibfield  {author} {\bibinfo {author} {\bibfnamefont {Y.~N.}\ \bibnamefont {{Gnedin}}}, \bibinfo {author} {\bibfnamefont {G.~G.}\ \bibnamefont {{Pavlov}}}, \ and\ \bibinfo {author} {\bibfnamefont {Y.~A.}\ \bibnamefont {{Shibanov}}},\ }\href@noop {} {\bibfield  {journal} {\bibinfo  {journal} {Soviet Astronomy Letters}\ }\textbf {\bibinfo {volume} {4}},\ \bibinfo {pages} {117} (\bibinfo {year} {1978})}\BibitemShut {NoStop}%
\bibitem [{\citenamefont {{Heyl}}\ and\ \citenamefont {{Hernquist}}(1997)}]{1997JPhA...30.6485H}%
  \BibitemOpen
  \bibfield  {author} {\bibinfo {author} {\bibfnamefont {J.~S.}\ \bibnamefont {{Heyl}}}\ and\ \bibinfo {author} {\bibfnamefont {L.}~\bibnamefont {{Hernquist}}},\ }\href {\doibase 10.1088/0305-4470/30/18/022} {\bibfield  {journal} {\bibinfo  {journal} {Journal of Physics A Mathematical General}\ }\textbf {\bibinfo {volume} {30}},\ \bibinfo {pages} {6485} (\bibinfo {year} {1997})},\ \Eprint {http://arxiv.org/abs/hep-ph/9705367} {arXiv:hep-ph/9705367 [astro-ph]} \BibitemShut {NoStop}%
\bibitem [{\citenamefont {{Ejlli}}\ \emph {et~al.}(2020)\citenamefont {{Ejlli}}, \citenamefont {{Della Valle}}, \citenamefont {{Gastaldi}}, \citenamefont {{Messineo}}, \citenamefont {{Pengo}}, \citenamefont {{Ruoso}},\ and\ \citenamefont {{Zavattini}}}]{2020PhR...871....1E}%
  \BibitemOpen
  \bibfield  {author} {\bibinfo {author} {\bibfnamefont {A.}~\bibnamefont {{Ejlli}}}, \bibinfo {author} {\bibfnamefont {F.}~\bibnamefont {{Della Valle}}}, \bibinfo {author} {\bibfnamefont {U.}~\bibnamefont {{Gastaldi}}}, \bibinfo {author} {\bibfnamefont {G.}~\bibnamefont {{Messineo}}}, \bibinfo {author} {\bibfnamefont {R.}~\bibnamefont {{Pengo}}}, \bibinfo {author} {\bibfnamefont {G.}~\bibnamefont {{Ruoso}}}, \ and\ \bibinfo {author} {\bibfnamefont {G.}~\bibnamefont {{Zavattini}}},\ }\href {\doibase 10.1016/j.physrep.2020.06.001} {\bibfield  {journal} {\bibinfo  {journal} {Physics Reports}\ }\textbf {\bibinfo {volume} {871}},\ \bibinfo {pages} {1} (\bibinfo {year} {2020})},\ \Eprint {http://arxiv.org/abs/2005.12913} {arXiv:2005.12913 [physics.optics]} \BibitemShut {NoStop}%
\bibitem [{\citenamefont {{Fern{\'a}ndez}}\ and\ \citenamefont {{Davis}}(2011)}]{2011ApJ...730..131F}%
  \BibitemOpen
  \bibfield  {author} {\bibinfo {author} {\bibfnamefont {R.}~\bibnamefont {{Fern{\'a}ndez}}}\ and\ \bibinfo {author} {\bibfnamefont {S.~W.}\ \bibnamefont {{Davis}}},\ }\href {\doibase 10.1088/0004-637X/730/2/131} {\bibfield  {journal} {\bibinfo  {journal} {\apj}\ }\textbf {\bibinfo {volume} {730}},\ \bibinfo {eid} {131} (\bibinfo {year} {2011})},\ \Eprint {http://arxiv.org/abs/1101.0834} {arXiv:1101.0834 [astro-ph.HE]} \BibitemShut {NoStop}%
\bibitem [{\citenamefont {{Heyl}}\ \emph {et~al.}(2003)\citenamefont {{Heyl}}, \citenamefont {{Shaviv}},\ and\ \citenamefont {{Lloyd}}}]{2003MNRAS.342..134H}%
  \BibitemOpen
  \bibfield  {author} {\bibinfo {author} {\bibfnamefont {J.~S.}\ \bibnamefont {{Heyl}}}, \bibinfo {author} {\bibfnamefont {N.~J.}\ \bibnamefont {{Shaviv}}}, \ and\ \bibinfo {author} {\bibfnamefont {D.}~\bibnamefont {{Lloyd}}},\ }\href {\doibase 10.1046/j.1365-8711.2003.06521.x} {\bibfield  {journal} {\bibinfo  {journal} {\mnras}\ }\textbf {\bibinfo {volume} {342}},\ \bibinfo {pages} {134} (\bibinfo {year} {2003})},\ \Eprint {http://arxiv.org/abs/astro-ph/0302118} {arXiv:astro-ph/0302118 [astro-ph]} \BibitemShut {NoStop}%
\bibitem [{\citenamefont {{Taverna}}\ \emph {et~al.}(2015)\citenamefont {{Taverna}}, \citenamefont {{Turolla}}, \citenamefont {{Gonzalez Caniulef}}, \citenamefont {{Zane}}, \citenamefont {{Muleri}},\ and\ \citenamefont {{Soffitta}}}]{2015MNRAS.454.3254T}%
  \BibitemOpen
  \bibfield  {author} {\bibinfo {author} {\bibfnamefont {R.}~\bibnamefont {{Taverna}}}, \bibinfo {author} {\bibfnamefont {R.}~\bibnamefont {{Turolla}}}, \bibinfo {author} {\bibfnamefont {D.}~\bibnamefont {{Gonzalez Caniulef}}}, \bibinfo {author} {\bibfnamefont {S.}~\bibnamefont {{Zane}}}, \bibinfo {author} {\bibfnamefont {F.}~\bibnamefont {{Muleri}}}, \ and\ \bibinfo {author} {\bibfnamefont {P.}~\bibnamefont {{Soffitta}}},\ }\href {\doibase 10.1093/mnras/stv2168} {\bibfield  {journal} {\bibinfo  {journal} {\mnras}\ }\textbf {\bibinfo {volume} {454}},\ \bibinfo {pages} {3254} (\bibinfo {year} {2015})},\ \Eprint {http://arxiv.org/abs/1509.05023} {arXiv:1509.05023 [astro-ph.HE]} \BibitemShut {NoStop}%
\bibitem [{\citenamefont {{Gonz{\'a}lez Caniulef}}\ \emph {et~al.}(2016)\citenamefont {{Gonz{\'a}lez Caniulef}}, \citenamefont {{Zane}}, \citenamefont {{Taverna}}, \citenamefont {{Turolla}},\ and\ \citenamefont {{Wu}}}]{2016MNRAS.459.3585G}%
  \BibitemOpen
  \bibfield  {author} {\bibinfo {author} {\bibfnamefont {D.}~\bibnamefont {{Gonz{\'a}lez Caniulef}}}, \bibinfo {author} {\bibfnamefont {S.}~\bibnamefont {{Zane}}}, \bibinfo {author} {\bibfnamefont {R.}~\bibnamefont {{Taverna}}}, \bibinfo {author} {\bibfnamefont {R.}~\bibnamefont {{Turolla}}}, \ and\ \bibinfo {author} {\bibfnamefont {K.}~\bibnamefont {{Wu}}},\ }\href {\doibase 10.1093/mnras/stw804} {\bibfield  {journal} {\bibinfo  {journal} {\mnras}\ }\textbf {\bibinfo {volume} {459}},\ \bibinfo {pages} {3585} (\bibinfo {year} {2016})},\ \Eprint {http://arxiv.org/abs/1604.01552} {arXiv:1604.01552 [astro-ph.HE]} \BibitemShut {NoStop}%
\bibitem [{\citenamefont {{Olausen}}\ and\ \citenamefont {{Kaspi}}(2014)}]{2014ApJS..212....6O}%
  \BibitemOpen
  \bibfield  {author} {\bibinfo {author} {\bibfnamefont {S.~A.}\ \bibnamefont {{Olausen}}}\ and\ \bibinfo {author} {\bibfnamefont {V.~M.}\ \bibnamefont {{Kaspi}}},\ }\href {\doibase 10.1088/0067-0049/212/1/6} {\bibfield  {journal} {\bibinfo  {journal} {\apjs}\ }\textbf {\bibinfo {volume} {212}},\ \bibinfo {eid} {6} (\bibinfo {year} {2014})},\ \Eprint {http://arxiv.org/abs/1309.4167} {arXiv:1309.4167 [astro-ph.HE]} \BibitemShut {NoStop}%
\bibitem [{\citenamefont {{Turolla}}\ \emph {et~al.}(2015)\citenamefont {{Turolla}}, \citenamefont {{Zane}},\ and\ \citenamefont {{Watts}}}]{2015RPPh...78k6901T}%
  \BibitemOpen
  \bibfield  {author} {\bibinfo {author} {\bibfnamefont {R.}~\bibnamefont {{Turolla}}}, \bibinfo {author} {\bibfnamefont {S.}~\bibnamefont {{Zane}}}, \ and\ \bibinfo {author} {\bibfnamefont {A.~L.}\ \bibnamefont {{Watts}}},\ }\href {\doibase 10.1088/0034-4885/78/11/116901} {\bibfield  {journal} {\bibinfo  {journal} {Reports on Progress in Physics}\ }\textbf {\bibinfo {volume} {78}},\ \bibinfo {eid} {116901} (\bibinfo {year} {2015})},\ \Eprint {http://arxiv.org/abs/1507.02924} {arXiv:1507.02924 [astro-ph.HE]} \BibitemShut {NoStop}%
\bibitem [{\citenamefont {{Taverna}}\ \emph {et~al.}(2020)\citenamefont {{Taverna}}, \citenamefont {{Turolla}}, \citenamefont {{Suleimanov}}, \citenamefont {{Potekhin}},\ and\ \citenamefont {{Zane}}}]{2020MNRAS.492.5057T}%
  \BibitemOpen
  \bibfield  {author} {\bibinfo {author} {\bibfnamefont {R.}~\bibnamefont {{Taverna}}}, \bibinfo {author} {\bibfnamefont {R.}~\bibnamefont {{Turolla}}}, \bibinfo {author} {\bibfnamefont {V.}~\bibnamefont {{Suleimanov}}}, \bibinfo {author} {\bibfnamefont {A.~Y.}\ \bibnamefont {{Potekhin}}}, \ and\ \bibinfo {author} {\bibfnamefont {S.}~\bibnamefont {{Zane}}},\ }\href {\doibase 10.1093/mnras/staa204} {\bibfield  {journal} {\bibinfo  {journal} {\mnras}\ }\textbf {\bibinfo {volume} {492}},\ \bibinfo {pages} {5057} (\bibinfo {year} {2020})},\ \Eprint {http://arxiv.org/abs/2001.07663} {arXiv:2001.07663 [astro-ph.HE]} \BibitemShut {NoStop}%
\bibitem [{\citenamefont {{van Adelsberg}}\ and\ \citenamefont {{Perna}}(2009)}]{2009MNRAS.399.1523V}%
  \BibitemOpen
  \bibfield  {author} {\bibinfo {author} {\bibfnamefont {M.}~\bibnamefont {{van Adelsberg}}}\ and\ \bibinfo {author} {\bibfnamefont {R.}~\bibnamefont {{Perna}}},\ }\href {\doibase 10.1111/j.1365-2966.2009.15374.x} {\bibfield  {journal} {\bibinfo  {journal} {\mnras}\ }\textbf {\bibinfo {volume} {399}},\ \bibinfo {pages} {1523} (\bibinfo {year} {2009})},\ \Eprint {http://arxiv.org/abs/0907.3499} {arXiv:0907.3499 [astro-ph.HE]} \BibitemShut {NoStop}%
\bibitem [{\citenamefont {{Baldini}}\ \emph {et~al.}(2022)\citenamefont {{Baldini}}, \citenamefont {{Bucciantini}}, \citenamefont {{Lalla}}, \citenamefont {{Ehlert}}, \citenamefont {{Manfreda}}, \citenamefont {{Negro}}, \citenamefont {{Omodei}}, \citenamefont {{Pesce-Rollins}}, \citenamefont {{Sgr{\`o}}},\ and\ \citenamefont {{Silvestri}}}]{2022SoftX..1901194B}%
  \BibitemOpen
  \bibfield  {author} {\bibinfo {author} {\bibfnamefont {L.}~\bibnamefont {{Baldini}}}, \bibinfo {author} {\bibfnamefont {N.}~\bibnamefont {{Bucciantini}}}, \bibinfo {author} {\bibfnamefont {N.~D.}\ \bibnamefont {{Lalla}}}, \bibinfo {author} {\bibfnamefont {S.}~\bibnamefont {{Ehlert}}}, \bibinfo {author} {\bibfnamefont {A.}~\bibnamefont {{Manfreda}}}, \bibinfo {author} {\bibfnamefont {M.}~\bibnamefont {{Negro}}}, \bibinfo {author} {\bibfnamefont {N.}~\bibnamefont {{Omodei}}}, \bibinfo {author} {\bibfnamefont {M.}~\bibnamefont {{Pesce-Rollins}}}, \bibinfo {author} {\bibfnamefont {C.}~\bibnamefont {{Sgr{\`o}}}}, \ and\ \bibinfo {author} {\bibfnamefont {S.}~\bibnamefont {{Silvestri}}},\ }\href {\doibase 10.1016/j.softx.2022.101194} {\bibfield  {journal} {\bibinfo  {journal} {SoftwareX}\ }\textbf {\bibinfo {volume} {19}},\ \bibinfo {eid} {101194} (\bibinfo {year} {2022})},\ \Eprint {http://arxiv.org/abs/2203.06384} {arXiv:2203.06384 [astro-ph.IM]} \BibitemShut {NoStop}%
\bibitem [{\citenamefont {{Taverna}}\ \emph {et~al.}(2022)\citenamefont {{Taverna}}, \citenamefont {{Turolla}}, \citenamefont {{Muleri}}, \citenamefont {{Heyl}}, \citenamefont {{Zane}}, \citenamefont {{Baldini}}, \citenamefont {{Gonz{\'a}lez-Caniulef}}, \citenamefont {{Bachetti}}, \citenamefont {{Rankin}}, \citenamefont {{Caiazzo}}, \citenamefont {{Di Lalla}}, \citenamefont {{Doroshenko}}, \citenamefont {{Errando}}, \citenamefont {{Gau}}, \citenamefont {{K{\i}rm{\i}z{\i}bayrak}}, \citenamefont {{Krawczynski}}, \citenamefont {{Negro}}, \citenamefont {{Ng}}, \citenamefont {{Omodei}}, \citenamefont {{Possenti}}, \citenamefont {{Tamagawa}}, \citenamefont {{Uchiyama}}, \citenamefont {{Weisskopf}}, \citenamefont {{Agudo}}, \citenamefont {{Antonelli}}, \citenamefont {{Baumgartner}}, \citenamefont {{Bellazzini}}, \citenamefont {{Bianchi}}, \citenamefont {{Bongiorno}}, \citenamefont {{Bonino}}, \citenamefont {{Brez}}, \citenamefont {{Bucciantini}}, \citenamefont {{Capitanio}}, \citenamefont {{Castellano}},
  \citenamefont {{Cavazzuti}}, \citenamefont {{Ciprini}}, \citenamefont {{Costa}}, \citenamefont {{De Rosa}}, \citenamefont {{Del Monte}}, \citenamefont {{Di Gesu}}, \citenamefont {{Di Marco}}, \citenamefont {{Donnarumma}}, \citenamefont {{Dov{\v{c}}iak}}, \citenamefont {{Ehlert}}, \citenamefont {{Enoto}}, \citenamefont {{Evangelista}}, \citenamefont {{Fabiani}}, \citenamefont {{Ferrazzoli}}, \citenamefont {{Garcia}}, \citenamefont {{Gunji}}, \citenamefont {{Hayashida}}, \citenamefont {{Iwakiri}}, \citenamefont {{Jorstad}}, \citenamefont {{Karas}}, \citenamefont {{Kitaguchi}}, \citenamefont {{Kolodziejczak}}, \citenamefont {{La Monaca}}, \citenamefont {{Latronico}}, \citenamefont {{Liodakis}}, \citenamefont {{Maldera}}, \citenamefont {{Manfreda}}, \citenamefont {{Marin}}, \citenamefont {{Marinucci}}, \citenamefont {{Marscher}}, \citenamefont {{Marshall}}, \citenamefont {{Matt}}, \citenamefont {{Mitsuishi}}, \citenamefont {{Mizuno}}, \citenamefont {{Ng}}, \citenamefont {{O{\textquoteright}Dell}}, \citenamefont
  {{Oppedisano}}, \citenamefont {{Papitto}}, \citenamefont {{Pavlov}}, \citenamefont {{Peirson}}, \citenamefont {{Perri}}, \citenamefont {{Pesce-Rollins}}, \citenamefont {{Pilia}}, \citenamefont {{Poutanen}}, \citenamefont {{Puccetti}}, \citenamefont {{Ramsey}}, \citenamefont {{Ratheesh}}, \citenamefont {{Romani}}, \citenamefont {{Sgr{\`o}}}, \citenamefont {{Slane}}, \citenamefont {{Soffitta}}, \citenamefont {{Spandre}}, \citenamefont {{Tavecchio}}, \citenamefont {{Tawara}}, \citenamefont {{Tennant}}, \citenamefont {{Thomas}}, \citenamefont {{Tombesi}}, \citenamefont {{Trois}}, \citenamefont {{Tsygankov}}, \citenamefont {{Vink}}, \citenamefont {{Wu}},\ and\ \citenamefont {{Xie}}}]{2022Sci...378..646T}%
  \BibitemOpen
  \bibfield  {author} {\bibinfo {author} {\bibfnamefont {R.}~\bibnamefont {{Taverna}}}, \bibinfo {author} {\bibfnamefont {R.}~\bibnamefont {{Turolla}}}, \bibinfo {author} {\bibfnamefont {F.}~\bibnamefont {{Muleri}}}, \bibinfo {author} {\bibfnamefont {J.}~\bibnamefont {{Heyl}}}, \bibinfo {author} {\bibfnamefont {S.}~\bibnamefont {{Zane}}}, \bibinfo {author} {\bibfnamefont {L.}~\bibnamefont {{Baldini}}}, \bibinfo {author} {\bibfnamefont {D.}~\bibnamefont {{Gonz{\'a}lez-Caniulef}}}, \bibinfo {author} {\bibfnamefont {M.}~\bibnamefont {{Bachetti}}}, \bibinfo {author} {\bibfnamefont {J.}~\bibnamefont {{Rankin}}}, \bibinfo {author} {\bibfnamefont {I.}~\bibnamefont {{Caiazzo}}}, \bibinfo {author} {\bibfnamefont {N.}~\bibnamefont {{Di Lalla}}}, \bibinfo {author} {\bibfnamefont {V.}~\bibnamefont {{Doroshenko}}}, \bibinfo {author} {\bibfnamefont {M.}~\bibnamefont {{Errando}}}, \bibinfo {author} {\bibfnamefont {E.}~\bibnamefont {{Gau}}}, \bibinfo {author} {\bibfnamefont {D.}~\bibnamefont {{K{\i}rm{\i}z{\i}bayrak}}},
  \bibinfo {author} {\bibfnamefont {H.}~\bibnamefont {{Krawczynski}}}, \bibinfo {author} {\bibfnamefont {M.}~\bibnamefont {{Negro}}}, \bibinfo {author} {\bibfnamefont {M.}~\bibnamefont {{Ng}}}, \bibinfo {author} {\bibfnamefont {N.}~\bibnamefont {{Omodei}}}, \bibinfo {author} {\bibfnamefont {A.}~\bibnamefont {{Possenti}}}, \bibinfo {author} {\bibfnamefont {T.}~\bibnamefont {{Tamagawa}}}, \bibinfo {author} {\bibfnamefont {K.}~\bibnamefont {{Uchiyama}}}, \bibinfo {author} {\bibfnamefont {M.~C.}\ \bibnamefont {{Weisskopf}}}, \bibinfo {author} {\bibfnamefont {I.}~\bibnamefont {{Agudo}}}, \bibinfo {author} {\bibfnamefont {L.~A.}\ \bibnamefont {{Antonelli}}}, \bibinfo {author} {\bibfnamefont {W.~H.}\ \bibnamefont {{Baumgartner}}}, \bibinfo {author} {\bibfnamefont {R.}~\bibnamefont {{Bellazzini}}}, \bibinfo {author} {\bibfnamefont {S.}~\bibnamefont {{Bianchi}}}, \bibinfo {author} {\bibfnamefont {S.~D.}\ \bibnamefont {{Bongiorno}}}, \bibinfo {author} {\bibfnamefont {R.}~\bibnamefont {{Bonino}}}, \bibinfo {author}
  {\bibfnamefont {A.}~\bibnamefont {{Brez}}}, \bibinfo {author} {\bibfnamefont {N.}~\bibnamefont {{Bucciantini}}}, \bibinfo {author} {\bibfnamefont {F.}~\bibnamefont {{Capitanio}}}, \bibinfo {author} {\bibfnamefont {S.}~\bibnamefont {{Castellano}}}, \bibinfo {author} {\bibfnamefont {E.}~\bibnamefont {{Cavazzuti}}}, \bibinfo {author} {\bibfnamefont {S.}~\bibnamefont {{Ciprini}}}, \bibinfo {author} {\bibfnamefont {E.}~\bibnamefont {{Costa}}}, \bibinfo {author} {\bibfnamefont {A.}~\bibnamefont {{De Rosa}}}, \bibinfo {author} {\bibfnamefont {E.}~\bibnamefont {{Del Monte}}}, \bibinfo {author} {\bibfnamefont {L.}~\bibnamefont {{Di Gesu}}}, \bibinfo {author} {\bibfnamefont {A.}~\bibnamefont {{Di Marco}}}, \bibinfo {author} {\bibfnamefont {I.}~\bibnamefont {{Donnarumma}}}, \bibinfo {author} {\bibfnamefont {M.}~\bibnamefont {{Dov{\v{c}}iak}}}, \bibinfo {author} {\bibfnamefont {S.~R.}\ \bibnamefont {{Ehlert}}}, \bibinfo {author} {\bibfnamefont {T.}~\bibnamefont {{Enoto}}}, \bibinfo {author} {\bibfnamefont
  {Y.}~\bibnamefont {{Evangelista}}}, \bibinfo {author} {\bibfnamefont {S.}~\bibnamefont {{Fabiani}}}, \bibinfo {author} {\bibfnamefont {R.}~\bibnamefont {{Ferrazzoli}}}, \bibinfo {author} {\bibfnamefont {J.~A.}\ \bibnamefont {{Garcia}}}, \bibinfo {author} {\bibfnamefont {S.}~\bibnamefont {{Gunji}}}, \bibinfo {author} {\bibfnamefont {K.}~\bibnamefont {{Hayashida}}}, \bibinfo {author} {\bibfnamefont {W.}~\bibnamefont {{Iwakiri}}}, \bibinfo {author} {\bibfnamefont {S.~G.}\ \bibnamefont {{Jorstad}}}, \bibinfo {author} {\bibfnamefont {V.}~\bibnamefont {{Karas}}}, \bibinfo {author} {\bibfnamefont {T.}~\bibnamefont {{Kitaguchi}}}, \bibinfo {author} {\bibfnamefont {J.~J.}\ \bibnamefont {{Kolodziejczak}}}, \bibinfo {author} {\bibfnamefont {F.}~\bibnamefont {{La Monaca}}}, \bibinfo {author} {\bibfnamefont {L.}~\bibnamefont {{Latronico}}}, \bibinfo {author} {\bibfnamefont {I.}~\bibnamefont {{Liodakis}}}, \bibinfo {author} {\bibfnamefont {S.}~\bibnamefont {{Maldera}}}, \bibinfo {author} {\bibfnamefont {A.}~\bibnamefont
  {{Manfreda}}}, \bibinfo {author} {\bibfnamefont {F.}~\bibnamefont {{Marin}}}, \bibinfo {author} {\bibfnamefont {A.}~\bibnamefont {{Marinucci}}}, \bibinfo {author} {\bibfnamefont {A.~P.}\ \bibnamefont {{Marscher}}}, \bibinfo {author} {\bibfnamefont {H.~L.}\ \bibnamefont {{Marshall}}}, \bibinfo {author} {\bibfnamefont {G.}~\bibnamefont {{Matt}}}, \bibinfo {author} {\bibfnamefont {I.}~\bibnamefont {{Mitsuishi}}}, \bibinfo {author} {\bibfnamefont {T.}~\bibnamefont {{Mizuno}}}, \bibinfo {author} {\bibfnamefont {S.~C.~Y.}\ \bibnamefont {{Ng}}}, \bibinfo {author} {\bibfnamefont {S.~L.}\ \bibnamefont {{O{\textquoteright}Dell}}}, \bibinfo {author} {\bibfnamefont {C.}~\bibnamefont {{Oppedisano}}}, \bibinfo {author} {\bibfnamefont {A.}~\bibnamefont {{Papitto}}}, \bibinfo {author} {\bibfnamefont {G.~G.}\ \bibnamefont {{Pavlov}}}, \bibinfo {author} {\bibfnamefont {A.~L.}\ \bibnamefont {{Peirson}}}, \bibinfo {author} {\bibfnamefont {M.}~\bibnamefont {{Perri}}}, \bibinfo {author} {\bibfnamefont {M.}~\bibnamefont
  {{Pesce-Rollins}}}, \bibinfo {author} {\bibfnamefont {M.}~\bibnamefont {{Pilia}}}, \bibinfo {author} {\bibfnamefont {J.}~\bibnamefont {{Poutanen}}}, \bibinfo {author} {\bibfnamefont {S.}~\bibnamefont {{Puccetti}}}, \bibinfo {author} {\bibfnamefont {B.~D.}\ \bibnamefont {{Ramsey}}}, \bibinfo {author} {\bibfnamefont {A.}~\bibnamefont {{Ratheesh}}}, \bibinfo {author} {\bibfnamefont {R.~W.}\ \bibnamefont {{Romani}}}, \bibinfo {author} {\bibfnamefont {C.}~\bibnamefont {{Sgr{\`o}}}}, \bibinfo {author} {\bibfnamefont {P.}~\bibnamefont {{Slane}}}, \bibinfo {author} {\bibfnamefont {P.}~\bibnamefont {{Soffitta}}}, \bibinfo {author} {\bibfnamefont {G.}~\bibnamefont {{Spandre}}}, \bibinfo {author} {\bibfnamefont {F.}~\bibnamefont {{Tavecchio}}}, \bibinfo {author} {\bibfnamefont {Y.}~\bibnamefont {{Tawara}}}, \bibinfo {author} {\bibfnamefont {A.~F.}\ \bibnamefont {{Tennant}}}, \bibinfo {author} {\bibfnamefont {N.~E.}\ \bibnamefont {{Thomas}}}, \bibinfo {author} {\bibfnamefont {F.}~\bibnamefont {{Tombesi}}}, \bibinfo
  {author} {\bibfnamefont {A.}~\bibnamefont {{Trois}}}, \bibinfo {author} {\bibfnamefont {S.~S.}\ \bibnamefont {{Tsygankov}}}, \bibinfo {author} {\bibfnamefont {J.}~\bibnamefont {{Vink}}}, \bibinfo {author} {\bibfnamefont {K.}~\bibnamefont {{Wu}}}, \ and\ \bibinfo {author} {\bibfnamefont {F.}~\bibnamefont {{Xie}}},\ }\href {\doibase 10.1126/science.add0080} {\bibfield  {journal} {\bibinfo  {journal} {Science}\ }\textbf {\bibinfo {volume} {378}},\ \bibinfo {pages} {646} (\bibinfo {year} {2022})},\ \Eprint {http://arxiv.org/abs/2205.08898} {arXiv:2205.08898 [astro-ph.HE]} \BibitemShut {NoStop}%
\bibitem [{\citenamefont {{Radhakrishnan}}\ and\ \citenamefont {{Cooke}}(1969)}]{1969ApL.....3..225R}%
  \BibitemOpen
  \bibfield  {author} {\bibinfo {author} {\bibfnamefont {V.}~\bibnamefont {{Radhakrishnan}}}\ and\ \bibinfo {author} {\bibfnamefont {D.~J.}\ \bibnamefont {{Cooke}}},\ }\href@noop {} {\bibfield  {journal} {\bibinfo  {journal} {Astrophysical Letters}\ }\textbf {\bibinfo {volume} {3}},\ \bibinfo {pages} {225} (\bibinfo {year} {1969})}\BibitemShut {NoStop}%
\bibitem [{\citenamefont {{Poutanen}}(2020)}]{2020A&A...641A.166P}%
  \BibitemOpen
  \bibfield  {author} {\bibinfo {author} {\bibfnamefont {J.}~\bibnamefont {{Poutanen}}},\ }\href {\doibase 10.1051/0004-6361/202038689} {\bibfield  {journal} {\bibinfo  {journal} {\aap}\ }\textbf {\bibinfo {volume} {641}},\ \bibinfo {eid} {A166} (\bibinfo {year} {2020})},\ \Eprint {http://arxiv.org/abs/2006.10448} {arXiv:2006.10448 [astro-ph.HE]} \BibitemShut {NoStop}%
\bibitem [{\citenamefont {{De Grandis}}\ \emph {et~al.}(2021)\citenamefont {{De Grandis}}, \citenamefont {{Taverna}}, \citenamefont {{Turolla}}, \citenamefont {{Gnarini}}, \citenamefont {{Popov}}, \citenamefont {{Zane}},\ and\ \citenamefont {{Wood}}}]{2021ApJ...914..118D}%
  \BibitemOpen
  \bibfield  {author} {\bibinfo {author} {\bibfnamefont {D.}~\bibnamefont {{De Grandis}}}, \bibinfo {author} {\bibfnamefont {R.}~\bibnamefont {{Taverna}}}, \bibinfo {author} {\bibfnamefont {R.}~\bibnamefont {{Turolla}}}, \bibinfo {author} {\bibfnamefont {A.}~\bibnamefont {{Gnarini}}}, \bibinfo {author} {\bibfnamefont {S.~B.}\ \bibnamefont {{Popov}}}, \bibinfo {author} {\bibfnamefont {S.}~\bibnamefont {{Zane}}}, \ and\ \bibinfo {author} {\bibfnamefont {T.~S.}\ \bibnamefont {{Wood}}},\ }\href {\doibase 10.3847/1538-4357/abfdac} {\bibfield  {journal} {\bibinfo  {journal} {\apj}\ }\textbf {\bibinfo {volume} {914}},\ \bibinfo {eid} {118} (\bibinfo {year} {2021})},\ \Eprint {http://arxiv.org/abs/2105.00684} {arXiv:2105.00684 [astro-ph.HE]} \BibitemShut {NoStop}%
\bibitem [{\citenamefont {{Suvorov}}\ and\ \citenamefont {{Pons}}(2025)}]{2025MNRAS.539.3655S}%
  \BibitemOpen
  \bibfield  {author} {\bibinfo {author} {\bibfnamefont {A.~G.}\ \bibnamefont {{Suvorov}}}\ and\ \bibinfo {author} {\bibfnamefont {J.~A.}\ \bibnamefont {{Pons}}},\ }\href {\doibase 10.1093/mnras/staf704} {\bibfield  {journal} {\bibinfo  {journal} {\mnras}\ }\textbf {\bibinfo {volume} {539}},\ \bibinfo {pages} {3655} (\bibinfo {year} {2025})},\ \Eprint {http://arxiv.org/abs/2503.01409} {arXiv:2503.01409 [astro-ph.HE]} \BibitemShut {NoStop}%
\bibitem [{\citenamefont {{Tiengo}}\ \emph {et~al.}(2013)\citenamefont {{Tiengo}}, \citenamefont {{Esposito}}, \citenamefont {{Mereghetti}}, \citenamefont {{Turolla}}, \citenamefont {{Nobili}}, \citenamefont {{Gastaldello}}, \citenamefont {{G{\"o}tz}}, \citenamefont {{Israel}}, \citenamefont {{Rea}}, \citenamefont {{Stella}}, \citenamefont {{Zane}},\ and\ \citenamefont {{Bignami}}}]{2013Natur.500..312T}%
  \BibitemOpen
  \bibfield  {author} {\bibinfo {author} {\bibfnamefont {A.}~\bibnamefont {{Tiengo}}}, \bibinfo {author} {\bibfnamefont {P.}~\bibnamefont {{Esposito}}}, \bibinfo {author} {\bibfnamefont {S.}~\bibnamefont {{Mereghetti}}}, \bibinfo {author} {\bibfnamefont {R.}~\bibnamefont {{Turolla}}}, \bibinfo {author} {\bibfnamefont {L.}~\bibnamefont {{Nobili}}}, \bibinfo {author} {\bibfnamefont {F.}~\bibnamefont {{Gastaldello}}}, \bibinfo {author} {\bibfnamefont {D.}~\bibnamefont {{G{\"o}tz}}}, \bibinfo {author} {\bibfnamefont {G.~L.}\ \bibnamefont {{Israel}}}, \bibinfo {author} {\bibfnamefont {N.}~\bibnamefont {{Rea}}}, \bibinfo {author} {\bibfnamefont {L.}~\bibnamefont {{Stella}}}, \bibinfo {author} {\bibfnamefont {S.}~\bibnamefont {{Zane}}}, \ and\ \bibinfo {author} {\bibfnamefont {G.~F.}\ \bibnamefont {{Bignami}}},\ }\href {\doibase 10.1038/nature12386} {\bibfield  {journal} {\bibinfo  {journal} {\nat}\ }\textbf {\bibinfo {volume} {500}},\ \bibinfo {pages} {312} (\bibinfo {year} {2013})},\ \Eprint
  {http://arxiv.org/abs/1308.4987} {arXiv:1308.4987 [astro-ph.HE]} \BibitemShut {NoStop}%
\bibitem [{\citenamefont {{Pizzocaro}}\ \emph {et~al.}(2019)\citenamefont {{Pizzocaro}}, \citenamefont {{Tiengo}}, \citenamefont {{Mereghetti}}, \citenamefont {{Turolla}}, \citenamefont {{Esposito}}, \citenamefont {{Stella}}, \citenamefont {{Zane}}, \citenamefont {{Rea}}, \citenamefont {{Coti Zelati}},\ and\ \citenamefont {{Israel}}}]{2019A&A...626A..39P}%
  \BibitemOpen
  \bibfield  {author} {\bibinfo {author} {\bibfnamefont {D.}~\bibnamefont {{Pizzocaro}}}, \bibinfo {author} {\bibfnamefont {A.}~\bibnamefont {{Tiengo}}}, \bibinfo {author} {\bibfnamefont {S.}~\bibnamefont {{Mereghetti}}}, \bibinfo {author} {\bibfnamefont {R.}~\bibnamefont {{Turolla}}}, \bibinfo {author} {\bibfnamefont {P.}~\bibnamefont {{Esposito}}}, \bibinfo {author} {\bibfnamefont {L.}~\bibnamefont {{Stella}}}, \bibinfo {author} {\bibfnamefont {S.}~\bibnamefont {{Zane}}}, \bibinfo {author} {\bibfnamefont {N.}~\bibnamefont {{Rea}}}, \bibinfo {author} {\bibfnamefont {F.}~\bibnamefont {{Coti Zelati}}}, \ and\ \bibinfo {author} {\bibfnamefont {G.}~\bibnamefont {{Israel}}},\ }\href {\doibase 10.1051/0004-6361/201834784} {\bibfield  {journal} {\bibinfo  {journal} {\aap}\ }\textbf {\bibinfo {volume} {626}},\ \bibinfo {eid} {A39} (\bibinfo {year} {2019})},\ \Eprint {http://arxiv.org/abs/1904.07553} {arXiv:1904.07553 [astro-ph.HE]} \BibitemShut {NoStop}%
\bibitem [{\citenamefont {{Rodr{\'\i}guez Castillo}}\ \emph {et~al.}(2016)\citenamefont {{Rodr{\'\i}guez Castillo}}, \citenamefont {{Israel}}, \citenamefont {{Tiengo}}, \citenamefont {{Salvetti}}, \citenamefont {{Turolla}}, \citenamefont {{Zane}}, \citenamefont {{Rea}}, \citenamefont {{Esposito}}, \citenamefont {{Mereghetti}}, \citenamefont {{Perna}}, \citenamefont {{Stella}}, \citenamefont {{Pons}}, \citenamefont {{Campana}}, \citenamefont {{G{\"o}tz}},\ and\ \citenamefont {{Motta}}}]{2016MNRAS.456.4145R}%
  \BibitemOpen
  \bibfield  {author} {\bibinfo {author} {\bibfnamefont {G.~A.}\ \bibnamefont {{Rodr{\'\i}guez Castillo}}}, \bibinfo {author} {\bibfnamefont {G.~L.}\ \bibnamefont {{Israel}}}, \bibinfo {author} {\bibfnamefont {A.}~\bibnamefont {{Tiengo}}}, \bibinfo {author} {\bibfnamefont {D.}~\bibnamefont {{Salvetti}}}, \bibinfo {author} {\bibfnamefont {R.}~\bibnamefont {{Turolla}}}, \bibinfo {author} {\bibfnamefont {S.}~\bibnamefont {{Zane}}}, \bibinfo {author} {\bibfnamefont {N.}~\bibnamefont {{Rea}}}, \bibinfo {author} {\bibfnamefont {P.}~\bibnamefont {{Esposito}}}, \bibinfo {author} {\bibfnamefont {S.}~\bibnamefont {{Mereghetti}}}, \bibinfo {author} {\bibfnamefont {R.}~\bibnamefont {{Perna}}}, \bibinfo {author} {\bibfnamefont {L.}~\bibnamefont {{Stella}}}, \bibinfo {author} {\bibfnamefont {J.~A.}\ \bibnamefont {{Pons}}}, \bibinfo {author} {\bibfnamefont {S.}~\bibnamefont {{Campana}}}, \bibinfo {author} {\bibfnamefont {D.}~\bibnamefont {{G{\"o}tz}}}, \ and\ \bibinfo {author} {\bibfnamefont {S.}~\bibnamefont {{Motta}}},\
  }\href {\doibase 10.1093/mnras/stv2490} {\bibfield  {journal} {\bibinfo  {journal} {\mnras}\ }\textbf {\bibinfo {volume} {456}},\ \bibinfo {pages} {4145} (\bibinfo {year} {2016})},\ \Eprint {http://arxiv.org/abs/1510.09157} {arXiv:1510.09157 [astro-ph.HE]} \BibitemShut {NoStop}%
\bibitem [{\citenamefont {{Borghese}}\ \emph {et~al.}(2015)\citenamefont {{Borghese}}, \citenamefont {{Rea}}, \citenamefont {{Coti Zelati}}, \citenamefont {{Tiengo}},\ and\ \citenamefont {{Turolla}}}]{2015ApJ...807L..20B}%
  \BibitemOpen
  \bibfield  {author} {\bibinfo {author} {\bibfnamefont {A.}~\bibnamefont {{Borghese}}}, \bibinfo {author} {\bibfnamefont {N.}~\bibnamefont {{Rea}}}, \bibinfo {author} {\bibfnamefont {F.}~\bibnamefont {{Coti Zelati}}}, \bibinfo {author} {\bibfnamefont {A.}~\bibnamefont {{Tiengo}}}, \ and\ \bibinfo {author} {\bibfnamefont {R.}~\bibnamefont {{Turolla}}},\ }\href {\doibase 10.1088/2041-8205/807/1/L20} {\bibfield  {journal} {\bibinfo  {journal} {\apjl}\ }\textbf {\bibinfo {volume} {807}},\ \bibinfo {eid} {L20} (\bibinfo {year} {2015})},\ \Eprint {http://arxiv.org/abs/1506.04206} {arXiv:1506.04206 [astro-ph.HE]} \BibitemShut {NoStop}%
\bibitem [{\citenamefont {{Borghese}}\ \emph {et~al.}(2017)\citenamefont {{Borghese}}, \citenamefont {{Rea}}, \citenamefont {{Coti Zelati}}, \citenamefont {{Tiengo}}, \citenamefont {{Turolla}},\ and\ \citenamefont {{Zane}}}]{2017MNRAS.468.2975B}%
  \BibitemOpen
  \bibfield  {author} {\bibinfo {author} {\bibfnamefont {A.}~\bibnamefont {{Borghese}}}, \bibinfo {author} {\bibfnamefont {N.}~\bibnamefont {{Rea}}}, \bibinfo {author} {\bibfnamefont {F.}~\bibnamefont {{Coti Zelati}}}, \bibinfo {author} {\bibfnamefont {A.}~\bibnamefont {{Tiengo}}}, \bibinfo {author} {\bibfnamefont {R.}~\bibnamefont {{Turolla}}}, \ and\ \bibinfo {author} {\bibfnamefont {S.}~\bibnamefont {{Zane}}},\ }\href {\doibase 10.1093/mnras/stx632} {\bibfield  {journal} {\bibinfo  {journal} {\mnras}\ }\textbf {\bibinfo {volume} {468}},\ \bibinfo {pages} {2975} (\bibinfo {year} {2017})},\ \Eprint {http://arxiv.org/abs/1703.05336} {arXiv:1703.05336 [astro-ph.HE]} \BibitemShut {NoStop}%
\bibitem [{\citenamefont {{Staubert}}\ \emph {et~al.}(2019)\citenamefont {{Staubert}}, \citenamefont {{Tr{\"u}mper}}, \citenamefont {{Kendziorra}}, \citenamefont {{Klochkov}}, \citenamefont {{Postnov}}, \citenamefont {{Kretschmar}}, \citenamefont {{Pottschmidt}}, \citenamefont {{Haberl}}, \citenamefont {{Rothschild}}, \citenamefont {{Santangelo}}, \citenamefont {{Wilms}}, \citenamefont {{Kreykenbohm}},\ and\ \citenamefont {{F{\"u}rst}}}]{2019A&A...622A..61S}%
  \BibitemOpen
  \bibfield  {author} {\bibinfo {author} {\bibfnamefont {R.}~\bibnamefont {{Staubert}}}, \bibinfo {author} {\bibfnamefont {J.}~\bibnamefont {{Tr{\"u}mper}}}, \bibinfo {author} {\bibfnamefont {E.}~\bibnamefont {{Kendziorra}}}, \bibinfo {author} {\bibfnamefont {D.}~\bibnamefont {{Klochkov}}}, \bibinfo {author} {\bibfnamefont {K.}~\bibnamefont {{Postnov}}}, \bibinfo {author} {\bibfnamefont {P.}~\bibnamefont {{Kretschmar}}}, \bibinfo {author} {\bibfnamefont {K.}~\bibnamefont {{Pottschmidt}}}, \bibinfo {author} {\bibfnamefont {F.}~\bibnamefont {{Haberl}}}, \bibinfo {author} {\bibfnamefont {R.~E.}\ \bibnamefont {{Rothschild}}}, \bibinfo {author} {\bibfnamefont {A.}~\bibnamefont {{Santangelo}}}, \bibinfo {author} {\bibfnamefont {J.}~\bibnamefont {{Wilms}}}, \bibinfo {author} {\bibfnamefont {I.}~\bibnamefont {{Kreykenbohm}}}, \ and\ \bibinfo {author} {\bibfnamefont {F.}~\bibnamefont {{F{\"u}rst}}},\ }\href {\doibase 10.1051/0004-6361/201834479} {\bibfield  {journal} {\bibinfo  {journal} {\aap}\ }\textbf {\bibinfo
  {volume} {622}},\ \bibinfo {eid} {A61} (\bibinfo {year} {2019})},\ \Eprint {http://arxiv.org/abs/1812.03461} {arXiv:1812.03461 [astro-ph.HE]} \BibitemShut {NoStop}%
\bibitem [{\citenamefont {{Mushtukov}}\ and\ \citenamefont {{Tsygankov}}(2024)}]{Mushtukov2022}%
  \BibitemOpen
  \bibfield  {author} {\bibinfo {author} {\bibfnamefont {A.}~\bibnamefont {{Mushtukov}}}\ and\ \bibinfo {author} {\bibfnamefont {S.}~\bibnamefont {{Tsygankov}}},\ }\enquote {\bibinfo {title} {{Accreting strongly magnetised neutron stars: X-ray Pulsars}},}\ in\ \href {\doibase 10.1007/978-981-19-6960-7_104} {\emph {\bibinfo {booktitle} {Handbook of X-ray and Gamma-ray Astrophysics}}},\ \bibinfo {editor} {edited by\ \bibinfo {editor} {\bibfnamefont {C.}~\bibnamefont {{Bambi}}}\ and\ \bibinfo {editor} {\bibfnamefont {A.}~\bibnamefont {{Santangelo}}}}\ (\bibinfo  {publisher} {Springer},\ \bibinfo {address} {Singapore},\ \bibinfo {year} {2024})\ pp.\ \bibinfo {pages} {4105--4176},\ \Eprint {http://arxiv.org/abs/2204.14185} {arXiv:2204.14185 [astro-ph.HE]} \BibitemShut {NoStop}%
\bibitem [{\citenamefont {{Doroshenko}}\ \emph {et~al.}(2015)\citenamefont {{Doroshenko}}, \citenamefont {{Santangelo}}, \citenamefont {{Doroshenko}}, \citenamefont {{Suleimanov}},\ and\ \citenamefont {{Piraino}}}]{2015MNRAS.452.2490D}%
  \BibitemOpen
  \bibfield  {author} {\bibinfo {author} {\bibfnamefont {R.}~\bibnamefont {{Doroshenko}}}, \bibinfo {author} {\bibfnamefont {A.}~\bibnamefont {{Santangelo}}}, \bibinfo {author} {\bibfnamefont {V.}~\bibnamefont {{Doroshenko}}}, \bibinfo {author} {\bibfnamefont {V.}~\bibnamefont {{Suleimanov}}}, \ and\ \bibinfo {author} {\bibfnamefont {S.}~\bibnamefont {{Piraino}}},\ }\href {\doibase 10.1093/mnras/stv1418} {\bibfield  {journal} {\bibinfo  {journal} {\mnras}\ }\textbf {\bibinfo {volume} {452}},\ \bibinfo {pages} {2490} (\bibinfo {year} {2015})}\BibitemShut {NoStop}%
\bibitem [{\citenamefont {{Molkov}}\ \emph {et~al.}(2019)\citenamefont {{Molkov}}, \citenamefont {{Lutovinov}}, \citenamefont {{Tsygankov}}, \citenamefont {{Mereminskiy}},\ and\ \citenamefont {{Mushtukov}}}]{2019ApJ...883L..11M}%
  \BibitemOpen
  \bibfield  {author} {\bibinfo {author} {\bibfnamefont {S.}~\bibnamefont {{Molkov}}}, \bibinfo {author} {\bibfnamefont {A.}~\bibnamefont {{Lutovinov}}}, \bibinfo {author} {\bibfnamefont {S.}~\bibnamefont {{Tsygankov}}}, \bibinfo {author} {\bibfnamefont {I.}~\bibnamefont {{Mereminskiy}}}, \ and\ \bibinfo {author} {\bibfnamefont {A.}~\bibnamefont {{Mushtukov}}},\ }\href {\doibase 10.3847/2041-8213/ab3e4d} {\bibfield  {journal} {\bibinfo  {journal} {\apjl}\ }\textbf {\bibinfo {volume} {883}},\ \bibinfo {eid} {L11} (\bibinfo {year} {2019})},\ \Eprint {http://arxiv.org/abs/1909.09159} {arXiv:1909.09159 [astro-ph.HE]} \BibitemShut {NoStop}%
\bibitem [{\citenamefont {{Molkov}}\ \emph {et~al.}(2021)\citenamefont {{Molkov}}, \citenamefont {{Doroshenko}}, \citenamefont {{Lutovinov}}, \citenamefont {{Tsygankov}}, \citenamefont {{Santangelo}}, \citenamefont {{Mereminskiy}},\ and\ \citenamefont {{Semena}}}]{2021ApJ...915L..27M}%
  \BibitemOpen
  \bibfield  {author} {\bibinfo {author} {\bibfnamefont {S.}~\bibnamefont {{Molkov}}}, \bibinfo {author} {\bibfnamefont {V.}~\bibnamefont {{Doroshenko}}}, \bibinfo {author} {\bibfnamefont {A.}~\bibnamefont {{Lutovinov}}}, \bibinfo {author} {\bibfnamefont {S.}~\bibnamefont {{Tsygankov}}}, \bibinfo {author} {\bibfnamefont {A.}~\bibnamefont {{Santangelo}}}, \bibinfo {author} {\bibfnamefont {I.}~\bibnamefont {{Mereminskiy}}}, \ and\ \bibinfo {author} {\bibfnamefont {A.}~\bibnamefont {{Semena}}},\ }\href {\doibase 10.3847/2041-8213/ac0c15} {\bibfield  {journal} {\bibinfo  {journal} {\apjl}\ }\textbf {\bibinfo {volume} {915}},\ \bibinfo {eid} {L27} (\bibinfo {year} {2021})},\ \Eprint {http://arxiv.org/abs/2106.09514} {arXiv:2106.09514 [astro-ph.HE]} \BibitemShut {NoStop}%
\bibitem [{\citenamefont {{Mushtukov}}\ \emph {et~al.}(2021)\citenamefont {{Mushtukov}}, \citenamefont {{Suleimanov}}, \citenamefont {{Tsygankov}},\ and\ \citenamefont {{Portegies Zwart}}}]{Mushtukov21}%
  \BibitemOpen
  \bibfield  {author} {\bibinfo {author} {\bibfnamefont {A.~A.}\ \bibnamefont {{Mushtukov}}}, \bibinfo {author} {\bibfnamefont {V.~F.}\ \bibnamefont {{Suleimanov}}}, \bibinfo {author} {\bibfnamefont {S.~S.}\ \bibnamefont {{Tsygankov}}}, \ and\ \bibinfo {author} {\bibfnamefont {S.}~\bibnamefont {{Portegies Zwart}}},\ }\href {\doibase 10.1093/mnras/stab811} {\bibfield  {journal} {\bibinfo  {journal} {\mnras}\ }\textbf {\bibinfo {volume} {503}},\ \bibinfo {pages} {5193} (\bibinfo {year} {2021})},\ \Eprint {http://arxiv.org/abs/2006.13596} {arXiv:2006.13596 [astro-ph.HE]} \BibitemShut {NoStop}%
\bibitem [{\citenamefont {{Sokolova-Lapa}}\ \emph {et~al.}(2021)\citenamefont {{Sokolova-Lapa}}, \citenamefont {{Gornostaev}}, \citenamefont {{Wilms}}, \citenamefont {{Ballhausen}}, \citenamefont {{Falkner}}, \citenamefont {{Postnov}}, \citenamefont {{Thalhammer}}, \citenamefont {{F{\"u}rst}}, \citenamefont {{Garc{\'\i}a}}, \citenamefont {{Shakura}}, \citenamefont {{Becker}}, \citenamefont {{Wolff}}, \citenamefont {{Pottschmidt}}, \citenamefont {{H{\"a}rer}},\ and\ \citenamefont {{Malacaria}}}]{SokolovaLapa21}%
  \BibitemOpen
  \bibfield  {author} {\bibinfo {author} {\bibfnamefont {E.}~\bibnamefont {{Sokolova-Lapa}}}, \bibinfo {author} {\bibfnamefont {M.}~\bibnamefont {{Gornostaev}}}, \bibinfo {author} {\bibfnamefont {J.}~\bibnamefont {{Wilms}}}, \bibinfo {author} {\bibfnamefont {R.}~\bibnamefont {{Ballhausen}}}, \bibinfo {author} {\bibfnamefont {S.}~\bibnamefont {{Falkner}}}, \bibinfo {author} {\bibfnamefont {K.}~\bibnamefont {{Postnov}}}, \bibinfo {author} {\bibfnamefont {P.}~\bibnamefont {{Thalhammer}}}, \bibinfo {author} {\bibfnamefont {F.}~\bibnamefont {{F{\"u}rst}}}, \bibinfo {author} {\bibfnamefont {J.~A.}\ \bibnamefont {{Garc{\'\i}a}}}, \bibinfo {author} {\bibfnamefont {N.}~\bibnamefont {{Shakura}}}, \bibinfo {author} {\bibfnamefont {P.~A.}\ \bibnamefont {{Becker}}}, \bibinfo {author} {\bibfnamefont {M.~T.}\ \bibnamefont {{Wolff}}}, \bibinfo {author} {\bibfnamefont {K.}~\bibnamefont {{Pottschmidt}}}, \bibinfo {author} {\bibfnamefont {L.}~\bibnamefont {{H{\"a}rer}}}, \ and\ \bibinfo {author} {\bibfnamefont {C.}~\bibnamefont
  {{Malacaria}}},\ }\href {\doibase 10.1051/0004-6361/202040228} {\bibfield  {journal} {\bibinfo  {journal} {\aap}\ }\textbf {\bibinfo {volume} {651}},\ \bibinfo {eid} {A12} (\bibinfo {year} {2021})},\ \Eprint {http://arxiv.org/abs/2104.06802} {arXiv:2104.06802 [astro-ph.HE]} \BibitemShut {NoStop}%
\bibitem [{\citenamefont {{Becker}}\ \emph {et~al.}(2012{\natexlab{a}})\citenamefont {{Becker}}, \citenamefont {{Klochkov}}, \citenamefont {{Sch{\"o}nherr}}, \citenamefont {{Nishimura}}, \citenamefont {{Ferrigno}}, \citenamefont {{Caballero}}, \citenamefont {{Kretschmar}}, \citenamefont {{Wolff}}, \citenamefont {{Wilms}},\ and\ \citenamefont {{Staubert}}}]{Becker2012}%
  \BibitemOpen
  \bibfield  {author} {\bibinfo {author} {\bibfnamefont {P.~A.}\ \bibnamefont {{Becker}}}, \bibinfo {author} {\bibfnamefont {D.}~\bibnamefont {{Klochkov}}}, \bibinfo {author} {\bibfnamefont {G.}~\bibnamefont {{Sch{\"o}nherr}}}, \bibinfo {author} {\bibfnamefont {O.}~\bibnamefont {{Nishimura}}}, \bibinfo {author} {\bibfnamefont {C.}~\bibnamefont {{Ferrigno}}}, \bibinfo {author} {\bibfnamefont {I.}~\bibnamefont {{Caballero}}}, \bibinfo {author} {\bibfnamefont {P.}~\bibnamefont {{Kretschmar}}}, \bibinfo {author} {\bibfnamefont {M.~T.}\ \bibnamefont {{Wolff}}}, \bibinfo {author} {\bibfnamefont {J.}~\bibnamefont {{Wilms}}}, \ and\ \bibinfo {author} {\bibfnamefont {R.}~\bibnamefont {{Staubert}}},\ }\href {\doibase 10.1051/0004-6361/201219065} {\bibfield  {journal} {\bibinfo  {journal} {\aap}\ }\textbf {\bibinfo {volume} {544}},\ \bibinfo {eid} {A123} (\bibinfo {year} {2012}{\natexlab{a}})},\ \Eprint {http://arxiv.org/abs/1205.5316} {arXiv:1205.5316 [astro-ph.HE]} \BibitemShut {NoStop}%
\bibitem [{\citenamefont {{Poutanen}}\ \emph {et~al.}(2024{\natexlab{a}})\citenamefont {{Poutanen}}, \citenamefont {{Tsygankov}},\ and\ \citenamefont {{Forsblom}}}]{Poutanen2024}%
  \BibitemOpen
  \bibfield  {author} {\bibinfo {author} {\bibfnamefont {J.}~\bibnamefont {{Poutanen}}}, \bibinfo {author} {\bibfnamefont {S.~S.}\ \bibnamefont {{Tsygankov}}}, \ and\ \bibinfo {author} {\bibfnamefont {S.~V.}\ \bibnamefont {{Forsblom}}},\ }\href {\doibase 10.3390/galaxies12040046} {\bibfield  {journal} {\bibinfo  {journal} {Galaxies}\ }\textbf {\bibinfo {volume} {12}},\ \bibinfo {eid} {46} (\bibinfo {year} {2024}{\natexlab{a}})},\ \Eprint {http://arxiv.org/abs/2408.04431} {arXiv:2408.04431 [astro-ph.HE]} \BibitemShut {NoStop}%
\bibitem [{\citenamefont {{London}}\ \emph {et~al.}(1986)\citenamefont {{London}}, \citenamefont {{Taam}},\ and\ \citenamefont {{Howard}}}]{1986ApJ...306..170L}%
  \BibitemOpen
  \bibfield  {author} {\bibinfo {author} {\bibfnamefont {R.~A.}\ \bibnamefont {{London}}}, \bibinfo {author} {\bibfnamefont {R.~E.}\ \bibnamefont {{Taam}}}, \ and\ \bibinfo {author} {\bibfnamefont {W.~M.}\ \bibnamefont {{Howard}}},\ }\href {\doibase 10.1086/164330} {\bibfield  {journal} {\bibinfo  {journal} {\apj}\ }\textbf {\bibinfo {volume} {306}},\ \bibinfo {pages} {170} (\bibinfo {year} {1986})}\BibitemShut {NoStop}%
\bibitem [{\citenamefont {{Pavlov}}\ \emph {et~al.}(2004)\citenamefont {{Pavlov}}, \citenamefont {{Sanwal}},\ and\ \citenamefont {{Teter}}}]{2004IAUS..218..239P}%
  \BibitemOpen
  \bibfield  {author} {\bibinfo {author} {\bibfnamefont {G.~G.}\ \bibnamefont {{Pavlov}}}, \bibinfo {author} {\bibfnamefont {D.}~\bibnamefont {{Sanwal}}}, \ and\ \bibinfo {author} {\bibfnamefont {M.~A.}\ \bibnamefont {{Teter}}},\ }in\ \href {\doibase 10.48550/arXiv.astro-ph/0311526} {\emph {\bibinfo {booktitle} {Young Neutron Stars and Their Environments}}},\ \bibinfo {series} {IAU Symposium}, Vol.\ \bibinfo {volume} {218},\ \bibinfo {editor} {edited by\ \bibinfo {editor} {\bibfnamefont {F.}~\bibnamefont {{Camilo}}}\ and\ \bibinfo {editor} {\bibfnamefont {B.~M.}\ \bibnamefont {{Gaensler}}}}\ (\bibinfo {year} {2004})\ p.\ \bibinfo {pages} {239},\ \Eprint {http://arxiv.org/abs/astro-ph/0311526} {arXiv:astro-ph/0311526 [astro-ph]} \BibitemShut {NoStop}%
\bibitem [{\citenamefont {{De Luca}}(2017)}]{2017JPhCS.932a2006D}%
  \BibitemOpen
  \bibfield  {author} {\bibinfo {author} {\bibfnamefont {A.}~\bibnamefont {{De Luca}}},\ }in\ \href {\doibase 10.1088/1742-6596/932/1/012006} {\emph {\bibinfo {booktitle} {Journal of Physics Conference Series}}},\ \bibinfo {series} {Journal of Physics Conference Series}, Vol.\ \bibinfo {volume} {932}\ (\bibinfo  {publisher} {IOP},\ \bibinfo {year} {2017})\ p.\ \bibinfo {pages} {012006},\ \Eprint {http://arxiv.org/abs/1711.07210} {arXiv:1711.07210 [astro-ph.HE]} \BibitemShut {NoStop}%
\bibitem [{\citenamefont {{Shibanov}}\ \emph {et~al.}(1992)\citenamefont {{Shibanov}}, \citenamefont {{Zavlin}}, \citenamefont {{Pavlov}},\ and\ \citenamefont {{Ventura}}}]{1992A&A...266..313S}%
  \BibitemOpen
  \bibfield  {author} {\bibinfo {author} {\bibfnamefont {I.~A.}\ \bibnamefont {{Shibanov}}}, \bibinfo {author} {\bibfnamefont {V.~E.}\ \bibnamefont {{Zavlin}}}, \bibinfo {author} {\bibfnamefont {G.~G.}\ \bibnamefont {{Pavlov}}}, \ and\ \bibinfo {author} {\bibfnamefont {J.}~\bibnamefont {{Ventura}}},\ }\href@noop {} {\bibfield  {journal} {\bibinfo  {journal} {\aap}\ }\textbf {\bibinfo {volume} {266}},\ \bibinfo {pages} {313} (\bibinfo {year} {1992})}\BibitemShut {NoStop}%
\bibitem [{\citenamefont {{Pavlov}}\ \emph {et~al.}(1994)\citenamefont {{Pavlov}}, \citenamefont {{Shibanov}}, \citenamefont {{Ventura}},\ and\ \citenamefont {{Zavlin}}}]{1994A&A...289..837P}%
  \BibitemOpen
  \bibfield  {author} {\bibinfo {author} {\bibfnamefont {G.~G.}\ \bibnamefont {{Pavlov}}}, \bibinfo {author} {\bibfnamefont {Y.~A.}\ \bibnamefont {{Shibanov}}}, \bibinfo {author} {\bibfnamefont {J.}~\bibnamefont {{Ventura}}}, \ and\ \bibinfo {author} {\bibfnamefont {V.~E.}\ \bibnamefont {{Zavlin}}},\ }\href@noop {} {\bibfield  {journal} {\bibinfo  {journal} {\aap}\ }\textbf {\bibinfo {volume} {289}},\ \bibinfo {pages} {837} (\bibinfo {year} {1994})}\BibitemShut {NoStop}%
\bibitem [{\citenamefont {{Zane}}\ \emph {et~al.}(2000)\citenamefont {{Zane}}, \citenamefont {{Turolla}},\ and\ \citenamefont {{Treves}}}]{2000ApJ...537..387Z}%
  \BibitemOpen
  \bibfield  {author} {\bibinfo {author} {\bibfnamefont {S.}~\bibnamefont {{Zane}}}, \bibinfo {author} {\bibfnamefont {R.}~\bibnamefont {{Turolla}}}, \ and\ \bibinfo {author} {\bibfnamefont {A.}~\bibnamefont {{Treves}}},\ }\href {\doibase 10.1086/309027} {\bibfield  {journal} {\bibinfo  {journal} {\apj}\ }\textbf {\bibinfo {volume} {537}},\ \bibinfo {pages} {387} (\bibinfo {year} {2000})}\BibitemShut {NoStop}%
\bibitem [{\citenamefont {{Ho}}\ \emph {et~al.}(2003)\citenamefont {{Ho}}, \citenamefont {{Lai}}, \citenamefont {{Potekhin}},\ and\ \citenamefont {{Chabrier}}}]{2003ApJ...599.1293H}%
  \BibitemOpen
  \bibfield  {author} {\bibinfo {author} {\bibfnamefont {W.~C.~G.}\ \bibnamefont {{Ho}}}, \bibinfo {author} {\bibfnamefont {D.}~\bibnamefont {{Lai}}}, \bibinfo {author} {\bibfnamefont {A.~Y.}\ \bibnamefont {{Potekhin}}}, \ and\ \bibinfo {author} {\bibfnamefont {G.}~\bibnamefont {{Chabrier}}},\ }\href {\doibase 10.1086/379507} {\bibfield  {journal} {\bibinfo  {journal} {\apj}\ }\textbf {\bibinfo {volume} {599}},\ \bibinfo {pages} {1293} (\bibinfo {year} {2003})},\ \Eprint {http://arxiv.org/abs/astro-ph/0309261} {arXiv:astro-ph/0309261 [astro-ph]} \BibitemShut {NoStop}%
\bibitem [{\citenamefont {{Potekhin}}\ \emph {et~al.}(2004)\citenamefont {{Potekhin}}, \citenamefont {{Lai}}, \citenamefont {{Chabrier}},\ and\ \citenamefont {{Ho}}}]{2004ApJ...612.1034P}%
  \BibitemOpen
  \bibfield  {author} {\bibinfo {author} {\bibfnamefont {A.~Y.}\ \bibnamefont {{Potekhin}}}, \bibinfo {author} {\bibfnamefont {D.}~\bibnamefont {{Lai}}}, \bibinfo {author} {\bibfnamefont {G.}~\bibnamefont {{Chabrier}}}, \ and\ \bibinfo {author} {\bibfnamefont {W.~C.~G.}\ \bibnamefont {{Ho}}},\ }\href {\doibase 10.1086/422679} {\bibfield  {journal} {\bibinfo  {journal} {\apj}\ }\textbf {\bibinfo {volume} {612}},\ \bibinfo {pages} {1034} (\bibinfo {year} {2004})},\ \Eprint {http://arxiv.org/abs/astro-ph/0405383} {arXiv:astro-ph/0405383 [astro-ph]} \BibitemShut {NoStop}%
\bibitem [{\citenamefont {{Mori}}\ and\ \citenamefont {{Ho}}(2007)}]{2007MNRAS.377..905M}%
  \BibitemOpen
  \bibfield  {author} {\bibinfo {author} {\bibfnamefont {K.}~\bibnamefont {{Mori}}}\ and\ \bibinfo {author} {\bibfnamefont {W.~C.~G.}\ \bibnamefont {{Ho}}},\ }\href {\doibase 10.1111/j.1365-2966.2007.11663.x} {\bibfield  {journal} {\bibinfo  {journal} {\mnras}\ }\textbf {\bibinfo {volume} {377}},\ \bibinfo {pages} {905} (\bibinfo {year} {2007})},\ \Eprint {http://arxiv.org/abs/astro-ph/0611145} {arXiv:astro-ph/0611145 [astro-ph]} \BibitemShut {NoStop}%
\bibitem [{\citenamefont {{Ho}}\ \emph {et~al.}(2007)\citenamefont {{Ho}}, \citenamefont {{Kaplan}}, \citenamefont {{Chang}}, \citenamefont {{van Adelsberg}},\ and\ \citenamefont {{Potekhin}}}]{2007MNRAS.375..821H}%
  \BibitemOpen
  \bibfield  {author} {\bibinfo {author} {\bibfnamefont {W.~C.~G.}\ \bibnamefont {{Ho}}}, \bibinfo {author} {\bibfnamefont {D.~L.}\ \bibnamefont {{Kaplan}}}, \bibinfo {author} {\bibfnamefont {P.}~\bibnamefont {{Chang}}}, \bibinfo {author} {\bibfnamefont {M.}~\bibnamefont {{van Adelsberg}}}, \ and\ \bibinfo {author} {\bibfnamefont {A.~Y.}\ \bibnamefont {{Potekhin}}},\ }\href {\doibase 10.1111/j.1365-2966.2006.11376.x} {\bibfield  {journal} {\bibinfo  {journal} {\mnras}\ }\textbf {\bibinfo {volume} {375}},\ \bibinfo {pages} {821} (\bibinfo {year} {2007})},\ \Eprint {http://arxiv.org/abs/astro-ph/0612145} {arXiv:astro-ph/0612145 [astro-ph]} \BibitemShut {NoStop}%
\bibitem [{\citenamefont {{Suleimanov}}\ \emph {et~al.}(2009)\citenamefont {{Suleimanov}}, \citenamefont {{Potekhin}},\ and\ \citenamefont {{Werner}}}]{2009A&A...500..891S}%
  \BibitemOpen
  \bibfield  {author} {\bibinfo {author} {\bibfnamefont {V.}~\bibnamefont {{Suleimanov}}}, \bibinfo {author} {\bibfnamefont {A.~Y.}\ \bibnamefont {{Potekhin}}}, \ and\ \bibinfo {author} {\bibfnamefont {K.}~\bibnamefont {{Werner}}},\ }\href {\doibase 10.1051/0004-6361/200912121} {\bibfield  {journal} {\bibinfo  {journal} {\aap}\ }\textbf {\bibinfo {volume} {500}},\ \bibinfo {pages} {891} (\bibinfo {year} {2009})},\ \Eprint {http://arxiv.org/abs/0905.3276} {arXiv:0905.3276 [astro-ph.SR]} \BibitemShut {NoStop}%
\bibitem [{\citenamefont {{Suleimanov}}\ \emph {et~al.}(2012)\citenamefont {{Suleimanov}}, \citenamefont {{Pavlov}},\ and\ \citenamefont {{Werner}}}]{2012ApJ...751...15S}%
  \BibitemOpen
  \bibfield  {author} {\bibinfo {author} {\bibfnamefont {V.~F.}\ \bibnamefont {{Suleimanov}}}, \bibinfo {author} {\bibfnamefont {G.~G.}\ \bibnamefont {{Pavlov}}}, \ and\ \bibinfo {author} {\bibfnamefont {K.}~\bibnamefont {{Werner}}},\ }\href {\doibase 10.1088/0004-637X/751/1/15} {\bibfield  {journal} {\bibinfo  {journal} {\apj}\ }\textbf {\bibinfo {volume} {751}},\ \bibinfo {eid} {15} (\bibinfo {year} {2012})},\ \Eprint {http://arxiv.org/abs/1201.5527} {arXiv:1201.5527 [astro-ph.HE]} \BibitemShut {NoStop}%
\bibitem [{\citenamefont {{Ho}}\ and\ \citenamefont {{Lai}}(2001)}]{2001MNRAS.327.1081H}%
  \BibitemOpen
  \bibfield  {author} {\bibinfo {author} {\bibfnamefont {W.~C.~G.}\ \bibnamefont {{Ho}}}\ and\ \bibinfo {author} {\bibfnamefont {D.}~\bibnamefont {{Lai}}},\ }\href {\doibase 10.1046/j.1365-8711.2001.04801.x} {\bibfield  {journal} {\bibinfo  {journal} {\mnras}\ }\textbf {\bibinfo {volume} {327}},\ \bibinfo {pages} {1081} (\bibinfo {year} {2001})},\ \Eprint {http://arxiv.org/abs/astro-ph/0104199} {arXiv:astro-ph/0104199 [astro-ph]} \BibitemShut {NoStop}%
\bibitem [{\citenamefont {{{\"O}zel}}(2001)}]{2001ApJ...563..276O}%
  \BibitemOpen
  \bibfield  {author} {\bibinfo {author} {\bibfnamefont {F.}~\bibnamefont {{{\"O}zel}}},\ }\href {\doibase 10.1086/323851} {\bibfield  {journal} {\bibinfo  {journal} {\apj}\ }\textbf {\bibinfo {volume} {563}},\ \bibinfo {pages} {276} (\bibinfo {year} {2001})},\ \Eprint {http://arxiv.org/abs/astro-ph/0103227} {arXiv:astro-ph/0103227 [astro-ph]} \BibitemShut {NoStop}%
\bibitem [{\citenamefont {{Zane}}\ \emph {et~al.}(2001)\citenamefont {{Zane}}, \citenamefont {{Turolla}}, \citenamefont {{Stella}},\ and\ \citenamefont {{Treves}}}]{2001ApJ...560..384Z}%
  \BibitemOpen
  \bibfield  {author} {\bibinfo {author} {\bibfnamefont {S.}~\bibnamefont {{Zane}}}, \bibinfo {author} {\bibfnamefont {R.}~\bibnamefont {{Turolla}}}, \bibinfo {author} {\bibfnamefont {L.}~\bibnamefont {{Stella}}}, \ and\ \bibinfo {author} {\bibfnamefont {A.}~\bibnamefont {{Treves}}},\ }\href {\doibase 10.1086/322360} {\bibfield  {journal} {\bibinfo  {journal} {\apj}\ }\textbf {\bibinfo {volume} {560}},\ \bibinfo {pages} {384} (\bibinfo {year} {2001})},\ \Eprint {http://arxiv.org/abs/astro-ph/0103316} {arXiv:astro-ph/0103316 [astro-ph]} \BibitemShut {NoStop}%
\bibitem [{\citenamefont {{van Adelsberg}}\ and\ \citenamefont {{Lai}}(2006)}]{vAL06}%
  \BibitemOpen
  \bibfield  {author} {\bibinfo {author} {\bibfnamefont {M.}~\bibnamefont {{van Adelsberg}}}\ and\ \bibinfo {author} {\bibfnamefont {D.}~\bibnamefont {{Lai}}},\ }\href {\doibase 10.1111/j.1365-2966.2006.11098.x} {\bibfield  {journal} {\bibinfo  {journal} {\mnras}\ }\textbf {\bibinfo {volume} {373}},\ \bibinfo {pages} {1495} (\bibinfo {year} {2006})},\ \Eprint {http://arxiv.org/abs/astro-ph/0607168} {arXiv:astro-ph/0607168 [astro-ph]} \BibitemShut {NoStop}%
\bibitem [{\citenamefont {{Potekhin}}(2014)}]{2014PhyU...57..735P}%
  \BibitemOpen
  \bibfield  {author} {\bibinfo {author} {\bibfnamefont {A.~Y.}\ \bibnamefont {{Potekhin}}},\ }\href {\doibase 10.3367/UFNe.0184.201408a.0793} {\bibfield  {journal} {\bibinfo  {journal} {Physics Uspekhi}\ }\textbf {\bibinfo {volume} {57}},\ \bibinfo {eid} {735-770} (\bibinfo {year} {2014})},\ \Eprint {http://arxiv.org/abs/1403.0074} {arXiv:1403.0074 [astro-ph.SR]} \BibitemShut {NoStop}%
\bibitem [{\citenamefont {{Ventura}}(1979)}]{1979PhRvD..19.1684V}%
  \BibitemOpen
  \bibfield  {author} {\bibinfo {author} {\bibfnamefont {J.}~\bibnamefont {{Ventura}}},\ }\href {\doibase 10.1103/PhysRevD.19.1684} {\bibfield  {journal} {\bibinfo  {journal} {\prd}\ }\textbf {\bibinfo {volume} {19}},\ \bibinfo {pages} {1684} (\bibinfo {year} {1979})}\BibitemShut {NoStop}%
\bibitem [{\citenamefont {{Ho}}\ and\ \citenamefont {{Lai}}(2003)}]{2003MNRAS.338..233H}%
  \BibitemOpen
  \bibfield  {author} {\bibinfo {author} {\bibfnamefont {W.~C.~G.}\ \bibnamefont {{Ho}}}\ and\ \bibinfo {author} {\bibfnamefont {D.}~\bibnamefont {{Lai}}},\ }\href {\doibase 10.1046/j.1365-8711.2003.06047.x} {\bibfield  {journal} {\bibinfo  {journal} {\mnras}\ }\textbf {\bibinfo {volume} {338}},\ \bibinfo {pages} {233} (\bibinfo {year} {2003})},\ \Eprint {http://arxiv.org/abs/astro-ph/0201380} {arXiv:astro-ph/0201380 [astro-ph]} \BibitemShut {NoStop}%
\bibitem [{\citenamefont {{Harding}}\ and\ \citenamefont {{Lai}}(2006)}]{2006RPPh...69.2631H}%
  \BibitemOpen
  \bibfield  {author} {\bibinfo {author} {\bibfnamefont {A.~K.}\ \bibnamefont {{Harding}}}\ and\ \bibinfo {author} {\bibfnamefont {D.}~\bibnamefont {{Lai}}},\ }\href {\doibase 10.1088/0034-4885/69/9/R03} {\bibfield  {journal} {\bibinfo  {journal} {Reports on Progress in Physics}\ }\textbf {\bibinfo {volume} {69}},\ \bibinfo {pages} {2631} (\bibinfo {year} {2006})},\ \Eprint {http://arxiv.org/abs/astro-ph/0606674} {arXiv:astro-ph/0606674 [astro-ph]} \BibitemShut {NoStop}%
\bibitem [{\citenamefont {{Pavlov}}\ and\ \citenamefont {{Shibanov}}(1979)}]{1979JETP...49..741P}%
  \BibitemOpen
  \bibfield  {author} {\bibinfo {author} {\bibfnamefont {G.~G.}\ \bibnamefont {{Pavlov}}}\ and\ \bibinfo {author} {\bibfnamefont {Y.~A.}\ \bibnamefont {{Shibanov}}},\ }\href@noop {} {\bibfield  {journal} {\bibinfo  {journal} {Soviet Journal of Experimental and Theoretical Physics}\ }\textbf {\bibinfo {volume} {49}},\ \bibinfo {pages} {741} (\bibinfo {year} {1979})}\BibitemShut {NoStop}%
\bibitem [{\citenamefont {{Kelly}}\ \emph {et~al.}(2024{\natexlab{a}})\citenamefont {{Kelly}}, \citenamefont {{Zane}}, \citenamefont {{Turolla}},\ and\ \citenamefont {{Taverna}}}]{2024MNRAS.528.3927K}%
  \BibitemOpen
  \bibfield  {author} {\bibinfo {author} {\bibfnamefont {R.~M.~E.}\ \bibnamefont {{Kelly}}}, \bibinfo {author} {\bibfnamefont {S.}~\bibnamefont {{Zane}}}, \bibinfo {author} {\bibfnamefont {R.}~\bibnamefont {{Turolla}}}, \ and\ \bibinfo {author} {\bibfnamefont {R.}~\bibnamefont {{Taverna}}},\ }\href {\doibase 10.1093/mnras/stae159} {\bibfield  {journal} {\bibinfo  {journal} {\mnras}\ }\textbf {\bibinfo {volume} {528}},\ \bibinfo {pages} {3927} (\bibinfo {year} {2024}{\natexlab{a}})},\ \Eprint {http://arxiv.org/abs/2401.07618} {arXiv:2401.07618 [astro-ph.HE]} \BibitemShut {NoStop}%
\bibitem [{\citenamefont {{Gonz{\'a}lez-Caniulef}}\ \emph {et~al.}(2019)\citenamefont {{Gonz{\'a}lez-Caniulef}}, \citenamefont {{Zane}}, \citenamefont {{Turolla}},\ and\ \citenamefont {{Wu}}}]{2019MNRAS.483..599G}%
  \BibitemOpen
  \bibfield  {author} {\bibinfo {author} {\bibfnamefont {D.}~\bibnamefont {{Gonz{\'a}lez-Caniulef}}}, \bibinfo {author} {\bibfnamefont {S.}~\bibnamefont {{Zane}}}, \bibinfo {author} {\bibfnamefont {R.}~\bibnamefont {{Turolla}}}, \ and\ \bibinfo {author} {\bibfnamefont {K.}~\bibnamefont {{Wu}}},\ }\href {\doibase 10.1093/mnras/sty3159} {\bibfield  {journal} {\bibinfo  {journal} {\mnras}\ }\textbf {\bibinfo {volume} {483}},\ \bibinfo {pages} {599} (\bibinfo {year} {2019})},\ \Eprint {http://arxiv.org/abs/1811.08526} {arXiv:1811.08526 [astro-ph.HE]} \BibitemShut {NoStop}%
\bibitem [{\citenamefont {{Kelly}}\ \emph {et~al.}(2024{\natexlab{b}})\citenamefont {{Kelly}}, \citenamefont {{Gonz{\'a}lez-Caniulef}}, \citenamefont {{Zane}}, \citenamefont {{Turolla}},\ and\ \citenamefont {{Taverna}}}]{2024MNRAS.534.1355K}%
  \BibitemOpen
  \bibfield  {author} {\bibinfo {author} {\bibfnamefont {R.~M.~E.}\ \bibnamefont {{Kelly}}}, \bibinfo {author} {\bibfnamefont {D.}~\bibnamefont {{Gonz{\'a}lez-Caniulef}}}, \bibinfo {author} {\bibfnamefont {S.}~\bibnamefont {{Zane}}}, \bibinfo {author} {\bibfnamefont {R.}~\bibnamefont {{Turolla}}}, \ and\ \bibinfo {author} {\bibfnamefont {R.}~\bibnamefont {{Taverna}}},\ }\href {\doibase 10.1093/mnras/stae2163} {\bibfield  {journal} {\bibinfo  {journal} {\mnras}\ }\textbf {\bibinfo {volume} {534}},\ \bibinfo {pages} {1355} (\bibinfo {year} {2024}{\natexlab{b}})},\ \Eprint {http://arxiv.org/abs/2409.11523} {arXiv:2409.11523 [astro-ph.HE]} \BibitemShut {NoStop}%
\bibitem [{\citenamefont {{Lai}}\ and\ \citenamefont {{Salpeter}}(1997)}]{1997ApJ...491..270L}%
  \BibitemOpen
  \bibfield  {author} {\bibinfo {author} {\bibfnamefont {D.}~\bibnamefont {{Lai}}}\ and\ \bibinfo {author} {\bibfnamefont {E.~E.}\ \bibnamefont {{Salpeter}}},\ }\href {\doibase 10.1086/304937} {\bibfield  {journal} {\bibinfo  {journal} {\apj}\ }\textbf {\bibinfo {volume} {491}},\ \bibinfo {pages} {270} (\bibinfo {year} {1997})},\ \Eprint {http://arxiv.org/abs/astro-ph/9704130} {arXiv:astro-ph/9704130 [astro-ph]} \BibitemShut {NoStop}%
\bibitem [{\citenamefont {{Lai}}(2001)}]{2001RvMP...73..629L}%
  \BibitemOpen
  \bibfield  {author} {\bibinfo {author} {\bibfnamefont {D.}~\bibnamefont {{Lai}}},\ }\href {\doibase 10.1103/RevModPhys.73.629} {\bibfield  {journal} {\bibinfo  {journal} {Reviews of Modern Physics}\ }\textbf {\bibinfo {volume} {73}},\ \bibinfo {pages} {629} (\bibinfo {year} {2001})},\ \Eprint {http://arxiv.org/abs/astro-ph/0009333} {arXiv:astro-ph/0009333 [astro-ph]} \BibitemShut {NoStop}%
\bibitem [{\citenamefont {{Turolla}}\ \emph {et~al.}(2004)\citenamefont {{Turolla}}, \citenamefont {{Zane}},\ and\ \citenamefont {{Drake}}}]{2004ApJ...603..265T}%
  \BibitemOpen
  \bibfield  {author} {\bibinfo {author} {\bibfnamefont {R.}~\bibnamefont {{Turolla}}}, \bibinfo {author} {\bibfnamefont {S.}~\bibnamefont {{Zane}}}, \ and\ \bibinfo {author} {\bibfnamefont {J.~J.}\ \bibnamefont {{Drake}}},\ }\href {\doibase 10.1086/379113} {\bibfield  {journal} {\bibinfo  {journal} {\apj}\ }\textbf {\bibinfo {volume} {603}},\ \bibinfo {pages} {265} (\bibinfo {year} {2004})},\ \Eprint {http://arxiv.org/abs/astro-ph/0308326} {arXiv:astro-ph/0308326 [astro-ph]} \BibitemShut {NoStop}%
\bibitem [{\citenamefont {{Potekhin}}\ \emph {et~al.}(2012)\citenamefont {{Potekhin}}, \citenamefont {{Suleimanov}}, \citenamefont {{van Adelsberg}},\ and\ \citenamefont {{Werner}}}]{2012A&A...546A.121P}%
  \BibitemOpen
  \bibfield  {author} {\bibinfo {author} {\bibfnamefont {A.~Y.}\ \bibnamefont {{Potekhin}}}, \bibinfo {author} {\bibfnamefont {V.~F.}\ \bibnamefont {{Suleimanov}}}, \bibinfo {author} {\bibfnamefont {M.}~\bibnamefont {{van Adelsberg}}}, \ and\ \bibinfo {author} {\bibfnamefont {K.}~\bibnamefont {{Werner}}},\ }\href {\doibase 10.1051/0004-6361/201219747} {\bibfield  {journal} {\bibinfo  {journal} {\aap}\ }\textbf {\bibinfo {volume} {546}},\ \bibinfo {eid} {A121} (\bibinfo {year} {2012})},\ \Eprint {http://arxiv.org/abs/1208.6582} {arXiv:1208.6582 [astro-ph.HE]} \BibitemShut {NoStop}%
\bibitem [{\citenamefont {{Bogdanov}}\ and\ \citenamefont {{Ho}}(2024)}]{2024ApJ...969...53B}%
  \BibitemOpen
  \bibfield  {author} {\bibinfo {author} {\bibfnamefont {S.}~\bibnamefont {{Bogdanov}}}\ and\ \bibinfo {author} {\bibfnamefont {W.~C.~G.}\ \bibnamefont {{Ho}}},\ }\href {\doibase 10.3847/1538-4357/ad452b} {\bibfield  {journal} {\bibinfo  {journal} {\apj}\ }\textbf {\bibinfo {volume} {969}},\ \bibinfo {eid} {53} (\bibinfo {year} {2024})},\ \Eprint {http://arxiv.org/abs/2407.00275} {arXiv:2407.00275 [astro-ph.HE]} \BibitemShut {NoStop}%
\bibitem [{\citenamefont {{Pavlov}}\ and\ \citenamefont {{Luna}}(2009)}]{2009ApJ...703..910P}%
  \BibitemOpen
  \bibfield  {author} {\bibinfo {author} {\bibfnamefont {G.~G.}\ \bibnamefont {{Pavlov}}}\ and\ \bibinfo {author} {\bibfnamefont {G.~J.~M.}\ \bibnamefont {{Luna}}},\ }\href {\doibase 10.1088/0004-637X/703/1/910} {\bibfield  {journal} {\bibinfo  {journal} {\apj}\ }\textbf {\bibinfo {volume} {703}},\ \bibinfo {pages} {910} (\bibinfo {year} {2009})},\ \Eprint {http://arxiv.org/abs/0905.3190} {arXiv:0905.3190 [astro-ph.HE]} \BibitemShut {NoStop}%
\bibitem [{\citenamefont {{Klochkov}}\ \emph {et~al.}(2015)\citenamefont {{Klochkov}}, \citenamefont {{Suleimanov}}, \citenamefont {{P{\"u}hlhofer}}, \citenamefont {{Yakovlev}}, \citenamefont {{Santangelo}},\ and\ \citenamefont {{Werner}}}]{2015A&A...573A..53K}%
  \BibitemOpen
  \bibfield  {author} {\bibinfo {author} {\bibfnamefont {D.}~\bibnamefont {{Klochkov}}}, \bibinfo {author} {\bibfnamefont {V.}~\bibnamefont {{Suleimanov}}}, \bibinfo {author} {\bibfnamefont {G.}~\bibnamefont {{P{\"u}hlhofer}}}, \bibinfo {author} {\bibfnamefont {D.~G.}\ \bibnamefont {{Yakovlev}}}, \bibinfo {author} {\bibfnamefont {A.}~\bibnamefont {{Santangelo}}}, \ and\ \bibinfo {author} {\bibfnamefont {K.}~\bibnamefont {{Werner}}},\ }\href {\doibase 10.1051/0004-6361/201424683} {\bibfield  {journal} {\bibinfo  {journal} {\aap}\ }\textbf {\bibinfo {volume} {573}},\ \bibinfo {eid} {A53} (\bibinfo {year} {2015})},\ \Eprint {http://arxiv.org/abs/1410.1055} {arXiv:1410.1055 [astro-ph.HE]} \BibitemShut {NoStop}%
\bibitem [{\citenamefont {{Ho}}\ and\ \citenamefont {{Heinke}}(2009)}]{2009Natur.462...71H}%
  \BibitemOpen
  \bibfield  {author} {\bibinfo {author} {\bibfnamefont {W.~C.~G.}\ \bibnamefont {{Ho}}}\ and\ \bibinfo {author} {\bibfnamefont {C.~O.}\ \bibnamefont {{Heinke}}},\ }\href {\doibase 10.1038/nature08525} {\bibfield  {journal} {\bibinfo  {journal} {\nat}\ }\textbf {\bibinfo {volume} {462}},\ \bibinfo {pages} {71} (\bibinfo {year} {2009})},\ \Eprint {http://arxiv.org/abs/0911.0672} {arXiv:0911.0672 [astro-ph.HE]} \BibitemShut {NoStop}%
\bibitem [{\citenamefont {{Doroshenko}}\ \emph {et~al.}(2022{\natexlab{a}})\citenamefont {{Doroshenko}}, \citenamefont {{Suleimanov}}, \citenamefont {{P{\"u}hlhofer}},\ and\ \citenamefont {{Santangelo}}}]{2022NatAs...6.1444D}%
  \BibitemOpen
  \bibfield  {author} {\bibinfo {author} {\bibfnamefont {V.}~\bibnamefont {{Doroshenko}}}, \bibinfo {author} {\bibfnamefont {V.}~\bibnamefont {{Suleimanov}}}, \bibinfo {author} {\bibfnamefont {G.}~\bibnamefont {{P{\"u}hlhofer}}}, \ and\ \bibinfo {author} {\bibfnamefont {A.}~\bibnamefont {{Santangelo}}},\ }\href {\doibase 10.1038/s41550-022-01800-1} {\bibfield  {journal} {\bibinfo  {journal} {Nature Astronomy}\ }\textbf {\bibinfo {volume} {6}},\ \bibinfo {pages} {1444} (\bibinfo {year} {2022}{\natexlab{a}})}\BibitemShut {NoStop}%
\bibitem [{\citenamefont {{Potekhin}}\ \emph {et~al.}(1999)\citenamefont {{Potekhin}}, \citenamefont {{Baiko}}, \citenamefont {{Haensel}},\ and\ \citenamefont {{Yakovlev}}}]{1999A&A...346..345P}%
  \BibitemOpen
  \bibfield  {author} {\bibinfo {author} {\bibfnamefont {A.~Y.}\ \bibnamefont {{Potekhin}}}, \bibinfo {author} {\bibfnamefont {D.~A.}\ \bibnamefont {{Baiko}}}, \bibinfo {author} {\bibfnamefont {P.}~\bibnamefont {{Haensel}}}, \ and\ \bibinfo {author} {\bibfnamefont {D.~G.}\ \bibnamefont {{Yakovlev}}},\ }\href {\doibase 10.48550/arXiv.astro-ph/9903127} {\bibfield  {journal} {\bibinfo  {journal} {\aap}\ }\textbf {\bibinfo {volume} {346}},\ \bibinfo {pages} {345} (\bibinfo {year} {1999})},\ \Eprint {http://arxiv.org/abs/astro-ph/9903127} {arXiv:astro-ph/9903127 [astro-ph]} \BibitemShut {NoStop}%
\bibitem [{\citenamefont {{Potekhin}}\ \emph {et~al.}(2015)\citenamefont {{Potekhin}}, \citenamefont {{Pons}},\ and\ \citenamefont {{Page}}}]{2015SSRv..191..239P}%
  \BibitemOpen
  \bibfield  {author} {\bibinfo {author} {\bibfnamefont {A.~Y.}\ \bibnamefont {{Potekhin}}}, \bibinfo {author} {\bibfnamefont {J.~A.}\ \bibnamefont {{Pons}}}, \ and\ \bibinfo {author} {\bibfnamefont {D.}~\bibnamefont {{Page}}},\ }\href {\doibase 10.1007/s11214-015-0180-9} {\bibfield  {journal} {\bibinfo  {journal} {\ssr}\ }\textbf {\bibinfo {volume} {191}},\ \bibinfo {pages} {239} (\bibinfo {year} {2015})},\ \Eprint {http://arxiv.org/abs/1507.06186} {arXiv:1507.06186 [astro-ph.HE]} \BibitemShut {NoStop}%
\bibitem [{\citenamefont {{Bogdanov}}(2014)}]{2014ApJ...790...94B}%
  \BibitemOpen
  \bibfield  {author} {\bibinfo {author} {\bibfnamefont {S.}~\bibnamefont {{Bogdanov}}},\ }\href {\doibase 10.1088/0004-637X/790/2/94} {\bibfield  {journal} {\bibinfo  {journal} {\apj}\ }\textbf {\bibinfo {volume} {790}},\ \bibinfo {eid} {94} (\bibinfo {year} {2014})},\ \Eprint {http://arxiv.org/abs/1406.0515} {arXiv:1406.0515 [astro-ph.HE]} \BibitemShut {NoStop}%
\bibitem [{\citenamefont {{Suleimanov}}\ \emph {et~al.}(2017)\citenamefont {{Suleimanov}}, \citenamefont {{Klochkov}}, \citenamefont {{Poutanen}},\ and\ \citenamefont {{Werner}}}]{2017A&A...600A..43S}%
  \BibitemOpen
  \bibfield  {author} {\bibinfo {author} {\bibfnamefont {V.~F.}\ \bibnamefont {{Suleimanov}}}, \bibinfo {author} {\bibfnamefont {D.}~\bibnamefont {{Klochkov}}}, \bibinfo {author} {\bibfnamefont {J.}~\bibnamefont {{Poutanen}}}, \ and\ \bibinfo {author} {\bibfnamefont {K.}~\bibnamefont {{Werner}}},\ }\href {\doibase 10.1051/0004-6361/201630028} {\bibfield  {journal} {\bibinfo  {journal} {\aap}\ }\textbf {\bibinfo {volume} {600}},\ \bibinfo {eid} {A43} (\bibinfo {year} {2017})},\ \Eprint {http://arxiv.org/abs/1701.06417} {arXiv:1701.06417 [astro-ph.HE]} \BibitemShut {NoStop}%
\bibitem [{\citenamefont {{Suleimanov}}\ \emph {et~al.}(2023{\natexlab{a}})\citenamefont {{Suleimanov}}, \citenamefont {{Poutanen}}, \citenamefont {{Doroshenko}},\ and\ \citenamefont {{Werner}}}]{2023A&A...673A..15S}%
  \BibitemOpen
  \bibfield  {author} {\bibinfo {author} {\bibfnamefont {V.~F.}\ \bibnamefont {{Suleimanov}}}, \bibinfo {author} {\bibfnamefont {J.}~\bibnamefont {{Poutanen}}}, \bibinfo {author} {\bibfnamefont {V.}~\bibnamefont {{Doroshenko}}}, \ and\ \bibinfo {author} {\bibfnamefont {K.}~\bibnamefont {{Werner}}},\ }\href {\doibase 10.1051/0004-6361/202346092} {\bibfield  {journal} {\bibinfo  {journal} {\aap}\ }\textbf {\bibinfo {volume} {673}},\ \bibinfo {eid} {A15} (\bibinfo {year} {2023}{\natexlab{a}})},\ \Eprint {http://arxiv.org/abs/2303.01382} {arXiv:2303.01382 [astro-ph.HE]} \BibitemShut {NoStop}%
\bibitem [{\citenamefont {{Mereghetti}}(2008)}]{2008A&ARv..15..225M}%
  \BibitemOpen
  \bibfield  {author} {\bibinfo {author} {\bibfnamefont {S.}~\bibnamefont {{Mereghetti}}},\ }\href {\doibase 10.1007/s00159-008-0011-z} {\bibfield  {journal} {\bibinfo  {journal} {The Astronomy and Astrophysics Review}\ }\textbf {\bibinfo {volume} {15}},\ \bibinfo {pages} {225} (\bibinfo {year} {2008})},\ \Eprint {http://arxiv.org/abs/0804.0250} {arXiv:0804.0250 [astro-ph]} \BibitemShut {NoStop}%
\bibitem [{\citenamefont {{Thompson}}\ \emph {et~al.}(2002)\citenamefont {{Thompson}}, \citenamefont {{Lyutikov}},\ and\ \citenamefont {{Kulkarni}}}]{Thompson2002}%
  \BibitemOpen
  \bibfield  {author} {\bibinfo {author} {\bibfnamefont {C.}~\bibnamefont {{Thompson}}}, \bibinfo {author} {\bibfnamefont {M.}~\bibnamefont {{Lyutikov}}}, \ and\ \bibinfo {author} {\bibfnamefont {S.~R.}\ \bibnamefont {{Kulkarni}}},\ }\href {\doibase 10.1086/340586} {\bibfield  {journal} {\bibinfo  {journal} {\apj}\ }\textbf {\bibinfo {volume} {574}},\ \bibinfo {pages} {332} (\bibinfo {year} {2002})},\ \Eprint {http://arxiv.org/abs/astro-ph/0110677} {arXiv:astro-ph/0110677 [astro-ph]} \BibitemShut {NoStop}%
\bibitem [{\citenamefont {{Duncan}}\ and\ \citenamefont {{Thompson}}(1992)}]{1992ApJ...392L...9D}%
  \BibitemOpen
  \bibfield  {author} {\bibinfo {author} {\bibfnamefont {R.~C.}\ \bibnamefont {{Duncan}}}\ and\ \bibinfo {author} {\bibfnamefont {C.}~\bibnamefont {{Thompson}}},\ }\href {\doibase 10.1086/186413} {\bibfield  {journal} {\bibinfo  {journal} {\apjl}\ }\textbf {\bibinfo {volume} {392}},\ \bibinfo {pages} {L9} (\bibinfo {year} {1992})}\BibitemShut {NoStop}%
\bibitem [{\citenamefont {{Nobili}}\ \emph {et~al.}(2008)\citenamefont {{Nobili}}, \citenamefont {{Turolla}},\ and\ \citenamefont {{Zane}}}]{2008MNRAS.386.1527N}%
  \BibitemOpen
  \bibfield  {author} {\bibinfo {author} {\bibfnamefont {L.}~\bibnamefont {{Nobili}}}, \bibinfo {author} {\bibfnamefont {R.}~\bibnamefont {{Turolla}}}, \ and\ \bibinfo {author} {\bibfnamefont {S.}~\bibnamefont {{Zane}}},\ }\href {\doibase 10.1111/j.1365-2966.2008.13125.x} {\bibfield  {journal} {\bibinfo  {journal} {\mnras}\ }\textbf {\bibinfo {volume} {386}},\ \bibinfo {pages} {1527} (\bibinfo {year} {2008})},\ \Eprint {http://arxiv.org/abs/0802.2647} {arXiv:0802.2647 [astro-ph]} \BibitemShut {NoStop}%
\bibitem [{\citenamefont {{Taverna}}\ \emph {et~al.}(2014)\citenamefont {{Taverna}}, \citenamefont {{Muleri}}, \citenamefont {{Turolla}}, \citenamefont {{Soffitta}}, \citenamefont {{Fabiani}},\ and\ \citenamefont {{Nobili}}}]{2014MNRAS.438.1686T}%
  \BibitemOpen
  \bibfield  {author} {\bibinfo {author} {\bibfnamefont {R.}~\bibnamefont {{Taverna}}}, \bibinfo {author} {\bibfnamefont {F.}~\bibnamefont {{Muleri}}}, \bibinfo {author} {\bibfnamefont {R.}~\bibnamefont {{Turolla}}}, \bibinfo {author} {\bibfnamefont {P.}~\bibnamefont {{Soffitta}}}, \bibinfo {author} {\bibfnamefont {S.}~\bibnamefont {{Fabiani}}}, \ and\ \bibinfo {author} {\bibfnamefont {L.}~\bibnamefont {{Nobili}}},\ }\href {\doibase 10.1093/mnras/stt2310} {\bibfield  {journal} {\bibinfo  {journal} {\mnras}\ }\textbf {\bibinfo {volume} {438}},\ \bibinfo {pages} {1686} (\bibinfo {year} {2014})},\ \Eprint {http://arxiv.org/abs/1311.7500} {arXiv:1311.7500 [astro-ph.HE]} \BibitemShut {NoStop}%
\bibitem [{\citenamefont {{Lai}}(2023)}]{2023PNAS..12016534L}%
  \BibitemOpen
  \bibfield  {author} {\bibinfo {author} {\bibfnamefont {D.}~\bibnamefont {{Lai}}},\ }\href {\doibase 10.1073/pnas.2216534120} {\bibfield  {journal} {\bibinfo  {journal} {Proceedings of the National Academy of Science}\ }\textbf {\bibinfo {volume} {120}},\ \bibinfo {eid} {e2216534120} (\bibinfo {year} {2023})},\ \Eprint {http://arxiv.org/abs/2209.13640} {arXiv:2209.13640 [astro-ph.HE]} \BibitemShut {NoStop}%
\bibitem [{\citenamefont {{Johnston}}\ \emph {et~al.}(2023)\citenamefont {{Johnston}}, \citenamefont {{Kramer}}, \citenamefont {{Karastergiou}}, \citenamefont {{Keith}}, \citenamefont {{Oswald}}, \citenamefont {{Parthasarathy}},\ and\ \citenamefont {{Weltevrede}}}]{Johnson2023}%
  \BibitemOpen
  \bibfield  {author} {\bibinfo {author} {\bibfnamefont {S.}~\bibnamefont {{Johnston}}}, \bibinfo {author} {\bibfnamefont {M.}~\bibnamefont {{Kramer}}}, \bibinfo {author} {\bibfnamefont {A.}~\bibnamefont {{Karastergiou}}}, \bibinfo {author} {\bibfnamefont {M.~J.}\ \bibnamefont {{Keith}}}, \bibinfo {author} {\bibfnamefont {L.~S.}\ \bibnamefont {{Oswald}}}, \bibinfo {author} {\bibfnamefont {A.}~\bibnamefont {{Parthasarathy}}}, \ and\ \bibinfo {author} {\bibfnamefont {P.}~\bibnamefont {{Weltevrede}}},\ }\href {\doibase 10.1093/mnras/stac3636} {\bibfield  {journal} {\bibinfo  {journal} {\mnras}\ }\textbf {\bibinfo {volume} {520}},\ \bibinfo {pages} {4801} (\bibinfo {year} {2023})},\ \Eprint {http://arxiv.org/abs/2212.03988} {arXiv:2212.03988 [astro-ph.HE]} \BibitemShut {NoStop}%
\bibitem [{\citenamefont {{Heyl}}\ \emph {et~al.}(2024{\natexlab{b}})\citenamefont {{Heyl}}, \citenamefont {{Doroshenko}}, \citenamefont {{Gonz{\'a}lez-Caniulef}}, \citenamefont {{Caiazzo}}, \citenamefont {{Poutanen}}, \citenamefont {{Mushtukov}}, \citenamefont {{Tsygankov}}, \citenamefont {{Kirmizibayrak}}, \citenamefont {{Bachetti}}, \citenamefont {{Pavlov}}, \citenamefont {{Forsblom}}, \citenamefont {{Malacaria}}, \citenamefont {{Suleimanov}}, \citenamefont {{Agudo}}, \citenamefont {{Antonelli}}, \citenamefont {{Baldini}}, \citenamefont {{Baumgartner}}, \citenamefont {{Bellazzini}}, \citenamefont {{Bianchi}}, \citenamefont {{Bongiorno}}, \citenamefont {{Bonino}}, \citenamefont {{Brez}}, \citenamefont {{Bucciantini}}, \citenamefont {{Capitanio}}, \citenamefont {{Castellano}}, \citenamefont {{Cavazzuti}}, \citenamefont {{Chen}}, \citenamefont {{Ciprini}}, \citenamefont {{Costa}}, \citenamefont {{De Rosa}}, \citenamefont {{Del Monte}}, \citenamefont {{Di Gesu}}, \citenamefont {{Di Lalla}}, \citenamefont {{Di
  Marco}}, \citenamefont {{Donnarumma}}, \citenamefont {{Dov{\v{c}}iak}}, \citenamefont {{Ehlert}}, \citenamefont {{Enoto}}, \citenamefont {{Evangelista}}, \citenamefont {{Fabiani}}, \citenamefont {{Ferrazzoli}}, \citenamefont {{Garcia}}, \citenamefont {{Gunji}}, \citenamefont {{Hayashida}}, \citenamefont {{Iwakiri}}, \citenamefont {{Jorstad}}, \citenamefont {{Kaaret}}, \citenamefont {{Karas}}, \citenamefont {{Kislat}}, \citenamefont {{Kitaguchi}}, \citenamefont {{Kolodziejczak}}, \citenamefont {{Krawczynski}}, \citenamefont {{La Monaca}}, \citenamefont {{Latronico}}, \citenamefont {{Liodakis}}, \citenamefont {{Maldera}}, \citenamefont {{Manfreda}}, \citenamefont {{Marin}}, \citenamefont {{Marinucci}}, \citenamefont {{Marscher}}, \citenamefont {{Marshall}}, \citenamefont {{Massaro}}, \citenamefont {{Matt}}, \citenamefont {{Mitsuishi}}, \citenamefont {{Mizuno}}, \citenamefont {{Muleri}}, \citenamefont {{Negro}}, \citenamefont {{Ng}}, \citenamefont {{O'Dell}}, \citenamefont {{Omodei}}, \citenamefont
  {{Oppedisano}}, \citenamefont {{Papitto}}, \citenamefont {{Peirson}}, \citenamefont {{Perri}}, \citenamefont {{Pesce-Rollins}}, \citenamefont {{Petrucci}}, \citenamefont {{Pilia}}, \citenamefont {{Possenti}}, \citenamefont {{Puccetti}}, \citenamefont {{Ramsey}}, \citenamefont {{Rankin}}, \citenamefont {{Ratheesh}}, \citenamefont {{Roberts}}, \citenamefont {{Romani}}, \citenamefont {{Sgr{\`o}}}, \citenamefont {{Slane}}, \citenamefont {{Soffitta}}, \citenamefont {{Spandre}}, \citenamefont {{Swartz}}, \citenamefont {{Tamagawa}}, \citenamefont {{Tavecchio}}, \citenamefont {{Taverna}}, \citenamefont {{Tawara}}, \citenamefont {{Tennant}}, \citenamefont {{Thomas}}, \citenamefont {{Tombesi}}, \citenamefont {{Trois}}, \citenamefont {{Turolla}}, \citenamefont {{Vink}}, \citenamefont {{Weisskopf}}, \citenamefont {{Wu}}, \citenamefont {{Xie}},\ and\ \citenamefont {{Zane}}}]{Heyl_herX-1}%
  \BibitemOpen
  \bibfield  {author} {\bibinfo {author} {\bibfnamefont {J.}~\bibnamefont {{Heyl}}}, \bibinfo {author} {\bibfnamefont {V.}~\bibnamefont {{Doroshenko}}}, \bibinfo {author} {\bibfnamefont {D.}~\bibnamefont {{Gonz{\'a}lez-Caniulef}}}, \bibinfo {author} {\bibfnamefont {I.}~\bibnamefont {{Caiazzo}}}, \bibinfo {author} {\bibfnamefont {J.}~\bibnamefont {{Poutanen}}}, \bibinfo {author} {\bibfnamefont {A.}~\bibnamefont {{Mushtukov}}}, \bibinfo {author} {\bibfnamefont {S.~S.}\ \bibnamefont {{Tsygankov}}}, \bibinfo {author} {\bibfnamefont {D.}~\bibnamefont {{Kirmizibayrak}}}, \bibinfo {author} {\bibfnamefont {M.}~\bibnamefont {{Bachetti}}}, \bibinfo {author} {\bibfnamefont {G.~G.}\ \bibnamefont {{Pavlov}}}, \bibinfo {author} {\bibfnamefont {S.~V.}\ \bibnamefont {{Forsblom}}}, \bibinfo {author} {\bibfnamefont {C.}~\bibnamefont {{Malacaria}}}, \bibinfo {author} {\bibfnamefont {V.~F.}\ \bibnamefont {{Suleimanov}}}, \bibinfo {author} {\bibfnamefont {I.}~\bibnamefont {{Agudo}}}, \bibinfo {author} {\bibfnamefont {L.~A.}\
  \bibnamefont {{Antonelli}}}, \bibinfo {author} {\bibfnamefont {L.}~\bibnamefont {{Baldini}}}, \bibinfo {author} {\bibfnamefont {W.~H.}\ \bibnamefont {{Baumgartner}}}, \bibinfo {author} {\bibfnamefont {R.}~\bibnamefont {{Bellazzini}}}, \bibinfo {author} {\bibfnamefont {S.}~\bibnamefont {{Bianchi}}}, \bibinfo {author} {\bibfnamefont {S.~D.}\ \bibnamefont {{Bongiorno}}}, \bibinfo {author} {\bibfnamefont {R.}~\bibnamefont {{Bonino}}}, \bibinfo {author} {\bibfnamefont {A.}~\bibnamefont {{Brez}}}, \bibinfo {author} {\bibfnamefont {N.}~\bibnamefont {{Bucciantini}}}, \bibinfo {author} {\bibfnamefont {F.}~\bibnamefont {{Capitanio}}}, \bibinfo {author} {\bibfnamefont {S.}~\bibnamefont {{Castellano}}}, \bibinfo {author} {\bibfnamefont {E.}~\bibnamefont {{Cavazzuti}}}, \bibinfo {author} {\bibfnamefont {C.-T.}\ \bibnamefont {{Chen}}}, \bibinfo {author} {\bibfnamefont {S.}~\bibnamefont {{Ciprini}}}, \bibinfo {author} {\bibfnamefont {E.}~\bibnamefont {{Costa}}}, \bibinfo {author} {\bibfnamefont {A.}~\bibnamefont {{De
  Rosa}}}, \bibinfo {author} {\bibfnamefont {E.}~\bibnamefont {{Del Monte}}}, \bibinfo {author} {\bibfnamefont {L.}~\bibnamefont {{Di Gesu}}}, \bibinfo {author} {\bibfnamefont {N.}~\bibnamefont {{Di Lalla}}}, \bibinfo {author} {\bibfnamefont {A.}~\bibnamefont {{Di Marco}}}, \bibinfo {author} {\bibfnamefont {I.}~\bibnamefont {{Donnarumma}}}, \bibinfo {author} {\bibfnamefont {M.}~\bibnamefont {{Dov{\v{c}}iak}}}, \bibinfo {author} {\bibfnamefont {S.~R.}\ \bibnamefont {{Ehlert}}}, \bibinfo {author} {\bibfnamefont {T.}~\bibnamefont {{Enoto}}}, \bibinfo {author} {\bibfnamefont {Y.}~\bibnamefont {{Evangelista}}}, \bibinfo {author} {\bibfnamefont {S.}~\bibnamefont {{Fabiani}}}, \bibinfo {author} {\bibfnamefont {R.}~\bibnamefont {{Ferrazzoli}}}, \bibinfo {author} {\bibfnamefont {J.~A.}\ \bibnamefont {{Garcia}}}, \bibinfo {author} {\bibfnamefont {S.}~\bibnamefont {{Gunji}}}, \bibinfo {author} {\bibfnamefont {K.}~\bibnamefont {{Hayashida}}}, \bibinfo {author} {\bibfnamefont {W.}~\bibnamefont {{Iwakiri}}}, \bibinfo
  {author} {\bibfnamefont {S.~G.}\ \bibnamefont {{Jorstad}}}, \bibinfo {author} {\bibfnamefont {P.}~\bibnamefont {{Kaaret}}}, \bibinfo {author} {\bibfnamefont {V.}~\bibnamefont {{Karas}}}, \bibinfo {author} {\bibfnamefont {F.}~\bibnamefont {{Kislat}}}, \bibinfo {author} {\bibfnamefont {T.}~\bibnamefont {{Kitaguchi}}}, \bibinfo {author} {\bibfnamefont {J.~J.}\ \bibnamefont {{Kolodziejczak}}}, \bibinfo {author} {\bibfnamefont {H.}~\bibnamefont {{Krawczynski}}}, \bibinfo {author} {\bibfnamefont {F.}~\bibnamefont {{La Monaca}}}, \bibinfo {author} {\bibfnamefont {L.}~\bibnamefont {{Latronico}}}, \bibinfo {author} {\bibfnamefont {I.}~\bibnamefont {{Liodakis}}}, \bibinfo {author} {\bibfnamefont {S.}~\bibnamefont {{Maldera}}}, \bibinfo {author} {\bibfnamefont {A.}~\bibnamefont {{Manfreda}}}, \bibinfo {author} {\bibfnamefont {F.}~\bibnamefont {{Marin}}}, \bibinfo {author} {\bibfnamefont {A.}~\bibnamefont {{Marinucci}}}, \bibinfo {author} {\bibfnamefont {A.~P.}\ \bibnamefont {{Marscher}}}, \bibinfo {author}
  {\bibfnamefont {H.~L.}\ \bibnamefont {{Marshall}}}, \bibinfo {author} {\bibfnamefont {F.}~\bibnamefont {{Massaro}}}, \bibinfo {author} {\bibfnamefont {G.}~\bibnamefont {{Matt}}}, \bibinfo {author} {\bibfnamefont {I.}~\bibnamefont {{Mitsuishi}}}, \bibinfo {author} {\bibfnamefont {T.}~\bibnamefont {{Mizuno}}}, \bibinfo {author} {\bibfnamefont {F.}~\bibnamefont {{Muleri}}}, \bibinfo {author} {\bibfnamefont {M.}~\bibnamefont {{Negro}}}, \bibinfo {author} {\bibfnamefont {C.~Y.}\ \bibnamefont {{Ng}}}, \bibinfo {author} {\bibfnamefont {S.~L.}\ \bibnamefont {{O'Dell}}}, \bibinfo {author} {\bibfnamefont {N.}~\bibnamefont {{Omodei}}}, \bibinfo {author} {\bibfnamefont {C.}~\bibnamefont {{Oppedisano}}}, \bibinfo {author} {\bibfnamefont {A.}~\bibnamefont {{Papitto}}}, \bibinfo {author} {\bibfnamefont {A.~L.}\ \bibnamefont {{Peirson}}}, \bibinfo {author} {\bibfnamefont {M.}~\bibnamefont {{Perri}}}, \bibinfo {author} {\bibfnamefont {M.}~\bibnamefont {{Pesce-Rollins}}}, \bibinfo {author} {\bibfnamefont {P.-O.}\
  \bibnamefont {{Petrucci}}}, \bibinfo {author} {\bibfnamefont {M.}~\bibnamefont {{Pilia}}}, \bibinfo {author} {\bibfnamefont {A.}~\bibnamefont {{Possenti}}}, \bibinfo {author} {\bibfnamefont {S.}~\bibnamefont {{Puccetti}}}, \bibinfo {author} {\bibfnamefont {B.~D.}\ \bibnamefont {{Ramsey}}}, \bibinfo {author} {\bibfnamefont {J.}~\bibnamefont {{Rankin}}}, \bibinfo {author} {\bibfnamefont {A.}~\bibnamefont {{Ratheesh}}}, \bibinfo {author} {\bibfnamefont {O.~J.}\ \bibnamefont {{Roberts}}}, \bibinfo {author} {\bibfnamefont {R.~W.}\ \bibnamefont {{Romani}}}, \bibinfo {author} {\bibfnamefont {C.}~\bibnamefont {{Sgr{\`o}}}}, \bibinfo {author} {\bibfnamefont {P.}~\bibnamefont {{Slane}}}, \bibinfo {author} {\bibfnamefont {P.}~\bibnamefont {{Soffitta}}}, \bibinfo {author} {\bibfnamefont {G.}~\bibnamefont {{Spandre}}}, \bibinfo {author} {\bibfnamefont {D.~A.}\ \bibnamefont {{Swartz}}}, \bibinfo {author} {\bibfnamefont {T.}~\bibnamefont {{Tamagawa}}}, \bibinfo {author} {\bibfnamefont {F.}~\bibnamefont {{Tavecchio}}},
  \bibinfo {author} {\bibfnamefont {R.}~\bibnamefont {{Taverna}}}, \bibinfo {author} {\bibfnamefont {Y.}~\bibnamefont {{Tawara}}}, \bibinfo {author} {\bibfnamefont {A.~F.}\ \bibnamefont {{Tennant}}}, \bibinfo {author} {\bibfnamefont {N.~E.}\ \bibnamefont {{Thomas}}}, \bibinfo {author} {\bibfnamefont {F.}~\bibnamefont {{Tombesi}}}, \bibinfo {author} {\bibfnamefont {A.}~\bibnamefont {{Trois}}}, \bibinfo {author} {\bibfnamefont {R.}~\bibnamefont {{Turolla}}}, \bibinfo {author} {\bibfnamefont {J.}~\bibnamefont {{Vink}}}, \bibinfo {author} {\bibfnamefont {M.~C.}\ \bibnamefont {{Weisskopf}}}, \bibinfo {author} {\bibfnamefont {K.}~\bibnamefont {{Wu}}}, \bibinfo {author} {\bibfnamefont {F.}~\bibnamefont {{Xie}}}, \ and\ \bibinfo {author} {\bibfnamefont {S.}~\bibnamefont {{Zane}}},\ }\href {\doibase 10.1038/s41550-024-02295-8} {\bibfield  {journal} {\bibinfo  {journal} {Nature Astronomy}\ }\textbf {\bibinfo {volume} {8}},\ \bibinfo {pages} {1047} (\bibinfo {year} {2024}{\natexlab{b}})}\BibitemShut {NoStop}%
\bibitem [{\citenamefont {{Kramer}}\ \emph {et~al.}(2007)\citenamefont {{Kramer}}, \citenamefont {{Stappers}}, \citenamefont {{Jessner}}, \citenamefont {{Lyne}},\ and\ \citenamefont {{Jordan}}}]{Kramer2007}%
  \BibitemOpen
  \bibfield  {author} {\bibinfo {author} {\bibfnamefont {M.}~\bibnamefont {{Kramer}}}, \bibinfo {author} {\bibfnamefont {B.~W.}\ \bibnamefont {{Stappers}}}, \bibinfo {author} {\bibfnamefont {A.}~\bibnamefont {{Jessner}}}, \bibinfo {author} {\bibfnamefont {A.~G.}\ \bibnamefont {{Lyne}}}, \ and\ \bibinfo {author} {\bibfnamefont {C.~A.}\ \bibnamefont {{Jordan}}},\ }\href {\doibase 10.1111/j.1365-2966.2007.11622.x} {\bibfield  {journal} {\bibinfo  {journal} {\mnras}\ }\textbf {\bibinfo {volume} {377}},\ \bibinfo {pages} {107} (\bibinfo {year} {2007})},\ \Eprint {http://arxiv.org/abs/astro-ph/0702365} {arXiv:astro-ph/0702365 [astro-ph]} \BibitemShut {NoStop}%
\bibitem [{\citenamefont {{Levin}}\ \emph {et~al.}(2012)\citenamefont {{Levin}}, \citenamefont {{Bailes}}, \citenamefont {{Bates}}, \citenamefont {{Bhat}}, \citenamefont {{Burgay}}, \citenamefont {{Burke-Spolaor}}, \citenamefont {{D'Amico}}, \citenamefont {{Johnston}}, \citenamefont {{Keith}}, \citenamefont {{Kramer}}, \citenamefont {{Milia}}, \citenamefont {{Possenti}}, \citenamefont {{Stappers}},\ and\ \citenamefont {{van Straten}}}]{Levin2012}%
  \BibitemOpen
  \bibfield  {author} {\bibinfo {author} {\bibfnamefont {L.}~\bibnamefont {{Levin}}}, \bibinfo {author} {\bibfnamefont {M.}~\bibnamefont {{Bailes}}}, \bibinfo {author} {\bibfnamefont {S.~D.}\ \bibnamefont {{Bates}}}, \bibinfo {author} {\bibfnamefont {N.~D.~R.}\ \bibnamefont {{Bhat}}}, \bibinfo {author} {\bibfnamefont {M.}~\bibnamefont {{Burgay}}}, \bibinfo {author} {\bibfnamefont {S.}~\bibnamefont {{Burke-Spolaor}}}, \bibinfo {author} {\bibfnamefont {N.}~\bibnamefont {{D'Amico}}}, \bibinfo {author} {\bibfnamefont {S.}~\bibnamefont {{Johnston}}}, \bibinfo {author} {\bibfnamefont {M.~J.}\ \bibnamefont {{Keith}}}, \bibinfo {author} {\bibfnamefont {M.}~\bibnamefont {{Kramer}}}, \bibinfo {author} {\bibfnamefont {S.}~\bibnamefont {{Milia}}}, \bibinfo {author} {\bibfnamefont {A.}~\bibnamefont {{Possenti}}}, \bibinfo {author} {\bibfnamefont {B.}~\bibnamefont {{Stappers}}}, \ and\ \bibinfo {author} {\bibfnamefont {W.}~\bibnamefont {{van Straten}}},\ }\href {\doibase 10.1111/j.1365-2966.2012.20807.x} {\bibfield
  {journal} {\bibinfo  {journal} {\mnras}\ }\textbf {\bibinfo {volume} {422}},\ \bibinfo {pages} {2489} (\bibinfo {year} {2012})},\ \Eprint {http://arxiv.org/abs/1204.2045} {arXiv:1204.2045 [astro-ph.HE]} \BibitemShut {NoStop}%
\bibitem [{\citenamefont {{Liu}}\ \emph {et~al.}(2025)\citenamefont {{Liu}}, \citenamefont {{Xu}}, \citenamefont {{Niu}}, \citenamefont {{Zhang}}, \citenamefont {{Jiang}}, \citenamefont {{Zhou}}, \citenamefont {{Han}}, \citenamefont {{Zhu}}, \citenamefont {{Lee}}, \citenamefont {{Li}}, \citenamefont {{Wang}}, \citenamefont {{Zhang}}, \citenamefont {{Chen}}, \citenamefont {{Luo}}, \citenamefont {{Luo}}, \citenamefont {{Niu}}, \citenamefont {{Qu}}, \citenamefont {{Wang}}, \citenamefont {{Wang}}, \citenamefont {{Wang}}, \citenamefont {{Wang}}, \citenamefont {{Wu}}, \citenamefont {{Wu}}, \citenamefont {{Xu}}, \citenamefont {{Yang}},\ and\ \citenamefont {{Zhang}}}]{Liu2025}%
  \BibitemOpen
  \bibfield  {author} {\bibinfo {author} {\bibfnamefont {X.}~\bibnamefont {{Liu}}}, \bibinfo {author} {\bibfnamefont {H.}~\bibnamefont {{Xu}}}, \bibinfo {author} {\bibfnamefont {J.}~\bibnamefont {{Niu}}}, \bibinfo {author} {\bibfnamefont {Y.}~\bibnamefont {{Zhang}}}, \bibinfo {author} {\bibfnamefont {J.}~\bibnamefont {{Jiang}}}, \bibinfo {author} {\bibfnamefont {D.}~\bibnamefont {{Zhou}}}, \bibinfo {author} {\bibfnamefont {J.}~\bibnamefont {{Han}}}, \bibinfo {author} {\bibfnamefont {W.}~\bibnamefont {{Zhu}}}, \bibinfo {author} {\bibfnamefont {K.}~\bibnamefont {{Lee}}}, \bibinfo {author} {\bibfnamefont {D.}~\bibnamefont {{Li}}}, \bibinfo {author} {\bibfnamefont {W.-Y.}\ \bibnamefont {{Wang}}}, \bibinfo {author} {\bibfnamefont {B.}~\bibnamefont {{Zhang}}}, \bibinfo {author} {\bibfnamefont {X.}~\bibnamefont {{Chen}}}, \bibinfo {author} {\bibfnamefont {J.-W.}\ \bibnamefont {{Luo}}}, \bibinfo {author} {\bibfnamefont {R.}~\bibnamefont {{Luo}}}, \bibinfo {author} {\bibfnamefont {C.}~\bibnamefont {{Niu}}}, \bibinfo
  {author} {\bibfnamefont {Y.}~\bibnamefont {{Qu}}}, \bibinfo {author} {\bibfnamefont {B.}~\bibnamefont {{Wang}}}, \bibinfo {author} {\bibfnamefont {F.}~\bibnamefont {{Wang}}}, \bibinfo {author} {\bibfnamefont {P.}~\bibnamefont {{Wang}}}, \bibinfo {author} {\bibfnamefont {T.}~\bibnamefont {{Wang}}}, \bibinfo {author} {\bibfnamefont {Q.}~\bibnamefont {{Wu}}}, \bibinfo {author} {\bibfnamefont {Z.}~\bibnamefont {{Wu}}}, \bibinfo {author} {\bibfnamefont {J.}~\bibnamefont {{Xu}}}, \bibinfo {author} {\bibfnamefont {Y.-P.}\ \bibnamefont {{Yang}}}, \ and\ \bibinfo {author} {\bibfnamefont {J.-S.}\ \bibnamefont {{Zhang}}},\ }\href {\doibase 10.3847/1538-4357/ade689} {\bibfield  {journal} {\bibinfo  {journal} {\apj}\ }\textbf {\bibinfo {volume} {988}},\ \bibinfo {eid} {175} (\bibinfo {year} {2025})},\ \Eprint {http://arxiv.org/abs/2504.00391} {arXiv:2504.00391 [astro-ph.HE]} \BibitemShut {NoStop}%
\bibitem [{\citenamefont {{Beloborodov}}(2009)}]{Beloborodov2009}%
  \BibitemOpen
  \bibfield  {author} {\bibinfo {author} {\bibfnamefont {A.~M.}\ \bibnamefont {{Beloborodov}}},\ }\href {\doibase 10.1088/0004-637X/703/1/1044} {\bibfield  {journal} {\bibinfo  {journal} {\apj}\ }\textbf {\bibinfo {volume} {703}},\ \bibinfo {pages} {1044} (\bibinfo {year} {2009})},\ \Eprint {http://arxiv.org/abs/0812.4873} {arXiv:0812.4873 [astro-ph]} \BibitemShut {NoStop}%
\bibitem [{\citenamefont {{Tong}}(2019)}]{Tong2019}%
  \BibitemOpen
  \bibfield  {author} {\bibinfo {author} {\bibfnamefont {H.}~\bibnamefont {{Tong}}},\ }\href {\doibase 10.1093/mnras/stz2438} {\bibfield  {journal} {\bibinfo  {journal} {\mnras}\ }\textbf {\bibinfo {volume} {489}},\ \bibinfo {pages} {3769} (\bibinfo {year} {2019})},\ \Eprint {http://arxiv.org/abs/1905.03476} {arXiv:1905.03476 [astro-ph.HE]} \BibitemShut {NoStop}%
\bibitem [{\citenamefont {{Tong}}\ \emph {et~al.}(2021)\citenamefont {{Tong}}, \citenamefont {{Wang}}, \citenamefont {{Wang}},\ and\ \citenamefont {{Yan}}}]{Tong2021}%
  \BibitemOpen
  \bibfield  {author} {\bibinfo {author} {\bibfnamefont {H.}~\bibnamefont {{Tong}}}, \bibinfo {author} {\bibfnamefont {P.~F.}\ \bibnamefont {{Wang}}}, \bibinfo {author} {\bibfnamefont {H.~G.}\ \bibnamefont {{Wang}}}, \ and\ \bibinfo {author} {\bibfnamefont {Z.}~\bibnamefont {{Yan}}},\ }\href {\doibase 10.1093/mnras/stab108} {\bibfield  {journal} {\bibinfo  {journal} {\mnras}\ }\textbf {\bibinfo {volume} {502}},\ \bibinfo {pages} {1549} (\bibinfo {year} {2021})},\ \Eprint {http://arxiv.org/abs/2101.04504} {arXiv:2101.04504 [astro-ph.HE]} \BibitemShut {NoStop}%
\bibitem [{\citenamefont {{Makishima}}\ \emph {et~al.}(2014)\citenamefont {{Makishima}}, \citenamefont {{Enoto}}, \citenamefont {{Hiraga}}, \citenamefont {{Nakano}}, \citenamefont {{Nakazawa}}, \citenamefont {{Sakurai}}, \citenamefont {{Sasano}},\ and\ \citenamefont {{Murakami}}}]{2014PhRvL.112q1102M}%
  \BibitemOpen
  \bibfield  {author} {\bibinfo {author} {\bibfnamefont {K.}~\bibnamefont {{Makishima}}}, \bibinfo {author} {\bibfnamefont {T.}~\bibnamefont {{Enoto}}}, \bibinfo {author} {\bibfnamefont {J.~S.}\ \bibnamefont {{Hiraga}}}, \bibinfo {author} {\bibfnamefont {T.}~\bibnamefont {{Nakano}}}, \bibinfo {author} {\bibfnamefont {K.}~\bibnamefont {{Nakazawa}}}, \bibinfo {author} {\bibfnamefont {S.}~\bibnamefont {{Sakurai}}}, \bibinfo {author} {\bibfnamefont {M.}~\bibnamefont {{Sasano}}}, \ and\ \bibinfo {author} {\bibfnamefont {H.}~\bibnamefont {{Murakami}}},\ }\href {\doibase 10.1103/PhysRevLett.112.171102} {\bibfield  {journal} {\bibinfo  {journal} {Phys. Rev. Lett.}\ }\textbf {\bibinfo {volume} {112}},\ \bibinfo {eid} {171102} (\bibinfo {year} {2014})},\ \Eprint {http://arxiv.org/abs/1404.3705} {arXiv:1404.3705 [astro-ph.HE]} \BibitemShut {NoStop}%
\bibitem [{\citenamefont {{Makishima}}\ \emph {et~al.}(2021{\natexlab{a}})\citenamefont {{Makishima}}, \citenamefont {{Enoto}}, \citenamefont {{Yoneda}},\ and\ \citenamefont {{Odaka}}}]{2021MNRAS.502.2266M}%
  \BibitemOpen
  \bibfield  {author} {\bibinfo {author} {\bibfnamefont {K.}~\bibnamefont {{Makishima}}}, \bibinfo {author} {\bibfnamefont {T.}~\bibnamefont {{Enoto}}}, \bibinfo {author} {\bibfnamefont {H.}~\bibnamefont {{Yoneda}}}, \ and\ \bibinfo {author} {\bibfnamefont {H.}~\bibnamefont {{Odaka}}},\ }\href {\doibase 10.1093/mnras/stab149} {\bibfield  {journal} {\bibinfo  {journal} {\mnras}\ }\textbf {\bibinfo {volume} {502}},\ \bibinfo {pages} {2266} (\bibinfo {year} {2021}{\natexlab{a}})},\ \Eprint {http://arxiv.org/abs/2102.00153} {arXiv:2102.00153 [astro-ph.HE]} \BibitemShut {NoStop}%
\bibitem [{\citenamefont {{Makishima}}\ \emph {et~al.}(2021{\natexlab{b}})\citenamefont {{Makishima}}, \citenamefont {{Tamba}}, \citenamefont {{Aizawa}}, \citenamefont {{Odaka}}, \citenamefont {{Yoneda}}, \citenamefont {{Enoto}},\ and\ \citenamefont {{Suzuki}}}]{2021ApJ...923...63M}%
  \BibitemOpen
  \bibfield  {author} {\bibinfo {author} {\bibfnamefont {K.}~\bibnamefont {{Makishima}}}, \bibinfo {author} {\bibfnamefont {T.}~\bibnamefont {{Tamba}}}, \bibinfo {author} {\bibfnamefont {Y.}~\bibnamefont {{Aizawa}}}, \bibinfo {author} {\bibfnamefont {H.}~\bibnamefont {{Odaka}}}, \bibinfo {author} {\bibfnamefont {H.}~\bibnamefont {{Yoneda}}}, \bibinfo {author} {\bibfnamefont {T.}~\bibnamefont {{Enoto}}}, \ and\ \bibinfo {author} {\bibfnamefont {H.}~\bibnamefont {{Suzuki}}},\ }\href {\doibase 10.3847/1538-4357/ac28fd} {\bibfield  {journal} {\bibinfo  {journal} {\apj}\ }\textbf {\bibinfo {volume} {923}},\ \bibinfo {eid} {63} (\bibinfo {year} {2021}{\natexlab{b}})},\ \Eprint {http://arxiv.org/abs/2109.11150} {arXiv:2109.11150 [astro-ph.HE]} \BibitemShut {NoStop}%
\bibitem [{\citenamefont {{Makishima}}\ \emph {et~al.}(2024)\citenamefont {{Makishima}}, \citenamefont {{Uchida}},\ and\ \citenamefont {{Enoto}}}]{2024PASJ...76..688M}%
  \BibitemOpen
  \bibfield  {author} {\bibinfo {author} {\bibfnamefont {K.}~\bibnamefont {{Makishima}}}, \bibinfo {author} {\bibfnamefont {N.}~\bibnamefont {{Uchida}}}, \ and\ \bibinfo {author} {\bibfnamefont {T.}~\bibnamefont {{Enoto}}},\ }\href {\doibase 10.1093/pasj/psae040} {\bibfield  {journal} {\bibinfo  {journal} {\pasj}\ }\textbf {\bibinfo {volume} {76}},\ \bibinfo {pages} {688} (\bibinfo {year} {2024})},\ \Eprint {http://arxiv.org/abs/2404.13799} {arXiv:2404.13799 [astro-ph.HE]} \BibitemShut {NoStop}%
\bibitem [{\citenamefont {{Desvignes}}\ \emph {et~al.}(2024)\citenamefont {{Desvignes}}, \citenamefont {{Weltevrede}}, \citenamefont {{Gao}}, \citenamefont {{Jones}}, \citenamefont {{Kramer}}, \citenamefont {{Caleb}}, \citenamefont {{Karuppusamy}}, \citenamefont {{Levin}}, \citenamefont {{Liu}}, \citenamefont {{Lyne}}, \citenamefont {{Shao}}, \citenamefont {{Stappers}},\ and\ \citenamefont {{P{\'e}tri}}}]{Desvignes2024}%
  \BibitemOpen
  \bibfield  {author} {\bibinfo {author} {\bibfnamefont {G.}~\bibnamefont {{Desvignes}}}, \bibinfo {author} {\bibfnamefont {P.}~\bibnamefont {{Weltevrede}}}, \bibinfo {author} {\bibfnamefont {Y.}~\bibnamefont {{Gao}}}, \bibinfo {author} {\bibfnamefont {D.~I.}\ \bibnamefont {{Jones}}}, \bibinfo {author} {\bibfnamefont {M.}~\bibnamefont {{Kramer}}}, \bibinfo {author} {\bibfnamefont {M.}~\bibnamefont {{Caleb}}}, \bibinfo {author} {\bibfnamefont {R.}~\bibnamefont {{Karuppusamy}}}, \bibinfo {author} {\bibfnamefont {L.}~\bibnamefont {{Levin}}}, \bibinfo {author} {\bibfnamefont {K.}~\bibnamefont {{Liu}}}, \bibinfo {author} {\bibfnamefont {A.~G.}\ \bibnamefont {{Lyne}}}, \bibinfo {author} {\bibfnamefont {L.}~\bibnamefont {{Shao}}}, \bibinfo {author} {\bibfnamefont {B.}~\bibnamefont {{Stappers}}}, \ and\ \bibinfo {author} {\bibfnamefont {J.}~\bibnamefont {{P{\'e}tri}}},\ }\href {\doibase 10.1038/s41550-024-02226-7} {\bibfield  {journal} {\bibinfo  {journal} {Nature Astronomy}\ }\textbf {\bibinfo {volume} {8}},\
  \bibinfo {pages} {617} (\bibinfo {year} {2024})}\BibitemShut {NoStop}%
\bibitem [{\citenamefont {{Cutler}}(2002)}]{2002PhRvD..66h4025C}%
  \BibitemOpen
  \bibfield  {author} {\bibinfo {author} {\bibfnamefont {C.}~\bibnamefont {{Cutler}}},\ }\href {\doibase 10.1103/PhysRevD.66.084025} {\bibfield  {journal} {\bibinfo  {journal} {Phys. Rev. D}\ }\textbf {\bibinfo {volume} {66}},\ \bibinfo {eid} {084025} (\bibinfo {year} {2002})},\ \Eprint {http://arxiv.org/abs/gr-qc/0206051} {arXiv:gr-qc/0206051 [gr-qc]} \BibitemShut {NoStop}%
\bibitem [{\citenamefont {{Dall'Osso}}\ \emph {et~al.}(2009)\citenamefont {{Dall'Osso}}, \citenamefont {{Shore}},\ and\ \citenamefont {{Stella}}}]{2009MNRAS.398.1869D}%
  \BibitemOpen
  \bibfield  {author} {\bibinfo {author} {\bibfnamefont {S.}~\bibnamefont {{Dall'Osso}}}, \bibinfo {author} {\bibfnamefont {S.~N.}\ \bibnamefont {{Shore}}}, \ and\ \bibinfo {author} {\bibfnamefont {L.}~\bibnamefont {{Stella}}},\ }\href {\doibase 10.1111/j.1365-2966.2008.14054.x} {\bibfield  {journal} {\bibinfo  {journal} {\mnras}\ }\textbf {\bibinfo {volume} {398}},\ \bibinfo {pages} {1869} (\bibinfo {year} {2009})},\ \Eprint {http://arxiv.org/abs/0811.4311} {arXiv:0811.4311 [astro-ph]} \BibitemShut {NoStop}%
\bibitem [{\citenamefont {{Cheng}}\ \emph {et~al.}(2019)\citenamefont {{Cheng}}, \citenamefont {{Zhang}}, \citenamefont {{Zheng}},\ and\ \citenamefont {{Fan}}}]{2019PhRvD..99h3011C}%
  \BibitemOpen
  \bibfield  {author} {\bibinfo {author} {\bibfnamefont {Q.}~\bibnamefont {{Cheng}}}, \bibinfo {author} {\bibfnamefont {S.-N.}\ \bibnamefont {{Zhang}}}, \bibinfo {author} {\bibfnamefont {X.-P.}\ \bibnamefont {{Zheng}}}, \ and\ \bibinfo {author} {\bibfnamefont {X.-L.}\ \bibnamefont {{Fan}}},\ }\href {\doibase 10.1103/PhysRevD.99.083011} {\bibfield  {journal} {\bibinfo  {journal} {Phys. Rev. D}\ }\textbf {\bibinfo {volume} {99}},\ \bibinfo {eid} {083011} (\bibinfo {year} {2019})},\ \Eprint {http://arxiv.org/abs/1904.06570} {arXiv:1904.06570 [astro-ph.HE]} \BibitemShut {NoStop}%
\bibitem [{\citenamefont {{Yan}}\ \emph {et~al.}(2024{\natexlab{a}})\citenamefont {{Yan}}, \citenamefont {{Cheng}},\ and\ \citenamefont {{Zheng}}}]{2024SCPMA..6729513Y}%
  \BibitemOpen
  \bibfield  {author} {\bibinfo {author} {\bibfnamefont {Y.-L.}\ \bibnamefont {{Yan}}}, \bibinfo {author} {\bibfnamefont {Q.}~\bibnamefont {{Cheng}}}, \ and\ \bibinfo {author} {\bibfnamefont {X.-P.}\ \bibnamefont {{Zheng}}},\ }\href {\doibase 10.1007/s11433-024-2473-6} {\bibfield  {journal} {\bibinfo  {journal} {Science China Physics, Mechanics, and Astronomy}\ }\textbf {\bibinfo {volume} {67}},\ \bibinfo {eid} {129513} (\bibinfo {year} {2024}{\natexlab{a}})},\ \Eprint {http://arxiv.org/abs/2410.23612} {arXiv:2410.23612 [astro-ph.HE]} \BibitemShut {NoStop}%
\bibitem [{\citenamefont {{Hu}}\ \emph {et~al.}(2023)\citenamefont {{Hu}}, \citenamefont {{Cheng}}, \citenamefont {{Zheng}}, \citenamefont {{Wang}}, \citenamefont {{Yan}}, \citenamefont {{Wang}},\ and\ \citenamefont {{Luo}}}]{2023RAA....23e5020H}%
  \BibitemOpen
  \bibfield  {author} {\bibinfo {author} {\bibfnamefont {F.-Y.}\ \bibnamefont {{Hu}}}, \bibinfo {author} {\bibfnamefont {Q.}~\bibnamefont {{Cheng}}}, \bibinfo {author} {\bibfnamefont {X.-P.}\ \bibnamefont {{Zheng}}}, \bibinfo {author} {\bibfnamefont {J.-Q.}\ \bibnamefont {{Wang}}}, \bibinfo {author} {\bibfnamefont {Y.-L.}\ \bibnamefont {{Yan}}}, \bibinfo {author} {\bibfnamefont {J.-Y.}\ \bibnamefont {{Wang}}}, \ and\ \bibinfo {author} {\bibfnamefont {T.-Y.}\ \bibnamefont {{Luo}}},\ }\href {\doibase 10.1088/1674-4527/accb7b} {\bibfield  {journal} {\bibinfo  {journal} {Research in Astronomy and Astrophysics}\ }\textbf {\bibinfo {volume} {23}},\ \bibinfo {eid} {055020} (\bibinfo {year} {2023})}\BibitemShut {NoStop}%
\bibitem [{\citenamefont {{Shu}}\ \emph {et~al.}(2025)\citenamefont {{Shu}}, \citenamefont {{Cheng}},\ and\ \citenamefont {{Zheng}}}]{2025PhRvD.111b3038S}%
  \BibitemOpen
  \bibfield  {author} {\bibinfo {author} {\bibfnamefont {J.}~\bibnamefont {{Shu}}}, \bibinfo {author} {\bibfnamefont {Q.}~\bibnamefont {{Cheng}}}, \ and\ \bibinfo {author} {\bibfnamefont {X.-P.}\ \bibnamefont {{Zheng}}},\ }\href {\doibase 10.1103/PhysRevD.111.023038} {\bibfield  {journal} {\bibinfo  {journal} {Phys. Rev. D}\ }\textbf {\bibinfo {volume} {111}},\ \bibinfo {eid} {023038} (\bibinfo {year} {2025})},\ \Eprint {http://arxiv.org/abs/2501.02887} {arXiv:2501.02887 [astro-ph.HE]} \BibitemShut {NoStop}%
\bibitem [{\citenamefont {{Yan}}\ \emph {et~al.}(2024{\natexlab{b}})\citenamefont {{Yan}}, \citenamefont {{Cheng}}, \citenamefont {{Zheng}},\ and\ \citenamefont {{Ouyang}}}]{2024EPJC...84.1043Y}%
  \BibitemOpen
  \bibfield  {author} {\bibinfo {author} {\bibfnamefont {Y.-L.}\ \bibnamefont {{Yan}}}, \bibinfo {author} {\bibfnamefont {Q.}~\bibnamefont {{Cheng}}}, \bibinfo {author} {\bibfnamefont {X.-P.}\ \bibnamefont {{Zheng}}}, \ and\ \bibinfo {author} {\bibfnamefont {X.-X.}\ \bibnamefont {{Ouyang}}},\ }\href {\doibase 10.1140/epjc/s10052-024-13406-0} {\bibfield  {journal} {\bibinfo  {journal} {European Physical Journal C}\ }\textbf {\bibinfo {volume} {84}},\ \bibinfo {eid} {1043} (\bibinfo {year} {2024}{\natexlab{b}})},\ \Eprint {http://arxiv.org/abs/2410.23576} {arXiv:2410.23576 [astro-ph.HE]} \BibitemShut {NoStop}%
\bibitem [{\citenamefont {{Ruderman}}(1969)}]{1969Natur.223..597R}%
  \BibitemOpen
  \bibfield  {author} {\bibinfo {author} {\bibfnamefont {M.}~\bibnamefont {{Ruderman}}},\ }\href {\doibase 10.1038/223597b0} {\bibfield  {journal} {\bibinfo  {journal} {\nat}\ }\textbf {\bibinfo {volume} {223}},\ \bibinfo {pages} {597} (\bibinfo {year} {1969})}\BibitemShut {NoStop}%
\bibitem [{\citenamefont {{Baym}}\ and\ \citenamefont {{Pines}}(1971)}]{1971AnPhy..66..816B}%
  \BibitemOpen
  \bibfield  {author} {\bibinfo {author} {\bibfnamefont {G.}~\bibnamefont {{Baym}}}\ and\ \bibinfo {author} {\bibfnamefont {D.}~\bibnamefont {{Pines}}},\ }\href {\doibase 10.1016/0003-4916(71)90084-4} {\bibfield  {journal} {\bibinfo  {journal} {Annals of Physics}\ }\textbf {\bibinfo {volume} {66}},\ \bibinfo {pages} {816} (\bibinfo {year} {1971})}\BibitemShut {NoStop}%
\bibitem [{\citenamefont {{Xu}}(2003)}]{2003ApJ...596L..59X}%
  \BibitemOpen
  \bibfield  {author} {\bibinfo {author} {\bibfnamefont {R.~X.}\ \bibnamefont {{Xu}}},\ }\href {\doibase 10.1086/379209} {\bibfield  {journal} {\bibinfo  {journal} {\apjl}\ }\textbf {\bibinfo {volume} {596}},\ \bibinfo {pages} {L59} (\bibinfo {year} {2003})},\ \Eprint {http://arxiv.org/abs/astro-ph/0302165} {arXiv:astro-ph/0302165 [astro-ph]} \BibitemShut {NoStop}%
\bibitem [{\citenamefont {{Baym}}\ \emph {et~al.}(1969)\citenamefont {{Baym}}, \citenamefont {{Pethick}},\ and\ \citenamefont {{Pines}}}]{1969Natur.224..673B}%
  \BibitemOpen
  \bibfield  {author} {\bibinfo {author} {\bibfnamefont {G.}~\bibnamefont {{Baym}}}, \bibinfo {author} {\bibfnamefont {C.}~\bibnamefont {{Pethick}}}, \ and\ \bibinfo {author} {\bibfnamefont {D.}~\bibnamefont {{Pines}}},\ }\href {\doibase 10.1038/224673a0} {\bibfield  {journal} {\bibinfo  {journal} {\nat}\ }\textbf {\bibinfo {volume} {224}},\ \bibinfo {pages} {673} (\bibinfo {year} {1969})}\BibitemShut {NoStop}%
\bibitem [{\citenamefont {{Anderson}}\ and\ \citenamefont {{Itoh}}(1975)}]{1975Natur.256...25A}%
  \BibitemOpen
  \bibfield  {author} {\bibinfo {author} {\bibfnamefont {P.~W.}\ \bibnamefont {{Anderson}}}\ and\ \bibinfo {author} {\bibfnamefont {N.}~\bibnamefont {{Itoh}}},\ }\href {\doibase 10.1038/256025a0} {\bibfield  {journal} {\bibinfo  {journal} {\nat}\ }\textbf {\bibinfo {volume} {256}},\ \bibinfo {pages} {25} (\bibinfo {year} {1975})}\BibitemShut {NoStop}%
\bibitem [{\citenamefont {{Antonopoulou}}\ \emph {et~al.}(2022)\citenamefont {{Antonopoulou}}, \citenamefont {{Haskell}},\ and\ \citenamefont {{Espinoza}}}]{2022RPPh...85l6901A}%
  \BibitemOpen
  \bibfield  {author} {\bibinfo {author} {\bibfnamefont {D.}~\bibnamefont {{Antonopoulou}}}, \bibinfo {author} {\bibfnamefont {B.}~\bibnamefont {{Haskell}}}, \ and\ \bibinfo {author} {\bibfnamefont {C.~M.}\ \bibnamefont {{Espinoza}}},\ }\href {\doibase 10.1088/1361-6633/ac9ced} {\bibfield  {journal} {\bibinfo  {journal} {Reports on Progress in Physics}\ }\textbf {\bibinfo {volume} {85}},\ \bibinfo {eid} {126901} (\bibinfo {year} {2022})}\BibitemShut {NoStop}%
\bibitem [{\citenamefont {{Zhou}}\ \emph {et~al.}(2022)\citenamefont {{Zhou}}, \citenamefont {{G{\"u}gercino{\u{g}}lu}}, \citenamefont {{Yuan}}, \citenamefont {{Ge}},\ and\ \citenamefont {{Yu}}}]{2022Univ....8..641Z}%
  \BibitemOpen
  \bibfield  {author} {\bibinfo {author} {\bibfnamefont {S.}~\bibnamefont {{Zhou}}}, \bibinfo {author} {\bibfnamefont {E.}~\bibnamefont {{G{\"u}gercino{\u{g}}lu}}}, \bibinfo {author} {\bibfnamefont {J.}~\bibnamefont {{Yuan}}}, \bibinfo {author} {\bibfnamefont {M.}~\bibnamefont {{Ge}}}, \ and\ \bibinfo {author} {\bibfnamefont {C.}~\bibnamefont {{Yu}}},\ }\href {\doibase 10.3390/universe8120641} {\bibfield  {journal} {\bibinfo  {journal} {Universe}\ }\textbf {\bibinfo {volume} {8}},\ \bibinfo {eid} {641} (\bibinfo {year} {2022})},\ \Eprint {http://arxiv.org/abs/2211.13885} {arXiv:2211.13885 [astro-ph.HE]} \BibitemShut {NoStop}%
\bibitem [{\citenamefont {{Pletsch}}\ \emph {et~al.}(2013)\citenamefont {{Pletsch}}, \citenamefont {{Guillemot}}, \citenamefont {{Allen}}, \citenamefont {{Anderson}}, \citenamefont {{Aulbert}}, \citenamefont {{Bock}}, \citenamefont {{Champion}}, \citenamefont {{Eggenstein}}, \citenamefont {{Fehrmann}}, \citenamefont {{Hammer}}, \citenamefont {{Karuppusamy}}, \citenamefont {{Keith}}, \citenamefont {{Kramer}}, \citenamefont {{Machenschalk}}, \citenamefont {{Ng}}, \citenamefont {{Papa}}, \citenamefont {{Ray}},\ and\ \citenamefont {{Siemens}}}]{2013ApJ...779L..11P}%
  \BibitemOpen
  \bibfield  {author} {\bibinfo {author} {\bibfnamefont {H.~J.}\ \bibnamefont {{Pletsch}}}, \bibinfo {author} {\bibfnamefont {L.}~\bibnamefont {{Guillemot}}}, \bibinfo {author} {\bibfnamefont {B.}~\bibnamefont {{Allen}}}, \bibinfo {author} {\bibfnamefont {D.}~\bibnamefont {{Anderson}}}, \bibinfo {author} {\bibfnamefont {C.}~\bibnamefont {{Aulbert}}}, \bibinfo {author} {\bibfnamefont {O.}~\bibnamefont {{Bock}}}, \bibinfo {author} {\bibfnamefont {D.~J.}\ \bibnamefont {{Champion}}}, \bibinfo {author} {\bibfnamefont {H.~B.}\ \bibnamefont {{Eggenstein}}}, \bibinfo {author} {\bibfnamefont {H.}~\bibnamefont {{Fehrmann}}}, \bibinfo {author} {\bibfnamefont {D.}~\bibnamefont {{Hammer}}}, \bibinfo {author} {\bibfnamefont {R.}~\bibnamefont {{Karuppusamy}}}, \bibinfo {author} {\bibfnamefont {M.}~\bibnamefont {{Keith}}}, \bibinfo {author} {\bibfnamefont {M.}~\bibnamefont {{Kramer}}}, \bibinfo {author} {\bibfnamefont {B.}~\bibnamefont {{Machenschalk}}}, \bibinfo {author} {\bibfnamefont {C.}~\bibnamefont {{Ng}}}, \bibinfo
  {author} {\bibfnamefont {M.~A.}\ \bibnamefont {{Papa}}}, \bibinfo {author} {\bibfnamefont {P.~S.}\ \bibnamefont {{Ray}}}, \ and\ \bibinfo {author} {\bibfnamefont {X.}~\bibnamefont {{Siemens}}},\ }\href {\doibase 10.1088/2041-8205/779/1/L11} {\bibfield  {journal} {\bibinfo  {journal} {\apjl}\ }\textbf {\bibinfo {volume} {779}},\ \bibinfo {eid} {L11} (\bibinfo {year} {2013})},\ \Eprint {http://arxiv.org/abs/1311.6427} {arXiv:1311.6427 [astro-ph.HE]} \BibitemShut {NoStop}%
\bibitem [{\citenamefont {{Dai}}\ \emph {et~al.}(2018)\citenamefont {{Dai}}, \citenamefont {{Johnston}}, \citenamefont {{Weltevrede}}, \citenamefont {{Kerr}}, \citenamefont {{Burgay}}, \citenamefont {{Esposito}}, \citenamefont {{Israel}}, \citenamefont {{Possenti}}, \citenamefont {{Rea}},\ and\ \citenamefont {{Sarkissian}}}]{2018MNRAS.480.3584D}%
  \BibitemOpen
  \bibfield  {author} {\bibinfo {author} {\bibfnamefont {S.}~\bibnamefont {{Dai}}}, \bibinfo {author} {\bibfnamefont {S.}~\bibnamefont {{Johnston}}}, \bibinfo {author} {\bibfnamefont {P.}~\bibnamefont {{Weltevrede}}}, \bibinfo {author} {\bibfnamefont {M.}~\bibnamefont {{Kerr}}}, \bibinfo {author} {\bibfnamefont {M.}~\bibnamefont {{Burgay}}}, \bibinfo {author} {\bibfnamefont {P.}~\bibnamefont {{Esposito}}}, \bibinfo {author} {\bibfnamefont {G.}~\bibnamefont {{Israel}}}, \bibinfo {author} {\bibfnamefont {A.}~\bibnamefont {{Possenti}}}, \bibinfo {author} {\bibfnamefont {N.}~\bibnamefont {{Rea}}}, \ and\ \bibinfo {author} {\bibfnamefont {J.}~\bibnamefont {{Sarkissian}}},\ }\href {\doibase 10.1093/mnras/sty2063} {\bibfield  {journal} {\bibinfo  {journal} {\mnras}\ }\textbf {\bibinfo {volume} {480}},\ \bibinfo {pages} {3584} (\bibinfo {year} {2018})},\ \Eprint {http://arxiv.org/abs/1806.05064} {arXiv:1806.05064 [astro-ph.HE]} \BibitemShut {NoStop}%
\bibitem [{\citenamefont {{Archibald}}\ \emph {et~al.}(2018)\citenamefont {{Archibald}}, \citenamefont {{Kaspi}}, \citenamefont {{Tendulkar}},\ and\ \citenamefont {{Scholz}}}]{2018ApJ...869..180A}%
  \BibitemOpen
  \bibfield  {author} {\bibinfo {author} {\bibfnamefont {R.~F.}\ \bibnamefont {{Archibald}}}, \bibinfo {author} {\bibfnamefont {V.~M.}\ \bibnamefont {{Kaspi}}}, \bibinfo {author} {\bibfnamefont {S.~P.}\ \bibnamefont {{Tendulkar}}}, \ and\ \bibinfo {author} {\bibfnamefont {P.}~\bibnamefont {{Scholz}}},\ }\href {\doibase 10.3847/1538-4357/aaee73} {\bibfield  {journal} {\bibinfo  {journal} {\apj}\ }\textbf {\bibinfo {volume} {869}},\ \bibinfo {eid} {180} (\bibinfo {year} {2018})},\ \Eprint {http://arxiv.org/abs/1806.01414} {arXiv:1806.01414 [astro-ph.HE]} \BibitemShut {NoStop}%
\bibitem [{\citenamefont {{Archibald}}\ \emph {et~al.}(2013)\citenamefont {{Archibald}}, \citenamefont {{Kaspi}}, \citenamefont {{Ng}}, \citenamefont {{Gourgouliatos}}, \citenamefont {{Tsang}}, \citenamefont {{Scholz}}, \citenamefont {{Beardmore}}, \citenamefont {{Gehrels}},\ and\ \citenamefont {{Kennea}}}]{2013Natur.497..591A}%
  \BibitemOpen
  \bibfield  {author} {\bibinfo {author} {\bibfnamefont {R.~F.}\ \bibnamefont {{Archibald}}}, \bibinfo {author} {\bibfnamefont {V.~M.}\ \bibnamefont {{Kaspi}}}, \bibinfo {author} {\bibfnamefont {C.~Y.}\ \bibnamefont {{Ng}}}, \bibinfo {author} {\bibfnamefont {K.~N.}\ \bibnamefont {{Gourgouliatos}}}, \bibinfo {author} {\bibfnamefont {D.}~\bibnamefont {{Tsang}}}, \bibinfo {author} {\bibfnamefont {P.}~\bibnamefont {{Scholz}}}, \bibinfo {author} {\bibfnamefont {A.~P.}\ \bibnamefont {{Beardmore}}}, \bibinfo {author} {\bibfnamefont {N.}~\bibnamefont {{Gehrels}}}, \ and\ \bibinfo {author} {\bibfnamefont {J.~A.}\ \bibnamefont {{Kennea}}},\ }\href {\doibase 10.1038/nature12159} {\bibfield  {journal} {\bibinfo  {journal} {\nat}\ }\textbf {\bibinfo {volume} {497}},\ \bibinfo {pages} {591} (\bibinfo {year} {2013})},\ \Eprint {http://arxiv.org/abs/1305.6894} {arXiv:1305.6894 [astro-ph.HE]} \BibitemShut {NoStop}%
\bibitem [{\citenamefont {{Younes}}\ \emph {et~al.}(2020)\citenamefont {{Younes}}, \citenamefont {{Ray}}, \citenamefont {{Baring}}, \citenamefont {{Kouveliotou}}, \citenamefont {{Fletcher}}, \citenamefont {{Wadiasingh}}, \citenamefont {{Harding}},\ and\ \citenamefont {{Goldstein}}}]{2020ApJ...896L..42Y}%
  \BibitemOpen
  \bibfield  {author} {\bibinfo {author} {\bibfnamefont {G.}~\bibnamefont {{Younes}}}, \bibinfo {author} {\bibfnamefont {P.~S.}\ \bibnamefont {{Ray}}}, \bibinfo {author} {\bibfnamefont {M.~G.}\ \bibnamefont {{Baring}}}, \bibinfo {author} {\bibfnamefont {C.}~\bibnamefont {{Kouveliotou}}}, \bibinfo {author} {\bibfnamefont {C.}~\bibnamefont {{Fletcher}}}, \bibinfo {author} {\bibfnamefont {Z.}~\bibnamefont {{Wadiasingh}}}, \bibinfo {author} {\bibfnamefont {A.~K.}\ \bibnamefont {{Harding}}}, \ and\ \bibinfo {author} {\bibfnamefont {A.}~\bibnamefont {{Goldstein}}},\ }\href {\doibase 10.3847/2041-8213/ab9a48} {\bibfield  {journal} {\bibinfo  {journal} {\apjl}\ }\textbf {\bibinfo {volume} {896}},\ \bibinfo {eid} {L42} (\bibinfo {year} {2020})},\ \Eprint {http://arxiv.org/abs/2006.04854} {arXiv:2006.04854 [astro-ph.HE]} \BibitemShut {NoStop}%
\bibitem [{\citenamefont {{{\c{S}}a{\c{s}}maz Mu{\c{s}}}}\ \emph {et~al.}(2014)\citenamefont {{{\c{S}}a{\c{s}}maz Mu{\c{s}}}}, \citenamefont {{Ayd{\i}n}},\ and\ \citenamefont {{G{\"o}{\u{g}}{\"u}{\c{s}}}}}]{2014MNRAS.440.2916S}%
  \BibitemOpen
  \bibfield  {author} {\bibinfo {author} {\bibfnamefont {S.}~\bibnamefont {{{\c{S}}a{\c{s}}maz Mu{\c{s}}}}}, \bibinfo {author} {\bibfnamefont {B.}~\bibnamefont {{Ayd{\i}n}}}, \ and\ \bibinfo {author} {\bibfnamefont {E.}~\bibnamefont {{G{\"o}{\u{g}}{\"u}{\c{s}}}}},\ }\href {\doibase 10.1093/mnras/stu436} {\bibfield  {journal} {\bibinfo  {journal} {\mnras}\ }\textbf {\bibinfo {volume} {440}},\ \bibinfo {pages} {2916} (\bibinfo {year} {2014})},\ \Eprint {http://arxiv.org/abs/1411.5274} {arXiv:1411.5274 [astro-ph.HE]} \BibitemShut {NoStop}%
\bibitem [{\citenamefont {{Ray}}\ \emph {et~al.}(2019)\citenamefont {{Ray}}, \citenamefont {{Guillot}}, \citenamefont {{Ho}}, \citenamefont {{Kerr}}, \citenamefont {{Enoto}}, \citenamefont {{Gendreau}}, \citenamefont {{Arzoumanian}}, \citenamefont {{Altamirano}}, \citenamefont {{Bogdanov}}, \citenamefont {{Campion}}, \citenamefont {{Chakrabarty}}, \citenamefont {{Deneva}}, \citenamefont {{Jaisawal}}, \citenamefont {{Kozon}}, \citenamefont {{Malacaria}}, \citenamefont {{Strohmayer}},\ and\ \citenamefont {{Wolff}}}]{2019ApJ...879..130R}%
  \BibitemOpen
  \bibfield  {author} {\bibinfo {author} {\bibfnamefont {P.~S.}\ \bibnamefont {{Ray}}}, \bibinfo {author} {\bibfnamefont {S.}~\bibnamefont {{Guillot}}}, \bibinfo {author} {\bibfnamefont {W.~C.~G.}\ \bibnamefont {{Ho}}}, \bibinfo {author} {\bibfnamefont {M.}~\bibnamefont {{Kerr}}}, \bibinfo {author} {\bibfnamefont {T.}~\bibnamefont {{Enoto}}}, \bibinfo {author} {\bibfnamefont {K.~C.}\ \bibnamefont {{Gendreau}}}, \bibinfo {author} {\bibfnamefont {Z.}~\bibnamefont {{Arzoumanian}}}, \bibinfo {author} {\bibfnamefont {D.}~\bibnamefont {{Altamirano}}}, \bibinfo {author} {\bibfnamefont {S.}~\bibnamefont {{Bogdanov}}}, \bibinfo {author} {\bibfnamefont {R.}~\bibnamefont {{Campion}}}, \bibinfo {author} {\bibfnamefont {D.}~\bibnamefont {{Chakrabarty}}}, \bibinfo {author} {\bibfnamefont {J.~S.}\ \bibnamefont {{Deneva}}}, \bibinfo {author} {\bibfnamefont {G.~K.}\ \bibnamefont {{Jaisawal}}}, \bibinfo {author} {\bibfnamefont {R.}~\bibnamefont {{Kozon}}}, \bibinfo {author} {\bibfnamefont {C.}~\bibnamefont {{Malacaria}}},
  \bibinfo {author} {\bibfnamefont {T.~E.}\ \bibnamefont {{Strohmayer}}}, \ and\ \bibinfo {author} {\bibfnamefont {M.~T.}\ \bibnamefont {{Wolff}}},\ }\href {\doibase 10.3847/1538-4357/ab24d8} {\bibfield  {journal} {\bibinfo  {journal} {\apj}\ }\textbf {\bibinfo {volume} {879}},\ \bibinfo {eid} {130} (\bibinfo {year} {2019})},\ \Eprint {http://arxiv.org/abs/1811.09218} {arXiv:1811.09218 [astro-ph.HE]} \BibitemShut {NoStop}%
\bibitem [{\citenamefont {{Tuo}}\ \emph {et~al.}(2024)\citenamefont {{Tuo}}, \citenamefont {{Serim}}, \citenamefont {{Antonelli}}, \citenamefont {{Ducci}}, \citenamefont {{Vahdat}}, \citenamefont {{Ge}}, \citenamefont {{Santangelo}},\ and\ \citenamefont {{Xie}}}]{2024ApJ...967L..13T}%
  \BibitemOpen
  \bibfield  {author} {\bibinfo {author} {\bibfnamefont {Y.}~\bibnamefont {{Tuo}}}, \bibinfo {author} {\bibfnamefont {M.~M.}\ \bibnamefont {{Serim}}}, \bibinfo {author} {\bibfnamefont {M.}~\bibnamefont {{Antonelli}}}, \bibinfo {author} {\bibfnamefont {L.}~\bibnamefont {{Ducci}}}, \bibinfo {author} {\bibfnamefont {A.}~\bibnamefont {{Vahdat}}}, \bibinfo {author} {\bibfnamefont {M.}~\bibnamefont {{Ge}}}, \bibinfo {author} {\bibfnamefont {A.}~\bibnamefont {{Santangelo}}}, \ and\ \bibinfo {author} {\bibfnamefont {F.}~\bibnamefont {{Xie}}},\ }\href {\doibase 10.3847/2041-8213/ad4488} {\bibfield  {journal} {\bibinfo  {journal} {\apjl}\ }\textbf {\bibinfo {volume} {967}},\ \bibinfo {eid} {L13} (\bibinfo {year} {2024})},\ \Eprint {http://arxiv.org/abs/2404.18158} {arXiv:2404.18158 [astro-ph.HE]} \BibitemShut {NoStop}%
\bibitem [{\citenamefont {{Kantor}}\ and\ \citenamefont {{Gusakov}}(2014)}]{2014ApJ...797L...4K}%
  \BibitemOpen
  \bibfield  {author} {\bibinfo {author} {\bibfnamefont {E.~M.}\ \bibnamefont {{Kantor}}}\ and\ \bibinfo {author} {\bibfnamefont {M.~E.}\ \bibnamefont {{Gusakov}}},\ }\href {\doibase 10.1088/2041-8205/797/1/L4} {\bibfield  {journal} {\bibinfo  {journal} {\apjl}\ }\textbf {\bibinfo {volume} {797}},\ \bibinfo {eid} {L4} (\bibinfo {year} {2014})},\ \Eprint {http://arxiv.org/abs/1411.2777} {arXiv:1411.2777 [astro-ph.SR]} \BibitemShut {NoStop}%
\bibitem [{\citenamefont {{Howitt}}\ and\ \citenamefont {{Melatos}}(2022)}]{2022MNRAS.514..863H}%
  \BibitemOpen
  \bibfield  {author} {\bibinfo {author} {\bibfnamefont {G.}~\bibnamefont {{Howitt}}}\ and\ \bibinfo {author} {\bibfnamefont {A.}~\bibnamefont {{Melatos}}},\ }\href {\doibase 10.1093/mnras/stac1358} {\bibfield  {journal} {\bibinfo  {journal} {\mnras}\ }\textbf {\bibinfo {volume} {514}},\ \bibinfo {pages} {863} (\bibinfo {year} {2022})},\ \Eprint {http://arxiv.org/abs/2205.05896} {arXiv:2205.05896 [astro-ph.HE]} \BibitemShut {NoStop}%
\bibitem [{\citenamefont {{Zhou}}\ \emph {et~al.}(2024)\citenamefont {{Zhou}}, \citenamefont {{Ye}}, \citenamefont {{Ge}}, \citenamefont {{G{\"u}gercino{\u{g}}lu}}, \citenamefont {{Zheng}}, \citenamefont {{Yu}}, \citenamefont {{Yuan}},\ and\ \citenamefont {{Zhang}}}]{2024ApJ...977..243Z}%
  \BibitemOpen
  \bibfield  {author} {\bibinfo {author} {\bibfnamefont {S.~Q.}\ \bibnamefont {{Zhou}}}, \bibinfo {author} {\bibfnamefont {W.~T.}\ \bibnamefont {{Ye}}}, \bibinfo {author} {\bibfnamefont {M.~Y.}\ \bibnamefont {{Ge}}}, \bibinfo {author} {\bibfnamefont {E.}~\bibnamefont {{G{\"u}gercino{\u{g}}lu}}}, \bibinfo {author} {\bibfnamefont {S.~J.}\ \bibnamefont {{Zheng}}}, \bibinfo {author} {\bibfnamefont {C.}~\bibnamefont {{Yu}}}, \bibinfo {author} {\bibfnamefont {J.~P.}\ \bibnamefont {{Yuan}}}, \ and\ \bibinfo {author} {\bibfnamefont {J.}~\bibnamefont {{Zhang}}},\ }\href {\doibase 10.3847/1538-4357/ad938d} {\bibfield  {journal} {\bibinfo  {journal} {\apj}\ }\textbf {\bibinfo {volume} {977}},\ \bibinfo {eid} {243} (\bibinfo {year} {2024})},\ \Eprint {http://arxiv.org/abs/2408.09204} {arXiv:2408.09204 [astro-ph.HE]} \BibitemShut {NoStop}%
\bibitem [{\citenamefont {{Huang}}\ and\ \citenamefont {{Geng}}(2014)}]{2014ApJ...782L..20H}%
  \BibitemOpen
  \bibfield  {author} {\bibinfo {author} {\bibfnamefont {Y.~F.}\ \bibnamefont {{Huang}}}\ and\ \bibinfo {author} {\bibfnamefont {J.~J.}\ \bibnamefont {{Geng}}},\ }\href {\doibase 10.1088/2041-8205/782/2/L20} {\bibfield  {journal} {\bibinfo  {journal} {\apjl}\ }\textbf {\bibinfo {volume} {782}},\ \bibinfo {eid} {L20} (\bibinfo {year} {2014})},\ \Eprint {http://arxiv.org/abs/1310.3324} {arXiv:1310.3324 [astro-ph.HE]} \BibitemShut {NoStop}%
\bibitem [{\citenamefont {{Harding}}\ \emph {et~al.}(1999)\citenamefont {{Harding}}, \citenamefont {{Contopoulos}},\ and\ \citenamefont {{Kazanas}}}]{1999ApJ...525L.125H}%
  \BibitemOpen
  \bibfield  {author} {\bibinfo {author} {\bibfnamefont {A.~K.}\ \bibnamefont {{Harding}}}, \bibinfo {author} {\bibfnamefont {I.}~\bibnamefont {{Contopoulos}}}, \ and\ \bibinfo {author} {\bibfnamefont {D.}~\bibnamefont {{Kazanas}}},\ }\href {\doibase 10.1086/312339} {\bibfield  {journal} {\bibinfo  {journal} {\apjl}\ }\textbf {\bibinfo {volume} {525}},\ \bibinfo {pages} {L125} (\bibinfo {year} {1999})},\ \Eprint {http://arxiv.org/abs/astro-ph/9908279} {arXiv:astro-ph/9908279 [astro-ph]} \BibitemShut {NoStop}%
\bibitem [{\citenamefont {{Tong}}(2014)}]{2014ApJ...784...86T}%
  \BibitemOpen
  \bibfield  {author} {\bibinfo {author} {\bibfnamefont {H.}~\bibnamefont {{Tong}}},\ }\href {\doibase 10.1088/0004-637X/784/2/86} {\bibfield  {journal} {\bibinfo  {journal} {\apj}\ }\textbf {\bibinfo {volume} {784}},\ \bibinfo {eid} {86} (\bibinfo {year} {2014})},\ \Eprint {http://arxiv.org/abs/1306.2445} {arXiv:1306.2445 [astro-ph.HE]} \BibitemShut {NoStop}%
\bibitem [{\citenamefont {{Lyutikov}}(2013)}]{2013arXiv1306.2264L}%
  \BibitemOpen
  \bibfield  {author} {\bibinfo {author} {\bibfnamefont {M.}~\bibnamefont {{Lyutikov}}},\ }\href {\doibase 10.48550/arXiv.1306.2264} {\bibfield  {journal} {\bibinfo  {journal} {arXiv e-prints}\ ,\ \bibinfo {eid} {arXiv:1306.2264}} (\bibinfo {year} {2013})},\ \Eprint {http://arxiv.org/abs/1306.2264} {arXiv:1306.2264 [astro-ph.HE]} \BibitemShut {NoStop}%
\bibitem [{\citenamefont {{Garcia}}\ and\ \citenamefont {{Ranea-Sandoval}}(2015)}]{2015MNRAS.449L..73G}%
  \BibitemOpen
  \bibfield  {author} {\bibinfo {author} {\bibfnamefont {F.}~\bibnamefont {{Garcia}}}\ and\ \bibinfo {author} {\bibfnamefont {I.~F.}\ \bibnamefont {{Ranea-Sandoval}}},\ }\href {\doibase 10.1093/mnrasl/slv019} {\bibfield  {journal} {\bibinfo  {journal} {\mnras}\ }\textbf {\bibinfo {volume} {449}},\ \bibinfo {pages} {L73} (\bibinfo {year} {2015})},\ \Eprint {http://arxiv.org/abs/1402.0848} {arXiv:1402.0848 [astro-ph.HE]} \BibitemShut {NoStop}%
\bibitem [{\citenamefont {{Marshall}}\ \emph {et~al.}(2015)\citenamefont {{Marshall}}, \citenamefont {{Guillemot}}, \citenamefont {{Harding}}, \citenamefont {{Martin}},\ and\ \citenamefont {{Smith}}}]{2015ApJ...807L..27M}%
  \BibitemOpen
  \bibfield  {author} {\bibinfo {author} {\bibfnamefont {F.~E.}\ \bibnamefont {{Marshall}}}, \bibinfo {author} {\bibfnamefont {L.}~\bibnamefont {{Guillemot}}}, \bibinfo {author} {\bibfnamefont {A.~K.}\ \bibnamefont {{Harding}}}, \bibinfo {author} {\bibfnamefont {P.}~\bibnamefont {{Martin}}}, \ and\ \bibinfo {author} {\bibfnamefont {D.~A.}\ \bibnamefont {{Smith}}},\ }\href {\doibase 10.1088/2041-8205/807/2/L27} {\bibfield  {journal} {\bibinfo  {journal} {\apjl}\ }\textbf {\bibinfo {volume} {807}},\ \bibinfo {eid} {L27} (\bibinfo {year} {2015})},\ \Eprint {http://arxiv.org/abs/1506.05765} {arXiv:1506.05765 [astro-ph.HE]} \BibitemShut {NoStop}%
\bibitem [{\citenamefont {{Zubieta}}\ \emph {et~al.}(2025)\citenamefont {{Zubieta}}, \citenamefont {{Garc{\'\i}a}}, \citenamefont {{del Palacio}}, \citenamefont {{Espinoza}}, \citenamefont {{Araujo Furlan}}, \citenamefont {{Gancio}}, \citenamefont {{Lousto}}, \citenamefont {{Combi}},\ and\ \citenamefont {{G{\"u}gercino{\u{g}}lu}}}]{2025A&A...694A.124Z}%
  \BibitemOpen
  \bibfield  {author} {\bibinfo {author} {\bibfnamefont {E.}~\bibnamefont {{Zubieta}}}, \bibinfo {author} {\bibfnamefont {F.}~\bibnamefont {{Garc{\'\i}a}}}, \bibinfo {author} {\bibfnamefont {S.}~\bibnamefont {{del Palacio}}}, \bibinfo {author} {\bibfnamefont {C.~M.}\ \bibnamefont {{Espinoza}}}, \bibinfo {author} {\bibfnamefont {S.~B.}\ \bibnamefont {{Araujo Furlan}}}, \bibinfo {author} {\bibfnamefont {G.}~\bibnamefont {{Gancio}}}, \bibinfo {author} {\bibfnamefont {C.~O.}\ \bibnamefont {{Lousto}}}, \bibinfo {author} {\bibfnamefont {J.~A.}\ \bibnamefont {{Combi}}}, \ and\ \bibinfo {author} {\bibfnamefont {E.}~\bibnamefont {{G{\"u}gercino{\u{g}}lu}}},\ }\href {\doibase 10.1051/0004-6361/202452693} {\bibfield  {journal} {\bibinfo  {journal} {\aap}\ }\textbf {\bibinfo {volume} {694}},\ \bibinfo {eid} {A124} (\bibinfo {year} {2025})},\ \Eprint {http://arxiv.org/abs/2412.17766} {arXiv:2412.17766 [astro-ph.HE]} \BibitemShut {NoStop}%
\bibitem [{\citenamefont {{Liu}}\ \emph {et~al.}(2022)\citenamefont {{Liu}}, \citenamefont {{Wang}}, \citenamefont {{Shen}}, \citenamefont {{Yan}}, \citenamefont {{Tong}}, \citenamefont {{Huang}},\ and\ \citenamefont {{Zhao}}}]{2022ApJ...931..103L}%
  \BibitemOpen
  \bibfield  {author} {\bibinfo {author} {\bibfnamefont {J.}~\bibnamefont {{Liu}}}, \bibinfo {author} {\bibfnamefont {H.-G.}\ \bibnamefont {{Wang}}}, \bibinfo {author} {\bibfnamefont {Z.-Q.}\ \bibnamefont {{Shen}}}, \bibinfo {author} {\bibfnamefont {Z.}~\bibnamefont {{Yan}}}, \bibinfo {author} {\bibfnamefont {H.}~\bibnamefont {{Tong}}}, \bibinfo {author} {\bibfnamefont {Z.-P.}\ \bibnamefont {{Huang}}}, \ and\ \bibinfo {author} {\bibfnamefont {R.-S.}\ \bibnamefont {{Zhao}}},\ }\href {\doibase 10.3847/1538-4357/ac6bf7} {\bibfield  {journal} {\bibinfo  {journal} {\apj}\ }\textbf {\bibinfo {volume} {931}},\ \bibinfo {eid} {103} (\bibinfo {year} {2022})}\BibitemShut {NoStop}%
\bibitem [{\citenamefont {{Palfreyman}}\ \emph {et~al.}(2018)\citenamefont {{Palfreyman}}, \citenamefont {{Dickey}}, \citenamefont {{Hotan}}, \citenamefont {{Ellingsen}},\ and\ \citenamefont {{van Straten}}}]{2018Natur.556..219P}%
  \BibitemOpen
  \bibfield  {author} {\bibinfo {author} {\bibfnamefont {J.}~\bibnamefont {{Palfreyman}}}, \bibinfo {author} {\bibfnamefont {J.~M.}\ \bibnamefont {{Dickey}}}, \bibinfo {author} {\bibfnamefont {A.}~\bibnamefont {{Hotan}}}, \bibinfo {author} {\bibfnamefont {S.}~\bibnamefont {{Ellingsen}}}, \ and\ \bibinfo {author} {\bibfnamefont {W.}~\bibnamefont {{van Straten}}},\ }\href {\doibase 10.1038/s41586-018-0001-x} {\bibfield  {journal} {\bibinfo  {journal} {\nat}\ }\textbf {\bibinfo {volume} {556}},\ \bibinfo {pages} {219} (\bibinfo {year} {2018})}\BibitemShut {NoStop}%
\bibitem [{\citenamefont {{Link}}\ \emph {et~al.}(1999)\citenamefont {{Link}}, \citenamefont {{Epstein}},\ and\ \citenamefont {{Lattimer}}}]{1999PhRvL..83.3362L}%
  \BibitemOpen
  \bibfield  {author} {\bibinfo {author} {\bibfnamefont {B.}~\bibnamefont {{Link}}}, \bibinfo {author} {\bibfnamefont {R.~I.}\ \bibnamefont {{Epstein}}}, \ and\ \bibinfo {author} {\bibfnamefont {J.~M.}\ \bibnamefont {{Lattimer}}},\ }\href {\doibase 10.1103/PhysRevLett.83.3362} {\bibfield  {journal} {\bibinfo  {journal} {Physical Review Letters}\ }\textbf {\bibinfo {volume} {83}},\ \bibinfo {pages} {3362} (\bibinfo {year} {1999})},\ \Eprint {http://arxiv.org/abs/astro-ph/9909146} {arXiv:astro-ph/9909146 [astro-ph]} \BibitemShut {NoStop}%
\bibitem [{\citenamefont {{Dib}}\ and\ \citenamefont {{Kaspi}}(2014)}]{2014ApJ...784...37D}%
  \BibitemOpen
  \bibfield  {author} {\bibinfo {author} {\bibfnamefont {R.}~\bibnamefont {{Dib}}}\ and\ \bibinfo {author} {\bibfnamefont {V.~M.}\ \bibnamefont {{Kaspi}}},\ }\href {\doibase 10.1088/0004-637X/784/1/37} {\bibfield  {journal} {\bibinfo  {journal} {\apj}\ }\textbf {\bibinfo {volume} {784}},\ \bibinfo {eid} {37} (\bibinfo {year} {2014})},\ \Eprint {http://arxiv.org/abs/1401.3085} {arXiv:1401.3085 [astro-ph.HE]} \BibitemShut {NoStop}%
\bibitem [{\citenamefont {{Akbal}}\ \emph {et~al.}(2015)\citenamefont {{Akbal}}, \citenamefont {{G{\"u}gercino{\u{g}}lu}}, \citenamefont {{{\c{S}}a{\c{s}}maz Mu{\c{s}}}},\ and\ \citenamefont {{Alpar}}}]{2015MNRAS.449..933A}%
  \BibitemOpen
  \bibfield  {author} {\bibinfo {author} {\bibfnamefont {O.}~\bibnamefont {{Akbal}}}, \bibinfo {author} {\bibfnamefont {E.}~\bibnamefont {{G{\"u}gercino{\u{g}}lu}}}, \bibinfo {author} {\bibfnamefont {S.}~\bibnamefont {{{\c{S}}a{\c{s}}maz Mu{\c{s}}}}}, \ and\ \bibinfo {author} {\bibfnamefont {M.~A.}\ \bibnamefont {{Alpar}}},\ }\href {\doibase 10.1093/mnras/stv322} {\bibfield  {journal} {\bibinfo  {journal} {\mnras}\ }\textbf {\bibinfo {volume} {449}},\ \bibinfo {pages} {933} (\bibinfo {year} {2015})},\ \Eprint {http://arxiv.org/abs/1502.03786} {arXiv:1502.03786 [astro-ph.HE]} \BibitemShut {NoStop}%
\bibitem [{\citenamefont {{Hu}}\ and\ \citenamefont {{Ng}}(2019)}]{2019AN....340..340H}%
  \BibitemOpen
  \bibfield  {author} {\bibinfo {author} {\bibfnamefont {C.~P.}\ \bibnamefont {{Hu}}}\ and\ \bibinfo {author} {\bibfnamefont {C.~Y.}\ \bibnamefont {{Ng}}},\ }\href {\doibase 10.1002/asna.201913621} {\bibfield  {journal} {\bibinfo  {journal} {Astronomische Nachrichten}\ }\textbf {\bibinfo {volume} {340}},\ \bibinfo {pages} {340} (\bibinfo {year} {2019})},\ \Eprint {http://arxiv.org/abs/1903.09736} {arXiv:1903.09736 [astro-ph.HE]} \BibitemShut {NoStop}%
\bibitem [{\citenamefont {{Fuentes}}\ \emph {et~al.}(2017)\citenamefont {{Fuentes}}, \citenamefont {{Espinoza}}, \citenamefont {{Reisenegger}}, \citenamefont {{Shaw}}, \citenamefont {{Stappers}},\ and\ \citenamefont {{Lyne}}}]{2017A&A...608A.131F}%
  \BibitemOpen
  \bibfield  {author} {\bibinfo {author} {\bibfnamefont {J.~R.}\ \bibnamefont {{Fuentes}}}, \bibinfo {author} {\bibfnamefont {C.~M.}\ \bibnamefont {{Espinoza}}}, \bibinfo {author} {\bibfnamefont {A.}~\bibnamefont {{Reisenegger}}}, \bibinfo {author} {\bibfnamefont {B.}~\bibnamefont {{Shaw}}}, \bibinfo {author} {\bibfnamefont {B.~W.}\ \bibnamefont {{Stappers}}}, \ and\ \bibinfo {author} {\bibfnamefont {A.~G.}\ \bibnamefont {{Lyne}}},\ }\href {\doibase 10.1051/0004-6361/201731519} {\bibfield  {journal} {\bibinfo  {journal} {\aap}\ }\textbf {\bibinfo {volume} {608}},\ \bibinfo {eid} {A131} (\bibinfo {year} {2017})},\ \Eprint {http://arxiv.org/abs/1710.00952} {arXiv:1710.00952 [astro-ph.HE]} \BibitemShut {NoStop}%
\bibitem [{\citenamefont {{Ruderman}}(1991)}]{1991ApJ...382..587R}%
  \BibitemOpen
  \bibfield  {author} {\bibinfo {author} {\bibfnamefont {M.}~\bibnamefont {{Ruderman}}},\ }\href {\doibase 10.1086/170745} {\bibfield  {journal} {\bibinfo  {journal} {\apj}\ }\textbf {\bibinfo {volume} {382}},\ \bibinfo {pages} {587} (\bibinfo {year} {1991})}\BibitemShut {NoStop}%
\bibitem [{\citenamefont {{Suvorov}}\ and\ \citenamefont {{Kokkotas}}(2019)}]{2019MNRAS.488.5887S}%
  \BibitemOpen
  \bibfield  {author} {\bibinfo {author} {\bibfnamefont {A.~G.}\ \bibnamefont {{Suvorov}}}\ and\ \bibinfo {author} {\bibfnamefont {K.~D.}\ \bibnamefont {{Kokkotas}}},\ }\href {\doibase 10.1093/mnras/stz2052} {\bibfield  {journal} {\bibinfo  {journal} {\mnras}\ }\textbf {\bibinfo {volume} {488}},\ \bibinfo {pages} {5887} (\bibinfo {year} {2019})},\ \Eprint {http://arxiv.org/abs/1907.10394} {arXiv:1907.10394 [astro-ph.HE]} \BibitemShut {NoStop}%
\bibitem [{\citenamefont {{Wang}}\ \emph {et~al.}(2018)\citenamefont {{Wang}}, \citenamefont {{Luo}}, \citenamefont {{Yue}}, \citenamefont {{Chen}}, \citenamefont {{Lee}},\ and\ \citenamefont {{Xu}}}]{2018ApJ...852..140W}%
  \BibitemOpen
  \bibfield  {author} {\bibinfo {author} {\bibfnamefont {W.}~\bibnamefont {{Wang}}}, \bibinfo {author} {\bibfnamefont {R.}~\bibnamefont {{Luo}}}, \bibinfo {author} {\bibfnamefont {H.}~\bibnamefont {{Yue}}}, \bibinfo {author} {\bibfnamefont {X.}~\bibnamefont {{Chen}}}, \bibinfo {author} {\bibfnamefont {K.}~\bibnamefont {{Lee}}}, \ and\ \bibinfo {author} {\bibfnamefont {R.}~\bibnamefont {{Xu}}},\ }\href {\doibase 10.3847/1538-4357/aaa025} {\bibfield  {journal} {\bibinfo  {journal} {\apj}\ }\textbf {\bibinfo {volume} {852}},\ \bibinfo {eid} {140} (\bibinfo {year} {2018})},\ \Eprint {http://arxiv.org/abs/1710.00541} {arXiv:1710.00541 [astro-ph.HE]} \BibitemShut {NoStop}%
\bibitem [{\citenamefont {{Wadiasingh}}\ and\ \citenamefont {{Timokhin}}(2019)}]{2019ApJ...879....4W}%
  \BibitemOpen
  \bibfield  {author} {\bibinfo {author} {\bibfnamefont {Z.}~\bibnamefont {{Wadiasingh}}}\ and\ \bibinfo {author} {\bibfnamefont {A.}~\bibnamefont {{Timokhin}}},\ }\href {\doibase 10.3847/1538-4357/ab2240} {\bibfield  {journal} {\bibinfo  {journal} {\apj}\ }\textbf {\bibinfo {volume} {879}},\ \bibinfo {eid} {4} (\bibinfo {year} {2019})},\ \Eprint {http://arxiv.org/abs/1904.12036} {arXiv:1904.12036 [astro-ph.HE]} \BibitemShut {NoStop}%
\bibitem [{\citenamefont {{Wasserman}}\ \emph {et~al.}(2022)\citenamefont {{Wasserman}}, \citenamefont {{Cordes}}, \citenamefont {{Chatterjee}},\ and\ \citenamefont {{Batra}}}]{2022ApJ...928...53W}%
  \BibitemOpen
  \bibfield  {author} {\bibinfo {author} {\bibfnamefont {I.}~\bibnamefont {{Wasserman}}}, \bibinfo {author} {\bibfnamefont {J.~M.}\ \bibnamefont {{Cordes}}}, \bibinfo {author} {\bibfnamefont {S.}~\bibnamefont {{Chatterjee}}}, \ and\ \bibinfo {author} {\bibfnamefont {G.}~\bibnamefont {{Batra}}},\ }\href {\doibase 10.3847/1538-4357/ac38a6} {\bibfield  {journal} {\bibinfo  {journal} {\apj}\ }\textbf {\bibinfo {volume} {928}},\ \bibinfo {eid} {53} (\bibinfo {year} {2022})},\ \Eprint {http://arxiv.org/abs/2107.12911} {arXiv:2107.12911 [astro-ph.HE]} \BibitemShut {NoStop}%
\bibitem [{\citenamefont {{Totani}}\ and\ \citenamefont {{Tsuzuki}}(2023)}]{2023MNRAS.526.2795T}%
  \BibitemOpen
  \bibfield  {author} {\bibinfo {author} {\bibfnamefont {T.}~\bibnamefont {{Totani}}}\ and\ \bibinfo {author} {\bibfnamefont {Y.}~\bibnamefont {{Tsuzuki}}},\ }\href {\doibase 10.1093/mnras/stad2532} {\bibfield  {journal} {\bibinfo  {journal} {\mnras}\ }\textbf {\bibinfo {volume} {526}},\ \bibinfo {pages} {2795} (\bibinfo {year} {2023})},\ \Eprint {http://arxiv.org/abs/2306.13612} {arXiv:2306.13612 [astro-ph.HE]} \BibitemShut {NoStop}%
\bibitem [{\citenamefont {{Wu}}\ \emph {et~al.}(2025)\citenamefont {{Wu}}, \citenamefont {{Wang}}, \citenamefont {{Zhao}}, \citenamefont {{Wang}}, \citenamefont {{Xu}}, \citenamefont {{Zhang}}, \citenamefont {{Zhou}}, \citenamefont {{Niu}}, \citenamefont {{Wang}}, \citenamefont {{Yi}}, \citenamefont {{Hua}}, \citenamefont {{Zhang}}, \citenamefont {{Han}}, \citenamefont {{Zhu}}, \citenamefont {{Lee}}, \citenamefont {{Li}}, \citenamefont {{Wu}}, \citenamefont {{Dai}},\ and\ \citenamefont {{Zhang}}}]{2025ApJ...979L..42W}%
  \BibitemOpen
  \bibfield  {author} {\bibinfo {author} {\bibfnamefont {Q.}~\bibnamefont {{Wu}}}, \bibinfo {author} {\bibfnamefont {F.~Y.}\ \bibnamefont {{Wang}}}, \bibinfo {author} {\bibfnamefont {Z.~Y.}\ \bibnamefont {{Zhao}}}, \bibinfo {author} {\bibfnamefont {P.}~\bibnamefont {{Wang}}}, \bibinfo {author} {\bibfnamefont {H.}~\bibnamefont {{Xu}}}, \bibinfo {author} {\bibfnamefont {Y.~K.}\ \bibnamefont {{Zhang}}}, \bibinfo {author} {\bibfnamefont {D.~J.}\ \bibnamefont {{Zhou}}}, \bibinfo {author} {\bibfnamefont {J.~R.}\ \bibnamefont {{Niu}}}, \bibinfo {author} {\bibfnamefont {W.~Y.}\ \bibnamefont {{Wang}}}, \bibinfo {author} {\bibfnamefont {S.~X.}\ \bibnamefont {{Yi}}}, \bibinfo {author} {\bibfnamefont {Z.~Q.}\ \bibnamefont {{Hua}}}, \bibinfo {author} {\bibfnamefont {S.~B.}\ \bibnamefont {{Zhang}}}, \bibinfo {author} {\bibfnamefont {J.~L.}\ \bibnamefont {{Han}}}, \bibinfo {author} {\bibfnamefont {W.~W.}\ \bibnamefont {{Zhu}}}, \bibinfo {author} {\bibfnamefont {K.~J.}\ \bibnamefont {{Lee}}}, \bibinfo {author} {\bibfnamefont
  {D.}~\bibnamefont {{Li}}}, \bibinfo {author} {\bibfnamefont {X.~F.}\ \bibnamefont {{Wu}}}, \bibinfo {author} {\bibfnamefont {Z.~G.}\ \bibnamefont {{Dai}}}, \ and\ \bibinfo {author} {\bibfnamefont {B.}~\bibnamefont {{Zhang}}},\ }\href {\doibase 10.3847/2041-8213/adaa7f} {\bibfield  {journal} {\bibinfo  {journal} {\apjl}\ }\textbf {\bibinfo {volume} {979}},\ \bibinfo {eid} {L42} (\bibinfo {year} {2025})},\ \Eprint {http://arxiv.org/abs/2501.09248} {arXiv:2501.09248 [astro-ph.HE]} \BibitemShut {NoStop}%
\bibitem [{\citenamefont {{Yang}}\ and\ \citenamefont {{Zhang}}(2021)}]{2021ApJ...919...89Y}%
  \BibitemOpen
  \bibfield  {author} {\bibinfo {author} {\bibfnamefont {Y.-P.}\ \bibnamefont {{Yang}}}\ and\ \bibinfo {author} {\bibfnamefont {B.}~\bibnamefont {{Zhang}}},\ }\href {\doibase 10.3847/1538-4357/ac14b5} {\bibfield  {journal} {\bibinfo  {journal} {\apj}\ }\textbf {\bibinfo {volume} {919}},\ \bibinfo {eid} {89} (\bibinfo {year} {2021})},\ \Eprint {http://arxiv.org/abs/2104.01925} {arXiv:2104.01925 [astro-ph.HE]} \BibitemShut {NoStop}%
\bibitem [{\citenamefont {{Li}}\ \emph {et~al.}(2022)\citenamefont {{Li}}, \citenamefont {{Ge}}, \citenamefont {{Lin}}, \citenamefont {{Zhang}}, \citenamefont {{Song}}, \citenamefont {{Cao}}, \citenamefont {{Zhang}}, \citenamefont {{Lu}}, \citenamefont {{Xu}}, \citenamefont {{Xiong}}, \citenamefont {{Tuo}}, \citenamefont {{Tan}}, \citenamefont {{Jiang}}, \citenamefont {{Qu}}, \citenamefont {{Zhang}}, \citenamefont {{Wang}}, \citenamefont {{Wang}}, \citenamefont {{Zhang}}, \citenamefont {{Zhang}}, \citenamefont {{Li}}, \citenamefont {{Liu}}, \citenamefont {{Li}}, \citenamefont {{Bu}}, \citenamefont {{Cai}}, \citenamefont {{Chen}}, \citenamefont {{Chen}}, \citenamefont {{Chang}}, \citenamefont {{Chen}}, \citenamefont {{Chen}}, \citenamefont {{Chen}}, \citenamefont {{Cui}}, \citenamefont {{Du}}, \citenamefont {{Gao}}, \citenamefont {{Gao}}, \citenamefont {{Gu}}, \citenamefont {{Guan}}, \citenamefont {{Guo}}, \citenamefont {{Han}}, \citenamefont {{Huang}}, \citenamefont {{Huo}}, \citenamefont {{Jia}},
  \citenamefont {{Jin}}, \citenamefont {{Kong}}, \citenamefont {{Li}}, \citenamefont {{Li}}, \citenamefont {{Li}}, \citenamefont {{Li}}, \citenamefont {{Li}}, \citenamefont {{Li}}, \citenamefont {{Liang}}, \citenamefont {{Liao}}, \citenamefont {{Liu}}, \citenamefont {{Liu}}, \citenamefont {{Liu}}, \citenamefont {{Lu}}, \citenamefont {{Luo}}, \citenamefont {{Luo}}, \citenamefont {{Ma}}, \citenamefont {{Ma}}, \citenamefont {{Ma}}, \citenamefont {{Meng}}, \citenamefont {{Nang}}, \citenamefont {{Nie}}, \citenamefont {{Ou}}, \citenamefont {{Ren}}, \citenamefont {{Sai}}, \citenamefont {{Song}}, \citenamefont {{Sun}}, \citenamefont {{Tao}}, \citenamefont {{Wang}}, \citenamefont {{Wang}}, \citenamefont {{Wang}}, \citenamefont {{Wang}}, \citenamefont {{Wen}}, \citenamefont {{Wu}}, \citenamefont {{Wu}}, \citenamefont {{Wu}}, \citenamefont {{Xiao}}, \citenamefont {{Yang}}, \citenamefont {{Yang}}, \citenamefont {{Yi}}, \citenamefont {{Yin}}, \citenamefont {{You}}, \citenamefont {{Yu}}, \citenamefont {{Zhang}},
  \citenamefont {{Zhang}}, \citenamefont {{Zhang}}, \citenamefont {{Zhang}}, \citenamefont {{Zhang}}, \citenamefont {{Zhang}}, \citenamefont {{Zhang}}, \citenamefont {{Zhao}}, \citenamefont {{Zhao}}, \citenamefont {{Zheng}},\ and\ \citenamefont {{Zhou}}}]{2022ApJ...931...56L}%
  \BibitemOpen
  \bibfield  {author} {\bibinfo {author} {\bibfnamefont {X.}~\bibnamefont {{Li}}}, \bibinfo {author} {\bibfnamefont {M.}~\bibnamefont {{Ge}}}, \bibinfo {author} {\bibfnamefont {L.}~\bibnamefont {{Lin}}}, \bibinfo {author} {\bibfnamefont {S.-N.}\ \bibnamefont {{Zhang}}}, \bibinfo {author} {\bibfnamefont {L.}~\bibnamefont {{Song}}}, \bibinfo {author} {\bibfnamefont {X.}~\bibnamefont {{Cao}}}, \bibinfo {author} {\bibfnamefont {B.}~\bibnamefont {{Zhang}}}, \bibinfo {author} {\bibfnamefont {F.}~\bibnamefont {{Lu}}}, \bibinfo {author} {\bibfnamefont {Y.}~\bibnamefont {{Xu}}}, \bibinfo {author} {\bibfnamefont {S.}~\bibnamefont {{Xiong}}}, \bibinfo {author} {\bibfnamefont {Y.}~\bibnamefont {{Tuo}}}, \bibinfo {author} {\bibfnamefont {Y.}~\bibnamefont {{Tan}}}, \bibinfo {author} {\bibfnamefont {W.}~\bibnamefont {{Jiang}}}, \bibinfo {author} {\bibfnamefont {J.}~\bibnamefont {{Qu}}}, \bibinfo {author} {\bibfnamefont {S.}~\bibnamefont {{Zhang}}}, \bibinfo {author} {\bibfnamefont {L.}~\bibnamefont {{Wang}}}, \bibinfo
  {author} {\bibfnamefont {J.}~\bibnamefont {{Wang}}}, \bibinfo {author} {\bibfnamefont {B.}~\bibnamefont {{Zhang}}}, \bibinfo {author} {\bibfnamefont {P.}~\bibnamefont {{Zhang}}}, \bibinfo {author} {\bibfnamefont {C.}~\bibnamefont {{Li}}}, \bibinfo {author} {\bibfnamefont {C.}~\bibnamefont {{Liu}}}, \bibinfo {author} {\bibfnamefont {T.}~\bibnamefont {{Li}}}, \bibinfo {author} {\bibfnamefont {Q.}~\bibnamefont {{Bu}}}, \bibinfo {author} {\bibfnamefont {C.}~\bibnamefont {{Cai}}}, \bibinfo {author} {\bibfnamefont {Y.}~\bibnamefont {{Chen}}}, \bibinfo {author} {\bibfnamefont {Y.}~\bibnamefont {{Chen}}}, \bibinfo {author} {\bibfnamefont {Z.}~\bibnamefont {{Chang}}}, \bibinfo {author} {\bibfnamefont {L.}~\bibnamefont {{Chen}}}, \bibinfo {author} {\bibfnamefont {T.}~\bibnamefont {{Chen}}}, \bibinfo {author} {\bibfnamefont {Y.}~\bibnamefont {{Chen}}}, \bibinfo {author} {\bibfnamefont {W.}~\bibnamefont {{Cui}}}, \bibinfo {author} {\bibfnamefont {Y.}~\bibnamefont {{Du}}}, \bibinfo {author} {\bibfnamefont
  {G.}~\bibnamefont {{Gao}}}, \bibinfo {author} {\bibfnamefont {H.}~\bibnamefont {{Gao}}}, \bibinfo {author} {\bibfnamefont {Y.}~\bibnamefont {{Gu}}}, \bibinfo {author} {\bibfnamefont {J.}~\bibnamefont {{Guan}}}, \bibinfo {author} {\bibfnamefont {C.}~\bibnamefont {{Guo}}}, \bibinfo {author} {\bibfnamefont {D.}~\bibnamefont {{Han}}}, \bibinfo {author} {\bibfnamefont {Y.}~\bibnamefont {{Huang}}}, \bibinfo {author} {\bibfnamefont {J.}~\bibnamefont {{Huo}}}, \bibinfo {author} {\bibfnamefont {S.}~\bibnamefont {{Jia}}}, \bibinfo {author} {\bibfnamefont {J.}~\bibnamefont {{Jin}}}, \bibinfo {author} {\bibfnamefont {L.}~\bibnamefont {{Kong}}}, \bibinfo {author} {\bibfnamefont {B.}~\bibnamefont {{Li}}}, \bibinfo {author} {\bibfnamefont {G.}~\bibnamefont {{Li}}}, \bibinfo {author} {\bibfnamefont {W.}~\bibnamefont {{Li}}}, \bibinfo {author} {\bibfnamefont {X.}~\bibnamefont {{Li}}}, \bibinfo {author} {\bibfnamefont {X.}~\bibnamefont {{Li}}}, \bibinfo {author} {\bibfnamefont {Z.}~\bibnamefont {{Li}}}, \bibinfo {author}
  {\bibfnamefont {X.}~\bibnamefont {{Liang}}}, \bibinfo {author} {\bibfnamefont {J.}~\bibnamefont {{Liao}}}, \bibinfo {author} {\bibfnamefont {H.}~\bibnamefont {{Liu}}}, \bibinfo {author} {\bibfnamefont {H.}~\bibnamefont {{Liu}}}, \bibinfo {author} {\bibfnamefont {X.}~\bibnamefont {{Liu}}}, \bibinfo {author} {\bibfnamefont {X.}~\bibnamefont {{Lu}}}, \bibinfo {author} {\bibfnamefont {Q.}~\bibnamefont {{Luo}}}, \bibinfo {author} {\bibfnamefont {T.}~\bibnamefont {{Luo}}}, \bibinfo {author} {\bibfnamefont {B.}~\bibnamefont {{Ma}}}, \bibinfo {author} {\bibfnamefont {R.}~\bibnamefont {{Ma}}}, \bibinfo {author} {\bibfnamefont {X.}~\bibnamefont {{Ma}}}, \bibinfo {author} {\bibfnamefont {B.}~\bibnamefont {{Meng}}}, \bibinfo {author} {\bibfnamefont {Y.}~\bibnamefont {{Nang}}}, \bibinfo {author} {\bibfnamefont {J.}~\bibnamefont {{Nie}}}, \bibinfo {author} {\bibfnamefont {G.}~\bibnamefont {{Ou}}}, \bibinfo {author} {\bibfnamefont {X.}~\bibnamefont {{Ren}}}, \bibinfo {author} {\bibfnamefont {N.}~\bibnamefont {{Sai}}},
  \bibinfo {author} {\bibfnamefont {X.}~\bibnamefont {{Song}}}, \bibinfo {author} {\bibfnamefont {L.}~\bibnamefont {{Sun}}}, \bibinfo {author} {\bibfnamefont {L.}~\bibnamefont {{Tao}}}, \bibinfo {author} {\bibfnamefont {C.}~\bibnamefont {{Wang}}}, \bibinfo {author} {\bibfnamefont {P.}~\bibnamefont {{Wang}}}, \bibinfo {author} {\bibfnamefont {W.}~\bibnamefont {{Wang}}}, \bibinfo {author} {\bibfnamefont {Y.}~\bibnamefont {{Wang}}}, \bibinfo {author} {\bibfnamefont {X.}~\bibnamefont {{Wen}}}, \bibinfo {author} {\bibfnamefont {B.}~\bibnamefont {{Wu}}}, \bibinfo {author} {\bibfnamefont {B.}~\bibnamefont {{Wu}}}, \bibinfo {author} {\bibfnamefont {M.}~\bibnamefont {{Wu}}}, \bibinfo {author} {\bibfnamefont {S.}~\bibnamefont {{Xiao}}}, \bibinfo {author} {\bibfnamefont {S.}~\bibnamefont {{Yang}}}, \bibinfo {author} {\bibfnamefont {Y.}~\bibnamefont {{Yang}}}, \bibinfo {author} {\bibfnamefont {Q.}~\bibnamefont {{Yi}}}, \bibinfo {author} {\bibfnamefont {Q.}~\bibnamefont {{Yin}}}, \bibinfo {author} {\bibfnamefont
  {Y.}~\bibnamefont {{You}}}, \bibinfo {author} {\bibfnamefont {W.}~\bibnamefont {{Yu}}}, \bibinfo {author} {\bibfnamefont {F.}~\bibnamefont {{Zhang}}}, \bibinfo {author} {\bibfnamefont {H.}~\bibnamefont {{Zhang}}}, \bibinfo {author} {\bibfnamefont {J.}~\bibnamefont {{Zhang}}}, \bibinfo {author} {\bibfnamefont {W.}~\bibnamefont {{Zhang}}}, \bibinfo {author} {\bibfnamefont {W.}~\bibnamefont {{Zhang}}}, \bibinfo {author} {\bibfnamefont {Y.}~\bibnamefont {{Zhang}}}, \bibinfo {author} {\bibfnamefont {Y.}~\bibnamefont {{Zhang}}}, \bibinfo {author} {\bibfnamefont {H.}~\bibnamefont {{Zhao}}}, \bibinfo {author} {\bibfnamefont {X.}~\bibnamefont {{Zhao}}}, \bibinfo {author} {\bibfnamefont {S.}~\bibnamefont {{Zheng}}}, \ and\ \bibinfo {author} {\bibfnamefont {D.}~\bibnamefont {{Zhou}}},\ }\href {\doibase 10.3847/1538-4357/ac6587} {\bibfield  {journal} {\bibinfo  {journal} {\apj}\ }\textbf {\bibinfo {volume} {931}},\ \bibinfo {eid} {56} (\bibinfo {year} {2022})},\ \Eprint {http://arxiv.org/abs/2204.03253}
  {arXiv:2204.03253 [astro-ph.HE]} \BibitemShut {NoStop}%
\bibitem [{\citenamefont {{Ge}}\ \emph {et~al.}(2024)\citenamefont {{Ge}}, \citenamefont {{Yang}}, \citenamefont {{Lu}}, \citenamefont {{Zhou}}, \citenamefont {{Ji}}, \citenamefont {{Zhang}}, \citenamefont {{Zhang}}, \citenamefont {{Zhang}}, \citenamefont {{Wang}}, \citenamefont {{Lee}}, \citenamefont {{Zhu}}, \citenamefont {{Li}}, \citenamefont {{Hou}},\ and\ \citenamefont {{Li}}}]{2024RAA....24a5016G}%
  \BibitemOpen
  \bibfield  {author} {\bibinfo {author} {\bibfnamefont {M.-Y.}\ \bibnamefont {{Ge}}}, \bibinfo {author} {\bibfnamefont {Y.-P.}\ \bibnamefont {{Yang}}}, \bibinfo {author} {\bibfnamefont {F.-J.}\ \bibnamefont {{Lu}}}, \bibinfo {author} {\bibfnamefont {S.-Q.}\ \bibnamefont {{Zhou}}}, \bibinfo {author} {\bibfnamefont {L.}~\bibnamefont {{Ji}}}, \bibinfo {author} {\bibfnamefont {S.-N.}\ \bibnamefont {{Zhang}}}, \bibinfo {author} {\bibfnamefont {B.}~\bibnamefont {{Zhang}}}, \bibinfo {author} {\bibfnamefont {L.}~\bibnamefont {{Zhang}}}, \bibinfo {author} {\bibfnamefont {P.}~\bibnamefont {{Wang}}}, \bibinfo {author} {\bibfnamefont {K.}~\bibnamefont {{Lee}}}, \bibinfo {author} {\bibfnamefont {W.}~\bibnamefont {{Zhu}}}, \bibinfo {author} {\bibfnamefont {J.}~\bibnamefont {{Li}}}, \bibinfo {author} {\bibfnamefont {X.}~\bibnamefont {{Hou}}}, \ and\ \bibinfo {author} {\bibfnamefont {Q.-C.}\ \bibnamefont {{Li}}},\ }\href {\doibase 10.1088/1674-4527/ad0f0c} {\bibfield  {journal} {\bibinfo  {journal} {Research in Astronomy
  and Astrophysics}\ }\textbf {\bibinfo {volume} {24}},\ \bibinfo {eid} {015016} (\bibinfo {year} {2024})}\BibitemShut {NoStop}%
\bibitem [{\citenamefont {{Shaw}}\ \emph {et~al.}(2018)\citenamefont {{Shaw}}, \citenamefont {{Lyne}}, \citenamefont {{Stappers}}, \citenamefont {{Weltevrede}}, \citenamefont {{Bassa}}, \citenamefont {{Lien}}, \citenamefont {{Mickaliger}}, \citenamefont {{Breton}}, \citenamefont {{Jordan}}, \citenamefont {{Keith}},\ and\ \citenamefont {{Krimm}}}]{2018MNRAS.478.3832S}%
  \BibitemOpen
  \bibfield  {author} {\bibinfo {author} {\bibfnamefont {B.}~\bibnamefont {{Shaw}}}, \bibinfo {author} {\bibfnamefont {A.~G.}\ \bibnamefont {{Lyne}}}, \bibinfo {author} {\bibfnamefont {B.~W.}\ \bibnamefont {{Stappers}}}, \bibinfo {author} {\bibfnamefont {P.}~\bibnamefont {{Weltevrede}}}, \bibinfo {author} {\bibfnamefont {C.~G.}\ \bibnamefont {{Bassa}}}, \bibinfo {author} {\bibfnamefont {A.~Y.}\ \bibnamefont {{Lien}}}, \bibinfo {author} {\bibfnamefont {M.~B.}\ \bibnamefont {{Mickaliger}}}, \bibinfo {author} {\bibfnamefont {R.~P.}\ \bibnamefont {{Breton}}}, \bibinfo {author} {\bibfnamefont {C.~A.}\ \bibnamefont {{Jordan}}}, \bibinfo {author} {\bibfnamefont {M.~J.}\ \bibnamefont {{Keith}}}, \ and\ \bibinfo {author} {\bibfnamefont {H.~A.}\ \bibnamefont {{Krimm}}},\ }\href {\doibase 10.1093/mnras/sty1294} {\bibfield  {journal} {\bibinfo  {journal} {\mnras}\ }\textbf {\bibinfo {volume} {478}},\ \bibinfo {pages} {3832} (\bibinfo {year} {2018})},\ \Eprint {http://arxiv.org/abs/1805.05110} {arXiv:1805.05110
  [astro-ph.HE]} \BibitemShut {NoStop}%
\bibitem [{\citenamefont {{Hu}}\ \emph {et~al.}(2024)\citenamefont {{Hu}}, \citenamefont {{Narita}}, \citenamefont {{Enoto}}, \citenamefont {{Younes}}, \citenamefont {{Wadiasingh}}, \citenamefont {{Baring}}, \citenamefont {{Ho}}, \citenamefont {{Guillot}}, \citenamefont {{Ray}}, \citenamefont {{G{\"u}ver}}, \citenamefont {{Rajwade}}, \citenamefont {{Arzoumanian}}, \citenamefont {{Kouveliotou}}, \citenamefont {{Harding}},\ and\ \citenamefont {{Gendreau}}}]{2024Natur.626..500H}%
  \BibitemOpen
  \bibfield  {author} {\bibinfo {author} {\bibfnamefont {C.-P.}\ \bibnamefont {{Hu}}}, \bibinfo {author} {\bibfnamefont {T.}~\bibnamefont {{Narita}}}, \bibinfo {author} {\bibfnamefont {T.}~\bibnamefont {{Enoto}}}, \bibinfo {author} {\bibfnamefont {G.}~\bibnamefont {{Younes}}}, \bibinfo {author} {\bibfnamefont {Z.}~\bibnamefont {{Wadiasingh}}}, \bibinfo {author} {\bibfnamefont {M.~G.}\ \bibnamefont {{Baring}}}, \bibinfo {author} {\bibfnamefont {W.~C.~G.}\ \bibnamefont {{Ho}}}, \bibinfo {author} {\bibfnamefont {S.}~\bibnamefont {{Guillot}}}, \bibinfo {author} {\bibfnamefont {P.~S.}\ \bibnamefont {{Ray}}}, \bibinfo {author} {\bibfnamefont {T.}~\bibnamefont {{G{\"u}ver}}}, \bibinfo {author} {\bibfnamefont {K.}~\bibnamefont {{Rajwade}}}, \bibinfo {author} {\bibfnamefont {Z.}~\bibnamefont {{Arzoumanian}}}, \bibinfo {author} {\bibfnamefont {C.}~\bibnamefont {{Kouveliotou}}}, \bibinfo {author} {\bibfnamefont {A.~K.}\ \bibnamefont {{Harding}}}, \ and\ \bibinfo {author} {\bibfnamefont {K.~C.}\ \bibnamefont
  {{Gendreau}}},\ }\href {\doibase 10.1038/s41586-023-07012-5} {\bibfield  {journal} {\bibinfo  {journal} {\nat}\ }\textbf {\bibinfo {volume} {626}},\ \bibinfo {pages} {500} (\bibinfo {year} {2024})},\ \Eprint {http://arxiv.org/abs/2402.09291} {arXiv:2402.09291 [astro-ph.HE]} \BibitemShut {NoStop}%
\bibitem [{\citenamefont {{Younes}}\ \emph {et~al.}(2023)\citenamefont {{Younes}}, \citenamefont {{Baring}}, \citenamefont {{Harding}}, \citenamefont {{Enoto}}, \citenamefont {{Wadiasingh}}, \citenamefont {{Pearlman}}, \citenamefont {{Ho}}, \citenamefont {{Guillot}}, \citenamefont {{Arzoumanian}}, \citenamefont {{Borghese}}, \citenamefont {{Gendreau}}, \citenamefont {{G{\"o}{\v{g}}{\"u}{\c{s}}}}, \citenamefont {{G{\"u}ver}}, \citenamefont {{van der Horst}}, \citenamefont {{Hu}}, \citenamefont {{Jaisawal}}, \citenamefont {{Kouveliotou}}, \citenamefont {{Lin}},\ and\ \citenamefont {{Majid}}}]{2023NatAs...7..339Y}%
  \BibitemOpen
  \bibfield  {author} {\bibinfo {author} {\bibfnamefont {G.}~\bibnamefont {{Younes}}}, \bibinfo {author} {\bibfnamefont {M.~G.}\ \bibnamefont {{Baring}}}, \bibinfo {author} {\bibfnamefont {A.~K.}\ \bibnamefont {{Harding}}}, \bibinfo {author} {\bibfnamefont {T.}~\bibnamefont {{Enoto}}}, \bibinfo {author} {\bibfnamefont {Z.}~\bibnamefont {{Wadiasingh}}}, \bibinfo {author} {\bibfnamefont {A.~B.}\ \bibnamefont {{Pearlman}}}, \bibinfo {author} {\bibfnamefont {W.~C.~G.}\ \bibnamefont {{Ho}}}, \bibinfo {author} {\bibfnamefont {S.}~\bibnamefont {{Guillot}}}, \bibinfo {author} {\bibfnamefont {Z.}~\bibnamefont {{Arzoumanian}}}, \bibinfo {author} {\bibfnamefont {A.}~\bibnamefont {{Borghese}}}, \bibinfo {author} {\bibfnamefont {K.}~\bibnamefont {{Gendreau}}}, \bibinfo {author} {\bibfnamefont {E.}~\bibnamefont {{G{\"o}{\v{g}}{\"u}{\c{s}}}}}, \bibinfo {author} {\bibfnamefont {T.}~\bibnamefont {{G{\"u}ver}}}, \bibinfo {author} {\bibfnamefont {A.~J.}\ \bibnamefont {{van der Horst}}}, \bibinfo {author} {\bibfnamefont {C.~P.}\
  \bibnamefont {{Hu}}}, \bibinfo {author} {\bibfnamefont {G.~K.}\ \bibnamefont {{Jaisawal}}}, \bibinfo {author} {\bibfnamefont {C.}~\bibnamefont {{Kouveliotou}}}, \bibinfo {author} {\bibfnamefont {L.}~\bibnamefont {{Lin}}}, \ and\ \bibinfo {author} {\bibfnamefont {W.~A.}\ \bibnamefont {{Majid}}},\ }\href {\doibase 10.1038/s41550-022-01865-y} {\bibfield  {journal} {\bibinfo  {journal} {Nature Astronomy}\ }\textbf {\bibinfo {volume} {7}},\ \bibinfo {pages} {339} (\bibinfo {year} {2023})},\ \Eprint {http://arxiv.org/abs/2210.11518} {arXiv:2210.11518 [astro-ph.HE]} \BibitemShut {NoStop}%
\bibitem [{\citenamefont {{Wu}}\ \emph {et~al.}(2023)\citenamefont {{Wu}}, \citenamefont {{Zhao}},\ and\ \citenamefont {{Wang}}}]{2023MNRAS.523.2732W}%
  \BibitemOpen
  \bibfield  {author} {\bibinfo {author} {\bibfnamefont {Q.}~\bibnamefont {{Wu}}}, \bibinfo {author} {\bibfnamefont {Z.-Y.}\ \bibnamefont {{Zhao}}}, \ and\ \bibinfo {author} {\bibfnamefont {F.-Y.}\ \bibnamefont {{Wang}}},\ }\href {\doibase 10.1093/mnras/stad1585} {\bibfield  {journal} {\bibinfo  {journal} {\mnras}\ }\textbf {\bibinfo {volume} {523}},\ \bibinfo {pages} {2732} (\bibinfo {year} {2023})},\ \Eprint {http://arxiv.org/abs/2305.17316} {arXiv:2305.17316 [astro-ph.HE]} \BibitemShut {NoStop}%
\bibitem [{\citenamefont {{Wang}}\ \emph {et~al.}(2021)\citenamefont {{Wang}}, \citenamefont {{Xu}}, \citenamefont {{Wang}}, \citenamefont {{Du}}, \citenamefont {{Cheng}}, \citenamefont {{Zheng}},\ and\ \citenamefont {{Xu}}}]{2021MNRAS.507.2208W}%
  \BibitemOpen
  \bibfield  {author} {\bibinfo {author} {\bibfnamefont {W.-H.}\ \bibnamefont {{Wang}}}, \bibinfo {author} {\bibfnamefont {H.}~\bibnamefont {{Xu}}}, \bibinfo {author} {\bibfnamefont {W.-Y.}\ \bibnamefont {{Wang}}}, \bibinfo {author} {\bibfnamefont {S.}~\bibnamefont {{Du}}}, \bibinfo {author} {\bibfnamefont {Q.}~\bibnamefont {{Cheng}}}, \bibinfo {author} {\bibfnamefont {X.-P.}\ \bibnamefont {{Zheng}}}, \ and\ \bibinfo {author} {\bibfnamefont {R.-X.}\ \bibnamefont {{Xu}}},\ }\href {\doibase 10.1093/mnras/stab2213} {\bibfield  {journal} {\bibinfo  {journal} {\mnras}\ }\textbf {\bibinfo {volume} {507}},\ \bibinfo {pages} {2208} (\bibinfo {year} {2021})},\ \Eprint {http://arxiv.org/abs/2107.13725} {arXiv:2107.13725 [astro-ph.HE]} \BibitemShut {NoStop}%
\bibitem [{\citenamefont {{Juett}}\ \emph {et~al.}(2002)\citenamefont {{Juett}}, \citenamefont {{Marshall}}, \citenamefont {{Chakrabarty}},\ and\ \citenamefont {{Schulz}}}]{2002ApJ...568L..31J}%
  \BibitemOpen
  \bibfield  {author} {\bibinfo {author} {\bibfnamefont {A.~M.}\ \bibnamefont {{Juett}}}, \bibinfo {author} {\bibfnamefont {H.~L.}\ \bibnamefont {{Marshall}}}, \bibinfo {author} {\bibfnamefont {D.}~\bibnamefont {{Chakrabarty}}}, \ and\ \bibinfo {author} {\bibfnamefont {N.~S.}\ \bibnamefont {{Schulz}}},\ }\href {\doibase 10.1086/340273} {\bibfield  {journal} {\bibinfo  {journal} {\apjl}\ }\textbf {\bibinfo {volume} {568}},\ \bibinfo {pages} {L31} (\bibinfo {year} {2002})},\ \Eprint {http://arxiv.org/abs/astro-ph/0202304} {arXiv:astro-ph/0202304 [astro-ph]} \BibitemShut {NoStop}%
\bibitem [{\citenamefont {{Hobbs}}\ \emph {et~al.}(2006)\citenamefont {{Hobbs}}, \citenamefont {{Edwards}},\ and\ \citenamefont {{Manchester}}}]{2006MNRAS.369..655H}%
  \BibitemOpen
  \bibfield  {author} {\bibinfo {author} {\bibfnamefont {G.~B.}\ \bibnamefont {{Hobbs}}}, \bibinfo {author} {\bibfnamefont {R.~T.}\ \bibnamefont {{Edwards}}}, \ and\ \bibinfo {author} {\bibfnamefont {R.~N.}\ \bibnamefont {{Manchester}}},\ }\href {\doibase 10.1111/j.1365-2966.2006.10302.x} {\bibfield  {journal} {\bibinfo  {journal} {\mnras}\ }\textbf {\bibinfo {volume} {369}},\ \bibinfo {pages} {655} (\bibinfo {year} {2006})},\ \Eprint {http://arxiv.org/abs/astro-ph/0603381} {arXiv:astro-ph/0603381 [astro-ph]} \BibitemShut {NoStop}%
\bibitem [{\citenamefont {{Mazets}}\ \emph {et~al.}(1979)\citenamefont {{Mazets}}, \citenamefont {{Golentskii}}, \citenamefont {{Ilinskii}}, \citenamefont {{Aptekar}},\ and\ \citenamefont {{Guryan}}}]{1979Natur.282..587M}%
  \BibitemOpen
  \bibfield  {author} {\bibinfo {author} {\bibfnamefont {E.~P.}\ \bibnamefont {{Mazets}}}, \bibinfo {author} {\bibfnamefont {S.~V.}\ \bibnamefont {{Golentskii}}}, \bibinfo {author} {\bibfnamefont {V.~N.}\ \bibnamefont {{Ilinskii}}}, \bibinfo {author} {\bibfnamefont {R.~L.}\ \bibnamefont {{Aptekar}}}, \ and\ \bibinfo {author} {\bibfnamefont {I.~A.}\ \bibnamefont {{Guryan}}},\ }\href {\doibase 10.1038/282587a0} {\bibfield  {journal} {\bibinfo  {journal} {\nat}\ }\textbf {\bibinfo {volume} {282}},\ \bibinfo {pages} {587} (\bibinfo {year} {1979})}\BibitemShut {NoStop}%
\bibitem [{\citenamefont {{Hurley}}\ \emph {et~al.}(1999)\citenamefont {{Hurley}}, \citenamefont {{Cline}}, \citenamefont {{Mazets}}, \citenamefont {{Barthelmy}}, \citenamefont {{Butterworth}}, \citenamefont {{Marshall}}, \citenamefont {{Palmer}}, \citenamefont {{Aptekar}}, \citenamefont {{Golenetskii}}, \citenamefont {{Il'Inskii}}, \citenamefont {{Frederiks}}, \citenamefont {{McTiernan}}, \citenamefont {{Gold}},\ and\ \citenamefont {{Trombka}}}]{1999Natur.397...41H}%
  \BibitemOpen
  \bibfield  {author} {\bibinfo {author} {\bibfnamefont {K.}~\bibnamefont {{Hurley}}}, \bibinfo {author} {\bibfnamefont {T.}~\bibnamefont {{Cline}}}, \bibinfo {author} {\bibfnamefont {E.}~\bibnamefont {{Mazets}}}, \bibinfo {author} {\bibfnamefont {S.}~\bibnamefont {{Barthelmy}}}, \bibinfo {author} {\bibfnamefont {P.}~\bibnamefont {{Butterworth}}}, \bibinfo {author} {\bibfnamefont {F.}~\bibnamefont {{Marshall}}}, \bibinfo {author} {\bibfnamefont {D.}~\bibnamefont {{Palmer}}}, \bibinfo {author} {\bibfnamefont {R.}~\bibnamefont {{Aptekar}}}, \bibinfo {author} {\bibfnamefont {S.}~\bibnamefont {{Golenetskii}}}, \bibinfo {author} {\bibfnamefont {V.}~\bibnamefont {{Il'Inskii}}}, \bibinfo {author} {\bibfnamefont {D.}~\bibnamefont {{Frederiks}}}, \bibinfo {author} {\bibfnamefont {J.}~\bibnamefont {{McTiernan}}}, \bibinfo {author} {\bibfnamefont {R.}~\bibnamefont {{Gold}}}, \ and\ \bibinfo {author} {\bibfnamefont {J.}~\bibnamefont {{Trombka}}},\ }\href {\doibase 10.1038/16199} {\bibfield  {journal} {\bibinfo  {journal}
  {\nat}\ }\textbf {\bibinfo {volume} {397}},\ \bibinfo {pages} {41} (\bibinfo {year} {1999})},\ \Eprint {http://arxiv.org/abs/astro-ph/9811443} {arXiv:astro-ph/9811443 [astro-ph]} \BibitemShut {NoStop}%
\bibitem [{\citenamefont {{Hurley}}\ \emph {et~al.}(2005)\citenamefont {{Hurley}}, \citenamefont {{Boggs}}, \citenamefont {{Smith}}, \citenamefont {{Duncan}}, \citenamefont {{Lin}}, \citenamefont {{Zoglauer}}, \citenamefont {{Krucker}}, \citenamefont {{Hurford}}, \citenamefont {{Hudson}}, \citenamefont {{Wigger}}, \citenamefont {{Hajdas}}, \citenamefont {{Thompson}}, \citenamefont {{Mitrofanov}}, \citenamefont {{Sanin}}, \citenamefont {{Boynton}}, \citenamefont {{Fellows}}, \citenamefont {{von Kienlin}}, \citenamefont {{Lichti}}, \citenamefont {{Rau}},\ and\ \citenamefont {{Cline}}}]{2005Natur.434.1098H}%
  \BibitemOpen
  \bibfield  {author} {\bibinfo {author} {\bibfnamefont {K.}~\bibnamefont {{Hurley}}}, \bibinfo {author} {\bibfnamefont {S.~E.}\ \bibnamefont {{Boggs}}}, \bibinfo {author} {\bibfnamefont {D.~M.}\ \bibnamefont {{Smith}}}, \bibinfo {author} {\bibfnamefont {R.~C.}\ \bibnamefont {{Duncan}}}, \bibinfo {author} {\bibfnamefont {R.}~\bibnamefont {{Lin}}}, \bibinfo {author} {\bibfnamefont {A.}~\bibnamefont {{Zoglauer}}}, \bibinfo {author} {\bibfnamefont {S.}~\bibnamefont {{Krucker}}}, \bibinfo {author} {\bibfnamefont {G.}~\bibnamefont {{Hurford}}}, \bibinfo {author} {\bibfnamefont {H.}~\bibnamefont {{Hudson}}}, \bibinfo {author} {\bibfnamefont {C.}~\bibnamefont {{Wigger}}}, \bibinfo {author} {\bibfnamefont {W.}~\bibnamefont {{Hajdas}}}, \bibinfo {author} {\bibfnamefont {C.}~\bibnamefont {{Thompson}}}, \bibinfo {author} {\bibfnamefont {I.}~\bibnamefont {{Mitrofanov}}}, \bibinfo {author} {\bibfnamefont {A.}~\bibnamefont {{Sanin}}}, \bibinfo {author} {\bibfnamefont {W.}~\bibnamefont {{Boynton}}}, \bibinfo {author}
  {\bibfnamefont {C.}~\bibnamefont {{Fellows}}}, \bibinfo {author} {\bibfnamefont {A.}~\bibnamefont {{von Kienlin}}}, \bibinfo {author} {\bibfnamefont {G.}~\bibnamefont {{Lichti}}}, \bibinfo {author} {\bibfnamefont {A.}~\bibnamefont {{Rau}}}, \ and\ \bibinfo {author} {\bibfnamefont {T.}~\bibnamefont {{Cline}}},\ }\href {\doibase 10.1038/nature03519} {\bibfield  {journal} {\bibinfo  {journal} {\nat}\ }\textbf {\bibinfo {volume} {434}},\ \bibinfo {pages} {1098} (\bibinfo {year} {2005})},\ \Eprint {http://arxiv.org/abs/astro-ph/0502329} {arXiv:astro-ph/0502329 [astro-ph]} \BibitemShut {NoStop}%
\bibitem [{\citenamefont {{Palmer}}\ \emph {et~al.}(2005)\citenamefont {{Palmer}}, \citenamefont {{Barthelmy}}, \citenamefont {{Gehrels}}, \citenamefont {{Kippen}}, \citenamefont {{Cayton}}, \citenamefont {{Kouveliotou}}, \citenamefont {{Eichler}}, \citenamefont {{Wijers}}, \citenamefont {{Woods}}, \citenamefont {{Granot}}, \citenamefont {{Lyubarsky}}, \citenamefont {{Ramirez-Ruiz}}, \citenamefont {{Barbier}}, \citenamefont {{Chester}}, \citenamefont {{Cummings}}, \citenamefont {{Fenimore}}, \citenamefont {{Finger}}, \citenamefont {{Gaensler}}, \citenamefont {{Hullinger}}, \citenamefont {{Krimm}}, \citenamefont {{Markwardt}}, \citenamefont {{Nousek}}, \citenamefont {{Parsons}}, \citenamefont {{Patel}}, \citenamefont {{Sakamoto}}, \citenamefont {{Sato}}, \citenamefont {{Suzuki}},\ and\ \citenamefont {{Tueller}}}]{2005Natur.434.1107P}%
  \BibitemOpen
  \bibfield  {author} {\bibinfo {author} {\bibfnamefont {D.~M.}\ \bibnamefont {{Palmer}}}, \bibinfo {author} {\bibfnamefont {S.}~\bibnamefont {{Barthelmy}}}, \bibinfo {author} {\bibfnamefont {N.}~\bibnamefont {{Gehrels}}}, \bibinfo {author} {\bibfnamefont {R.~M.}\ \bibnamefont {{Kippen}}}, \bibinfo {author} {\bibfnamefont {T.}~\bibnamefont {{Cayton}}}, \bibinfo {author} {\bibfnamefont {C.}~\bibnamefont {{Kouveliotou}}}, \bibinfo {author} {\bibfnamefont {D.}~\bibnamefont {{Eichler}}}, \bibinfo {author} {\bibfnamefont {R.~A.~M.~J.}\ \bibnamefont {{Wijers}}}, \bibinfo {author} {\bibfnamefont {P.~M.}\ \bibnamefont {{Woods}}}, \bibinfo {author} {\bibfnamefont {J.}~\bibnamefont {{Granot}}}, \bibinfo {author} {\bibfnamefont {Y.~E.}\ \bibnamefont {{Lyubarsky}}}, \bibinfo {author} {\bibfnamefont {E.}~\bibnamefont {{Ramirez-Ruiz}}}, \bibinfo {author} {\bibfnamefont {L.}~\bibnamefont {{Barbier}}}, \bibinfo {author} {\bibfnamefont {M.}~\bibnamefont {{Chester}}}, \bibinfo {author} {\bibfnamefont {J.}~\bibnamefont
  {{Cummings}}}, \bibinfo {author} {\bibfnamefont {E.~E.}\ \bibnamefont {{Fenimore}}}, \bibinfo {author} {\bibfnamefont {M.~H.}\ \bibnamefont {{Finger}}}, \bibinfo {author} {\bibfnamefont {B.~M.}\ \bibnamefont {{Gaensler}}}, \bibinfo {author} {\bibfnamefont {D.}~\bibnamefont {{Hullinger}}}, \bibinfo {author} {\bibfnamefont {H.}~\bibnamefont {{Krimm}}}, \bibinfo {author} {\bibfnamefont {C.~B.}\ \bibnamefont {{Markwardt}}}, \bibinfo {author} {\bibfnamefont {J.~A.}\ \bibnamefont {{Nousek}}}, \bibinfo {author} {\bibfnamefont {A.}~\bibnamefont {{Parsons}}}, \bibinfo {author} {\bibfnamefont {S.}~\bibnamefont {{Patel}}}, \bibinfo {author} {\bibfnamefont {T.}~\bibnamefont {{Sakamoto}}}, \bibinfo {author} {\bibfnamefont {G.}~\bibnamefont {{Sato}}}, \bibinfo {author} {\bibfnamefont {M.}~\bibnamefont {{Suzuki}}}, \ and\ \bibinfo {author} {\bibfnamefont {J.}~\bibnamefont {{Tueller}}},\ }\href {\doibase 10.1038/nature03525} {\bibfield  {journal} {\bibinfo  {journal} {\nat}\ }\textbf {\bibinfo {volume} {434}},\ \bibinfo
  {pages} {1107} (\bibinfo {year} {2005})},\ \Eprint {http://arxiv.org/abs/astro-ph/0503030} {arXiv:astro-ph/0503030 [astro-ph]} \BibitemShut {NoStop}%
\bibitem [{\citenamefont {{Svinkin}}\ \emph {et~al.}(2021)\citenamefont {{Svinkin}}, \citenamefont {{Frederiks}}, \citenamefont {{Hurley}}, \citenamefont {{Aptekar}}, \citenamefont {{Golenetskii}}, \citenamefont {{Lysenko}}, \citenamefont {{Ridnaia}}, \citenamefont {{Tsvetkova}}, \citenamefont {{Ulanov}}, \citenamefont {{Cline}}, \citenamefont {{Mitrofanov}}, \citenamefont {{Golovin}}, \citenamefont {{Kozyrev}}, \citenamefont {{Litvak}}, \citenamefont {{Sanin}}, \citenamefont {{Goldstein}}, \citenamefont {{Briggs}}, \citenamefont {{Wilson-Hodge}}, \citenamefont {{von Kienlin}}, \citenamefont {{Zhang}}, \citenamefont {{Rau}}, \citenamefont {{Savchenko}}, \citenamefont {{Bozzo}}, \citenamefont {{Ferrigno}}, \citenamefont {{Ubertini}}, \citenamefont {{Bazzano}}, \citenamefont {{Rodi}}, \citenamefont {{Barthelmy}}, \citenamefont {{Cummings}}, \citenamefont {{Krimm}}, \citenamefont {{Palmer}}, \citenamefont {{Boynton}}, \citenamefont {{Fellows}}, \citenamefont {{Harshman}}, \citenamefont {{Enos}},\ and\
  \citenamefont {{Starr}}}]{Svinkin2021Natur}%
  \BibitemOpen
  \bibfield  {author} {\bibinfo {author} {\bibfnamefont {D.}~\bibnamefont {{Svinkin}}}, \bibinfo {author} {\bibfnamefont {D.}~\bibnamefont {{Frederiks}}}, \bibinfo {author} {\bibfnamefont {K.}~\bibnamefont {{Hurley}}}, \bibinfo {author} {\bibfnamefont {R.}~\bibnamefont {{Aptekar}}}, \bibinfo {author} {\bibfnamefont {S.}~\bibnamefont {{Golenetskii}}}, \bibinfo {author} {\bibfnamefont {A.}~\bibnamefont {{Lysenko}}}, \bibinfo {author} {\bibfnamefont {A.~V.}\ \bibnamefont {{Ridnaia}}}, \bibinfo {author} {\bibfnamefont {A.}~\bibnamefont {{Tsvetkova}}}, \bibinfo {author} {\bibfnamefont {M.}~\bibnamefont {{Ulanov}}}, \bibinfo {author} {\bibfnamefont {T.~L.}\ \bibnamefont {{Cline}}}, \bibinfo {author} {\bibfnamefont {I.}~\bibnamefont {{Mitrofanov}}}, \bibinfo {author} {\bibfnamefont {D.}~\bibnamefont {{Golovin}}}, \bibinfo {author} {\bibfnamefont {A.}~\bibnamefont {{Kozyrev}}}, \bibinfo {author} {\bibfnamefont {M.}~\bibnamefont {{Litvak}}}, \bibinfo {author} {\bibfnamefont {A.}~\bibnamefont {{Sanin}}}, \bibinfo
  {author} {\bibfnamefont {A.}~\bibnamefont {{Goldstein}}}, \bibinfo {author} {\bibfnamefont {M.~S.}\ \bibnamefont {{Briggs}}}, \bibinfo {author} {\bibfnamefont {C.}~\bibnamefont {{Wilson-Hodge}}}, \bibinfo {author} {\bibfnamefont {A.}~\bibnamefont {{von Kienlin}}}, \bibinfo {author} {\bibfnamefont {X.~L.}\ \bibnamefont {{Zhang}}}, \bibinfo {author} {\bibfnamefont {A.}~\bibnamefont {{Rau}}}, \bibinfo {author} {\bibfnamefont {V.}~\bibnamefont {{Savchenko}}}, \bibinfo {author} {\bibfnamefont {E.}~\bibnamefont {{Bozzo}}}, \bibinfo {author} {\bibfnamefont {C.}~\bibnamefont {{Ferrigno}}}, \bibinfo {author} {\bibfnamefont {P.}~\bibnamefont {{Ubertini}}}, \bibinfo {author} {\bibfnamefont {A.}~\bibnamefont {{Bazzano}}}, \bibinfo {author} {\bibfnamefont {J.~C.}\ \bibnamefont {{Rodi}}}, \bibinfo {author} {\bibfnamefont {S.}~\bibnamefont {{Barthelmy}}}, \bibinfo {author} {\bibfnamefont {J.}~\bibnamefont {{Cummings}}}, \bibinfo {author} {\bibfnamefont {H.}~\bibnamefont {{Krimm}}}, \bibinfo {author} {\bibfnamefont
  {D.~M.}\ \bibnamefont {{Palmer}}}, \bibinfo {author} {\bibfnamefont {W.}~\bibnamefont {{Boynton}}}, \bibinfo {author} {\bibfnamefont {C.~W.}\ \bibnamefont {{Fellows}}}, \bibinfo {author} {\bibfnamefont {K.~P.}\ \bibnamefont {{Harshman}}}, \bibinfo {author} {\bibfnamefont {H.}~\bibnamefont {{Enos}}}, \ and\ \bibinfo {author} {\bibfnamefont {R.}~\bibnamefont {{Starr}}},\ }\href {\doibase 10.1038/s41586-020-03076-9} {\bibfield  {journal} {\bibinfo  {journal} {\nat}\ }\textbf {\bibinfo {volume} {589}},\ \bibinfo {pages} {211} (\bibinfo {year} {2021})},\ \Eprint {http://arxiv.org/abs/2101.05104} {arXiv:2101.05104 [astro-ph.HE]} \BibitemShut {NoStop}%
\bibitem [{\citenamefont {{Coti Zelati}}\ \emph {et~al.}(2018)\citenamefont {{Coti Zelati}}, \citenamefont {{Rea}}, \citenamefont {{Pons}}, \citenamefont {{Campana}},\ and\ \citenamefont {{Esposito}}}]{CotiZelati2018}%
  \BibitemOpen
  \bibfield  {author} {\bibinfo {author} {\bibfnamefont {F.}~\bibnamefont {{Coti Zelati}}}, \bibinfo {author} {\bibfnamefont {N.}~\bibnamefont {{Rea}}}, \bibinfo {author} {\bibfnamefont {J.~A.}\ \bibnamefont {{Pons}}}, \bibinfo {author} {\bibfnamefont {S.}~\bibnamefont {{Campana}}}, \ and\ \bibinfo {author} {\bibfnamefont {P.}~\bibnamefont {{Esposito}}},\ }\href {\doibase 10.1093/mnras/stx2679} {\bibfield  {journal} {\bibinfo  {journal} {\mnras}\ }\textbf {\bibinfo {volume} {474}},\ \bibinfo {pages} {961} (\bibinfo {year} {2018})},\ \Eprint {http://arxiv.org/abs/1710.04671} {arXiv:1710.04671 [astro-ph.HE]} \BibitemShut {NoStop}%
\bibitem [{\citenamefont {{Yuan}}\ \emph {et~al.}(2025)\citenamefont {{Yuan}}, \citenamefont {{Dai}}, \citenamefont {{Feng}}, \citenamefont {{Jin}}, \citenamefont {{Jonker}}, \citenamefont {{Kuulkers}}, \citenamefont {{Liu}}, \citenamefont {{Nandra}}, \citenamefont {{O'Brien}}, \citenamefont {{Piro}}, \citenamefont {{Rau}}, \citenamefont {{Rea}}, \citenamefont {{Sanders}}, \citenamefont {{Tao}}, \citenamefont {{Wang}}, \citenamefont {{Wu}}, \citenamefont {{Zhang}}, \citenamefont {{Zhang}}, \citenamefont {{Ai}}, \citenamefont {{Buchner}}, \citenamefont {{Bulbul}}, \citenamefont {{Chen}}, \citenamefont {{Chen}}, \citenamefont {{Chen}}, \citenamefont {{Chen}}, \citenamefont {{Coleiro}}, \citenamefont {{Coti Zelati}}, \citenamefont {{Dai}}, \citenamefont {{Fan}}, \citenamefont {{Fan}}, \citenamefont {{Friedrich}}, \citenamefont {{Gao}}, \citenamefont {{Ge}}, \citenamefont {{Ge}}, \citenamefont {{Geng}}, \citenamefont {{Ghirlanda}}, \citenamefont {{Gianfagna}}, \citenamefont {{Gou}}, \citenamefont {{Guillot}},
  \citenamefont {{Hou}}, \citenamefont {{Hu}}, \citenamefont {{Huang}}, \citenamefont {{Ji}}, \citenamefont {{Jia}}, \citenamefont {{Komossa}}, \citenamefont {{Kong}}, \citenamefont {{Lan}}, \citenamefont {{Li}}, \citenamefont {{Li}}, \citenamefont {{Li}}, \citenamefont {{Li}}, \citenamefont {{Li}}, \citenamefont {{Li}}, \citenamefont {{Ling}}, \citenamefont {{Liu}}, \citenamefont {{Liu}}, \citenamefont {{Liu}}, \citenamefont {{Liu}}, \citenamefont {{Luo}}, \citenamefont {{Ma}}, \citenamefont {{Maggi}}, \citenamefont {{Maitra}}, \citenamefont {{Marino}}, \citenamefont {{Chi-Yung Ng}}, \citenamefont {{Pan}}, \citenamefont {{Rukdee}}, \citenamefont {{Soria}}, \citenamefont {{Sun}}, \citenamefont {{Tam}}, \citenamefont {{Linesh Thakur}}, \citenamefont {{Tian}}, \citenamefont {{Troja}}, \citenamefont {{Wang}}, \citenamefont {{Wang}}, \citenamefont {{Wang}}, \citenamefont {{Wei}}, \citenamefont {{Wen}}, \citenamefont {{Wu}}, \citenamefont {{Wu}}, \citenamefont {{Xiao}}, \citenamefont {{Xu}}, \citenamefont {{Xu}},
  \citenamefont {{Xu}}, \citenamefont {{Xu}}, \citenamefont {{Yang}}, \citenamefont {{You}}, \citenamefont {{Yu}}, \citenamefont {{Yu}}, \citenamefont {{Zhang}}, \citenamefont {{Zhang}}, \citenamefont {{Zhang}}, \citenamefont {{Zhang}}, \citenamefont {{Zhang}}, \citenamefont {{Zhang}}, \citenamefont {{Zhou}},\ and\ \citenamefont {{Zou}}}]{Yuan2025EP}%
  \BibitemOpen
  \bibfield  {author} {\bibinfo {author} {\bibfnamefont {W.}~\bibnamefont {{Yuan}}}, \bibinfo {author} {\bibfnamefont {L.}~\bibnamefont {{Dai}}}, \bibinfo {author} {\bibfnamefont {H.}~\bibnamefont {{Feng}}}, \bibinfo {author} {\bibfnamefont {C.}~\bibnamefont {{Jin}}}, \bibinfo {author} {\bibfnamefont {P.}~\bibnamefont {{Jonker}}}, \bibinfo {author} {\bibfnamefont {E.}~\bibnamefont {{Kuulkers}}}, \bibinfo {author} {\bibfnamefont {Y.}~\bibnamefont {{Liu}}}, \bibinfo {author} {\bibfnamefont {K.}~\bibnamefont {{Nandra}}}, \bibinfo {author} {\bibfnamefont {P.}~\bibnamefont {{O'Brien}}}, \bibinfo {author} {\bibfnamefont {L.}~\bibnamefont {{Piro}}}, \bibinfo {author} {\bibfnamefont {A.}~\bibnamefont {{Rau}}}, \bibinfo {author} {\bibfnamefont {N.}~\bibnamefont {{Rea}}}, \bibinfo {author} {\bibfnamefont {J.}~\bibnamefont {{Sanders}}}, \bibinfo {author} {\bibfnamefont {L.}~\bibnamefont {{Tao}}}, \bibinfo {author} {\bibfnamefont {J.}~\bibnamefont {{Wang}}}, \bibinfo {author} {\bibfnamefont {X.}~\bibnamefont {{Wu}}},
  \bibinfo {author} {\bibfnamefont {B.}~\bibnamefont {{Zhang}}}, \bibinfo {author} {\bibfnamefont {S.}~\bibnamefont {{Zhang}}}, \bibinfo {author} {\bibfnamefont {S.}~\bibnamefont {{Ai}}}, \bibinfo {author} {\bibfnamefont {J.}~\bibnamefont {{Buchner}}}, \bibinfo {author} {\bibfnamefont {E.}~\bibnamefont {{Bulbul}}}, \bibinfo {author} {\bibfnamefont {H.}~\bibnamefont {{Chen}}}, \bibinfo {author} {\bibfnamefont {M.}~\bibnamefont {{Chen}}}, \bibinfo {author} {\bibfnamefont {Y.}~\bibnamefont {{Chen}}}, \bibinfo {author} {\bibfnamefont {Y.-P.}\ \bibnamefont {{Chen}}}, \bibinfo {author} {\bibfnamefont {A.}~\bibnamefont {{Coleiro}}}, \bibinfo {author} {\bibfnamefont {F.}~\bibnamefont {{Coti Zelati}}}, \bibinfo {author} {\bibfnamefont {Z.}~\bibnamefont {{Dai}}}, \bibinfo {author} {\bibfnamefont {X.}~\bibnamefont {{Fan}}}, \bibinfo {author} {\bibfnamefont {Z.}~\bibnamefont {{Fan}}}, \bibinfo {author} {\bibfnamefont {S.}~\bibnamefont {{Friedrich}}}, \bibinfo {author} {\bibfnamefont {H.}~\bibnamefont {{Gao}}}, \bibinfo
  {author} {\bibfnamefont {C.}~\bibnamefont {{Ge}}}, \bibinfo {author} {\bibfnamefont {M.}~\bibnamefont {{Ge}}}, \bibinfo {author} {\bibfnamefont {J.}~\bibnamefont {{Geng}}}, \bibinfo {author} {\bibfnamefont {G.}~\bibnamefont {{Ghirlanda}}}, \bibinfo {author} {\bibfnamefont {G.}~\bibnamefont {{Gianfagna}}}, \bibinfo {author} {\bibfnamefont {L.}~\bibnamefont {{Gou}}}, \bibinfo {author} {\bibfnamefont {S.}~\bibnamefont {{Guillot}}}, \bibinfo {author} {\bibfnamefont {X.}~\bibnamefont {{Hou}}}, \bibinfo {author} {\bibfnamefont {J.}~\bibnamefont {{Hu}}}, \bibinfo {author} {\bibfnamefont {Y.}~\bibnamefont {{Huang}}}, \bibinfo {author} {\bibfnamefont {L.}~\bibnamefont {{Ji}}}, \bibinfo {author} {\bibfnamefont {S.}~\bibnamefont {{Jia}}}, \bibinfo {author} {\bibfnamefont {S.}~\bibnamefont {{Komossa}}}, \bibinfo {author} {\bibfnamefont {A.~K.~H.}\ \bibnamefont {{Kong}}}, \bibinfo {author} {\bibfnamefont {L.}~\bibnamefont {{Lan}}}, \bibinfo {author} {\bibfnamefont {A.}~\bibnamefont {{Li}}}, \bibinfo {author}
  {\bibfnamefont {A.}~\bibnamefont {{Li}}}, \bibinfo {author} {\bibfnamefont {C.}~\bibnamefont {{Li}}}, \bibinfo {author} {\bibfnamefont {D.}~\bibnamefont {{Li}}}, \bibinfo {author} {\bibfnamefont {J.}~\bibnamefont {{Li}}}, \bibinfo {author} {\bibfnamefont {Z.}~\bibnamefont {{Li}}}, \bibinfo {author} {\bibfnamefont {Z.}~\bibnamefont {{Ling}}}, \bibinfo {author} {\bibfnamefont {A.}~\bibnamefont {{Liu}}}, \bibinfo {author} {\bibfnamefont {J.}~\bibnamefont {{Liu}}}, \bibinfo {author} {\bibfnamefont {L.}~\bibnamefont {{Liu}}}, \bibinfo {author} {\bibfnamefont {Z.}~\bibnamefont {{Liu}}}, \bibinfo {author} {\bibfnamefont {J.}~\bibnamefont {{Luo}}}, \bibinfo {author} {\bibfnamefont {R.}~\bibnamefont {{Ma}}}, \bibinfo {author} {\bibfnamefont {P.}~\bibnamefont {{Maggi}}}, \bibinfo {author} {\bibfnamefont {C.}~\bibnamefont {{Maitra}}}, \bibinfo {author} {\bibfnamefont {A.}~\bibnamefont {{Marino}}}, \bibinfo {author} {\bibfnamefont {S.}~\bibnamefont {{Chi-Yung Ng}}}, \bibinfo {author} {\bibfnamefont {H.}~\bibnamefont
  {{Pan}}}, \bibinfo {author} {\bibfnamefont {S.}~\bibnamefont {{Rukdee}}}, \bibinfo {author} {\bibfnamefont {R.}~\bibnamefont {{Soria}}}, \bibinfo {author} {\bibfnamefont {H.}~\bibnamefont {{Sun}}}, \bibinfo {author} {\bibfnamefont {P.-H.~T.}\ \bibnamefont {{Tam}}}, \bibinfo {author} {\bibfnamefont {A.}~\bibnamefont {{Linesh Thakur}}}, \bibinfo {author} {\bibfnamefont {H.}~\bibnamefont {{Tian}}}, \bibinfo {author} {\bibfnamefont {E.}~\bibnamefont {{Troja}}}, \bibinfo {author} {\bibfnamefont {W.}~\bibnamefont {{Wang}}}, \bibinfo {author} {\bibfnamefont {X.}~\bibnamefont {{Wang}}}, \bibinfo {author} {\bibfnamefont {Y.}~\bibnamefont {{Wang}}}, \bibinfo {author} {\bibfnamefont {J.}~\bibnamefont {{Wei}}}, \bibinfo {author} {\bibfnamefont {S.}~\bibnamefont {{Wen}}}, \bibinfo {author} {\bibfnamefont {J.}~\bibnamefont {{Wu}}}, \bibinfo {author} {\bibfnamefont {T.}~\bibnamefont {{Wu}}}, \bibinfo {author} {\bibfnamefont {D.}~\bibnamefont {{Xiao}}}, \bibinfo {author} {\bibfnamefont {D.}~\bibnamefont {{Xu}}}, \bibinfo
  {author} {\bibfnamefont {R.}~\bibnamefont {{Xu}}}, \bibinfo {author} {\bibfnamefont {Y.}~\bibnamefont {{Xu}}}, \bibinfo {author} {\bibfnamefont {Y.}~\bibnamefont {{Xu}}}, \bibinfo {author} {\bibfnamefont {H.}~\bibnamefont {{Yang}}}, \bibinfo {author} {\bibfnamefont {B.}~\bibnamefont {{You}}}, \bibinfo {author} {\bibfnamefont {H.}~\bibnamefont {{Yu}}}, \bibinfo {author} {\bibfnamefont {Y.}~\bibnamefont {{Yu}}}, \bibinfo {author} {\bibfnamefont {B.}~\bibnamefont {{Zhang}}}, \bibinfo {author} {\bibfnamefont {C.}~\bibnamefont {{Zhang}}}, \bibinfo {author} {\bibfnamefont {G.}~\bibnamefont {{Zhang}}}, \bibinfo {author} {\bibfnamefont {L.}~\bibnamefont {{Zhang}}}, \bibinfo {author} {\bibfnamefont {W.}~\bibnamefont {{Zhang}}}, \bibinfo {author} {\bibfnamefont {Y.}~\bibnamefont {{Zhang}}}, \bibinfo {author} {\bibfnamefont {P.}~\bibnamefont {{Zhou}}}, \ and\ \bibinfo {author} {\bibfnamefont {Z.}~\bibnamefont {{Zou}}},\ }\href {\doibase 10.48550/arXiv.2501.07362} {\bibfield  {journal} {\bibinfo  {journal} {arXiv
  e-prints}\ ,\ \bibinfo {eid} {arXiv:2501.07362}} (\bibinfo {year} {2025})},\ \Eprint {http://arxiv.org/abs/2501.07362} {arXiv:2501.07362 [astro-ph.HE]} \BibitemShut {NoStop}%
\bibitem [{\citenamefont {{Wei}}\ \emph {et~al.}(2016)\citenamefont {{Wei}}, \citenamefont {{Cordier}}, \citenamefont {{Antier}}, \citenamefont {{Antilogus}}, \citenamefont {{Atteia}}, \citenamefont {{Bajat}}, \citenamefont {{Basa}}, \citenamefont {{Beckmann}}, \citenamefont {{Bernardini}}, \citenamefont {{Boissier}}, \citenamefont {{Bouchet}}, \citenamefont {{Burwitz}}, \citenamefont {{Claret}}, \citenamefont {{Dai}}, \citenamefont {{Daigne}}, \citenamefont {{Deng}}, \citenamefont {{Dornic}}, \citenamefont {{Feng}}, \citenamefont {{Foglizzo}}, \citenamefont {{Gao}}, \citenamefont {{Gehrels}}, \citenamefont {{Godet}}, \citenamefont {{Goldwurm}}, \citenamefont {{Gonzalez}}, \citenamefont {{Gosset}}, \citenamefont {{G{\"o}tz}}, \citenamefont {{Gouiffes}}, \citenamefont {{Grise}}, \citenamefont {{Gros}}, \citenamefont {{Guilet}}, \citenamefont {{Han}}, \citenamefont {{Huang}}, \citenamefont {{Huang}}, \citenamefont {{Jouret}}, \citenamefont {{Klotz}}, \citenamefont {{La Marle}}, \citenamefont {{Lachaud}},
  \citenamefont {{Le Floch}}, \citenamefont {{Lee}}, \citenamefont {{Leroy}}, \citenamefont {{Li}}, \citenamefont {{Li}}, \citenamefont {{Li}}, \citenamefont {{Liang}}, \citenamefont {{Lyu}}, \citenamefont {{Mercier}}, \citenamefont {{Migliori}}, \citenamefont {{Mochkovitch}}, \citenamefont {{O'Brien}}, \citenamefont {{Osborne}}, \citenamefont {{Paul}}, \citenamefont {{Perinati}}, \citenamefont {{Petitjean}}, \citenamefont {{Piron}}, \citenamefont {{Qiu}}, \citenamefont {{Rau}}, \citenamefont {{Rodriguez}}, \citenamefont {{Schanne}}, \citenamefont {{Tanvir}}, \citenamefont {{Vangioni}}, \citenamefont {{Vergani}}, \citenamefont {{Wang}}, \citenamefont {{Wang}}, \citenamefont {{Wang}}, \citenamefont {{Wang}}, \citenamefont {{Watson}}, \citenamefont {{Webb}}, \citenamefont {{Wei}}, \citenamefont {{Willingale}}, \citenamefont {{Wu}}, \citenamefont {{Wu}}, \citenamefont {{Xin}}, \citenamefont {{Xu}}, \citenamefont {{Yu}}, \citenamefont {{Yu}}, \citenamefont {{Yu}}, \citenamefont {{Zhang}}, \citenamefont {{Zhang}},
  \citenamefont {{Zhang}},\ and\ \citenamefont {{Zhou}}}]{Wei2016SVOM}%
  \BibitemOpen
  \bibfield  {author} {\bibinfo {author} {\bibfnamefont {J.}~\bibnamefont {{Wei}}}, \bibinfo {author} {\bibfnamefont {B.}~\bibnamefont {{Cordier}}}, \bibinfo {author} {\bibfnamefont {S.}~\bibnamefont {{Antier}}}, \bibinfo {author} {\bibfnamefont {P.}~\bibnamefont {{Antilogus}}}, \bibinfo {author} {\bibfnamefont {J.~L.}\ \bibnamefont {{Atteia}}}, \bibinfo {author} {\bibfnamefont {A.}~\bibnamefont {{Bajat}}}, \bibinfo {author} {\bibfnamefont {S.}~\bibnamefont {{Basa}}}, \bibinfo {author} {\bibfnamefont {V.}~\bibnamefont {{Beckmann}}}, \bibinfo {author} {\bibfnamefont {M.~G.}\ \bibnamefont {{Bernardini}}}, \bibinfo {author} {\bibfnamefont {S.}~\bibnamefont {{Boissier}}}, \bibinfo {author} {\bibfnamefont {L.}~\bibnamefont {{Bouchet}}}, \bibinfo {author} {\bibfnamefont {V.}~\bibnamefont {{Burwitz}}}, \bibinfo {author} {\bibfnamefont {A.}~\bibnamefont {{Claret}}}, \bibinfo {author} {\bibfnamefont {Z.~G.}\ \bibnamefont {{Dai}}}, \bibinfo {author} {\bibfnamefont {F.}~\bibnamefont {{Daigne}}}, \bibinfo {author}
  {\bibfnamefont {J.}~\bibnamefont {{Deng}}}, \bibinfo {author} {\bibfnamefont {D.}~\bibnamefont {{Dornic}}}, \bibinfo {author} {\bibfnamefont {H.}~\bibnamefont {{Feng}}}, \bibinfo {author} {\bibfnamefont {T.}~\bibnamefont {{Foglizzo}}}, \bibinfo {author} {\bibfnamefont {H.}~\bibnamefont {{Gao}}}, \bibinfo {author} {\bibfnamefont {N.}~\bibnamefont {{Gehrels}}}, \bibinfo {author} {\bibfnamefont {O.}~\bibnamefont {{Godet}}}, \bibinfo {author} {\bibfnamefont {A.}~\bibnamefont {{Goldwurm}}}, \bibinfo {author} {\bibfnamefont {F.}~\bibnamefont {{Gonzalez}}}, \bibinfo {author} {\bibfnamefont {L.}~\bibnamefont {{Gosset}}}, \bibinfo {author} {\bibfnamefont {D.}~\bibnamefont {{G{\"o}tz}}}, \bibinfo {author} {\bibfnamefont {C.}~\bibnamefont {{Gouiffes}}}, \bibinfo {author} {\bibfnamefont {F.}~\bibnamefont {{Grise}}}, \bibinfo {author} {\bibfnamefont {A.}~\bibnamefont {{Gros}}}, \bibinfo {author} {\bibfnamefont {J.}~\bibnamefont {{Guilet}}}, \bibinfo {author} {\bibfnamefont {X.}~\bibnamefont {{Han}}}, \bibinfo {author}
  {\bibfnamefont {M.}~\bibnamefont {{Huang}}}, \bibinfo {author} {\bibfnamefont {Y.~F.}\ \bibnamefont {{Huang}}}, \bibinfo {author} {\bibfnamefont {M.}~\bibnamefont {{Jouret}}}, \bibinfo {author} {\bibfnamefont {A.}~\bibnamefont {{Klotz}}}, \bibinfo {author} {\bibfnamefont {O.}~\bibnamefont {{La Marle}}}, \bibinfo {author} {\bibfnamefont {C.}~\bibnamefont {{Lachaud}}}, \bibinfo {author} {\bibfnamefont {E.}~\bibnamefont {{Le Floch}}}, \bibinfo {author} {\bibfnamefont {W.}~\bibnamefont {{Lee}}}, \bibinfo {author} {\bibfnamefont {N.}~\bibnamefont {{Leroy}}}, \bibinfo {author} {\bibfnamefont {L.~X.}\ \bibnamefont {{Li}}}, \bibinfo {author} {\bibfnamefont {S.~C.}\ \bibnamefont {{Li}}}, \bibinfo {author} {\bibfnamefont {Z.}~\bibnamefont {{Li}}}, \bibinfo {author} {\bibfnamefont {E.~W.}\ \bibnamefont {{Liang}}}, \bibinfo {author} {\bibfnamefont {H.}~\bibnamefont {{Lyu}}}, \bibinfo {author} {\bibfnamefont {K.}~\bibnamefont {{Mercier}}}, \bibinfo {author} {\bibfnamefont {G.}~\bibnamefont {{Migliori}}}, \bibinfo
  {author} {\bibfnamefont {R.}~\bibnamefont {{Mochkovitch}}}, \bibinfo {author} {\bibfnamefont {P.}~\bibnamefont {{O'Brien}}}, \bibinfo {author} {\bibfnamefont {J.}~\bibnamefont {{Osborne}}}, \bibinfo {author} {\bibfnamefont {J.}~\bibnamefont {{Paul}}}, \bibinfo {author} {\bibfnamefont {E.}~\bibnamefont {{Perinati}}}, \bibinfo {author} {\bibfnamefont {P.}~\bibnamefont {{Petitjean}}}, \bibinfo {author} {\bibfnamefont {F.}~\bibnamefont {{Piron}}}, \bibinfo {author} {\bibfnamefont {Y.}~\bibnamefont {{Qiu}}}, \bibinfo {author} {\bibfnamefont {A.}~\bibnamefont {{Rau}}}, \bibinfo {author} {\bibfnamefont {J.}~\bibnamefont {{Rodriguez}}}, \bibinfo {author} {\bibfnamefont {S.}~\bibnamefont {{Schanne}}}, \bibinfo {author} {\bibfnamefont {N.}~\bibnamefont {{Tanvir}}}, \bibinfo {author} {\bibfnamefont {E.}~\bibnamefont {{Vangioni}}}, \bibinfo {author} {\bibfnamefont {S.}~\bibnamefont {{Vergani}}}, \bibinfo {author} {\bibfnamefont {F.~Y.}\ \bibnamefont {{Wang}}}, \bibinfo {author} {\bibfnamefont {J.}~\bibnamefont
  {{Wang}}}, \bibinfo {author} {\bibfnamefont {X.~G.}\ \bibnamefont {{Wang}}}, \bibinfo {author} {\bibfnamefont {X.~Y.}\ \bibnamefont {{Wang}}}, \bibinfo {author} {\bibfnamefont {A.}~\bibnamefont {{Watson}}}, \bibinfo {author} {\bibfnamefont {N.}~\bibnamefont {{Webb}}}, \bibinfo {author} {\bibfnamefont {J.~J.}\ \bibnamefont {{Wei}}}, \bibinfo {author} {\bibfnamefont {R.}~\bibnamefont {{Willingale}}}, \bibinfo {author} {\bibfnamefont {C.}~\bibnamefont {{Wu}}}, \bibinfo {author} {\bibfnamefont {X.~F.}\ \bibnamefont {{Wu}}}, \bibinfo {author} {\bibfnamefont {L.~P.}\ \bibnamefont {{Xin}}}, \bibinfo {author} {\bibfnamefont {D.}~\bibnamefont {{Xu}}}, \bibinfo {author} {\bibfnamefont {S.}~\bibnamefont {{Yu}}}, \bibinfo {author} {\bibfnamefont {W.~F.}\ \bibnamefont {{Yu}}}, \bibinfo {author} {\bibfnamefont {Y.~W.}\ \bibnamefont {{Yu}}}, \bibinfo {author} {\bibfnamefont {B.}~\bibnamefont {{Zhang}}}, \bibinfo {author} {\bibfnamefont {S.~N.}\ \bibnamefont {{Zhang}}}, \bibinfo {author} {\bibfnamefont {Y.}~\bibnamefont
  {{Zhang}}}, \ and\ \bibinfo {author} {\bibfnamefont {X.~L.}\ \bibnamefont {{Zhou}}},\ }\href {\doibase 10.48550/arXiv.1610.06892} {\bibfield  {journal} {\bibinfo  {journal} {arXiv e-prints}\ ,\ \bibinfo {eid} {arXiv:1610.06892}} (\bibinfo {year} {2016})},\ \Eprint {http://arxiv.org/abs/1610.06892} {arXiv:1610.06892 [astro-ph.IM]} \BibitemShut {NoStop}%
\bibitem [{\citenamefont {{Sturrock}}\ \emph {et~al.}(1989)\citenamefont {{Sturrock}}, \citenamefont {{Harding}},\ and\ \citenamefont {{Daugherty}}}]{1989BAAS...21..768S}%
  \BibitemOpen
  \bibfield  {author} {\bibinfo {author} {\bibfnamefont {P.~A.}\ \bibnamefont {{Sturrock}}}, \bibinfo {author} {\bibfnamefont {A.~K.}\ \bibnamefont {{Harding}}}, \ and\ \bibinfo {author} {\bibfnamefont {J.~K.}\ \bibnamefont {{Daugherty}}},\ }in\ \href@noop {} {\emph {\bibinfo {booktitle} {Bulletin of the American Astronomical Society}}},\ Vol.~\bibinfo {volume} {21}\ (\bibinfo {year} {1989})\ p.\ \bibinfo {pages} {768}\BibitemShut {NoStop}%
\bibitem [{\citenamefont {{Thompson}}\ and\ \citenamefont {{Duncan}}(1995)}]{1995MNRAS.275..255T}%
  \BibitemOpen
  \bibfield  {author} {\bibinfo {author} {\bibfnamefont {C.}~\bibnamefont {{Thompson}}}\ and\ \bibinfo {author} {\bibfnamefont {R.~C.}\ \bibnamefont {{Duncan}}},\ }\href {\doibase 10.1093/mnras/275.2.255} {\bibfield  {journal} {\bibinfo  {journal} {\mnras}\ }\textbf {\bibinfo {volume} {275}},\ \bibinfo {pages} {255} (\bibinfo {year} {1995})}\BibitemShut {NoStop}%
\bibitem [{\citenamefont {{van Putten}}\ \emph {et~al.}(2016)\citenamefont {{van Putten}}, \citenamefont {{Watts}}, \citenamefont {{Baring}},\ and\ \citenamefont {{Wijers}}}]{2016MNRAS.461..877V}%
  \BibitemOpen
  \bibfield  {author} {\bibinfo {author} {\bibfnamefont {T.}~\bibnamefont {{van Putten}}}, \bibinfo {author} {\bibfnamefont {A.~L.}\ \bibnamefont {{Watts}}}, \bibinfo {author} {\bibfnamefont {M.~G.}\ \bibnamefont {{Baring}}}, \ and\ \bibinfo {author} {\bibfnamefont {R.~A.~M.~J.}\ \bibnamefont {{Wijers}}},\ }\href {\doibase 10.1093/mnras/stw1279} {\bibfield  {journal} {\bibinfo  {journal} {\mnras}\ }\textbf {\bibinfo {volume} {461}},\ \bibinfo {pages} {877} (\bibinfo {year} {2016})},\ \Eprint {http://arxiv.org/abs/1605.08022} {arXiv:1605.08022 [astro-ph.HE]} \BibitemShut {NoStop}%
\bibitem [{\citenamefont {{Taverna}}\ and\ \citenamefont {{Turolla}}(2017)}]{2017MNRAS.469.3610T}%
  \BibitemOpen
  \bibfield  {author} {\bibinfo {author} {\bibfnamefont {R.}~\bibnamefont {{Taverna}}}\ and\ \bibinfo {author} {\bibfnamefont {R.}~\bibnamefont {{Turolla}}},\ }\href {\doibase 10.1093/mnras/stx1086} {\bibfield  {journal} {\bibinfo  {journal} {\mnras}\ }\textbf {\bibinfo {volume} {469}},\ \bibinfo {pages} {3610} (\bibinfo {year} {2017})},\ \Eprint {http://arxiv.org/abs/1705.01130} {arXiv:1705.01130 [astro-ph.HE]} \BibitemShut {NoStop}%
\bibitem [{\citenamefont {{Israel}}\ \emph {et~al.}(2008)\citenamefont {{Israel}}, \citenamefont {{Romano}}, \citenamefont {{Mangano}}, \citenamefont {{Dall'Osso}}, \citenamefont {{Chincarini}}, \citenamefont {{Stella}}, \citenamefont {{Campana}}, \citenamefont {{Belloni}}, \citenamefont {{Tagliaferri}}, \citenamefont {{Blustin}}, \citenamefont {{Sakamoto}}, \citenamefont {{Hurley}}, \citenamefont {{Zane}}, \citenamefont {{Moretti}}, \citenamefont {{Palmer}}, \citenamefont {{Guidorzi}}, \citenamefont {{Burrows}}, \citenamefont {{Gehrels}},\ and\ \citenamefont {{Krimm}}}]{2008ApJ...685.1114I}%
  \BibitemOpen
  \bibfield  {author} {\bibinfo {author} {\bibfnamefont {G.~L.}\ \bibnamefont {{Israel}}}, \bibinfo {author} {\bibfnamefont {P.}~\bibnamefont {{Romano}}}, \bibinfo {author} {\bibfnamefont {V.}~\bibnamefont {{Mangano}}}, \bibinfo {author} {\bibfnamefont {S.}~\bibnamefont {{Dall'Osso}}}, \bibinfo {author} {\bibfnamefont {G.}~\bibnamefont {{Chincarini}}}, \bibinfo {author} {\bibfnamefont {L.}~\bibnamefont {{Stella}}}, \bibinfo {author} {\bibfnamefont {S.}~\bibnamefont {{Campana}}}, \bibinfo {author} {\bibfnamefont {T.}~\bibnamefont {{Belloni}}}, \bibinfo {author} {\bibfnamefont {G.}~\bibnamefont {{Tagliaferri}}}, \bibinfo {author} {\bibfnamefont {A.~J.}\ \bibnamefont {{Blustin}}}, \bibinfo {author} {\bibfnamefont {T.}~\bibnamefont {{Sakamoto}}}, \bibinfo {author} {\bibfnamefont {K.}~\bibnamefont {{Hurley}}}, \bibinfo {author} {\bibfnamefont {S.}~\bibnamefont {{Zane}}}, \bibinfo {author} {\bibfnamefont {A.}~\bibnamefont {{Moretti}}}, \bibinfo {author} {\bibfnamefont {D.}~\bibnamefont {{Palmer}}}, \bibinfo
  {author} {\bibfnamefont {C.}~\bibnamefont {{Guidorzi}}}, \bibinfo {author} {\bibfnamefont {D.~N.}\ \bibnamefont {{Burrows}}}, \bibinfo {author} {\bibfnamefont {N.}~\bibnamefont {{Gehrels}}}, \ and\ \bibinfo {author} {\bibfnamefont {H.~A.}\ \bibnamefont {{Krimm}}},\ }\href {\doibase 10.1086/590486} {\bibfield  {journal} {\bibinfo  {journal} {\apj}\ }\textbf {\bibinfo {volume} {685}},\ \bibinfo {pages} {1114} (\bibinfo {year} {2008})},\ \Eprint {http://arxiv.org/abs/0805.3919} {arXiv:0805.3919 [astro-ph]} \BibitemShut {NoStop}%
\bibitem [{\citenamefont {{Rea}}\ and\ \citenamefont {{Esposito}}(2011)}]{2011ASSP...21..247R}%
  \BibitemOpen
  \bibfield  {author} {\bibinfo {author} {\bibfnamefont {N.}~\bibnamefont {{Rea}}}\ and\ \bibinfo {author} {\bibfnamefont {P.}~\bibnamefont {{Esposito}}},\ }in\ \href {\doibase 10.1007/978-3-642-17251-9_21} {\emph {\bibinfo {booktitle} {High-Energy Emission from Pulsars and their Systems}}},\ \bibinfo {series} {Astrophysics and Space Science Proceedings}, Vol.~\bibinfo {volume} {21},\ \bibinfo {editor} {edited by\ \bibinfo {editor} {\bibfnamefont {D.~F.}\ \bibnamefont {{Torres}}}\ and\ \bibinfo {editor} {\bibfnamefont {N.}~\bibnamefont {{Rea}}}}\ (\bibinfo {year} {2011})\ p.\ \bibinfo {pages} {247},\ \Eprint {http://arxiv.org/abs/1101.4472} {arXiv:1101.4472 [astro-ph.GA]} \BibitemShut {NoStop}%
\bibitem [{\citenamefont {{Mereghetti}}\ \emph {et~al.}(2015)\citenamefont {{Mereghetti}}, \citenamefont {{Pons}},\ and\ \citenamefont {{Melatos}}}]{2015SSRv..191..315M}%
  \BibitemOpen
  \bibfield  {author} {\bibinfo {author} {\bibfnamefont {S.}~\bibnamefont {{Mereghetti}}}, \bibinfo {author} {\bibfnamefont {J.~A.}\ \bibnamefont {{Pons}}}, \ and\ \bibinfo {author} {\bibfnamefont {A.}~\bibnamefont {{Melatos}}},\ }\href {\doibase 10.1007/s11214-015-0146-y} {\bibfield  {journal} {\bibinfo  {journal} {\ssr}\ }\textbf {\bibinfo {volume} {191}},\ \bibinfo {pages} {315} (\bibinfo {year} {2015})},\ \Eprint {http://arxiv.org/abs/1503.06313} {arXiv:1503.06313 [astro-ph.HE]} \BibitemShut {NoStop}%
\bibitem [{\citenamefont {{Enoto}}\ \emph {et~al.}(2017)\citenamefont {{Enoto}}, \citenamefont {{Shibata}}, \citenamefont {{Kitaguchi}}, \citenamefont {{Suwa}}, \citenamefont {{Uchide}}, \citenamefont {{Nishioka}}, \citenamefont {{Kisaka}}, \citenamefont {{Nakano}}, \citenamefont {{Murakami}},\ and\ \citenamefont {{Makishima}}}]{2017ApJS..231....8E}%
  \BibitemOpen
  \bibfield  {author} {\bibinfo {author} {\bibfnamefont {T.}~\bibnamefont {{Enoto}}}, \bibinfo {author} {\bibfnamefont {S.}~\bibnamefont {{Shibata}}}, \bibinfo {author} {\bibfnamefont {T.}~\bibnamefont {{Kitaguchi}}}, \bibinfo {author} {\bibfnamefont {Y.}~\bibnamefont {{Suwa}}}, \bibinfo {author} {\bibfnamefont {T.}~\bibnamefont {{Uchide}}}, \bibinfo {author} {\bibfnamefont {H.}~\bibnamefont {{Nishioka}}}, \bibinfo {author} {\bibfnamefont {S.}~\bibnamefont {{Kisaka}}}, \bibinfo {author} {\bibfnamefont {T.}~\bibnamefont {{Nakano}}}, \bibinfo {author} {\bibfnamefont {H.}~\bibnamefont {{Murakami}}}, \ and\ \bibinfo {author} {\bibfnamefont {K.}~\bibnamefont {{Makishima}}},\ }\href {\doibase 10.3847/1538-4365/aa6f0a} {\bibfield  {journal} {\bibinfo  {journal} {\apjs}\ }\textbf {\bibinfo {volume} {231}},\ \bibinfo {eid} {8} (\bibinfo {year} {2017})},\ \Eprint {http://arxiv.org/abs/1704.07018} {arXiv:1704.07018 [astro-ph.HE]} \BibitemShut {NoStop}%
\bibitem [{\citenamefont {{Yu-Cong}}\ \emph {et~al.}(2025)\citenamefont {{Yu-Cong}}, \citenamefont {{Lin}}, \citenamefont {{Ming-Yu}}, \citenamefont {{Teruaki}}, \citenamefont {{Chin-Ping}}, \citenamefont {{George}}, \citenamefont {{Ersin}},\ and\ \citenamefont {{Christian}}}]{2025ApJ...980...99Y}%
  \BibitemOpen
  \bibfield  {author} {\bibinfo {author} {\bibfnamefont {F.}~\bibnamefont {{Yu-Cong}}}, \bibinfo {author} {\bibfnamefont {L.}~\bibnamefont {{Lin}}}, \bibinfo {author} {\bibfnamefont {G.}~\bibnamefont {{Ming-Yu}}}, \bibinfo {author} {\bibfnamefont {E.}~\bibnamefont {{Teruaki}}}, \bibinfo {author} {\bibfnamefont {H.}~\bibnamefont {{Chin-Ping}}}, \bibinfo {author} {\bibfnamefont {Y.}~\bibnamefont {{George}}}, \bibinfo {author} {\bibfnamefont {G.}~\bibnamefont {{Ersin}}}, \ and\ \bibinfo {author} {\bibfnamefont {M.}~\bibnamefont {{Christian}}},\ }\href {\doibase 10.3847/1538-4357/ada936} {\bibfield  {journal} {\bibinfo  {journal} {\apj}\ }\textbf {\bibinfo {volume} {980}},\ \bibinfo {eid} {99} (\bibinfo {year} {2025})},\ \Eprint {http://arxiv.org/abs/2501.07049} {arXiv:2501.07049 [astro-ph.HE]} \BibitemShut {NoStop}%
\bibitem [{\citenamefont {{Gotthelf}}\ \emph {et~al.}(2019)\citenamefont {{Gotthelf}}, \citenamefont {{Halpern}}, \citenamefont {{Alford}}, \citenamefont {{Mihara}}, \citenamefont {{Negoro}}, \citenamefont {{Kawai}}, \citenamefont {{Dai}}, \citenamefont {{Lower}}, \citenamefont {{Johnston}}, \citenamefont {{Bailes}}, \citenamefont {{Os{\l}owski}}, \citenamefont {{Camilo}}, \citenamefont {{Miyasaka}},\ and\ \citenamefont {{Madsen}}}]{2019ApJ...874L..25G}%
  \BibitemOpen
  \bibfield  {author} {\bibinfo {author} {\bibfnamefont {E.~V.}\ \bibnamefont {{Gotthelf}}}, \bibinfo {author} {\bibfnamefont {J.~P.}\ \bibnamefont {{Halpern}}}, \bibinfo {author} {\bibfnamefont {J.~A.~J.}\ \bibnamefont {{Alford}}}, \bibinfo {author} {\bibfnamefont {T.}~\bibnamefont {{Mihara}}}, \bibinfo {author} {\bibfnamefont {H.}~\bibnamefont {{Negoro}}}, \bibinfo {author} {\bibfnamefont {N.}~\bibnamefont {{Kawai}}}, \bibinfo {author} {\bibfnamefont {S.}~\bibnamefont {{Dai}}}, \bibinfo {author} {\bibfnamefont {M.~E.}\ \bibnamefont {{Lower}}}, \bibinfo {author} {\bibfnamefont {S.}~\bibnamefont {{Johnston}}}, \bibinfo {author} {\bibfnamefont {M.}~\bibnamefont {{Bailes}}}, \bibinfo {author} {\bibfnamefont {S.}~\bibnamefont {{Os{\l}owski}}}, \bibinfo {author} {\bibfnamefont {F.}~\bibnamefont {{Camilo}}}, \bibinfo {author} {\bibfnamefont {H.}~\bibnamefont {{Miyasaka}}}, \ and\ \bibinfo {author} {\bibfnamefont {K.~K.}\ \bibnamefont {{Madsen}}},\ }\href {\doibase 10.3847/2041-8213/ab101a} {\bibfield  {journal}
  {\bibinfo  {journal} {\apjl}\ }\textbf {\bibinfo {volume} {874}},\ \bibinfo {eid} {L25} (\bibinfo {year} {2019})},\ \Eprint {http://arxiv.org/abs/1902.08358} {arXiv:1902.08358 [astro-ph.HE]} \BibitemShut {NoStop}%
\bibitem [{\citenamefont {{Borghese}}\ \emph {et~al.}(2021)\citenamefont {{Borghese}}, \citenamefont {{Rea}}, \citenamefont {{Turolla}}, \citenamefont {{Rigoselli}}, \citenamefont {{Alford}}, \citenamefont {{Gotthelf}}, \citenamefont {{Burgay}}, \citenamefont {{Possenti}}, \citenamefont {{Zane}}, \citenamefont {{Coti Zelati}}, \citenamefont {{Perna}}, \citenamefont {{Esposito}}, \citenamefont {{Mereghetti}}, \citenamefont {{Vigan{\`o}}}, \citenamefont {{Tiengo}}, \citenamefont {{G{\"o}tz}}, \citenamefont {{Ibrahim}}, \citenamefont {{Israel}}, \citenamefont {{Pons}},\ and\ \citenamefont {{Sathyaprakash}}}]{2021MNRAS.504.5244B}%
  \BibitemOpen
  \bibfield  {author} {\bibinfo {author} {\bibfnamefont {A.}~\bibnamefont {{Borghese}}}, \bibinfo {author} {\bibfnamefont {N.}~\bibnamefont {{Rea}}}, \bibinfo {author} {\bibfnamefont {R.}~\bibnamefont {{Turolla}}}, \bibinfo {author} {\bibfnamefont {M.}~\bibnamefont {{Rigoselli}}}, \bibinfo {author} {\bibfnamefont {J.~A.~J.}\ \bibnamefont {{Alford}}}, \bibinfo {author} {\bibfnamefont {E.~V.}\ \bibnamefont {{Gotthelf}}}, \bibinfo {author} {\bibfnamefont {M.}~\bibnamefont {{Burgay}}}, \bibinfo {author} {\bibfnamefont {A.}~\bibnamefont {{Possenti}}}, \bibinfo {author} {\bibfnamefont {S.}~\bibnamefont {{Zane}}}, \bibinfo {author} {\bibfnamefont {F.}~\bibnamefont {{Coti Zelati}}}, \bibinfo {author} {\bibfnamefont {R.}~\bibnamefont {{Perna}}}, \bibinfo {author} {\bibfnamefont {P.}~\bibnamefont {{Esposito}}}, \bibinfo {author} {\bibfnamefont {S.}~\bibnamefont {{Mereghetti}}}, \bibinfo {author} {\bibfnamefont {D.}~\bibnamefont {{Vigan{\`o}}}}, \bibinfo {author} {\bibfnamefont {A.}~\bibnamefont {{Tiengo}}}, \bibinfo
  {author} {\bibfnamefont {D.}~\bibnamefont {{G{\"o}tz}}}, \bibinfo {author} {\bibfnamefont {A.}~\bibnamefont {{Ibrahim}}}, \bibinfo {author} {\bibfnamefont {G.~L.}\ \bibnamefont {{Israel}}}, \bibinfo {author} {\bibfnamefont {J.}~\bibnamefont {{Pons}}}, \ and\ \bibinfo {author} {\bibfnamefont {R.}~\bibnamefont {{Sathyaprakash}}},\ }\href {\doibase 10.1093/mnras/stab1236} {\bibfield  {journal} {\bibinfo  {journal} {\mnras}\ }\textbf {\bibinfo {volume} {504}},\ \bibinfo {pages} {5244} (\bibinfo {year} {2021})},\ \Eprint {http://arxiv.org/abs/2104.11083} {arXiv:2104.11083 [astro-ph.HE]} \BibitemShut {NoStop}%
\bibitem [{\citenamefont {{Bochenek}}\ \emph {et~al.}(2020)\citenamefont {{Bochenek}}, \citenamefont {{Ravi}}, \citenamefont {{Belov}}, \citenamefont {{Hallinan}}, \citenamefont {{Kocz}}, \citenamefont {{Kulkarni}},\ and\ \citenamefont {{McKenna}}}]{Bochenek20}%
  \BibitemOpen
  \bibfield  {author} {\bibinfo {author} {\bibfnamefont {C.~D.}\ \bibnamefont {{Bochenek}}}, \bibinfo {author} {\bibfnamefont {V.}~\bibnamefont {{Ravi}}}, \bibinfo {author} {\bibfnamefont {K.~V.}\ \bibnamefont {{Belov}}}, \bibinfo {author} {\bibfnamefont {G.}~\bibnamefont {{Hallinan}}}, \bibinfo {author} {\bibfnamefont {J.}~\bibnamefont {{Kocz}}}, \bibinfo {author} {\bibfnamefont {S.~R.}\ \bibnamefont {{Kulkarni}}}, \ and\ \bibinfo {author} {\bibfnamefont {D.~L.}\ \bibnamefont {{McKenna}}},\ }\href {\doibase 10.1038/s41586-020-2872-x} {\bibfield  {journal} {\bibinfo  {journal} {\nat}\ }\textbf {\bibinfo {volume} {587}},\ \bibinfo {pages} {59} (\bibinfo {year} {2020})},\ \Eprint {http://arxiv.org/abs/2005.10828} {arXiv:2005.10828 [astro-ph.HE]} \BibitemShut {NoStop}%
\bibitem [{\citenamefont {{CHIME/FRB Collaboration}}\ \emph {et~al.}(2020)\citenamefont {{CHIME/FRB Collaboration}}, \citenamefont {{Andersen}}, \citenamefont {{Bandura}}, \citenamefont {{Bhardwaj}}, \citenamefont {{Bij}}, \citenamefont {{Boyce}}, \citenamefont {{Boyle}}, \citenamefont {{Brar}}, \citenamefont {{Cassanelli}}, \citenamefont {{Chawla}}, \citenamefont {{Chen}}, \citenamefont {{Cliche}}, \citenamefont {{Cook}}, \citenamefont {{Cubranic}}, \citenamefont {{Curtin}}, \citenamefont {{Denman}}, \citenamefont {{Dobbs}}, \citenamefont {{Dong}}, \citenamefont {{Fandino}}, \citenamefont {{Fonseca}}, \citenamefont {{Gaensler}}, \citenamefont {{Giri}}, \citenamefont {{Good}}, \citenamefont {{Halpern}}, \citenamefont {{Hill}}, \citenamefont {{Hinshaw}}, \citenamefont {{H{\"o}fer}}, \citenamefont {{Josephy}}, \citenamefont {{Kania}}, \citenamefont {{Kaspi}}, \citenamefont {{Landecker}}, \citenamefont {{Leung}}, \citenamefont {{Li}}, \citenamefont {{Lin}}, \citenamefont {{Masui}}, \citenamefont {{McKinven}},
  \citenamefont {{Mena-Parra}}, \citenamefont {{Merryfield}}, \citenamefont {{Meyers}}, \citenamefont {{Michilli}}, \citenamefont {{Milutinovic}}, \citenamefont {{Mirhosseini}}, \citenamefont {{M{\"u}nchmeyer}}, \citenamefont {{Naidu}}, \citenamefont {{Newburgh}}, \citenamefont {{Ng}}, \citenamefont {{Patel}}, \citenamefont {{Pen}}, \citenamefont {{Pinsonneault-Marotte}}, \citenamefont {{Pleunis}}, \citenamefont {{Quine}}, \citenamefont {{Rafiei-Ravandi}}, \citenamefont {{Rahman}}, \citenamefont {{Ransom}}, \citenamefont {{Renard}}, \citenamefont {{Sanghavi}}, \citenamefont {{Scholz}}, \citenamefont {{Shaw}}, \citenamefont {{Shin}}, \citenamefont {{Siegel}}, \citenamefont {{Singh}}, \citenamefont {{Smegal}}, \citenamefont {{Smith}}, \citenamefont {{Stairs}}, \citenamefont {{Tan}}, \citenamefont {{Tendulkar}}, \citenamefont {{Tretyakov}}, \citenamefont {{Vanderlinde}}, \citenamefont {{Wang}}, \citenamefont {{Wulf}},\ and\ \citenamefont {{Zwaniga}}}]{Andersen2020Natur}%
  \BibitemOpen
  \bibfield  {author} {\bibinfo {author} {\bibnamefont {{CHIME/FRB Collaboration}}}, \bibinfo {author} {\bibfnamefont {B.~C.}\ \bibnamefont {{Andersen}}}, \bibinfo {author} {\bibfnamefont {K.~M.}\ \bibnamefont {{Bandura}}}, \bibinfo {author} {\bibfnamefont {M.}~\bibnamefont {{Bhardwaj}}}, \bibinfo {author} {\bibfnamefont {A.}~\bibnamefont {{Bij}}}, \bibinfo {author} {\bibfnamefont {M.~M.}\ \bibnamefont {{Boyce}}}, \bibinfo {author} {\bibfnamefont {P.~J.}\ \bibnamefont {{Boyle}}}, \bibinfo {author} {\bibfnamefont {C.}~\bibnamefont {{Brar}}}, \bibinfo {author} {\bibfnamefont {T.}~\bibnamefont {{Cassanelli}}}, \bibinfo {author} {\bibfnamefont {P.}~\bibnamefont {{Chawla}}}, \bibinfo {author} {\bibfnamefont {T.}~\bibnamefont {{Chen}}}, \bibinfo {author} {\bibfnamefont {J.~F.}\ \bibnamefont {{Cliche}}}, \bibinfo {author} {\bibfnamefont {A.}~\bibnamefont {{Cook}}}, \bibinfo {author} {\bibfnamefont {D.}~\bibnamefont {{Cubranic}}}, \bibinfo {author} {\bibfnamefont {A.~P.}\ \bibnamefont {{Curtin}}}, \bibinfo {author}
  {\bibfnamefont {N.~T.}\ \bibnamefont {{Denman}}}, \bibinfo {author} {\bibfnamefont {M.}~\bibnamefont {{Dobbs}}}, \bibinfo {author} {\bibfnamefont {F.~Q.}\ \bibnamefont {{Dong}}}, \bibinfo {author} {\bibfnamefont {M.}~\bibnamefont {{Fandino}}}, \bibinfo {author} {\bibfnamefont {E.}~\bibnamefont {{Fonseca}}}, \bibinfo {author} {\bibfnamefont {B.~M.}\ \bibnamefont {{Gaensler}}}, \bibinfo {author} {\bibfnamefont {U.}~\bibnamefont {{Giri}}}, \bibinfo {author} {\bibfnamefont {D.~C.}\ \bibnamefont {{Good}}}, \bibinfo {author} {\bibfnamefont {M.}~\bibnamefont {{Halpern}}}, \bibinfo {author} {\bibfnamefont {A.~S.}\ \bibnamefont {{Hill}}}, \bibinfo {author} {\bibfnamefont {G.~F.}\ \bibnamefont {{Hinshaw}}}, \bibinfo {author} {\bibfnamefont {C.}~\bibnamefont {{H{\"o}fer}}}, \bibinfo {author} {\bibfnamefont {A.}~\bibnamefont {{Josephy}}}, \bibinfo {author} {\bibfnamefont {J.~W.}\ \bibnamefont {{Kania}}}, \bibinfo {author} {\bibfnamefont {V.~M.}\ \bibnamefont {{Kaspi}}}, \bibinfo {author} {\bibfnamefont {T.~L.}\
  \bibnamefont {{Landecker}}}, \bibinfo {author} {\bibfnamefont {C.}~\bibnamefont {{Leung}}}, \bibinfo {author} {\bibfnamefont {D.~Z.}\ \bibnamefont {{Li}}}, \bibinfo {author} {\bibfnamefont {H.~H.}\ \bibnamefont {{Lin}}}, \bibinfo {author} {\bibfnamefont {K.~W.}\ \bibnamefont {{Masui}}}, \bibinfo {author} {\bibfnamefont {R.}~\bibnamefont {{McKinven}}}, \bibinfo {author} {\bibfnamefont {J.}~\bibnamefont {{Mena-Parra}}}, \bibinfo {author} {\bibfnamefont {M.}~\bibnamefont {{Merryfield}}}, \bibinfo {author} {\bibfnamefont {B.~W.}\ \bibnamefont {{Meyers}}}, \bibinfo {author} {\bibfnamefont {D.}~\bibnamefont {{Michilli}}}, \bibinfo {author} {\bibfnamefont {N.}~\bibnamefont {{Milutinovic}}}, \bibinfo {author} {\bibfnamefont {A.}~\bibnamefont {{Mirhosseini}}}, \bibinfo {author} {\bibfnamefont {M.}~\bibnamefont {{M{\"u}nchmeyer}}}, \bibinfo {author} {\bibfnamefont {A.}~\bibnamefont {{Naidu}}}, \bibinfo {author} {\bibfnamefont {L.~B.}\ \bibnamefont {{Newburgh}}}, \bibinfo {author} {\bibfnamefont {C.}~\bibnamefont
  {{Ng}}}, \bibinfo {author} {\bibfnamefont {C.}~\bibnamefont {{Patel}}}, \bibinfo {author} {\bibfnamefont {U.~L.}\ \bibnamefont {{Pen}}}, \bibinfo {author} {\bibfnamefont {T.}~\bibnamefont {{Pinsonneault-Marotte}}}, \bibinfo {author} {\bibfnamefont {Z.}~\bibnamefont {{Pleunis}}}, \bibinfo {author} {\bibfnamefont {B.~M.}\ \bibnamefont {{Quine}}}, \bibinfo {author} {\bibfnamefont {M.}~\bibnamefont {{Rafiei-Ravandi}}}, \bibinfo {author} {\bibfnamefont {M.}~\bibnamefont {{Rahman}}}, \bibinfo {author} {\bibfnamefont {S.~M.}\ \bibnamefont {{Ransom}}}, \bibinfo {author} {\bibfnamefont {A.}~\bibnamefont {{Renard}}}, \bibinfo {author} {\bibfnamefont {P.}~\bibnamefont {{Sanghavi}}}, \bibinfo {author} {\bibfnamefont {P.}~\bibnamefont {{Scholz}}}, \bibinfo {author} {\bibfnamefont {J.~R.}\ \bibnamefont {{Shaw}}}, \bibinfo {author} {\bibfnamefont {K.}~\bibnamefont {{Shin}}}, \bibinfo {author} {\bibfnamefont {S.~R.}\ \bibnamefont {{Siegel}}}, \bibinfo {author} {\bibfnamefont {S.}~\bibnamefont {{Singh}}}, \bibinfo {author}
  {\bibfnamefont {R.~J.}\ \bibnamefont {{Smegal}}}, \bibinfo {author} {\bibfnamefont {K.~M.}\ \bibnamefont {{Smith}}}, \bibinfo {author} {\bibfnamefont {I.~H.}\ \bibnamefont {{Stairs}}}, \bibinfo {author} {\bibfnamefont {C.~M.}\ \bibnamefont {{Tan}}}, \bibinfo {author} {\bibfnamefont {S.~P.}\ \bibnamefont {{Tendulkar}}}, \bibinfo {author} {\bibfnamefont {I.}~\bibnamefont {{Tretyakov}}}, \bibinfo {author} {\bibfnamefont {K.}~\bibnamefont {{Vanderlinde}}}, \bibinfo {author} {\bibfnamefont {H.}~\bibnamefont {{Wang}}}, \bibinfo {author} {\bibfnamefont {D.}~\bibnamefont {{Wulf}}}, \ and\ \bibinfo {author} {\bibfnamefont {A.~V.}\ \bibnamefont {{Zwaniga}}},\ }\href {\doibase 10.1038/s41586-020-2863-y} {\bibfield  {journal} {\bibinfo  {journal} {\nat}\ }\textbf {\bibinfo {volume} {587}},\ \bibinfo {pages} {54} (\bibinfo {year} {2020})},\ \Eprint {http://arxiv.org/abs/2005.10324} {arXiv:2005.10324 [astro-ph.HE]} \BibitemShut {NoStop}%
\bibitem [{\citenamefont {{Li}}\ \emph {et~al.}(2021)\citenamefont {{Li}}, \citenamefont {{Lin}}, \citenamefont {{Xiong}}, \citenamefont {{Ge}}, \citenamefont {{Li}}, \citenamefont {{Li}}, \citenamefont {{Lu}}, \citenamefont {{Zhang}}, \citenamefont {{Tuo}}, \citenamefont {{Nang}}, \citenamefont {{Zhang}}, \citenamefont {{Xiao}}, \citenamefont {{Chen}}, \citenamefont {{Song}}, \citenamefont {{Xu}}, \citenamefont {{Liu}}, \citenamefont {{Jia}}, \citenamefont {{Cao}}, \citenamefont {{Qu}}, \citenamefont {{Zhang}}, \citenamefont {{Gu}}, \citenamefont {{Liao}}, \citenamefont {{Zhao}}, \citenamefont {{Tan}}, \citenamefont {{Nie}}, \citenamefont {{Zhao}}, \citenamefont {{Zheng}}, \citenamefont {{Zheng}}, \citenamefont {{Luo}}, \citenamefont {{Cai}}, \citenamefont {{Li}}, \citenamefont {{Xue}}, \citenamefont {{Bu}}, \citenamefont {{Chang}}, \citenamefont {{Chen}}, \citenamefont {{Chen}}, \citenamefont {{Chen}}, \citenamefont {{Chen}}, \citenamefont {{Chen}}, \citenamefont {{Cui}}, \citenamefont {{Cui}},
  \citenamefont {{Deng}}, \citenamefont {{Dong}}, \citenamefont {{Du}}, \citenamefont {{Fu}}, \citenamefont {{Gao}}, \citenamefont {{Gao}}, \citenamefont {{Gao}}, \citenamefont {{Gu}}, \citenamefont {{Guan}}, \citenamefont {{Guo}}, \citenamefont {{Han}}, \citenamefont {{Huang}}, \citenamefont {{Huo}}, \citenamefont {{Jiang}}, \citenamefont {{Jiang}}, \citenamefont {{Jin}}, \citenamefont {{Jin}}, \citenamefont {{Kong}}, \citenamefont {{Li}}, \citenamefont {{Li}}, \citenamefont {{Li}}, \citenamefont {{Li}}, \citenamefont {{Li}}, \citenamefont {{Li}}, \citenamefont {{Li}}, \citenamefont {{Liang}}, \citenamefont {{Liu}}, \citenamefont {{Liu}}, \citenamefont {{Liu}}, \citenamefont {{Liu}}, \citenamefont {{Liu}}, \citenamefont {{Lu}}, \citenamefont {{Lu}}, \citenamefont {{Luo}}, \citenamefont {{Ma}}, \citenamefont {{Meng}}, \citenamefont {{Ou}}, \citenamefont {{Sai}}, \citenamefont {{Shang}}, \citenamefont {{Song}}, \citenamefont {{Sun}}, \citenamefont {{Tao}}, \citenamefont {{Wang}}, \citenamefont {{Wang}},
  \citenamefont {{Wang}}, \citenamefont {{Wang}}, \citenamefont {{Wang}}, \citenamefont {{Wen}}, \citenamefont {{Wu}}, \citenamefont {{Wu}}, \citenamefont {{Wu}}, \citenamefont {{Xiao}}, \citenamefont {{Xu}}, \citenamefont {{Yang}}, \citenamefont {{Yang}}, \citenamefont {{Yang}}, \citenamefont {{Yang}}, \citenamefont {{Yi}}, \citenamefont {{Yin}}, \citenamefont {{You}}, \citenamefont {{Zhang}}, \citenamefont {{Zhang}}, \citenamefont {{Zhang}}, \citenamefont {{Zhang}}, \citenamefont {{Zhang}}, \citenamefont {{Zhang}}, \citenamefont {{Zhang}}, \citenamefont {{Zhang}}, \citenamefont {{Zhang}}, \citenamefont {{Zhang}}, \citenamefont {{Zhang}}, \citenamefont {{Zhang}}, \citenamefont {{Zhang}}, \citenamefont {{Zhang}}, \citenamefont {{Zhang}}, \citenamefont {{Zhang}}, \citenamefont {{Zhou}}, \citenamefont {{Zhou}}, \citenamefont {{Zhu}}, \citenamefont {{Zhu}},\ and\ \citenamefont {{Zhuang}}}]{LCK21}%
  \BibitemOpen
  \bibfield  {author} {\bibinfo {author} {\bibfnamefont {C.~K.}\ \bibnamefont {{Li}}}, \bibinfo {author} {\bibfnamefont {L.}~\bibnamefont {{Lin}}}, \bibinfo {author} {\bibfnamefont {S.~L.}\ \bibnamefont {{Xiong}}}, \bibinfo {author} {\bibfnamefont {M.~Y.}\ \bibnamefont {{Ge}}}, \bibinfo {author} {\bibfnamefont {X.~B.}\ \bibnamefont {{Li}}}, \bibinfo {author} {\bibfnamefont {T.~P.}\ \bibnamefont {{Li}}}, \bibinfo {author} {\bibfnamefont {F.~J.}\ \bibnamefont {{Lu}}}, \bibinfo {author} {\bibfnamefont {S.~N.}\ \bibnamefont {{Zhang}}}, \bibinfo {author} {\bibfnamefont {Y.~L.}\ \bibnamefont {{Tuo}}}, \bibinfo {author} {\bibfnamefont {Y.}~\bibnamefont {{Nang}}}, \bibinfo {author} {\bibfnamefont {B.}~\bibnamefont {{Zhang}}}, \bibinfo {author} {\bibfnamefont {S.}~\bibnamefont {{Xiao}}}, \bibinfo {author} {\bibfnamefont {Y.}~\bibnamefont {{Chen}}}, \bibinfo {author} {\bibfnamefont {L.~M.}\ \bibnamefont {{Song}}}, \bibinfo {author} {\bibfnamefont {Y.~P.}\ \bibnamefont {{Xu}}}, \bibinfo {author} {\bibfnamefont {C.~Z.}\
  \bibnamefont {{Liu}}}, \bibinfo {author} {\bibfnamefont {S.~M.}\ \bibnamefont {{Jia}}}, \bibinfo {author} {\bibfnamefont {X.~L.}\ \bibnamefont {{Cao}}}, \bibinfo {author} {\bibfnamefont {J.~L.}\ \bibnamefont {{Qu}}}, \bibinfo {author} {\bibfnamefont {S.}~\bibnamefont {{Zhang}}}, \bibinfo {author} {\bibfnamefont {Y.~D.}\ \bibnamefont {{Gu}}}, \bibinfo {author} {\bibfnamefont {J.~Y.}\ \bibnamefont {{Liao}}}, \bibinfo {author} {\bibfnamefont {X.~F.}\ \bibnamefont {{Zhao}}}, \bibinfo {author} {\bibfnamefont {Y.}~\bibnamefont {{Tan}}}, \bibinfo {author} {\bibfnamefont {J.~Y.}\ \bibnamefont {{Nie}}}, \bibinfo {author} {\bibfnamefont {H.~S.}\ \bibnamefont {{Zhao}}}, \bibinfo {author} {\bibfnamefont {S.~J.}\ \bibnamefont {{Zheng}}}, \bibinfo {author} {\bibfnamefont {Y.~G.}\ \bibnamefont {{Zheng}}}, \bibinfo {author} {\bibfnamefont {Q.}~\bibnamefont {{Luo}}}, \bibinfo {author} {\bibfnamefont {C.}~\bibnamefont {{Cai}}}, \bibinfo {author} {\bibfnamefont {B.}~\bibnamefont {{Li}}}, \bibinfo {author} {\bibfnamefont
  {W.~C.}\ \bibnamefont {{Xue}}}, \bibinfo {author} {\bibfnamefont {Q.~C.}\ \bibnamefont {{Bu}}}, \bibinfo {author} {\bibfnamefont {Z.}~\bibnamefont {{Chang}}}, \bibinfo {author} {\bibfnamefont {G.}~\bibnamefont {{Chen}}}, \bibinfo {author} {\bibfnamefont {L.}~\bibnamefont {{Chen}}}, \bibinfo {author} {\bibfnamefont {T.~X.}\ \bibnamefont {{Chen}}}, \bibinfo {author} {\bibfnamefont {Y.~B.}\ \bibnamefont {{Chen}}}, \bibinfo {author} {\bibfnamefont {Y.~P.}\ \bibnamefont {{Chen}}}, \bibinfo {author} {\bibfnamefont {W.}~\bibnamefont {{Cui}}}, \bibinfo {author} {\bibfnamefont {W.~W.}\ \bibnamefont {{Cui}}}, \bibinfo {author} {\bibfnamefont {J.~K.}\ \bibnamefont {{Deng}}}, \bibinfo {author} {\bibfnamefont {Y.~W.}\ \bibnamefont {{Dong}}}, \bibinfo {author} {\bibfnamefont {Y.~Y.}\ \bibnamefont {{Du}}}, \bibinfo {author} {\bibfnamefont {M.~X.}\ \bibnamefont {{Fu}}}, \bibinfo {author} {\bibfnamefont {G.~H.}\ \bibnamefont {{Gao}}}, \bibinfo {author} {\bibfnamefont {H.}~\bibnamefont {{Gao}}}, \bibinfo {author}
  {\bibfnamefont {M.}~\bibnamefont {{Gao}}}, \bibinfo {author} {\bibfnamefont {Y.~D.}\ \bibnamefont {{Gu}}}, \bibinfo {author} {\bibfnamefont {J.}~\bibnamefont {{Guan}}}, \bibinfo {author} {\bibfnamefont {C.~C.}\ \bibnamefont {{Guo}}}, \bibinfo {author} {\bibfnamefont {D.~W.}\ \bibnamefont {{Han}}}, \bibinfo {author} {\bibfnamefont {Y.}~\bibnamefont {{Huang}}}, \bibinfo {author} {\bibfnamefont {J.}~\bibnamefont {{Huo}}}, \bibinfo {author} {\bibfnamefont {L.~H.}\ \bibnamefont {{Jiang}}}, \bibinfo {author} {\bibfnamefont {W.~C.}\ \bibnamefont {{Jiang}}}, \bibinfo {author} {\bibfnamefont {J.}~\bibnamefont {{Jin}}}, \bibinfo {author} {\bibfnamefont {Y.~J.}\ \bibnamefont {{Jin}}}, \bibinfo {author} {\bibfnamefont {L.~D.}\ \bibnamefont {{Kong}}}, \bibinfo {author} {\bibfnamefont {G.}~\bibnamefont {{Li}}}, \bibinfo {author} {\bibfnamefont {M.~S.}\ \bibnamefont {{Li}}}, \bibinfo {author} {\bibfnamefont {W.}~\bibnamefont {{Li}}}, \bibinfo {author} {\bibfnamefont {X.}~\bibnamefont {{Li}}}, \bibinfo {author}
  {\bibfnamefont {X.~F.}\ \bibnamefont {{Li}}}, \bibinfo {author} {\bibfnamefont {Y.~G.}\ \bibnamefont {{Li}}}, \bibinfo {author} {\bibfnamefont {Z.~W.}\ \bibnamefont {{Li}}}, \bibinfo {author} {\bibfnamefont {X.~H.}\ \bibnamefont {{Liang}}}, \bibinfo {author} {\bibfnamefont {B.~S.}\ \bibnamefont {{Liu}}}, \bibinfo {author} {\bibfnamefont {G.~Q.}\ \bibnamefont {{Liu}}}, \bibinfo {author} {\bibfnamefont {H.~W.}\ \bibnamefont {{Liu}}}, \bibinfo {author} {\bibfnamefont {X.~J.}\ \bibnamefont {{Liu}}}, \bibinfo {author} {\bibfnamefont {Y.~N.}\ \bibnamefont {{Liu}}}, \bibinfo {author} {\bibfnamefont {B.}~\bibnamefont {{Lu}}}, \bibinfo {author} {\bibfnamefont {X.~F.}\ \bibnamefont {{Lu}}}, \bibinfo {author} {\bibfnamefont {T.}~\bibnamefont {{Luo}}}, \bibinfo {author} {\bibfnamefont {X.}~\bibnamefont {{Ma}}}, \bibinfo {author} {\bibfnamefont {B.}~\bibnamefont {{Meng}}}, \bibinfo {author} {\bibfnamefont {G.}~\bibnamefont {{Ou}}}, \bibinfo {author} {\bibfnamefont {N.}~\bibnamefont {{Sai}}}, \bibinfo {author}
  {\bibfnamefont {R.~C.}\ \bibnamefont {{Shang}}}, \bibinfo {author} {\bibfnamefont {X.~Y.}\ \bibnamefont {{Song}}}, \bibinfo {author} {\bibfnamefont {L.}~\bibnamefont {{Sun}}}, \bibinfo {author} {\bibfnamefont {L.}~\bibnamefont {{Tao}}}, \bibinfo {author} {\bibfnamefont {C.}~\bibnamefont {{Wang}}}, \bibinfo {author} {\bibfnamefont {G.~F.}\ \bibnamefont {{Wang}}}, \bibinfo {author} {\bibfnamefont {J.}~\bibnamefont {{Wang}}}, \bibinfo {author} {\bibfnamefont {W.~S.}\ \bibnamefont {{Wang}}}, \bibinfo {author} {\bibfnamefont {Y.~S.}\ \bibnamefont {{Wang}}}, \bibinfo {author} {\bibfnamefont {X.~Y.}\ \bibnamefont {{Wen}}}, \bibinfo {author} {\bibfnamefont {B.~B.}\ \bibnamefont {{Wu}}}, \bibinfo {author} {\bibfnamefont {B.~Y.}\ \bibnamefont {{Wu}}}, \bibinfo {author} {\bibfnamefont {M.}~\bibnamefont {{Wu}}}, \bibinfo {author} {\bibfnamefont {G.~C.}\ \bibnamefont {{Xiao}}}, \bibinfo {author} {\bibfnamefont {H.}~\bibnamefont {{Xu}}}, \bibinfo {author} {\bibfnamefont {J.~W.}\ \bibnamefont {{Yang}}}, \bibinfo {author}
  {\bibfnamefont {S.}~\bibnamefont {{Yang}}}, \bibinfo {author} {\bibfnamefont {Y.~J.}\ \bibnamefont {{Yang}}}, \bibinfo {author} {\bibfnamefont {Y.-J.}\ \bibnamefont {{Yang}}}, \bibinfo {author} {\bibfnamefont {Q.~B.}\ \bibnamefont {{Yi}}}, \bibinfo {author} {\bibfnamefont {Q.~Q.}\ \bibnamefont {{Yin}}}, \bibinfo {author} {\bibfnamefont {Y.}~\bibnamefont {{You}}}, \bibinfo {author} {\bibfnamefont {A.~M.}\ \bibnamefont {{Zhang}}}, \bibinfo {author} {\bibfnamefont {C.~M.}\ \bibnamefont {{Zhang}}}, \bibinfo {author} {\bibfnamefont {F.}~\bibnamefont {{Zhang}}}, \bibinfo {author} {\bibfnamefont {H.~M.}\ \bibnamefont {{Zhang}}}, \bibinfo {author} {\bibfnamefont {J.}~\bibnamefont {{Zhang}}}, \bibinfo {author} {\bibfnamefont {T.}~\bibnamefont {{Zhang}}}, \bibinfo {author} {\bibfnamefont {W.}~\bibnamefont {{Zhang}}}, \bibinfo {author} {\bibfnamefont {W.~C.}\ \bibnamefont {{Zhang}}}, \bibinfo {author} {\bibfnamefont {W.~Z.}\ \bibnamefont {{Zhang}}}, \bibinfo {author} {\bibfnamefont {Y.}~\bibnamefont {{Zhang}}},
  \bibinfo {author} {\bibfnamefont {Y.}~\bibnamefont {{Zhang}}}, \bibinfo {author} {\bibfnamefont {Y.~F.}\ \bibnamefont {{Zhang}}}, \bibinfo {author} {\bibfnamefont {Y.~J.}\ \bibnamefont {{Zhang}}}, \bibinfo {author} {\bibfnamefont {Z.}~\bibnamefont {{Zhang}}}, \bibinfo {author} {\bibfnamefont {Z.}~\bibnamefont {{Zhang}}}, \bibinfo {author} {\bibfnamefont {Z.~L.}\ \bibnamefont {{Zhang}}}, \bibinfo {author} {\bibfnamefont {D.~K.}\ \bibnamefont {{Zhou}}}, \bibinfo {author} {\bibfnamefont {J.~F.}\ \bibnamefont {{Zhou}}}, \bibinfo {author} {\bibfnamefont {Y.}~\bibnamefont {{Zhu}}}, \bibinfo {author} {\bibfnamefont {Y.~X.}\ \bibnamefont {{Zhu}}}, \ and\ \bibinfo {author} {\bibfnamefont {R.~L.}\ \bibnamefont {{Zhuang}}},\ }\href {\doibase 10.1038/s41550-021-01302-6} {\bibfield  {journal} {\bibinfo  {journal} {Nature Astronomy}\ }\textbf {\bibinfo {volume} {5}},\ \bibinfo {pages} {378} (\bibinfo {year} {2021})},\ \Eprint {http://arxiv.org/abs/2005.11071} {arXiv:2005.11071 [astro-ph.HE]} \BibitemShut {NoStop}%
\bibitem [{\citenamefont {{Mereghetti}}\ \emph {et~al.}(2020)\citenamefont {{Mereghetti}}, \citenamefont {{Savchenko}}, \citenamefont {{Ferrigno}}, \citenamefont {{G{\"o}tz}}, \citenamefont {{Rigoselli}}, \citenamefont {{Tiengo}}, \citenamefont {{Bazzano}}, \citenamefont {{Bozzo}}, \citenamefont {{Coleiro}}, \citenamefont {{Courvoisier}}, \citenamefont {{Doyle}}, \citenamefont {{Goldwurm}}, \citenamefont {{Hanlon}}, \citenamefont {{Jourdain}}, \citenamefont {{von Kienlin}}, \citenamefont {{Lutovinov}}, \citenamefont {{Martin-Carrillo}}, \citenamefont {{Molkov}}, \citenamefont {{Natalucci}}, \citenamefont {{Onori}}, \citenamefont {{Panessa}}, \citenamefont {{Rodi}}, \citenamefont {{Rodriguez}}, \citenamefont {{S{\'a}nchez-Fern{\'a}ndez}}, \citenamefont {{Sunyaev}},\ and\ \citenamefont {{Ubertini}}}]{Mereghetti20}%
  \BibitemOpen
  \bibfield  {author} {\bibinfo {author} {\bibfnamefont {S.}~\bibnamefont {{Mereghetti}}}, \bibinfo {author} {\bibfnamefont {V.}~\bibnamefont {{Savchenko}}}, \bibinfo {author} {\bibfnamefont {C.}~\bibnamefont {{Ferrigno}}}, \bibinfo {author} {\bibfnamefont {D.}~\bibnamefont {{G{\"o}tz}}}, \bibinfo {author} {\bibfnamefont {M.}~\bibnamefont {{Rigoselli}}}, \bibinfo {author} {\bibfnamefont {A.}~\bibnamefont {{Tiengo}}}, \bibinfo {author} {\bibfnamefont {A.}~\bibnamefont {{Bazzano}}}, \bibinfo {author} {\bibfnamefont {E.}~\bibnamefont {{Bozzo}}}, \bibinfo {author} {\bibfnamefont {A.}~\bibnamefont {{Coleiro}}}, \bibinfo {author} {\bibfnamefont {T.~J.~L.}\ \bibnamefont {{Courvoisier}}}, \bibinfo {author} {\bibfnamefont {M.}~\bibnamefont {{Doyle}}}, \bibinfo {author} {\bibfnamefont {A.}~\bibnamefont {{Goldwurm}}}, \bibinfo {author} {\bibfnamefont {L.}~\bibnamefont {{Hanlon}}}, \bibinfo {author} {\bibfnamefont {E.}~\bibnamefont {{Jourdain}}}, \bibinfo {author} {\bibfnamefont {A.}~\bibnamefont {{von Kienlin}}},
  \bibinfo {author} {\bibfnamefont {A.}~\bibnamefont {{Lutovinov}}}, \bibinfo {author} {\bibfnamefont {A.}~\bibnamefont {{Martin-Carrillo}}}, \bibinfo {author} {\bibfnamefont {S.}~\bibnamefont {{Molkov}}}, \bibinfo {author} {\bibfnamefont {L.}~\bibnamefont {{Natalucci}}}, \bibinfo {author} {\bibfnamefont {F.}~\bibnamefont {{Onori}}}, \bibinfo {author} {\bibfnamefont {F.}~\bibnamefont {{Panessa}}}, \bibinfo {author} {\bibfnamefont {J.}~\bibnamefont {{Rodi}}}, \bibinfo {author} {\bibfnamefont {J.}~\bibnamefont {{Rodriguez}}}, \bibinfo {author} {\bibfnamefont {C.}~\bibnamefont {{S{\'a}nchez-Fern{\'a}ndez}}}, \bibinfo {author} {\bibfnamefont {R.}~\bibnamefont {{Sunyaev}}}, \ and\ \bibinfo {author} {\bibfnamefont {P.}~\bibnamefont {{Ubertini}}},\ }\href {\doibase 10.3847/2041-8213/aba2cf} {\bibfield  {journal} {\bibinfo  {journal} {\apjl}\ }\textbf {\bibinfo {volume} {898}},\ \bibinfo {eid} {L29} (\bibinfo {year} {2020})},\ \Eprint {http://arxiv.org/abs/2005.06335} {arXiv:2005.06335 [astro-ph.HE]} \BibitemShut
  {NoStop}%
\bibitem [{\citenamefont {{Tavani}}\ \emph {et~al.}(2021)\citenamefont {{Tavani}}, \citenamefont {{Casentini}}, \citenamefont {{Ursi}}, \citenamefont {{Verrecchia}}, \citenamefont {{Addis}}, \citenamefont {{Antonelli}}, \citenamefont {{Argan}}, \citenamefont {{Barbiellini}}, \citenamefont {{Baroncelli}}, \citenamefont {{Bernardi}}, \citenamefont {{Bianchi}}, \citenamefont {{Bulgarelli}}, \citenamefont {{Caraveo}}, \citenamefont {{Cardillo}}, \citenamefont {{Cattaneo}}, \citenamefont {{Chen}}, \citenamefont {{Costa}}, \citenamefont {{Del Monte}}, \citenamefont {{Di Cocco}}, \citenamefont {{Di Persio}}, \citenamefont {{Donnarumma}}, \citenamefont {{Evangelista}}, \citenamefont {{Feroci}}, \citenamefont {{Ferrari}}, \citenamefont {{Fioretti}}, \citenamefont {{Fuschino}}, \citenamefont {{Galli}}, \citenamefont {{Gianotti}}, \citenamefont {{Giuliani}}, \citenamefont {{Labanti}}, \citenamefont {{Lazzarotto}}, \citenamefont {{Lipari}}, \citenamefont {{Longo}}, \citenamefont {{Lucarelli}}, \citenamefont {{Magro}},
  \citenamefont {{Marisaldi}}, \citenamefont {{Mereghetti}}, \citenamefont {{Morelli}}, \citenamefont {{Morselli}}, \citenamefont {{Naldi}}, \citenamefont {{Pacciani}}, \citenamefont {{Parmiggiani}}, \citenamefont {{Paoletti}}, \citenamefont {{Pellizzoni}}, \citenamefont {{Perri}}, \citenamefont {{Perotti}}, \citenamefont {{Piano}}, \citenamefont {{Picozza}}, \citenamefont {{Pilia}}, \citenamefont {{Pittori}}, \citenamefont {{Puccetti}}, \citenamefont {{Pupillo}}, \citenamefont {{Rapisarda}}, \citenamefont {{Rappoldi}}, \citenamefont {{Rubini}}, \citenamefont {{Setti}}, \citenamefont {{Soffitta}}, \citenamefont {{Trifoglio}}, \citenamefont {{Trois}}, \citenamefont {{Vercellone}}, \citenamefont {{Vittorini}}, \citenamefont {{Giommi}},\ and\ \citenamefont {{D'Amico}}}]{Tavani21}%
  \BibitemOpen
  \bibfield  {author} {\bibinfo {author} {\bibfnamefont {M.}~\bibnamefont {{Tavani}}}, \bibinfo {author} {\bibfnamefont {C.}~\bibnamefont {{Casentini}}}, \bibinfo {author} {\bibfnamefont {A.}~\bibnamefont {{Ursi}}}, \bibinfo {author} {\bibfnamefont {F.}~\bibnamefont {{Verrecchia}}}, \bibinfo {author} {\bibfnamefont {A.}~\bibnamefont {{Addis}}}, \bibinfo {author} {\bibfnamefont {L.~A.}\ \bibnamefont {{Antonelli}}}, \bibinfo {author} {\bibfnamefont {A.}~\bibnamefont {{Argan}}}, \bibinfo {author} {\bibfnamefont {G.}~\bibnamefont {{Barbiellini}}}, \bibinfo {author} {\bibfnamefont {L.}~\bibnamefont {{Baroncelli}}}, \bibinfo {author} {\bibfnamefont {G.}~\bibnamefont {{Bernardi}}}, \bibinfo {author} {\bibfnamefont {G.}~\bibnamefont {{Bianchi}}}, \bibinfo {author} {\bibfnamefont {A.}~\bibnamefont {{Bulgarelli}}}, \bibinfo {author} {\bibfnamefont {P.}~\bibnamefont {{Caraveo}}}, \bibinfo {author} {\bibfnamefont {M.}~\bibnamefont {{Cardillo}}}, \bibinfo {author} {\bibfnamefont {P.~W.}\ \bibnamefont {{Cattaneo}}},
  \bibinfo {author} {\bibfnamefont {A.~W.}\ \bibnamefont {{Chen}}}, \bibinfo {author} {\bibfnamefont {E.}~\bibnamefont {{Costa}}}, \bibinfo {author} {\bibfnamefont {E.}~\bibnamefont {{Del Monte}}}, \bibinfo {author} {\bibfnamefont {G.}~\bibnamefont {{Di Cocco}}}, \bibinfo {author} {\bibfnamefont {G.}~\bibnamefont {{Di Persio}}}, \bibinfo {author} {\bibfnamefont {I.}~\bibnamefont {{Donnarumma}}}, \bibinfo {author} {\bibfnamefont {Y.}~\bibnamefont {{Evangelista}}}, \bibinfo {author} {\bibfnamefont {M.}~\bibnamefont {{Feroci}}}, \bibinfo {author} {\bibfnamefont {A.}~\bibnamefont {{Ferrari}}}, \bibinfo {author} {\bibfnamefont {V.}~\bibnamefont {{Fioretti}}}, \bibinfo {author} {\bibfnamefont {F.}~\bibnamefont {{Fuschino}}}, \bibinfo {author} {\bibfnamefont {M.}~\bibnamefont {{Galli}}}, \bibinfo {author} {\bibfnamefont {F.}~\bibnamefont {{Gianotti}}}, \bibinfo {author} {\bibfnamefont {A.}~\bibnamefont {{Giuliani}}}, \bibinfo {author} {\bibfnamefont {C.}~\bibnamefont {{Labanti}}}, \bibinfo {author} {\bibfnamefont
  {F.}~\bibnamefont {{Lazzarotto}}}, \bibinfo {author} {\bibfnamefont {P.}~\bibnamefont {{Lipari}}}, \bibinfo {author} {\bibfnamefont {F.}~\bibnamefont {{Longo}}}, \bibinfo {author} {\bibfnamefont {F.}~\bibnamefont {{Lucarelli}}}, \bibinfo {author} {\bibfnamefont {A.}~\bibnamefont {{Magro}}}, \bibinfo {author} {\bibfnamefont {M.}~\bibnamefont {{Marisaldi}}}, \bibinfo {author} {\bibfnamefont {S.}~\bibnamefont {{Mereghetti}}}, \bibinfo {author} {\bibfnamefont {E.}~\bibnamefont {{Morelli}}}, \bibinfo {author} {\bibfnamefont {A.}~\bibnamefont {{Morselli}}}, \bibinfo {author} {\bibfnamefont {G.}~\bibnamefont {{Naldi}}}, \bibinfo {author} {\bibfnamefont {L.}~\bibnamefont {{Pacciani}}}, \bibinfo {author} {\bibfnamefont {N.}~\bibnamefont {{Parmiggiani}}}, \bibinfo {author} {\bibfnamefont {F.}~\bibnamefont {{Paoletti}}}, \bibinfo {author} {\bibfnamefont {A.}~\bibnamefont {{Pellizzoni}}}, \bibinfo {author} {\bibfnamefont {M.}~\bibnamefont {{Perri}}}, \bibinfo {author} {\bibfnamefont {F.}~\bibnamefont {{Perotti}}},
  \bibinfo {author} {\bibfnamefont {G.}~\bibnamefont {{Piano}}}, \bibinfo {author} {\bibfnamefont {P.}~\bibnamefont {{Picozza}}}, \bibinfo {author} {\bibfnamefont {M.}~\bibnamefont {{Pilia}}}, \bibinfo {author} {\bibfnamefont {C.}~\bibnamefont {{Pittori}}}, \bibinfo {author} {\bibfnamefont {S.}~\bibnamefont {{Puccetti}}}, \bibinfo {author} {\bibfnamefont {G.}~\bibnamefont {{Pupillo}}}, \bibinfo {author} {\bibfnamefont {M.}~\bibnamefont {{Rapisarda}}}, \bibinfo {author} {\bibfnamefont {A.}~\bibnamefont {{Rappoldi}}}, \bibinfo {author} {\bibfnamefont {A.}~\bibnamefont {{Rubini}}}, \bibinfo {author} {\bibfnamefont {G.}~\bibnamefont {{Setti}}}, \bibinfo {author} {\bibfnamefont {P.}~\bibnamefont {{Soffitta}}}, \bibinfo {author} {\bibfnamefont {M.}~\bibnamefont {{Trifoglio}}}, \bibinfo {author} {\bibfnamefont {A.}~\bibnamefont {{Trois}}}, \bibinfo {author} {\bibfnamefont {S.}~\bibnamefont {{Vercellone}}}, \bibinfo {author} {\bibfnamefont {V.}~\bibnamefont {{Vittorini}}}, \bibinfo {author} {\bibfnamefont
  {P.}~\bibnamefont {{Giommi}}}, \ and\ \bibinfo {author} {\bibfnamefont {F.}~\bibnamefont {{D'Amico}}},\ }\href {\doibase 10.1038/s41550-020-01276-x} {\bibfield  {journal} {\bibinfo  {journal} {Nature Astronomy}\ }\textbf {\bibinfo {volume} {5}},\ \bibinfo {pages} {401} (\bibinfo {year} {2021})},\ \Eprint {http://arxiv.org/abs/2005.12164} {arXiv:2005.12164 [astro-ph.HE]} \BibitemShut {NoStop}%
\bibitem [{\citenamefont {{Ridnaia}}\ \emph {et~al.}(2021)\citenamefont {{Ridnaia}}, \citenamefont {{Svinkin}}, \citenamefont {{Frederiks}}, \citenamefont {{Bykov}}, \citenamefont {{Popov}}, \citenamefont {{Aptekar}}, \citenamefont {{Golenetskii}}, \citenamefont {{Lysenko}}, \citenamefont {{Tsvetkova}}, \citenamefont {{Ulanov}},\ and\ \citenamefont {{Cline}}}]{Ridnaia21}%
  \BibitemOpen
  \bibfield  {author} {\bibinfo {author} {\bibfnamefont {A.}~\bibnamefont {{Ridnaia}}}, \bibinfo {author} {\bibfnamefont {D.}~\bibnamefont {{Svinkin}}}, \bibinfo {author} {\bibfnamefont {D.}~\bibnamefont {{Frederiks}}}, \bibinfo {author} {\bibfnamefont {A.}~\bibnamefont {{Bykov}}}, \bibinfo {author} {\bibfnamefont {S.}~\bibnamefont {{Popov}}}, \bibinfo {author} {\bibfnamefont {R.}~\bibnamefont {{Aptekar}}}, \bibinfo {author} {\bibfnamefont {S.}~\bibnamefont {{Golenetskii}}}, \bibinfo {author} {\bibfnamefont {A.}~\bibnamefont {{Lysenko}}}, \bibinfo {author} {\bibfnamefont {A.}~\bibnamefont {{Tsvetkova}}}, \bibinfo {author} {\bibfnamefont {M.}~\bibnamefont {{Ulanov}}}, \ and\ \bibinfo {author} {\bibfnamefont {T.~L.}\ \bibnamefont {{Cline}}},\ }\href {\doibase 10.1038/s41550-020-01265-0} {\bibfield  {journal} {\bibinfo  {journal} {Nature Astronomy}\ }\textbf {\bibinfo {volume} {5}},\ \bibinfo {pages} {372} (\bibinfo {year} {2021})},\ \Eprint {http://arxiv.org/abs/2005.11178} {arXiv:2005.11178 [astro-ph.HE]}
  \BibitemShut {NoStop}%
\bibitem [{\citenamefont {{Xie}}\ \emph {et~al.}(2023)\citenamefont {{Xie}}, \citenamefont {{Geng}}, \citenamefont {{Zhu}}, \citenamefont {{Zhao}}, \citenamefont {{Lei}}, \citenamefont {{Yuan}}, \citenamefont {{Zhao}}, \citenamefont {{Wu}},\ and\ \citenamefont {{Qiao}}}]{Xieyu23}%
  \BibitemOpen
  \bibfield  {author} {\bibinfo {author} {\bibfnamefont {Y.}~\bibnamefont {{Xie}}}, \bibinfo {author} {\bibfnamefont {J.-J.}\ \bibnamefont {{Geng}}}, \bibinfo {author} {\bibfnamefont {X.-W.}\ \bibnamefont {{Zhu}}}, \bibinfo {author} {\bibfnamefont {Z.-H.}\ \bibnamefont {{Zhao}}}, \bibinfo {author} {\bibfnamefont {Z.}~\bibnamefont {{Lei}}}, \bibinfo {author} {\bibfnamefont {W.-Q.}\ \bibnamefont {{Yuan}}}, \bibinfo {author} {\bibfnamefont {G.}~\bibnamefont {{Zhao}}}, \bibinfo {author} {\bibfnamefont {X.-F.}\ \bibnamefont {{Wu}}}, \ and\ \bibinfo {author} {\bibfnamefont {B.}~\bibnamefont {{Qiao}}},\ }\href {\doibase 10.1016/j.scib.2023.06.005} {\bibfield  {journal} {\bibinfo  {journal} {Science Bulletin}\ }\textbf {\bibinfo {volume} {68}},\ \bibinfo {pages} {1857} (\bibinfo {year} {2023})}\BibitemShut {NoStop}%
\bibitem [{\citenamefont {{Tsygankov}}\ \emph {et~al.}(2022)\citenamefont {{Tsygankov}}, \citenamefont {{Doroshenko}}, \citenamefont {{Poutanen}}, \citenamefont {{Heyl}}, \citenamefont {{Mushtukov}}, \citenamefont {{Caiazzo}}, \citenamefont {{Di Marco}}, \citenamefont {{Forsblom}}, \citenamefont {{Gonz{\'a}lez-Caniulef}}, \citenamefont {{Klawin}}, \citenamefont {{La Monaca}}, \citenamefont {{Malacaria}}, \citenamefont {{Marshall}}, \citenamefont {{Muleri}}, \citenamefont {{Ng}}, \citenamefont {{Suleimanov}}, \citenamefont {{Sunyaev}}, \citenamefont {{Turolla}}, \citenamefont {{Agudo}}, \citenamefont {{Antonelli}}, \citenamefont {{Bachetti}}, \citenamefont {{Baldini}}, \citenamefont {{Baumgartner}}, \citenamefont {{Bellazzini}}, \citenamefont {{Bianchi}}, \citenamefont {{Bongiorno}}, \citenamefont {{Bonino}}, \citenamefont {{Brez}}, \citenamefont {{Bucciantini}}, \citenamefont {{Capitanio}}, \citenamefont {{Castellano}}, \citenamefont {{Cavazzuti}}, \citenamefont {{Ciprini}}, \citenamefont {{Costa}},
  \citenamefont {{De Rosa}}, \citenamefont {{Del Monte}}, \citenamefont {{Di Gesu}}, \citenamefont {{Di Lalla}}, \citenamefont {{Donnarumma}}, \citenamefont {{Dov{\v{c}}iak}}, \citenamefont {{Ehlert}}, \citenamefont {{Enoto}}, \citenamefont {{Evangelista}}, \citenamefont {{Fabiani}}, \citenamefont {{Ferrazzoli}}, \citenamefont {{Garcia}}, \citenamefont {{Gunji}}, \citenamefont {{Hayashida}}, \citenamefont {{Iwakiri}}, \citenamefont {{Jorstad}}, \citenamefont {{Karas}}, \citenamefont {{Kitaguchi}}, \citenamefont {{Kolodziejczak}}, \citenamefont {{Krawczynski}}, \citenamefont {{Latronico}}, \citenamefont {{Liodakis}}, \citenamefont {{Maldera}}, \citenamefont {{Manfreda}}, \citenamefont {{Marin}}, \citenamefont {{Marinucci}}, \citenamefont {{Marscher}}, \citenamefont {{Matt}}, \citenamefont {{Mitsuishi}}, \citenamefont {{Mizuno}}, \citenamefont {{Ng}}, \citenamefont {{O'Dell}}, \citenamefont {{Omodei}}, \citenamefont {{Oppedisano}}, \citenamefont {{Papitto}}, \citenamefont {{Pavlov}}, \citenamefont {{Peirson}},
  \citenamefont {{Perri}}, \citenamefont {{Pesce-Rollins}}, \citenamefont {{Petrucci}}, \citenamefont {{Pilia}}, \citenamefont {{Possenti}}, \citenamefont {{Puccetti}}, \citenamefont {{Ramsey}}, \citenamefont {{Rankin}}, \citenamefont {{Ratheesh}}, \citenamefont {{Romani}}, \citenamefont {{Sgr{\`o}}}, \citenamefont {{Slane}}, \citenamefont {{Soffitta}}, \citenamefont {{Spandre}}, \citenamefont {{Tamagawa}}, \citenamefont {{Tavecchio}}, \citenamefont {{Taverna}}, \citenamefont {{Tawara}}, \citenamefont {{Tennant}}, \citenamefont {{Thomas}}, \citenamefont {{Tombesi}}, \citenamefont {{Trois}}, \citenamefont {{Vink}}, \citenamefont {{Weisskopf}}, \citenamefont {{Wu}}, \citenamefont {{Xie}}, \citenamefont {{Zane}},\ and\ \citenamefont {{IXPE Collaboration}}}]{Tsygankov_etal_2022_cenx-3}%
  \BibitemOpen
  \bibfield  {author} {\bibinfo {author} {\bibfnamefont {S.~S.}\ \bibnamefont {{Tsygankov}}}, \bibinfo {author} {\bibfnamefont {V.}~\bibnamefont {{Doroshenko}}}, \bibinfo {author} {\bibfnamefont {J.}~\bibnamefont {{Poutanen}}}, \bibinfo {author} {\bibfnamefont {J.}~\bibnamefont {{Heyl}}}, \bibinfo {author} {\bibfnamefont {A.~A.}\ \bibnamefont {{Mushtukov}}}, \bibinfo {author} {\bibfnamefont {I.}~\bibnamefont {{Caiazzo}}}, \bibinfo {author} {\bibfnamefont {A.}~\bibnamefont {{Di Marco}}}, \bibinfo {author} {\bibfnamefont {S.~V.}\ \bibnamefont {{Forsblom}}}, \bibinfo {author} {\bibfnamefont {D.}~\bibnamefont {{Gonz{\'a}lez-Caniulef}}}, \bibinfo {author} {\bibfnamefont {M.}~\bibnamefont {{Klawin}}}, \bibinfo {author} {\bibfnamefont {F.}~\bibnamefont {{La Monaca}}}, \bibinfo {author} {\bibfnamefont {C.}~\bibnamefont {{Malacaria}}}, \bibinfo {author} {\bibfnamefont {H.~L.}\ \bibnamefont {{Marshall}}}, \bibinfo {author} {\bibfnamefont {F.}~\bibnamefont {{Muleri}}}, \bibinfo {author} {\bibfnamefont {M.}~\bibnamefont
  {{Ng}}}, \bibinfo {author} {\bibfnamefont {V.~F.}\ \bibnamefont {{Suleimanov}}}, \bibinfo {author} {\bibfnamefont {R.~A.}\ \bibnamefont {{Sunyaev}}}, \bibinfo {author} {\bibfnamefont {R.}~\bibnamefont {{Turolla}}}, \bibinfo {author} {\bibfnamefont {I.}~\bibnamefont {{Agudo}}}, \bibinfo {author} {\bibfnamefont {L.~A.}\ \bibnamefont {{Antonelli}}}, \bibinfo {author} {\bibfnamefont {M.}~\bibnamefont {{Bachetti}}}, \bibinfo {author} {\bibfnamefont {L.}~\bibnamefont {{Baldini}}}, \bibinfo {author} {\bibfnamefont {W.~H.}\ \bibnamefont {{Baumgartner}}}, \bibinfo {author} {\bibfnamefont {R.}~\bibnamefont {{Bellazzini}}}, \bibinfo {author} {\bibfnamefont {S.}~\bibnamefont {{Bianchi}}}, \bibinfo {author} {\bibfnamefont {S.~D.}\ \bibnamefont {{Bongiorno}}}, \bibinfo {author} {\bibfnamefont {R.}~\bibnamefont {{Bonino}}}, \bibinfo {author} {\bibfnamefont {A.}~\bibnamefont {{Brez}}}, \bibinfo {author} {\bibfnamefont {N.}~\bibnamefont {{Bucciantini}}}, \bibinfo {author} {\bibfnamefont {F.}~\bibnamefont {{Capitanio}}},
  \bibinfo {author} {\bibfnamefont {S.}~\bibnamefont {{Castellano}}}, \bibinfo {author} {\bibfnamefont {E.}~\bibnamefont {{Cavazzuti}}}, \bibinfo {author} {\bibfnamefont {S.}~\bibnamefont {{Ciprini}}}, \bibinfo {author} {\bibfnamefont {E.}~\bibnamefont {{Costa}}}, \bibinfo {author} {\bibfnamefont {A.}~\bibnamefont {{De Rosa}}}, \bibinfo {author} {\bibfnamefont {E.}~\bibnamefont {{Del Monte}}}, \bibinfo {author} {\bibfnamefont {L.}~\bibnamefont {{Di Gesu}}}, \bibinfo {author} {\bibfnamefont {N.}~\bibnamefont {{Di Lalla}}}, \bibinfo {author} {\bibfnamefont {I.}~\bibnamefont {{Donnarumma}}}, \bibinfo {author} {\bibfnamefont {M.}~\bibnamefont {{Dov{\v{c}}iak}}}, \bibinfo {author} {\bibfnamefont {S.~R.}\ \bibnamefont {{Ehlert}}}, \bibinfo {author} {\bibfnamefont {T.}~\bibnamefont {{Enoto}}}, \bibinfo {author} {\bibfnamefont {Y.}~\bibnamefont {{Evangelista}}}, \bibinfo {author} {\bibfnamefont {S.}~\bibnamefont {{Fabiani}}}, \bibinfo {author} {\bibfnamefont {R.}~\bibnamefont {{Ferrazzoli}}}, \bibinfo {author}
  {\bibfnamefont {J.~A.}\ \bibnamefont {{Garcia}}}, \bibinfo {author} {\bibfnamefont {S.}~\bibnamefont {{Gunji}}}, \bibinfo {author} {\bibfnamefont {K.}~\bibnamefont {{Hayashida}}}, \bibinfo {author} {\bibfnamefont {W.}~\bibnamefont {{Iwakiri}}}, \bibinfo {author} {\bibfnamefont {S.~G.}\ \bibnamefont {{Jorstad}}}, \bibinfo {author} {\bibfnamefont {V.}~\bibnamefont {{Karas}}}, \bibinfo {author} {\bibfnamefont {T.}~\bibnamefont {{Kitaguchi}}}, \bibinfo {author} {\bibfnamefont {J.~J.}\ \bibnamefont {{Kolodziejczak}}}, \bibinfo {author} {\bibfnamefont {H.}~\bibnamefont {{Krawczynski}}}, \bibinfo {author} {\bibfnamefont {L.}~\bibnamefont {{Latronico}}}, \bibinfo {author} {\bibfnamefont {I.}~\bibnamefont {{Liodakis}}}, \bibinfo {author} {\bibfnamefont {S.}~\bibnamefont {{Maldera}}}, \bibinfo {author} {\bibfnamefont {A.}~\bibnamefont {{Manfreda}}}, \bibinfo {author} {\bibfnamefont {F.}~\bibnamefont {{Marin}}}, \bibinfo {author} {\bibfnamefont {A.}~\bibnamefont {{Marinucci}}}, \bibinfo {author} {\bibfnamefont
  {A.~P.}\ \bibnamefont {{Marscher}}}, \bibinfo {author} {\bibfnamefont {G.}~\bibnamefont {{Matt}}}, \bibinfo {author} {\bibfnamefont {I.}~\bibnamefont {{Mitsuishi}}}, \bibinfo {author} {\bibfnamefont {T.}~\bibnamefont {{Mizuno}}}, \bibinfo {author} {\bibfnamefont {C.-Y.}\ \bibnamefont {{Ng}}}, \bibinfo {author} {\bibfnamefont {S.~L.}\ \bibnamefont {{O'Dell}}}, \bibinfo {author} {\bibfnamefont {N.}~\bibnamefont {{Omodei}}}, \bibinfo {author} {\bibfnamefont {C.}~\bibnamefont {{Oppedisano}}}, \bibinfo {author} {\bibfnamefont {A.}~\bibnamefont {{Papitto}}}, \bibinfo {author} {\bibfnamefont {G.~G.}\ \bibnamefont {{Pavlov}}}, \bibinfo {author} {\bibfnamefont {A.~L.}\ \bibnamefont {{Peirson}}}, \bibinfo {author} {\bibfnamefont {M.}~\bibnamefont {{Perri}}}, \bibinfo {author} {\bibfnamefont {M.}~\bibnamefont {{Pesce-Rollins}}}, \bibinfo {author} {\bibfnamefont {P.-O.}\ \bibnamefont {{Petrucci}}}, \bibinfo {author} {\bibfnamefont {M.}~\bibnamefont {{Pilia}}}, \bibinfo {author} {\bibfnamefont {A.}~\bibnamefont
  {{Possenti}}}, \bibinfo {author} {\bibfnamefont {S.}~\bibnamefont {{Puccetti}}}, \bibinfo {author} {\bibfnamefont {B.~D.}\ \bibnamefont {{Ramsey}}}, \bibinfo {author} {\bibfnamefont {J.}~\bibnamefont {{Rankin}}}, \bibinfo {author} {\bibfnamefont {A.}~\bibnamefont {{Ratheesh}}}, \bibinfo {author} {\bibfnamefont {R.~W.}\ \bibnamefont {{Romani}}}, \bibinfo {author} {\bibfnamefont {C.}~\bibnamefont {{Sgr{\`o}}}}, \bibinfo {author} {\bibfnamefont {P.}~\bibnamefont {{Slane}}}, \bibinfo {author} {\bibfnamefont {P.}~\bibnamefont {{Soffitta}}}, \bibinfo {author} {\bibfnamefont {G.}~\bibnamefont {{Spandre}}}, \bibinfo {author} {\bibfnamefont {T.}~\bibnamefont {{Tamagawa}}}, \bibinfo {author} {\bibfnamefont {F.}~\bibnamefont {{Tavecchio}}}, \bibinfo {author} {\bibfnamefont {R.}~\bibnamefont {{Taverna}}}, \bibinfo {author} {\bibfnamefont {Y.}~\bibnamefont {{Tawara}}}, \bibinfo {author} {\bibfnamefont {A.~F.}\ \bibnamefont {{Tennant}}}, \bibinfo {author} {\bibfnamefont {N.~E.}\ \bibnamefont {{Thomas}}}, \bibinfo
  {author} {\bibfnamefont {F.}~\bibnamefont {{Tombesi}}}, \bibinfo {author} {\bibfnamefont {A.}~\bibnamefont {{Trois}}}, \bibinfo {author} {\bibfnamefont {J.}~\bibnamefont {{Vink}}}, \bibinfo {author} {\bibfnamefont {M.~C.}\ \bibnamefont {{Weisskopf}}}, \bibinfo {author} {\bibfnamefont {K.}~\bibnamefont {{Wu}}}, \bibinfo {author} {\bibfnamefont {F.}~\bibnamefont {{Xie}}}, \bibinfo {author} {\bibfnamefont {S.}~\bibnamefont {{Zane}}}, \ and\ \bibinfo {author} {\bibnamefont {{IXPE Collaboration}}},\ }\href {\doibase 10.3847/2041-8213/aca486} {\bibfield  {journal} {\bibinfo  {journal} {\apjl}\ }\textbf {\bibinfo {volume} {941}},\ \bibinfo {eid} {L14} (\bibinfo {year} {2022})},\ \Eprint {http://arxiv.org/abs/2209.02447} {arXiv:2209.02447 [astro-ph.HE]} \BibitemShut {NoStop}%
\bibitem [{\citenamefont {{Weng}}\ and\ \citenamefont {{Ji}}(2024)}]{Weng2024}%
  \BibitemOpen
  \bibfield  {author} {\bibinfo {author} {\bibfnamefont {S.-S.}\ \bibnamefont {{Weng}}}\ and\ \bibinfo {author} {\bibfnamefont {L.}~\bibnamefont {{Ji}}},\ }\href {\doibase 10.3390/universe10120453} {\bibfield  {journal} {\bibinfo  {journal} {Universe}\ }\textbf {\bibinfo {volume} {10}},\ \bibinfo {eid} {453} (\bibinfo {year} {2024})},\ \Eprint {http://arxiv.org/abs/2412.17275} {arXiv:2412.17275 [astro-ph.HE]} \BibitemShut {NoStop}%
\bibitem [{\citenamefont {{Basko}}\ and\ \citenamefont {{Sunyaev}}(1976)}]{Basko1976}%
  \BibitemOpen
  \bibfield  {author} {\bibinfo {author} {\bibfnamefont {M.~M.}\ \bibnamefont {{Basko}}}\ and\ \bibinfo {author} {\bibfnamefont {R.~A.}\ \bibnamefont {{Sunyaev}}},\ }\href {\doibase 10.1093/mnras/175.2.395} {\bibfield  {journal} {\bibinfo  {journal} {\mnras}\ }\textbf {\bibinfo {volume} {175}},\ \bibinfo {pages} {395} (\bibinfo {year} {1976})}\BibitemShut {NoStop}%
\bibitem [{\citenamefont {{Mushtukov}}\ \emph {et~al.}(2015{\natexlab{a}})\citenamefont {{Mushtukov}}, \citenamefont {{Suleimanov}}, \citenamefont {{Tsygankov}},\ and\ \citenamefont {{Poutanen}}}]{Mushtukov2015}%
  \BibitemOpen
  \bibfield  {author} {\bibinfo {author} {\bibfnamefont {A.~A.}\ \bibnamefont {{Mushtukov}}}, \bibinfo {author} {\bibfnamefont {V.~F.}\ \bibnamefont {{Suleimanov}}}, \bibinfo {author} {\bibfnamefont {S.~S.}\ \bibnamefont {{Tsygankov}}}, \ and\ \bibinfo {author} {\bibfnamefont {J.}~\bibnamefont {{Poutanen}}},\ }\href {\doibase 10.1093/mnras/stu2484} {\bibfield  {journal} {\bibinfo  {journal} {\mnras}\ }\textbf {\bibinfo {volume} {447}},\ \bibinfo {pages} {1847} (\bibinfo {year} {2015}{\natexlab{a}})},\ \Eprint {http://arxiv.org/abs/1409.6457} {arXiv:1409.6457 [astro-ph.HE]} \BibitemShut {NoStop}%
\bibitem [{\citenamefont {{Becker}}\ \emph {et~al.}(2012{\natexlab{b}})\citenamefont {{Becker}}, \citenamefont {{Klochkov}}, \citenamefont {{Sch{\"o}nherr}}, \citenamefont {{Nishimura}}, \citenamefont {{Ferrigno}}, \citenamefont {{Caballero}}, \citenamefont {{Kretschmar}}, \citenamefont {{Wolff}}, \citenamefont {{Wilms}},\ and\ \citenamefont {{Staubert}}}]{Becker2012A&A...544A.123B}%
  \BibitemOpen
  \bibfield  {author} {\bibinfo {author} {\bibfnamefont {P.~A.}\ \bibnamefont {{Becker}}}, \bibinfo {author} {\bibfnamefont {D.}~\bibnamefont {{Klochkov}}}, \bibinfo {author} {\bibfnamefont {G.}~\bibnamefont {{Sch{\"o}nherr}}}, \bibinfo {author} {\bibfnamefont {O.}~\bibnamefont {{Nishimura}}}, \bibinfo {author} {\bibfnamefont {C.}~\bibnamefont {{Ferrigno}}}, \bibinfo {author} {\bibfnamefont {I.}~\bibnamefont {{Caballero}}}, \bibinfo {author} {\bibfnamefont {P.}~\bibnamefont {{Kretschmar}}}, \bibinfo {author} {\bibfnamefont {M.~T.}\ \bibnamefont {{Wolff}}}, \bibinfo {author} {\bibfnamefont {J.}~\bibnamefont {{Wilms}}}, \ and\ \bibinfo {author} {\bibfnamefont {R.}~\bibnamefont {{Staubert}}},\ }\href {\doibase 10.1051/0004-6361/201219065} {\bibfield  {journal} {\bibinfo  {journal} {\aap}\ }\textbf {\bibinfo {volume} {544}},\ \bibinfo {eid} {A123} (\bibinfo {year} {2012}{\natexlab{b}})},\ \Eprint {http://arxiv.org/abs/1205.5316} {arXiv:1205.5316 [astro-ph.HE]} \BibitemShut {NoStop}%
\bibitem [{\citenamefont {{Meszaros}}\ \emph {et~al.}(1988)\citenamefont {{Meszaros}}, \citenamefont {{Novick}}, \citenamefont {{Szentgyorgyi}}, \citenamefont {{Chanan}},\ and\ \citenamefont {{Weisskopf}}}]{Meszaros1988}%
  \BibitemOpen
  \bibfield  {author} {\bibinfo {author} {\bibfnamefont {P.}~\bibnamefont {{Meszaros}}}, \bibinfo {author} {\bibfnamefont {R.}~\bibnamefont {{Novick}}}, \bibinfo {author} {\bibfnamefont {A.}~\bibnamefont {{Szentgyorgyi}}}, \bibinfo {author} {\bibfnamefont {G.~A.}\ \bibnamefont {{Chanan}}}, \ and\ \bibinfo {author} {\bibfnamefont {M.~C.}\ \bibnamefont {{Weisskopf}}},\ }\href {\doibase 10.1086/165962} {\bibfield  {journal} {\bibinfo  {journal} {\apj}\ }\textbf {\bibinfo {volume} {324}},\ \bibinfo {pages} {1056} (\bibinfo {year} {1988})}\BibitemShut {NoStop}%
\bibitem [{\citenamefont {{Weisskopf}}\ \emph {et~al.}(2022)\citenamefont {{Weisskopf}}, \citenamefont {{Soffitta}}, \citenamefont {{Baldini}}, \citenamefont {{Ramsey}}, \citenamefont {{O'Dell}}, \citenamefont {{Romani}}, \citenamefont {{Matt}}, \citenamefont {{Deininger}}, \citenamefont {{Baumgartner}}, \citenamefont {{Bellazzini}}, \citenamefont {{Costa}}, \citenamefont {{Kolodziejczak}}, \citenamefont {{Latronico}}, \citenamefont {{Marshall}}, \citenamefont {{Muleri}}, \citenamefont {{Bongiorno}}, \citenamefont {{Tennant}}, \citenamefont {{Bucciantini}}, \citenamefont {{Dovciak}}, \citenamefont {{Marin}}, \citenamefont {{Marscher}}, \citenamefont {{Poutanen}}, \citenamefont {{Slane}}, \citenamefont {{Turolla}}, \citenamefont {{Kalinowski}}, \citenamefont {{Di Marco}}, \citenamefont {{Fabiani}}, \citenamefont {{Minuti}}, \citenamefont {{La Monaca}}, \citenamefont {{Pinchera}}, \citenamefont {{Rankin}}, \citenamefont {{Sgro'}}, \citenamefont {{Trois}}, \citenamefont {{Xie}}, \citenamefont {{Alexander}},
  \citenamefont {{Allen}}, \citenamefont {{Amici}}, \citenamefont {{Andersen}}, \citenamefont {{Antonelli}}, \citenamefont {{Antoniak}}, \citenamefont {{Attina'}}, \citenamefont {{Barbanera}}, \citenamefont {{Bachetti}}, \citenamefont {{Baggett}}, \citenamefont {{Bladt}}, \citenamefont {{Brez}}, \citenamefont {{Bonino}}, \citenamefont {{Boree}}, \citenamefont {{Borotto}}, \citenamefont {{Breeding}}, \citenamefont {{Brienza}}, \citenamefont {{Bygott}}, \citenamefont {{Caporale}}, \citenamefont {{Cardelli}}, \citenamefont {{Carpentiero}}, \citenamefont {{Castellano}}, \citenamefont {{Castronuovo}}, \citenamefont {{Cavalli}}, \citenamefont {{Cavazzuti}}, \citenamefont {{Ceccanti}}, \citenamefont {{Centrone}}, \citenamefont {{Citraro}}, \citenamefont {{D' Amico}}, \citenamefont {{D'Alba}}, \citenamefont {{Di Gesu}}, \citenamefont {{Del Monte}}, \citenamefont {{Dietz}}, \citenamefont {{Di Lalla}}, \citenamefont {{Di Persio}}, \citenamefont {{Dolan}}, \citenamefont {{Donnarumma}}, \citenamefont {{Evangelista}},
  \citenamefont {{Ferrant}}, \citenamefont {{Ferrazzoli}}, \citenamefont {{Ferrie}}, \citenamefont {{Footdale}}, \citenamefont {{Forsyth}}, \citenamefont {{Foster}}, \citenamefont {{Garelick}}, \citenamefont {{Gunji}}, \citenamefont {{Gurnee}}, \citenamefont {{Head}}, \citenamefont {{Hibbard}}, \citenamefont {{Johnson}}, \citenamefont {{Kelly}}, \citenamefont {{Kilaru}}, \citenamefont {{Lefevre}}, \citenamefont {{Le Roy}}, \citenamefont {{Loffredo}}, \citenamefont {{Lorenzi}}, \citenamefont {{Lucchesi}}, \citenamefont {{Maddox}}, \citenamefont {{Magazzu}}, \citenamefont {{Maldera}}, \citenamefont {{Manfreda}}, \citenamefont {{Mangraviti}}, \citenamefont {{Marengo}}, \citenamefont {{Marrocchesi}}, \citenamefont {{Massaro}}, \citenamefont {{Mauger}}, \citenamefont {{McCracken}}, \citenamefont {{McEachen}}, \citenamefont {{Mize}}, \citenamefont {{Mereu}}, \citenamefont {{Mitchell}}, \citenamefont {{Mitsuishi}}, \citenamefont {{Morbidini}}, \citenamefont {{Mosti}}, \citenamefont {{Nasimi}}, \citenamefont
  {{Negri}}, \citenamefont {{Negro}}, \citenamefont {{Nguyen}}, \citenamefont {{Nitschke}}, \citenamefont {{Nuti}}, \citenamefont {{Onizuka}}, \citenamefont {{Oppedisano}}, \citenamefont {{Orsini}}, \citenamefont {{Osborne}}, \citenamefont {{Pacheco}}, \citenamefont {{Paggi}}, \citenamefont {{Painter}}, \citenamefont {{Pavelitz}}, \citenamefont {{Pentz}}, \citenamefont {{Piazzolla}}, \citenamefont {{Perri}}, \citenamefont {{Pesce-Rollins}}, \citenamefont {{Peterson}}, \citenamefont {{Pilia}}, \citenamefont {{Profeti}}, \citenamefont {{Puccetti}}, \citenamefont {{Ranganathan}}, \citenamefont {{Ratheesh}}, \citenamefont {{Reedy}}, \citenamefont {{Root}}, \citenamefont {{Rubini}}, \citenamefont {{Ruswick}}, \citenamefont {{Sanchez}}, \citenamefont {{Sarra}}, \citenamefont {{Santoli}}, \citenamefont {{Scalise}}, \citenamefont {{Sciortino}}, \citenamefont {{Schroeder}}, \citenamefont {{Seek}}, \citenamefont {{Sosdian}}, \citenamefont {{Spandre}}, \citenamefont {{Speegle}}, \citenamefont {{Tamagawa}}, \citenamefont
  {{Tardiola}}, \citenamefont {{Tobia}}, \citenamefont {{Thomas}}, \citenamefont {{Valerie}}, \citenamefont {{Vimercati}}, \citenamefont {{Walden}}, \citenamefont {{Weddendorf}}, \citenamefont {{Wedmore}}, \citenamefont {{Welch}}, \citenamefont {{Zanetti}},\ and\ \citenamefont {{Zanetti}}}]{Weisskopf2022}%
  \BibitemOpen
  \bibfield  {author} {\bibinfo {author} {\bibfnamefont {M.~C.}\ \bibnamefont {{Weisskopf}}}, \bibinfo {author} {\bibfnamefont {P.}~\bibnamefont {{Soffitta}}}, \bibinfo {author} {\bibfnamefont {L.}~\bibnamefont {{Baldini}}}, \bibinfo {author} {\bibfnamefont {B.~D.}\ \bibnamefont {{Ramsey}}}, \bibinfo {author} {\bibfnamefont {S.~L.}\ \bibnamefont {{O'Dell}}}, \bibinfo {author} {\bibfnamefont {R.~W.}\ \bibnamefont {{Romani}}}, \bibinfo {author} {\bibfnamefont {G.}~\bibnamefont {{Matt}}}, \bibinfo {author} {\bibfnamefont {W.~D.}\ \bibnamefont {{Deininger}}}, \bibinfo {author} {\bibfnamefont {W.~H.}\ \bibnamefont {{Baumgartner}}}, \bibinfo {author} {\bibfnamefont {R.}~\bibnamefont {{Bellazzini}}}, \bibinfo {author} {\bibfnamefont {E.}~\bibnamefont {{Costa}}}, \bibinfo {author} {\bibfnamefont {J.~J.}\ \bibnamefont {{Kolodziejczak}}}, \bibinfo {author} {\bibfnamefont {L.}~\bibnamefont {{Latronico}}}, \bibinfo {author} {\bibfnamefont {H.~L.}\ \bibnamefont {{Marshall}}}, \bibinfo {author} {\bibfnamefont
  {F.}~\bibnamefont {{Muleri}}}, \bibinfo {author} {\bibfnamefont {S.~D.}\ \bibnamefont {{Bongiorno}}}, \bibinfo {author} {\bibfnamefont {A.}~\bibnamefont {{Tennant}}}, \bibinfo {author} {\bibfnamefont {N.}~\bibnamefont {{Bucciantini}}}, \bibinfo {author} {\bibfnamefont {M.}~\bibnamefont {{Dovciak}}}, \bibinfo {author} {\bibfnamefont {F.}~\bibnamefont {{Marin}}}, \bibinfo {author} {\bibfnamefont {A.}~\bibnamefont {{Marscher}}}, \bibinfo {author} {\bibfnamefont {J.}~\bibnamefont {{Poutanen}}}, \bibinfo {author} {\bibfnamefont {P.}~\bibnamefont {{Slane}}}, \bibinfo {author} {\bibfnamefont {R.}~\bibnamefont {{Turolla}}}, \bibinfo {author} {\bibfnamefont {W.}~\bibnamefont {{Kalinowski}}}, \bibinfo {author} {\bibfnamefont {A.}~\bibnamefont {{Di Marco}}}, \bibinfo {author} {\bibfnamefont {S.}~\bibnamefont {{Fabiani}}}, \bibinfo {author} {\bibfnamefont {M.}~\bibnamefont {{Minuti}}}, \bibinfo {author} {\bibfnamefont {F.}~\bibnamefont {{La Monaca}}}, \bibinfo {author} {\bibfnamefont {M.}~\bibnamefont {{Pinchera}}},
  \bibinfo {author} {\bibfnamefont {J.}~\bibnamefont {{Rankin}}}, \bibinfo {author} {\bibfnamefont {C.}~\bibnamefont {{Sgro'}}}, \bibinfo {author} {\bibfnamefont {A.}~\bibnamefont {{Trois}}}, \bibinfo {author} {\bibfnamefont {F.}~\bibnamefont {{Xie}}}, \bibinfo {author} {\bibfnamefont {C.}~\bibnamefont {{Alexander}}}, \bibinfo {author} {\bibfnamefont {D.~Z.}\ \bibnamefont {{Allen}}}, \bibinfo {author} {\bibfnamefont {F.}~\bibnamefont {{Amici}}}, \bibinfo {author} {\bibfnamefont {J.}~\bibnamefont {{Andersen}}}, \bibinfo {author} {\bibfnamefont {A.}~\bibnamefont {{Antonelli}}}, \bibinfo {author} {\bibfnamefont {S.}~\bibnamefont {{Antoniak}}}, \bibinfo {author} {\bibfnamefont {P.}~\bibnamefont {{Attina'}}}, \bibinfo {author} {\bibfnamefont {M.}~\bibnamefont {{Barbanera}}}, \bibinfo {author} {\bibfnamefont {M.}~\bibnamefont {{Bachetti}}}, \bibinfo {author} {\bibfnamefont {R.~M.}\ \bibnamefont {{Baggett}}}, \bibinfo {author} {\bibfnamefont {J.}~\bibnamefont {{Bladt}}}, \bibinfo {author} {\bibfnamefont
  {A.}~\bibnamefont {{Brez}}}, \bibinfo {author} {\bibfnamefont {R.}~\bibnamefont {{Bonino}}}, \bibinfo {author} {\bibfnamefont {C.}~\bibnamefont {{Boree}}}, \bibinfo {author} {\bibfnamefont {F.}~\bibnamefont {{Borotto}}}, \bibinfo {author} {\bibfnamefont {S.}~\bibnamefont {{Breeding}}}, \bibinfo {author} {\bibfnamefont {D.}~\bibnamefont {{Brienza}}}, \bibinfo {author} {\bibfnamefont {H.~K.}\ \bibnamefont {{Bygott}}}, \bibinfo {author} {\bibfnamefont {C.}~\bibnamefont {{Caporale}}}, \bibinfo {author} {\bibfnamefont {C.}~\bibnamefont {{Cardelli}}}, \bibinfo {author} {\bibfnamefont {R.}~\bibnamefont {{Carpentiero}}}, \bibinfo {author} {\bibfnamefont {S.}~\bibnamefont {{Castellano}}}, \bibinfo {author} {\bibfnamefont {M.}~\bibnamefont {{Castronuovo}}}, \bibinfo {author} {\bibfnamefont {L.}~\bibnamefont {{Cavalli}}}, \bibinfo {author} {\bibfnamefont {E.}~\bibnamefont {{Cavazzuti}}}, \bibinfo {author} {\bibfnamefont {M.}~\bibnamefont {{Ceccanti}}}, \bibinfo {author} {\bibfnamefont {M.}~\bibnamefont {{Centrone}}},
  \bibinfo {author} {\bibfnamefont {S.}~\bibnamefont {{Citraro}}}, \bibinfo {author} {\bibfnamefont {F.}~\bibnamefont {{D' Amico}}}, \bibinfo {author} {\bibfnamefont {E.}~\bibnamefont {{D'Alba}}}, \bibinfo {author} {\bibfnamefont {L.}~\bibnamefont {{Di Gesu}}}, \bibinfo {author} {\bibfnamefont {E.}~\bibnamefont {{Del Monte}}}, \bibinfo {author} {\bibfnamefont {K.~L.}\ \bibnamefont {{Dietz}}}, \bibinfo {author} {\bibfnamefont {N.}~\bibnamefont {{Di Lalla}}}, \bibinfo {author} {\bibfnamefont {G.}~\bibnamefont {{Di Persio}}}, \bibinfo {author} {\bibfnamefont {D.}~\bibnamefont {{Dolan}}}, \bibinfo {author} {\bibfnamefont {I.}~\bibnamefont {{Donnarumma}}}, \bibinfo {author} {\bibfnamefont {Y.}~\bibnamefont {{Evangelista}}}, \bibinfo {author} {\bibfnamefont {K.}~\bibnamefont {{Ferrant}}}, \bibinfo {author} {\bibfnamefont {R.}~\bibnamefont {{Ferrazzoli}}}, \bibinfo {author} {\bibfnamefont {M.}~\bibnamefont {{Ferrie}}}, \bibinfo {author} {\bibfnamefont {J.}~\bibnamefont {{Footdale}}}, \bibinfo {author} {\bibfnamefont
  {B.}~\bibnamefont {{Forsyth}}}, \bibinfo {author} {\bibfnamefont {M.}~\bibnamefont {{Foster}}}, \bibinfo {author} {\bibfnamefont {B.}~\bibnamefont {{Garelick}}}, \bibinfo {author} {\bibfnamefont {S.}~\bibnamefont {{Gunji}}}, \bibinfo {author} {\bibfnamefont {E.}~\bibnamefont {{Gurnee}}}, \bibinfo {author} {\bibfnamefont {M.}~\bibnamefont {{Head}}}, \bibinfo {author} {\bibfnamefont {G.}~\bibnamefont {{Hibbard}}}, \bibinfo {author} {\bibfnamefont {S.}~\bibnamefont {{Johnson}}}, \bibinfo {author} {\bibfnamefont {E.}~\bibnamefont {{Kelly}}}, \bibinfo {author} {\bibfnamefont {K.}~\bibnamefont {{Kilaru}}}, \bibinfo {author} {\bibfnamefont {C.}~\bibnamefont {{Lefevre}}}, \bibinfo {author} {\bibfnamefont {S.}~\bibnamefont {{Le Roy}}}, \bibinfo {author} {\bibfnamefont {P.}~\bibnamefont {{Loffredo}}}, \bibinfo {author} {\bibfnamefont {P.}~\bibnamefont {{Lorenzi}}}, \bibinfo {author} {\bibfnamefont {L.}~\bibnamefont {{Lucchesi}}}, \bibinfo {author} {\bibfnamefont {T.}~\bibnamefont {{Maddox}}}, \bibinfo {author}
  {\bibfnamefont {G.}~\bibnamefont {{Magazzu}}}, \bibinfo {author} {\bibfnamefont {S.}~\bibnamefont {{Maldera}}}, \bibinfo {author} {\bibfnamefont {A.}~\bibnamefont {{Manfreda}}}, \bibinfo {author} {\bibfnamefont {E.}~\bibnamefont {{Mangraviti}}}, \bibinfo {author} {\bibfnamefont {M.}~\bibnamefont {{Marengo}}}, \bibinfo {author} {\bibfnamefont {A.}~\bibnamefont {{Marrocchesi}}}, \bibinfo {author} {\bibfnamefont {F.}~\bibnamefont {{Massaro}}}, \bibinfo {author} {\bibfnamefont {D.}~\bibnamefont {{Mauger}}}, \bibinfo {author} {\bibfnamefont {J.}~\bibnamefont {{McCracken}}}, \bibinfo {author} {\bibfnamefont {M.}~\bibnamefont {{McEachen}}}, \bibinfo {author} {\bibfnamefont {R.}~\bibnamefont {{Mize}}}, \bibinfo {author} {\bibfnamefont {P.}~\bibnamefont {{Mereu}}}, \bibinfo {author} {\bibfnamefont {S.}~\bibnamefont {{Mitchell}}}, \bibinfo {author} {\bibfnamefont {I.}~\bibnamefont {{Mitsuishi}}}, \bibinfo {author} {\bibfnamefont {A.}~\bibnamefont {{Morbidini}}}, \bibinfo {author} {\bibfnamefont {F.}~\bibnamefont
  {{Mosti}}}, \bibinfo {author} {\bibfnamefont {H.}~\bibnamefont {{Nasimi}}}, \bibinfo {author} {\bibfnamefont {B.}~\bibnamefont {{Negri}}}, \bibinfo {author} {\bibfnamefont {M.}~\bibnamefont {{Negro}}}, \bibinfo {author} {\bibfnamefont {T.}~\bibnamefont {{Nguyen}}}, \bibinfo {author} {\bibfnamefont {I.}~\bibnamefont {{Nitschke}}}, \bibinfo {author} {\bibfnamefont {A.}~\bibnamefont {{Nuti}}}, \bibinfo {author} {\bibfnamefont {M.}~\bibnamefont {{Onizuka}}}, \bibinfo {author} {\bibfnamefont {C.}~\bibnamefont {{Oppedisano}}}, \bibinfo {author} {\bibfnamefont {L.}~\bibnamefont {{Orsini}}}, \bibinfo {author} {\bibfnamefont {D.}~\bibnamefont {{Osborne}}}, \bibinfo {author} {\bibfnamefont {R.}~\bibnamefont {{Pacheco}}}, \bibinfo {author} {\bibfnamefont {A.}~\bibnamefont {{Paggi}}}, \bibinfo {author} {\bibfnamefont {W.}~\bibnamefont {{Painter}}}, \bibinfo {author} {\bibfnamefont {S.~D.}\ \bibnamefont {{Pavelitz}}}, \bibinfo {author} {\bibfnamefont {C.}~\bibnamefont {{Pentz}}}, \bibinfo {author} {\bibfnamefont
  {R.}~\bibnamefont {{Piazzolla}}}, \bibinfo {author} {\bibfnamefont {M.}~\bibnamefont {{Perri}}}, \bibinfo {author} {\bibfnamefont {M.}~\bibnamefont {{Pesce-Rollins}}}, \bibinfo {author} {\bibfnamefont {C.}~\bibnamefont {{Peterson}}}, \bibinfo {author} {\bibfnamefont {M.}~\bibnamefont {{Pilia}}}, \bibinfo {author} {\bibfnamefont {A.}~\bibnamefont {{Profeti}}}, \bibinfo {author} {\bibfnamefont {S.}~\bibnamefont {{Puccetti}}}, \bibinfo {author} {\bibfnamefont {J.}~\bibnamefont {{Ranganathan}}}, \bibinfo {author} {\bibfnamefont {A.}~\bibnamefont {{Ratheesh}}}, \bibinfo {author} {\bibfnamefont {L.}~\bibnamefont {{Reedy}}}, \bibinfo {author} {\bibfnamefont {N.}~\bibnamefont {{Root}}}, \bibinfo {author} {\bibfnamefont {A.}~\bibnamefont {{Rubini}}}, \bibinfo {author} {\bibfnamefont {S.}~\bibnamefont {{Ruswick}}}, \bibinfo {author} {\bibfnamefont {J.}~\bibnamefont {{Sanchez}}}, \bibinfo {author} {\bibfnamefont {P.}~\bibnamefont {{Sarra}}}, \bibinfo {author} {\bibfnamefont {F.}~\bibnamefont {{Santoli}}}, \bibinfo
  {author} {\bibfnamefont {E.}~\bibnamefont {{Scalise}}}, \bibinfo {author} {\bibfnamefont {A.}~\bibnamefont {{Sciortino}}}, \bibinfo {author} {\bibfnamefont {C.}~\bibnamefont {{Schroeder}}}, \bibinfo {author} {\bibfnamefont {T.}~\bibnamefont {{Seek}}}, \bibinfo {author} {\bibfnamefont {K.}~\bibnamefont {{Sosdian}}}, \bibinfo {author} {\bibfnamefont {G.}~\bibnamefont {{Spandre}}}, \bibinfo {author} {\bibfnamefont {C.~O.}\ \bibnamefont {{Speegle}}}, \bibinfo {author} {\bibfnamefont {T.}~\bibnamefont {{Tamagawa}}}, \bibinfo {author} {\bibfnamefont {M.}~\bibnamefont {{Tardiola}}}, \bibinfo {author} {\bibfnamefont {A.}~\bibnamefont {{Tobia}}}, \bibinfo {author} {\bibfnamefont {N.~E.}\ \bibnamefont {{Thomas}}}, \bibinfo {author} {\bibfnamefont {R.}~\bibnamefont {{Valerie}}}, \bibinfo {author} {\bibfnamefont {M.}~\bibnamefont {{Vimercati}}}, \bibinfo {author} {\bibfnamefont {A.~L.}\ \bibnamefont {{Walden}}}, \bibinfo {author} {\bibfnamefont {B.}~\bibnamefont {{Weddendorf}}}, \bibinfo {author} {\bibfnamefont
  {J.}~\bibnamefont {{Wedmore}}}, \bibinfo {author} {\bibfnamefont {D.}~\bibnamefont {{Welch}}}, \bibinfo {author} {\bibfnamefont {D.}~\bibnamefont {{Zanetti}}}, \ and\ \bibinfo {author} {\bibfnamefont {F.}~\bibnamefont {{Zanetti}}},\ }\href {\doibase 10.1117/1.JATIS.8.2.026002} {\bibfield  {journal} {\bibinfo  {journal} {JATIS}\ }\textbf {\bibinfo {volume} {8}},\ \bibinfo {pages} {026002} (\bibinfo {year} {2022})},\ \Eprint {http://arxiv.org/abs/2112.01269} {arXiv:2112.01269 [astro-ph.IM]} \BibitemShut {NoStop}%
\bibitem [{\citenamefont {{Neumann}}\ \emph {et~al.}(2023)\citenamefont {{Neumann}}, \citenamefont {{Avakyan}}, \citenamefont {{Doroshenko}},\ and\ \citenamefont {{Santangelo}}}]{Neumann2023}%
  \BibitemOpen
  \bibfield  {author} {\bibinfo {author} {\bibfnamefont {M.}~\bibnamefont {{Neumann}}}, \bibinfo {author} {\bibfnamefont {A.}~\bibnamefont {{Avakyan}}}, \bibinfo {author} {\bibfnamefont {V.}~\bibnamefont {{Doroshenko}}}, \ and\ \bibinfo {author} {\bibfnamefont {A.}~\bibnamefont {{Santangelo}}},\ }\href {\doibase 10.1051/0004-6361/202245728} {\bibfield  {journal} {\bibinfo  {journal} {\aap}\ }\textbf {\bibinfo {volume} {677}},\ \bibinfo {eid} {A134} (\bibinfo {year} {2023})},\ \Eprint {http://arxiv.org/abs/2303.16137} {arXiv:2303.16137 [astro-ph.HE]} \BibitemShut {NoStop}%
\bibitem [{\citenamefont {{Doroshenko}}\ \emph {et~al.}(2022{\natexlab{b}})\citenamefont {{Doroshenko}}, \citenamefont {{Poutanen}}, \citenamefont {{Tsygankov}}, \citenamefont {{Suleimanov}}, \citenamefont {{Bachetti}}, \citenamefont {{Caiazzo}}, \citenamefont {{Costa}}, \citenamefont {{Di Marco}}, \citenamefont {{Heyl}}, \citenamefont {{La Monaca}},\ and\ \citenamefont {et~al.}}]{Doroshenko_etal_2022_herx-1}%
  \BibitemOpen
  \bibfield  {author} {\bibinfo {author} {\bibfnamefont {V.}~\bibnamefont {{Doroshenko}}}, \bibinfo {author} {\bibfnamefont {J.}~\bibnamefont {{Poutanen}}}, \bibinfo {author} {\bibfnamefont {S.~S.}\ \bibnamefont {{Tsygankov}}}, \bibinfo {author} {\bibfnamefont {V.~F.}\ \bibnamefont {{Suleimanov}}}, \bibinfo {author} {\bibfnamefont {M.}~\bibnamefont {{Bachetti}}}, \bibinfo {author} {\bibfnamefont {I.}~\bibnamefont {{Caiazzo}}}, \bibinfo {author} {\bibfnamefont {E.}~\bibnamefont {{Costa}}}, \bibinfo {author} {\bibfnamefont {A.}~\bibnamefont {{Di Marco}}}, \bibinfo {author} {\bibfnamefont {J.}~\bibnamefont {{Heyl}}}, \bibinfo {author} {\bibfnamefont {F.}~\bibnamefont {{La Monaca}}}, \ and\ \bibinfo {author} {\bibnamefont {et~al.}},\ }\href {\doibase 10.1038/s41550-022-01799-5} {\bibfield  {journal} {\bibinfo  {journal} {Nature Astronomy}\ }\textbf {\bibinfo {volume} {6}},\ \bibinfo {pages} {1433} (\bibinfo {year} {2022}{\natexlab{b}})},\ \Eprint {http://arxiv.org/abs/2206.07138} {arXiv:2206.07138 [astro-ph.HE]}
  \BibitemShut {NoStop}%
\bibitem [{\citenamefont {{Garg}}\ \emph {et~al.}(2023)\citenamefont {{Garg}}, \citenamefont {{Rawat}}, \citenamefont {{Bhargava}}, \citenamefont {{M{\'e}ndez}},\ and\ \citenamefont {{Bhattacharyya}}}]{Garg_herx-1}%
  \BibitemOpen
  \bibfield  {author} {\bibinfo {author} {\bibfnamefont {A.}~\bibnamefont {{Garg}}}, \bibinfo {author} {\bibfnamefont {D.}~\bibnamefont {{Rawat}}}, \bibinfo {author} {\bibfnamefont {Y.}~\bibnamefont {{Bhargava}}}, \bibinfo {author} {\bibfnamefont {M.}~\bibnamefont {{M{\'e}ndez}}}, \ and\ \bibinfo {author} {\bibfnamefont {S.}~\bibnamefont {{Bhattacharyya}}},\ }\href {\doibase 10.3847/2041-8213/acccfa} {\bibfield  {journal} {\bibinfo  {journal} {\apjl}\ }\textbf {\bibinfo {volume} {948}},\ \bibinfo {eid} {L10} (\bibinfo {year} {2023})}\BibitemShut {NoStop}%
\bibitem [{\citenamefont {{Zhao}}\ \emph {et~al.}(2024)\citenamefont {{Zhao}}, \citenamefont {{Li}}, \citenamefont {{Tao}}, \citenamefont {{Feng}}, \citenamefont {{Zhang}}, \citenamefont {{Walter}}, \citenamefont {{Ge}}, \citenamefont {{Tong}}, \citenamefont {{Ji}}, \citenamefont {{Zhang}}, \citenamefont {{Qu}}, \citenamefont {{Huang}}, \citenamefont {{Ma}}, \citenamefont {{Zhang}}, \citenamefont {{Yin}}, \citenamefont {{Yin}}, \citenamefont {{Ma}}, \citenamefont {{Zhao}}, \citenamefont {{Li}}, \citenamefont {{Yang}}, \citenamefont {{Liu}}, \citenamefont {{Yu}}, \citenamefont {{Huang}}, \citenamefont {{Li}}, \citenamefont {{Li}}, \citenamefont {{Xiao}},\ and\ \citenamefont {{Zhao}}}]{Zhao_herx-1}%
  \BibitemOpen
  \bibfield  {author} {\bibinfo {author} {\bibfnamefont {Q.~C.}\ \bibnamefont {{Zhao}}}, \bibinfo {author} {\bibfnamefont {H.~C.}\ \bibnamefont {{Li}}}, \bibinfo {author} {\bibfnamefont {L.}~\bibnamefont {{Tao}}}, \bibinfo {author} {\bibfnamefont {H.}~\bibnamefont {{Feng}}}, \bibinfo {author} {\bibfnamefont {S.~N.}\ \bibnamefont {{Zhang}}}, \bibinfo {author} {\bibfnamefont {R.}~\bibnamefont {{Walter}}}, \bibinfo {author} {\bibfnamefont {M.~Y.}\ \bibnamefont {{Ge}}}, \bibinfo {author} {\bibfnamefont {H.}~\bibnamefont {{Tong}}}, \bibinfo {author} {\bibfnamefont {L.}~\bibnamefont {{Ji}}}, \bibinfo {author} {\bibfnamefont {L.}~\bibnamefont {{Zhang}}}, \bibinfo {author} {\bibfnamefont {J.~L.}\ \bibnamefont {{Qu}}}, \bibinfo {author} {\bibfnamefont {Y.}~\bibnamefont {{Huang}}}, \bibinfo {author} {\bibfnamefont {X.}~\bibnamefont {{Ma}}}, \bibinfo {author} {\bibfnamefont {S.}~\bibnamefont {{Zhang}}}, \bibinfo {author} {\bibfnamefont {Q.~Q.}\ \bibnamefont {{Yin}}}, \bibinfo {author} {\bibfnamefont {H.~X.}\
  \bibnamefont {{Yin}}}, \bibinfo {author} {\bibfnamefont {R.~C.}\ \bibnamefont {{Ma}}}, \bibinfo {author} {\bibfnamefont {S.~J.}\ \bibnamefont {{Zhao}}}, \bibinfo {author} {\bibfnamefont {P.~P.}\ \bibnamefont {{Li}}}, \bibinfo {author} {\bibfnamefont {Z.~X.}\ \bibnamefont {{Yang}}}, \bibinfo {author} {\bibfnamefont {H.~X.}\ \bibnamefont {{Liu}}}, \bibinfo {author} {\bibfnamefont {W.}~\bibnamefont {{Yu}}}, \bibinfo {author} {\bibfnamefont {Y.~M.}\ \bibnamefont {{Huang}}}, \bibinfo {author} {\bibfnamefont {Z.~X.}\ \bibnamefont {{Li}}}, \bibinfo {author} {\bibfnamefont {Y.~J.}\ \bibnamefont {{Li}}}, \bibinfo {author} {\bibfnamefont {J.~Y.}\ \bibnamefont {{Xiao}}}, \ and\ \bibinfo {author} {\bibfnamefont {K.}~\bibnamefont {{Zhao}}},\ }\href {\doibase 10.1093/mnras/stae1173} {\bibfield  {journal} {\bibinfo  {journal} {\mnras}\ }\textbf {\bibinfo {volume} {531}},\ \bibinfo {pages} {3935} (\bibinfo {year} {2024})},\ \Eprint {http://arxiv.org/abs/2405.00509} {arXiv:2405.00509 [astro-ph.HE]} \BibitemShut {NoStop}%
\bibitem [{\citenamefont {{Tsygankov}}\ \emph {et~al.}(2023)\citenamefont {{Tsygankov}}, \citenamefont {{Doroshenko}}, \citenamefont {{Mushtukov}}, \citenamefont {{Poutanen}}, \citenamefont {{Di Marco}}, \citenamefont {{Heyl}}, \citenamefont {{La Monaca}}, \citenamefont {{Forsblom}}, \citenamefont {{Malacaria}}, \citenamefont {{Marshall}}, \citenamefont {{Suleimanov}}, \citenamefont {{Svoboda}}, \citenamefont {{Taverna}}, \citenamefont {{Ursini}}, \citenamefont {{Agudo}}, \citenamefont {{Antonelli}}, \citenamefont {{Bachetti}}, \citenamefont {{Baldini}}, \citenamefont {{Baumgartner}}, \citenamefont {{Bellazzini}}, \citenamefont {{Bianchi}}, \citenamefont {{Bongiorno}}, \citenamefont {{Bonino}}, \citenamefont {{Brez}}, \citenamefont {{Bucciantini}}, \citenamefont {{Capitanio}}, \citenamefont {{Castellano}}, \citenamefont {{Cavazzuti}}, \citenamefont {{Chen}}, \citenamefont {{Ciprini}}, \citenamefont {{Costa}}, \citenamefont {{De Rosa}}, \citenamefont {{Del Monte}}, \citenamefont {{Di Gesu}}, \citenamefont {{Di
  Lalla}}, \citenamefont {{Donnarumma}}, \citenamefont {{Dov{\v{c}}iak}}, \citenamefont {{Ehlert}}, \citenamefont {{Enoto}}, \citenamefont {{Evangelista}}, \citenamefont {{Fabiani}}, \citenamefont {{Ferrazzoli}}, \citenamefont {{Garcia}}, \citenamefont {{Gunji}}, \citenamefont {{Hayashida}}, \citenamefont {{Iwakiri}}, \citenamefont {{Jorstad}}, \citenamefont {{Kaaret}}, \citenamefont {{Karas}}, \citenamefont {{Kislat}}, \citenamefont {{Kitaguchi}}, \citenamefont {{Kolodziejczak}}, \citenamefont {{Krawczynski}}, \citenamefont {{Latronico}}, \citenamefont {{Liodakis}}, \citenamefont {{Maldera}}, \citenamefont {{Manfreda}}, \citenamefont {{Marin}}, \citenamefont {{Marinucci}}, \citenamefont {{Marscher}}, \citenamefont {{Massaro}}, \citenamefont {{Matt}}, \citenamefont {{Mitsuishi}}, \citenamefont {{Mizuno}}, \citenamefont {{Muleri}}, \citenamefont {{Negro}}, \citenamefont {{Ng}}, \citenamefont {{O'Dell}}, \citenamefont {{Omodei}}, \citenamefont {{Oppedisano}}, \citenamefont {{Papitto}}, \citenamefont {{Pavlov}},
  \citenamefont {{Peirson}}, \citenamefont {{Perri}}, \citenamefont {{Pesce-Rollins}}, \citenamefont {{Petrucci}}, \citenamefont {{Pilia}}, \citenamefont {{Possenti}}, \citenamefont {{Puccetti}}, \citenamefont {{Ramsey}}, \citenamefont {{Rankin}}, \citenamefont {{Ratheesh}}, \citenamefont {{Roberts}}, \citenamefont {{Romani}}, \citenamefont {{Sgr{\`o}}}, \citenamefont {{Slane}}, \citenamefont {{Soffitta}}, \citenamefont {{Spandre}}, \citenamefont {{Swartz}}, \citenamefont {{Tamagawa}}, \citenamefont {{Tavecchio}}, \citenamefont {{Tawara}}, \citenamefont {{Tennant}}, \citenamefont {{Thomas}}, \citenamefont {{Tombesi}}, \citenamefont {{Trois}}, \citenamefont {{Turolla}}, \citenamefont {{Vink}}, \citenamefont {{Weisskopf}}, \citenamefont {{Wu}}, \citenamefont {{Xie}},\ and\ \citenamefont {{Zane}}}]{Tsygankov_groj1008}%
  \BibitemOpen
  \bibfield  {author} {\bibinfo {author} {\bibfnamefont {S.~S.}\ \bibnamefont {{Tsygankov}}}, \bibinfo {author} {\bibfnamefont {V.}~\bibnamefont {{Doroshenko}}}, \bibinfo {author} {\bibfnamefont {A.~A.}\ \bibnamefont {{Mushtukov}}}, \bibinfo {author} {\bibfnamefont {J.}~\bibnamefont {{Poutanen}}}, \bibinfo {author} {\bibfnamefont {A.}~\bibnamefont {{Di Marco}}}, \bibinfo {author} {\bibfnamefont {J.}~\bibnamefont {{Heyl}}}, \bibinfo {author} {\bibfnamefont {F.}~\bibnamefont {{La Monaca}}}, \bibinfo {author} {\bibfnamefont {S.~V.}\ \bibnamefont {{Forsblom}}}, \bibinfo {author} {\bibfnamefont {C.}~\bibnamefont {{Malacaria}}}, \bibinfo {author} {\bibfnamefont {H.~L.}\ \bibnamefont {{Marshall}}}, \bibinfo {author} {\bibfnamefont {V.~F.}\ \bibnamefont {{Suleimanov}}}, \bibinfo {author} {\bibfnamefont {J.}~\bibnamefont {{Svoboda}}}, \bibinfo {author} {\bibfnamefont {R.}~\bibnamefont {{Taverna}}}, \bibinfo {author} {\bibfnamefont {F.}~\bibnamefont {{Ursini}}}, \bibinfo {author} {\bibfnamefont {I.}~\bibnamefont
  {{Agudo}}}, \bibinfo {author} {\bibfnamefont {L.~A.}\ \bibnamefont {{Antonelli}}}, \bibinfo {author} {\bibfnamefont {M.}~\bibnamefont {{Bachetti}}}, \bibinfo {author} {\bibfnamefont {L.}~\bibnamefont {{Baldini}}}, \bibinfo {author} {\bibfnamefont {W.~H.}\ \bibnamefont {{Baumgartner}}}, \bibinfo {author} {\bibfnamefont {R.}~\bibnamefont {{Bellazzini}}}, \bibinfo {author} {\bibfnamefont {S.}~\bibnamefont {{Bianchi}}}, \bibinfo {author} {\bibfnamefont {S.~D.}\ \bibnamefont {{Bongiorno}}}, \bibinfo {author} {\bibfnamefont {R.}~\bibnamefont {{Bonino}}}, \bibinfo {author} {\bibfnamefont {A.}~\bibnamefont {{Brez}}}, \bibinfo {author} {\bibfnamefont {N.}~\bibnamefont {{Bucciantini}}}, \bibinfo {author} {\bibfnamefont {F.}~\bibnamefont {{Capitanio}}}, \bibinfo {author} {\bibfnamefont {S.}~\bibnamefont {{Castellano}}}, \bibinfo {author} {\bibfnamefont {E.}~\bibnamefont {{Cavazzuti}}}, \bibinfo {author} {\bibfnamefont {C.-T.}\ \bibnamefont {{Chen}}}, \bibinfo {author} {\bibfnamefont {S.}~\bibnamefont {{Ciprini}}},
  \bibinfo {author} {\bibfnamefont {E.}~\bibnamefont {{Costa}}}, \bibinfo {author} {\bibfnamefont {A.}~\bibnamefont {{De Rosa}}}, \bibinfo {author} {\bibfnamefont {E.}~\bibnamefont {{Del Monte}}}, \bibinfo {author} {\bibfnamefont {L.}~\bibnamefont {{Di Gesu}}}, \bibinfo {author} {\bibfnamefont {N.}~\bibnamefont {{Di Lalla}}}, \bibinfo {author} {\bibfnamefont {I.}~\bibnamefont {{Donnarumma}}}, \bibinfo {author} {\bibfnamefont {M.}~\bibnamefont {{Dov{\v{c}}iak}}}, \bibinfo {author} {\bibfnamefont {S.~R.}\ \bibnamefont {{Ehlert}}}, \bibinfo {author} {\bibfnamefont {T.}~\bibnamefont {{Enoto}}}, \bibinfo {author} {\bibfnamefont {Y.}~\bibnamefont {{Evangelista}}}, \bibinfo {author} {\bibfnamefont {S.}~\bibnamefont {{Fabiani}}}, \bibinfo {author} {\bibfnamefont {R.}~\bibnamefont {{Ferrazzoli}}}, \bibinfo {author} {\bibfnamefont {J.~A.}\ \bibnamefont {{Garcia}}}, \bibinfo {author} {\bibfnamefont {S.}~\bibnamefont {{Gunji}}}, \bibinfo {author} {\bibfnamefont {K.}~\bibnamefont {{Hayashida}}}, \bibinfo {author}
  {\bibfnamefont {W.}~\bibnamefont {{Iwakiri}}}, \bibinfo {author} {\bibfnamefont {S.~G.}\ \bibnamefont {{Jorstad}}}, \bibinfo {author} {\bibfnamefont {P.}~\bibnamefont {{Kaaret}}}, \bibinfo {author} {\bibfnamefont {V.}~\bibnamefont {{Karas}}}, \bibinfo {author} {\bibfnamefont {F.}~\bibnamefont {{Kislat}}}, \bibinfo {author} {\bibfnamefont {T.}~\bibnamefont {{Kitaguchi}}}, \bibinfo {author} {\bibfnamefont {J.~J.}\ \bibnamefont {{Kolodziejczak}}}, \bibinfo {author} {\bibfnamefont {H.}~\bibnamefont {{Krawczynski}}}, \bibinfo {author} {\bibfnamefont {L.}~\bibnamefont {{Latronico}}}, \bibinfo {author} {\bibfnamefont {I.}~\bibnamefont {{Liodakis}}}, \bibinfo {author} {\bibfnamefont {S.}~\bibnamefont {{Maldera}}}, \bibinfo {author} {\bibfnamefont {A.}~\bibnamefont {{Manfreda}}}, \bibinfo {author} {\bibfnamefont {F.}~\bibnamefont {{Marin}}}, \bibinfo {author} {\bibfnamefont {A.}~\bibnamefont {{Marinucci}}}, \bibinfo {author} {\bibfnamefont {A.~P.}\ \bibnamefont {{Marscher}}}, \bibinfo {author} {\bibfnamefont
  {F.}~\bibnamefont {{Massaro}}}, \bibinfo {author} {\bibfnamefont {G.}~\bibnamefont {{Matt}}}, \bibinfo {author} {\bibfnamefont {I.}~\bibnamefont {{Mitsuishi}}}, \bibinfo {author} {\bibfnamefont {T.}~\bibnamefont {{Mizuno}}}, \bibinfo {author} {\bibfnamefont {F.}~\bibnamefont {{Muleri}}}, \bibinfo {author} {\bibfnamefont {M.}~\bibnamefont {{Negro}}}, \bibinfo {author} {\bibfnamefont {C.-Y.}\ \bibnamefont {{Ng}}}, \bibinfo {author} {\bibfnamefont {S.~L.}\ \bibnamefont {{O'Dell}}}, \bibinfo {author} {\bibfnamefont {N.}~\bibnamefont {{Omodei}}}, \bibinfo {author} {\bibfnamefont {C.}~\bibnamefont {{Oppedisano}}}, \bibinfo {author} {\bibfnamefont {A.}~\bibnamefont {{Papitto}}}, \bibinfo {author} {\bibfnamefont {G.~G.}\ \bibnamefont {{Pavlov}}}, \bibinfo {author} {\bibfnamefont {A.~L.}\ \bibnamefont {{Peirson}}}, \bibinfo {author} {\bibfnamefont {M.}~\bibnamefont {{Perri}}}, \bibinfo {author} {\bibfnamefont {M.}~\bibnamefont {{Pesce-Rollins}}}, \bibinfo {author} {\bibfnamefont {P.-O.}\ \bibnamefont {{Petrucci}}},
  \bibinfo {author} {\bibfnamefont {M.}~\bibnamefont {{Pilia}}}, \bibinfo {author} {\bibfnamefont {A.}~\bibnamefont {{Possenti}}}, \bibinfo {author} {\bibfnamefont {S.}~\bibnamefont {{Puccetti}}}, \bibinfo {author} {\bibfnamefont {B.~D.}\ \bibnamefont {{Ramsey}}}, \bibinfo {author} {\bibfnamefont {J.}~\bibnamefont {{Rankin}}}, \bibinfo {author} {\bibfnamefont {A.}~\bibnamefont {{Ratheesh}}}, \bibinfo {author} {\bibfnamefont {O.~J.}\ \bibnamefont {{Roberts}}}, \bibinfo {author} {\bibfnamefont {R.~W.}\ \bibnamefont {{Romani}}}, \bibinfo {author} {\bibfnamefont {C.}~\bibnamefont {{Sgr{\`o}}}}, \bibinfo {author} {\bibfnamefont {P.}~\bibnamefont {{Slane}}}, \bibinfo {author} {\bibfnamefont {P.}~\bibnamefont {{Soffitta}}}, \bibinfo {author} {\bibfnamefont {G.}~\bibnamefont {{Spandre}}}, \bibinfo {author} {\bibfnamefont {D.~A.}\ \bibnamefont {{Swartz}}}, \bibinfo {author} {\bibfnamefont {T.}~\bibnamefont {{Tamagawa}}}, \bibinfo {author} {\bibfnamefont {F.}~\bibnamefont {{Tavecchio}}}, \bibinfo {author}
  {\bibfnamefont {Y.}~\bibnamefont {{Tawara}}}, \bibinfo {author} {\bibfnamefont {A.~F.}\ \bibnamefont {{Tennant}}}, \bibinfo {author} {\bibfnamefont {N.~E.}\ \bibnamefont {{Thomas}}}, \bibinfo {author} {\bibfnamefont {F.}~\bibnamefont {{Tombesi}}}, \bibinfo {author} {\bibfnamefont {A.}~\bibnamefont {{Trois}}}, \bibinfo {author} {\bibfnamefont {R.}~\bibnamefont {{Turolla}}}, \bibinfo {author} {\bibfnamefont {J.}~\bibnamefont {{Vink}}}, \bibinfo {author} {\bibfnamefont {M.~C.}\ \bibnamefont {{Weisskopf}}}, \bibinfo {author} {\bibfnamefont {K.}~\bibnamefont {{Wu}}}, \bibinfo {author} {\bibfnamefont {F.}~\bibnamefont {{Xie}}}, \ and\ \bibinfo {author} {\bibfnamefont {S.}~\bibnamefont {{Zane}}},\ }\href {\doibase 10.1051/0004-6361/202346134} {\bibfield  {journal} {\bibinfo  {journal} {\aap}\ }\textbf {\bibinfo {volume} {675}},\ \bibinfo {eid} {A48} (\bibinfo {year} {2023})},\ \Eprint {http://arxiv.org/abs/2302.06680} {arXiv:2302.06680 [astro-ph.HE]} \BibitemShut {NoStop}%
\bibitem [{\citenamefont {{Marshall}}\ \emph {et~al.}(2022)\citenamefont {{Marshall}}, \citenamefont {{Ng}}, \citenamefont {{Rogantini}}, \citenamefont {{Heyl}}, \citenamefont {{Tsygankov}}, \citenamefont {{Poutanen}}, \citenamefont {{Costa}}, \citenamefont {{Zane}}, \citenamefont {{Malacaria}}, \citenamefont {{Agudo}}, \citenamefont {{Antonelli}}, \citenamefont {{Bachetti}}, \citenamefont {{Baldini}}, \citenamefont {{Baumgartner}}, \citenamefont {{Bellazzini}}, \citenamefont {{Bianchi}}, \citenamefont {{Bongiorno}}, \citenamefont {{Bonino}}, \citenamefont {{Brez}}, \citenamefont {{Bucciantini}}, \citenamefont {{Capitanio}}, \citenamefont {{Castellano}}, \citenamefont {{Cavazzuti}}, \citenamefont {{Ciprini}}, \citenamefont {{De Rosa}}, \citenamefont {{Del Monte}}, \citenamefont {{Di Gesu}}, \citenamefont {{Di Lalla}}, \citenamefont {{Di Marco}}, \citenamefont {{Donnarumma}}, \citenamefont {{Doroshenko}}, \citenamefont {{Dov{\v{c}}iak}}, \citenamefont {{Ehlert}}, \citenamefont {{Enoto}}, \citenamefont
  {{Evangelista}}, \citenamefont {{Fabiani}}, \citenamefont {{Ferrazzoli}}, \citenamefont {{Garcia}}, \citenamefont {{Gunji}}, \citenamefont {{Hayashida}}, \citenamefont {{Iwakiri}}, \citenamefont {{Jorstad}}, \citenamefont {{Karas}}, \citenamefont {{Kitaguchi}}, \citenamefont {{Kolodziejczak}}, \citenamefont {{Krawczynski}}, \citenamefont {{La Monaca}}, \citenamefont {{Latronico}}, \citenamefont {{Liodakis}}, \citenamefont {{Maldera}}, \citenamefont {{Manfreda}}, \citenamefont {{Marin}}, \citenamefont {{Marinucci}}, \citenamefont {{Marscher}}, \citenamefont {{Matt}}, \citenamefont {{Mitsuishi}}, \citenamefont {{Mizuno}}, \citenamefont {{Muleri}}, \citenamefont {{Ng}}, \citenamefont {{O'Dell}}, \citenamefont {{Omodei}}, \citenamefont {{Oppedisano}}, \citenamefont {{Papitto}}, \citenamefont {{Pavlov}}, \citenamefont {{Peirson}}, \citenamefont {{Perri}}, \citenamefont {{Pesce-Rollins}}, \citenamefont {{Petrucci}}, \citenamefont {{Pilia}}, \citenamefont {{Possenti}}, \citenamefont {{Puccetti}}, \citenamefont
  {{Ramsey}}, \citenamefont {{Rankin}}, \citenamefont {{Ratheesh}}, \citenamefont {{Romani}}, \citenamefont {{Sgr{\`o}}}, \citenamefont {{Slane}}, \citenamefont {{Soffitta}}, \citenamefont {{Spandre}}, \citenamefont {{Tamagawa}}, \citenamefont {{Tavecchio}}, \citenamefont {{Taverna}}, \citenamefont {{Tawara}}, \citenamefont {{Tennant}}, \citenamefont {{Thomas}}, \citenamefont {{Tombesi}}, \citenamefont {{Trois}}, \citenamefont {{Turolla}}, \citenamefont {{Vink}}, \citenamefont {{Weisskopf}}, \citenamefont {{Wu}}, \citenamefont {{Xie}}, \citenamefont {{IXPE Collaboration}}, \citenamefont {{Schulz}},\ and\ \citenamefont {{Chakrabarty}}}]{Marshall_4U1627}%
  \BibitemOpen
  \bibfield  {author} {\bibinfo {author} {\bibfnamefont {H.~L.}\ \bibnamefont {{Marshall}}}, \bibinfo {author} {\bibfnamefont {M.}~\bibnamefont {{Ng}}}, \bibinfo {author} {\bibfnamefont {D.}~\bibnamefont {{Rogantini}}}, \bibinfo {author} {\bibfnamefont {J.}~\bibnamefont {{Heyl}}}, \bibinfo {author} {\bibfnamefont {S.~S.}\ \bibnamefont {{Tsygankov}}}, \bibinfo {author} {\bibfnamefont {J.}~\bibnamefont {{Poutanen}}}, \bibinfo {author} {\bibfnamefont {E.}~\bibnamefont {{Costa}}}, \bibinfo {author} {\bibfnamefont {S.}~\bibnamefont {{Zane}}}, \bibinfo {author} {\bibfnamefont {C.}~\bibnamefont {{Malacaria}}}, \bibinfo {author} {\bibfnamefont {I.}~\bibnamefont {{Agudo}}}, \bibinfo {author} {\bibfnamefont {L.~A.}\ \bibnamefont {{Antonelli}}}, \bibinfo {author} {\bibfnamefont {M.}~\bibnamefont {{Bachetti}}}, \bibinfo {author} {\bibfnamefont {L.}~\bibnamefont {{Baldini}}}, \bibinfo {author} {\bibfnamefont {W.~H.}\ \bibnamefont {{Baumgartner}}}, \bibinfo {author} {\bibfnamefont {R.}~\bibnamefont {{Bellazzini}}},
  \bibinfo {author} {\bibfnamefont {S.}~\bibnamefont {{Bianchi}}}, \bibinfo {author} {\bibfnamefont {S.~D.}\ \bibnamefont {{Bongiorno}}}, \bibinfo {author} {\bibfnamefont {R.}~\bibnamefont {{Bonino}}}, \bibinfo {author} {\bibfnamefont {A.}~\bibnamefont {{Brez}}}, \bibinfo {author} {\bibfnamefont {N.}~\bibnamefont {{Bucciantini}}}, \bibinfo {author} {\bibfnamefont {F.}~\bibnamefont {{Capitanio}}}, \bibinfo {author} {\bibfnamefont {S.}~\bibnamefont {{Castellano}}}, \bibinfo {author} {\bibfnamefont {E.}~\bibnamefont {{Cavazzuti}}}, \bibinfo {author} {\bibfnamefont {S.}~\bibnamefont {{Ciprini}}}, \bibinfo {author} {\bibfnamefont {A.}~\bibnamefont {{De Rosa}}}, \bibinfo {author} {\bibfnamefont {E.}~\bibnamefont {{Del Monte}}}, \bibinfo {author} {\bibfnamefont {L.}~\bibnamefont {{Di Gesu}}}, \bibinfo {author} {\bibfnamefont {N.}~\bibnamefont {{Di Lalla}}}, \bibinfo {author} {\bibfnamefont {A.}~\bibnamefont {{Di Marco}}}, \bibinfo {author} {\bibfnamefont {I.}~\bibnamefont {{Donnarumma}}}, \bibinfo {author}
  {\bibfnamefont {V.}~\bibnamefont {{Doroshenko}}}, \bibinfo {author} {\bibfnamefont {M.}~\bibnamefont {{Dov{\v{c}}iak}}}, \bibinfo {author} {\bibfnamefont {S.~R.}\ \bibnamefont {{Ehlert}}}, \bibinfo {author} {\bibfnamefont {T.}~\bibnamefont {{Enoto}}}, \bibinfo {author} {\bibfnamefont {Y.}~\bibnamefont {{Evangelista}}}, \bibinfo {author} {\bibfnamefont {S.}~\bibnamefont {{Fabiani}}}, \bibinfo {author} {\bibfnamefont {R.}~\bibnamefont {{Ferrazzoli}}}, \bibinfo {author} {\bibfnamefont {J.~A.}\ \bibnamefont {{Garcia}}}, \bibinfo {author} {\bibfnamefont {S.}~\bibnamefont {{Gunji}}}, \bibinfo {author} {\bibfnamefont {K.}~\bibnamefont {{Hayashida}}}, \bibinfo {author} {\bibfnamefont {W.}~\bibnamefont {{Iwakiri}}}, \bibinfo {author} {\bibfnamefont {S.~G.}\ \bibnamefont {{Jorstad}}}, \bibinfo {author} {\bibfnamefont {V.}~\bibnamefont {{Karas}}}, \bibinfo {author} {\bibfnamefont {T.}~\bibnamefont {{Kitaguchi}}}, \bibinfo {author} {\bibfnamefont {J.~J.}\ \bibnamefont {{Kolodziejczak}}}, \bibinfo {author}
  {\bibfnamefont {H.}~\bibnamefont {{Krawczynski}}}, \bibinfo {author} {\bibfnamefont {F.}~\bibnamefont {{La Monaca}}}, \bibinfo {author} {\bibfnamefont {L.}~\bibnamefont {{Latronico}}}, \bibinfo {author} {\bibfnamefont {I.}~\bibnamefont {{Liodakis}}}, \bibinfo {author} {\bibfnamefont {S.}~\bibnamefont {{Maldera}}}, \bibinfo {author} {\bibfnamefont {A.}~\bibnamefont {{Manfreda}}}, \bibinfo {author} {\bibfnamefont {F.}~\bibnamefont {{Marin}}}, \bibinfo {author} {\bibfnamefont {A.}~\bibnamefont {{Marinucci}}}, \bibinfo {author} {\bibfnamefont {A.~P.}\ \bibnamefont {{Marscher}}}, \bibinfo {author} {\bibfnamefont {G.}~\bibnamefont {{Matt}}}, \bibinfo {author} {\bibfnamefont {I.}~\bibnamefont {{Mitsuishi}}}, \bibinfo {author} {\bibfnamefont {T.}~\bibnamefont {{Mizuno}}}, \bibinfo {author} {\bibfnamefont {F.}~\bibnamefont {{Muleri}}}, \bibinfo {author} {\bibfnamefont {C.~Y.}\ \bibnamefont {{Ng}}}, \bibinfo {author} {\bibfnamefont {S.~L.}\ \bibnamefont {{O'Dell}}}, \bibinfo {author} {\bibfnamefont {N.}~\bibnamefont
  {{Omodei}}}, \bibinfo {author} {\bibfnamefont {C.}~\bibnamefont {{Oppedisano}}}, \bibinfo {author} {\bibfnamefont {A.}~\bibnamefont {{Papitto}}}, \bibinfo {author} {\bibfnamefont {G.~G.}\ \bibnamefont {{Pavlov}}}, \bibinfo {author} {\bibfnamefont {A.~L.}\ \bibnamefont {{Peirson}}}, \bibinfo {author} {\bibfnamefont {M.}~\bibnamefont {{Perri}}}, \bibinfo {author} {\bibfnamefont {M.}~\bibnamefont {{Pesce-Rollins}}}, \bibinfo {author} {\bibfnamefont {P.-O.}\ \bibnamefont {{Petrucci}}}, \bibinfo {author} {\bibfnamefont {M.}~\bibnamefont {{Pilia}}}, \bibinfo {author} {\bibfnamefont {A.}~\bibnamefont {{Possenti}}}, \bibinfo {author} {\bibfnamefont {S.}~\bibnamefont {{Puccetti}}}, \bibinfo {author} {\bibfnamefont {B.~D.}\ \bibnamefont {{Ramsey}}}, \bibinfo {author} {\bibfnamefont {J.}~\bibnamefont {{Rankin}}}, \bibinfo {author} {\bibfnamefont {A.}~\bibnamefont {{Ratheesh}}}, \bibinfo {author} {\bibfnamefont {R.~W.}\ \bibnamefont {{Romani}}}, \bibinfo {author} {\bibfnamefont {C.}~\bibnamefont {{Sgr{\`o}}}}, \bibinfo
  {author} {\bibfnamefont {P.}~\bibnamefont {{Slane}}}, \bibinfo {author} {\bibfnamefont {P.}~\bibnamefont {{Soffitta}}}, \bibinfo {author} {\bibfnamefont {G.}~\bibnamefont {{Spandre}}}, \bibinfo {author} {\bibfnamefont {T.}~\bibnamefont {{Tamagawa}}}, \bibinfo {author} {\bibfnamefont {F.}~\bibnamefont {{Tavecchio}}}, \bibinfo {author} {\bibfnamefont {R.}~\bibnamefont {{Taverna}}}, \bibinfo {author} {\bibfnamefont {Y.}~\bibnamefont {{Tawara}}}, \bibinfo {author} {\bibfnamefont {A.~F.}\ \bibnamefont {{Tennant}}}, \bibinfo {author} {\bibfnamefont {N.~E.}\ \bibnamefont {{Thomas}}}, \bibinfo {author} {\bibfnamefont {F.}~\bibnamefont {{Tombesi}}}, \bibinfo {author} {\bibfnamefont {A.}~\bibnamefont {{Trois}}}, \bibinfo {author} {\bibfnamefont {R.}~\bibnamefont {{Turolla}}}, \bibinfo {author} {\bibfnamefont {J.}~\bibnamefont {{Vink}}}, \bibinfo {author} {\bibfnamefont {M.~C.}\ \bibnamefont {{Weisskopf}}}, \bibinfo {author} {\bibfnamefont {K.}~\bibnamefont {{Wu}}}, \bibinfo {author} {\bibfnamefont {F.}~\bibnamefont
  {{Xie}}}, \bibinfo {author} {\bibnamefont {{IXPE Collaboration}}}, \bibinfo {author} {\bibfnamefont {N.~S.}\ \bibnamefont {{Schulz}}}, \ and\ \bibinfo {author} {\bibfnamefont {D.}~\bibnamefont {{Chakrabarty}}},\ }\href {\doibase 10.3847/1538-4357/ac98c2} {\bibfield  {journal} {\bibinfo  {journal} {\apj}\ }\textbf {\bibinfo {volume} {940}},\ \bibinfo {eid} {70} (\bibinfo {year} {2022})},\ \Eprint {http://arxiv.org/abs/2210.03194} {arXiv:2210.03194 [astro-ph.HE]} \BibitemShut {NoStop}%
\bibitem [{\citenamefont {{Mushtukov}}\ \emph {et~al.}(2023)\citenamefont {{Mushtukov}}, \citenamefont {{Tsygankov}}, \citenamefont {{Poutanen}}, \citenamefont {{Doroshenko}}, \citenamefont {{Salganik}}, \citenamefont {{Costa}}, \citenamefont {{Marco}}, \citenamefont {{Heyl}}, \citenamefont {{Monaca}}, \citenamefont {{Lutovinov}}, \citenamefont {{Mereminsky}}, \citenamefont {{Papitto}}, \citenamefont {{Semena}}, \citenamefont {{Shtykovsky}}, \citenamefont {{Suleimanov}}, \citenamefont {{Forsblom}}, \citenamefont {{Gonz{\'a}lez-Caniulef}}, \citenamefont {{Malacaria}}, \citenamefont {{Sunyaev}}, \citenamefont {{Agudo}}, \citenamefont {{Antonelli}}, \citenamefont {{Bachetti}}, \citenamefont {{Baldini}}, \citenamefont {{Baumgartner}}, \citenamefont {{Bellazzini}}, \citenamefont {{Bianchi}}, \citenamefont {{Bongiorno}}, \citenamefont {{Bonino}}, \citenamefont {{Brez}}, \citenamefont {{Bucciantini}}, \citenamefont {{Capitanio}}, \citenamefont {{Castellano}}, \citenamefont {{Cavazzuti}}, \citenamefont {{Chen}},
  \citenamefont {{Ciprini}}, \citenamefont {{De Rosa}}, \citenamefont {{Del Monte}}, \citenamefont {{Gesu}}, \citenamefont {{Lalla}}, \citenamefont {{Donnarumma}}, \citenamefont {{Dov{\v{c}}iak}}, \citenamefont {{Ehlert}}, \citenamefont {{Enoto}}, \citenamefont {{Evangelista}}, \citenamefont {{Fabiani}}, \citenamefont {{Ferrazzoli}}, \citenamefont {{Garcia}}, \citenamefont {{Gunji}}, \citenamefont {{Hayashida}}, \citenamefont {{Iwakiri}}, \citenamefont {{Jorstad}}, \citenamefont {{Kaaret}}, \citenamefont {{Karas}}, \citenamefont {{Kislat}}, \citenamefont {{Kitaguchi}}, \citenamefont {{Kolodziejczak}}, \citenamefont {{Krawczynski}}, \citenamefont {{Latronico}}, \citenamefont {{Liodakis}}, \citenamefont {{Maldera}}, \citenamefont {{Manfreda}}, \citenamefont {{Marin}}, \citenamefont {{Marscher}}, \citenamefont {{Marshall}}, \citenamefont {{Massaro}}, \citenamefont {{Matt}}, \citenamefont {{Mitsuishi}}, \citenamefont {{Mizuno}}, \citenamefont {{Muleri}}, \citenamefont {{Negro}}, \citenamefont {{Ng}},
  \citenamefont {{O'Dell}}, \citenamefont {{Omodei}}, \citenamefont {{Oppedisano}}, \citenamefont {{Pavlov}}, \citenamefont {{Peirson}}, \citenamefont {{Perri}}, \citenamefont {{Pesce-Rollins}}, \citenamefont {{Petrucci}}, \citenamefont {{Pilia}}, \citenamefont {{Possenti}}, \citenamefont {{Puccetti}}, \citenamefont {{Ramsey}}, \citenamefont {{Rankin}}, \citenamefont {{Ratheesh}}, \citenamefont {{Roberts}}, \citenamefont {{Romani}}, \citenamefont {{Sgr{\`o}}}, \citenamefont {{Slane}}, \citenamefont {{Soffitta}}, \citenamefont {{Spandre}}, \citenamefont {{Swartz}}, \citenamefont {{Tamagawa}}, \citenamefont {{Tavecchio}}, \citenamefont {{Taverna}}, \citenamefont {{Tawara}}, \citenamefont {{Tennant}}, \citenamefont {{Thomas}}, \citenamefont {{Tombesi}}, \citenamefont {{Trois}}, \citenamefont {{Turolla}}, \citenamefont {{Vink}}, \citenamefont {{Weisskopf}}, \citenamefont {{Wu}}, \citenamefont {{Xie}},\ and\ \citenamefont {{Zane}}}]{Mushtukov_xpeisei}%
  \BibitemOpen
  \bibfield  {author} {\bibinfo {author} {\bibfnamefont {A.~A.}\ \bibnamefont {{Mushtukov}}}, \bibinfo {author} {\bibfnamefont {S.~S.}\ \bibnamefont {{Tsygankov}}}, \bibinfo {author} {\bibfnamefont {J.}~\bibnamefont {{Poutanen}}}, \bibinfo {author} {\bibfnamefont {V.}~\bibnamefont {{Doroshenko}}}, \bibinfo {author} {\bibfnamefont {A.}~\bibnamefont {{Salganik}}}, \bibinfo {author} {\bibfnamefont {E.}~\bibnamefont {{Costa}}}, \bibinfo {author} {\bibfnamefont {A.~D.}\ \bibnamefont {{Marco}}}, \bibinfo {author} {\bibfnamefont {J.}~\bibnamefont {{Heyl}}}, \bibinfo {author} {\bibfnamefont {F.~L.}\ \bibnamefont {{Monaca}}}, \bibinfo {author} {\bibfnamefont {A.~A.}\ \bibnamefont {{Lutovinov}}}, \bibinfo {author} {\bibfnamefont {I.~A.}\ \bibnamefont {{Mereminsky}}}, \bibinfo {author} {\bibfnamefont {A.}~\bibnamefont {{Papitto}}}, \bibinfo {author} {\bibfnamefont {A.~N.}\ \bibnamefont {{Semena}}}, \bibinfo {author} {\bibfnamefont {A.~E.}\ \bibnamefont {{Shtykovsky}}}, \bibinfo {author} {\bibfnamefont {V.~F.}\
  \bibnamefont {{Suleimanov}}}, \bibinfo {author} {\bibfnamefont {S.~V.}\ \bibnamefont {{Forsblom}}}, \bibinfo {author} {\bibfnamefont {D.}~\bibnamefont {{Gonz{\'a}lez-Caniulef}}}, \bibinfo {author} {\bibfnamefont {C.}~\bibnamefont {{Malacaria}}}, \bibinfo {author} {\bibfnamefont {R.~A.}\ \bibnamefont {{Sunyaev}}}, \bibinfo {author} {\bibfnamefont {I.}~\bibnamefont {{Agudo}}}, \bibinfo {author} {\bibfnamefont {L.~A.}\ \bibnamefont {{Antonelli}}}, \bibinfo {author} {\bibfnamefont {M.}~\bibnamefont {{Bachetti}}}, \bibinfo {author} {\bibfnamefont {L.}~\bibnamefont {{Baldini}}}, \bibinfo {author} {\bibfnamefont {W.~H.}\ \bibnamefont {{Baumgartner}}}, \bibinfo {author} {\bibfnamefont {R.}~\bibnamefont {{Bellazzini}}}, \bibinfo {author} {\bibfnamefont {S.}~\bibnamefont {{Bianchi}}}, \bibinfo {author} {\bibfnamefont {S.~D.}\ \bibnamefont {{Bongiorno}}}, \bibinfo {author} {\bibfnamefont {R.}~\bibnamefont {{Bonino}}}, \bibinfo {author} {\bibfnamefont {A.}~\bibnamefont {{Brez}}}, \bibinfo {author} {\bibfnamefont
  {N.}~\bibnamefont {{Bucciantini}}}, \bibinfo {author} {\bibfnamefont {F.}~\bibnamefont {{Capitanio}}}, \bibinfo {author} {\bibfnamefont {S.}~\bibnamefont {{Castellano}}}, \bibinfo {author} {\bibfnamefont {E.}~\bibnamefont {{Cavazzuti}}}, \bibinfo {author} {\bibfnamefont {C.~T.}\ \bibnamefont {{Chen}}}, \bibinfo {author} {\bibfnamefont {S.}~\bibnamefont {{Ciprini}}}, \bibinfo {author} {\bibfnamefont {A.}~\bibnamefont {{De Rosa}}}, \bibinfo {author} {\bibfnamefont {E.}~\bibnamefont {{Del Monte}}}, \bibinfo {author} {\bibfnamefont {L.~D.}\ \bibnamefont {{Gesu}}}, \bibinfo {author} {\bibfnamefont {N.~D.}\ \bibnamefont {{Lalla}}}, \bibinfo {author} {\bibfnamefont {I.}~\bibnamefont {{Donnarumma}}}, \bibinfo {author} {\bibfnamefont {M.}~\bibnamefont {{Dov{\v{c}}iak}}}, \bibinfo {author} {\bibfnamefont {S.~R.}\ \bibnamefont {{Ehlert}}}, \bibinfo {author} {\bibfnamefont {T.}~\bibnamefont {{Enoto}}}, \bibinfo {author} {\bibfnamefont {Y.}~\bibnamefont {{Evangelista}}}, \bibinfo {author} {\bibfnamefont
  {S.}~\bibnamefont {{Fabiani}}}, \bibinfo {author} {\bibfnamefont {R.}~\bibnamefont {{Ferrazzoli}}}, \bibinfo {author} {\bibfnamefont {J.~A.}\ \bibnamefont {{Garcia}}}, \bibinfo {author} {\bibfnamefont {S.}~\bibnamefont {{Gunji}}}, \bibinfo {author} {\bibfnamefont {K.}~\bibnamefont {{Hayashida}}}, \bibinfo {author} {\bibfnamefont {W.}~\bibnamefont {{Iwakiri}}}, \bibinfo {author} {\bibfnamefont {S.~G.}\ \bibnamefont {{Jorstad}}}, \bibinfo {author} {\bibfnamefont {P.}~\bibnamefont {{Kaaret}}}, \bibinfo {author} {\bibfnamefont {V.}~\bibnamefont {{Karas}}}, \bibinfo {author} {\bibfnamefont {F.}~\bibnamefont {{Kislat}}}, \bibinfo {author} {\bibfnamefont {T.}~\bibnamefont {{Kitaguchi}}}, \bibinfo {author} {\bibfnamefont {J.~J.}\ \bibnamefont {{Kolodziejczak}}}, \bibinfo {author} {\bibfnamefont {H.}~\bibnamefont {{Krawczynski}}}, \bibinfo {author} {\bibfnamefont {L.}~\bibnamefont {{Latronico}}}, \bibinfo {author} {\bibfnamefont {I.}~\bibnamefont {{Liodakis}}}, \bibinfo {author} {\bibfnamefont {S.}~\bibnamefont
  {{Maldera}}}, \bibinfo {author} {\bibfnamefont {A.}~\bibnamefont {{Manfreda}}}, \bibinfo {author} {\bibfnamefont {F.}~\bibnamefont {{Marin}}}, \bibinfo {author} {\bibfnamefont {A.~P.}\ \bibnamefont {{Marscher}}}, \bibinfo {author} {\bibfnamefont {H.~L.}\ \bibnamefont {{Marshall}}}, \bibinfo {author} {\bibfnamefont {F.}~\bibnamefont {{Massaro}}}, \bibinfo {author} {\bibfnamefont {G.}~\bibnamefont {{Matt}}}, \bibinfo {author} {\bibfnamefont {I.}~\bibnamefont {{Mitsuishi}}}, \bibinfo {author} {\bibfnamefont {T.}~\bibnamefont {{Mizuno}}}, \bibinfo {author} {\bibfnamefont {F.}~\bibnamefont {{Muleri}}}, \bibinfo {author} {\bibfnamefont {M.}~\bibnamefont {{Negro}}}, \bibinfo {author} {\bibfnamefont {C.~Y.}\ \bibnamefont {{Ng}}}, \bibinfo {author} {\bibfnamefont {S.~L.}\ \bibnamefont {{O'Dell}}}, \bibinfo {author} {\bibfnamefont {N.}~\bibnamefont {{Omodei}}}, \bibinfo {author} {\bibfnamefont {C.}~\bibnamefont {{Oppedisano}}}, \bibinfo {author} {\bibfnamefont {G.~G.}\ \bibnamefont {{Pavlov}}}, \bibinfo {author}
  {\bibfnamefont {A.~L.}\ \bibnamefont {{Peirson}}}, \bibinfo {author} {\bibfnamefont {M.}~\bibnamefont {{Perri}}}, \bibinfo {author} {\bibfnamefont {M.}~\bibnamefont {{Pesce-Rollins}}}, \bibinfo {author} {\bibfnamefont {P.~O.}\ \bibnamefont {{Petrucci}}}, \bibinfo {author} {\bibfnamefont {M.}~\bibnamefont {{Pilia}}}, \bibinfo {author} {\bibfnamefont {A.}~\bibnamefont {{Possenti}}}, \bibinfo {author} {\bibfnamefont {S.}~\bibnamefont {{Puccetti}}}, \bibinfo {author} {\bibfnamefont {B.~D.}\ \bibnamefont {{Ramsey}}}, \bibinfo {author} {\bibfnamefont {J.}~\bibnamefont {{Rankin}}}, \bibinfo {author} {\bibfnamefont {A.}~\bibnamefont {{Ratheesh}}}, \bibinfo {author} {\bibfnamefont {O.~J.}\ \bibnamefont {{Roberts}}}, \bibinfo {author} {\bibfnamefont {R.~W.}\ \bibnamefont {{Romani}}}, \bibinfo {author} {\bibfnamefont {C.}~\bibnamefont {{Sgr{\`o}}}}, \bibinfo {author} {\bibfnamefont {P.}~\bibnamefont {{Slane}}}, \bibinfo {author} {\bibfnamefont {P.}~\bibnamefont {{Soffitta}}}, \bibinfo {author} {\bibfnamefont
  {G.}~\bibnamefont {{Spandre}}}, \bibinfo {author} {\bibfnamefont {D.~A.}\ \bibnamefont {{Swartz}}}, \bibinfo {author} {\bibfnamefont {T.}~\bibnamefont {{Tamagawa}}}, \bibinfo {author} {\bibfnamefont {F.}~\bibnamefont {{Tavecchio}}}, \bibinfo {author} {\bibfnamefont {R.}~\bibnamefont {{Taverna}}}, \bibinfo {author} {\bibfnamefont {Y.}~\bibnamefont {{Tawara}}}, \bibinfo {author} {\bibfnamefont {A.~F.}\ \bibnamefont {{Tennant}}}, \bibinfo {author} {\bibfnamefont {N.~E.}\ \bibnamefont {{Thomas}}}, \bibinfo {author} {\bibfnamefont {F.}~\bibnamefont {{Tombesi}}}, \bibinfo {author} {\bibfnamefont {A.}~\bibnamefont {{Trois}}}, \bibinfo {author} {\bibfnamefont {R.}~\bibnamefont {{Turolla}}}, \bibinfo {author} {\bibfnamefont {J.}~\bibnamefont {{Vink}}}, \bibinfo {author} {\bibfnamefont {M.~C.}\ \bibnamefont {{Weisskopf}}}, \bibinfo {author} {\bibfnamefont {K.}~\bibnamefont {{Wu}}}, \bibinfo {author} {\bibfnamefont {F.}~\bibnamefont {{Xie}}}, \ and\ \bibinfo {author} {\bibfnamefont {S.}~\bibnamefont {{Zane}}},\ }\href
  {\doibase 10.1093/mnras/stad1961} {\bibfield  {journal} {\bibinfo  {journal} {\mnras}\ }\textbf {\bibinfo {volume} {524}},\ \bibinfo {pages} {2004} (\bibinfo {year} {2023})},\ \Eprint {http://arxiv.org/abs/2303.17325} {arXiv:2303.17325 [astro-ph.HE]} \BibitemShut {NoStop}%
\bibitem [{\citenamefont {{Forsblom}}\ \emph {et~al.}(2023)\citenamefont {{Forsblom}}, \citenamefont {{Poutanen}}, \citenamefont {{Tsygankov}}, \citenamefont {{Bachetti}}, \citenamefont {{Di Marco}}, \citenamefont {{Doroshenko}}, \citenamefont {{Heyl}}, \citenamefont {{La Monaca}}, \citenamefont {{Malacaria}}, \citenamefont {{Marshall}}, \citenamefont {{Muleri}}, \citenamefont {{Mushtukov}}, \citenamefont {{Pilia}}, \citenamefont {{Rogantini}}, \citenamefont {{Suleimanov}}, \citenamefont {{Taverna}}, \citenamefont {{Xie}}, \citenamefont {{Agudo}}, \citenamefont {{Antonelli}}, \citenamefont {{Baldini}}, \citenamefont {{Baumgartner}}, \citenamefont {{Bellazzini}}, \citenamefont {{Bianchi}}, \citenamefont {{Bongiorno}}, \citenamefont {{Bonino}}, \citenamefont {{Brez}}, \citenamefont {{Bucciantini}}, \citenamefont {{Capitanio}}, \citenamefont {{Castellano}}, \citenamefont {{Cavazzuti}}, \citenamefont {{Chen}}, \citenamefont {{Ciprini}}, \citenamefont {{Costa}}, \citenamefont {{De Rosa}}, \citenamefont {{Del
  Monte}}, \citenamefont {{Di Gesu}}, \citenamefont {{Di Lalla}}, \citenamefont {{Donnarumma}}, \citenamefont {{Dov{\v{c}}iak}}, \citenamefont {{Ehlert}}, \citenamefont {{Enoto}}, \citenamefont {{Evangelista}}, \citenamefont {{Fabiani}}, \citenamefont {{Ferrazzoli}}, \citenamefont {{Garcia}}, \citenamefont {{Gunji}}, \citenamefont {{Hayashida}}, \citenamefont {{Iwakiri}}, \citenamefont {{Jorstad}}, \citenamefont {{Kaaret}}, \citenamefont {{Karas}}, \citenamefont {{Kitaguchi}}, \citenamefont {{Kolodziejczak}}, \citenamefont {{Krawczynski}}, \citenamefont {{Latronico}}, \citenamefont {{Liodakis}}, \citenamefont {{Maldera}}, \citenamefont {{Manfreda}}, \citenamefont {{Marin}}, \citenamefont {{Marinucci}}, \citenamefont {{Marscher}}, \citenamefont {{Matt}}, \citenamefont {{Mitsuishi}}, \citenamefont {{Mizuno}}, \citenamefont {{Negro}}, \citenamefont {{Ng}}, \citenamefont {{O'Dell}}, \citenamefont {{Omodei}}, \citenamefont {{Oppedisano}}, \citenamefont {{Papitto}}, \citenamefont {{Pavlov}}, \citenamefont
  {{Peirson}}, \citenamefont {{Perri}}, \citenamefont {{Pesce-Rollins}}, \citenamefont {{Petrucci}}, \citenamefont {{Possenti}}, \citenamefont {{Puccetti}}, \citenamefont {{Ramsey}}, \citenamefont {{Rankin}}, \citenamefont {{Ratheesh}}, \citenamefont {{Roberts}}, \citenamefont {{Romani}}, \citenamefont {{Sgr{\`o}}}, \citenamefont {{Slane}}, \citenamefont {{Soffitta}}, \citenamefont {{Spandre}}, \citenamefont {{Sunyaev}}, \citenamefont {{Swartz}}, \citenamefont {{Tamagawa}}, \citenamefont {{Tavecchio}}, \citenamefont {{Tawara}}, \citenamefont {{Tennant}}, \citenamefont {{Thomas}}, \citenamefont {{Tombesi}}, \citenamefont {{Trois}}, \citenamefont {{Turolla}}, \citenamefont {{Vink}}, \citenamefont {{Weisskopf}}, \citenamefont {{Wu}}, \citenamefont {{Zane}},\ and\ \citenamefont {{IXPE Collaboration}}}]{Forsblom_velaX-1}%
  \BibitemOpen
  \bibfield  {author} {\bibinfo {author} {\bibfnamefont {S.~V.}\ \bibnamefont {{Forsblom}}}, \bibinfo {author} {\bibfnamefont {J.}~\bibnamefont {{Poutanen}}}, \bibinfo {author} {\bibfnamefont {S.~S.}\ \bibnamefont {{Tsygankov}}}, \bibinfo {author} {\bibfnamefont {M.}~\bibnamefont {{Bachetti}}}, \bibinfo {author} {\bibfnamefont {A.}~\bibnamefont {{Di Marco}}}, \bibinfo {author} {\bibfnamefont {V.}~\bibnamefont {{Doroshenko}}}, \bibinfo {author} {\bibfnamefont {J.}~\bibnamefont {{Heyl}}}, \bibinfo {author} {\bibfnamefont {F.}~\bibnamefont {{La Monaca}}}, \bibinfo {author} {\bibfnamefont {C.}~\bibnamefont {{Malacaria}}}, \bibinfo {author} {\bibfnamefont {H.~L.}\ \bibnamefont {{Marshall}}}, \bibinfo {author} {\bibfnamefont {F.}~\bibnamefont {{Muleri}}}, \bibinfo {author} {\bibfnamefont {A.~A.}\ \bibnamefont {{Mushtukov}}}, \bibinfo {author} {\bibfnamefont {M.}~\bibnamefont {{Pilia}}}, \bibinfo {author} {\bibfnamefont {D.}~\bibnamefont {{Rogantini}}}, \bibinfo {author} {\bibfnamefont {V.~F.}\ \bibnamefont
  {{Suleimanov}}}, \bibinfo {author} {\bibfnamefont {R.}~\bibnamefont {{Taverna}}}, \bibinfo {author} {\bibfnamefont {F.}~\bibnamefont {{Xie}}}, \bibinfo {author} {\bibfnamefont {I.}~\bibnamefont {{Agudo}}}, \bibinfo {author} {\bibfnamefont {L.~A.}\ \bibnamefont {{Antonelli}}}, \bibinfo {author} {\bibfnamefont {L.}~\bibnamefont {{Baldini}}}, \bibinfo {author} {\bibfnamefont {W.~H.}\ \bibnamefont {{Baumgartner}}}, \bibinfo {author} {\bibfnamefont {R.}~\bibnamefont {{Bellazzini}}}, \bibinfo {author} {\bibfnamefont {S.}~\bibnamefont {{Bianchi}}}, \bibinfo {author} {\bibfnamefont {S.~D.}\ \bibnamefont {{Bongiorno}}}, \bibinfo {author} {\bibfnamefont {R.}~\bibnamefont {{Bonino}}}, \bibinfo {author} {\bibfnamefont {A.}~\bibnamefont {{Brez}}}, \bibinfo {author} {\bibfnamefont {N.}~\bibnamefont {{Bucciantini}}}, \bibinfo {author} {\bibfnamefont {F.}~\bibnamefont {{Capitanio}}}, \bibinfo {author} {\bibfnamefont {S.}~\bibnamefont {{Castellano}}}, \bibinfo {author} {\bibfnamefont {E.}~\bibnamefont {{Cavazzuti}}},
  \bibinfo {author} {\bibfnamefont {C.-T.}\ \bibnamefont {{Chen}}}, \bibinfo {author} {\bibfnamefont {S.}~\bibnamefont {{Ciprini}}}, \bibinfo {author} {\bibfnamefont {E.}~\bibnamefont {{Costa}}}, \bibinfo {author} {\bibfnamefont {A.}~\bibnamefont {{De Rosa}}}, \bibinfo {author} {\bibfnamefont {E.}~\bibnamefont {{Del Monte}}}, \bibinfo {author} {\bibfnamefont {L.}~\bibnamefont {{Di Gesu}}}, \bibinfo {author} {\bibfnamefont {N.}~\bibnamefont {{Di Lalla}}}, \bibinfo {author} {\bibfnamefont {I.}~\bibnamefont {{Donnarumma}}}, \bibinfo {author} {\bibfnamefont {M.}~\bibnamefont {{Dov{\v{c}}iak}}}, \bibinfo {author} {\bibfnamefont {S.~R.}\ \bibnamefont {{Ehlert}}}, \bibinfo {author} {\bibfnamefont {T.}~\bibnamefont {{Enoto}}}, \bibinfo {author} {\bibfnamefont {Y.}~\bibnamefont {{Evangelista}}}, \bibinfo {author} {\bibfnamefont {S.}~\bibnamefont {{Fabiani}}}, \bibinfo {author} {\bibfnamefont {R.}~\bibnamefont {{Ferrazzoli}}}, \bibinfo {author} {\bibfnamefont {J.~A.}\ \bibnamefont {{Garcia}}}, \bibinfo {author}
  {\bibfnamefont {S.}~\bibnamefont {{Gunji}}}, \bibinfo {author} {\bibfnamefont {K.}~\bibnamefont {{Hayashida}}}, \bibinfo {author} {\bibfnamefont {W.}~\bibnamefont {{Iwakiri}}}, \bibinfo {author} {\bibfnamefont {S.~G.}\ \bibnamefont {{Jorstad}}}, \bibinfo {author} {\bibfnamefont {P.}~\bibnamefont {{Kaaret}}}, \bibinfo {author} {\bibfnamefont {V.}~\bibnamefont {{Karas}}}, \bibinfo {author} {\bibfnamefont {T.}~\bibnamefont {{Kitaguchi}}}, \bibinfo {author} {\bibfnamefont {J.~J.}\ \bibnamefont {{Kolodziejczak}}}, \bibinfo {author} {\bibfnamefont {H.}~\bibnamefont {{Krawczynski}}}, \bibinfo {author} {\bibfnamefont {L.}~\bibnamefont {{Latronico}}}, \bibinfo {author} {\bibfnamefont {I.}~\bibnamefont {{Liodakis}}}, \bibinfo {author} {\bibfnamefont {S.}~\bibnamefont {{Maldera}}}, \bibinfo {author} {\bibfnamefont {A.}~\bibnamefont {{Manfreda}}}, \bibinfo {author} {\bibfnamefont {F.}~\bibnamefont {{Marin}}}, \bibinfo {author} {\bibfnamefont {A.}~\bibnamefont {{Marinucci}}}, \bibinfo {author} {\bibfnamefont {A.~P.}\
  \bibnamefont {{Marscher}}}, \bibinfo {author} {\bibfnamefont {G.}~\bibnamefont {{Matt}}}, \bibinfo {author} {\bibfnamefont {I.}~\bibnamefont {{Mitsuishi}}}, \bibinfo {author} {\bibfnamefont {T.}~\bibnamefont {{Mizuno}}}, \bibinfo {author} {\bibfnamefont {M.}~\bibnamefont {{Negro}}}, \bibinfo {author} {\bibfnamefont {C.-Y.}\ \bibnamefont {{Ng}}}, \bibinfo {author} {\bibfnamefont {S.~L.}\ \bibnamefont {{O'Dell}}}, \bibinfo {author} {\bibfnamefont {N.}~\bibnamefont {{Omodei}}}, \bibinfo {author} {\bibfnamefont {C.}~\bibnamefont {{Oppedisano}}}, \bibinfo {author} {\bibfnamefont {A.}~\bibnamefont {{Papitto}}}, \bibinfo {author} {\bibfnamefont {G.~G.}\ \bibnamefont {{Pavlov}}}, \bibinfo {author} {\bibfnamefont {A.~L.}\ \bibnamefont {{Peirson}}}, \bibinfo {author} {\bibfnamefont {M.}~\bibnamefont {{Perri}}}, \bibinfo {author} {\bibfnamefont {M.}~\bibnamefont {{Pesce-Rollins}}}, \bibinfo {author} {\bibfnamefont {P.-O.}\ \bibnamefont {{Petrucci}}}, \bibinfo {author} {\bibfnamefont {A.}~\bibnamefont {{Possenti}}},
  \bibinfo {author} {\bibfnamefont {S.}~\bibnamefont {{Puccetti}}}, \bibinfo {author} {\bibfnamefont {B.~D.}\ \bibnamefont {{Ramsey}}}, \bibinfo {author} {\bibfnamefont {J.}~\bibnamefont {{Rankin}}}, \bibinfo {author} {\bibfnamefont {A.}~\bibnamefont {{Ratheesh}}}, \bibinfo {author} {\bibfnamefont {O.~J.}\ \bibnamefont {{Roberts}}}, \bibinfo {author} {\bibfnamefont {R.~W.}\ \bibnamefont {{Romani}}}, \bibinfo {author} {\bibfnamefont {C.}~\bibnamefont {{Sgr{\`o}}}}, \bibinfo {author} {\bibfnamefont {P.}~\bibnamefont {{Slane}}}, \bibinfo {author} {\bibfnamefont {P.}~\bibnamefont {{Soffitta}}}, \bibinfo {author} {\bibfnamefont {G.}~\bibnamefont {{Spandre}}}, \bibinfo {author} {\bibfnamefont {R.~A.}\ \bibnamefont {{Sunyaev}}}, \bibinfo {author} {\bibfnamefont {D.~A.}\ \bibnamefont {{Swartz}}}, \bibinfo {author} {\bibfnamefont {T.}~\bibnamefont {{Tamagawa}}}, \bibinfo {author} {\bibfnamefont {F.}~\bibnamefont {{Tavecchio}}}, \bibinfo {author} {\bibfnamefont {Y.}~\bibnamefont {{Tawara}}}, \bibinfo {author}
  {\bibfnamefont {A.~F.}\ \bibnamefont {{Tennant}}}, \bibinfo {author} {\bibfnamefont {N.~E.}\ \bibnamefont {{Thomas}}}, \bibinfo {author} {\bibfnamefont {F.}~\bibnamefont {{Tombesi}}}, \bibinfo {author} {\bibfnamefont {A.}~\bibnamefont {{Trois}}}, \bibinfo {author} {\bibfnamefont {R.}~\bibnamefont {{Turolla}}}, \bibinfo {author} {\bibfnamefont {J.}~\bibnamefont {{Vink}}}, \bibinfo {author} {\bibfnamefont {M.~C.}\ \bibnamefont {{Weisskopf}}}, \bibinfo {author} {\bibfnamefont {K.}~\bibnamefont {{Wu}}}, \bibinfo {author} {\bibfnamefont {S.}~\bibnamefont {{Zane}}}, \ and\ \bibinfo {author} {\bibnamefont {{IXPE Collaboration}}},\ }\href {\doibase 10.3847/2041-8213/acc391} {\bibfield  {journal} {\bibinfo  {journal} {\apjl}\ }\textbf {\bibinfo {volume} {947}},\ \bibinfo {eid} {L20} (\bibinfo {year} {2023})},\ \Eprint {http://arxiv.org/abs/2303.01800} {arXiv:2303.01800 [astro-ph.HE]} \BibitemShut {NoStop}%
\bibitem [{\citenamefont {{Forsblom}}\ \emph {et~al.}(2025)\citenamefont {{Forsblom}}, \citenamefont {{Tsygankov}}, \citenamefont {{Suleimanov}}, \citenamefont {{Mushtukov}},\ and\ \citenamefont {{Poutanen}}}]{Forsblom2025}%
  \BibitemOpen
  \bibfield  {author} {\bibinfo {author} {\bibfnamefont {S.~V.}\ \bibnamefont {{Forsblom}}}, \bibinfo {author} {\bibfnamefont {S.~S.}\ \bibnamefont {{Tsygankov}}}, \bibinfo {author} {\bibfnamefont {V.~F.}\ \bibnamefont {{Suleimanov}}}, \bibinfo {author} {\bibfnamefont {A.~A.}\ \bibnamefont {{Mushtukov}}}, \ and\ \bibinfo {author} {\bibfnamefont {J.}~\bibnamefont {{Poutanen}}},\ }\href {\doibase 10.48550/arXiv.2501.14324} {\bibfield  {journal} {\bibinfo  {journal} {\aap, in press}\ ,\ \bibinfo {eid} {arXiv:2501.14324}} (\bibinfo {year} {2025})},\ \Eprint {http://arxiv.org/abs/2501.14324} {arXiv:2501.14324 [astro-ph.HE]} \BibitemShut {NoStop}%
\bibitem [{\citenamefont {{Malacaria}}\ \emph {et~al.}(2023)\citenamefont {{Malacaria}}, \citenamefont {{Heyl}}, \citenamefont {{Doroshenko}}, \citenamefont {{Tsygankov}}, \citenamefont {{Poutanen}}, \citenamefont {{Forsblom}}, \citenamefont {{Capitanio}}, \citenamefont {{Di Marco}}, \citenamefont {{Du}}, \citenamefont {{Ducci}}, \citenamefont {{La Monaca}}, \citenamefont {{Lutovinov}}, \citenamefont {{Marshall}}, \citenamefont {{Mereminskiy}}, \citenamefont {{Molkov}}, \citenamefont {{Mushtukov}}, \citenamefont {{Ng}}, \citenamefont {{Petrucci}}, \citenamefont {{Santangelo}}, \citenamefont {{Shtykovsky}}, \citenamefont {{Suleimanov}}, \citenamefont {{Agudo}}, \citenamefont {{Antonelli}}, \citenamefont {{Bachetti}}, \citenamefont {{Baldini}}, \citenamefont {{Baumgartner}}, \citenamefont {{Bellazzini}}, \citenamefont {{Bianchi}}, \citenamefont {{Bongiorno}}, \citenamefont {{Bonino}}, \citenamefont {{Brez}}, \citenamefont {{Bucciantini}}, \citenamefont {{Castellano}}, \citenamefont {{Cavazzuti}}, \citenamefont
  {{Chen}}, \citenamefont {{Ciprini}}, \citenamefont {{Costa}}, \citenamefont {{De Rosa}}, \citenamefont {{Del Monte}}, \citenamefont {{Di Gesu}}, \citenamefont {{Di Lalla}}, \citenamefont {{Donnarumma}}, \citenamefont {{Dov{\v{c}}iak}}, \citenamefont {{Ehlert}}, \citenamefont {{Enoto}}, \citenamefont {{Evangelista}}, \citenamefont {{Fabiani}}, \citenamefont {{Ferrazzoli}}, \citenamefont {{Garcia}}, \citenamefont {{Gunji}}, \citenamefont {{Hayashida}}, \citenamefont {{Iwakiri}}, \citenamefont {{Jorstad}}, \citenamefont {{Kaaret}}, \citenamefont {{Karas}}, \citenamefont {{Kislat}}, \citenamefont {{Kitaguchi}}, \citenamefont {{Kolodziejczak}}, \citenamefont {{Krawczynski}}, \citenamefont {{Latronico}}, \citenamefont {{Liodakis}}, \citenamefont {{Maldera}}, \citenamefont {{Manfreda}}, \citenamefont {{Marin}}, \citenamefont {{Marinucci}}, \citenamefont {{Marscher}}, \citenamefont {{Massaro}}, \citenamefont {{Matt}}, \citenamefont {{Mitsuishi}}, \citenamefont {{Mizuno}}, \citenamefont {{Muleri}}, \citenamefont
  {{Negro}}, \citenamefont {{Ng}}, \citenamefont {{O'Dell}}, \citenamefont {{Omodei}}, \citenamefont {{Oppedisano}}, \citenamefont {{Papitto}}, \citenamefont {{Pavlov}}, \citenamefont {{Peirson}}, \citenamefont {{Perri}}, \citenamefont {{Pesce-Rollins}}, \citenamefont {{Pilia}}, \citenamefont {{Possenti}}, \citenamefont {{Puccetti}}, \citenamefont {{Ramsey}}, \citenamefont {{Rankin}}, \citenamefont {{Ratheesh}}, \citenamefont {{Roberts}}, \citenamefont {{Romani}}, \citenamefont {{Sgr{\`o}}}, \citenamefont {{Slane}}, \citenamefont {{Soffitta}}, \citenamefont {{Spandre}}, \citenamefont {{Swartz}}, \citenamefont {{Tamagawa}}, \citenamefont {{Tavecchio}}, \citenamefont {{Taverna}}, \citenamefont {{Tawara}}, \citenamefont {{Tennant}}, \citenamefont {{Thomas}}, \citenamefont {{Tombesi}}, \citenamefont {{Trois}}, \citenamefont {{Turolla}}, \citenamefont {{Vink}}, \citenamefont {{Weisskopf}}, \citenamefont {{Wu}}, \citenamefont {{Xie}},\ and\ \citenamefont {{Zane}}}]{Malacaria_exo2030}%
  \BibitemOpen
  \bibfield  {author} {\bibinfo {author} {\bibfnamefont {C.}~\bibnamefont {{Malacaria}}}, \bibinfo {author} {\bibfnamefont {J.}~\bibnamefont {{Heyl}}}, \bibinfo {author} {\bibfnamefont {V.}~\bibnamefont {{Doroshenko}}}, \bibinfo {author} {\bibfnamefont {S.~S.}\ \bibnamefont {{Tsygankov}}}, \bibinfo {author} {\bibfnamefont {J.}~\bibnamefont {{Poutanen}}}, \bibinfo {author} {\bibfnamefont {S.~V.}\ \bibnamefont {{Forsblom}}}, \bibinfo {author} {\bibfnamefont {F.}~\bibnamefont {{Capitanio}}}, \bibinfo {author} {\bibfnamefont {A.}~\bibnamefont {{Di Marco}}}, \bibinfo {author} {\bibfnamefont {Y.}~\bibnamefont {{Du}}}, \bibinfo {author} {\bibfnamefont {L.}~\bibnamefont {{Ducci}}}, \bibinfo {author} {\bibfnamefont {F.}~\bibnamefont {{La Monaca}}}, \bibinfo {author} {\bibfnamefont {A.~A.}\ \bibnamefont {{Lutovinov}}}, \bibinfo {author} {\bibfnamefont {H.~L.}\ \bibnamefont {{Marshall}}}, \bibinfo {author} {\bibfnamefont {I.~A.}\ \bibnamefont {{Mereminskiy}}}, \bibinfo {author} {\bibfnamefont {S.~V.}\ \bibnamefont
  {{Molkov}}}, \bibinfo {author} {\bibfnamefont {A.~A.}\ \bibnamefont {{Mushtukov}}}, \bibinfo {author} {\bibfnamefont {M.}~\bibnamefont {{Ng}}}, \bibinfo {author} {\bibfnamefont {P.-O.}\ \bibnamefont {{Petrucci}}}, \bibinfo {author} {\bibfnamefont {A.}~\bibnamefont {{Santangelo}}}, \bibinfo {author} {\bibfnamefont {A.~E.}\ \bibnamefont {{Shtykovsky}}}, \bibinfo {author} {\bibfnamefont {V.~F.}\ \bibnamefont {{Suleimanov}}}, \bibinfo {author} {\bibfnamefont {I.}~\bibnamefont {{Agudo}}}, \bibinfo {author} {\bibfnamefont {L.~A.}\ \bibnamefont {{Antonelli}}}, \bibinfo {author} {\bibfnamefont {M.}~\bibnamefont {{Bachetti}}}, \bibinfo {author} {\bibfnamefont {L.}~\bibnamefont {{Baldini}}}, \bibinfo {author} {\bibfnamefont {W.~H.}\ \bibnamefont {{Baumgartner}}}, \bibinfo {author} {\bibfnamefont {R.}~\bibnamefont {{Bellazzini}}}, \bibinfo {author} {\bibfnamefont {S.}~\bibnamefont {{Bianchi}}}, \bibinfo {author} {\bibfnamefont {S.~D.}\ \bibnamefont {{Bongiorno}}}, \bibinfo {author} {\bibfnamefont {R.}~\bibnamefont
  {{Bonino}}}, \bibinfo {author} {\bibfnamefont {A.}~\bibnamefont {{Brez}}}, \bibinfo {author} {\bibfnamefont {N.}~\bibnamefont {{Bucciantini}}}, \bibinfo {author} {\bibfnamefont {S.}~\bibnamefont {{Castellano}}}, \bibinfo {author} {\bibfnamefont {E.}~\bibnamefont {{Cavazzuti}}}, \bibinfo {author} {\bibfnamefont {C.-T.}\ \bibnamefont {{Chen}}}, \bibinfo {author} {\bibfnamefont {S.}~\bibnamefont {{Ciprini}}}, \bibinfo {author} {\bibfnamefont {E.}~\bibnamefont {{Costa}}}, \bibinfo {author} {\bibfnamefont {A.}~\bibnamefont {{De Rosa}}}, \bibinfo {author} {\bibfnamefont {E.}~\bibnamefont {{Del Monte}}}, \bibinfo {author} {\bibfnamefont {L.}~\bibnamefont {{Di Gesu}}}, \bibinfo {author} {\bibfnamefont {N.}~\bibnamefont {{Di Lalla}}}, \bibinfo {author} {\bibfnamefont {I.}~\bibnamefont {{Donnarumma}}}, \bibinfo {author} {\bibfnamefont {M.}~\bibnamefont {{Dov{\v{c}}iak}}}, \bibinfo {author} {\bibfnamefont {S.~R.}\ \bibnamefont {{Ehlert}}}, \bibinfo {author} {\bibfnamefont {T.}~\bibnamefont {{Enoto}}}, \bibinfo
  {author} {\bibfnamefont {Y.}~\bibnamefont {{Evangelista}}}, \bibinfo {author} {\bibfnamefont {S.}~\bibnamefont {{Fabiani}}}, \bibinfo {author} {\bibfnamefont {R.}~\bibnamefont {{Ferrazzoli}}}, \bibinfo {author} {\bibfnamefont {J.~A.}\ \bibnamefont {{Garcia}}}, \bibinfo {author} {\bibfnamefont {S.}~\bibnamefont {{Gunji}}}, \bibinfo {author} {\bibfnamefont {K.}~\bibnamefont {{Hayashida}}}, \bibinfo {author} {\bibfnamefont {W.}~\bibnamefont {{Iwakiri}}}, \bibinfo {author} {\bibfnamefont {S.~G.}\ \bibnamefont {{Jorstad}}}, \bibinfo {author} {\bibfnamefont {P.}~\bibnamefont {{Kaaret}}}, \bibinfo {author} {\bibfnamefont {V.}~\bibnamefont {{Karas}}}, \bibinfo {author} {\bibfnamefont {F.}~\bibnamefont {{Kislat}}}, \bibinfo {author} {\bibfnamefont {T.}~\bibnamefont {{Kitaguchi}}}, \bibinfo {author} {\bibfnamefont {J.~J.}\ \bibnamefont {{Kolodziejczak}}}, \bibinfo {author} {\bibfnamefont {H.}~\bibnamefont {{Krawczynski}}}, \bibinfo {author} {\bibfnamefont {L.}~\bibnamefont {{Latronico}}}, \bibinfo {author}
  {\bibfnamefont {I.}~\bibnamefont {{Liodakis}}}, \bibinfo {author} {\bibfnamefont {S.}~\bibnamefont {{Maldera}}}, \bibinfo {author} {\bibfnamefont {A.}~\bibnamefont {{Manfreda}}}, \bibinfo {author} {\bibfnamefont {F.}~\bibnamefont {{Marin}}}, \bibinfo {author} {\bibfnamefont {A.}~\bibnamefont {{Marinucci}}}, \bibinfo {author} {\bibfnamefont {A.~P.}\ \bibnamefont {{Marscher}}}, \bibinfo {author} {\bibfnamefont {F.}~\bibnamefont {{Massaro}}}, \bibinfo {author} {\bibfnamefont {G.}~\bibnamefont {{Matt}}}, \bibinfo {author} {\bibfnamefont {I.}~\bibnamefont {{Mitsuishi}}}, \bibinfo {author} {\bibfnamefont {T.}~\bibnamefont {{Mizuno}}}, \bibinfo {author} {\bibfnamefont {F.}~\bibnamefont {{Muleri}}}, \bibinfo {author} {\bibfnamefont {M.}~\bibnamefont {{Negro}}}, \bibinfo {author} {\bibfnamefont {C.-Y.}\ \bibnamefont {{Ng}}}, \bibinfo {author} {\bibfnamefont {S.~L.}\ \bibnamefont {{O'Dell}}}, \bibinfo {author} {\bibfnamefont {N.}~\bibnamefont {{Omodei}}}, \bibinfo {author} {\bibfnamefont {C.}~\bibnamefont
  {{Oppedisano}}}, \bibinfo {author} {\bibfnamefont {A.}~\bibnamefont {{Papitto}}}, \bibinfo {author} {\bibfnamefont {G.~G.}\ \bibnamefont {{Pavlov}}}, \bibinfo {author} {\bibfnamefont {A.~L.}\ \bibnamefont {{Peirson}}}, \bibinfo {author} {\bibfnamefont {M.}~\bibnamefont {{Perri}}}, \bibinfo {author} {\bibfnamefont {M.}~\bibnamefont {{Pesce-Rollins}}}, \bibinfo {author} {\bibfnamefont {M.}~\bibnamefont {{Pilia}}}, \bibinfo {author} {\bibfnamefont {A.}~\bibnamefont {{Possenti}}}, \bibinfo {author} {\bibfnamefont {S.}~\bibnamefont {{Puccetti}}}, \bibinfo {author} {\bibfnamefont {B.~D.}\ \bibnamefont {{Ramsey}}}, \bibinfo {author} {\bibfnamefont {J.}~\bibnamefont {{Rankin}}}, \bibinfo {author} {\bibfnamefont {A.}~\bibnamefont {{Ratheesh}}}, \bibinfo {author} {\bibfnamefont {O.~J.}\ \bibnamefont {{Roberts}}}, \bibinfo {author} {\bibfnamefont {R.~W.}\ \bibnamefont {{Romani}}}, \bibinfo {author} {\bibfnamefont {C.}~\bibnamefont {{Sgr{\`o}}}}, \bibinfo {author} {\bibfnamefont {P.}~\bibnamefont {{Slane}}}, \bibinfo
  {author} {\bibfnamefont {P.}~\bibnamefont {{Soffitta}}}, \bibinfo {author} {\bibfnamefont {G.}~\bibnamefont {{Spandre}}}, \bibinfo {author} {\bibfnamefont {D.~A.}\ \bibnamefont {{Swartz}}}, \bibinfo {author} {\bibfnamefont {T.}~\bibnamefont {{Tamagawa}}}, \bibinfo {author} {\bibfnamefont {F.}~\bibnamefont {{Tavecchio}}}, \bibinfo {author} {\bibfnamefont {R.}~\bibnamefont {{Taverna}}}, \bibinfo {author} {\bibfnamefont {Y.}~\bibnamefont {{Tawara}}}, \bibinfo {author} {\bibfnamefont {A.~F.}\ \bibnamefont {{Tennant}}}, \bibinfo {author} {\bibfnamefont {N.~E.}\ \bibnamefont {{Thomas}}}, \bibinfo {author} {\bibfnamefont {F.}~\bibnamefont {{Tombesi}}}, \bibinfo {author} {\bibfnamefont {A.}~\bibnamefont {{Trois}}}, \bibinfo {author} {\bibfnamefont {R.}~\bibnamefont {{Turolla}}}, \bibinfo {author} {\bibfnamefont {J.}~\bibnamefont {{Vink}}}, \bibinfo {author} {\bibfnamefont {M.~C.}\ \bibnamefont {{Weisskopf}}}, \bibinfo {author} {\bibfnamefont {K.}~\bibnamefont {{Wu}}}, \bibinfo {author} {\bibfnamefont
  {F.}~\bibnamefont {{Xie}}}, \ and\ \bibinfo {author} {\bibfnamefont {S.}~\bibnamefont {{Zane}}},\ }\href {\doibase 10.1051/0004-6361/202346581} {\bibfield  {journal} {\bibinfo  {journal} {\aap}\ }\textbf {\bibinfo {volume} {675}},\ \bibinfo {eid} {A29} (\bibinfo {year} {2023})},\ \Eprint {http://arxiv.org/abs/2304.00925} {arXiv:2304.00925 [astro-ph.HE]} \BibitemShut {NoStop}%
\bibitem [{\citenamefont {{Suleimanov}}\ \emph {et~al.}(2023{\natexlab{b}})\citenamefont {{Suleimanov}}, \citenamefont {{Forsblom}}, \citenamefont {{Tsygankov}}, \citenamefont {{Poutanen}}, \citenamefont {{Doroshenko}}, \citenamefont {{Doroshenko}}, \citenamefont {{Capitanio}}, \citenamefont {{Di Marco}}, \citenamefont {{Gonz{\'a}lez-Caniulef}}, \citenamefont {{Heyl}}, \citenamefont {{La Monaca}}, \citenamefont {{Lutovinov}}, \citenamefont {{Molkov}}, \citenamefont {{Malacaria}}, \citenamefont {{Mushtukov}}, \citenamefont {{Shtykovsky}}, \citenamefont {{Agudo}}, \citenamefont {{Antonelli}}, \citenamefont {{Bachetti}}, \citenamefont {{Baldini}}, \citenamefont {{Baumgartner}}, \citenamefont {{Bellazzini}}, \citenamefont {{Bianchi}}, \citenamefont {{Bongiorno}}, \citenamefont {{Bonino}}, \citenamefont {{Brez}}, \citenamefont {{Bucciantini}}, \citenamefont {{Castellano}}, \citenamefont {{Cavazzuti}}, \citenamefont {{Chen}}, \citenamefont {{Ciprini}}, \citenamefont {{Costa}}, \citenamefont {{De Rosa}},
  \citenamefont {{Del Monte}}, \citenamefont {{Di Gesu}}, \citenamefont {{Di Lalla}}, \citenamefont {{Donnarumma}}, \citenamefont {{Dov{\v{c}}iak}}, \citenamefont {{Ehlert}}, \citenamefont {{Enoto}}, \citenamefont {{Evangelista}}, \citenamefont {{Fabiani}}, \citenamefont {{Ferrazzoli}}, \citenamefont {{Garcia}}, \citenamefont {{Gunji}}, \citenamefont {{Hayashida}}, \citenamefont {{Iwakiri}}, \citenamefont {{Jorstad}}, \citenamefont {{Kaaret}}, \citenamefont {{Karas}}, \citenamefont {{Kislat}}, \citenamefont {{Kitaguchi}}, \citenamefont {{Kolodziejczak}}, \citenamefont {{Krawczynski}}, \citenamefont {{Latronico}}, \citenamefont {{Liodakis}}, \citenamefont {{Maldera}}, \citenamefont {{Manfreda}}, \citenamefont {{Marin}}, \citenamefont {{Marinucci}}, \citenamefont {{Marscher}}, \citenamefont {{Marshall}}, \citenamefont {{Massaro}}, \citenamefont {{Matt}}, \citenamefont {{Mitsuishi}}, \citenamefont {{Mizuno}}, \citenamefont {{Muleri}}, \citenamefont {{Negro}}, \citenamefont {{Ng}}, \citenamefont {{O'Dell}},
  \citenamefont {{Omodei}}, \citenamefont {{Oppedisano}}, \citenamefont {{Papitto}}, \citenamefont {{Pavlov}}, \citenamefont {{Peirson}}, \citenamefont {{Perri}}, \citenamefont {{Pesce-Rollins}}, \citenamefont {{Petrucci}}, \citenamefont {{Pilia}}, \citenamefont {{Possenti}}, \citenamefont {{Puccetti}}, \citenamefont {{Ramsey}}, \citenamefont {{Rankin}}, \citenamefont {{Ratheesh}}, \citenamefont {{Roberts}}, \citenamefont {{Romani}}, \citenamefont {{Sgr{\`o}}}, \citenamefont {{Slane}}, \citenamefont {{Soffitta}}, \citenamefont {{Spandre}}, \citenamefont {{Swartz}}, \citenamefont {{Tamagawa}}, \citenamefont {{Tavecchio}}, \citenamefont {{Taverna}}, \citenamefont {{Tawara}}, \citenamefont {{Tennant}}, \citenamefont {{Thomas}}, \citenamefont {{Tombesi}}, \citenamefont {{Trois}}, \citenamefont {{Turolla}}, \citenamefont {{Vink}}, \citenamefont {{Weisskopf}}, \citenamefont {{Wu}}, \citenamefont {{Xie}},\ and\ \citenamefont {{Zane}}}]{Suleimanov_gx301}%
  \BibitemOpen
  \bibfield  {author} {\bibinfo {author} {\bibfnamefont {V.~F.}\ \bibnamefont {{Suleimanov}}}, \bibinfo {author} {\bibfnamefont {S.~V.}\ \bibnamefont {{Forsblom}}}, \bibinfo {author} {\bibfnamefont {S.~S.}\ \bibnamefont {{Tsygankov}}}, \bibinfo {author} {\bibfnamefont {J.}~\bibnamefont {{Poutanen}}}, \bibinfo {author} {\bibfnamefont {V.}~\bibnamefont {{Doroshenko}}}, \bibinfo {author} {\bibfnamefont {R.}~\bibnamefont {{Doroshenko}}}, \bibinfo {author} {\bibfnamefont {F.}~\bibnamefont {{Capitanio}}}, \bibinfo {author} {\bibfnamefont {A.}~\bibnamefont {{Di Marco}}}, \bibinfo {author} {\bibfnamefont {D.}~\bibnamefont {{Gonz{\'a}lez-Caniulef}}}, \bibinfo {author} {\bibfnamefont {J.}~\bibnamefont {{Heyl}}}, \bibinfo {author} {\bibfnamefont {F.}~\bibnamefont {{La Monaca}}}, \bibinfo {author} {\bibfnamefont {A.~A.}\ \bibnamefont {{Lutovinov}}}, \bibinfo {author} {\bibfnamefont {S.~V.}\ \bibnamefont {{Molkov}}}, \bibinfo {author} {\bibfnamefont {C.}~\bibnamefont {{Malacaria}}}, \bibinfo {author} {\bibfnamefont
  {A.~A.}\ \bibnamefont {{Mushtukov}}}, \bibinfo {author} {\bibfnamefont {A.~E.}\ \bibnamefont {{Shtykovsky}}}, \bibinfo {author} {\bibfnamefont {I.}~\bibnamefont {{Agudo}}}, \bibinfo {author} {\bibfnamefont {L.~A.}\ \bibnamefont {{Antonelli}}}, \bibinfo {author} {\bibfnamefont {M.}~\bibnamefont {{Bachetti}}}, \bibinfo {author} {\bibfnamefont {L.}~\bibnamefont {{Baldini}}}, \bibinfo {author} {\bibfnamefont {W.~H.}\ \bibnamefont {{Baumgartner}}}, \bibinfo {author} {\bibfnamefont {R.}~\bibnamefont {{Bellazzini}}}, \bibinfo {author} {\bibfnamefont {S.}~\bibnamefont {{Bianchi}}}, \bibinfo {author} {\bibfnamefont {S.~D.}\ \bibnamefont {{Bongiorno}}}, \bibinfo {author} {\bibfnamefont {R.}~\bibnamefont {{Bonino}}}, \bibinfo {author} {\bibfnamefont {A.}~\bibnamefont {{Brez}}}, \bibinfo {author} {\bibfnamefont {N.}~\bibnamefont {{Bucciantini}}}, \bibinfo {author} {\bibfnamefont {S.}~\bibnamefont {{Castellano}}}, \bibinfo {author} {\bibfnamefont {E.}~\bibnamefont {{Cavazzuti}}}, \bibinfo {author} {\bibfnamefont
  {C.-T.}\ \bibnamefont {{Chen}}}, \bibinfo {author} {\bibfnamefont {S.}~\bibnamefont {{Ciprini}}}, \bibinfo {author} {\bibfnamefont {E.}~\bibnamefont {{Costa}}}, \bibinfo {author} {\bibfnamefont {A.}~\bibnamefont {{De Rosa}}}, \bibinfo {author} {\bibfnamefont {E.}~\bibnamefont {{Del Monte}}}, \bibinfo {author} {\bibfnamefont {L.}~\bibnamefont {{Di Gesu}}}, \bibinfo {author} {\bibfnamefont {N.}~\bibnamefont {{Di Lalla}}}, \bibinfo {author} {\bibfnamefont {I.}~\bibnamefont {{Donnarumma}}}, \bibinfo {author} {\bibfnamefont {M.}~\bibnamefont {{Dov{\v{c}}iak}}}, \bibinfo {author} {\bibfnamefont {S.~R.}\ \bibnamefont {{Ehlert}}}, \bibinfo {author} {\bibfnamefont {T.}~\bibnamefont {{Enoto}}}, \bibinfo {author} {\bibfnamefont {Y.}~\bibnamefont {{Evangelista}}}, \bibinfo {author} {\bibfnamefont {S.}~\bibnamefont {{Fabiani}}}, \bibinfo {author} {\bibfnamefont {R.}~\bibnamefont {{Ferrazzoli}}}, \bibinfo {author} {\bibfnamefont {J.~A.}\ \bibnamefont {{Garcia}}}, \bibinfo {author} {\bibfnamefont {S.}~\bibnamefont
  {{Gunji}}}, \bibinfo {author} {\bibfnamefont {K.}~\bibnamefont {{Hayashida}}}, \bibinfo {author} {\bibfnamefont {W.}~\bibnamefont {{Iwakiri}}}, \bibinfo {author} {\bibfnamefont {S.~G.}\ \bibnamefont {{Jorstad}}}, \bibinfo {author} {\bibfnamefont {P.}~\bibnamefont {{Kaaret}}}, \bibinfo {author} {\bibfnamefont {V.}~\bibnamefont {{Karas}}}, \bibinfo {author} {\bibfnamefont {F.}~\bibnamefont {{Kislat}}}, \bibinfo {author} {\bibfnamefont {T.}~\bibnamefont {{Kitaguchi}}}, \bibinfo {author} {\bibfnamefont {J.~J.}\ \bibnamefont {{Kolodziejczak}}}, \bibinfo {author} {\bibfnamefont {H.}~\bibnamefont {{Krawczynski}}}, \bibinfo {author} {\bibfnamefont {L.}~\bibnamefont {{Latronico}}}, \bibinfo {author} {\bibfnamefont {I.}~\bibnamefont {{Liodakis}}}, \bibinfo {author} {\bibfnamefont {S.}~\bibnamefont {{Maldera}}}, \bibinfo {author} {\bibfnamefont {A.}~\bibnamefont {{Manfreda}}}, \bibinfo {author} {\bibfnamefont {F.}~\bibnamefont {{Marin}}}, \bibinfo {author} {\bibfnamefont {A.}~\bibnamefont {{Marinucci}}}, \bibinfo
  {author} {\bibfnamefont {A.~P.}\ \bibnamefont {{Marscher}}}, \bibinfo {author} {\bibfnamefont {H.~L.}\ \bibnamefont {{Marshall}}}, \bibinfo {author} {\bibfnamefont {F.}~\bibnamefont {{Massaro}}}, \bibinfo {author} {\bibfnamefont {G.}~\bibnamefont {{Matt}}}, \bibinfo {author} {\bibfnamefont {I.}~\bibnamefont {{Mitsuishi}}}, \bibinfo {author} {\bibfnamefont {T.}~\bibnamefont {{Mizuno}}}, \bibinfo {author} {\bibfnamefont {F.}~\bibnamefont {{Muleri}}}, \bibinfo {author} {\bibfnamefont {M.}~\bibnamefont {{Negro}}}, \bibinfo {author} {\bibfnamefont {C.-Y.}\ \bibnamefont {{Ng}}}, \bibinfo {author} {\bibfnamefont {S.~L.}\ \bibnamefont {{O'Dell}}}, \bibinfo {author} {\bibfnamefont {N.}~\bibnamefont {{Omodei}}}, \bibinfo {author} {\bibfnamefont {C.}~\bibnamefont {{Oppedisano}}}, \bibinfo {author} {\bibfnamefont {A.}~\bibnamefont {{Papitto}}}, \bibinfo {author} {\bibfnamefont {G.~G.}\ \bibnamefont {{Pavlov}}}, \bibinfo {author} {\bibfnamefont {A.~L.}\ \bibnamefont {{Peirson}}}, \bibinfo {author} {\bibfnamefont
  {M.}~\bibnamefont {{Perri}}}, \bibinfo {author} {\bibfnamefont {M.}~\bibnamefont {{Pesce-Rollins}}}, \bibinfo {author} {\bibfnamefont {P.-O.}\ \bibnamefont {{Petrucci}}}, \bibinfo {author} {\bibfnamefont {M.}~\bibnamefont {{Pilia}}}, \bibinfo {author} {\bibfnamefont {A.}~\bibnamefont {{Possenti}}}, \bibinfo {author} {\bibfnamefont {S.}~\bibnamefont {{Puccetti}}}, \bibinfo {author} {\bibfnamefont {B.~D.}\ \bibnamefont {{Ramsey}}}, \bibinfo {author} {\bibfnamefont {J.}~\bibnamefont {{Rankin}}}, \bibinfo {author} {\bibfnamefont {A.}~\bibnamefont {{Ratheesh}}}, \bibinfo {author} {\bibfnamefont {O.~J.}\ \bibnamefont {{Roberts}}}, \bibinfo {author} {\bibfnamefont {R.~W.}\ \bibnamefont {{Romani}}}, \bibinfo {author} {\bibfnamefont {C.}~\bibnamefont {{Sgr{\`o}}}}, \bibinfo {author} {\bibfnamefont {P.}~\bibnamefont {{Slane}}}, \bibinfo {author} {\bibfnamefont {P.}~\bibnamefont {{Soffitta}}}, \bibinfo {author} {\bibfnamefont {G.}~\bibnamefont {{Spandre}}}, \bibinfo {author} {\bibfnamefont {D.~A.}\ \bibnamefont
  {{Swartz}}}, \bibinfo {author} {\bibfnamefont {T.}~\bibnamefont {{Tamagawa}}}, \bibinfo {author} {\bibfnamefont {F.}~\bibnamefont {{Tavecchio}}}, \bibinfo {author} {\bibfnamefont {R.}~\bibnamefont {{Taverna}}}, \bibinfo {author} {\bibfnamefont {Y.}~\bibnamefont {{Tawara}}}, \bibinfo {author} {\bibfnamefont {A.~F.}\ \bibnamefont {{Tennant}}}, \bibinfo {author} {\bibfnamefont {N.~E.}\ \bibnamefont {{Thomas}}}, \bibinfo {author} {\bibfnamefont {F.}~\bibnamefont {{Tombesi}}}, \bibinfo {author} {\bibfnamefont {A.}~\bibnamefont {{Trois}}}, \bibinfo {author} {\bibfnamefont {R.}~\bibnamefont {{Turolla}}}, \bibinfo {author} {\bibfnamefont {J.}~\bibnamefont {{Vink}}}, \bibinfo {author} {\bibfnamefont {M.~C.}\ \bibnamefont {{Weisskopf}}}, \bibinfo {author} {\bibfnamefont {K.}~\bibnamefont {{Wu}}}, \bibinfo {author} {\bibfnamefont {F.}~\bibnamefont {{Xie}}}, \ and\ \bibinfo {author} {\bibfnamefont {S.}~\bibnamefont {{Zane}}},\ }\href {\doibase 10.1051/0004-6361/202346994} {\bibfield  {journal} {\bibinfo  {journal}
  {\aap}\ }\textbf {\bibinfo {volume} {678}},\ \bibinfo {eid} {A119} (\bibinfo {year} {2023}{\natexlab{b}})},\ \Eprint {http://arxiv.org/abs/2305.15309} {arXiv:2305.15309 [astro-ph.HE]} \BibitemShut {NoStop}%
\bibitem [{\citenamefont {{Doroshenko}}\ \emph {et~al.}(2023)\citenamefont {{Doroshenko}}, \citenamefont {{Poutanen}}, \citenamefont {{Heyl}}, \citenamefont {{Tsygankov}}, \citenamefont {{Caiazzo}}, \citenamefont {{Turolla}}, \citenamefont {{Veledina}}, \citenamefont {{Weisskopf}}, \citenamefont {{Forsblom}}, \citenamefont {{Gonz{\'a}lez-Caniulef}}, \citenamefont {{Loktev}}, \citenamefont {{Malacaria}}, \citenamefont {{Mushtukov}}, \citenamefont {{Suleimanov}}, \citenamefont {{Lutovinov}}, \citenamefont {{Mereminskiy}}, \citenamefont {{Molkov}}, \citenamefont {{Salganik}}, \citenamefont {{Santangelo}}, \citenamefont {{Berdyugin}}, \citenamefont {{Kravtsov}}, \citenamefont {{Nitindala}}, \citenamefont {{Agudo}}, \citenamefont {{Antonelli}}, \citenamefont {{Bachetti}}, \citenamefont {{Baldini}}, \citenamefont {{Baumgartner}}, \citenamefont {{Bellazzini}}, \citenamefont {{Bianchi}}, \citenamefont {{Bongiorno}}, \citenamefont {{Bonino}}, \citenamefont {{Brez}}, \citenamefont {{Bucciantini}}, \citenamefont
  {{Capitanio}}, \citenamefont {{Castellano}}, \citenamefont {{Cavazzuti}}, \citenamefont {{Chen}}, \citenamefont {{Ciprini}}, \citenamefont {{Costa}}, \citenamefont {{De Rosa}}, \citenamefont {{Del Monte}}, \citenamefont {{Di Gesu}}, \citenamefont {{Di Lalla}}, \citenamefont {{Di Marco}}, \citenamefont {{Donnarumma}}, \citenamefont {{Dov{\v{c}}iak}}, \citenamefont {{Ehlert}}, \citenamefont {{Enoto}}, \citenamefont {{Evangelista}}, \citenamefont {{Fabiani}}, \citenamefont {{Ferrazzoli}}, \citenamefont {{Garc{\'\i}a}}, \citenamefont {{Gunji}}, \citenamefont {{Hayashida}}, \citenamefont {{Iwakiri}}, \citenamefont {{Jorstad}}, \citenamefont {{Kaaret}}, \citenamefont {{Karas}}, \citenamefont {{Kislat}}, \citenamefont {{Kitaguchi}}, \citenamefont {{Kolodziejczak}}, \citenamefont {{Krawczynski}}, \citenamefont {{La Monaca}}, \citenamefont {{Latronico}}, \citenamefont {{Liodakis}}, \citenamefont {{Maldera}}, \citenamefont {{Manfreda}}, \citenamefont {{Marin}}, \citenamefont {{Marinucci}}, \citenamefont {{Marscher}},
  \citenamefont {{Marshall}}, \citenamefont {{Massaro}}, \citenamefont {{Matt}}, \citenamefont {{Mitsuishi}}, \citenamefont {{Mizuno}}, \citenamefont {{Muleri}}, \citenamefont {{Negro}}, \citenamefont {{Ng}}, \citenamefont {{O'Dell}}, \citenamefont {{Omodei}}, \citenamefont {{Oppedisano}}, \citenamefont {{Papitto}}, \citenamefont {{Pavlov}}, \citenamefont {{Peirson}}, \citenamefont {{Perri}}, \citenamefont {{Pesce-Rollins}}, \citenamefont {{Petrucci}}, \citenamefont {{Pilia}}, \citenamefont {{Possenti}}, \citenamefont {{Puccetti}}, \citenamefont {{Ramsey}}, \citenamefont {{Rankin}}, \citenamefont {{Ratheesh}}, \citenamefont {{Roberts}}, \citenamefont {{Romani}}, \citenamefont {{Sgr{\`o}}}, \citenamefont {{Slane}}, \citenamefont {{Soffitta}}, \citenamefont {{Spandre}}, \citenamefont {{Swartz}}, \citenamefont {{Tamagawa}}, \citenamefont {{Tavecchio}}, \citenamefont {{Taverna}}, \citenamefont {{Tawara}}, \citenamefont {{Tennant}}, \citenamefont {{Thomas}}, \citenamefont {{Tombesi}}, \citenamefont {{Trois}},
  \citenamefont {{Vink}}, \citenamefont {{Wu}}, \citenamefont {{Xie}},\ and\ \citenamefont {{Zane}}}]{Victor_etal_2023}%
  \BibitemOpen
  \bibfield  {author} {\bibinfo {author} {\bibfnamefont {V.}~\bibnamefont {{Doroshenko}}}, \bibinfo {author} {\bibfnamefont {J.}~\bibnamefont {{Poutanen}}}, \bibinfo {author} {\bibfnamefont {J.}~\bibnamefont {{Heyl}}}, \bibinfo {author} {\bibfnamefont {S.~S.}\ \bibnamefont {{Tsygankov}}}, \bibinfo {author} {\bibfnamefont {I.}~\bibnamefont {{Caiazzo}}}, \bibinfo {author} {\bibfnamefont {R.}~\bibnamefont {{Turolla}}}, \bibinfo {author} {\bibfnamefont {A.}~\bibnamefont {{Veledina}}}, \bibinfo {author} {\bibfnamefont {M.~C.}\ \bibnamefont {{Weisskopf}}}, \bibinfo {author} {\bibfnamefont {S.~V.}\ \bibnamefont {{Forsblom}}}, \bibinfo {author} {\bibfnamefont {D.}~\bibnamefont {{Gonz{\'a}lez-Caniulef}}}, \bibinfo {author} {\bibfnamefont {V.}~\bibnamefont {{Loktev}}}, \bibinfo {author} {\bibfnamefont {C.}~\bibnamefont {{Malacaria}}}, \bibinfo {author} {\bibfnamefont {A.~A.}\ \bibnamefont {{Mushtukov}}}, \bibinfo {author} {\bibfnamefont {V.~F.}\ \bibnamefont {{Suleimanov}}}, \bibinfo {author} {\bibfnamefont {A.~A.}\
  \bibnamefont {{Lutovinov}}}, \bibinfo {author} {\bibfnamefont {I.~A.}\ \bibnamefont {{Mereminskiy}}}, \bibinfo {author} {\bibfnamefont {S.~V.}\ \bibnamefont {{Molkov}}}, \bibinfo {author} {\bibfnamefont {A.}~\bibnamefont {{Salganik}}}, \bibinfo {author} {\bibfnamefont {A.}~\bibnamefont {{Santangelo}}}, \bibinfo {author} {\bibfnamefont {A.~V.}\ \bibnamefont {{Berdyugin}}}, \bibinfo {author} {\bibfnamefont {V.}~\bibnamefont {{Kravtsov}}}, \bibinfo {author} {\bibfnamefont {A.~P.}\ \bibnamefont {{Nitindala}}}, \bibinfo {author} {\bibfnamefont {I.}~\bibnamefont {{Agudo}}}, \bibinfo {author} {\bibfnamefont {L.~A.}\ \bibnamefont {{Antonelli}}}, \bibinfo {author} {\bibfnamefont {M.}~\bibnamefont {{Bachetti}}}, \bibinfo {author} {\bibfnamefont {L.}~\bibnamefont {{Baldini}}}, \bibinfo {author} {\bibfnamefont {W.~H.}\ \bibnamefont {{Baumgartner}}}, \bibinfo {author} {\bibfnamefont {R.}~\bibnamefont {{Bellazzini}}}, \bibinfo {author} {\bibfnamefont {S.}~\bibnamefont {{Bianchi}}}, \bibinfo {author} {\bibfnamefont
  {S.~D.}\ \bibnamefont {{Bongiorno}}}, \bibinfo {author} {\bibfnamefont {R.}~\bibnamefont {{Bonino}}}, \bibinfo {author} {\bibfnamefont {A.}~\bibnamefont {{Brez}}}, \bibinfo {author} {\bibfnamefont {N.}~\bibnamefont {{Bucciantini}}}, \bibinfo {author} {\bibfnamefont {F.}~\bibnamefont {{Capitanio}}}, \bibinfo {author} {\bibfnamefont {S.}~\bibnamefont {{Castellano}}}, \bibinfo {author} {\bibfnamefont {E.}~\bibnamefont {{Cavazzuti}}}, \bibinfo {author} {\bibfnamefont {C.-T.}\ \bibnamefont {{Chen}}}, \bibinfo {author} {\bibfnamefont {S.}~\bibnamefont {{Ciprini}}}, \bibinfo {author} {\bibfnamefont {E.}~\bibnamefont {{Costa}}}, \bibinfo {author} {\bibfnamefont {A.}~\bibnamefont {{De Rosa}}}, \bibinfo {author} {\bibfnamefont {E.}~\bibnamefont {{Del Monte}}}, \bibinfo {author} {\bibfnamefont {L.}~\bibnamefont {{Di Gesu}}}, \bibinfo {author} {\bibfnamefont {N.}~\bibnamefont {{Di Lalla}}}, \bibinfo {author} {\bibfnamefont {A.}~\bibnamefont {{Di Marco}}}, \bibinfo {author} {\bibfnamefont {I.}~\bibnamefont
  {{Donnarumma}}}, \bibinfo {author} {\bibfnamefont {M.}~\bibnamefont {{Dov{\v{c}}iak}}}, \bibinfo {author} {\bibfnamefont {S.~R.}\ \bibnamefont {{Ehlert}}}, \bibinfo {author} {\bibfnamefont {T.}~\bibnamefont {{Enoto}}}, \bibinfo {author} {\bibfnamefont {Y.}~\bibnamefont {{Evangelista}}}, \bibinfo {author} {\bibfnamefont {S.}~\bibnamefont {{Fabiani}}}, \bibinfo {author} {\bibfnamefont {R.}~\bibnamefont {{Ferrazzoli}}}, \bibinfo {author} {\bibfnamefont {J.~A.}\ \bibnamefont {{Garc{\'\i}a}}}, \bibinfo {author} {\bibfnamefont {S.}~\bibnamefont {{Gunji}}}, \bibinfo {author} {\bibfnamefont {K.}~\bibnamefont {{Hayashida}}}, \bibinfo {author} {\bibfnamefont {W.}~\bibnamefont {{Iwakiri}}}, \bibinfo {author} {\bibfnamefont {S.~G.}\ \bibnamefont {{Jorstad}}}, \bibinfo {author} {\bibfnamefont {P.}~\bibnamefont {{Kaaret}}}, \bibinfo {author} {\bibfnamefont {V.}~\bibnamefont {{Karas}}}, \bibinfo {author} {\bibfnamefont {F.}~\bibnamefont {{Kislat}}}, \bibinfo {author} {\bibfnamefont {T.}~\bibnamefont {{Kitaguchi}}},
  \bibinfo {author} {\bibfnamefont {J.~J.}\ \bibnamefont {{Kolodziejczak}}}, \bibinfo {author} {\bibfnamefont {H.}~\bibnamefont {{Krawczynski}}}, \bibinfo {author} {\bibfnamefont {F.}~\bibnamefont {{La Monaca}}}, \bibinfo {author} {\bibfnamefont {L.}~\bibnamefont {{Latronico}}}, \bibinfo {author} {\bibfnamefont {I.}~\bibnamefont {{Liodakis}}}, \bibinfo {author} {\bibfnamefont {S.}~\bibnamefont {{Maldera}}}, \bibinfo {author} {\bibfnamefont {A.}~\bibnamefont {{Manfreda}}}, \bibinfo {author} {\bibfnamefont {F.}~\bibnamefont {{Marin}}}, \bibinfo {author} {\bibfnamefont {A.}~\bibnamefont {{Marinucci}}}, \bibinfo {author} {\bibfnamefont {A.~P.}\ \bibnamefont {{Marscher}}}, \bibinfo {author} {\bibfnamefont {H.~L.}\ \bibnamefont {{Marshall}}}, \bibinfo {author} {\bibfnamefont {F.}~\bibnamefont {{Massaro}}}, \bibinfo {author} {\bibfnamefont {G.}~\bibnamefont {{Matt}}}, \bibinfo {author} {\bibfnamefont {I.}~\bibnamefont {{Mitsuishi}}}, \bibinfo {author} {\bibfnamefont {T.}~\bibnamefont {{Mizuno}}}, \bibinfo {author}
  {\bibfnamefont {F.}~\bibnamefont {{Muleri}}}, \bibinfo {author} {\bibfnamefont {M.}~\bibnamefont {{Negro}}}, \bibinfo {author} {\bibfnamefont {C.-Y.}\ \bibnamefont {{Ng}}}, \bibinfo {author} {\bibfnamefont {S.~L.}\ \bibnamefont {{O'Dell}}}, \bibinfo {author} {\bibfnamefont {N.}~\bibnamefont {{Omodei}}}, \bibinfo {author} {\bibfnamefont {C.}~\bibnamefont {{Oppedisano}}}, \bibinfo {author} {\bibfnamefont {A.}~\bibnamefont {{Papitto}}}, \bibinfo {author} {\bibfnamefont {G.~G.}\ \bibnamefont {{Pavlov}}}, \bibinfo {author} {\bibfnamefont {A.~L.}\ \bibnamefont {{Peirson}}}, \bibinfo {author} {\bibfnamefont {M.}~\bibnamefont {{Perri}}}, \bibinfo {author} {\bibfnamefont {M.}~\bibnamefont {{Pesce-Rollins}}}, \bibinfo {author} {\bibfnamefont {P.-O.}\ \bibnamefont {{Petrucci}}}, \bibinfo {author} {\bibfnamefont {M.}~\bibnamefont {{Pilia}}}, \bibinfo {author} {\bibfnamefont {A.}~\bibnamefont {{Possenti}}}, \bibinfo {author} {\bibfnamefont {S.}~\bibnamefont {{Puccetti}}}, \bibinfo {author} {\bibfnamefont {B.~D.}\
  \bibnamefont {{Ramsey}}}, \bibinfo {author} {\bibfnamefont {J.}~\bibnamefont {{Rankin}}}, \bibinfo {author} {\bibfnamefont {A.}~\bibnamefont {{Ratheesh}}}, \bibinfo {author} {\bibfnamefont {O.~J.}\ \bibnamefont {{Roberts}}}, \bibinfo {author} {\bibfnamefont {R.~W.}\ \bibnamefont {{Romani}}}, \bibinfo {author} {\bibfnamefont {C.}~\bibnamefont {{Sgr{\`o}}}}, \bibinfo {author} {\bibfnamefont {P.}~\bibnamefont {{Slane}}}, \bibinfo {author} {\bibfnamefont {P.}~\bibnamefont {{Soffitta}}}, \bibinfo {author} {\bibfnamefont {G.}~\bibnamefont {{Spandre}}}, \bibinfo {author} {\bibfnamefont {D.~A.}\ \bibnamefont {{Swartz}}}, \bibinfo {author} {\bibfnamefont {T.}~\bibnamefont {{Tamagawa}}}, \bibinfo {author} {\bibfnamefont {F.}~\bibnamefont {{Tavecchio}}}, \bibinfo {author} {\bibfnamefont {R.}~\bibnamefont {{Taverna}}}, \bibinfo {author} {\bibfnamefont {Y.}~\bibnamefont {{Tawara}}}, \bibinfo {author} {\bibfnamefont {A.~F.}\ \bibnamefont {{Tennant}}}, \bibinfo {author} {\bibfnamefont {N.~E.}\ \bibnamefont {{Thomas}}},
  \bibinfo {author} {\bibfnamefont {F.}~\bibnamefont {{Tombesi}}}, \bibinfo {author} {\bibfnamefont {A.}~\bibnamefont {{Trois}}}, \bibinfo {author} {\bibfnamefont {J.}~\bibnamefont {{Vink}}}, \bibinfo {author} {\bibfnamefont {K.}~\bibnamefont {{Wu}}}, \bibinfo {author} {\bibfnamefont {F.}~\bibnamefont {{Xie}}}, \ and\ \bibinfo {author} {\bibfnamefont {S.}~\bibnamefont {{Zane}}},\ }\href {\doibase 10.1051/0004-6361/202347088} {\bibfield  {journal} {\bibinfo  {journal} {\aap}\ }\textbf {\bibinfo {volume} {677}},\ \bibinfo {eid} {A57} (\bibinfo {year} {2023})},\ \Eprint {http://arxiv.org/abs/2306.02116} {arXiv:2306.02116 [astro-ph.HE]} \BibitemShut {NoStop}%
\bibitem [{\citenamefont {{Zhao}}\ \emph {et~al.}(2025)\citenamefont {{Zhao}}, \citenamefont {{Tao}}, \citenamefont {{Tsygankov}}, \citenamefont {{Mushtukov}}, \citenamefont {{Feng}}, \citenamefont {{Ge}}, \citenamefont {{Li}}, \citenamefont {{Zhang}},\ and\ \citenamefont {{Zhang}}}]{Zhao_RXJ0440}%
  \BibitemOpen
  \bibfield  {author} {\bibinfo {author} {\bibfnamefont {Q.~C.}\ \bibnamefont {{Zhao}}}, \bibinfo {author} {\bibfnamefont {L.}~\bibnamefont {{Tao}}}, \bibinfo {author} {\bibfnamefont {S.~S.}\ \bibnamefont {{Tsygankov}}}, \bibinfo {author} {\bibfnamefont {A.~A.}\ \bibnamefont {{Mushtukov}}}, \bibinfo {author} {\bibfnamefont {H.}~\bibnamefont {{Feng}}}, \bibinfo {author} {\bibfnamefont {M.~Y.}\ \bibnamefont {{Ge}}}, \bibinfo {author} {\bibfnamefont {H.~C.}\ \bibnamefont {{Li}}}, \bibinfo {author} {\bibfnamefont {S.~N.}\ \bibnamefont {{Zhang}}}, \ and\ \bibinfo {author} {\bibfnamefont {L.}~\bibnamefont {{Zhang}}},\ }\href {\doibase 10.1051/0004-6361/202452872} {\bibfield  {journal} {\bibinfo  {journal} {\aap}\ }\textbf {\bibinfo {volume} {693}},\ \bibinfo {eid} {A241} (\bibinfo {year} {2025})},\ \Eprint {http://arxiv.org/abs/2412.15955} {arXiv:2412.15955 [astro-ph.HE]} \BibitemShut {NoStop}%
\bibitem [{\citenamefont {{Majumder}}\ \emph {et~al.}(2024)\citenamefont {{Majumder}}, \citenamefont {{Chatterjee}}, \citenamefont {{Jayasurya}}, \citenamefont {{Das}},\ and\ \citenamefont {{Nandi}}}]{SwiftJ0243_Majumder}%
  \BibitemOpen
  \bibfield  {author} {\bibinfo {author} {\bibfnamefont {S.}~\bibnamefont {{Majumder}}}, \bibinfo {author} {\bibfnamefont {R.}~\bibnamefont {{Chatterjee}}}, \bibinfo {author} {\bibfnamefont {K.~M.}\ \bibnamefont {{Jayasurya}}}, \bibinfo {author} {\bibfnamefont {S.}~\bibnamefont {{Das}}}, \ and\ \bibinfo {author} {\bibfnamefont {A.}~\bibnamefont {{Nandi}}},\ }\href {\doibase 10.3847/2041-8213/ad67e5} {\bibfield  {journal} {\bibinfo  {journal} {\apjl}\ }\textbf {\bibinfo {volume} {971}},\ \bibinfo {eid} {L21} (\bibinfo {year} {2024})},\ \Eprint {http://arxiv.org/abs/2402.11602} {arXiv:2402.11602 [astro-ph.HE]} \BibitemShut {NoStop}%
\bibitem [{\citenamefont {{Poutanen}}\ \emph {et~al.}(2024{\natexlab{b}})\citenamefont {{Poutanen}}, \citenamefont {{Tsygankov}}, \citenamefont {{Doroshenko}}, \citenamefont {{Forsblom}}, \citenamefont {{Jenke}}, \citenamefont {{Kaaret}}, \citenamefont {{Berdyugin}}, \citenamefont {{Blinov}}, \citenamefont {{Kravtsov}}, \citenamefont {{Liodakis}}, \citenamefont {{Tzouvanou}}, \citenamefont {{Di Marco}}, \citenamefont {{Heyl}}, \citenamefont {{La Monaca}}, \citenamefont {{Mushtukov}}, \citenamefont {{Pavlov}}, \citenamefont {{Salganik}}, \citenamefont {{Veledina}}, \citenamefont {{Weisskopf}}, \citenamefont {{Zane}}, \citenamefont {{Loktev}}, \citenamefont {{Suleimanov}}, \citenamefont {{Wilson-Hodge}}, \citenamefont {{Berdyugina}}, \citenamefont {{Kagitani}}, \citenamefont {{Piirola}}, \citenamefont {{Sakanoi}}, \citenamefont {{Agudo}}, \citenamefont {{Antonelli}}, \citenamefont {{Bachetti}}, \citenamefont {{Baldini}}, \citenamefont {{Baumgartner}}, \citenamefont {{Bellazzini}}, \citenamefont {{Bianchi}},
  \citenamefont {{Bongiorno}}, \citenamefont {{Bonino}}, \citenamefont {{Brez}}, \citenamefont {{Bucciantini}}, \citenamefont {{Capitanio}}, \citenamefont {{Castellano}}, \citenamefont {{Cavazzuti}}, \citenamefont {{Chen}}, \citenamefont {{Ciprini}}, \citenamefont {{Costa}}, \citenamefont {{De Rosa}}, \citenamefont {{Del Monte}}, \citenamefont {{Di Gesu}}, \citenamefont {{Di Lalla}}, \citenamefont {{Donnarumma}}, \citenamefont {{Dov{\v{c}}iak}}, \citenamefont {{Ehlert}}, \citenamefont {{Enoto}}, \citenamefont {{Evangelista}}, \citenamefont {{Fabiani}}, \citenamefont {{Ferrazzoli}}, \citenamefont {{Garcia}}, \citenamefont {{Gunji}}, \citenamefont {{Hayashida}}, \citenamefont {{Iwakiri}}, \citenamefont {{Jorstad}}, \citenamefont {{Karas}}, \citenamefont {{Kislat}}, \citenamefont {{Kitaguchi}}, \citenamefont {{Kolodziejczak}}, \citenamefont {{Latronico}}, \citenamefont {{Maldera}}, \citenamefont {{Manfreda}}, \citenamefont {{Marin}}, \citenamefont {{Marinucci}}, \citenamefont {{Marscher}}, \citenamefont
  {{Marshall}}, \citenamefont {{Massaro}}, \citenamefont {{Matt}}, \citenamefont {{Mitsuishi}}, \citenamefont {{Mizuno}}, \citenamefont {{Muleri}}, \citenamefont {{Negro}}, \citenamefont {{Ng}}, \citenamefont {{O'Dell}}, \citenamefont {{Omodei}}, \citenamefont {{Oppedisano}}, \citenamefont {{Papitto}}, \citenamefont {{Peirson}}, \citenamefont {{Perri}}, \citenamefont {{Pesce-Rollins}}, \citenamefont {{Petrucci}}, \citenamefont {{Pilia}}, \citenamefont {{Possenti}}, \citenamefont {{Puccetti}}, \citenamefont {{Ramsey}}, \citenamefont {{Rankin}}, \citenamefont {{Ratheesh}}, \citenamefont {{Roberts}}, \citenamefont {{Romani}}, \citenamefont {{Sgr{\`o}}}, \citenamefont {{Slane}}, \citenamefont {{Soffitta}}, \citenamefont {{Spandre}}, \citenamefont {{Swartz}}, \citenamefont {{Tamagawa}}, \citenamefont {{Tavecchio}}, \citenamefont {{Taverna}}, \citenamefont {{Tawara}}, \citenamefont {{Tennant}}, \citenamefont {{Thomas}}, \citenamefont {{Tombesi}}, \citenamefont {{Trois}}, \citenamefont {{Turolla}}, \citenamefont
  {{Vink}}, \citenamefont {{Wu}},\ and\ \citenamefont {{Xie}}}]{SwiftJ0243_Poutanen}%
  \BibitemOpen
  \bibfield  {author} {\bibinfo {author} {\bibfnamefont {J.}~\bibnamefont {{Poutanen}}}, \bibinfo {author} {\bibfnamefont {S.~S.}\ \bibnamefont {{Tsygankov}}}, \bibinfo {author} {\bibfnamefont {V.}~\bibnamefont {{Doroshenko}}}, \bibinfo {author} {\bibfnamefont {S.~V.}\ \bibnamefont {{Forsblom}}}, \bibinfo {author} {\bibfnamefont {P.}~\bibnamefont {{Jenke}}}, \bibinfo {author} {\bibfnamefont {P.}~\bibnamefont {{Kaaret}}}, \bibinfo {author} {\bibfnamefont {A.~V.}\ \bibnamefont {{Berdyugin}}}, \bibinfo {author} {\bibfnamefont {D.}~\bibnamefont {{Blinov}}}, \bibinfo {author} {\bibfnamefont {V.}~\bibnamefont {{Kravtsov}}}, \bibinfo {author} {\bibfnamefont {I.}~\bibnamefont {{Liodakis}}}, \bibinfo {author} {\bibfnamefont {A.}~\bibnamefont {{Tzouvanou}}}, \bibinfo {author} {\bibfnamefont {A.}~\bibnamefont {{Di Marco}}}, \bibinfo {author} {\bibfnamefont {J.}~\bibnamefont {{Heyl}}}, \bibinfo {author} {\bibfnamefont {F.}~\bibnamefont {{La Monaca}}}, \bibinfo {author} {\bibfnamefont {A.~A.}\ \bibnamefont {{Mushtukov}}},
  \bibinfo {author} {\bibfnamefont {G.~G.}\ \bibnamefont {{Pavlov}}}, \bibinfo {author} {\bibfnamefont {A.}~\bibnamefont {{Salganik}}}, \bibinfo {author} {\bibfnamefont {A.}~\bibnamefont {{Veledina}}}, \bibinfo {author} {\bibfnamefont {M.~C.}\ \bibnamefont {{Weisskopf}}}, \bibinfo {author} {\bibfnamefont {S.}~\bibnamefont {{Zane}}}, \bibinfo {author} {\bibfnamefont {V.}~\bibnamefont {{Loktev}}}, \bibinfo {author} {\bibfnamefont {V.~F.}\ \bibnamefont {{Suleimanov}}}, \bibinfo {author} {\bibfnamefont {C.}~\bibnamefont {{Wilson-Hodge}}}, \bibinfo {author} {\bibfnamefont {S.~V.}\ \bibnamefont {{Berdyugina}}}, \bibinfo {author} {\bibfnamefont {M.}~\bibnamefont {{Kagitani}}}, \bibinfo {author} {\bibfnamefont {V.}~\bibnamefont {{Piirola}}}, \bibinfo {author} {\bibfnamefont {T.}~\bibnamefont {{Sakanoi}}}, \bibinfo {author} {\bibfnamefont {I.}~\bibnamefont {{Agudo}}}, \bibinfo {author} {\bibfnamefont {L.~A.}\ \bibnamefont {{Antonelli}}}, \bibinfo {author} {\bibfnamefont {M.}~\bibnamefont {{Bachetti}}}, \bibinfo
  {author} {\bibfnamefont {L.}~\bibnamefont {{Baldini}}}, \bibinfo {author} {\bibfnamefont {W.~H.}\ \bibnamefont {{Baumgartner}}}, \bibinfo {author} {\bibfnamefont {R.}~\bibnamefont {{Bellazzini}}}, \bibinfo {author} {\bibfnamefont {S.}~\bibnamefont {{Bianchi}}}, \bibinfo {author} {\bibfnamefont {S.~D.}\ \bibnamefont {{Bongiorno}}}, \bibinfo {author} {\bibfnamefont {R.}~\bibnamefont {{Bonino}}}, \bibinfo {author} {\bibfnamefont {A.}~\bibnamefont {{Brez}}}, \bibinfo {author} {\bibfnamefont {N.}~\bibnamefont {{Bucciantini}}}, \bibinfo {author} {\bibfnamefont {F.}~\bibnamefont {{Capitanio}}}, \bibinfo {author} {\bibfnamefont {S.}~\bibnamefont {{Castellano}}}, \bibinfo {author} {\bibfnamefont {E.}~\bibnamefont {{Cavazzuti}}}, \bibinfo {author} {\bibfnamefont {C.-T.}\ \bibnamefont {{Chen}}}, \bibinfo {author} {\bibfnamefont {S.}~\bibnamefont {{Ciprini}}}, \bibinfo {author} {\bibfnamefont {E.}~\bibnamefont {{Costa}}}, \bibinfo {author} {\bibfnamefont {A.}~\bibnamefont {{De Rosa}}}, \bibinfo {author} {\bibfnamefont
  {E.}~\bibnamefont {{Del Monte}}}, \bibinfo {author} {\bibfnamefont {L.}~\bibnamefont {{Di Gesu}}}, \bibinfo {author} {\bibfnamefont {N.}~\bibnamefont {{Di Lalla}}}, \bibinfo {author} {\bibfnamefont {I.}~\bibnamefont {{Donnarumma}}}, \bibinfo {author} {\bibfnamefont {M.}~\bibnamefont {{Dov{\v{c}}iak}}}, \bibinfo {author} {\bibfnamefont {S.~R.}\ \bibnamefont {{Ehlert}}}, \bibinfo {author} {\bibfnamefont {T.}~\bibnamefont {{Enoto}}}, \bibinfo {author} {\bibfnamefont {Y.}~\bibnamefont {{Evangelista}}}, \bibinfo {author} {\bibfnamefont {S.}~\bibnamefont {{Fabiani}}}, \bibinfo {author} {\bibfnamefont {R.}~\bibnamefont {{Ferrazzoli}}}, \bibinfo {author} {\bibfnamefont {J.~A.}\ \bibnamefont {{Garcia}}}, \bibinfo {author} {\bibfnamefont {S.}~\bibnamefont {{Gunji}}}, \bibinfo {author} {\bibfnamefont {K.}~\bibnamefont {{Hayashida}}}, \bibinfo {author} {\bibfnamefont {W.}~\bibnamefont {{Iwakiri}}}, \bibinfo {author} {\bibfnamefont {S.~G.}\ \bibnamefont {{Jorstad}}}, \bibinfo {author} {\bibfnamefont {V.}~\bibnamefont
  {{Karas}}}, \bibinfo {author} {\bibfnamefont {F.}~\bibnamefont {{Kislat}}}, \bibinfo {author} {\bibfnamefont {T.}~\bibnamefont {{Kitaguchi}}}, \bibinfo {author} {\bibfnamefont {J.~J.}\ \bibnamefont {{Kolodziejczak}}}, \bibinfo {author} {\bibfnamefont {L.}~\bibnamefont {{Latronico}}}, \bibinfo {author} {\bibfnamefont {S.}~\bibnamefont {{Maldera}}}, \bibinfo {author} {\bibfnamefont {A.}~\bibnamefont {{Manfreda}}}, \bibinfo {author} {\bibfnamefont {F.}~\bibnamefont {{Marin}}}, \bibinfo {author} {\bibfnamefont {A.}~\bibnamefont {{Marinucci}}}, \bibinfo {author} {\bibfnamefont {A.~P.}\ \bibnamefont {{Marscher}}}, \bibinfo {author} {\bibfnamefont {H.~L.}\ \bibnamefont {{Marshall}}}, \bibinfo {author} {\bibfnamefont {F.}~\bibnamefont {{Massaro}}}, \bibinfo {author} {\bibfnamefont {G.}~\bibnamefont {{Matt}}}, \bibinfo {author} {\bibfnamefont {I.}~\bibnamefont {{Mitsuishi}}}, \bibinfo {author} {\bibfnamefont {T.}~\bibnamefont {{Mizuno}}}, \bibinfo {author} {\bibfnamefont {F.}~\bibnamefont {{Muleri}}}, \bibinfo
  {author} {\bibfnamefont {M.}~\bibnamefont {{Negro}}}, \bibinfo {author} {\bibfnamefont {C.-Y.}\ \bibnamefont {{Ng}}}, \bibinfo {author} {\bibfnamefont {S.~L.}\ \bibnamefont {{O'Dell}}}, \bibinfo {author} {\bibfnamefont {N.}~\bibnamefont {{Omodei}}}, \bibinfo {author} {\bibfnamefont {C.}~\bibnamefont {{Oppedisano}}}, \bibinfo {author} {\bibfnamefont {A.}~\bibnamefont {{Papitto}}}, \bibinfo {author} {\bibfnamefont {A.~L.}\ \bibnamefont {{Peirson}}}, \bibinfo {author} {\bibfnamefont {M.}~\bibnamefont {{Perri}}}, \bibinfo {author} {\bibfnamefont {M.}~\bibnamefont {{Pesce-Rollins}}}, \bibinfo {author} {\bibfnamefont {P.-O.}\ \bibnamefont {{Petrucci}}}, \bibinfo {author} {\bibfnamefont {M.}~\bibnamefont {{Pilia}}}, \bibinfo {author} {\bibfnamefont {A.}~\bibnamefont {{Possenti}}}, \bibinfo {author} {\bibfnamefont {S.}~\bibnamefont {{Puccetti}}}, \bibinfo {author} {\bibfnamefont {B.~D.}\ \bibnamefont {{Ramsey}}}, \bibinfo {author} {\bibfnamefont {J.}~\bibnamefont {{Rankin}}}, \bibinfo {author} {\bibfnamefont
  {A.}~\bibnamefont {{Ratheesh}}}, \bibinfo {author} {\bibfnamefont {O.~J.}\ \bibnamefont {{Roberts}}}, \bibinfo {author} {\bibfnamefont {R.~W.}\ \bibnamefont {{Romani}}}, \bibinfo {author} {\bibfnamefont {C.}~\bibnamefont {{Sgr{\`o}}}}, \bibinfo {author} {\bibfnamefont {P.}~\bibnamefont {{Slane}}}, \bibinfo {author} {\bibfnamefont {P.}~\bibnamefont {{Soffitta}}}, \bibinfo {author} {\bibfnamefont {G.}~\bibnamefont {{Spandre}}}, \bibinfo {author} {\bibfnamefont {D.~A.}\ \bibnamefont {{Swartz}}}, \bibinfo {author} {\bibfnamefont {T.}~\bibnamefont {{Tamagawa}}}, \bibinfo {author} {\bibfnamefont {F.}~\bibnamefont {{Tavecchio}}}, \bibinfo {author} {\bibfnamefont {R.}~\bibnamefont {{Taverna}}}, \bibinfo {author} {\bibfnamefont {Y.}~\bibnamefont {{Tawara}}}, \bibinfo {author} {\bibfnamefont {A.~F.}\ \bibnamefont {{Tennant}}}, \bibinfo {author} {\bibfnamefont {N.~E.}\ \bibnamefont {{Thomas}}}, \bibinfo {author} {\bibfnamefont {F.}~\bibnamefont {{Tombesi}}}, \bibinfo {author} {\bibfnamefont {A.}~\bibnamefont
  {{Trois}}}, \bibinfo {author} {\bibfnamefont {R.}~\bibnamefont {{Turolla}}}, \bibinfo {author} {\bibfnamefont {J.}~\bibnamefont {{Vink}}}, \bibinfo {author} {\bibfnamefont {K.}~\bibnamefont {{Wu}}}, \ and\ \bibinfo {author} {\bibfnamefont {F.}~\bibnamefont {{Xie}}},\ }\href {\doibase 10.1051/0004-6361/202450696} {\bibfield  {journal} {\bibinfo  {journal} {\aap}\ }\textbf {\bibinfo {volume} {691}},\ \bibinfo {eid} {A123} (\bibinfo {year} {2024}{\natexlab{b}})},\ \Eprint {http://arxiv.org/abs/2405.08107} {arXiv:2405.08107 [astro-ph.HE]} \BibitemShut {NoStop}%
\bibitem [{\citenamefont {{Forsblom}}\ \emph {et~al.}(2024)\citenamefont {{Forsblom}}, \citenamefont {{Tsygankov}}, \citenamefont {{Poutanen}}, \citenamefont {{Doroshenko}}, \citenamefont {{Mushtukov}}, \citenamefont {{Ng}}, \citenamefont {{Ravi}}, \citenamefont {{Marshall}}, \citenamefont {{Di Marco}}, \citenamefont {{La Monaca}}, \citenamefont {{Malacaria}}, \citenamefont {{Mastroserio}}, \citenamefont {{Loktev}}, \citenamefont {{Possenti}}, \citenamefont {{Suleimanov}}, \citenamefont {{Taverna}}, \citenamefont {{Agudo}}, \citenamefont {{Antonelli}}, \citenamefont {{Bachetti}}, \citenamefont {{Baldini}}, \citenamefont {{Baumgartner}}, \citenamefont {{Bellazzini}}, \citenamefont {{Bianchi}}, \citenamefont {{Bongiorno}}, \citenamefont {{Bonino}}, \citenamefont {{Brez}}, \citenamefont {{Bucciantini}}, \citenamefont {{Capitanio}}, \citenamefont {{Castellano}}, \citenamefont {{Cavazzuti}}, \citenamefont {{Chen}}, \citenamefont {{Ciprini}}, \citenamefont {{Costa}}, \citenamefont {{De Rosa}}, \citenamefont {{Del
  Monte}}, \citenamefont {{Di Gesu}}, \citenamefont {{Di Lalla}}, \citenamefont {{Donnarumma}}, \citenamefont {{Dov{\v{c}}iak}}, \citenamefont {{Ehlert}}, \citenamefont {{Enoto}}, \citenamefont {{Evangelista}}, \citenamefont {{Fabiani}}, \citenamefont {{Ferrazzoli}}, \citenamefont {{Garcia}}, \citenamefont {{Gunji}}, \citenamefont {{Hayashida}}, \citenamefont {{Heyl}}, \citenamefont {{Iwakiri}}, \citenamefont {{Jorstad}}, \citenamefont {{Kaaret}}, \citenamefont {{Karas}}, \citenamefont {{Kislat}}, \citenamefont {{Kitaguchi}}, \citenamefont {{Kolodziejczak}}, \citenamefont {{Krawczynski}}, \citenamefont {{Latronico}}, \citenamefont {{Liodakis}}, \citenamefont {{Maldera}}, \citenamefont {{Manfreda}}, \citenamefont {{Marin}}, \citenamefont {{Marinucci}}, \citenamefont {{Marscher}}, \citenamefont {{Massaro}}, \citenamefont {{Matt}}, \citenamefont {{Mitsuishi}}, \citenamefont {{Mizuno}}, \citenamefont {{Muleri}}, \citenamefont {{Negro}}, \citenamefont {{Ng}}, \citenamefont {{O'Dell}}, \citenamefont {{Omodei}},
  \citenamefont {{Oppedisano}}, \citenamefont {{Papitto}}, \citenamefont {{Pavlov}}, \citenamefont {{Peirson}}, \citenamefont {{Perri}}, \citenamefont {{Pesce-Rollins}}, \citenamefont {{Petrucci}}, \citenamefont {{Pilia}}, \citenamefont {{Puccetti}}, \citenamefont {{Ramsey}}, \citenamefont {{Rankin}}, \citenamefont {{Ratheesh}}, \citenamefont {{Roberts}}, \citenamefont {{Romani}}, \citenamefont {{Sgr{\`o}}}, \citenamefont {{Slane}}, \citenamefont {{Soffitta}}, \citenamefont {{Spandre}}, \citenamefont {{Swartz}}, \citenamefont {{Tamagawa}}, \citenamefont {{Tavecchio}}, \citenamefont {{Tawara}}, \citenamefont {{Tennant}}, \citenamefont {{Thomas}}, \citenamefont {{Tombesi}}, \citenamefont {{Trois}}, \citenamefont {{Turolla}}, \citenamefont {{Vink}}, \citenamefont {{Weisskopf}}, \citenamefont {{Wu}}, \citenamefont {{Xie}},\ and\ \citenamefont {{Zane}}}]{SMCX-1_Forsblom}%
  \BibitemOpen
  \bibfield  {author} {\bibinfo {author} {\bibfnamefont {S.~V.}\ \bibnamefont {{Forsblom}}}, \bibinfo {author} {\bibfnamefont {S.~S.}\ \bibnamefont {{Tsygankov}}}, \bibinfo {author} {\bibfnamefont {J.}~\bibnamefont {{Poutanen}}}, \bibinfo {author} {\bibfnamefont {V.}~\bibnamefont {{Doroshenko}}}, \bibinfo {author} {\bibfnamefont {A.~A.}\ \bibnamefont {{Mushtukov}}}, \bibinfo {author} {\bibfnamefont {M.}~\bibnamefont {{Ng}}}, \bibinfo {author} {\bibfnamefont {S.}~\bibnamefont {{Ravi}}}, \bibinfo {author} {\bibfnamefont {H.~L.}\ \bibnamefont {{Marshall}}}, \bibinfo {author} {\bibfnamefont {A.}~\bibnamefont {{Di Marco}}}, \bibinfo {author} {\bibfnamefont {F.}~\bibnamefont {{La Monaca}}}, \bibinfo {author} {\bibfnamefont {C.}~\bibnamefont {{Malacaria}}}, \bibinfo {author} {\bibfnamefont {G.}~\bibnamefont {{Mastroserio}}}, \bibinfo {author} {\bibfnamefont {V.}~\bibnamefont {{Loktev}}}, \bibinfo {author} {\bibfnamefont {A.}~\bibnamefont {{Possenti}}}, \bibinfo {author} {\bibfnamefont {V.~F.}\ \bibnamefont
  {{Suleimanov}}}, \bibinfo {author} {\bibfnamefont {R.}~\bibnamefont {{Taverna}}}, \bibinfo {author} {\bibfnamefont {I.}~\bibnamefont {{Agudo}}}, \bibinfo {author} {\bibfnamefont {L.~A.}\ \bibnamefont {{Antonelli}}}, \bibinfo {author} {\bibfnamefont {M.}~\bibnamefont {{Bachetti}}}, \bibinfo {author} {\bibfnamefont {L.}~\bibnamefont {{Baldini}}}, \bibinfo {author} {\bibfnamefont {W.~H.}\ \bibnamefont {{Baumgartner}}}, \bibinfo {author} {\bibfnamefont {R.}~\bibnamefont {{Bellazzini}}}, \bibinfo {author} {\bibfnamefont {S.}~\bibnamefont {{Bianchi}}}, \bibinfo {author} {\bibfnamefont {S.~D.}\ \bibnamefont {{Bongiorno}}}, \bibinfo {author} {\bibfnamefont {R.}~\bibnamefont {{Bonino}}}, \bibinfo {author} {\bibfnamefont {A.}~\bibnamefont {{Brez}}}, \bibinfo {author} {\bibfnamefont {N.}~\bibnamefont {{Bucciantini}}}, \bibinfo {author} {\bibfnamefont {F.}~\bibnamefont {{Capitanio}}}, \bibinfo {author} {\bibfnamefont {S.}~\bibnamefont {{Castellano}}}, \bibinfo {author} {\bibfnamefont {E.}~\bibnamefont {{Cavazzuti}}},
  \bibinfo {author} {\bibfnamefont {C.-T.}\ \bibnamefont {{Chen}}}, \bibinfo {author} {\bibfnamefont {S.}~\bibnamefont {{Ciprini}}}, \bibinfo {author} {\bibfnamefont {E.}~\bibnamefont {{Costa}}}, \bibinfo {author} {\bibfnamefont {A.}~\bibnamefont {{De Rosa}}}, \bibinfo {author} {\bibfnamefont {E.}~\bibnamefont {{Del Monte}}}, \bibinfo {author} {\bibfnamefont {L.}~\bibnamefont {{Di Gesu}}}, \bibinfo {author} {\bibfnamefont {N.}~\bibnamefont {{Di Lalla}}}, \bibinfo {author} {\bibfnamefont {I.}~\bibnamefont {{Donnarumma}}}, \bibinfo {author} {\bibfnamefont {M.}~\bibnamefont {{Dov{\v{c}}iak}}}, \bibinfo {author} {\bibfnamefont {S.~R.}\ \bibnamefont {{Ehlert}}}, \bibinfo {author} {\bibfnamefont {T.}~\bibnamefont {{Enoto}}}, \bibinfo {author} {\bibfnamefont {Y.}~\bibnamefont {{Evangelista}}}, \bibinfo {author} {\bibfnamefont {S.}~\bibnamefont {{Fabiani}}}, \bibinfo {author} {\bibfnamefont {R.}~\bibnamefont {{Ferrazzoli}}}, \bibinfo {author} {\bibfnamefont {J.~A.}\ \bibnamefont {{Garcia}}}, \bibinfo {author}
  {\bibfnamefont {S.}~\bibnamefont {{Gunji}}}, \bibinfo {author} {\bibfnamefont {K.}~\bibnamefont {{Hayashida}}}, \bibinfo {author} {\bibfnamefont {J.}~\bibnamefont {{Heyl}}}, \bibinfo {author} {\bibfnamefont {W.}~\bibnamefont {{Iwakiri}}}, \bibinfo {author} {\bibfnamefont {S.~G.}\ \bibnamefont {{Jorstad}}}, \bibinfo {author} {\bibfnamefont {P.}~\bibnamefont {{Kaaret}}}, \bibinfo {author} {\bibfnamefont {V.}~\bibnamefont {{Karas}}}, \bibinfo {author} {\bibfnamefont {F.}~\bibnamefont {{Kislat}}}, \bibinfo {author} {\bibfnamefont {T.}~\bibnamefont {{Kitaguchi}}}, \bibinfo {author} {\bibfnamefont {J.~J.}\ \bibnamefont {{Kolodziejczak}}}, \bibinfo {author} {\bibfnamefont {H.}~\bibnamefont {{Krawczynski}}}, \bibinfo {author} {\bibfnamefont {L.}~\bibnamefont {{Latronico}}}, \bibinfo {author} {\bibfnamefont {I.}~\bibnamefont {{Liodakis}}}, \bibinfo {author} {\bibfnamefont {S.}~\bibnamefont {{Maldera}}}, \bibinfo {author} {\bibfnamefont {A.}~\bibnamefont {{Manfreda}}}, \bibinfo {author} {\bibfnamefont
  {F.}~\bibnamefont {{Marin}}}, \bibinfo {author} {\bibfnamefont {A.}~\bibnamefont {{Marinucci}}}, \bibinfo {author} {\bibfnamefont {A.~P.}\ \bibnamefont {{Marscher}}}, \bibinfo {author} {\bibfnamefont {F.}~\bibnamefont {{Massaro}}}, \bibinfo {author} {\bibfnamefont {G.}~\bibnamefont {{Matt}}}, \bibinfo {author} {\bibfnamefont {I.}~\bibnamefont {{Mitsuishi}}}, \bibinfo {author} {\bibfnamefont {T.}~\bibnamefont {{Mizuno}}}, \bibinfo {author} {\bibfnamefont {F.}~\bibnamefont {{Muleri}}}, \bibinfo {author} {\bibfnamefont {M.}~\bibnamefont {{Negro}}}, \bibinfo {author} {\bibfnamefont {C.-Y.}\ \bibnamefont {{Ng}}}, \bibinfo {author} {\bibfnamefont {S.~L.}\ \bibnamefont {{O'Dell}}}, \bibinfo {author} {\bibfnamefont {N.}~\bibnamefont {{Omodei}}}, \bibinfo {author} {\bibfnamefont {C.}~\bibnamefont {{Oppedisano}}}, \bibinfo {author} {\bibfnamefont {A.}~\bibnamefont {{Papitto}}}, \bibinfo {author} {\bibfnamefont {G.~G.}\ \bibnamefont {{Pavlov}}}, \bibinfo {author} {\bibfnamefont {A.~L.}\ \bibnamefont {{Peirson}}},
  \bibinfo {author} {\bibfnamefont {M.}~\bibnamefont {{Perri}}}, \bibinfo {author} {\bibfnamefont {M.}~\bibnamefont {{Pesce-Rollins}}}, \bibinfo {author} {\bibfnamefont {P.-O.}\ \bibnamefont {{Petrucci}}}, \bibinfo {author} {\bibfnamefont {M.}~\bibnamefont {{Pilia}}}, \bibinfo {author} {\bibfnamefont {S.}~\bibnamefont {{Puccetti}}}, \bibinfo {author} {\bibfnamefont {B.~D.}\ \bibnamefont {{Ramsey}}}, \bibinfo {author} {\bibfnamefont {J.}~\bibnamefont {{Rankin}}}, \bibinfo {author} {\bibfnamefont {A.}~\bibnamefont {{Ratheesh}}}, \bibinfo {author} {\bibfnamefont {O.~J.}\ \bibnamefont {{Roberts}}}, \bibinfo {author} {\bibfnamefont {R.~W.}\ \bibnamefont {{Romani}}}, \bibinfo {author} {\bibfnamefont {C.}~\bibnamefont {{Sgr{\`o}}}}, \bibinfo {author} {\bibfnamefont {P.}~\bibnamefont {{Slane}}}, \bibinfo {author} {\bibfnamefont {P.}~\bibnamefont {{Soffitta}}}, \bibinfo {author} {\bibfnamefont {G.}~\bibnamefont {{Spandre}}}, \bibinfo {author} {\bibfnamefont {D.~A.}\ \bibnamefont {{Swartz}}}, \bibinfo {author}
  {\bibfnamefont {T.}~\bibnamefont {{Tamagawa}}}, \bibinfo {author} {\bibfnamefont {F.}~\bibnamefont {{Tavecchio}}}, \bibinfo {author} {\bibfnamefont {Y.}~\bibnamefont {{Tawara}}}, \bibinfo {author} {\bibfnamefont {A.~F.}\ \bibnamefont {{Tennant}}}, \bibinfo {author} {\bibfnamefont {N.~E.}\ \bibnamefont {{Thomas}}}, \bibinfo {author} {\bibfnamefont {F.}~\bibnamefont {{Tombesi}}}, \bibinfo {author} {\bibfnamefont {A.}~\bibnamefont {{Trois}}}, \bibinfo {author} {\bibfnamefont {R.}~\bibnamefont {{Turolla}}}, \bibinfo {author} {\bibfnamefont {J.}~\bibnamefont {{Vink}}}, \bibinfo {author} {\bibfnamefont {M.~C.}\ \bibnamefont {{Weisskopf}}}, \bibinfo {author} {\bibfnamefont {K.}~\bibnamefont {{Wu}}}, \bibinfo {author} {\bibfnamefont {F.}~\bibnamefont {{Xie}}}, \ and\ \bibinfo {author} {\bibfnamefont {S.}~\bibnamefont {{Zane}}},\ }\href {\doibase 10.1051/0004-6361/202450937} {\bibfield  {journal} {\bibinfo  {journal} {\aap}\ }\textbf {\bibinfo {volume} {691}},\ \bibinfo {eid} {A216} (\bibinfo {year} {2024})},\
  \Eprint {http://arxiv.org/abs/2406.08988} {arXiv:2406.08988 [astro-ph.HE]} \BibitemShut {NoStop}%
\bibitem [{\citenamefont {{Loktev}}\ \emph {et~al.}(2025)\citenamefont {{Loktev}}, \citenamefont {{Forsblom}}, \citenamefont {{Tsygankov}}, \citenamefont {{Poutanen}}, \citenamefont {{Mushtukov}}, \citenamefont {{Di Marco}}, \citenamefont {{Heyl}}, \citenamefont {{Kelly}}, \citenamefont {{La Monaca}}, \citenamefont {{Ng}}, \citenamefont {{Ravi}}, \citenamefont {{Salganik}}, \citenamefont {{Santangelo}}, \citenamefont {{Suleimanov}},\ and\ \citenamefont {{Zane}}}]{Loktev25}%
  \BibitemOpen
  \bibfield  {author} {\bibinfo {author} {\bibfnamefont {V.}~\bibnamefont {{Loktev}}}, \bibinfo {author} {\bibfnamefont {S.~V.}\ \bibnamefont {{Forsblom}}}, \bibinfo {author} {\bibfnamefont {S.~S.}\ \bibnamefont {{Tsygankov}}}, \bibinfo {author} {\bibfnamefont {J.}~\bibnamefont {{Poutanen}}}, \bibinfo {author} {\bibfnamefont {A.~A.}\ \bibnamefont {{Mushtukov}}}, \bibinfo {author} {\bibfnamefont {A.}~\bibnamefont {{Di Marco}}}, \bibinfo {author} {\bibfnamefont {J.}~\bibnamefont {{Heyl}}}, \bibinfo {author} {\bibfnamefont {R.~M.~E.}\ \bibnamefont {{Kelly}}}, \bibinfo {author} {\bibfnamefont {F.}~\bibnamefont {{La Monaca}}}, \bibinfo {author} {\bibfnamefont {M.}~\bibnamefont {{Ng}}}, \bibinfo {author} {\bibfnamefont {S.}~\bibnamefont {{Ravi}}}, \bibinfo {author} {\bibfnamefont {A.}~\bibnamefont {{Salganik}}}, \bibinfo {author} {\bibfnamefont {A.}~\bibnamefont {{Santangelo}}}, \bibinfo {author} {\bibfnamefont {V.~F.}\ \bibnamefont {{Suleimanov}}}, \ and\ \bibinfo {author} {\bibfnamefont {S.}~\bibnamefont
  {{Zane}}},\ }\href {\doibase 10.48550/arXiv.2503.13720} {\bibfield  {journal} {\bibinfo  {journal} {\aap, submitted}\ ,\ \bibinfo {eid} {arXiv:2503.13720}} (\bibinfo {year} {2025})},\ \Eprint {http://arxiv.org/abs/2503.13720} {arXiv:2503.13720 [astro-ph.HE]} \BibitemShut {NoStop}%
\bibitem [{\citenamefont {{Zhou}}\ \emph {et~al.}(2025)\citenamefont {{Zhou}}, \citenamefont {{Ducci}}, \citenamefont {{Liu}}, \citenamefont {{Tsygankov}}, \citenamefont {{Forsblom}}, \citenamefont {{Mushtukov}}, \citenamefont {{Suleimanov}}, \citenamefont {{Poutanen}}, \citenamefont {{Wang}}, \citenamefont {{Di Marco}}, \citenamefont {{Doroshenko}}, \citenamefont {{La Monaca}}, \citenamefont {{Loktev}}, \citenamefont {{Salganik}},\ and\ \citenamefont {{Santangelo}}}]{ZhouML2025}%
  \BibitemOpen
  \bibfield  {author} {\bibinfo {author} {\bibfnamefont {M.}~\bibnamefont {{Zhou}}}, \bibinfo {author} {\bibfnamefont {L.}~\bibnamefont {{Ducci}}}, \bibinfo {author} {\bibfnamefont {H.}~\bibnamefont {{Liu}}}, \bibinfo {author} {\bibfnamefont {S.~S.}\ \bibnamefont {{Tsygankov}}}, \bibinfo {author} {\bibfnamefont {S.~V.}\ \bibnamefont {{Forsblom}}}, \bibinfo {author} {\bibfnamefont {A.~A.}\ \bibnamefont {{Mushtukov}}}, \bibinfo {author} {\bibfnamefont {V.~F.}\ \bibnamefont {{Suleimanov}}}, \bibinfo {author} {\bibfnamefont {J.}~\bibnamefont {{Poutanen}}}, \bibinfo {author} {\bibfnamefont {P.}~\bibnamefont {{Wang}}}, \bibinfo {author} {\bibfnamefont {A.}~\bibnamefont {{Di Marco}}}, \bibinfo {author} {\bibfnamefont {V.}~\bibnamefont {{Doroshenko}}}, \bibinfo {author} {\bibfnamefont {F.}~\bibnamefont {{La Monaca}}}, \bibinfo {author} {\bibfnamefont {V.}~\bibnamefont {{Loktev}}}, \bibinfo {author} {\bibfnamefont {A.}~\bibnamefont {{Salganik}}}, \ and\ \bibinfo {author} {\bibfnamefont {A.}~\bibnamefont
  {{Santangelo}}},\ }\href {\doibase 10.48550/arXiv.2507.17873} {\bibfield  {journal} {\bibinfo  {journal} {arXiv e-prints}\ ,\ \bibinfo {eid} {arXiv:2507.17873}} (\bibinfo {year} {2025})},\ \Eprint {http://arxiv.org/abs/2507.17873} {arXiv:2507.17873 [astro-ph.HE]} \BibitemShut {NoStop}%
\bibitem [{\citenamefont {{Nitindala}}\ \emph {et~al.}(2025)\citenamefont {{Nitindala}}, \citenamefont {{Veledina}},\ and\ \citenamefont {{Poutanen}}}]{Nitindala25}%
  \BibitemOpen
  \bibfield  {author} {\bibinfo {author} {\bibfnamefont {A.~P.}\ \bibnamefont {{Nitindala}}}, \bibinfo {author} {\bibfnamefont {A.}~\bibnamefont {{Veledina}}}, \ and\ \bibinfo {author} {\bibfnamefont {J.}~\bibnamefont {{Poutanen}}},\ }\href {\doibase 10.1051/0004-6361/202453188} {\bibfield  {journal} {\bibinfo  {journal} {\aap}\ }\textbf {\bibinfo {volume} {694}},\ \bibinfo {eid} {A230} (\bibinfo {year} {2025})},\ \Eprint {http://arxiv.org/abs/2411.18299} {arXiv:2411.18299 [astro-ph.HE]} \BibitemShut {NoStop}%
\bibitem [{\citenamefont {{Tsygankov}}\ \emph {et~al.}(2016)\citenamefont {{Tsygankov}}, \citenamefont {{Lutovinov}}, \citenamefont {{Doroshenko}}, \citenamefont {{Mushtukov}}, \citenamefont {{Suleimanov}},\ and\ \citenamefont {{Poutanen}}}]{2016A&A...593A..16T}%
  \BibitemOpen
  \bibfield  {author} {\bibinfo {author} {\bibfnamefont {S.~S.}\ \bibnamefont {{Tsygankov}}}, \bibinfo {author} {\bibfnamefont {A.~A.}\ \bibnamefont {{Lutovinov}}}, \bibinfo {author} {\bibfnamefont {V.}~\bibnamefont {{Doroshenko}}}, \bibinfo {author} {\bibfnamefont {A.~A.}\ \bibnamefont {{Mushtukov}}}, \bibinfo {author} {\bibfnamefont {V.}~\bibnamefont {{Suleimanov}}}, \ and\ \bibinfo {author} {\bibfnamefont {J.}~\bibnamefont {{Poutanen}}},\ }\href {\doibase 10.1051/0004-6361/201628236} {\bibfield  {journal} {\bibinfo  {journal} {\aap}\ }\textbf {\bibinfo {volume} {593}},\ \bibinfo {eid} {A16} (\bibinfo {year} {2016})},\ \Eprint {http://arxiv.org/abs/1602.03177} {arXiv:1602.03177 [astro-ph.HE]} \BibitemShut {NoStop}%
\bibitem [{\citenamefont {{Doroshenko}}\ \emph {et~al.}(2020{\natexlab{a}})\citenamefont {{Doroshenko}}, \citenamefont {{Santangelo}}, \citenamefont {{Suleimanov}},\ and\ \citenamefont {{Tsygankov}}}]{Doroshenko2020b}%
  \BibitemOpen
  \bibfield  {author} {\bibinfo {author} {\bibfnamefont {V.}~\bibnamefont {{Doroshenko}}}, \bibinfo {author} {\bibfnamefont {A.}~\bibnamefont {{Santangelo}}}, \bibinfo {author} {\bibfnamefont {V.~F.}\ \bibnamefont {{Suleimanov}}}, \ and\ \bibinfo {author} {\bibfnamefont {S.~S.}\ \bibnamefont {{Tsygankov}}},\ }\href {\doibase 10.1051/0004-6361/202038948} {\bibfield  {journal} {\bibinfo  {journal} {\aap}\ }\textbf {\bibinfo {volume} {643}},\ \bibinfo {eid} {A173} (\bibinfo {year} {2020}{\natexlab{a}})},\ \Eprint {http://arxiv.org/abs/2009.14064} {arXiv:2009.14064 [astro-ph.HE]} \BibitemShut {NoStop}%
\bibitem [{\citenamefont {{Illarionov}}\ and\ \citenamefont {{Sunyaev}}(1975)}]{Illarionov1975}%
  \BibitemOpen
  \bibfield  {author} {\bibinfo {author} {\bibfnamefont {A.~F.}\ \bibnamefont {{Illarionov}}}\ and\ \bibinfo {author} {\bibfnamefont {R.~A.}\ \bibnamefont {{Sunyaev}}},\ }\href@noop {} {\bibfield  {journal} {\bibinfo  {journal} {\aap}\ }\textbf {\bibinfo {volume} {39}},\ \bibinfo {pages} {185} (\bibinfo {year} {1975})}\BibitemShut {NoStop}%
\bibitem [{\citenamefont {{Kong}}\ \emph {et~al.}(2022)\citenamefont {{Kong}}, \citenamefont {{Zhang}}, \citenamefont {{Zhang}}, \citenamefont {{Ji}}, \citenamefont {{Doroshenko}}, \citenamefont {{Santangelo}}, \citenamefont {{Chen}}, \citenamefont {{Lu}}, \citenamefont {{Ge}}, \citenamefont {{Wang}}, \citenamefont {{Tao}}, \citenamefont {{Qu}}, \citenamefont {{Li}}, \citenamefont {{Liu}}, \citenamefont {{Liao}}, \citenamefont {{Chang}}, \citenamefont {{Peng}},\ and\ \citenamefont {{Shui}}}]{Kong2022ApJ...933L...3K}%
  \BibitemOpen
  \bibfield  {author} {\bibinfo {author} {\bibfnamefont {L.-D.}\ \bibnamefont {{Kong}}}, \bibinfo {author} {\bibfnamefont {S.}~\bibnamefont {{Zhang}}}, \bibinfo {author} {\bibfnamefont {S.-N.}\ \bibnamefont {{Zhang}}}, \bibinfo {author} {\bibfnamefont {L.}~\bibnamefont {{Ji}}}, \bibinfo {author} {\bibfnamefont {V.}~\bibnamefont {{Doroshenko}}}, \bibinfo {author} {\bibfnamefont {A.}~\bibnamefont {{Santangelo}}}, \bibinfo {author} {\bibfnamefont {Y.-P.}\ \bibnamefont {{Chen}}}, \bibinfo {author} {\bibfnamefont {F.-J.}\ \bibnamefont {{Lu}}}, \bibinfo {author} {\bibfnamefont {M.-Y.}\ \bibnamefont {{Ge}}}, \bibinfo {author} {\bibfnamefont {P.-J.}\ \bibnamefont {{Wang}}}, \bibinfo {author} {\bibfnamefont {L.}~\bibnamefont {{Tao}}}, \bibinfo {author} {\bibfnamefont {J.-L.}\ \bibnamefont {{Qu}}}, \bibinfo {author} {\bibfnamefont {T.-P.}\ \bibnamefont {{Li}}}, \bibinfo {author} {\bibfnamefont {C.-Z.}\ \bibnamefont {{Liu}}}, \bibinfo {author} {\bibfnamefont {J.-Y.}\ \bibnamefont {{Liao}}}, \bibinfo {author}
  {\bibfnamefont {Z.}~\bibnamefont {{Chang}}}, \bibinfo {author} {\bibfnamefont {J.-Q.}\ \bibnamefont {{Peng}}}, \ and\ \bibinfo {author} {\bibfnamefont {Q.-C.}\ \bibnamefont {{Shui}}},\ }\href {\doibase 10.3847/2041-8213/ac7711} {\bibfield  {journal} {\bibinfo  {journal} {\apjl}\ }\textbf {\bibinfo {volume} {933}},\ \bibinfo {eid} {L3} (\bibinfo {year} {2022})},\ \Eprint {http://arxiv.org/abs/2206.04283} {arXiv:2206.04283 [astro-ph.HE]} \BibitemShut {NoStop}%
\bibitem [{\citenamefont {{Tsygankov}}\ \emph {et~al.}(2017)\citenamefont {{Tsygankov}}, \citenamefont {{Wijnands}}, \citenamefont {{Lutovinov}}, \citenamefont {{Degenaar}},\ and\ \citenamefont {{Poutanen}}}]{2017MNRAS.470..126T}%
  \BibitemOpen
  \bibfield  {author} {\bibinfo {author} {\bibfnamefont {S.~S.}\ \bibnamefont {{Tsygankov}}}, \bibinfo {author} {\bibfnamefont {R.}~\bibnamefont {{Wijnands}}}, \bibinfo {author} {\bibfnamefont {A.~A.}\ \bibnamefont {{Lutovinov}}}, \bibinfo {author} {\bibfnamefont {N.}~\bibnamefont {{Degenaar}}}, \ and\ \bibinfo {author} {\bibfnamefont {J.}~\bibnamefont {{Poutanen}}},\ }\href {\doibase 10.1093/mnras/stx1255} {\bibfield  {journal} {\bibinfo  {journal} {\mnras}\ }\textbf {\bibinfo {volume} {470}},\ \bibinfo {pages} {126} (\bibinfo {year} {2017})},\ \Eprint {http://arxiv.org/abs/1703.04634} {arXiv:1703.04634 [astro-ph.HE]} \BibitemShut {NoStop}%
\bibitem [{\citenamefont {{Tsygankov}}\ \emph {et~al.}(2019{\natexlab{a}})\citenamefont {{Tsygankov}}, \citenamefont {{Rouco Escorial}}, \citenamefont {{Suleimanov}}, \citenamefont {{Mushtukov}}, \citenamefont {{Doroshenko}}, \citenamefont {{Lutovinov}}, \citenamefont {{Wijnands}},\ and\ \citenamefont {{Poutanen}}}]{2019MNRAS.483L.144T}%
  \BibitemOpen
  \bibfield  {author} {\bibinfo {author} {\bibfnamefont {S.~S.}\ \bibnamefont {{Tsygankov}}}, \bibinfo {author} {\bibfnamefont {A.}~\bibnamefont {{Rouco Escorial}}}, \bibinfo {author} {\bibfnamefont {V.~F.}\ \bibnamefont {{Suleimanov}}}, \bibinfo {author} {\bibfnamefont {A.~A.}\ \bibnamefont {{Mushtukov}}}, \bibinfo {author} {\bibfnamefont {V.}~\bibnamefont {{Doroshenko}}}, \bibinfo {author} {\bibfnamefont {A.~A.}\ \bibnamefont {{Lutovinov}}}, \bibinfo {author} {\bibfnamefont {R.}~\bibnamefont {{Wijnands}}}, \ and\ \bibinfo {author} {\bibfnamefont {J.}~\bibnamefont {{Poutanen}}},\ }\href {\doibase 10.1093/mnrasl/sly236} {\bibfield  {journal} {\bibinfo  {journal} {\mnras}\ }\textbf {\bibinfo {volume} {483}},\ \bibinfo {pages} {L144} (\bibinfo {year} {2019}{\natexlab{a}})},\ \Eprint {http://arxiv.org/abs/1810.13307} {arXiv:1810.13307 [astro-ph.HE]} \BibitemShut {NoStop}%
\bibitem [{\citenamefont {{Tsygankov}}\ \emph {et~al.}(2019{\natexlab{b}})\citenamefont {{Tsygankov}}, \citenamefont {{Doroshenko}}, \citenamefont {{Mushtukov}}, \citenamefont {{Suleimanov}}, \citenamefont {{Lutovinov}},\ and\ \citenamefont {{Poutanen}}}]{2019MNRAS.487L..30T}%
  \BibitemOpen
  \bibfield  {author} {\bibinfo {author} {\bibfnamefont {S.~S.}\ \bibnamefont {{Tsygankov}}}, \bibinfo {author} {\bibfnamefont {V.}~\bibnamefont {{Doroshenko}}}, \bibinfo {author} {\bibfnamefont {A.~A.}\ \bibnamefont {{Mushtukov}}}, \bibinfo {author} {\bibfnamefont {V.~F.}\ \bibnamefont {{Suleimanov}}}, \bibinfo {author} {\bibfnamefont {A.~A.}\ \bibnamefont {{Lutovinov}}}, \ and\ \bibinfo {author} {\bibfnamefont {J.}~\bibnamefont {{Poutanen}}},\ }\href {\doibase 10.1093/mnrasl/slz079} {\bibfield  {journal} {\bibinfo  {journal} {\mnras}\ }\textbf {\bibinfo {volume} {487}},\ \bibinfo {pages} {L30} (\bibinfo {year} {2019}{\natexlab{b}})},\ \Eprint {http://arxiv.org/abs/1905.09496} {arXiv:1905.09496 [astro-ph.HE]} \BibitemShut {NoStop}%
\bibitem [{\citenamefont {{Suleimanov}}\ \emph {et~al.}(2018)\citenamefont {{Suleimanov}}, \citenamefont {{Poutanen}},\ and\ \citenamefont {{Werner}}}]{Suleimanov18}%
  \BibitemOpen
  \bibfield  {author} {\bibinfo {author} {\bibfnamefont {V.~F.}\ \bibnamefont {{Suleimanov}}}, \bibinfo {author} {\bibfnamefont {J.}~\bibnamefont {{Poutanen}}}, \ and\ \bibinfo {author} {\bibfnamefont {K.}~\bibnamefont {{Werner}}},\ }\href {\doibase 10.1051/0004-6361/201833581} {\bibfield  {journal} {\bibinfo  {journal} {\aap}\ }\textbf {\bibinfo {volume} {619}},\ \bibinfo {eid} {A114} (\bibinfo {year} {2018})},\ \Eprint {http://arxiv.org/abs/1808.10655} {arXiv:1808.10655 [astro-ph.HE]} \BibitemShut {NoStop}%
\bibitem [{\citenamefont {{Doroshenko}}\ \emph {et~al.}(2017)\citenamefont {{Doroshenko}}, \citenamefont {{Tsygankov}}, \citenamefont {{Mushtukov}}, \citenamefont {{Lutovinov}}, \citenamefont {{Santangelo}}, \citenamefont {{Suleimanov}},\ and\ \citenamefont {{Poutanen}}}]{Doroshenko2017}%
  \BibitemOpen
  \bibfield  {author} {\bibinfo {author} {\bibfnamefont {V.}~\bibnamefont {{Doroshenko}}}, \bibinfo {author} {\bibfnamefont {S.~S.}\ \bibnamefont {{Tsygankov}}}, \bibinfo {author} {\bibfnamefont {A.~A.}\ \bibnamefont {{Mushtukov}}}, \bibinfo {author} {\bibfnamefont {A.~A.}\ \bibnamefont {{Lutovinov}}}, \bibinfo {author} {\bibfnamefont {A.}~\bibnamefont {{Santangelo}}}, \bibinfo {author} {\bibfnamefont {V.~F.}\ \bibnamefont {{Suleimanov}}}, \ and\ \bibinfo {author} {\bibfnamefont {J.}~\bibnamefont {{Poutanen}}},\ }\href {\doibase 10.1093/mnras/stw3236} {\bibfield  {journal} {\bibinfo  {journal} {\mnras}\ }\textbf {\bibinfo {volume} {466}},\ \bibinfo {pages} {2143} (\bibinfo {year} {2017})},\ \Eprint {http://arxiv.org/abs/1607.03933} {arXiv:1607.03933 [astro-ph.HE]} \BibitemShut {NoStop}%
\bibitem [{\citenamefont {{Kong}}\ \emph {et~al.}(2021)\citenamefont {{Kong}}, \citenamefont {{Zhang}}, \citenamefont {{Ji}}, \citenamefont {{Reig}}, \citenamefont {{Doroshenko}}, \citenamefont {{Santangelo}}, \citenamefont {{Staubert}}, \citenamefont {{Zhang}}, \citenamefont {{Soria}}, \citenamefont {{Chang}}, \citenamefont {{Chen}}, \citenamefont {{Wang}}, \citenamefont {{Tao}},\ and\ \citenamefont {{Qu}}}]{Kong2021}%
  \BibitemOpen
  \bibfield  {author} {\bibinfo {author} {\bibfnamefont {L.~D.}\ \bibnamefont {{Kong}}}, \bibinfo {author} {\bibfnamefont {S.}~\bibnamefont {{Zhang}}}, \bibinfo {author} {\bibfnamefont {L.}~\bibnamefont {{Ji}}}, \bibinfo {author} {\bibfnamefont {P.}~\bibnamefont {{Reig}}}, \bibinfo {author} {\bibfnamefont {V.}~\bibnamefont {{Doroshenko}}}, \bibinfo {author} {\bibfnamefont {A.}~\bibnamefont {{Santangelo}}}, \bibinfo {author} {\bibfnamefont {R.}~\bibnamefont {{Staubert}}}, \bibinfo {author} {\bibfnamefont {S.~N.}\ \bibnamefont {{Zhang}}}, \bibinfo {author} {\bibfnamefont {R.}~\bibnamefont {{Soria}}}, \bibinfo {author} {\bibfnamefont {Z.}~\bibnamefont {{Chang}}}, \bibinfo {author} {\bibfnamefont {Y.~P.}\ \bibnamefont {{Chen}}}, \bibinfo {author} {\bibfnamefont {P.~J.}\ \bibnamefont {{Wang}}}, \bibinfo {author} {\bibfnamefont {L.}~\bibnamefont {{Tao}}}, \ and\ \bibinfo {author} {\bibfnamefont {J.~L.}\ \bibnamefont {{Qu}}},\ }\href {\doibase 10.3847/2041-8213/ac1ad3} {\bibfield  {journal} {\bibinfo  {journal}
  {\apjl}\ }\textbf {\bibinfo {volume} {917}},\ \bibinfo {eid} {L38} (\bibinfo {year} {2021})},\ \Eprint {http://arxiv.org/abs/2108.02485} {arXiv:2108.02485 [astro-ph.HE]} \BibitemShut {NoStop}%
\bibitem [{\citenamefont {{Wang}}\ \emph {et~al.}(2022)\citenamefont {{Wang}}, \citenamefont {{Kong}}, \citenamefont {{Zhang}}, \citenamefont {{Doroshenko}}, \citenamefont {{Santangelo}}, \citenamefont {{Ji}}, \citenamefont {{Yorgancioglu}}, \citenamefont {{Chen}}, \citenamefont {{Zhang}}, \citenamefont {{Qu}}, \citenamefont {{Ge}}, \citenamefont {{Li}}, \citenamefont {{Chang}}, \citenamefont {{Tao}}, \citenamefont {{Peng}},\ and\ \citenamefont {{Shui}}}]{WangPJ2022}%
  \BibitemOpen
  \bibfield  {author} {\bibinfo {author} {\bibfnamefont {P.~J.}\ \bibnamefont {{Wang}}}, \bibinfo {author} {\bibfnamefont {L.~D.}\ \bibnamefont {{Kong}}}, \bibinfo {author} {\bibfnamefont {S.}~\bibnamefont {{Zhang}}}, \bibinfo {author} {\bibfnamefont {V.}~\bibnamefont {{Doroshenko}}}, \bibinfo {author} {\bibfnamefont {A.}~\bibnamefont {{Santangelo}}}, \bibinfo {author} {\bibfnamefont {L.}~\bibnamefont {{Ji}}}, \bibinfo {author} {\bibfnamefont {E.~S.}\ \bibnamefont {{Yorgancioglu}}}, \bibinfo {author} {\bibfnamefont {Y.~P.}\ \bibnamefont {{Chen}}}, \bibinfo {author} {\bibfnamefont {S.~N.}\ \bibnamefont {{Zhang}}}, \bibinfo {author} {\bibfnamefont {J.~L.}\ \bibnamefont {{Qu}}}, \bibinfo {author} {\bibfnamefont {M.~Y.}\ \bibnamefont {{Ge}}}, \bibinfo {author} {\bibfnamefont {J.}~\bibnamefont {{Li}}}, \bibinfo {author} {\bibfnamefont {Z.}~\bibnamefont {{Chang}}}, \bibinfo {author} {\bibfnamefont {L.}~\bibnamefont {{Tao}}}, \bibinfo {author} {\bibfnamefont {J.~Q.}\ \bibnamefont {{Peng}}}, \ and\ \bibinfo {author}
  {\bibfnamefont {Q.~C.}\ \bibnamefont {{Shui}}},\ }\href {\doibase 10.3847/1538-4357/ac8230} {\bibfield  {journal} {\bibinfo  {journal} {\apj}\ }\textbf {\bibinfo {volume} {935}},\ \bibinfo {eid} {125} (\bibinfo {year} {2022})},\ \Eprint {http://arxiv.org/abs/2208.13340} {arXiv:2208.13340 [astro-ph.HE]} \BibitemShut {NoStop}%
\bibitem [{\citenamefont {{Postnov}}\ \emph {et~al.}(2015)\citenamefont {{Postnov}}, \citenamefont {{Gornostaev}}, \citenamefont {{Klochkov}}, \citenamefont {{Laplace}}, \citenamefont {{Lukin}},\ and\ \citenamefont {{Shakura}}}]{Postnov2015}%
  \BibitemOpen
  \bibfield  {author} {\bibinfo {author} {\bibfnamefont {K.~A.}\ \bibnamefont {{Postnov}}}, \bibinfo {author} {\bibfnamefont {M.~I.}\ \bibnamefont {{Gornostaev}}}, \bibinfo {author} {\bibfnamefont {D.}~\bibnamefont {{Klochkov}}}, \bibinfo {author} {\bibfnamefont {E.}~\bibnamefont {{Laplace}}}, \bibinfo {author} {\bibfnamefont {V.~V.}\ \bibnamefont {{Lukin}}}, \ and\ \bibinfo {author} {\bibfnamefont {N.~I.}\ \bibnamefont {{Shakura}}},\ }\href {\doibase 10.1093/mnras/stv1393} {\bibfield  {journal} {\bibinfo  {journal} {\mnras}\ }\textbf {\bibinfo {volume} {452}},\ \bibinfo {pages} {1601} (\bibinfo {year} {2015})},\ \Eprint {http://arxiv.org/abs/1506.07082} {arXiv:1506.07082 [astro-ph.HE]} \BibitemShut {NoStop}%
\bibitem [{\citenamefont {{Reig}}\ and\ \citenamefont {{Nespoli}}(2013)}]{Reig2013}%
  \BibitemOpen
  \bibfield  {author} {\bibinfo {author} {\bibfnamefont {P.}~\bibnamefont {{Reig}}}\ and\ \bibinfo {author} {\bibfnamefont {E.}~\bibnamefont {{Nespoli}}},\ }\href {\doibase 10.1051/0004-6361/201219806} {\bibfield  {journal} {\bibinfo  {journal} {\aap}\ }\textbf {\bibinfo {volume} {551}},\ \bibinfo {eid} {A1} (\bibinfo {year} {2013})},\ \Eprint {http://arxiv.org/abs/1212.5944} {arXiv:1212.5944 [astro-ph.HE]} \BibitemShut {NoStop}%
\bibitem [{\citenamefont {{Poutanen}}\ \emph {et~al.}(2013)\citenamefont {{Poutanen}}, \citenamefont {{Mushtukov}}, \citenamefont {{Suleimanov}}, \citenamefont {{Tsygankov}}, \citenamefont {{Nagirner}}, \citenamefont {{Doroshenko}},\ and\ \citenamefont {{Lutovinov}}}]{2013ApJ...777..115P}%
  \BibitemOpen
  \bibfield  {author} {\bibinfo {author} {\bibfnamefont {J.}~\bibnamefont {{Poutanen}}}, \bibinfo {author} {\bibfnamefont {A.~A.}\ \bibnamefont {{Mushtukov}}}, \bibinfo {author} {\bibfnamefont {V.~F.}\ \bibnamefont {{Suleimanov}}}, \bibinfo {author} {\bibfnamefont {S.~S.}\ \bibnamefont {{Tsygankov}}}, \bibinfo {author} {\bibfnamefont {D.~I.}\ \bibnamefont {{Nagirner}}}, \bibinfo {author} {\bibfnamefont {V.}~\bibnamefont {{Doroshenko}}}, \ and\ \bibinfo {author} {\bibfnamefont {A.~A.}\ \bibnamefont {{Lutovinov}}},\ }\href {\doibase 10.1088/0004-637X/777/2/115} {\bibfield  {journal} {\bibinfo  {journal} {\apj}\ }\textbf {\bibinfo {volume} {777}},\ \bibinfo {eid} {115} (\bibinfo {year} {2013})},\ \Eprint {http://arxiv.org/abs/1304.2633} {arXiv:1304.2633 [astro-ph.HE]} \BibitemShut {NoStop}%
\bibitem [{\citenamefont {{Fabbiano}}(1989)}]{Fabbiano1989}%
  \BibitemOpen
  \bibfield  {author} {\bibinfo {author} {\bibfnamefont {G.}~\bibnamefont {{Fabbiano}}},\ }\href {\doibase 10.1146/annurev.aa.27.090189.000511} {\bibfield  {journal} {\bibinfo  {journal} {\araa}\ }\textbf {\bibinfo {volume} {27}},\ \bibinfo {pages} {87} (\bibinfo {year} {1989})}\BibitemShut {NoStop}%
\bibitem [{\citenamefont {{Kaaret}}\ \emph {et~al.}(2017)\citenamefont {{Kaaret}}, \citenamefont {{Feng}},\ and\ \citenamefont {{Roberts}}}]{Kaaret2017}%
  \BibitemOpen
  \bibfield  {author} {\bibinfo {author} {\bibfnamefont {P.}~\bibnamefont {{Kaaret}}}, \bibinfo {author} {\bibfnamefont {H.}~\bibnamefont {{Feng}}}, \ and\ \bibinfo {author} {\bibfnamefont {T.~P.}\ \bibnamefont {{Roberts}}},\ }\href {\doibase 10.1146/annurev-astro-091916-055259} {\bibfield  {journal} {\bibinfo  {journal} {\araa}\ }\textbf {\bibinfo {volume} {55}},\ \bibinfo {pages} {303} (\bibinfo {year} {2017})},\ \Eprint {http://arxiv.org/abs/1703.10728} {arXiv:1703.10728 [astro-ph.HE]} \BibitemShut {NoStop}%
\bibitem [{\citenamefont {{Fabrika}}\ \emph {et~al.}(2021)\citenamefont {{Fabrika}}, \citenamefont {{Atapin}}, \citenamefont {{Vinokurov}},\ and\ \citenamefont {{Sholukhova}}}]{Fabrika2021}%
  \BibitemOpen
  \bibfield  {author} {\bibinfo {author} {\bibfnamefont {S.~N.}\ \bibnamefont {{Fabrika}}}, \bibinfo {author} {\bibfnamefont {K.~E.}\ \bibnamefont {{Atapin}}}, \bibinfo {author} {\bibfnamefont {A.~S.}\ \bibnamefont {{Vinokurov}}}, \ and\ \bibinfo {author} {\bibfnamefont {O.~N.}\ \bibnamefont {{Sholukhova}}},\ }\href {\doibase 10.1134/S1990341321010077} {\bibfield  {journal} {\bibinfo  {journal} {Astrophysical Bulletin}\ }\textbf {\bibinfo {volume} {76}},\ \bibinfo {pages} {6} (\bibinfo {year} {2021})},\ \Eprint {http://arxiv.org/abs/2105.10537} {arXiv:2105.10537 [astro-ph.GA]} \BibitemShut {NoStop}%
\bibitem [{\citenamefont {{King}}\ \emph {et~al.}(2023)\citenamefont {{King}}, \citenamefont {{Lasota}},\ and\ \citenamefont {{Middleton}}}]{King2023}%
  \BibitemOpen
  \bibfield  {author} {\bibinfo {author} {\bibfnamefont {A.}~\bibnamefont {{King}}}, \bibinfo {author} {\bibfnamefont {J.-P.}\ \bibnamefont {{Lasota}}}, \ and\ \bibinfo {author} {\bibfnamefont {M.}~\bibnamefont {{Middleton}}},\ }\href {\doibase 10.1016/j.newar.2022.101672} {\bibfield  {journal} {\bibinfo  {journal} {\nar}\ }\textbf {\bibinfo {volume} {96}},\ \bibinfo {eid} {101672} (\bibinfo {year} {2023})},\ \Eprint {http://arxiv.org/abs/2302.10605} {arXiv:2302.10605 [astro-ph.HE]} \BibitemShut {NoStop}%
\bibitem [{\citenamefont {{Pinto}}\ and\ \citenamefont {{Walton}}(2023)}]{Pinto2023}%
  \BibitemOpen
  \bibfield  {author} {\bibinfo {author} {\bibfnamefont {C.}~\bibnamefont {{Pinto}}}\ and\ \bibinfo {author} {\bibfnamefont {D.~J.}\ \bibnamefont {{Walton}}},\ }\enquote {\bibinfo {title} {{Ultra-Luminous X-Ray Sources: Extreme Accretion and Feedback}},}\ in\ \href {\doibase 10.1007/978-981-99-4409-5_12} {\emph {\bibinfo {booktitle} {High-Resolution X-ray Spectroscopy}}},\ \bibinfo {editor} {edited by\ \bibinfo {editor} {\bibfnamefont {C.}~\bibnamefont {{Bambi}}}\ and\ \bibinfo {editor} {\bibfnamefont {J.}~\bibnamefont {{Jiang}}}}\ (\bibinfo  {publisher} {Springer},\ \bibinfo {address} {Singapore},\ \bibinfo {year} {2023})\ p.\ \bibinfo {pages} {345–391},\ \Eprint {http://arxiv.org/abs/2302.00006} {arXiv:2302.00006 [astro-ph.HE]} \BibitemShut {NoStop}%
\bibitem [{\citenamefont {{Kovlakas}}\ \emph {et~al.}(2020)\citenamefont {{Kovlakas}}, \citenamefont {{Zezas}}, \citenamefont {{Andrews}}, \citenamefont {{Basu-Zych}}, \citenamefont {{Fragos}}, \citenamefont {{Hornschemeier}}, \citenamefont {{Lehmer}},\ and\ \citenamefont {{Ptak}}}]{Kovlakas2020_ULXcat}%
  \BibitemOpen
  \bibfield  {author} {\bibinfo {author} {\bibfnamefont {K.}~\bibnamefont {{Kovlakas}}}, \bibinfo {author} {\bibfnamefont {A.}~\bibnamefont {{Zezas}}}, \bibinfo {author} {\bibfnamefont {J.~J.}\ \bibnamefont {{Andrews}}}, \bibinfo {author} {\bibfnamefont {A.}~\bibnamefont {{Basu-Zych}}}, \bibinfo {author} {\bibfnamefont {T.}~\bibnamefont {{Fragos}}}, \bibinfo {author} {\bibfnamefont {A.}~\bibnamefont {{Hornschemeier}}}, \bibinfo {author} {\bibfnamefont {B.}~\bibnamefont {{Lehmer}}}, \ and\ \bibinfo {author} {\bibfnamefont {A.}~\bibnamefont {{Ptak}}},\ }\href {\doibase 10.1093/mnras/staa2481} {\bibfield  {journal} {\bibinfo  {journal} {\mnras}\ }\textbf {\bibinfo {volume} {498}},\ \bibinfo {pages} {4790} (\bibinfo {year} {2020})},\ \Eprint {http://arxiv.org/abs/2008.10572} {arXiv:2008.10572 [astro-ph.GA]} \BibitemShut {NoStop}%
\bibitem [{\citenamefont {{Walton}}\ \emph {et~al.}(2022)\citenamefont {{Walton}}, \citenamefont {{Mackenzie}}, \citenamefont {{Gully}}, \citenamefont {{Patel}}, \citenamefont {{Roberts}}, \citenamefont {{Earnshaw}},\ and\ \citenamefont {{Mateos}}}]{Walton2022catalogue}%
  \BibitemOpen
  \bibfield  {author} {\bibinfo {author} {\bibfnamefont {D.~J.}\ \bibnamefont {{Walton}}}, \bibinfo {author} {\bibfnamefont {A.~D.~A.}\ \bibnamefont {{Mackenzie}}}, \bibinfo {author} {\bibfnamefont {H.}~\bibnamefont {{Gully}}}, \bibinfo {author} {\bibfnamefont {N.~R.}\ \bibnamefont {{Patel}}}, \bibinfo {author} {\bibfnamefont {T.~P.}\ \bibnamefont {{Roberts}}}, \bibinfo {author} {\bibfnamefont {H.~P.}\ \bibnamefont {{Earnshaw}}}, \ and\ \bibinfo {author} {\bibfnamefont {S.}~\bibnamefont {{Mateos}}},\ }\href {\doibase 10.1093/mnras/stab3001} {\bibfield  {journal} {\bibinfo  {journal} {\mnras}\ }\textbf {\bibinfo {volume} {509}},\ \bibinfo {pages} {1587} (\bibinfo {year} {2022})},\ \Eprint {http://arxiv.org/abs/2110.07625} {arXiv:2110.07625 [astro-ph.HE]} \BibitemShut {NoStop}%
\bibitem [{\citenamefont {{Tranin}}\ \emph {et~al.}(2024)\citenamefont {{Tranin}}, \citenamefont {{Webb}}, \citenamefont {{Godet}},\ and\ \citenamefont {{Quintin}}}]{Tranin2024_cat}%
  \BibitemOpen
  \bibfield  {author} {\bibinfo {author} {\bibfnamefont {H.}~\bibnamefont {{Tranin}}}, \bibinfo {author} {\bibfnamefont {N.}~\bibnamefont {{Webb}}}, \bibinfo {author} {\bibfnamefont {O.}~\bibnamefont {{Godet}}}, \ and\ \bibinfo {author} {\bibfnamefont {E.}~\bibnamefont {{Quintin}}},\ }\href {\doibase 10.1051/0004-6361/202244952} {\bibfield  {journal} {\bibinfo  {journal} {\aap}\ }\textbf {\bibinfo {volume} {681}},\ \bibinfo {eid} {A16} (\bibinfo {year} {2024})},\ \Eprint {http://arxiv.org/abs/2304.11216} {arXiv:2304.11216 [astro-ph.HE]} \BibitemShut {NoStop}%
\bibitem [{\citenamefont {{Colbert}}\ and\ \citenamefont {{Mushotzky}}(1999)}]{Colbert1999}%
  \BibitemOpen
  \bibfield  {author} {\bibinfo {author} {\bibfnamefont {E.~J.~M.}\ \bibnamefont {{Colbert}}}\ and\ \bibinfo {author} {\bibfnamefont {R.~F.}\ \bibnamefont {{Mushotzky}}},\ }\href {\doibase 10.1086/307356} {\bibfield  {journal} {\bibinfo  {journal} {\apj}\ }\textbf {\bibinfo {volume} {519}},\ \bibinfo {pages} {89} (\bibinfo {year} {1999})},\ \Eprint {http://arxiv.org/abs/astro-ph/9901023} {arXiv:astro-ph/9901023 [astro-ph]} \BibitemShut {NoStop}%
\bibitem [{\citenamefont {{King}}\ \emph {et~al.}(2001)\citenamefont {{King}}, \citenamefont {{Davies}}, \citenamefont {{Ward}}, \citenamefont {{Fabbiano}},\ and\ \citenamefont {{Elvis}}}]{King2001}%
  \BibitemOpen
  \bibfield  {author} {\bibinfo {author} {\bibfnamefont {A.~R.}\ \bibnamefont {{King}}}, \bibinfo {author} {\bibfnamefont {M.~B.}\ \bibnamefont {{Davies}}}, \bibinfo {author} {\bibfnamefont {M.~J.}\ \bibnamefont {{Ward}}}, \bibinfo {author} {\bibfnamefont {G.}~\bibnamefont {{Fabbiano}}}, \ and\ \bibinfo {author} {\bibfnamefont {M.}~\bibnamefont {{Elvis}}},\ }\href {\doibase 10.1086/320343} {\bibfield  {journal} {\bibinfo  {journal} {\apjl}\ }\textbf {\bibinfo {volume} {552}},\ \bibinfo {pages} {L109} (\bibinfo {year} {2001})},\ \Eprint {http://arxiv.org/abs/astro-ph/0104333} {arXiv:astro-ph/0104333 [astro-ph]} \BibitemShut {NoStop}%
\bibitem [{\citenamefont {{Poutanen}}\ \emph {et~al.}(2007)\citenamefont {{Poutanen}}, \citenamefont {{Lipunova}}, \citenamefont {{Fabrika}}, \citenamefont {{Butkevich}},\ and\ \citenamefont {{Abolmasov}}}]{Poutanen2007}%
  \BibitemOpen
  \bibfield  {author} {\bibinfo {author} {\bibfnamefont {J.}~\bibnamefont {{Poutanen}}}, \bibinfo {author} {\bibfnamefont {G.}~\bibnamefont {{Lipunova}}}, \bibinfo {author} {\bibfnamefont {S.}~\bibnamefont {{Fabrika}}}, \bibinfo {author} {\bibfnamefont {A.~G.}\ \bibnamefont {{Butkevich}}}, \ and\ \bibinfo {author} {\bibfnamefont {P.}~\bibnamefont {{Abolmasov}}},\ }\href {\doibase 10.1111/j.1365-2966.2007.11668.x} {\bibfield  {journal} {\bibinfo  {journal} {\mnras}\ }\textbf {\bibinfo {volume} {377}},\ \bibinfo {pages} {1187} (\bibinfo {year} {2007})},\ \Eprint {http://arxiv.org/abs/astro-ph/0609274} {arXiv:astro-ph/0609274 [astro-ph]} \BibitemShut {NoStop}%
\bibitem [{\citenamefont {{Zampieri}}\ and\ \citenamefont {{Roberts}}(2009)}]{Zampieri2009}%
  \BibitemOpen
  \bibfield  {author} {\bibinfo {author} {\bibfnamefont {L.}~\bibnamefont {{Zampieri}}}\ and\ \bibinfo {author} {\bibfnamefont {T.~P.}\ \bibnamefont {{Roberts}}},\ }\href {\doibase 10.1111/j.1365-2966.2009.15509.x} {\bibfield  {journal} {\bibinfo  {journal} {\mnras}\ }\textbf {\bibinfo {volume} {400}},\ \bibinfo {pages} {677} (\bibinfo {year} {2009})},\ \Eprint {http://arxiv.org/abs/0909.1017} {arXiv:0909.1017 [astro-ph.HE]} \BibitemShut {NoStop}%
\bibitem [{\citenamefont {{Gladstone}}\ \emph {et~al.}(2009)\citenamefont {{Gladstone}}, \citenamefont {{Roberts}},\ and\ \citenamefont {{Done}}}]{Gladstone2009}%
  \BibitemOpen
  \bibfield  {author} {\bibinfo {author} {\bibfnamefont {J.~C.}\ \bibnamefont {{Gladstone}}}, \bibinfo {author} {\bibfnamefont {T.~P.}\ \bibnamefont {{Roberts}}}, \ and\ \bibinfo {author} {\bibfnamefont {C.}~\bibnamefont {{Done}}},\ }\href {\doibase 10.1111/j.1365-2966.2009.15123.x} {\bibfield  {journal} {\bibinfo  {journal} {\mnras}\ }\textbf {\bibinfo {volume} {397}},\ \bibinfo {pages} {1836} (\bibinfo {year} {2009})},\ \Eprint {http://arxiv.org/abs/0905.4076} {arXiv:0905.4076 [astro-ph.CO]} \BibitemShut {NoStop}%
\bibitem [{\citenamefont {{Sutton}}\ \emph {et~al.}(2013)\citenamefont {{Sutton}}, \citenamefont {{Roberts}},\ and\ \citenamefont {{Middleton}}}]{Sutton2013}%
  \BibitemOpen
  \bibfield  {author} {\bibinfo {author} {\bibfnamefont {A.~D.}\ \bibnamefont {{Sutton}}}, \bibinfo {author} {\bibfnamefont {T.~P.}\ \bibnamefont {{Roberts}}}, \ and\ \bibinfo {author} {\bibfnamefont {M.~J.}\ \bibnamefont {{Middleton}}},\ }\href {\doibase 10.1093/mnras/stt1419} {\bibfield  {journal} {\bibinfo  {journal} {\mnras}\ }\textbf {\bibinfo {volume} {435}},\ \bibinfo {pages} {1758} (\bibinfo {year} {2013})},\ \Eprint {http://arxiv.org/abs/1307.8044} {arXiv:1307.8044 [astro-ph.HE]} \BibitemShut {NoStop}%
\bibitem [{\citenamefont {{Pinto}}\ \emph {et~al.}(2016)\citenamefont {{Pinto}}, \citenamefont {{Middleton}},\ and\ \citenamefont {{Fabian}}}]{Pinto2016}%
  \BibitemOpen
  \bibfield  {author} {\bibinfo {author} {\bibfnamefont {C.}~\bibnamefont {{Pinto}}}, \bibinfo {author} {\bibfnamefont {M.~J.}\ \bibnamefont {{Middleton}}}, \ and\ \bibinfo {author} {\bibfnamefont {A.~C.}\ \bibnamefont {{Fabian}}},\ }\href {\doibase 10.1038/nature17417} {\bibfield  {journal} {\bibinfo  {journal} {\nat}\ }\textbf {\bibinfo {volume} {533}},\ \bibinfo {pages} {64} (\bibinfo {year} {2016})},\ \Eprint {http://arxiv.org/abs/1604.08593} {arXiv:1604.08593 [astro-ph.HE]} \BibitemShut {NoStop}%
\bibitem [{\citenamefont {{Kosec}}\ \emph {et~al.}(2021)\citenamefont {{Kosec}}, \citenamefont {{Pinto}}, \citenamefont {{Reynolds}}, \citenamefont {{Guainazzi}}, \citenamefont {{Kara}}, \citenamefont {{Walton}}, \citenamefont {{Fabian}}, \citenamefont {{Parker}},\ and\ \citenamefont {{Valtchanov}}}]{Kosec2021}%
  \BibitemOpen
  \bibfield  {author} {\bibinfo {author} {\bibfnamefont {P.}~\bibnamefont {{Kosec}}}, \bibinfo {author} {\bibfnamefont {C.}~\bibnamefont {{Pinto}}}, \bibinfo {author} {\bibfnamefont {C.~S.}\ \bibnamefont {{Reynolds}}}, \bibinfo {author} {\bibfnamefont {M.}~\bibnamefont {{Guainazzi}}}, \bibinfo {author} {\bibfnamefont {E.}~\bibnamefont {{Kara}}}, \bibinfo {author} {\bibfnamefont {D.~J.}\ \bibnamefont {{Walton}}}, \bibinfo {author} {\bibfnamefont {A.~C.}\ \bibnamefont {{Fabian}}}, \bibinfo {author} {\bibfnamefont {M.~L.}\ \bibnamefont {{Parker}}}, \ and\ \bibinfo {author} {\bibfnamefont {I.}~\bibnamefont {{Valtchanov}}},\ }\href {\doibase 10.1093/mnras/stab2856} {\bibfield  {journal} {\bibinfo  {journal} {\mnras}\ }\textbf {\bibinfo {volume} {508}},\ \bibinfo {pages} {3569} (\bibinfo {year} {2021})},\ \Eprint {http://arxiv.org/abs/2109.14683} {arXiv:2109.14683 [astro-ph.HE]} \BibitemShut {NoStop}%
\bibitem [{\citenamefont {{Bachetti}}\ \emph {et~al.}(2014)\citenamefont {{Bachetti}}, \citenamefont {{Harrison}}, \citenamefont {{Walton}}, \citenamefont {{Grefenstette}}, \citenamefont {{Chakrabarty}}, \citenamefont {{F{\"u}rst}}, \citenamefont {{Barret}}, \citenamefont {{Beloborodov}}, \citenamefont {{Boggs}}, \citenamefont {{Christensen}}, \citenamefont {{Craig}}, \citenamefont {{Fabian}}, \citenamefont {{Hailey}}, \citenamefont {{Hornschemeier}}, \citenamefont {{Kaspi}}, \citenamefont {{Kulkarni}}, \citenamefont {{Maccarone}}, \citenamefont {{Miller}}, \citenamefont {{Rana}}, \citenamefont {{Stern}}, \citenamefont {{Tendulkar}}, \citenamefont {{Tomsick}}, \citenamefont {{Webb}},\ and\ \citenamefont {{Zhang}}}]{Bachetti2014}%
  \BibitemOpen
  \bibfield  {author} {\bibinfo {author} {\bibfnamefont {M.}~\bibnamefont {{Bachetti}}}, \bibinfo {author} {\bibfnamefont {F.~A.}\ \bibnamefont {{Harrison}}}, \bibinfo {author} {\bibfnamefont {D.~J.}\ \bibnamefont {{Walton}}}, \bibinfo {author} {\bibfnamefont {B.~W.}\ \bibnamefont {{Grefenstette}}}, \bibinfo {author} {\bibfnamefont {D.}~\bibnamefont {{Chakrabarty}}}, \bibinfo {author} {\bibfnamefont {F.}~\bibnamefont {{F{\"u}rst}}}, \bibinfo {author} {\bibfnamefont {D.}~\bibnamefont {{Barret}}}, \bibinfo {author} {\bibfnamefont {A.}~\bibnamefont {{Beloborodov}}}, \bibinfo {author} {\bibfnamefont {S.~E.}\ \bibnamefont {{Boggs}}}, \bibinfo {author} {\bibfnamefont {F.~E.}\ \bibnamefont {{Christensen}}}, \bibinfo {author} {\bibfnamefont {W.~W.}\ \bibnamefont {{Craig}}}, \bibinfo {author} {\bibfnamefont {A.~C.}\ \bibnamefont {{Fabian}}}, \bibinfo {author} {\bibfnamefont {C.~J.}\ \bibnamefont {{Hailey}}}, \bibinfo {author} {\bibfnamefont {A.}~\bibnamefont {{Hornschemeier}}}, \bibinfo {author} {\bibfnamefont
  {V.}~\bibnamefont {{Kaspi}}}, \bibinfo {author} {\bibfnamefont {S.~R.}\ \bibnamefont {{Kulkarni}}}, \bibinfo {author} {\bibfnamefont {T.}~\bibnamefont {{Maccarone}}}, \bibinfo {author} {\bibfnamefont {J.~M.}\ \bibnamefont {{Miller}}}, \bibinfo {author} {\bibfnamefont {V.}~\bibnamefont {{Rana}}}, \bibinfo {author} {\bibfnamefont {D.}~\bibnamefont {{Stern}}}, \bibinfo {author} {\bibfnamefont {S.~P.}\ \bibnamefont {{Tendulkar}}}, \bibinfo {author} {\bibfnamefont {J.}~\bibnamefont {{Tomsick}}}, \bibinfo {author} {\bibfnamefont {N.~A.}\ \bibnamefont {{Webb}}}, \ and\ \bibinfo {author} {\bibfnamefont {W.~W.}\ \bibnamefont {{Zhang}}},\ }\href {\doibase 10.1038/nature13791} {\bibfield  {journal} {\bibinfo  {journal} {\nat}\ }\textbf {\bibinfo {volume} {514}},\ \bibinfo {pages} {202} (\bibinfo {year} {2014})},\ \Eprint {http://arxiv.org/abs/1410.3590} {arXiv:1410.3590 [astro-ph.HE]} \BibitemShut {NoStop}%
\bibitem [{\citenamefont {{F{\"u}rst}}\ \emph {et~al.}(2016)\citenamefont {{F{\"u}rst}}, \citenamefont {{Walton}}, \citenamefont {{Harrison}}, \citenamefont {{Stern}}, \citenamefont {{Barret}}, \citenamefont {{Brightman}}, \citenamefont {{Fabian}}, \citenamefont {{Grefenstette}}, \citenamefont {{Madsen}}, \citenamefont {{Middleton}}, \citenamefont {{Miller}}, \citenamefont {{Pottschmidt}}, \citenamefont {{Ptak}}, \citenamefont {{Rana}},\ and\ \citenamefont {{Webb}}}]{Furst2016}%
  \BibitemOpen
  \bibfield  {author} {\bibinfo {author} {\bibfnamefont {F.}~\bibnamefont {{F{\"u}rst}}}, \bibinfo {author} {\bibfnamefont {D.~J.}\ \bibnamefont {{Walton}}}, \bibinfo {author} {\bibfnamefont {F.~A.}\ \bibnamefont {{Harrison}}}, \bibinfo {author} {\bibfnamefont {D.}~\bibnamefont {{Stern}}}, \bibinfo {author} {\bibfnamefont {D.}~\bibnamefont {{Barret}}}, \bibinfo {author} {\bibfnamefont {M.}~\bibnamefont {{Brightman}}}, \bibinfo {author} {\bibfnamefont {A.~C.}\ \bibnamefont {{Fabian}}}, \bibinfo {author} {\bibfnamefont {B.}~\bibnamefont {{Grefenstette}}}, \bibinfo {author} {\bibfnamefont {K.~K.}\ \bibnamefont {{Madsen}}}, \bibinfo {author} {\bibfnamefont {M.~J.}\ \bibnamefont {{Middleton}}}, \bibinfo {author} {\bibfnamefont {J.~M.}\ \bibnamefont {{Miller}}}, \bibinfo {author} {\bibfnamefont {K.}~\bibnamefont {{Pottschmidt}}}, \bibinfo {author} {\bibfnamefont {A.}~\bibnamefont {{Ptak}}}, \bibinfo {author} {\bibfnamefont {V.}~\bibnamefont {{Rana}}}, \ and\ \bibinfo {author} {\bibfnamefont {N.}~\bibnamefont
  {{Webb}}},\ }\href {\doibase 10.3847/2041-8205/831/2/L14} {\bibfield  {journal} {\bibinfo  {journal} {\apjl}\ }\textbf {\bibinfo {volume} {831}},\ \bibinfo {eid} {L14} (\bibinfo {year} {2016})},\ \Eprint {http://arxiv.org/abs/1609.07129} {arXiv:1609.07129 [astro-ph.HE]} \BibitemShut {NoStop}%
\bibitem [{\citenamefont {{Israel}}\ \emph {et~al.}(2017{\natexlab{a}})\citenamefont {{Israel}}, \citenamefont {{Papitto}}, \citenamefont {{Esposito}}, \citenamefont {{Stella}}, \citenamefont {{Zampieri}}, \citenamefont {{Belfiore}}, \citenamefont {{Rodr{\'\i}guez Castillo}}, \citenamefont {{De Luca}}, \citenamefont {{Tiengo}}, \citenamefont {{Haberl}}, \citenamefont {{Greiner}}, \citenamefont {{Salvaterra}}, \citenamefont {{Sandrelli}},\ and\ \citenamefont {{Lisini}}}]{Israel2017}%
  \BibitemOpen
  \bibfield  {author} {\bibinfo {author} {\bibfnamefont {G.~L.}\ \bibnamefont {{Israel}}}, \bibinfo {author} {\bibfnamefont {A.}~\bibnamefont {{Papitto}}}, \bibinfo {author} {\bibfnamefont {P.}~\bibnamefont {{Esposito}}}, \bibinfo {author} {\bibfnamefont {L.}~\bibnamefont {{Stella}}}, \bibinfo {author} {\bibfnamefont {L.}~\bibnamefont {{Zampieri}}}, \bibinfo {author} {\bibfnamefont {A.}~\bibnamefont {{Belfiore}}}, \bibinfo {author} {\bibfnamefont {G.~A.}\ \bibnamefont {{Rodr{\'\i}guez Castillo}}}, \bibinfo {author} {\bibfnamefont {A.}~\bibnamefont {{De Luca}}}, \bibinfo {author} {\bibfnamefont {A.}~\bibnamefont {{Tiengo}}}, \bibinfo {author} {\bibfnamefont {F.}~\bibnamefont {{Haberl}}}, \bibinfo {author} {\bibfnamefont {J.}~\bibnamefont {{Greiner}}}, \bibinfo {author} {\bibfnamefont {R.}~\bibnamefont {{Salvaterra}}}, \bibinfo {author} {\bibfnamefont {S.}~\bibnamefont {{Sandrelli}}}, \ and\ \bibinfo {author} {\bibfnamefont {G.}~\bibnamefont {{Lisini}}},\ }\href {\doibase 10.1093/mnrasl/slw218} {\bibfield
  {journal} {\bibinfo  {journal} {\mnras}\ }\textbf {\bibinfo {volume} {466}},\ \bibinfo {pages} {L48} (\bibinfo {year} {2017}{\natexlab{a}})},\ \Eprint {http://arxiv.org/abs/1609.06538} {arXiv:1609.06538 [astro-ph.HE]} \BibitemShut {NoStop}%
\bibitem [{\citenamefont {{Israel}}\ \emph {et~al.}(2017{\natexlab{b}})\citenamefont {{Israel}}, \citenamefont {{Belfiore}}, \citenamefont {{Stella}}, \citenamefont {{Esposito}}, \citenamefont {{Casella}}, \citenamefont {{De Luca}}, \citenamefont {{Marelli}}, \citenamefont {{Papitto}}, \citenamefont {{Perri}}, \citenamefont {{Puccetti}}, \citenamefont {{Castillo}}, \citenamefont {{Salvetti}}, \citenamefont {{Tiengo}}, \citenamefont {{Zampieri}}, \citenamefont {{D'Agostino}}, \citenamefont {{Greiner}}, \citenamefont {{Haberl}}, \citenamefont {{Novara}}, \citenamefont {{Salvaterra}}, \citenamefont {{Turolla}}, \citenamefont {{Watson}}, \citenamefont {{Wilms}},\ and\ \citenamefont {{Wolter}}}]{Israel2017a}%
  \BibitemOpen
  \bibfield  {author} {\bibinfo {author} {\bibfnamefont {G.~L.}\ \bibnamefont {{Israel}}}, \bibinfo {author} {\bibfnamefont {A.}~\bibnamefont {{Belfiore}}}, \bibinfo {author} {\bibfnamefont {L.}~\bibnamefont {{Stella}}}, \bibinfo {author} {\bibfnamefont {P.}~\bibnamefont {{Esposito}}}, \bibinfo {author} {\bibfnamefont {P.}~\bibnamefont {{Casella}}}, \bibinfo {author} {\bibfnamefont {A.}~\bibnamefont {{De Luca}}}, \bibinfo {author} {\bibfnamefont {M.}~\bibnamefont {{Marelli}}}, \bibinfo {author} {\bibfnamefont {A.}~\bibnamefont {{Papitto}}}, \bibinfo {author} {\bibfnamefont {M.}~\bibnamefont {{Perri}}}, \bibinfo {author} {\bibfnamefont {S.}~\bibnamefont {{Puccetti}}}, \bibinfo {author} {\bibfnamefont {G.~A.~R.}\ \bibnamefont {{Castillo}}}, \bibinfo {author} {\bibfnamefont {D.}~\bibnamefont {{Salvetti}}}, \bibinfo {author} {\bibfnamefont {A.}~\bibnamefont {{Tiengo}}}, \bibinfo {author} {\bibfnamefont {L.}~\bibnamefont {{Zampieri}}}, \bibinfo {author} {\bibfnamefont {D.}~\bibnamefont {{D'Agostino}}}, \bibinfo
  {author} {\bibfnamefont {J.}~\bibnamefont {{Greiner}}}, \bibinfo {author} {\bibfnamefont {F.}~\bibnamefont {{Haberl}}}, \bibinfo {author} {\bibfnamefont {G.}~\bibnamefont {{Novara}}}, \bibinfo {author} {\bibfnamefont {R.}~\bibnamefont {{Salvaterra}}}, \bibinfo {author} {\bibfnamefont {R.}~\bibnamefont {{Turolla}}}, \bibinfo {author} {\bibfnamefont {M.}~\bibnamefont {{Watson}}}, \bibinfo {author} {\bibfnamefont {J.}~\bibnamefont {{Wilms}}}, \ and\ \bibinfo {author} {\bibfnamefont {A.}~\bibnamefont {{Wolter}}},\ }\href {\doibase 10.1126/science.aai8635} {\bibfield  {journal} {\bibinfo  {journal} {Science}\ }\textbf {\bibinfo {volume} {355}},\ \bibinfo {pages} {817} (\bibinfo {year} {2017}{\natexlab{b}})},\ \Eprint {http://arxiv.org/abs/1609.07375} {arXiv:1609.07375 [astro-ph.HE]} \BibitemShut {NoStop}%
\bibitem [{\citenamefont {{Carpano}}\ \emph {et~al.}(2018)\citenamefont {{Carpano}}, \citenamefont {{Haberl}}, \citenamefont {{Maitra}},\ and\ \citenamefont {{Vasilopoulos}}}]{Carpano2018}%
  \BibitemOpen
  \bibfield  {author} {\bibinfo {author} {\bibfnamefont {S.}~\bibnamefont {{Carpano}}}, \bibinfo {author} {\bibfnamefont {F.}~\bibnamefont {{Haberl}}}, \bibinfo {author} {\bibfnamefont {C.}~\bibnamefont {{Maitra}}}, \ and\ \bibinfo {author} {\bibfnamefont {G.}~\bibnamefont {{Vasilopoulos}}},\ }\href {\doibase 10.1093/mnrasl/sly030} {\bibfield  {journal} {\bibinfo  {journal} {\mnras}\ }\textbf {\bibinfo {volume} {476}},\ \bibinfo {pages} {L45} (\bibinfo {year} {2018})},\ \Eprint {http://arxiv.org/abs/1802.10341} {arXiv:1802.10341 [astro-ph.HE]} \BibitemShut {NoStop}%
\bibitem [{\citenamefont {{Sathyaprakash}}\ \emph {et~al.}(2019)\citenamefont {{Sathyaprakash}}, \citenamefont {{Roberts}}, \citenamefont {{Walton}}, \citenamefont {{Fuerst}}, \citenamefont {{Bachetti}}, \citenamefont {{Pinto}}, \citenamefont {{Alston}}, \citenamefont {{Earnshaw}}, \citenamefont {{Fabian}}, \citenamefont {{Middleton}},\ and\ \citenamefont {{Soria}}}]{Sathyaprakash2019}%
  \BibitemOpen
  \bibfield  {author} {\bibinfo {author} {\bibfnamefont {R.}~\bibnamefont {{Sathyaprakash}}}, \bibinfo {author} {\bibfnamefont {T.~P.}\ \bibnamefont {{Roberts}}}, \bibinfo {author} {\bibfnamefont {D.~J.}\ \bibnamefont {{Walton}}}, \bibinfo {author} {\bibfnamefont {F.}~\bibnamefont {{Fuerst}}}, \bibinfo {author} {\bibfnamefont {M.}~\bibnamefont {{Bachetti}}}, \bibinfo {author} {\bibfnamefont {C.}~\bibnamefont {{Pinto}}}, \bibinfo {author} {\bibfnamefont {W.~N.}\ \bibnamefont {{Alston}}}, \bibinfo {author} {\bibfnamefont {H.~P.}\ \bibnamefont {{Earnshaw}}}, \bibinfo {author} {\bibfnamefont {A.~C.}\ \bibnamefont {{Fabian}}}, \bibinfo {author} {\bibfnamefont {M.~J.}\ \bibnamefont {{Middleton}}}, \ and\ \bibinfo {author} {\bibfnamefont {R.}~\bibnamefont {{Soria}}},\ }\href {\doibase 10.1093/mnrasl/slz086} {\bibfield  {journal} {\bibinfo  {journal} {\mnras}\ }\textbf {\bibinfo {volume} {488}},\ \bibinfo {pages} {L35} (\bibinfo {year} {2019})},\ \Eprint {http://arxiv.org/abs/1906.00640} {arXiv:1906.00640
  [astro-ph.HE]} \BibitemShut {NoStop}%
\bibitem [{\citenamefont {{Rodr{\'\i}guez Castillo}}\ \emph {et~al.}(2020)\citenamefont {{Rodr{\'\i}guez Castillo}}, \citenamefont {{Israel}}, \citenamefont {{Belfiore}}, \citenamefont {{Bernardini}}, \citenamefont {{Esposito}}, \citenamefont {{Pintore}}, \citenamefont {{De Luca}}, \citenamefont {{Papitto}}, \citenamefont {{Stella}}, \citenamefont {{Tiengo}}, \citenamefont {{Zampieri}}, \citenamefont {{Bachetti}}, \citenamefont {{Brightman}}, \citenamefont {{Casella}}, \citenamefont {{D'Agostino}}, \citenamefont {{Dall'Osso}}, \citenamefont {{Earnshaw}}, \citenamefont {{F{\"u}rst}}, \citenamefont {{Haberl}}, \citenamefont {{Harrison}}, \citenamefont {{Mapelli}}, \citenamefont {{Marelli}}, \citenamefont {{Middleton}}, \citenamefont {{Pinto}}, \citenamefont {{Roberts}}, \citenamefont {{Salvaterra}}, \citenamefont {{Turolla}}, \citenamefont {{Walton}},\ and\ \citenamefont {{Wolter}}}]{RodriguezCastillo2020}%
  \BibitemOpen
  \bibfield  {author} {\bibinfo {author} {\bibfnamefont {G.~A.}\ \bibnamefont {{Rodr{\'\i}guez Castillo}}}, \bibinfo {author} {\bibfnamefont {G.~L.}\ \bibnamefont {{Israel}}}, \bibinfo {author} {\bibfnamefont {A.}~\bibnamefont {{Belfiore}}}, \bibinfo {author} {\bibfnamefont {F.}~\bibnamefont {{Bernardini}}}, \bibinfo {author} {\bibfnamefont {P.}~\bibnamefont {{Esposito}}}, \bibinfo {author} {\bibfnamefont {F.}~\bibnamefont {{Pintore}}}, \bibinfo {author} {\bibfnamefont {A.}~\bibnamefont {{De Luca}}}, \bibinfo {author} {\bibfnamefont {A.}~\bibnamefont {{Papitto}}}, \bibinfo {author} {\bibfnamefont {L.}~\bibnamefont {{Stella}}}, \bibinfo {author} {\bibfnamefont {A.}~\bibnamefont {{Tiengo}}}, \bibinfo {author} {\bibfnamefont {L.}~\bibnamefont {{Zampieri}}}, \bibinfo {author} {\bibfnamefont {M.}~\bibnamefont {{Bachetti}}}, \bibinfo {author} {\bibfnamefont {M.}~\bibnamefont {{Brightman}}}, \bibinfo {author} {\bibfnamefont {P.}~\bibnamefont {{Casella}}}, \bibinfo {author} {\bibfnamefont {D.}~\bibnamefont
  {{D'Agostino}}}, \bibinfo {author} {\bibfnamefont {S.}~\bibnamefont {{Dall'Osso}}}, \bibinfo {author} {\bibfnamefont {H.~P.}\ \bibnamefont {{Earnshaw}}}, \bibinfo {author} {\bibfnamefont {F.}~\bibnamefont {{F{\"u}rst}}}, \bibinfo {author} {\bibfnamefont {F.}~\bibnamefont {{Haberl}}}, \bibinfo {author} {\bibfnamefont {F.~A.}\ \bibnamefont {{Harrison}}}, \bibinfo {author} {\bibfnamefont {M.}~\bibnamefont {{Mapelli}}}, \bibinfo {author} {\bibfnamefont {M.}~\bibnamefont {{Marelli}}}, \bibinfo {author} {\bibfnamefont {M.}~\bibnamefont {{Middleton}}}, \bibinfo {author} {\bibfnamefont {C.}~\bibnamefont {{Pinto}}}, \bibinfo {author} {\bibfnamefont {T.~P.}\ \bibnamefont {{Roberts}}}, \bibinfo {author} {\bibfnamefont {R.}~\bibnamefont {{Salvaterra}}}, \bibinfo {author} {\bibfnamefont {R.}~\bibnamefont {{Turolla}}}, \bibinfo {author} {\bibfnamefont {D.~J.}\ \bibnamefont {{Walton}}}, \ and\ \bibinfo {author} {\bibfnamefont {A.}~\bibnamefont {{Wolter}}},\ }\href {\doibase 10.3847/1538-4357/ab8a44} {\bibfield  {journal}
  {\bibinfo  {journal} {\apj}\ }\textbf {\bibinfo {volume} {895}},\ \bibinfo {eid} {60} (\bibinfo {year} {2020})},\ \Eprint {http://arxiv.org/abs/1906.04791} {arXiv:1906.04791 [astro-ph.HE]} \BibitemShut {NoStop}%
\bibitem [{\citenamefont {{Bachetti}}\ \emph {et~al.}(2022)\citenamefont {{Bachetti}}, \citenamefont {{Heida}}, \citenamefont {{Maccarone}}, \citenamefont {{Huppenkothen}}, \citenamefont {{Israel}}, \citenamefont {{Barret}}, \citenamefont {{Brightman}}, \citenamefont {{Brumback}}, \citenamefont {{Earnshaw}}, \citenamefont {{Forster}}, \citenamefont {{F{\"u}rst}}, \citenamefont {{Grefenstette}}, \citenamefont {{Harrison}}, \citenamefont {{Jaodand}}, \citenamefont {{Madsen}}, \citenamefont {{Middleton}}, \citenamefont {{Pike}}, \citenamefont {{Pilia}}, \citenamefont {{Poutanen}}, \citenamefont {{Stern}}, \citenamefont {{Tomsick}}, \citenamefont {{Walton}}, \citenamefont {{Webb}},\ and\ \citenamefont {{Wilms}}}]{Bachetti2022}%
  \BibitemOpen
  \bibfield  {author} {\bibinfo {author} {\bibfnamefont {M.}~\bibnamefont {{Bachetti}}}, \bibinfo {author} {\bibfnamefont {M.}~\bibnamefont {{Heida}}}, \bibinfo {author} {\bibfnamefont {T.}~\bibnamefont {{Maccarone}}}, \bibinfo {author} {\bibfnamefont {D.}~\bibnamefont {{Huppenkothen}}}, \bibinfo {author} {\bibfnamefont {G.~L.}\ \bibnamefont {{Israel}}}, \bibinfo {author} {\bibfnamefont {D.}~\bibnamefont {{Barret}}}, \bibinfo {author} {\bibfnamefont {M.}~\bibnamefont {{Brightman}}}, \bibinfo {author} {\bibfnamefont {M.}~\bibnamefont {{Brumback}}}, \bibinfo {author} {\bibfnamefont {H.~P.}\ \bibnamefont {{Earnshaw}}}, \bibinfo {author} {\bibfnamefont {K.}~\bibnamefont {{Forster}}}, \bibinfo {author} {\bibfnamefont {F.}~\bibnamefont {{F{\"u}rst}}}, \bibinfo {author} {\bibfnamefont {B.~W.}\ \bibnamefont {{Grefenstette}}}, \bibinfo {author} {\bibfnamefont {F.~A.}\ \bibnamefont {{Harrison}}}, \bibinfo {author} {\bibfnamefont {A.~D.}\ \bibnamefont {{Jaodand}}}, \bibinfo {author} {\bibfnamefont {K.~K.}\ \bibnamefont
  {{Madsen}}}, \bibinfo {author} {\bibfnamefont {M.}~\bibnamefont {{Middleton}}}, \bibinfo {author} {\bibfnamefont {S.~N.}\ \bibnamefont {{Pike}}}, \bibinfo {author} {\bibfnamefont {M.}~\bibnamefont {{Pilia}}}, \bibinfo {author} {\bibfnamefont {J.}~\bibnamefont {{Poutanen}}}, \bibinfo {author} {\bibfnamefont {D.}~\bibnamefont {{Stern}}}, \bibinfo {author} {\bibfnamefont {J.~A.}\ \bibnamefont {{Tomsick}}}, \bibinfo {author} {\bibfnamefont {D.~J.}\ \bibnamefont {{Walton}}}, \bibinfo {author} {\bibfnamefont {N.}~\bibnamefont {{Webb}}}, \ and\ \bibinfo {author} {\bibfnamefont {J.}~\bibnamefont {{Wilms}}},\ }\href {\doibase 10.3847/1538-4357/ac8d67} {\bibfield  {journal} {\bibinfo  {journal} {\apj}\ }\textbf {\bibinfo {volume} {937}},\ \bibinfo {eid} {125} (\bibinfo {year} {2022})},\ \Eprint {http://arxiv.org/abs/2112.00339} {arXiv:2112.00339 [astro-ph.HE]} \BibitemShut {NoStop}%
\bibitem [{\citenamefont {{Liu}}(2024)}]{LiuJR24}%
  \BibitemOpen
  \bibfield  {author} {\bibinfo {author} {\bibfnamefont {J.}~\bibnamefont {{Liu}}},\ }\href {\doibase 10.3847/1538-4357/ad17c7} {\bibfield  {journal} {\bibinfo  {journal} {\apj}\ }\textbf {\bibinfo {volume} {961}},\ \bibinfo {eid} {196} (\bibinfo {year} {2024})},\ \Eprint {http://arxiv.org/abs/2312.16770} {arXiv:2312.16770 [astro-ph.HE]} \BibitemShut {NoStop}%
\bibitem [{\citenamefont {{Pintore}}\ \emph {et~al.}(2017)\citenamefont {{Pintore}}, \citenamefont {{Zampieri}}, \citenamefont {{Stella}}, \citenamefont {{Wolter}}, \citenamefont {{Mereghetti}},\ and\ \citenamefont {{Israel}}}]{Pintore2017}%
  \BibitemOpen
  \bibfield  {author} {\bibinfo {author} {\bibfnamefont {F.}~\bibnamefont {{Pintore}}}, \bibinfo {author} {\bibfnamefont {L.}~\bibnamefont {{Zampieri}}}, \bibinfo {author} {\bibfnamefont {L.}~\bibnamefont {{Stella}}}, \bibinfo {author} {\bibfnamefont {A.}~\bibnamefont {{Wolter}}}, \bibinfo {author} {\bibfnamefont {S.}~\bibnamefont {{Mereghetti}}}, \ and\ \bibinfo {author} {\bibfnamefont {G.~L.}\ \bibnamefont {{Israel}}},\ }\href {\doibase 10.3847/1538-4357/836/1/113} {\bibfield  {journal} {\bibinfo  {journal} {\apj}\ }\textbf {\bibinfo {volume} {836}},\ \bibinfo {eid} {113} (\bibinfo {year} {2017})},\ \Eprint {http://arxiv.org/abs/1701.03595} {arXiv:1701.03595 [astro-ph.HE]} \BibitemShut {NoStop}%
\bibitem [{\citenamefont {{G{\'u}rpide}}\ \emph {et~al.}(2021)\citenamefont {{G{\'u}rpide}}, \citenamefont {{Godet}}, \citenamefont {{Koliopanos}}, \citenamefont {{Webb}},\ and\ \citenamefont {{Olive}}}]{Gurpide2021a}%
  \BibitemOpen
  \bibfield  {author} {\bibinfo {author} {\bibfnamefont {A.}~\bibnamefont {{G{\'u}rpide}}}, \bibinfo {author} {\bibfnamefont {O.}~\bibnamefont {{Godet}}}, \bibinfo {author} {\bibfnamefont {F.}~\bibnamefont {{Koliopanos}}}, \bibinfo {author} {\bibfnamefont {N.}~\bibnamefont {{Webb}}}, \ and\ \bibinfo {author} {\bibfnamefont {J.~F.}\ \bibnamefont {{Olive}}},\ }\href {\doibase 10.1051/0004-6361/202039572} {\bibfield  {journal} {\bibinfo  {journal} {\aap}\ }\textbf {\bibinfo {volume} {649}},\ \bibinfo {eid} {A104} (\bibinfo {year} {2021})},\ \Eprint {http://arxiv.org/abs/2102.11159} {arXiv:2102.11159 [astro-ph.HE]} \BibitemShut {NoStop}%
\bibitem [{\citenamefont {{Walton}}\ \emph {et~al.}(2018)\citenamefont {{Walton}}, \citenamefont {{F{\"u}rst}}, \citenamefont {{Heida}}, \citenamefont {{Harrison}}, \citenamefont {{Barret}}, \citenamefont {{Stern}}, \citenamefont {{Bachetti}}, \citenamefont {{Brightman}}, \citenamefont {{Fabian}},\ and\ \citenamefont {{Middleton}}}]{Walton2018}%
  \BibitemOpen
  \bibfield  {author} {\bibinfo {author} {\bibfnamefont {D.~J.}\ \bibnamefont {{Walton}}}, \bibinfo {author} {\bibfnamefont {F.}~\bibnamefont {{F{\"u}rst}}}, \bibinfo {author} {\bibfnamefont {M.}~\bibnamefont {{Heida}}}, \bibinfo {author} {\bibfnamefont {F.~A.}\ \bibnamefont {{Harrison}}}, \bibinfo {author} {\bibfnamefont {D.}~\bibnamefont {{Barret}}}, \bibinfo {author} {\bibfnamefont {D.}~\bibnamefont {{Stern}}}, \bibinfo {author} {\bibfnamefont {M.}~\bibnamefont {{Bachetti}}}, \bibinfo {author} {\bibfnamefont {M.}~\bibnamefont {{Brightman}}}, \bibinfo {author} {\bibfnamefont {A.~C.}\ \bibnamefont {{Fabian}}}, \ and\ \bibinfo {author} {\bibfnamefont {M.~J.}\ \bibnamefont {{Middleton}}},\ }\href {\doibase 10.3847/1538-4357/aab610} {\bibfield  {journal} {\bibinfo  {journal} {\apj}\ }\textbf {\bibinfo {volume} {856}},\ \bibinfo {eid} {128} (\bibinfo {year} {2018})},\ \Eprint {http://arxiv.org/abs/1803.04424} {arXiv:1803.04424 [astro-ph.HE]} \BibitemShut {NoStop}%
\bibitem [{\citenamefont {{Feng}}\ \emph {et~al.}(2010)\citenamefont {{Feng}}, \citenamefont {{Rao}},\ and\ \citenamefont {{Kaaret}}}]{Feng2010}%
  \BibitemOpen
  \bibfield  {author} {\bibinfo {author} {\bibfnamefont {H.}~\bibnamefont {{Feng}}}, \bibinfo {author} {\bibfnamefont {F.}~\bibnamefont {{Rao}}}, \ and\ \bibinfo {author} {\bibfnamefont {P.}~\bibnamefont {{Kaaret}}},\ }\href {\doibase 10.1088/2041-8205/710/2/L137} {\bibfield  {journal} {\bibinfo  {journal} {\apjl}\ }\textbf {\bibinfo {volume} {710}},\ \bibinfo {pages} {L137} (\bibinfo {year} {2010})},\ \Eprint {http://arxiv.org/abs/1001.3456} {arXiv:1001.3456 [astro-ph.HE]} \BibitemShut {NoStop}%
\bibitem [{\citenamefont {{Imbrogno}}\ \emph {et~al.}(2024)\citenamefont {{Imbrogno}}, \citenamefont {{Motta}}, \citenamefont {{Amato}}, \citenamefont {{Israel}}, \citenamefont {{Rodr{\'\i}guez Castillo}}, \citenamefont {{Brightman}}, \citenamefont {{Casella}}, \citenamefont {{Bachetti}}, \citenamefont {{F{\"u}rst}}, \citenamefont {{Stella}}, \citenamefont {{Pinto}}, \citenamefont {{Pintore}}, \citenamefont {{Tombesi}}, \citenamefont {{G{\'u}rpide}}, \citenamefont {{Middleton}}, \citenamefont {{Salvaggio}}, \citenamefont {{Tiengo}}, \citenamefont {{Belfiore}}, \citenamefont {{De Luca}}, \citenamefont {{Esposito}}, \citenamefont {{Wolter}}, \citenamefont {{Earnshaw}}, \citenamefont {{Walton}}, \citenamefont {{Roberts}}, \citenamefont {{Zampieri}}, \citenamefont {{Marelli}},\ and\ \citenamefont {{Salvaterra}}}]{Imbrogno2024}%
  \BibitemOpen
  \bibfield  {author} {\bibinfo {author} {\bibfnamefont {M.}~\bibnamefont {{Imbrogno}}}, \bibinfo {author} {\bibfnamefont {S.~E.}\ \bibnamefont {{Motta}}}, \bibinfo {author} {\bibfnamefont {R.}~\bibnamefont {{Amato}}}, \bibinfo {author} {\bibfnamefont {G.~L.}\ \bibnamefont {{Israel}}}, \bibinfo {author} {\bibfnamefont {G.~A.}\ \bibnamefont {{Rodr{\'\i}guez Castillo}}}, \bibinfo {author} {\bibfnamefont {M.}~\bibnamefont {{Brightman}}}, \bibinfo {author} {\bibfnamefont {P.}~\bibnamefont {{Casella}}}, \bibinfo {author} {\bibfnamefont {M.}~\bibnamefont {{Bachetti}}}, \bibinfo {author} {\bibfnamefont {F.}~\bibnamefont {{F{\"u}rst}}}, \bibinfo {author} {\bibfnamefont {L.}~\bibnamefont {{Stella}}}, \bibinfo {author} {\bibfnamefont {C.}~\bibnamefont {{Pinto}}}, \bibinfo {author} {\bibfnamefont {F.}~\bibnamefont {{Pintore}}}, \bibinfo {author} {\bibfnamefont {F.}~\bibnamefont {{Tombesi}}}, \bibinfo {author} {\bibfnamefont {A.}~\bibnamefont {{G{\'u}rpide}}}, \bibinfo {author} {\bibfnamefont {M.~J.}\ \bibnamefont
  {{Middleton}}}, \bibinfo {author} {\bibfnamefont {C.}~\bibnamefont {{Salvaggio}}}, \bibinfo {author} {\bibfnamefont {A.}~\bibnamefont {{Tiengo}}}, \bibinfo {author} {\bibfnamefont {A.}~\bibnamefont {{Belfiore}}}, \bibinfo {author} {\bibfnamefont {A.}~\bibnamefont {{De Luca}}}, \bibinfo {author} {\bibfnamefont {P.}~\bibnamefont {{Esposito}}}, \bibinfo {author} {\bibfnamefont {A.}~\bibnamefont {{Wolter}}}, \bibinfo {author} {\bibfnamefont {H.~P.}\ \bibnamefont {{Earnshaw}}}, \bibinfo {author} {\bibfnamefont {D.~J.}\ \bibnamefont {{Walton}}}, \bibinfo {author} {\bibfnamefont {T.~P.}\ \bibnamefont {{Roberts}}}, \bibinfo {author} {\bibfnamefont {L.}~\bibnamefont {{Zampieri}}}, \bibinfo {author} {\bibfnamefont {M.}~\bibnamefont {{Marelli}}}, \ and\ \bibinfo {author} {\bibfnamefont {R.}~\bibnamefont {{Salvaterra}}},\ }\href {\doibase 10.1051/0004-6361/202450432} {\bibfield  {journal} {\bibinfo  {journal} {\aap}\ }\textbf {\bibinfo {volume} {689}},\ \bibinfo {eid} {A284} (\bibinfo {year} {2024})},\ \Eprint
  {http://arxiv.org/abs/2407.09240} {arXiv:2407.09240 [astro-ph.HE]} \BibitemShut {NoStop}%
\bibitem [{\citenamefont {{Dall'Osso}}\ \emph {et~al.}(2015)\citenamefont {{Dall'Osso}}, \citenamefont {{Perna}},\ and\ \citenamefont {{Stella}}}]{DallOsso2015}%
  \BibitemOpen
  \bibfield  {author} {\bibinfo {author} {\bibfnamefont {S.}~\bibnamefont {{Dall'Osso}}}, \bibinfo {author} {\bibfnamefont {R.}~\bibnamefont {{Perna}}}, \ and\ \bibinfo {author} {\bibfnamefont {L.}~\bibnamefont {{Stella}}},\ }\href {\doibase 10.1093/mnras/stv170} {\bibfield  {journal} {\bibinfo  {journal} {\mnras}\ }\textbf {\bibinfo {volume} {449}},\ \bibinfo {pages} {2144} (\bibinfo {year} {2015})},\ \Eprint {http://arxiv.org/abs/1412.1823} {arXiv:1412.1823 [astro-ph.HE]} \BibitemShut {NoStop}%
\bibitem [{\citenamefont {{Mushtukov}}\ \emph {et~al.}(2015{\natexlab{b}})\citenamefont {{Mushtukov}}, \citenamefont {{Suleimanov}}, \citenamefont {{Tsygankov}},\ and\ \citenamefont {{Poutanen}}}]{Mushtukov2015MNRAS.454.2539M}%
  \BibitemOpen
  \bibfield  {author} {\bibinfo {author} {\bibfnamefont {A.~A.}\ \bibnamefont {{Mushtukov}}}, \bibinfo {author} {\bibfnamefont {V.~F.}\ \bibnamefont {{Suleimanov}}}, \bibinfo {author} {\bibfnamefont {S.~S.}\ \bibnamefont {{Tsygankov}}}, \ and\ \bibinfo {author} {\bibfnamefont {J.}~\bibnamefont {{Poutanen}}},\ }\href {\doibase 10.1093/mnras/stv2087} {\bibfield  {journal} {\bibinfo  {journal} {\mnras}\ }\textbf {\bibinfo {volume} {454}},\ \bibinfo {pages} {2539} (\bibinfo {year} {2015}{\natexlab{b}})},\ \Eprint {http://arxiv.org/abs/1506.03600} {arXiv:1506.03600 [astro-ph.HE]} \BibitemShut {NoStop}%
\bibitem [{\citenamefont {{Tong}}(2015)}]{Tong2015}%
  \BibitemOpen
  \bibfield  {author} {\bibinfo {author} {\bibfnamefont {H.}~\bibnamefont {{Tong}}},\ }\href {\doibase 10.1002/asna.201512233} {\bibfield  {journal} {\bibinfo  {journal} {Astronomische Nachrichten}\ }\textbf {\bibinfo {volume} {336}},\ \bibinfo {pages} {835} (\bibinfo {year} {2015})},\ \Eprint {http://arxiv.org/abs/1508.03115} {arXiv:1508.03115 [astro-ph.HE]} \BibitemShut {NoStop}%
\bibitem [{\citenamefont {{King}}(2009)}]{King2009}%
  \BibitemOpen
  \bibfield  {author} {\bibinfo {author} {\bibfnamefont {A.~R.}\ \bibnamefont {{King}}},\ }\href {\doibase 10.1111/j.1745-3933.2008.00594.x} {\bibfield  {journal} {\bibinfo  {journal} {\mnras}\ }\textbf {\bibinfo {volume} {393}},\ \bibinfo {pages} {L41} (\bibinfo {year} {2009})},\ \Eprint {http://arxiv.org/abs/0811.1473} {arXiv:0811.1473 [astro-ph]} \BibitemShut {NoStop}%
\bibitem [{\citenamefont {{King}}\ \emph {et~al.}(2017)\citenamefont {{King}}, \citenamefont {{Lasota}},\ and\ \citenamefont {{Klu{\'z}niak}}}]{King2017}%
  \BibitemOpen
  \bibfield  {author} {\bibinfo {author} {\bibfnamefont {A.}~\bibnamefont {{King}}}, \bibinfo {author} {\bibfnamefont {J.-P.}\ \bibnamefont {{Lasota}}}, \ and\ \bibinfo {author} {\bibfnamefont {W.}~\bibnamefont {{Klu{\'z}niak}}},\ }\href {\doibase 10.1093/mnrasl/slx020} {\bibfield  {journal} {\bibinfo  {journal} {\mnras}\ }\textbf {\bibinfo {volume} {468}},\ \bibinfo {pages} {L59} (\bibinfo {year} {2017})},\ \Eprint {http://arxiv.org/abs/1702.00808} {arXiv:1702.00808 [astro-ph.HE]} \BibitemShut {NoStop}%
\bibitem [{\citenamefont {{King}}\ and\ \citenamefont {{Lasota}}(2020)}]{King2020}%
  \BibitemOpen
  \bibfield  {author} {\bibinfo {author} {\bibfnamefont {A.}~\bibnamefont {{King}}}\ and\ \bibinfo {author} {\bibfnamefont {J.-P.}\ \bibnamefont {{Lasota}}},\ }\href {\doibase 10.1093/mnras/staa930} {\bibfield  {journal} {\bibinfo  {journal} {\mnras}\ }\textbf {\bibinfo {volume} {494}},\ \bibinfo {pages} {3611} (\bibinfo {year} {2020})},\ \Eprint {http://arxiv.org/abs/2003.14019} {arXiv:2003.14019 [astro-ph.HE]} \BibitemShut {NoStop}%
\bibitem [{\citenamefont {{Brightman}}\ \emph {et~al.}(2018)\citenamefont {{Brightman}}, \citenamefont {{Harrison}}, \citenamefont {{F{\"u}rst}}, \citenamefont {{Middleton}}, \citenamefont {{Walton}}, \citenamefont {{Stern}}, \citenamefont {{Fabian}}, \citenamefont {{Heida}}, \citenamefont {{Barret}},\ and\ \citenamefont {{Bachetti}}}]{2018NatAs...2..312B}%
  \BibitemOpen
  \bibfield  {author} {\bibinfo {author} {\bibfnamefont {M.}~\bibnamefont {{Brightman}}}, \bibinfo {author} {\bibfnamefont {F.~A.}\ \bibnamefont {{Harrison}}}, \bibinfo {author} {\bibfnamefont {F.}~\bibnamefont {{F{\"u}rst}}}, \bibinfo {author} {\bibfnamefont {M.~J.}\ \bibnamefont {{Middleton}}}, \bibinfo {author} {\bibfnamefont {D.~J.}\ \bibnamefont {{Walton}}}, \bibinfo {author} {\bibfnamefont {D.}~\bibnamefont {{Stern}}}, \bibinfo {author} {\bibfnamefont {A.~C.}\ \bibnamefont {{Fabian}}}, \bibinfo {author} {\bibfnamefont {M.}~\bibnamefont {{Heida}}}, \bibinfo {author} {\bibfnamefont {D.}~\bibnamefont {{Barret}}}, \ and\ \bibinfo {author} {\bibfnamefont {M.}~\bibnamefont {{Bachetti}}},\ }\href {\doibase 10.1038/s41550-018-0391-6} {\bibfield  {journal} {\bibinfo  {journal} {Nature Astronomy}\ }\textbf {\bibinfo {volume} {2}},\ \bibinfo {pages} {312} (\bibinfo {year} {2018})},\ \Eprint {http://arxiv.org/abs/1803.02376} {arXiv:1803.02376 [astro-ph.HE]} \BibitemShut {NoStop}%
\bibitem [{\citenamefont {{Doroshenko}}\ \emph {et~al.}(2020{\natexlab{b}})\citenamefont {{Doroshenko}}, \citenamefont {{Zhang}}, \citenamefont {{Santangelo}}, \citenamefont {{Ji}}, \citenamefont {{Tsygankov}}, \citenamefont {{Mushtukov}}, \citenamefont {{Qu}}, \citenamefont {{Zhang}}, \citenamefont {{Ge}}, \citenamefont {{Chen}}, \citenamefont {{Bu}}, \citenamefont {{Cao}}, \citenamefont {{Chang}}, \citenamefont {{Chen}}, \citenamefont {{Chen}}, \citenamefont {{Chen}}, \citenamefont {{Chen}}, \citenamefont {{Chen}}, \citenamefont {{Cui}}, \citenamefont {{Cui}}, \citenamefont {{Deng}}, \citenamefont {{Dong}}, \citenamefont {{Du}}, \citenamefont {{Fu}}, \citenamefont {{Gao}}, \citenamefont {{Gao}}, \citenamefont {{Gao}}, \citenamefont {{Gu}}, \citenamefont {{Guan}}, \citenamefont {{Guo}}, \citenamefont {{Han}}, \citenamefont {{Hu}}, \citenamefont {{Huang}}, \citenamefont {{Huo}}, \citenamefont {{Jia}}, \citenamefont {{Jiang}}, \citenamefont {{Jiang}}, \citenamefont {{Jin}}, \citenamefont {{Jin}}, \citenamefont
  {{Kong}}, \citenamefont {{Li}}, \citenamefont {{Li}}, \citenamefont {{Li}}, \citenamefont {{Li}}, \citenamefont {{Li}}, \citenamefont {{Li}}, \citenamefont {{Li}}, \citenamefont {{Li}}, \citenamefont {{Li}}, \citenamefont {{Li}}, \citenamefont {{Li}}, \citenamefont {{Li}}, \citenamefont {{Liang}}, \citenamefont {{Liao}}, \citenamefont {{Liu}}, \citenamefont {{Liu}}, \citenamefont {{Liu}}, \citenamefont {{Liu}}, \citenamefont {{Liu}}, \citenamefont {{Liu}}, \citenamefont {{Liu}}, \citenamefont {{Lu}}, \citenamefont {{Lu}}, \citenamefont {{Lu}}, \citenamefont {{Luo}}, \citenamefont {{Ma}}, \citenamefont {{Meng}}, \citenamefont {{Nang}}, \citenamefont {{Nie}}, \citenamefont {{Ou}}, \citenamefont {{Sai}}, \citenamefont {{Shang}}, \citenamefont {{Song}}, \citenamefont {{Song}}, \citenamefont {{Sun}}, \citenamefont {{Tan}}, \citenamefont {{Tao}}, \citenamefont {{Tuo}}, \citenamefont {{Wang}}, \citenamefont {{Wang}}, \citenamefont {{Wang}}, \citenamefont {{Wang}}, \citenamefont {{Wen}}, \citenamefont {{Wu}},
  \citenamefont {{Wu}}, \citenamefont {{Xiao}}, \citenamefont {{Xiong}}, \citenamefont {{Xu}}, \citenamefont {{Xu}}, \citenamefont {{Yang}}, \citenamefont {{Yang}}, \citenamefont {{Yang}}, \citenamefont {{Yang}}, \citenamefont {{Zhang}}, \citenamefont {{Zhang}}, \citenamefont {{Zhang}}, \citenamefont {{Zhang}}, \citenamefont {{Zhang}}, \citenamefont {{Zhang}}, \citenamefont {{Zhang}}, \citenamefont {{Zhang}}, \citenamefont {{Zhang}}, \citenamefont {{Zhang}}, \citenamefont {{Zhang}}, \citenamefont {{Zhang}}, \citenamefont {{Zhang}}, \citenamefont {{Zhang}}, \citenamefont {{Zhang}}, \citenamefont {{Zhang}}, \citenamefont {{Zhang}}, \citenamefont {{Zhao}}, \citenamefont {{Zhao}}, \citenamefont {{Zhao}}, \citenamefont {{Zheng}}, \citenamefont {{Zhu}}, \citenamefont {{Zhu}}, \citenamefont {{Zou}},\ and\ \citenamefont {{Zhang}}}]{Doroshenko2020}%
  \BibitemOpen
  \bibfield  {author} {\bibinfo {author} {\bibfnamefont {V.}~\bibnamefont {{Doroshenko}}}, \bibinfo {author} {\bibfnamefont {S.~N.}\ \bibnamefont {{Zhang}}}, \bibinfo {author} {\bibfnamefont {A.}~\bibnamefont {{Santangelo}}}, \bibinfo {author} {\bibfnamefont {L.}~\bibnamefont {{Ji}}}, \bibinfo {author} {\bibfnamefont {S.}~\bibnamefont {{Tsygankov}}}, \bibinfo {author} {\bibfnamefont {A.}~\bibnamefont {{Mushtukov}}}, \bibinfo {author} {\bibfnamefont {L.~J.}\ \bibnamefont {{Qu}}}, \bibinfo {author} {\bibfnamefont {S.}~\bibnamefont {{Zhang}}}, \bibinfo {author} {\bibfnamefont {M.~Y.}\ \bibnamefont {{Ge}}}, \bibinfo {author} {\bibfnamefont {Y.~P.}\ \bibnamefont {{Chen}}}, \bibinfo {author} {\bibfnamefont {Q.~C.}\ \bibnamefont {{Bu}}}, \bibinfo {author} {\bibfnamefont {X.~L.}\ \bibnamefont {{Cao}}}, \bibinfo {author} {\bibfnamefont {Z.}~\bibnamefont {{Chang}}}, \bibinfo {author} {\bibfnamefont {G.}~\bibnamefont {{Chen}}}, \bibinfo {author} {\bibfnamefont {L.}~\bibnamefont {{Chen}}}, \bibinfo {author}
  {\bibfnamefont {T.~X.}\ \bibnamefont {{Chen}}}, \bibinfo {author} {\bibfnamefont {Y.}~\bibnamefont {{Chen}}}, \bibinfo {author} {\bibfnamefont {Y.~B.}\ \bibnamefont {{Chen}}}, \bibinfo {author} {\bibfnamefont {W.}~\bibnamefont {{Cui}}}, \bibinfo {author} {\bibfnamefont {W.~W.}\ \bibnamefont {{Cui}}}, \bibinfo {author} {\bibfnamefont {J.~K.}\ \bibnamefont {{Deng}}}, \bibinfo {author} {\bibfnamefont {Y.~W.}\ \bibnamefont {{Dong}}}, \bibinfo {author} {\bibfnamefont {Y.~Y.}\ \bibnamefont {{Du}}}, \bibinfo {author} {\bibfnamefont {M.~X.}\ \bibnamefont {{Fu}}}, \bibinfo {author} {\bibfnamefont {G.~H.}\ \bibnamefont {{Gao}}}, \bibinfo {author} {\bibfnamefont {H.}~\bibnamefont {{Gao}}}, \bibinfo {author} {\bibfnamefont {M.}~\bibnamefont {{Gao}}}, \bibinfo {author} {\bibfnamefont {Y.~D.}\ \bibnamefont {{Gu}}}, \bibinfo {author} {\bibfnamefont {J.}~\bibnamefont {{Guan}}}, \bibinfo {author} {\bibfnamefont {C.~C.}\ \bibnamefont {{Guo}}}, \bibinfo {author} {\bibfnamefont {D.~W.}\ \bibnamefont {{Han}}}, \bibinfo {author}
  {\bibfnamefont {W.}~\bibnamefont {{Hu}}}, \bibinfo {author} {\bibfnamefont {Y.}~\bibnamefont {{Huang}}}, \bibinfo {author} {\bibfnamefont {J.}~\bibnamefont {{Huo}}}, \bibinfo {author} {\bibfnamefont {S.~M.}\ \bibnamefont {{Jia}}}, \bibinfo {author} {\bibfnamefont {L.~H.}\ \bibnamefont {{Jiang}}}, \bibinfo {author} {\bibfnamefont {W.~C.}\ \bibnamefont {{Jiang}}}, \bibinfo {author} {\bibfnamefont {J.}~\bibnamefont {{Jin}}}, \bibinfo {author} {\bibfnamefont {Y.~J.}\ \bibnamefont {{Jin}}}, \bibinfo {author} {\bibfnamefont {L.~D.}\ \bibnamefont {{Kong}}}, \bibinfo {author} {\bibfnamefont {B.}~\bibnamefont {{Li}}}, \bibinfo {author} {\bibfnamefont {C.~K.}\ \bibnamefont {{Li}}}, \bibinfo {author} {\bibfnamefont {G.}~\bibnamefont {{Li}}}, \bibinfo {author} {\bibfnamefont {M.~S.}\ \bibnamefont {{Li}}}, \bibinfo {author} {\bibfnamefont {T.~P.}\ \bibnamefont {{Li}}}, \bibinfo {author} {\bibfnamefont {W.}~\bibnamefont {{Li}}}, \bibinfo {author} {\bibfnamefont {X.}~\bibnamefont {{Li}}}, \bibinfo {author} {\bibfnamefont
  {X.~B.}\ \bibnamefont {{Li}}}, \bibinfo {author} {\bibfnamefont {X.~F.}\ \bibnamefont {{Li}}}, \bibinfo {author} {\bibfnamefont {Y.~G.}\ \bibnamefont {{Li}}}, \bibinfo {author} {\bibfnamefont {Z.~J.}\ \bibnamefont {{Li}}}, \bibinfo {author} {\bibfnamefont {Z.~W.}\ \bibnamefont {{Li}}}, \bibinfo {author} {\bibfnamefont {X.~H.}\ \bibnamefont {{Liang}}}, \bibinfo {author} {\bibfnamefont {J.~Y.}\ \bibnamefont {{Liao}}}, \bibinfo {author} {\bibfnamefont {C.~Z.}\ \bibnamefont {{Liu}}}, \bibinfo {author} {\bibfnamefont {G.~Q.}\ \bibnamefont {{Liu}}}, \bibinfo {author} {\bibfnamefont {H.~W.}\ \bibnamefont {{Liu}}}, \bibinfo {author} {\bibfnamefont {S.~Z.}\ \bibnamefont {{Liu}}}, \bibinfo {author} {\bibfnamefont {X.~J.}\ \bibnamefont {{Liu}}}, \bibinfo {author} {\bibfnamefont {Y.}~\bibnamefont {{Liu}}}, \bibinfo {author} {\bibfnamefont {Y.~N.}\ \bibnamefont {{Liu}}}, \bibinfo {author} {\bibfnamefont {B.}~\bibnamefont {{Lu}}}, \bibinfo {author} {\bibfnamefont {F.~J.}\ \bibnamefont {{Lu}}}, \bibinfo {author}
  {\bibfnamefont {X.~F.}\ \bibnamefont {{Lu}}}, \bibinfo {author} {\bibfnamefont {T.}~\bibnamefont {{Luo}}}, \bibinfo {author} {\bibfnamefont {X.}~\bibnamefont {{Ma}}}, \bibinfo {author} {\bibfnamefont {B.}~\bibnamefont {{Meng}}}, \bibinfo {author} {\bibfnamefont {Y.}~\bibnamefont {{Nang}}}, \bibinfo {author} {\bibfnamefont {J.~Y.}\ \bibnamefont {{Nie}}}, \bibinfo {author} {\bibfnamefont {G.}~\bibnamefont {{Ou}}}, \bibinfo {author} {\bibfnamefont {N.}~\bibnamefont {{Sai}}}, \bibinfo {author} {\bibfnamefont {R.~C.}\ \bibnamefont {{Shang}}}, \bibinfo {author} {\bibfnamefont {L.~M.}\ \bibnamefont {{Song}}}, \bibinfo {author} {\bibfnamefont {X.~Y.}\ \bibnamefont {{Song}}}, \bibinfo {author} {\bibfnamefont {L.}~\bibnamefont {{Sun}}}, \bibinfo {author} {\bibfnamefont {Y.}~\bibnamefont {{Tan}}}, \bibinfo {author} {\bibfnamefont {L.}~\bibnamefont {{Tao}}}, \bibinfo {author} {\bibfnamefont {Y.~L.}\ \bibnamefont {{Tuo}}}, \bibinfo {author} {\bibfnamefont {G.~F.}\ \bibnamefont {{Wang}}}, \bibinfo {author} {\bibfnamefont
  {J.}~\bibnamefont {{Wang}}}, \bibinfo {author} {\bibfnamefont {W.~S.}\ \bibnamefont {{Wang}}}, \bibinfo {author} {\bibfnamefont {Y.~S.}\ \bibnamefont {{Wang}}}, \bibinfo {author} {\bibfnamefont {X.~Y.}\ \bibnamefont {{Wen}}}, \bibinfo {author} {\bibfnamefont {B.~B.}\ \bibnamefont {{Wu}}}, \bibinfo {author} {\bibfnamefont {M.}~\bibnamefont {{Wu}}}, \bibinfo {author} {\bibfnamefont {G.~C.}\ \bibnamefont {{Xiao}}}, \bibinfo {author} {\bibfnamefont {S.~L.}\ \bibnamefont {{Xiong}}}, \bibinfo {author} {\bibfnamefont {H.}~\bibnamefont {{Xu}}}, \bibinfo {author} {\bibfnamefont {Y.~P.}\ \bibnamefont {{Xu}}}, \bibinfo {author} {\bibfnamefont {Y.~R.}\ \bibnamefont {{Yang}}}, \bibinfo {author} {\bibfnamefont {J.~W.}\ \bibnamefont {{Yang}}}, \bibinfo {author} {\bibfnamefont {S.}~\bibnamefont {{Yang}}}, \bibinfo {author} {\bibfnamefont {Y.~J.}\ \bibnamefont {{Yang}}}, \bibinfo {author} {\bibfnamefont {A.~M.}\ \bibnamefont {{Zhang}}}, \bibinfo {author} {\bibfnamefont {C.~L.}\ \bibnamefont {{Zhang}}}, \bibinfo {author}
  {\bibfnamefont {C.~M.}\ \bibnamefont {{Zhang}}}, \bibinfo {author} {\bibfnamefont {F.}~\bibnamefont {{Zhang}}}, \bibinfo {author} {\bibfnamefont {H.~M.}\ \bibnamefont {{Zhang}}}, \bibinfo {author} {\bibfnamefont {J.}~\bibnamefont {{Zhang}}}, \bibinfo {author} {\bibfnamefont {Q.}~\bibnamefont {{Zhang}}}, \bibinfo {author} {\bibfnamefont {T.}~\bibnamefont {{Zhang}}}, \bibinfo {author} {\bibfnamefont {W.}~\bibnamefont {{Zhang}}}, \bibinfo {author} {\bibfnamefont {W.~C.}\ \bibnamefont {{Zhang}}}, \bibinfo {author} {\bibfnamefont {W.~Z.}\ \bibnamefont {{Zhang}}}, \bibinfo {author} {\bibfnamefont {Y.}~\bibnamefont {{Zhang}}}, \bibinfo {author} {\bibfnamefont {Y.}~\bibnamefont {{Zhang}}}, \bibinfo {author} {\bibfnamefont {Y.~F.}\ \bibnamefont {{Zhang}}}, \bibinfo {author} {\bibfnamefont {Y.~J.}\ \bibnamefont {{Zhang}}}, \bibinfo {author} {\bibfnamefont {Z.}~\bibnamefont {{Zhang}}}, \bibinfo {author} {\bibfnamefont {Z.~L.}\ \bibnamefont {{Zhang}}}, \bibinfo {author} {\bibfnamefont {H.~S.}\ \bibnamefont {{Zhao}}},
  \bibinfo {author} {\bibfnamefont {J.~L.}\ \bibnamefont {{Zhao}}}, \bibinfo {author} {\bibfnamefont {X.~F.}\ \bibnamefont {{Zhao}}}, \bibinfo {author} {\bibfnamefont {S.~J.}\ \bibnamefont {{Zheng}}}, \bibinfo {author} {\bibfnamefont {Y.}~\bibnamefont {{Zhu}}}, \bibinfo {author} {\bibfnamefont {Y.~X.}\ \bibnamefont {{Zhu}}}, \bibinfo {author} {\bibfnamefont {C.~L.}\ \bibnamefont {{Zou}}}, \ and\ \bibinfo {author} {\bibfnamefont {R.~L.}\ \bibnamefont {{Zhang}}},\ }\href {\doibase 10.1093/mnras/stz2879} {\bibfield  {journal} {\bibinfo  {journal} {\mnras}\ }\textbf {\bibinfo {volume} {491}},\ \bibinfo {pages} {1857} (\bibinfo {year} {2020}{\natexlab{b}})},\ \Eprint {http://arxiv.org/abs/1909.12614} {arXiv:1909.12614 [astro-ph.HE]} \BibitemShut {NoStop}%
\bibitem [{\citenamefont {{Cui}}\ \emph {et~al.}(2023)\citenamefont {{Cui}}, \citenamefont {{Wang}}, \citenamefont {{Zhao}}, \citenamefont {{Zhang}}, \citenamefont {{Meidinger}}, \citenamefont {{Yang}}, \citenamefont {{Keil}}, \citenamefont {{Zhang}}, \citenamefont {{Huo}}, \citenamefont {{Wang}}, \citenamefont {{Song}}, \citenamefont {{Lu}}, \citenamefont {{Ma}}, \citenamefont {{Wang}}, \citenamefont {{Xu}}, \citenamefont {{Zhu}}, \citenamefont {{Li}}, \citenamefont {{Li}}, \citenamefont {{Luo}}, \citenamefont {{Han}}, \citenamefont {{Zhao}}, \citenamefont {{Hou}}, \citenamefont {{Yang}}, \citenamefont {{Geng}}, \citenamefont {{Li}}, \citenamefont {{Chen}}, \citenamefont {{Tang}}, \citenamefont {{Chen}},\ and\ \citenamefont {{Chen}}}]{2023ExA....55..603C}%
  \BibitemOpen
  \bibfield  {author} {\bibinfo {author} {\bibfnamefont {W.}~\bibnamefont {{Cui}}}, \bibinfo {author} {\bibfnamefont {H.}~\bibnamefont {{Wang}}}, \bibinfo {author} {\bibfnamefont {X.}~\bibnamefont {{Zhao}}}, \bibinfo {author} {\bibfnamefont {J.}~\bibnamefont {{Zhang}}}, \bibinfo {author} {\bibfnamefont {N.}~\bibnamefont {{Meidinger}}}, \bibinfo {author} {\bibfnamefont {Y.}~\bibnamefont {{Yang}}}, \bibinfo {author} {\bibfnamefont {I.}~\bibnamefont {{Keil}}}, \bibinfo {author} {\bibfnamefont {Z.}~\bibnamefont {{Zhang}}}, \bibinfo {author} {\bibfnamefont {J.}~\bibnamefont {{Huo}}}, \bibinfo {author} {\bibfnamefont {J.}~\bibnamefont {{Wang}}}, \bibinfo {author} {\bibfnamefont {Z.}~\bibnamefont {{Song}}}, \bibinfo {author} {\bibfnamefont {F.}~\bibnamefont {{Lu}}}, \bibinfo {author} {\bibfnamefont {J.}~\bibnamefont {{Ma}}}, \bibinfo {author} {\bibfnamefont {Y.}~\bibnamefont {{Wang}}}, \bibinfo {author} {\bibfnamefont {J.}~\bibnamefont {{Xu}}}, \bibinfo {author} {\bibfnamefont {Y.}~\bibnamefont {{Zhu}}}, \bibinfo
  {author} {\bibfnamefont {T.}~\bibnamefont {{Li}}}, \bibinfo {author} {\bibfnamefont {W.}~\bibnamefont {{Li}}}, \bibinfo {author} {\bibfnamefont {L.}~\bibnamefont {{Luo}}}, \bibinfo {author} {\bibfnamefont {D.}~\bibnamefont {{Han}}}, \bibinfo {author} {\bibfnamefont {Z.}~\bibnamefont {{Zhao}}}, \bibinfo {author} {\bibfnamefont {D.}~\bibnamefont {{Hou}}}, \bibinfo {author} {\bibfnamefont {X.}~\bibnamefont {{Yang}}}, \bibinfo {author} {\bibfnamefont {H.}~\bibnamefont {{Geng}}}, \bibinfo {author} {\bibfnamefont {S.}~\bibnamefont {{Li}}}, \bibinfo {author} {\bibfnamefont {H.}~\bibnamefont {{Chen}}}, \bibinfo {author} {\bibfnamefont {Q.}~\bibnamefont {{Tang}}}, \bibinfo {author} {\bibfnamefont {Y.}~\bibnamefont {{Chen}}}, \ and\ \bibinfo {author} {\bibfnamefont {Y.}~\bibnamefont {{Chen}}},\ }\href {\doibase 10.1007/s10686-023-09891-y} {\bibfield  {journal} {\bibinfo  {journal} {Experimental Astronomy}\ }\textbf {\bibinfo {volume} {55}},\ \bibinfo {pages} {603} (\bibinfo {year} {2023})}\BibitemShut {NoStop}%
\end{thebibliography}%
%\begin{thebibliography}{99}
%\bibitem {ref1} Exmple: M. Aspelmeyer, T. J. Kippenberg, and F. Marquardt, Rev. Mod. Phys. 86, 1391 (2014).

%\bibitem {ref2} Example: R. J. Hunter, \textit{Zeta Potential in Colloid Science} (Academic, New York, 1981), p. 120 %%Books

%\end{thebibliography}

%%%%%%%%%%%%%%%%%%%%%%%%%%%%%%%%%%%%%%%%%%%%%%%%%%%%%%%
%%% Appendix sections. ??????, ????
%%%%%%%%%%%%%%%%%%%%%%%%%%%%%%%%%%%%%%%%%%%%%%%%%%%%%%%
%\begin{appendix}
%\section{Name}

%\end{appendix}

%\begin{appendices}
%\section{Appendix}
%\end{appendices}
%\appendix

%\appendix

%\renewcommand{\thesection}{Appendix}

%\section{}

%\def\thesubsection{\Arabic.\alph{section}}

%\section{}

%\end{appendices}
%\end{appendix}

\end{multicols}
\end{document}